\pdfoutput=1
\documentclass[reprint,twocolumn,longbibliography,superscriptaddress,
showpacs,preprintnumbers,
notitlepage,amsmath,amssymb,
aps,
]{revtex4-1}

\usepackage{graphicx}
\usepackage{dcolumn}
\usepackage{bm}
\usepackage{multirow}
\usepackage{bbm}
\usepackage{ulem}
\usepackage{diagbox}

\usepackage{graphicx}
\usepackage{enumerate}
\usepackage{subfigure}
\usepackage{epstopdf}
\usepackage[colorlinks,
linkcolor=red,
anchorcolor=black,
citecolor=blue]{hyperref}
\usepackage{amssymb}
\usepackage{framed}
\usepackage{tabularx}
\usepackage{booktabs}
\usepackage{float}
\usepackage[table]{xcolor}
\usepackage{array}
\usepackage{tikz}
\hypersetup{colorlinks=true}

\date{\today}
\begin{document}
\title{Construction and classification of crystalline topological superconductor and insulators in three-dimensional interacting fermion systems}
\author{Jian-Hao Zhang}
\affiliation{Department of Physics, The Chinese University of Hong Kong, Shatin, New Territories, Hong Kong, China}
\author{Shang-Qiang Ning}
\affiliation{Department of Physics, The Chinese University of Hong Kong, Shatin, New Territories, Hong Kong, China}
\author{Yang Qi}
\email{qiyang@fudan.edu.cn}
\affiliation{State Key Laboratory of Surface Physics, Fudan University, Shanghai 200433, China}
\affiliation{Center for Field Theory and Particle Physics, Department of Physics, Fudan University, Shanghai 200433, China}
\affiliation{Collaborative Innovation Center of Advanced Microstructures, Nanjing 210093, China}
\author{Zheng-Cheng Gu}
\email{zcgu@phy.cuhk.edu.hk}
\affiliation{Department of Physics, The Chinese University of Hong Kong, Shatin, New Territories, Hong Kong, China}

\begin{abstract}
The natural existence of crystalline symmetry in real materials manifests the importance of understanding crystalline symmetry-protected topological (SPT) phases, especially for interacting systems. In this paper, we systematically construct and classify all the crystalline topological superconductors and insulators in three-dimensional (3D) interacting fermion systems using the novel concept of topological crystal. The corresponding higher-order topological surface theory can also be systematically studied via higher-order bulk-boundary correspondence. 
In particular, we discover an intriguing fact that almost all topological crystals with nontrivial 2D block states are intrinsically interacting topological phases that cannot be realized in any free-fermion systems. Moreover, the crystalline equivalence principle for 3D interacting fermionic systems is also verified, with an additional subtle ``twist'' on the spin of fermions. 
\end{abstract}

\maketitle
\tableofcontents
\newcommand{\lra}{\longrightarrow}
\newcommand{\xra}{\xrightarrow}
\newcommand{\ra}{\rightarrow}
\newcommand{\bs}{\boldsymbol}
\newcommand{\ul}{\underline}
\newcommand{\dif}{\mathrm{d}}
\newcommand{\Z}{\mathbb{Z}}
\newcommand{\1}{\text{\uppercase\expandafter{\romannumeral1}}}
\newcommand{\2}{\text{\uppercase\expandafter{\romannumeral2}}}
\newcommand{\3}{\text{\uppercase\expandafter{\romannumeral3}}}
\newcommand{\4}{\text{\uppercase\expandafter{\romannumeral4}}}
\newcommand{\5}{\text{\uppercase\expandafter{\romannumeral5}}}
\newcommand{\6}{\text{\uppercase\expandafter{\romannumeral6}}}

\section{Introduction}
\subsection{The goal of this paper}
Topological phases of quantum matter have become a fascinating subject of condensed matter physics during the past few decades. In particular, the patterns of long-range entanglement provide us a systematic way of understanding intrinsic topological order~\cite{entanglement}. Furthermore, the interplay between symmetry and topology plays a central role in 
recent years. In particular, the so-called symmetry-protected topological (SPT) phases have been systematically constructed and classified for interacting bosonic and fermionic systems \cite{ZCGu2009,chen11a,XieChenScience,cohomology,Senthil_2015,E8,Lu12,invertible2,invertible3,special,general1,general2,Kapustin2014,Kapustin2015,Kapustin2017,Gu-Levin,gauging1,gauging3,dimensionalreduction,gauging2,2DFSPT,braiding}.
In general, by ``gauging'' the internal (unitary) symmetry~\cite{LevinGu,Gu-Levin, gauging1,threeloop,ran14,wangj15,wangcj15,lin15,gauging3,dimensionalreduction,gauging2,2DFSPT,braiding} and investigating the braiding statistics of the corresponding gauge fluxes/flux lines, different SPT phases can be uniquely identified.
Moreover, the gapless edge states or anomalous surface topological orders have been proposed as another very powerful way to characterize different SPT phases in interacting systems~\cite{Ashvin2013,ChongWang2013,XieChen2015,ChenjieWang2016,XLQi2013,Senthil2013,Lukasz2013,XieChen2014,ChongWang2014,Ning21a}. A well known example of SPT phases first proposed in non-interacting fermion systems are topological insulator, protected by time-reversal and charge-conservation symmetry~\cite{KaneRMP,ZhangRMP}. Recently, crystalline SPT phases protected by space group symmetry have been intensively studied theoretically\cite{TCI,Fu2012,ITCI,reduction,building,correspondence,SET,230,BCSPT,Jiang2017,Kane2017,Shiozaki2018,ZDSong2018,defect,realspace,KenX,rotation,dihedral,LuX,YMLu2018,Cheng2018,Hermele2018,Po2020,Huang2020PRR,Huang2021PRR,wallpaper,PEPS, SQ2105,Maissam2020,Maissam2021,Max18,Meng18fLSM,JosephMeng19,indicator3,indicator5,indicator1,indicator2} and experimentally~\cite{TCIrealization1,TCIrealization2,TCIrealization3,TCIrealization4}. In particular, very different from SPT phases protected by internal symmetry, the boundaries of $d$-dimensional crystalline SPT phases are typically gapped but with protected $(d-n)$-dimensional gapless modes ($1<n\leq d$) at the hinges or corners. This type of topological phases are dubbed \textit{higher-order (HO) topological phases} \cite{Wang2018,Yan2018,Nori2018,Wangyuxuan2018,Ryu2018,Zhang2019,Hsu2018,bultinck2019three,Roy_2020,Roy_2021,Laubscher_2019,Laubscher_2020,JHZhang2022,May-Mann2022}. 

So far, The study of HO topological phases mainly focuses on non-interacting or weakly interacting fermion systems. Due to the limitation of analytical methods, only numerical simulations on finite systems can provide us with some insights into the nature of HO topological phases in strongly correlated electron systems. However, in large classes of real materials, strong electronic interactions practically play a significant role and cannot be avoided. Therefore, the complete construction and classification of crystalline SPT phases for interacting fermion systems become an important but challenging problem. Very recently, a real-space construction of topological crystals has been established to construct and classify all topological crystalline phases in interacting bosonic and non-interacting fermionic systems \cite{Hermele2018,realspace}. For two-dimensional (2D) interacting fermion systems, the systematic constructions and classifications have already been established  \cite{rotation,dihedral,wallpaper} for crystalline topological superconductor (TSC) and topological insulator (TI). Furthermore, it was pointed out that there are profound relationships between classifications of crystalline SPT and internal SPT phases. In Ref. \cite{correspondence}, a ``\textit{crystalline equivalence principle}'' was conjectured: crystalline SPT phases with space group $G$ are in one-to-one correspondence with SPT phases protected by internal symmetry with an identical abstract group structure, but acting in a twisted way, where if an element of $G$ is a mirror reflection as an orientation reserving symmetry operation, it should be regarded as an antiunitary time-reversal symmetry operation. This conjecture is rigorously proven in interacting bosonic systems \cite{realspace}. In interacting fermionic systems, it has been conjectured that the crystalline equivalence principle should be applied in a twisted way: a spinless (spin-1/2) fermions should be mapped into spin-1/2 (spinless) fermions  \cite{rotation,dihedral,wallpaper}. Besides the correspondence of the bulk classification, this principle also implies the correspondence of their boundaries\cite{Zhang2112,Zhang2309}.

In this paper, we systematically construct and classify all crystalline TSC and TI for 3D interacting fermion systems protected by crystalline point group symmetry. (In real materials, the translation symmetry is typically broken by disorder and there is no need to consider the full space group symmetry in general \footnote{For optical lattice or other artificial lattice systems, the full space group symmetry can be preserved and our method is easy to be generalized for those cases.}.)  
We also study the corresponding HO topological surface theories via higher-order bulk-boundary correspondence, for both spinless and spin-1/2 fermions. 
We conclude that each 2D block state has a second-order topological surface theory, each 1D block state has a third-order topological surface theory, and all 0D block states do not have nontrivial HO topological surface theory. In particular, we discover an intriguing fact that almost all 2D block states are intrinsically interacting topological phases. The systematic investigation of HO topological surface theory serves a significant experimental relevance of crystalline SPT phases, because tunneling spectroscopy and transport experiments may directly probe the topological surface theory. 
Finally, by comparing our results with the classifications of 3D fermionic SPT (fSPT) phases protected by internal symmetries with an identical abstract group structure, we confirm the crystalline equivalence principle for generic 3D interacting fermionic systems, which requires the following twists on the symmetry group:
\begin{enumerate}[1.]
\item orientation-reserving operations, such as the Mirror reflection symmetry, should be treated as antiunitary time reversal symmetry operations;
\item The systems with spinless fermions should be mapped into the systems with spin-1/2 fermions, and vice versa.
Here, the property of spin-1/2 is formulated mathematically as a particular 2-cocycle described in Sec.~\ref{sec:pgs12}, which can be extracted from the structure of the double group associated with the point group~\cite{Elcoro:ks5574}.
In particular, for point groups $S_2$, $S_4$ and $S_6$ with even-fold roto-reflections as their generators, the $\mathbb{Z}_2^f$ extensions are trivial for both spinless and spin-1/2 fermions.
\end{enumerate}
We will demonstrate the second issue in Sec. \ref{inversion} by the simplest example: 3D inversion-symmetric ($S_2$-symmetric) lattice, with a pictorial interpretation of this subtlety. 

Our general paradigm of constructing topological crystals includes the following steps:
\begin{description}
\item[Cell decomposition]For a 3D lattice with a specific point group symmetry, we divide the whole system into an assembly of lower-dimensional blocks.
\item[block state decoration]With concrete cell decomposition, we can decorate proper lower-dimensional block states on blocks with different dimensions. The \textit{no-open-edge condition} requires that an \textit{obstruction-free} block state should have a fully-gapped bulk.
\item[Bubble equivalence]For a specific obstruction-free block state, we need to further seek if it can be trivialized by \textit{bubbles} at higher-dimensional blocks (``bubble'' on $d$D block $\zeta$ is defined as a $(d-1)$D state respecting the symmetry of $\zeta$, which can be shrunken to a point and trivialized). An obstruction and trivialization-free block state corresponds to a 3D nontrivial crystalline fSPT phase.
\item[Higher-order bulk edge correspondence]With a nontrivial $d$D block state ($d=1,2$), the $(d-1)$D topological edge theories of block states will reside on the corresponding higher-order surface. A $d$D block state corresponds to a $(4-d)^{\mathrm{th}}$-order surface theory.
\end{description}
In addition, we should further investigate the group structure of the classification by nontrivial stacking between different block states: nontrivial stacking implicates a nontrivial group structure of the classification.

\subsection{Point group symmetry for spinless and spin-1/2 fermions}
\label{sec:pgs12}
We should clarify the precise meaning of ``spinless'' and ``spin-1/2'' fermions for systems with and without $U^f(1)$ charge conservation. 

A fermionic system always has a fermion parity symmetry $\mathbb{Z}_2^f=\{1,P_f=(-1)^F\}$, where $F$ is the total number of fermions. And $\mathbb{Z}_2^f$ is the center of the total symmetry $G_f$ because all physical symmetries commute with $P_f$. In particular, for fermionic systems without $U^f(1)$ charge conservation, we can define a physical (bosonic) symmetry group by a quotient group $G_b=G_f/\mathbb{Z}_2^f$. Reversely, for a given physical symmetry group $G_b$, there are several choices of the total symmetry group $G_f$ as different central extensions of $G_b$ by $\mathbb{Z}_2^f$, described by the following short exact sequence:
\begin{align}
0\rightarrow\mathbb{Z}_2^f\rightarrow G_f\rightarrow G_b\rightarrow0
\end{align}
and different extensions $G_f$ are characterized by different factor systems of the above short exact sequence that are 2-cocycles $\omega_2\in\mathcal{H}^2(G_b,\mathbb{Z}_2^f)$. We phrase $G_f=\mathbb{Z}_2^f\times_{\omega_2}G_b$.

For the systems with $U^f(1)$ charge conservation, the group element is $U_\theta=e^{i\theta F}$, and aforementioned fermion parity operator $P_f=U_\pi$ is the order 2 element of $U^f(1)$. $U^f(1)$ charge conservation is a normal subgroup of the total symmetry group $G_f$, which can be expressed by the following short exact sequence:
\begin{align}
0\rightarrow U^f(1)\rightarrow G_f\rightarrow G\rightarrow0
\end{align}
where we have defined the physical symmetry group $G:=G_f/U^f(1)$. Reversely, for a given physical symmetry group $G$, we can define the total symmetry group $G_f=U^f(1)\rtimes_{\omega_2}G$. Here $\omega_2\in\mathbb{R}/\mathbb{Z}=[0,1)$ as a $U(1)$ phase, which is related to above short exact sequence. The multiplication of the total symmetry group $G_f$ is defined as:
\begin{align}
(1,g)\times(1,h)=\left(e^{2\pi i\omega_2(g,h)F},gh\right)\in G_f
\end{align}
where $g,h\in G$. Hence $\omega_2\in\mathcal{H}^2(G,\mathbb{R}/\mathbb{Z})$ is a 2-cocycle, characterizing different total symmetry groups.

The spin of fermions (spinless or spin-1/2 fermions) is characterized by different $\omega_2$ for systems with and without $U^f(1)$ charge conservation: the spinless fermions correspond to a trivial $\omega_2$ while spin-1/2 fermions correspond to a specific choice of nontrivial $\omega_2$.
For a given point group $G$, this specific $\omega_2$ can be computed from the structure of the corresponding double group ${}^dG$~\cite{Elcoro:ks5574}:
Each group element $g\in G$ is represented by a O(3) matrix, which can be written as $g=\pm\phi_g$, where $\phi_g$ is an SO(3) matrix.
It is well known that there is a two-to-one map from SU(2) to SO(3), and we can (arbitrarily) choose an SU(2) matrix $\tilde\phi_g$ from the two preimages of $\phi_g$.
Once we fix the choice of $\tilde\phi_g$ for each $g$, they form a projective representation of $G$, and $\omega_2$ is given by its factors:
\begin{equation}
    \label{eq:w2-proj}
    \tilde\phi_g\cdot\tilde\phi_h=\omega_2(g, h)\tilde\phi_{gh}.
\end{equation}

We give a pictorial argument of the spin of fermions: a point group symmetry operation $O$ rotates a fermion by $2\pi/n$ if $O^n=1$, equivalently, if we take $O$ by $n$ times, a fermion will be rotated by a round. Rotate a spinless fermion leads to nothing, but rotate a spin-1/2 fermion leads to an additional $-1$ phase of the wavefunction of each fermion. The factor system $\omega_2$ of aforementioned short exact sequences characterize the phase caused by rotating a fermion by a round.
Furthermore, we notice that the possible minus sign in $g=\pm\phi_g$ is discarded in the construction of $\omega_2$.
This is because the inversion symmetry $S_1$ extends trivially over $\mathbb Z_2^f$ even for spin-1/2 fermions. (In other words, $S_1^2=+1$.)
This can be understood intuitively from the fact that inversion acts trivially on the electron spin.

\begin{table}[!htbp]
\renewcommand\arraystretch{1.2}
  \centering
  \begin{tabular}{|c|c|c|c|c|c|c|c|c|}
   \hline
   \multirow{2}*{\diagbox[innerwidth=1.4cm]{~$G_b$}{spin~}}& \multicolumn{4}{c|}{spinless}&\multicolumn{4}{c|}{spin-1/2} \\
   \cline{2-9}
{}&$E_{0}^{\mathrm{2D}}$&$E_{0}^{\mathrm{1D}}$&$E_{0}^{\mathrm{0D}}$&~$\mathcal{G}_{0}$~&$E_{1/2}^{\mathrm{2D}}$&$E_{1/2}^{\mathrm{1D}}$&$E_{1/2}^{\mathrm{0D}}$&~$\mathcal{G}_{1/2}$~\\
\hline
$C_1$&$\mathbb{Z}_1$&$\mathbb{Z}_1$&$\mathbb{Z}_1$&$\mathbb{Z}_1$&$\mathbb{Z}_1$&$\mathbb{Z}_1$&$\mathbb{Z}_1$&$\mathbb{Z}_1$\\
\hline
$C_i=S_2$&$\mathbb{Z}_1$&$\mathbb{Z}_1$&$\mathbb{Z}_1$&$\mathbb{Z}_1$&$\mathbb{Z}_1$&$\mathbb{Z}_1$&$\mathbb{Z}_1$&$\mathbb{Z}_{1}$\\
\hline
$C_2$&$\mathbb{Z}_1$&$\mathbb{Z}_1$&$\mathbb{Z}_1$&$\mathbb{Z}_1$&$\mathbb{Z}_1$&$\mathbb{Z}_1$&$\mathbb{Z}_1$&$\mathbb{Z}_1$\\
\hline
$C_s=C_{1h}$&$\mathbb{Z}_{16}$&$\mathbb{Z}_1$&$\mathbb{Z}_1$&$\mathbb{Z}_{16}$&$\mathbb{Z}_{1}$&$\mathbb{Z}_1$&$\mathbb{Z}_1$&$\mathbb{Z}_1$\\
\hline
$C_{2h}$&$\mathbb{Z}_{8}$&$\mathbb{Z}_1$&$\mathbb{Z}_1$&$\mathbb{Z}_{8}$&$\mathbb{Z}_1$&$\mathbb{Z}_1$&$\mathbb{Z}_1$&$\mathbb{Z}_1$\\
\hline
$D_2=V$&$\mathbb{Z}_{1}$&$\mathbb{Z}_1$&$\mathbb{Z}_1$&$\mathbb{Z}_1$&$\mathbb{Z}_{1}$&$\mathbb{Z}_1$&$\mathbb{Z}_2^2$&$\mathbb{Z}_2^2$\\
\hline
$C_{2v}$&$\mathbb{Z}_2^2$&$\mathbb{Z}_2$&$\mathbb{Z}_1$&$\mathbb{Z}_2^3$&$\mathbb{Z}_1$&$\mathbb{Z}_1$&$\mathbb{Z}_1$&$\mathbb{Z}_1$\\
\hline
$D_{2h}=V_h$&$\mathbb{Z}_2^3$&$\mathbb{Z}_2$&$\mathbb{Z}_2$&$\mathbb{Z}_2^5$&$\mathbb{Z}_1$&$\mathbb{Z}_1$&$\mathbb{Z}_2^3$&$\mathbb{Z}_2^3$\\
\hline
$C_4$&$\mathbb{Z}_1$&$\mathbb{Z}_1$&$\mathbb{Z}_1$&$\mathbb{Z}_1$&$\mathbb{Z}_1$&$\mathbb{Z}_1$&$\mathbb{Z}_1$&$\mathbb{Z}_1$\\
\hline
$S_4$&$\mathbb{Z}_1$&$\mathbb{Z}_2$&$\mathbb{Z}_2$&$\mathbb{Z}_2^2$&$\mathbb{Z}_2$&$\mathbb{Z}_1$&$\mathbb{Z}_2$&$\mathbb{Z}_2^2$\\
\hline
$C_{4h}$&$\mathbb{Z}_{8}$&$\mathbb{Z}_1$&$\mathbb{Z}_2$&$\mathbb{Z}_{8}\times\mathbb{Z}_2$&$\mathbb{Z}_{1}$&$\mathbb{Z}_1$&$\mathbb{Z}_2$&$\mathbb{Z}_2$\\
\hline
$D_4$&$\mathbb{Z}_1$&$\mathbb{Z}_2$&$\mathbb{Z}_1$&$\mathbb{Z}_2$&$\mathbb{Z}_1$&$\mathbb{Z}_1$&$\mathbb{Z}_2^2$&$\mathbb{Z}_2^2$\\
\hline
$C_{4v}$&$\mathbb{Z}_2^2$&$\mathbb{Z}_2^2$&$\mathbb{Z}_1$&$\mathbb{Z}_2^4$&$\mathbb{Z}_1$&$\mathbb{Z}_1$&$\mathbb{Z}_1$&$\mathbb{Z}_1$\\
\hline
$D_{2d}=V_d$&$\mathbb{Z}_2$&$\mathbb{Z}_2$&$\mathbb{Z}_2$&$\mathbb{Z}_2^3$&$\mathbb{Z}_1$&$\mathbb{Z}_1$&$\mathbb{Z}_2$&$\mathbb{Z}_2$\\
\hline
$D_{4h}$&$\mathbb{Z}_2^3$&$\mathbb{Z}_2^2$&$\mathbb{Z}_2$&$\mathbb{Z}_2^6$&$\mathbb{Z}_1$&$\mathbb{Z}_1$&$\mathbb{Z}_2^3$&$\mathbb{Z}_2^3$\\
\hline
$C_3$&$\mathbb{Z}_1$&$\mathbb{Z}_1$&$\mathbb{Z}_1$&$\mathbb{Z}_1$&$\mathbb{Z}_1$&$\mathbb{Z}_1$&$\mathbb{Z}_1$&$\mathbb{Z}_1$\\
\hline
$S_6$&$\mathbb{Z}_1$&$\mathbb{Z}_1$&$\mathbb{Z}_1$&$\mathbb{Z}_1$&$\mathbb{Z}_1$&$\mathbb{Z}_1$&$\mathbb{Z}_1$&$\mathbb{Z}_1$\\
\hline
$D_3$&$\mathbb{Z}_1$&$\mathbb{Z}_1$&$\mathbb{Z}_1$&$\mathbb{Z}_1$&$\mathbb{Z}_1$&$\mathbb{Z}_1$&$\mathbb{Z}_1$&$\mathbb{Z}_1$\\
\hline
$C_{3v}$&$\mathbb{Z}_{16}$&$\mathbb{Z}_1$&$\mathbb{Z}_1$&$\mathbb{Z}_{16}$&$\mathbb{Z}_{1}$&$\mathbb{Z}_1$&$\mathbb{Z}_1$&$\mathbb{Z}_{1}$\\
\hline
$D_{3d}$&$\mathbb{Z}_2$&$\mathbb{Z}_2$&$\mathbb{Z}_2$&$\mathbb{Z}_2^3$&$\mathbb{Z}_1$&$\mathbb{Z}_1$&$\mathbb{Z}_1$&$\mathbb{Z}_1$\\
\hline
$C_6$&$\mathbb{Z}_1$&$\mathbb{Z}_1$&$\mathbb{Z}_1$&$\mathbb{Z}_1$&$\mathbb{Z}_1$&$\mathbb{Z}_1$&$\mathbb{Z}_1$&$\mathbb{Z}_1$\\
\hline
$C_{3h}$&$\mathbb{Z}_{8}$&$\mathbb{Z}_1$&$\mathbb{Z}_1$&$\mathbb{Z}_{8}$&$\mathbb{Z}_{1}$&$\mathbb{Z}_1$&$\mathbb{Z}_1$&$\mathbb{Z}_{1}$\\
\hline
$C_{6h}$&$\mathbb{Z}_{8}$&$\mathbb{Z}_1$&$\mathbb{Z}_1$&$\mathbb{Z}_{8}$&$\mathbb{Z}_{1}$&$\mathbb{Z}_1$&$\mathbb{Z}_1$&$\mathbb{Z}_1$\\
\hline
$D_{6}$&$\mathbb{Z}_{1}$&$\mathbb{Z}_1$&$\mathbb{Z}_1$&$\mathbb{Z}_{1}$&$\mathbb{Z}_{1}$&$\mathbb{Z}_1$&$\mathbb{Z}_2^2$&$\mathbb{Z}_2^2$\\
\hline
$C_{6v}$&$\mathbb{Z}_{2}^2$&$\mathbb{Z}_2$&$\mathbb{Z}_1$&$\mathbb{Z}_2^3$&$\mathbb{Z}_1$&$\mathbb{Z}_1$&$\mathbb{Z}_1$&$\mathbb{Z}_1$\\
\hline
$D_{3h}$&$\mathbb{Z}_{2}^2$&$\mathbb{Z}_2$&$\mathbb{Z}_1$&$\mathbb{Z}_2^3$&$\mathbb{Z}_1$&$\mathbb{Z}_1$&$\mathbb{Z}_1$&$\mathbb{Z}_1$\\
\hline
$D_{6h}$&$\mathbb{Z}_2^3$&$\mathbb{Z}_2$&$\mathbb{Z}_2$&$\mathbb{Z}_2^5$&$\mathbb{Z}_1$&$\mathbb{Z}_1$&$\mathbb{Z}_2^3$&$\mathbb{Z}_2^3$\\
\hline
$T$&$\mathbb{Z}_1$&$\mathbb{Z}_1$&$\mathbb{Z}_1$&$\mathbb{Z}_1$&$\mathbb{Z}_1$&$\mathbb{Z}_1$&$\mathbb{Z}_1$&$\mathbb{Z}_1$\\
\hline
$T_h$&$\mathbb{Z}_2$&$\mathbb{Z}_2$&$\mathbb{Z}_2$&$\mathbb{Z}_2^3$&$\mathbb{Z}_1$&$\mathbb{Z}_1$&$\mathbb{Z}_2$&$\mathbb{Z}_2$\\
\hline
$T_d$&$\mathbb{Z}_2$&$\mathbb{Z}_1$&$\mathbb{Z}_2^2$&$\mathbb{Z}_2^3$&$\mathbb{Z}_1$&$\mathbb{Z}_1$&$\mathbb{Z}_1$&$\mathbb{Z}_1$\\
\hline
$O$&$\mathbb{Z}_1$&$\mathbb{Z}_2$&$\mathbb{Z}_1$&$\mathbb{Z}_2$&$\mathbb{Z}_1$&$\mathbb{Z}_1$&$\mathbb{Z}_2$&$\mathbb{Z}_2$\\
\hline
$O_h$&$\mathbb{Z}_2^2$&$\mathbb{Z}_2^2$&$\mathbb{Z}_2$&$\mathbb{Z}_2^5$&$\mathbb{Z}_1$&$\mathbb{Z}_1$&$\mathbb{Z}_2^2$&$\mathbb{Z}_2^2$\\
\hline
\end{tabular}
\caption{Classifications of 3D point group symmetric TSC phases for both spinless and spin-1/2 fermions, with classification data $E_{0}^{d\mathrm{D}}$ and $E_{1/2}^{d\mathrm{D}}$ for block states with all dimensions $d=0,1,2$.}
\label{TSC}
\end{table}

\begin{table*}[!htbp]
\renewcommand\arraystretch{1.2}
  \centering
  \begin{tabular}{|c|c|c|c|c|c|c|c|c|}
   \hline
   \multirow{2}*{\diagbox[innerwidth=1.5cm]{~$G_b$}{spin~}}& \multicolumn{4}{c|}{spinless}&\multicolumn{4}{c|}{spin-1/2} \\
   \cline{2-9}
{}&$E_{0,U(1)}^{\mathrm{2D}}$&$E_{0,U(1)}^{\mathrm{1D}}$&$E_{0,U(1)}^{\mathrm{0D}}$&~$\mathcal{G}_{0,U(1)}$~&$E_{1/2,U(1)}^{\mathrm{2D}}$&$E_{1/2,U(1)}^{\mathrm{1D}}$&$E_{1/2,U(1)}^{\mathrm{0D}}$&~$\mathcal{G}_{1/2,U(1)}$~\\
\hline
$C_1$&$\mathbb{Z}_1$&$\mathbb{Z}_1$&$\mathbb{Z}_1$&$\mathbb{Z}_1$&$\mathbb{Z}_1$&$\mathbb{Z}_1$&$\mathbb{Z}_1$&$\mathbb{Z}_1$\\
\hline
$C_i=S_2$&$\mathbb{Z}_2^2$&$\mathbb{Z}_1$&$\mathbb{Z}_4$&$\mathbb{Z}_8\times\mathbb{Z}_2$&$\mathbb{Z}_2^2$&$\mathbb{Z}_1$&$\mathbb{Z}_4$&$\mathbb{Z}_8\times\mathbb{Z}_2$\\
\hline
$C_2$&$\mathbb{Z}_1$&$\mathbb{Z}_1$&$\mathbb{Z}_1$&$\mathbb{Z}_1$&$\mathbb{Z}_1$&$\mathbb{Z}_1$&$\mathbb{Z}_1$&$\mathbb{Z}_1$\\
\hline
$C_s=C_{1h}$&$\mathbb{Z}_8\times\mathbb{Z}_2$&$\mathbb{Z}_1$&$\mathbb{Z}_1$&$\mathbb{Z}_8\times\mathbb{Z}_2$&$\mathbb{Z}_8\times\mathbb{Z}_2$&$\mathbb{Z}_1$&$\mathbb{Z}_1$&$\mathbb{Z}_8\times\mathbb{Z}_2$\\
\hline
$C_{2h}$&$\mathbb{Z}_8\times\mathbb{Z}_2^2$&$\mathbb{Z}_1$&$\mathbb{Z}_2$&$\mathbb{Z}_8\times\mathbb{Z}_4\times\mathbb{Z}_2$&$\mathbb{Z}_8\times\mathbb{Z}_2^2$&$\mathbb{Z}_1$&$\mathbb{Z}_2$&$\mathbb{Z}_8\times\mathbb{Z}_4\times\mathbb{Z}_2$\\
\hline
$D_2=V$&$\mathbb{Z}_{1}$&$\mathbb{Z}_1$&$\mathbb{Z}_1$&$\mathbb{Z}_1$&$\mathbb{Z}_{1}$&$\mathbb{Z}_1$&$\mathbb{Z}_2^3$&$\mathbb{Z}_2^3$\\
\hline
$C_{2v}$&$\mathbb{Z}_2^3$&$\mathbb{Z}_2$&$\mathbb{Z}_1$&$\mathbb{Z}_2^4$&$\mathbb{Z}_8\times\mathbb{Z}_4\times\mathbb{Z}_2$&$\mathbb{Z}_1$&$\mathbb{Z}_1$&$\mathbb{Z}_8\times\mathbb{Z}_4\times\mathbb{Z}_2$\\
\hline
$D_{2h}=V_h$&$\mathbb{Z}_2^4$&$\mathbb{Z}_2^3$&$\mathbb{Z}_4$&$\mathbb{Z}_4\times\mathbb{Z}_2^7$&$\mathbb{Z}_8\times\mathbb{Z}_4^2\times\mathbb{Z}_2$&$\mathbb{Z}_1$&$\mathbb{Z}_2^4$&$\mathbb{Z}_8\times\mathbb{Z}_4^2\times\mathbb{Z}_2^5$\\
\hline
$C_4$&$\mathbb{Z}_1$&$\mathbb{Z}_1$&$\mathbb{Z}_1$&$\mathbb{Z}_1$&$\mathbb{Z}_1$&$\mathbb{Z}_1$&$\mathbb{Z}_1$&$\mathbb{Z}_1$\\
\hline
$S_4$&$\mathbb{Z}_2^2$&$\mathbb{Z}_1$&$\mathbb{Z}_2^2$&$\mathbb{Z}_2^4$&$\mathbb{Z}_2^2$&$\mathbb{Z}_1$&$\mathbb{Z}_2^2$&$\mathbb{Z}_2^4$\\
\hline
$C_{4h}$&$\mathbb{Z}_8\times\mathbb{Z}_2$&$\mathbb{Z}_1$&$\mathbb{Z}_4\times\mathbb{Z}_2$&$\mathbb{Z}_8\times\mathbb{Z}_4\times\mathbb{Z}_2^2$&$\mathbb{Z}_8\times\mathbb{Z}_2$&$\mathbb{Z}_1$&$\mathbb{Z}_4\times\mathbb{Z}_2$&$\mathbb{Z}_8\times\mathbb{Z}_4\times\mathbb{Z}_2^2$\\
\hline
$D_4$&$\mathbb{Z}_1$&$\mathbb{Z}_1$&$\mathbb{Z}_1$&$\mathbb{Z}_1$&$\mathbb{Z}_1$&$\mathbb{Z}_1$&$\mathbb{Z}_2^2$&$\mathbb{Z}_2^2$\\
\hline
$C_{4v}$&$\mathbb{Z}_2^3$&$\mathbb{Z}_2$&$\mathbb{Z}_1$&$\mathbb{Z}_2^4$&$\mathbb{Z}_8\times\mathbb{Z}_4\times\mathbb{Z}_2$&$\mathbb{Z}_1$&$\mathbb{Z}_1$&$\mathbb{Z}_8\times\mathbb{Z}_4\times\mathbb{Z}_2$\\
\hline
$D_{2d}=V_d$&$\mathbb{Z}_2^2$&$\mathbb{Z}_2$&$\mathbb{Z}_4$&$\mathbb{Z}_4\times\mathbb{Z}_2^3$&$\mathbb{Z}_8\times\mathbb{Z}_2^2$&$\mathbb{Z}_1$&$\mathbb{Z}_2^3$&$\mathbb{Z}_8\times\mathbb{Z}_2^5$\\
\hline
$D_{4h}$&$\mathbb{Z}_2^4$&$\mathbb{Z}_2^3$&$\mathbb{Z}_4$&$\mathbb{Z}_4\times\mathbb{Z}_2^7$&$\mathbb{Z}_8\times\mathbb{Z}_4^2\times\mathbb{Z}_2$&$\mathbb{Z}_1$&$\mathbb{Z}_2^4$&$\mathbb{Z}_8\times\mathbb{Z}_4^2\times\mathbb{Z}_2^5$\\
\hline
$C_3$&$\mathbb{Z}_1$&$\mathbb{Z}_1$&$\mathbb{Z}_1$&$\mathbb{Z}_1$&$\mathbb{Z}_1$&$\mathbb{Z}_1$&$\mathbb{Z}_1$&$\mathbb{Z}_1$\\
\hline
$S_6$&$\mathbb{Z}_2^2$&$\mathbb{Z}_1$&$\mathbb{Z}_4$&$\mathbb{Z}_4\times\mathbb{Z}_2^2$&$\mathbb{Z}_2^2$&$\mathbb{Z}_1$&$\mathbb{Z}_4$&$\mathbb{Z}_4\times\mathbb{Z}_2^2$\\
\hline
$D_3$&$\mathbb{Z}_1$&$\mathbb{Z}_1$&$\mathbb{Z}_1$&$\mathbb{Z}_1$&$\mathbb{Z}_1$&$\mathbb{Z}_1$&$\mathbb{Z}_1$&$\mathbb{Z}_1$\\
\hline
$C_{3v}$&$\mathbb{Z}_8\times\mathbb{Z}_2$&$\mathbb{Z}_1$&$\mathbb{Z}_1$&$\mathbb{Z}_8\times\mathbb{Z}_2$&$\mathbb{Z}_8\times\mathbb{Z}_2$&$\mathbb{Z}_1$&$\mathbb{Z}_1$&$\mathbb{Z}_8\times\mathbb{Z}_2$\\
\hline
$D_{3d}$&$\mathbb{Z}_2^2$&$\mathbb{Z}_1$&$\mathbb{Z}_4$&$\mathbb{Z}_4\times\mathbb{Z}_2^2$&$\mathbb{Z}_8\times\mathbb{Z}_2$&$\mathbb{Z}_1$&$\mathbb{Z}_2^2$&$\mathbb{Z}_8\times\mathbb{Z}_2^3$\\
\hline
$C_6$&$\mathbb{Z}_1$&$\mathbb{Z}_1$&$\mathbb{Z}_1$&$\mathbb{Z}_1$&$\mathbb{Z}_1$&$\mathbb{Z}_1$&$\mathbb{Z}_1$&$\mathbb{Z}_1$\\
\hline
$C_{3h}$&$\mathbb{Z}_8\times\mathbb{Z}_2$&$\mathbb{Z}_1$&$\mathbb{Z}_1$&$\mathbb{Z}_8\times\mathbb{Z}_2$&$\mathbb{Z}_8\times\mathbb{Z}_2$&$\mathbb{Z}_1$&$\mathbb{Z}_1$&$\mathbb{Z}_8\times\mathbb{Z}_2$\\
\hline
$C_{6h}$&$\mathbb{Z}_8\times\mathbb{Z}_2$&$\mathbb{Z}_1$&$\mathbb{Z}_4$&$\mathbb{Z}_8\times\mathbb{Z}_4\times\mathbb{Z}_2$&$\mathbb{Z}_8\times\mathbb{Z}_2$&$\mathbb{Z}_1$&$\mathbb{Z}_4$&$\mathbb{Z}_8\times\mathbb{Z}_4\times\mathbb{Z}_2$\\
\hline
$D_{6}$&$\mathbb{Z}_{1}$&$\mathbb{Z}_1$&$\mathbb{Z}_1$&$\mathbb{Z}_1$&$\mathbb{Z}_{1}$&$\mathbb{Z}_1$&$\mathbb{Z}_2^3$&$\mathbb{Z}_2^3$\\
\hline
$C_{6v}$&$\mathbb{Z}_2^3$&$\mathbb{Z}_2$&$\mathbb{Z}_1$&$\mathbb{Z}_2^4$&$\mathbb{Z}_8\times\mathbb{Z}_4\times\mathbb{Z}_2$&$\mathbb{Z}_1$&$\mathbb{Z}_1$&$\mathbb{Z}_8\times\mathbb{Z}_4\times\mathbb{Z}_2$\\
\hline
$D_{3h}$&$\mathbb{Z}_2^3$&$\mathbb{Z}_2$&$\mathbb{Z}_2$&$\mathbb{Z}_2^5$&$\mathbb{Z}_8\times\mathbb{Z}_4\times\mathbb{Z}_2$&$\mathbb{Z}_1$&$\mathbb{Z}_2^2$&$\mathbb{Z}_8\times\mathbb{Z}_4\times\mathbb{Z}_2^3$\\
\hline
$D_{6h}$&$\mathbb{Z}_2^4$&$\mathbb{Z}_2^3$&$\mathbb{Z}_4$&$\mathbb{Z}_4\times\mathbb{Z}_2^7$&$\mathbb{Z}_8\times\mathbb{Z}_4^2\times\mathbb{Z}_2$&$\mathbb{Z}_1$&$\mathbb{Z}_2^4$&$\mathbb{Z}_8\times\mathbb{Z}_4^2\times\mathbb{Z}_2^5$\\
\hline
$T$&$\mathbb{Z}_1$&$\mathbb{Z}_1$&$\mathbb{Z}_1$&$\mathbb{Z}_1$&$\mathbb{Z}_1$&$\mathbb{Z}_1$&$\mathbb{Z}_1$&$\mathbb{Z}_1$\\
\hline
$T_h$&$\mathbb{Z}_2^2$&$\mathbb{Z}_2$&$\mathbb{Z}_4$&$\mathbb{Z}_4\times\mathbb{Z}_2^3$&$\mathbb{Z}_8\times\mathbb{Z}_2$&$\mathbb{Z}_1$&$\mathbb{Z}_2^2$&$\mathbb{Z}_8\times\mathbb{Z}_2^3$\\
\hline
$T_d$&$\mathbb{Z}_2^2$&$\mathbb{Z}_2$&$\mathbb{Z}_2$&$\mathbb{Z}_2^4$&$\mathbb{Z}_8\times\mathbb{Z}_2$&$\mathbb{Z}_1$&$\mathbb{Z}_2$&$\mathbb{Z}_8\times\mathbb{Z}_2^2$\\
\hline
$O$&$\mathbb{Z}_1$&$\mathbb{Z}_1$&$\mathbb{Z}_1$&$\mathbb{Z}_1$&$\mathbb{Z}_1$&$\mathbb{Z}_1$&$\mathbb{Z}_2$&$\mathbb{Z}_2$\\
\hline
$O_h$&$\mathbb{Z}_2^3$&$\mathbb{Z}_2$&$\mathbb{Z}_4$&$\mathbb{Z}_4\times\mathbb{Z}_2^4$&$\mathbb{Z}_8\times\mathbb{Z}_4\times\mathbb{Z}_2$&$\mathbb{Z}_1$&$\mathbb{Z}_2^3$&$\mathbb{Z}_8\times\mathbb{Z}_4\times\mathbb{Z}_2^4$\\
\hline
\end{tabular}
\caption{Classifications of 3D point group symmetric TI phases for both spinless and spin-1/2 fermions, with classification data $E_{0,U(1)}^{d\mathrm{D}}$ and $E_{1/2,U(1)}^{d\mathrm{D}}$ for block states with all dimensions $d=0,1,2$.}
\label{TI}
\end{table*}

\subsection{Summary of main results}
Now we first summarize the classification results of crystalline TSC and TI protected by all 32 point groups, for 3D systems with both spinless and spin-1/2 fermions layer by layer (i.e., classification contributed by 0D/1D/2D block state decorations, respectively). We label the classification attributed to $p$-dimensional block states by $E^{d\mathrm{D}}$, and the subscript depicts the spin of fermions and the systems with/without $U^f(1)$ charge conservation. For crystalline TSC with spinless/spin-1/2 fermions, all classification results are summarized in Table. \ref{TSC}; for crystalline TI with spinless/spin-1/2 fermions, all classification results are summarized in Table. \ref{TI}. Furthermore, the results indicate that there are several nontrivial group structures of classification results, which are attributed to stacking several copies of block states that will be deformed to another block state.

The rest of the paper is organized as follows: we introduce the general paradigm of constructing three-dimensional topological crystals and classifying crystalline SPT phases protected by point group symmetry in Sec. \ref{general} and demonstrate that almost all 2D block states represent intriguing interacting crystalline SPT phases. In Sec. \ref{TSC classification}, we construct and classify the crystalline TSC in 3D interacting fermionic systems, for both spinless and spin-1/2 fermions. We also construct and classify the crystalline TI in 3D interacting fermionic systems in Sec. \ref{TI classification} for spinless and spin-1/2 fermions, and stress the corresponding HO topological surface theory by higher-order bulk-boundary correspondence for each nontrivial block state of different symmetry classes. 
Compare all classification results calculated in this paper to the classifications of 3D fSPT phases protected by corresponding internal symmetry groups, including results in Refs. \cite{mathematical,wang2021exactly} and that we compute for $\omega_2=0$ using the formulas in Refs.~\cite{general1,general2} and the algorithm in Ref.~\cite{resolution}, we confirm the \textit{crystalline equivalence principle} for 3D interacting fermionic systems. Finally, in Sec. \ref{conclusion}, conclusions of main results and discussions about further applications of topological crystals and experimental relevances of HO topological surface theories are presented. In Supplementary Materials, we first review the $K$-matrix formalism of interacting fSPT phases, including both bulk Chern-Simons theory and edge nonchiral Luttinger liquid theory; then we rigorously prove the equivalence of Kitaev's $E_8$ state and 16 layers of $(p+ip)$-wave superconductors [$(p+ip)$-SCs] in 2D interacting fermionic systems without $U^f(1)$ charge conservation, which interprets that why we do not treat the Kitaev's $E_8$ state as an independent 2D root phase for crystalline TSC but treat it as an independent 2D root phase for crystalline TI; finally, we summarize crystalline fSPT phases protected by all other 3D point groups in the main text, including full classification results, concrete block state as their root phases, and the corresponding HO topological surface theories \cite{supplementary}.

\section{General paradigm of real-space construction\label{general}}
In this section, we outline the general approach for constructing topological crystalline phases in 3D interacting fermionic systems through real-space methods. Firstly, we break down the system into lower-dimensional blocks. Then, we introduce appropriate lower-dimensional block states and assess their validity. Specifically, we determine if the bulk state of a block state construction can be completely gapped. If it cannot, we classify such decoration as \textit{obstructed}. Finally, we utilize the concept of bubble equivalence to explore all possible ways of \textit{trivializing} the system. A decoration that is free from obstructions and trivializations corresponds to a nontrivial crystalline symmetry-protected topological (SPT) phase. To illustrate these procedures comprehensively, we provide a detailed demonstration using the 3D $T$-symmetric cubic lattice as an example.

\subsection{Cell decomposition\label{Seccell}}
\begin{figure*}
\begin{tikzpicture}[scale=0.78]
\tikzstyle{sergio}=[rectangle,draw=none]
\draw[thick] (-1,1.5) -- (3,1.5);
\draw[thick] (-1,1.5) -- (-3,0);
\draw[thick] (3,1.5) -- (1,0);
\draw[thick] (-3,0) -- (1,0);
\draw[thick] (-3,0) -- (-3,-4);
\draw[thick] (1,-4) -- (-3,-4);
\draw[thick] (1,-4) -- (1,0);
\draw[thick] (3,-2.5) -- (3,1.5);
\draw[thick] (3,-2.5) -- (1,-4);
\draw[thick,color=green] (0,-1.25) -- (0,0.75);
\draw[thick,color=green] (0,-1.25) -- (0,-3.25);
\draw[thick,color=green] (0,-1.25) -- (-2,-1.25);
\draw[thick,color=green] (0,-1.25) -- (2,-1.25);
\draw[thick,color=green] (0,-1.25) -- (1,-0.5);
\draw[thick,color=green] (0,-1.25) -- (-1,-2);
\draw[densely dashed] (-1,-2.5) -- (-1,1.5);
\draw[densely dashed] (-1,-2.5) -- (3,-2.5);
\draw[densely dashed] (-1,-2.5) -- (-3,-4);
\draw[densely dashed,color=red] (3,1.5) -- (-3,-4);
\draw[thick,color=violet] (-1,1.5) -- (0,-1.25);
\draw[densely dashed,color=red] (-3,0) -- (3,-2.5);
\draw[thick,color=violet] (0,-1.25) -- (1,0);
\path (-3.25,-4.25) node [style=sergio] {$C_3^1$};
\path (3.25,-2.75) node [style=sergio] {$C_3^2$};
\path (0.75,0.25) node [style=sergio] {$C_3^3$};
\path (-1.25,1.75) node [style=sergio] {$C_3^4$};
\draw[thick,color=violet] (0,-1.25) -- (-3,-4);
\draw[densely dashed,color=red] (0,-1.25) -- (-1,-2.5);
\draw[thick,color=violet] (0,-1.25) -- (3,-2.5);
\draw[densely dashed,color=red] (0,-1.25) -- (1,-4);
\draw[thick] (3,-2.5) -- (1,-4);
\draw[thick] (-3,-4) -- (1,-4);
\path (-2,-0.75) node [style=sergio] {$\tau_1$};
\path (-0.25,-2) node [style=sergio] {$\tau_1$};
\path (0.4,-3) node [style=sergio] {$\tau_1$};
\path (1.75,0) node [style=sergio] {$\tau_1$};
\path (0,1) node [style=sergio] {$\tau_3,C_2^1$};
\path (0,-3.5) node [style=sergio] {$\tau_3$};
\path (-2.25,-1.25) node [style=sergio] {$\tau_3$};
\path (2.5,-1.5) node [style=sergio] {$\tau_3,C_2^3$};
\path (-1.25,-2) node [style=sergio] {$\tau_3$};
\path (1.5,-0.5) node [style=sergio] {$\tau_3,C_2^2$};
\path (1.5,-1.65) node [style=sergio] {$\tau_2$};
\path (-2,-2.75) node [style=sergio] {$\tau_2$};
\path (-0.5,-0.5) node [style=sergio] {$\tau_2$};
\path (0.3,-0.5) node [style=sergio] {$\tau_2$};
\path (-0.3,-1.3) node [style=sergio] {$\mu$};
\draw[thick] (4.5,-2.25) -- (8.5,-2.25);
\draw[densely dashed,thick] (6.5,-0.75) -- (10.5,-0.75);
\draw[densely dashed,thick] (6.5,-0.75) -- (4.5,-2.25);
\draw[thick] (10.5,-0.75) -- (8.5,-2.25);
\draw[densely dashed,color=red] (6.5,-0.75) -- (7.5,0.5);
\draw[densely dashed,color=red] (8.5,-2.25) -- (7.5,0.5);
\draw[thick,color=violet] (4.5,-2.25) -- (7.5,0.5);
\draw[thick,color=violet] (10.5,-0.75) -- (7.5,0.5);
\draw[thick,color=green] (7.5,-1.5) -- (7.5,0.5);
\filldraw[fill=black, draw=black] (7.5,0.5)circle (3pt);
\draw[densely dashed,thick] (6.5,-0.75) -- (8.5,-2.25);
\draw[densely dashed,thick] (10.5,-0.75) -- (4.5,-2.25);
\draw[thick] (11.5,-2.25) -- (15.5,-2.25);
\draw[densely dashed,thick] (14.5,-1.5) -- (11.5,-2.25);
\draw[color=green,thick] (14.5,0.5) -- (14.5,-1.5);
\draw[color=violet,thick] (14.5,0.5) -- (11.5,-2.25);
\draw[thick] (17.5,-0.75) -- (15.5,-2.25);
\draw[densely dashed,thick] (14.5,-1.5) -- (17.5,-0.75);
\draw[color=violet,thick] (14.5,0.5) -- (17.5,-0.75);
\path (14.5,-2.5) node [style=sergio] {$\lambda$};
\path (16.75,0.25) node [style=sergio] {$\sigma_2$};
\draw[thick,->] (13.5,-1.25) -- (13,-0.25);
\draw[thick,->] (16,-0.75) -- (16.5,0.25);
\draw[densely dashed,color=red] (15.5,-2.25) -- (14.5,0.5);
\path (12.75,-0.25) node [style=sergio] {$\sigma_2$};
\path (14,-0.75) node [style=sergio] {$\sigma_1$};
\path (15.5,-0.5) node [style=sergio] {$\sigma_1$};
\filldraw[fill=black, draw=black] (0,-1.25)circle (3pt);
\end{tikzpicture}
\caption{The cell decomposition of the 3D $T$-symmetric lattice. There are four axes of 3-fold rotation symmetry across the center of the system as indicated by the solid dot $\mu$, and labeled by $C_3^j$ ($j=1,2,3,4$); and three axes of 2-fold rotation symmetry across the center, labeled by $C_2^{1,2,3}$. The medium panel illustrates the bottom rectangular pyramid of the left cubic; the right panel illustrates two independent triangular pyramids, $\lambda_{1,2}$, of the upper rectangular pyramid, where $\sigma_{1,2}$ are two independent 2D blocks in the system.}
\label{T cell decomposition}
\end{figure*}
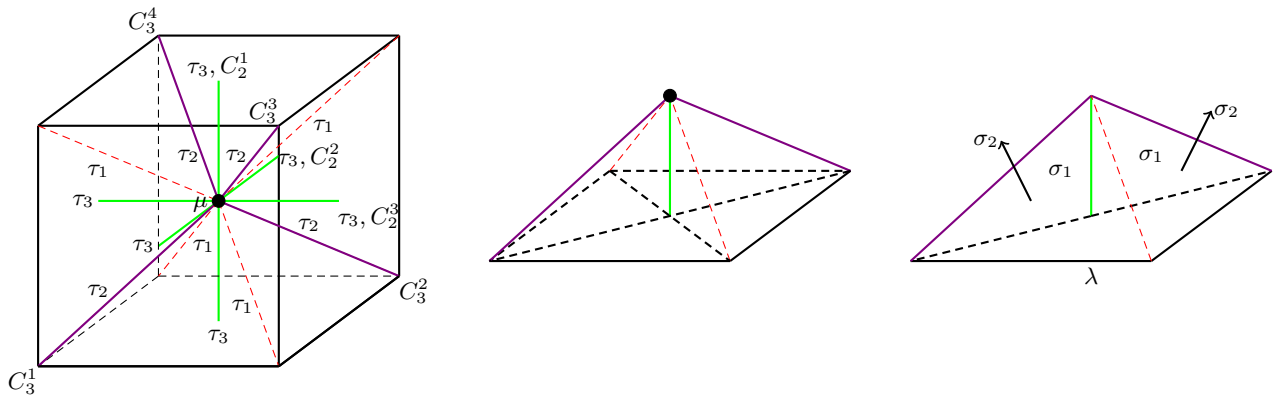

For 3D $T$-symmetric cubic lattice, we decompose the wavefunction of the whole system to a direct product states of the wavefunctions of different blocks: Suppose $|\psi\rangle$ is an SPT state that cannot be trivialized by a symmetric local unitary transformation $O^{\mathrm{loc}}$. Nevertheless, we can still define an alternative local unitary to \textit{extensively trivialize} the original wavefunction $|\psi\rangle$. First we can trivialize the region $\lambda$ located downward in the cubic (see Fig. \ref{T cell decomposition}) and restrict $O^{\mathrm{loc}}$ to $\lambda$ as $O^{\mathrm{loc}}_{\lambda}$ and act it on the wavefunction $|\psi\rangle$:
\begin{align}
O_{\lambda}^{\mathrm{loc}}|\psi\rangle=|T_{\lambda}\rangle\otimes|\psi^{\bar{\lambda}}\rangle
\end{align}
where the regime $\lambda$ is in the product state $|T_{\lambda}\rangle$ and the remainder of the system $\bar\lambda$ is in the state $|\psi^{\bar\lambda}\rangle$ and all nontrivial topological properties are encoded in this wavefunction. Furthermore, to trivialize the system in a symmetric way, we denote that $V_gO^{\mathrm{loc}}_{\lambda} V_g^{-1}$ trivializes the region $g\lambda$ ($g\in T$). Therefore, we act on the original wavefunction $|\psi\rangle$ with:
\begin{align}
O^{\mathrm{loc}}_{\mathrm{3D}}=\bigotimes_{g\in T}V_gO^{\mathrm{loc}}_{\lambda} V_g^{-1}
\end{align}
which results in an extensively trivialized wavefunction:
\begin{align}
|\psi'\rangle=O_{\mathrm{3D}}^{\mathrm{loc}}|\psi\rangle=\bigotimes\limits_{g\in T}|T_{g\lambda}\rangle\otimes|\phi\rangle
\end{align}
where $|T_{g\lambda}\rangle$ represents the deformed wavefunction of the 3D block labeled by $g\lambda$ ($g\in T$), and $|\phi\rangle$ represents the wavefunction of all lower-dimensional blocks. 

Subsequently we trivialize the region $\sigma_j$ as indicated in Fig. \ref{T cell decomposition}: repeatedly restrict $O^{\mathrm{loc}}$ to $\sigma_j$ ($j=1,2$) as $O^{\mathrm{loc}}_{\sigma_j}$ and act it on the wavefunction $|\phi\rangle$:
\begin{align}
O^{\mathrm{loc}}_{\sigma_j}|\phi\rangle=|T_{\sigma_j}\rangle\otimes|\phi^{\bar\sigma_j}\rangle
\end{align}
where the regime $\sigma_j$ is in the product state/invertible topological state $|T_{\sigma_j}\rangle$ and the remainder of the lower-dimensional blocks are in the state $|\phi^{\bar\sigma_j}\rangle$, all nontrivial topological properties (except the decorations of invertible topological phases) are encoded in this wavefunction. Again, to trivialize the system in a symmetric way, we denote that $V_gO_{\sigma_j}^{\mathrm{loc}}V_g^{-1}$ trivializes the 2D block $g\sigma_j$ ($g\in T$). Therefore, we act on the wavefunction $|\phi\rangle$ with:
\begin{align}
O^{\mathrm{loc}}_{\mathrm{2D}}=\bigotimes_{j=1}^2\bigotimes_{g\in T}V_gO_{\sigma_j}^{\mathrm{loc}}V_g^{-1}
\end{align}
which results in an extensively trivialized wavefunction:
\begin{align}
|\phi'\rangle=O^{\mathrm{loc}}_{\mathrm{2D}}|\phi\rangle=\bigotimes_{j=1}^2\bigotimes_{g\in A_4}|T_{g\sigma_j}\rangle\otimes|\eta\rangle
\end{align}
where $|T_{g\sigma_j}\rangle$ represents the deformed wavefunction of the 2D block labeled by $g\sigma_j$, and $|\eta\rangle$ represents the wavefunction of the remaining 1D and 0D blocks. 

Next we decompose the 1D and 0D block states: define a local unitary operator $W$ that is $\mathbb{Z}_3$ symmetric, hence it is only well-defined on 1D blocks. Restrict $W$ to $\tau_{1,2}$ and act it on the wavefunction $|\eta\rangle$:
\begin{align}
W_{\tau_{1,2}}|\eta\rangle=|\beta_{\tau_{1,2}}\rangle\otimes|\eta^{\bar{\tau}_{1,2}}\rangle
\end{align}
the regime $\tau_{1,2}$ is in the $\mathbb{Z}_3$-symmetric state $|\beta_{\tau_{1,2}}\rangle$ and the remainder of the 1D and 0D blocks is in the state $|\eta^{\bar{\tau}_{1,2}}\rangle$. To deform the system in a symmetric way, we denote that $V_gW_{\tau_{1,2}}V_g^{-1}$ deforms the 1D block $g\tau_{1,2}$ ($g\in T$) and define:
\begin{align}
W_{\mathrm{1D}}=\bigotimes_{j=1}^2\bigotimes_{g\in T}V_gW_{\tau_j}V_g^{-1}
\end{align}
Similarly, define a $\mathbb{Z}_2$-symmetric local unitary operator $X_{\mathrm{1D}}$ on 1D blocks $g\tau_3$ ($g\in T$). Now we act on the wavefunction $|\eta\rangle$ with $W_{\mathrm{1D}}$ and $X_{\mathrm{1D}}$:
\begin{align}
W_{\mathrm{1D}}X_{\mathrm{1D}}|\eta\rangle=\bigotimes_{j=1}^3\bigotimes_{g\in T}|\beta_{g\tau_{j}}\rangle\otimes|\alpha_\mu\rangle
\end{align}
where $|\beta_{g\tau_{1,2}}\rangle$ represents the 1D $\mathbb{Z}_3$-symmetric fSPT wavefunction of the 1D block $g\tau_{1,2}$, $|\beta_{g\tau_{3}}\rangle$ represents the 1D $\mathbb{Z}_2$-symmetric fSPT wavefunction of the 1D block $g\tau_{3}$, and $|\alpha_{\mu}\rangle$ represents the wavefunction of the 0D block $\mu$. Now the topological properties of the system are encoded in the block states with different dimensions. 

\subsection{Block state decoration}
\label{sec:blockdecoration}
With cell decomposition, we decorate some lower-dimensional block states on the corresponding blocks, in analogue to the domain wall decoration for internal SPT \cite{Chen2014NC,QR2104,QRPRX20}. We still consider the 3D $T$-symmetric cubic lattice (see Fig. \ref{T cell decomposition}). All $d$D block states form the group $\{\mathrm{BS}\}^{d\mathrm{D}}$, and all block states form the group:
\begin{align}
\{\mathrm{BS}\}=\bigotimes_{d=0}^3\{\mathrm{BS}\}^{d\mathrm{D}}
\end{align}
Furthermore, the decorated states should respect the \textit{no-open-edge condition}\cite{ZDSong2018}. Once we decorate some lower-dimensional block states, they might leave several gapless modes on the edge of the corresponding blocks, and there are several gapless edge modes coinciding near the blocks with lower dimension. In order to contribute to an SPT state, the bulk of the system should be fully gapped, hence these gapless modes should be gapped out (by some proper interactions, mass terms, entanglement pairs, etc.) in a symmetric way. If the bulk of the system cannot be symmetrically fully gapped, we call the corresponding block state is \textit{obstructed}. Equivalently, an obstruction-free decoration should satisfy the no-open-edge condition. 

For 2D blocks, the decorated block states will leave several 1D gapless modes on 1D blocks as their shared border, at which several symmetry actions (forming the group $G_{\mathrm{1D}}$ as a subgroup of total 3D point group) act as internal symmetries. 
A necessary condition of yielding a fully gapped 1D block is that the corresponding 1D gapless modes should be nonchiral, i.e., the total chiral central charge of 1D gapless modes should vanish. Furthermore,  1D gapless nonchiral modes of invertible phases can be formulated by a nonchiral topological Luttinger liquid with the Lagrangian \cite{supplementary}:
\begin{align}
\mathcal{L}_{1\mathrm{D}}=\frac{K_{IJ}}{4\pi}\left(\partial_x\phi^I\right)\left(\partial_t\phi^J\right)+\frac{V_{IJ}}{8\pi}\left(\partial_x\phi^I\right)\left(\partial_x\phi^J\right)
\label{Luttinger}
\end{align}
where $\phi=(\phi^1,\cdot\cdot\cdot,\phi^{2n})^T$ is a $2n$-component bosonic field. Such theory can be viewed as topological boundary theory of   2D  Abelian Chern-Simons theory with $K_{IJ}$ as the coefficient of $U(1)$ dynamical gauge fields $\{a^I\}$  which are related to $\phi^I$ field on the boundary by $a^I_\mu=\partial_\mu \phi^I$\cite{Lu12,Ning21a}.  For fSPT, the determinant of $K$ must be $\pm1$ and further all such $K$ with $2n\times 2n$ dimension can be equivalently transformed into the canonical form as direct sum of $n$ copies of $\sigma_z$ by unimodular transformation\cite{Lu12}.   A symmetry transformation $S\in G_{\mathrm{1D}}$ on 1D nonchiral Luttinger liquid (\ref{Luttinger}) is defined as:
\begin{align}
S:\phi(x)\mapsto W^S\phi(x)+\delta\phi^S 
\label{K-matrix symmetry}
\end{align}
such that $W^SK(W^S)^T=K$ for unitary symmetry element $S$.   $\delta\phi^S$ is spatial-independent constant phase taking value in $[0,2\pi)$  and  depends on the symmetry element $S \in G_{1D}$. In particular, the fermion parity  can be simply realized as
 $W^{\mathbb{Z}_2^f}=\mathbbm{1}_{n\times n}$ and $\delta\phi^{\mathbb{Z}_2^f}=\pi(1,\cdot\cdot\cdot,1)^T$ (mod $2\pi$) if we transform $K$ into the canonical form. Moreover,  the $U^f(1)$ charge conservation can be characterized by $W^{U^f(1)}=\mathbbm{1}_{n\times n}$ and $\delta\phi^{U^f(1)}=\theta(\pm1,\cdot\cdot\cdot,\pm1)^T$ for $n$-component bosonic field $\phi$.

 To see whether these gapless modes can be further symmetrically gapped out for a crystalline symmetric state, we  try to construct ``backscattering" terms that gap out the edge without breaking symmetry, neither explicitly nor spontaneously. Such backscattering terms are in the form:
\begin{align}
U=\sum\limits_{k}U(\Lambda_k)=\sum\limits_{k}U(x)\cos\left[\Lambda_k^TK\phi-\alpha(x)\right]
\label{backscattering}
\end{align}
which are necessary to be symmetric under Eq.(\ref{K-matrix symmetry}) for all $S\in G_{\text{1D}}$.
The backscattering term (\ref{backscattering}) can gap out the edge as long as the two conditions are both satisfied: (1) there are $n$ such terms $\Lambda_i$ ($i=1,...,n$) for $2n$ number bosonic fields, (2) the vectors $\{\Lambda_k\}$ satisfy the ``null-vector'' conditions \cite{Haldane1995} for $\forall i,j$:
\begin{align}
\Lambda_i^TK\Lambda_j=0
\label{null-vector}
\end{align}
In general, such gapped terms may cause spontaneous symmetry breaking  (SSB) of Eq.(\ref{K-matrix symmetry}) which is not expected for constructing SPT state. To check whether the gapped terms result in SSB or not, one can use the primitivity criterion that all the $n\times n$  minors of matrix $\Lambda$ which is $n\times 2n$ dimension with the $i$th column given by $\Lambda_i$  are coprime \cite{Levin12, Chenjie13}. If they are coprime, there is no SSB, otherwise there is.

The no-open-edge condition requires that for bosonic field $\phi$ with $2n$ components, we need at least $n$ independent scattering terms satisfying Eq. (\ref{null-vector}) and also primitivity criterion to fully gap out the 1D nonchiral Luttinger liquid described by Lagrangian (\ref{Luttinger}) without SSB. For such purpose, we can also stack such bosonic fields with some extra fields corresponding to trivial 2D bulk so that they together can be symmetrically gapped out.  If there is no way to symmetrically gap out the gapless modes, we call the corresponding 2D block state \textit{obstructed}. As the gapping scattering terms are constructive,  we can also define one topological invariant that can assess whether such a symmetric gapping process exists or not, called the \textit{anomaly indicator}. For all the cases we consider in this paper, the 1D blocks are with $\mathbb{Z}_2$ or $\mathbb{Z}_2\times \mathbb{Z}_2$ symmetry.  As   $\mathbb{Z}_2\times \mathbb{Z}_2$ FSPT are classified by $(Z_8)^2\times Z_4$ and the three root states are protected by three $Z_2$ subgroup alone, so it is sufficient for us to consider the $Z_2$ symmetry anomaly indicator. (For more detailed analysis of Luttinger liquid boundaries $\mathbb{Z}_2^f\times \mathbb{Z}_2\times \mathbb{Z}_2$ FSPT can be referred to Ref. \cite{Ning21a}.) For a $\mathbb{Z}_2$ symmetry operation, the $(K, W, \delta\phi)$ can be reformulated to the following canonical form by some proper unimodular transformations \cite{Heinrich_2018}, as
\begin{align}
W=\left(
\begin{array}{cccc}
-\mathbbm{1}_{n_--m} & 0 & 0 & 0\\
0 & \mathbbm{1}_{n_+-m} & 0 & 0\\
0 & 0 & 0 & \mathbbm{1}_m\\
0 & 0 & \mathbbm{1}_m & 0
\end{array}
\right)
\end{align}
\begin{align}
K=\left(
\begin{array}{cccc}
A & 0 & B & -B\\
0 & C & D & D\\
B^T & D^T & E & F\\
-B^T & D^T & F^T & E
\end{array}
\right),~\delta\phi=\left(
\begin{array}{cccc}
0\\
\chi_2\\
0\\
0
\end{array}
\right)
\end{align}
where $\mathbbm{1}_{m}$ depicts an $m\times m$ identity matrix, $n_-$, $m$, and $n_+$ are non-negative integers satisfying $n_++n_-=N$ and $m\leq n_\pm$. Based on the canonical form of $(K, W, \delta\chi)$, we further define the \textit{auxiliary vector} $\chi_+$ as
\begin{align}
\chi_+=\left(
\begin{array}{cccc}
0\\
\chi_2+2a\\
\mathrm{diag}(E+F)/2+b\\
\mathrm{diag}(E+F)/2+b
\end{array}
\right)
\end{align}
for $\forall a, b\in\mathbb{Z}$. The anomaly indicator $\nu$ is defined by
\begin{align}
\nu\equiv\frac{1}{2}\chi_+^TK^{-1}\chi_++\frac{1}{4}\mathrm{sig}(K(1-W))~(\mathrm{mod}~2)
\label{anomaly indicator}
\end{align}
where ``sig'' denotes the signature of the matrix. For more detailed about this indicator, one can refer the Ref.\cite{Heinrich_2018}. One important fact here for us is that the anomaly-free criterion of $(K, W, \delta\phi)$ based on the anomaly indicator $\nu$ is
\begin{align}
\nu=0~(\mathrm{mod}~2)
\end{align}
The most important application of this anomaly-free criterion is that it promises the existence of such scattering terms, with extra bosonic fields corresponding to trivial bulk if necessary, that can symmetrically gap out the gapless Luttinger liquids without causing SSB. We would use this criterion to assess whether we can gap out the gapless Luttinger liquids without causing SSB or not.

For 1D blocks of crystalline TSC, the decorated block states leave several 0D gapless modes that can be formulated in terms of Majorana zero modes (e.g., a complex fermion can be reformulated to two Majorana zero modes, a spin-1/2 degree of freedom can be reformulated to four Majorana zero modes) at the center of the 3D point group as their shared border, at which the point group symmetry acting internally. Together with the $\Z_2^f$ fermion parity, we label the total effective on-site symmetry group by $G_{\mathrm{0D}}$. Suppose there are $2n$ Majorana zero modes $\gamma_j$ and $\gamma_j'$ ($j=1,\cdot\cdot\cdot,n$) near the 0D block, they can be reformulated to $n$ complex fermions: $c_j^\dag=\left(\gamma_j+i\gamma_j'\right)/2$, and they span a $2^n$-dimensional Hilbert space. An arbitrary symmetry action $g\in G_{\mathrm{0D}}$ permutes the Majorana zero modes, hence it can be phrased by a $2^n\times2^n$ matrix $A(g)$ on the Hilbert space. All $A(g)$ form a matrix representation of $G_{\mathrm{0D}}$. For $\forall g_1,g_2\in G_{\mathrm{0D}}$, the matrix representation $A$ satisfies:
\begin{align}
A(g_1g_2)=\omega_2(g_1,g_2)A(g_1)A(g_2)
\end{align}
where $\omega_2(g_1,g_2)\in\mathcal{H}^2[G,U(1)]$ satisfies the 2-cocycle condition:
\begin{align}
{\omega_2(g_2,g_3)\omega_2(g_1g_2,g_3)}/{\omega_2(g_1,g_2g_3)\omega_2(g_2,g_3)}=1
\end{align}
A sufficient and necessary condition of fully gapping out the $2n$ Majorana zero modes is that $A$ should be a linear representation of $G_{\mathrm{0D}}$ without breaking symmetry, according to the wisdom of 1D SPT. Equivalently, $\omega_2$ should be a 2-coboundary with $U(1)$ coefficient: for $\forall g_1,g_2\in G_{\mathrm{0D}}$, there exists a 1-cochain $\nu(g)$ as a linear map from $G_{\mathrm{0D}}$ to $U(1)$ that $\omega_2$ can be rephrased in terms of $\nu_1$:
\begin{align}
\omega_2(g_1,g_2)=\mathrm{d}\nu_1(g_1,g_2)={\nu_1(g_1)\nu_1(g_2)}/{\nu_1(g_1g_2)}
\end{align}
If not, $A$ will be a projective representation of $G_{\mathrm{0D}}$, and the $2n$ Majorana zero modes cannot be gapped out in a $G_{\mathrm{0D}}$-symmetric way. We call the corresponding 1D block state is \textit{obstructed}.

For 1D blocks of crystalline TI, as $U^f(1)$ is always incompatible with a single Majorana chain \cite{fidkowski11, Ning23PRR}, we only need to consider decorated 1D SPT on the 1D blocks.  We will argue that the only possible root phase is the Haldane chain. For the 1D blocks as the axis of $n$-fold rotation, the corresponding total symmetry is $U^f(1)\times\mathbb{Z}_n$ for both spinless and spin-1/2 fermions. The corresponding 1D fSPT phases are classified by group 2-cohomology:
\begin{align}
\mathcal{H}^2\left[U^f(1)\times\mathbb{Z}_n,U(1)\right]=\mathbb{Z}_1
\end{align}
i.e., there is no nontrivial block state. For the 1D blocks as the axis of $3$-fold dihedral group, the corresponding total symmetry is $U^f(1)\times(\mathbb{Z}_3\rtimes\mathbb{Z}_2)$ for both spinless and spin-1/2 fermions. The corresponding 1D fSPT phases are classified by group 2-cohomology:
\begin{align}
\mathcal{H}^2\left[U^f(1)\times(\mathbb{Z}_3\rtimes\mathbb{Z}_2),U(1)\right]=\mathbb{Z}_1
\end{align}
i.e., there is no nontrivial block state. For 1D blocks as the axis of $2k$-fold dihedral group ($k=1,2,3$), the corresponding total symmetry is $U^f(1)\times(\mathbb{Z}_{2k}\rtimes\mathbb{Z}_2)/U^f(1)\rtimes_{\omega_2}(\mathbb{Z}_{2k}\rtimes\mathbb{Z}_2)$ for spinless/spin-1/2 fermions. The corresponding 1D fSPT phases are classified by group 2-cohomologies:
\begin{align}
\begin{aligned}
&\mathcal{H}^2\left[U^f(1)\times(\mathbb{Z}_{2k}\rtimes\mathbb{Z}_2),U(1)\right]=\mathbb{Z}_2\\
&\mathcal{H}^2\left[U^f(1)\rtimes_{\omega_2}(\mathbb{Z}_{2k}\rtimes\mathbb{Z}_2),U(1)\right]=\mathbb{Z}_1
\end{aligned}
\end{align}
i.e., for spinless fermions, the only possible 1D root phase is the Haldane chain; for spin-1/2 fermions, there is no nontrivial 1D root phase. 

For the 0D block, i.e., the center of the system, which has the total point group symmetry (i.e., $G_{0D}$) as internal symmetry, we can decorate different 1D representations of the point group symmetry, classified by $\mathcal{H}^1(G_{0D}, U(1))$, while  higher dimensional irreducible representations will cause state degeneracy, and then obstructed. 

To summarize, all obstruction-free $d$D block states form the group $\{\mathrm{OFBS}\}^{d\mathrm{D}}$  which is a subset of $\{\mathrm{BS}\}^{d\mathrm{D}}$, and all obstruction-free block states formally form the group:
\begin{align}
\{\mathrm{OFBS}\}=\bigotimes_{d=0}^3\{\mathrm{OFBS}\}^{d\mathrm{D}}\subset\{\mathrm{BS}\}
\end{align}
Each element in $\{\mathrm{OFBS}\}$ contribute to a construction of valid symmetric and fully gapped state. However, not every nontrivial elements of its would contribute to a nontrivial FSPT in the classification sense, that is means some of its nontrivial elements may just results in a trivial state by the above decorated block state procedures. To see such a equivalence of seeming nontrivial decorated block states and trivial state, which we call it \textit{trivialization}, one can use the so-called bubble equivalence as below, which in fact is a local unitary  transformation similar to the $O^{loc}$ in Sec.\ref{Seccell}. 

\subsection{Bubble equivalence}
In order to obtain a nontrivial SPT state from obstruction-free block state decorations, we should further consider possible trivializations. For blocks with a dimension larger than 0, we can further decorate some codimension 1 degree of freedom that could be trivialized when they shrink to a point. This construction is called \textit{bubble equivalence}, and we demonstrate it for different dimensions. The bubble equivalence are widely used in the bosonic crystalline SPT \cite{reduction}, free fermion SPT state \cite{ZDSong2018} and 2D interacting fermionic crystalline SPT \cite{wallpaper}  and  we generalize it to the 3D interacting fermionic crystalline SPT.

\paragraph{3D bubble equivalence}For 3D blocks, we can consider a 2D sphere that can be shrunk to a point inside each 3D block, and there is no on-site symmetry on them for all possible cases. For crystalline TSC, the only possible state we can decorate is $(p+ip)$-SC. In particular, the 3D ``$(p+ip)$-SC'' bubble does not change the parity of number of $(p+ip)$-SCs on 2D blocks: each 2D block is the shared border of two nearby 3D blocks, hence the number of $(p+ip)$-SCs on this 2D block can only be changed by 0 or 2 by 3D ``$(p+ip)$-SC'' bubble, depending on the orientation of the bubbles. For crystalline TI, the possible states we can decorate are the Chern insulator and Kitaev's $E_8$ state, whose number can be changed by $0$ or 2 similarly to the $p+ip$-SC bubbles.

\paragraph{2D bubble equivalence}For 2D blocks, we can consider a 1D sphere which can be shrunk to a point inside each 2D block. By reflection symmetry acting internally, the only possible effective ``on-site'' symmetry of 2D block is $\mathbb{Z}_2$. For crystalline TSC with spinless fermions, there are two possible states we can decorate:
\begin{enumerate}[1.]
\item Majorana chain with anti-periodic boundary condition (anti-PBC);
\item 1D fSPT phase, composed of double Majorana chains.
\end{enumerate}
For crystalline TSC with spin-1/2 fermions and crystalline TI, there is no possible state we can decorate because of the trivial classification of the corresponding 1D fSPT phases. 

\paragraph{1D bubble equivalence}For 1D blocks, we can consider two 1D irreducible representations of the corresponding total on-site symmetry of the 1D blocks that should be trivialized if they shrunk to a point. There are two possibilities: the first one is a fermionic 1D bubble: consider two complex fermions with the following geometry:
\begin{align}
\begin{tikzpicture}
\tikzstyle{sergio}=[rectangle,draw=none]
\draw[thick] (-2.5,2)--(-1.5,2);
\filldraw[fill=red, draw=red] (-2.5,2)circle (2pt);
\filldraw[fill=yellow, draw=yellow] (-1.5,2)circle (2pt);
\path (-3.1,2) node [style=sergio] {$a_l^\dag/b_l^\dag$};
\path (-0.9,2) node [style=sergio] {$a_r^\dag/b_r^\dag$};
\end{tikzpicture}
\label{1D bubble}
\end{align}
Where yellow and red dots represent two complex fermions $a_l^\dag$ and $a_r^\dag$ which are trivialized when they are fused, i.e., $a_l^\dag a_r^\dag|0\rangle$ is a trivial atomic insulating state with even fermion parity. We demonstrate that this 1D bubble can be shrunk to a point and trivialized by a finite-depth circuit: if we decorate a 1D bubble, we can enclose $a_l^\dag$ and $a_r^\dag$ by an LU transformation. Repeatedly applying this LU transformation, we can shrink these two modes to a point. Therefore, the creation operator of fermionic 1D bubbles in the entire lattice is:
\begin{align}
B_j^f=\bigotimes_{\tau}(a_l^\tau)^\dag(a_r^\tau)^\dag
\end{align}

Another one is a bosonic 1D bubble: consider  the geometry indicated in Eq. (\ref{1D bubble}), where yellow and red dots represent two bosons $b_l^\dag$ and $b_r^\dag$ that carry 1D irreducible representations of the physical symmetry group (total symmetry group quotient by fermion parity $\mathbb{Z}_2^f$) of corresponding 1D blocks. They should be trivialized by shrinking them to a point: $b_l^\dag b_r^\dag|0\rangle$ carries a trivial 1D irreducible representation of the physical symmetry group. The creation operator of bosonic 1D bubbles in the entire lattice is:
\begin{align}
B_j^b=\bigotimes_{\tau}(b_l^\tau)^\dag(b_r^\tau)^\dag
\end{align}
And the creation operator of general 1D bubbles is:
\begin{align}
B_j=B_j^f\otimes B_j^b
\end{align}
Enlarge these bubbles and proximate to the nearby lower-dimensional blocks, the FSPT phases decorated on the bubble can be fused with the original states on the nearby lower-dimensional blocks, which leads to some possible \textit{trivializations} of lower-dimensional block state decorations. All trivialized $d$D block states form the group $\{\mathrm{TBS}\}^{d\mathrm{D}}\subset\{\mathrm{OFBS}\}^{d\mathrm{D}}$, and all trivialized block states form the group:
\begin{align}
\{\mathrm{TBS}\}=\bigotimes_{d=0}^3\{\mathrm{TBS}\}^{d\mathrm{D}}\subset\{\mathrm{OFBS}\}
\end{align}
Therefore, an obstruction and trivialization-free block state can be labeled by a group element of the following quotient group:
\begin{align}
\mathcal{G}=\{\mathrm{OFBS}\}/\{\mathrm{TBS}\}
\end{align}
and all group elements in $\mathcal{G}$ are not equivalent. Equivalently, group $G$ gives the classifications of the corresponding crystalline topological phases. In particular, we note that the block states are constructed layer by layer. Therefore, we should specify the $d$-dimensional obstruction-free and trivialization-free block states:
\begin{align}
E^{d\mathrm{D}}=\{\mathrm{OFBS}\}^{d\mathrm{D}}/\{\mathrm{TBS}\}^{d\mathrm{D}}
\end{align}
We should note that $E^{d\mathrm{D}}$ is not a group in the sense of SPT classification, because to obtain the ultimate classification of SPT phases, we should further consider the possible stacking between block states with different dimensions. $E^{d\mathrm{D}}$ can only be treated as a group only in the sense of $d$D block states. 

With all obstruction and trivialization, free block states with different dimensions, the ultimate classification with accurate group structure of 2D crystalline fSPT phases is an extension between $E^{2\mathrm{D}}$, $E^{1\mathrm{D}}$ and $E^{0\mathrm{D}}$:
\begin{align}
\mathcal{G}=\left(E^{2\mathrm{D}}\times_{\omega_2}E^{1\mathrm{D}}\right)\times_{\omega_2}E^{0\mathrm{D}}
\end{align}
here the symbol $\times_{\omega_2}$ depicts the possible extensions of $E^{2\mathrm{D}}$, $E^{1\mathrm{D}}$ and $E^{0\mathrm{D}}$ that is characterized by following short exact sequence:
\begin{align}
\begin{aligned}
0&\rightarrow E^{2\mathrm{D}}\rightarrow E^{2\mathrm{D}}\times_{\omega_2}E^{1\mathrm{D}}\rightarrow E^{1\mathrm{D}}\rightarrow0\\
&0\rightarrow E^{2\mathrm{D}}\times_{\omega_2}E^{1\mathrm{D}}\rightarrow\mathcal{G}\rightarrow E^{0\mathrm{D}}\rightarrow0
\end{aligned}
\end{align}
In particular, we elucidate that numerous 2D block states correspond to intriguingly interacting topological crystalline phases. The backscattering terms as the solutions to the ``null-vector'' problem (\ref{null-vector}) including more than two non-vanishing elements correspond to interactions of the 1D (nonchiral) Luttinger liquids at the 1D blocks as the shared border of the nearby 2D blocks.

\subsection{Higher-order topological surface theory}
By considering obstruction and trivialization-free nontrivial block states, we proceed to analyze their topological surface theories. Notably, all the procedures described in Section \ref{Seccell} involve smooth deformations, which means they do not alter the topological properties of the robust surface theory. As a result, we can directly examine the topological surface theory of lower-dimensional block states by assigning them to the 3D open system and obtaining the corresponding surface theories by truncating the block states on the boundary.

However, it is important to note that a nontrivial topological crystalline phase does not guarantee a higher-order (HO) topological surface state. Specifically, 2D block states give rise to second-order topological surface theories featuring 1D gapless modes, while 1D block states result in third-order topological surface theories with 0D gapless modes. On the other hand, 0D block states do not lead to an HO topological surface theory.

Furthermore, we previously mentioned that two block states are considered ``equivalent" if they can be smoothly deformed into each other through bubble constructions in bulk. Similarly, their topological surface theories can also be smoothly transformed into each other through ``plates" constructions on the boundary. This demonstrates the direct relationship between the bulk and boundary in 3D crystalline fractal symmetry-protected topological (fSPT) phases, known as the bulk-boundary correspondence.

In the following sections, we explicitly apply these procedures to calculate the classification of crystalline TSC and TI and the corresponding HO topological surface theory, by several representative examples for each crystallographic system.

\section{Construction and classification of crystalline TSC\label{TSC classification}}
In this section, we describe the details of real-space construction for crystalline TSC protected by point groups in 3D interacting fermionic systems by analyzing several typical examples. It is well known that all 32 point groups in 3D systems can be divided into seven different crystallographic systems:
\begin{description}
\item[Triclinic lattice]$C_1$, $S_2$.
\item[Monoclinic lattice]$C_2$, $C_{1h}~(C_s)$, $C_{2h}$.
\item[Orthorhombic lattice]$D_2$, $C_{2v}$, $D_{2h}$.
\item[Tetragonal lattice]$C_4$, $S_4$, $C_{4h}$, $D_4$, $C_{4v}$, $D_{2d}$, $D_{4h}$.
\item[Trigonal lattice]$C_3$, $S_6$, $D_3$, $C_{3v}$, $D_{3d}$.
\item[Hexagonal lattice]$C_6$, $C_{3h}$, $C_{6h}$, $D_6$, $C_{6v}$, $D_{3h}$, $D_{6h}$.
\item[Cubic lattice]$T$, $T_h$, $T_d$, $O$, $O_h$.
\end{description}
In particular, we apply the general paradigm of real-space construction to investigate five representative cases that belong to cubic lattice after a pedagogical introduction of an inversion-symmetric system.

\subsection{Inversion-symmetric lattice\label{inversion}}
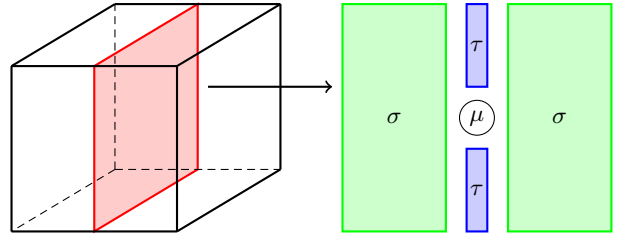
\begin{figure}
\begin{tikzpicture}[scale=0.55]
\tikzstyle{sergio}=[rectangle,draw=none]
\filldraw[fill=red!20, draw=red, thick] (-1,0)--(1.5,1.5)--(1.5,-2.5)--(-1,-4)--cycle;
\draw[thick] (-0.5,1.5) -- (3.5,1.5);
\draw[thick] (-0.5,1.5) -- (-3,0);
\draw[thick] (3.5,1.5) -- (1,0);
\draw[thick] (-3,0) -- (1,0);
\draw[thick] (-3,0) -- (-3,-4);
\draw[thick] (1,-4) -- (-3,-4);
\draw[thick] (1,-4) -- (1,0);
\draw[thick] (3.5,-2.5) -- (3.5,1.5);
\draw[thick] (3.5,-2.5) -- (1,-4);
\draw[densely dashed] (-0.5,-2.5) -- (-0.5,1.5);
\draw[densely dashed] (-0.5,-2.5) -- (3.5,-2.5);
\draw[densely dashed] (-0.5,-2.5) -- (-3,-4);
\filldraw[fill=green!20, draw=green, thick] (5,1.5)--(7.5,1.5)--(7.5,-4)--(5,-4)--cycle;
\filldraw[fill=green!20, draw=green, thick] (9,1.5)--(11.5,1.5)--(11.5,-4)--(9,-4)--cycle;
\filldraw[fill=blue!20, draw=blue, thick] (8.5,1.5)--(8,1.5)--(8,-0.5)--(8.5,-0.5)--cycle;
\filldraw[fill=blue!20, draw=blue, thick] (8.5,-2)--(8,-2)--(8,-4)--(8.5,-4)--cycle;
\filldraw[fill=white, draw=black] (8.25,-1.25)circle (12pt);
\path (6.25,-1.25) node [style=sergio] {$\sigma$};
\path (10.25,-1.25) node [style=sergio] {$\sigma$};
\path (8.25,-1.25) node [style=sergio] {$\mu$};
\path (8.25,0.5) node [style=sergio] {$\tau$};
\path (8.25,-3) node [style=sergio] {$\tau$};
\draw[thick,->] (1.75,-0.5) -- (4.75,-0.5);
\end{tikzpicture}
\caption{The cell decomposition of the 3D inversion-symmetric lattice. The red plate in the left panel is the equator including the center of inversion; the right panel shows the lower-dimensional blocks included in the equator, where green plates depict 2D blocks, blue segments depict 1D blocks and the black circle depicts 0D blocks as the center of inversion.}
\label{inversion cell decomposition}
\end{figure}

For inversion-symmetric lattice with the cell decomposition in Fig. \ref{inversion cell decomposition}, the ground-state wavefunction of the system can be decomposed to the direct products of wavefunctions of lower-dimensional blocks as:
\begin{align}
|\Psi\rangle=\bigotimes_{g\in S_2}|T_{g\lambda}\rangle\otimes|T_{g\sigma}\rangle\otimes|T_{g\tau}\rangle\otimes|\alpha_\mu\rangle
\label{Inversion wavefunction}
\end{align}
where $|T_{g(\lambda,\sigma,\tau)}\rangle$ is the wavefunction of $d$D block state on $g(\lambda,\sigma,\tau)$ ($d=1,2,3$), which is topological trivial or invertible topological phase; $|\alpha_\mu\rangle$ is the wavefunction of 0D block state which is $\mathbb{Z}_2$ symmetric. 

We argue that for 3D inversion-symmetric lattice, both spinless and spin-1/2 fermions correspond to trivial group extension with fermion parity $\mathbb{Z}_2^f$: for spinless fermions, it is obvious; for spin-1/2 fermions, we can treat an inversion $\bs{I}$ operation in three spatial dimensions as the composition of a 2-fold rotation $\bs{R}_z$ with respect to $z$-axis and a reflection $\bs{M}_{xy}$ with respect to $xy$-plane. The spin-1/2 nature of fermions requires the relations $\bs{R}_z^2=\bs{M}_{xy}^2=-1$, hence $\bs{I}^2=1$ for either spinless or spin-1/2 fermions, which corresponds to the trivial $\mathbb{Z}_2^f$ extension of the inversion. 

Subsequently, we consider the lower-dimensional block state decorations. For 2D blocks, the only possible nontrivial root phase is  $(p+ip)$-SC, leaving two chiral Majorana modes $\gamma_{1,2}$ near 1D block $\tau$, with opposite chiralities, described by Hamiltonian:
\begin{align}
H_0=\int\mathrm{d}x\cdot\gamma^T\left(i\sigma^3\partial_x\right)\gamma,~\gamma=(\gamma_1,\gamma_2)^T
\end{align}
The inversion transforms the Majorana modes $\gamma_1$ and $\gamma_2$ as:
\[
\gamma_1(x)\mapsto\gamma_2(-x),~~\gamma_2(x)\mapsto\gamma_1(-x).
\]
The only possible mass term 
\begin{align}
i m(x)\gamma_1(x)\gamma_2(x)  \mapsto - i m(x)\gamma_1(-x)\gamma_2(-x).
\end{align}
For  symmetric mass, it is necessary that $m(x)=-m(x)$ and in particular, $m(0)=0$, which is the famous domain wall structure of the nonzero $m(x)$ ($x>0$) for majorna fermion and would trap the single majorana zero\cite{Jackiw1976}. Therefore, the single layer $p+ip$-SC decorated state is obstructed.
On the other hand, bilayer $(p+ip)$-SCs can be trivialized by 3D bubble equivalence: 3D ``$p+ip$'' bubble changes the layers of $(p+ip)$-SCs by two, equivalent, bilayer $(p+ip)$-SCs are trivialized by 3D ``$p+ip$'' bubble. 

For 1D blocks, the only possible nontrivial root phase is Kitaev's Majorana chain, leaving two dangling Majorana zero modes near the inversion center. These two Majorana zero modes cannot be gapped out in a symmetric way, since $i\gamma_1 \gamma_2\mapsto i\gamma_2\gamma_1=-i\gamma_1\gamma_2$ does not preserve the inversion symmetry: $\gamma_{1,2}\mapsto \gamma_{2,1}$. Similar situation also occurs in 2D case \cite{dihedral}.

For 0D blocks, all possible root phases form the $\mathbb{Z}_2^2$ group, with two root phases: complex fermion and eigenvalues $\pm$ of $\mathbb{Z}_2$ by inversion symmetry acting internally. From Ref. \cite{dihedral}, all these root phases are trivialized. Therefore, the classification of 3D inversion-symmetric TSCs are trivial, for both spinless and spin-1/2 fermions, and there is no nontrivial HO topological surface theory.

\subsection{$T$-symmetric lattice\label{T}}
For $T$-symmetric cubic, by cell decomposition as illustrated in Fig. \ref{T cell decomposition}, the ground-state wavefunction of the system can be decomposed to the direct products of wavefunctions of lower-dimensional blocks as:
\begin{align}
|\Psi\rangle=\bigotimes_{g\in T}|T_{g\lambda}\rangle\otimes\bigotimes\limits_{k=1}^2|T_{g\sigma_k}\rangle\otimes \bigotimes\limits_{l=1}^3|\beta_{g\tau_l}\rangle\otimes|\alpha_\mu\rangle
\label{T wavefunction}
\end{align}
where $|T_{g\lambda}\rangle$ is the wavefunction of 3D block state labeled by $g\lambda$ which is topologically trivial, $|T_{g\sigma_{1,2}}\rangle$ is the wavefunction of 2D block state labeled by $g\sigma_{1,2}$; $|\beta_{g\tau_{1,2}}\rangle$ is the wavefunction of 1D block state labeled by $g\tau_{1,2}$ which is $\mathbb{Z}_3$-symmetric, and $|\beta_{g\tau_3}\rangle$ is the wavefunction of 1D block state labeled by $g\tau_3$ which is $\mathbb{Z}_2$-symmetric; $|\alpha_\mu\rangle$ is the wavefunction of 0D block state labeled by $\mu$ which is $A_4$-symmetric.

With topological crystals, we decorate the lower-dimensional block states and investigate the possible \textit{obstructions} and \textit{trivializations}.

\subsubsection{2D block states}
There is no effective ``on-site'' symmetry on all 2D blocks. The only possible root phase is $(p+ip)$-SC with gapless chiral Majorana modes as its edge mode. If we decorate a 2D $(p+ip)$-SC with quantum number $n_k\in\mathbb{Z}$ on each 2D block $\sigma_k$ ($k=1,2$), the chiral central charges on the 1D blocks $\tau_1$, $\tau_2$ and $\tau_3$ are $\frac{3n_2}{2}$, $\frac{3n_1+3n_2}{2}$ and $n_1$, respectively. Fully gapped bulk requires that all of these three quantities should vanish:
\[
3n_2=3n_1+3n_2=n_1=0
\]
There is no non-vanishing solution, hence all nontrivial 2D block state decorations are \textit{obstructed}.

\subsubsection{1D block states\label{1D block}}
The effective ``on-site'' symmetry $G_{\mathrm{1D}}$ on the 1D block labeled by $\tau_{1,2}$ is $\mathbb{Z}_3$ by 3-fold rotation symmetry acting internally. Hence the candidate 1D block states are classified by group supercohomology \cite{special,general1,general2}:
\begin{align}
\begin{aligned}
&n_0\in\mathcal{H}^0(G_{\mathrm{1D}},\mathbb{Z}_2)=\mathbb{Z}_2\\
&n_1\in\mathcal{H}^1(G_{\mathrm{1D}},\mathbb{Z}_2)=\mathbb{Z}_1\\
&\nu_2\in\mathcal{H}^2\left[G_{\mathrm{1D}},U(1)\right]=\mathbb{Z}_1
\end{aligned}
\label{2-supercohomology}
\end{align}
where $n_0$ depicts the Majorana chain, $n_1$ depicts the complex fermion decoration  and $\nu_2$ depicts the 1D bosonic SPT (bSPT), with the twisted cocycle conditions ($\forall g_1,g_2,g_3\in G_{\mathrm{1D}}$):
\begin{align}
\begin{aligned}
&n_1(g_1)+n_1(g_2)-n_1(g_1,g_2)=\omega_2\smile n_0(g_1,g_2)\\
&\frac{\nu_2(g_1,g_2g_3)\nu_2(g_2,g_3)}{\nu_2(g_1,g_2)\nu_2(g_1g_2,g_3)}=(-1)^{\omega_2\smile{n_1}(g_1,g_2,g_3)}
\end{aligned}
\label{2-twisted}
\end{align}
Here $\omega_2\in\mathcal{H}^2(G_{\mathrm{1D}},\mathbb{Z}_2)$ characterizes the spin of fermions. For 1D blocks $\tau_{1,2}$ with $\mathbb{Z}_3$ on-site symmetry, $\mathcal{H}^2(\mathbb{Z}_3,\mathbb{Z}_2)=\mathbb{Z}_1$, hence the spin of fermions is irrelevant. Eqs. (\ref{2-supercohomology}) and (\ref{2-twisted}) indicate that the only possible nontrivial 1D block state is Majorana chain characterized by nonzero $n_0$ in Eq. (\ref{2-supercohomology}). 

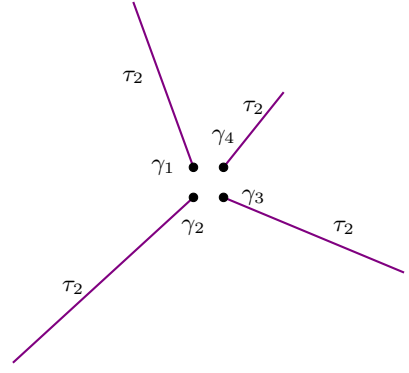
\begin{figure}
\begin{tikzpicture}[scale=0.8]
\tikzstyle{sergio}=[rectangle,draw=none]
\path (-1,0) node [style=sergio] {$\tau_2$};
\path (-2,-3.5) node [style=sergio] {$\tau_2$};
\path (1,-0.5) node [style=sergio] {$\tau_2$};
\path (2.5,-2.5) node [style=sergio] {$\tau_2$};
\draw[thick,color=violet] (-1,1.25) -- (0,-1.5);
\draw[thick,color=violet] (0,-2) -- (-3,-4.75);
\draw[thick,color=violet] (0.5,-1.5) -- (1.5,-0.25);
\draw[thick,color=violet] (0.5,-2) -- (3.5,-3.25);
\filldraw[fill=black, draw=black] (0,-1.5)circle (2pt);
\filldraw[fill=black, draw=black] (0.5,-1.5)circle (2pt);
\filldraw[fill=black, draw=black] (0,-2)circle (2pt);
\filldraw[fill=black, draw=black] (0.5,-2)circle (2pt);
\path (-0.5,-1.5) node [style=sergio] {$\gamma_1$};
\path (0,-2.5) node [style=sergio] {$\gamma_2$};
\path (1,-2) node [style=sergio] {$\gamma_3$};
\path (0.5,-1) node [style=sergio] {$\gamma_4$};
\end{tikzpicture}
\caption{Majorana chain decorations on 1D blocks $\tau_2$, leaving 4 dangling Majorana zero modes $\gamma_j$ at $\mu$.}
\label{tau2}
\end{figure}

Subsequently, we consider the Majorana chain decorations on 1D blocks labeled by $\tau_2$ (see Fig. \ref{tau2}) that leave 4 dangling Majorana zero modes $\gamma_j$ ($j=1,2,3,4$). The symmetry properties $\bs{R}_j\in C_3^j$ ($j=1,2,3,4$) of these Majorana zero modes are:
\begin{align}
\begin{aligned}
&\bs{R}_1:\left(\gamma_1,\gamma_2,\gamma_3,\gamma_4\right)\mapsto\left(\gamma_4,\gamma_2,\gamma_1,\gamma_3\right)\\
&\bs{R}_2:\left(\gamma_1,\gamma_2,\gamma_3,\gamma_4\right)\mapsto\left(\gamma_2,\gamma_4,\gamma_3,\gamma_1\right)\\
&\bs{R}_3:\left(\gamma_1,\gamma_2,\gamma_3,\gamma_4\right)\mapsto\left(\gamma_2,\gamma_3,\gamma_1,\gamma_4\right)\\
&\bs{R}_4:\left(\gamma_1,\gamma_2,\gamma_3,\gamma_4\right)\mapsto\left(\gamma_1,\gamma_3,\gamma_4,\gamma_2\right)
\end{aligned}
\end{align}
To get a gapped SPT state, we should investigate whether Majorana zero modes $\gamma_j$ ($j=1,2,3,4$) can be gapped out symmetrically. Firstly we introduce the interacting Hamiltonian:
\begin{align}
H_U=U\gamma_1\gamma_2\gamma_3\gamma_4,~U>0
\label{HU}
\end{align}
This Hamiltonian provides an energy gap but leaves ground-state degeneracy (GSD). The ground state corresponds to the constraint $\gamma_1\gamma_2\gamma_3\gamma_4=-1$, with two possibilities:
\begin{align}
\left\{
\begin{aligned}
&-i\gamma_1\gamma_2=-1\\
&-i\gamma_3\gamma_4=1
\end{aligned}
\right.,~~\left\{
\begin{aligned}
&-i\gamma_1\gamma_2=1\\
&-i\gamma_3\gamma_4=-1
\end{aligned}
\right.
\end{align}
To lift the GSD, we should further introduce some symmetric mass terms  like $i\gamma_1\gamma_2$, i.e., Majorana pairs. We list all possible Majorana pairs as:
\begin{align}
i\gamma_1\gamma_2,~i\gamma_1\gamma_3,~i\gamma_1\gamma_4,~i\gamma_2\gamma_3,~i\gamma_2\gamma_4,~i\gamma_3\gamma_4
\label{mass term}
\end{align}
Consider $i\gamma_1\gamma_2$ as an example. Under $\bs{R}_1$, it will be transformed as:
\begin{align}
\bs{R}_1:~i\gamma_1\gamma_2\mapsto i\gamma_4\gamma_2\mapsto i\gamma_3\gamma_2\mapsto i\gamma_1\gamma_2
\end{align}
Hence if the symmetric mass term includes $i\gamma_1\gamma_2$, the terms $-i\gamma_2\gamma_4$ and $-i\gamma_2\gamma_3$ should also be included. Under $\bs{R}_2$, $i\gamma_1\gamma_2$ will be transformed as:
\begin{align}
\bs{R}_2:~i\gamma_1\gamma_2\mapsto i\gamma_2\gamma_4\mapsto i\gamma_4\gamma_1\mapsto i\gamma_1\gamma_2
\end{align}
Hence if the symmetric mass term includes $i\gamma_1\gamma_2$, the terms $i\gamma_2\gamma_4$ and $-i\gamma_1\gamma_4$ should also be included that is in contradiction to $\bs{R}_1$-symmtry requirement. As the consequence, the term $i\gamma_1\gamma_2$ cannot be included in the symmetric mass term. Similar to all other terms in Eq. (\ref{mass term}), we conclude that all terms in Eq. (\ref{mass term}) break symmetry, and the GSD cannot be lifted. As the consequence, the Majorana chain decoration on 1D blocks labeled by $\tau_2$ is obstructed. Similar for the 1D blocks labeled by $\tau_1$.

\begin{figure}
\begin{tikzpicture}[scale=0.8]
\tikzstyle{sergio}=[rectangle,draw=none]
\path (-1,0.5) node [style=sergio] {$\tau_2$};
\path (-2,-4) node [style=sergio] {$\tau_2$};
\path (1,0) node [style=sergio] {$\tau_2$};
\path (2.5,-3) node [style=sergio] {$\tau_2$};
\draw[thick,color=violet] (-1,1.75) -- (0,-1);
\draw[thick,color=violet] (0,-2.5) -- (-3,-5.25);
\draw[thick,color=violet] (0.5,-1) -- (1.5,0.25);
\draw[thick,color=violet] (0.5,-2.5) -- (3.5,-3.75);
\filldraw[fill=black, draw=black] (0,-1)circle (2pt);
\filldraw[fill=black, draw=black] (0.5,-1)circle (2pt);
\filldraw[fill=black, draw=black] (0,-2.5)circle (2pt);
\filldraw[fill=black, draw=black] (0.5,-2.5)circle (2pt);
\path (-0.5,-1) node [style=sergio] {$\gamma_1$};
\path (0,-3) node [style=sergio] {$\gamma_2$};
\path (0.5,-3) node [style=sergio] {$\gamma_3$};
\path (0.5,-0.5) node [style=sergio] {$\gamma_4$};
\draw[densely dashed,color=red] (-4,0) -- (-0.5,-1.5);
\path (-1,-1.6) node [style=sergio] {$\gamma_1'$};
\filldraw[fill=black, draw=black] (-0.5,-1.5)circle (2pt);
\draw[densely dashed,color=red] (-0.5,-2) -- (-1.5,-3.25);
\path (-0.5,-2.5) node [style=sergio] {$\gamma_2'$};
\filldraw[fill=black, draw=black] (-0.5,-2)circle (2pt);
\draw[densely dashed,color=red] (3.5,0.5) -- (1,-1.5);
\path (1.5,-2) node [style=sergio] {$\gamma_3'$};
\filldraw[fill=black, draw=black] (1,-2)circle (2pt);
\draw[densely dashed,color=red] (1,-2) -- (2,-4.75);
\path (1,-1.1) node [style=sergio] {$\gamma_4'$};
\filldraw[fill=black, draw=black] (1,-1.5)circle (2pt);
\end{tikzpicture}
\caption{Majorana chain decorations on 1D blocks $\tau_1$ and $\tau_2$, leaving 8 dangling Majorana zero modes $\gamma_j$ and $\gamma_j'$ at $\mu$.}
\label{together}
\end{figure}
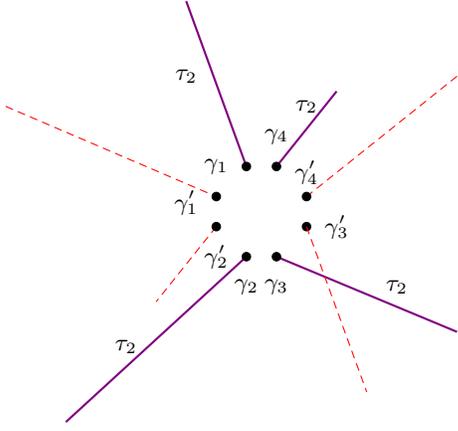

There is one exception: If we consider all 1D blocks $\tau_1$ and $\tau_2$ together and decorate a Majorana chain on each of them (see Fig. \ref{together}), there are 8 dangling Majorana zero modes $\gamma_j$ and $\gamma_j'$ ($j=1,2,3,4$). The symmetry properties $\bs{R}_j\in C_3^j$ of these Majorana zero modes are:
\begin{align}
\begin{aligned}
&\bs{R}_1:\begin{aligned}
&(\gamma_1,\gamma_2,\gamma_3,\gamma_4)\mapsto(\gamma_4,\gamma_2,\gamma_1,\gamma_3)\\
&(\gamma_1',\gamma_2',\gamma_3',\gamma_4')\mapsto(\gamma_3',\gamma_1',\gamma_2',\gamma_4')
\end{aligned}\\
&\bs{R}_2:\begin{aligned}
&(\gamma_1,\gamma_2,\gamma_3,\gamma_4)\mapsto(\gamma_2,\gamma_4,\gamma_3,\gamma_1)\\
&(\gamma_1',\gamma_2',\gamma_3',\gamma_4')\mapsto(\gamma_1',\gamma_3',\gamma_4',\gamma_2')
\end{aligned}\\
&\bs{R}_3:\begin{aligned}
&(\gamma_1,\gamma_2,\gamma_3,\gamma_4)\mapsto(\gamma_2,\gamma_3,\gamma_1,\gamma_4)\\
&(\gamma_1',\gamma_2',\gamma_3',\gamma_4')\mapsto(\gamma_3',\gamma_2',\gamma_4',\gamma_1')
\end{aligned}\\
&\bs{R}_4:\begin{aligned}
&(\gamma_1,\gamma_2,\gamma_3,\gamma_4)\mapsto(\gamma_1,\gamma_3,\gamma_4,\gamma_2)\\
&(\gamma_1',\gamma_2',\gamma_3',\gamma_4')\mapsto(\gamma_2',\gamma_4',\gamma_3',\gamma_1')
\end{aligned}
\end{aligned}
\end{align}
It is straightforwardly to verify that the Hamiltonian with 4 Majorana entanglement pairs elucidated as following is symmetric under $R_i$ and can gap out all dangling Majorana zero modes $\gamma_j$ and $\gamma_j'$ ($j=1,2,3,4$):
\begin{align}
H=i\gamma_1\gamma_3'+i\gamma_2\gamma_4'+i\gamma_3\gamma_1'+i\gamma_4\gamma_2'
\end{align}

Nevertheless, we demonstrate that Majorana chain decorations on both $\tau_1$ and $\tau_2$ can be trivialized: consider the system with a cubic-shaped boundary as illustrated in Fig. \ref{T cell decomposition}. Hence each Majorana chain decorated on the 1D block has an open boundary condition (OBC) and leaves a dangling Majorana zero mode ($\xi_1$) on each vertex of the cubic. Put a Majorana chain with OBC on each hinge of the cubic, this procedure further leaves 3 dangling Majorana zero modes on each vertex of the cubic ($\xi_2$, $\xi_3$ and $\xi_4$). We note that each vertex of the cubic is aligned on a 3-fold rotation axis, hence above 4 Majorana zero modes $\xi_j$ ($j=1,2,3,4$) have the following symmetry properties:
\begin{align}
\bs{R}\in C_3:~\left(\xi_1,\xi_2,\xi_3,\xi_4\right)\mapsto\left(\xi_1,\xi_3,\xi_4,\xi_2\right)
\end{align}
We can introduce the following $C_3$-symmetric mass term to gap out these 4 Majorana zero modes:
\begin{align}
H_m=i\xi_1\xi_2+i\xi_1\xi_3+i\xi_1\xi_4
\end{align}
In addition to the fact that there is a trivial vacuum state on each surface of the cubic, we know that this case indeed corresponds to a trivial bulk because a gapped, short-range entangled symmetric boundary termination is obtained. Equivalently, the case of Majorana chain decorations on 1D blocks $\tau_1$ and $\tau_2$ is trivialized.

The effective ``on-site'' symmetry on the 1D block labeled by $\tau_3$ is $\mathbb{Z}_2$ by 2-fold rotation symmetry acting internally. Hence the candidate 1D block states are classified by group supercohomology (\ref{2-supercohomology}) with twisted cocycle conditions (\ref{2-twisted}) \cite{special,general1,general2}, with $G_{\mathrm{1D}}=\mathbb{Z}_2$. For 1D block $\tau_3$, $\omega_2\in\mathcal{H}^2(\mathbb{Z}_2,\mathbb{Z}_2)=\mathbb{Z}_2$, so we should discuss the spinless and spin-1/2 fermions separately. 

\paragraph{Spinless fermions}For spinless fermions, there are two root phases on the 1D block $\tau_3$:
\begin{enumerate}[1.]
\item Majorana chain;
\item 1D fSPT phase, formed by double Majorana chains; $\mathbb{Z}_2$ symmetry action exchanges two Majorana chains.
\end{enumerate}

\begin{figure}
\begin{tikzpicture}
\tikzstyle{sergio}=[rectangle,draw=none]
\draw[thick,color=green] (0,-1.25) -- (0,0.75);
\draw[thick,color=green] (0,-1.25) -- (0,-3.25);
\draw[thick,color=green] (0,-1.25) -- (-2,-1.25);
\draw[thick,color=green] (0,-1.25) -- (2,-1.25);
\draw[thick,color=green] (0,-1.25) -- (1,-0.5);
\draw[thick,color=green] (0,-1.25) -- (-1,-2);
\filldraw[fill=black, draw=black] (-0.5524,-1.6852)circle (2pt);
\filldraw[fill=black, draw=black] (0.0116,-0.5702)circle (2pt);
\filldraw[fill=black, draw=black] (0.6281,-0.7932)circle (2pt);
\filldraw[fill=black, draw=black] (-0.5904,-1.2594)circle (2pt);
\filldraw[fill=black, draw=black] (0.6003,-1.2494)circle (2pt);
\filldraw[fill=black, draw=black] (0,-2)circle (2pt);
\path (-0.2625,-0.267) node [style=sergio] {$\gamma_1$};
\path (0.3003,-2.2448) node [style=sergio] {$\gamma_1'$};
\path (0.5,-0.5) node [style=sergio] {$\gamma_2$};
\path (-0.5,-2) node [style=sergio] {$\gamma_2'$};
\path (-0.7902,-0.9623) node [style=sergio] {$\gamma_3$};
\path (0.8673,-1.615) node [style=sergio] {$\gamma_3'$};
\end{tikzpicture}
\caption{Majorana chain decoration on 1D blocks $\tau_3$, leaving 6 Majorana zero modes $\gamma_{1,2,3}$ and $\gamma_{1,2,3}'$ near the 0D block $\mu$.}
\label{tau3M}
\end{figure}
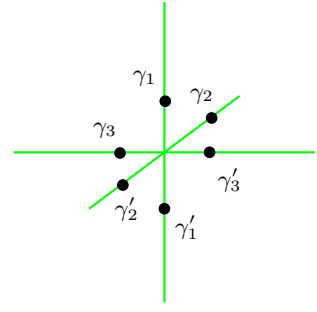

Majorana chain decoration on each 1D block $\tau_3$ leaves 6 dangling Majorana zero modes $\gamma_j$ and $\gamma_j'$ ($j=1,2,3$) near the 0D block $\mu$ (see Fig. \ref{tau3M}), forming 3 complex fermions:
\begin{align}
c_j^\dag=\frac{1}{2}(\gamma_j+i\gamma_j'),~~j=1,2,3.
\end{align}
which span an 8-dimensional Hilbert space. In this Hilbert space, Majorana zero modes $\gamma_{1,2,3}$ and $\gamma_{1,2,3}'$ can be represented as $8\times8$ matrices. Furthermore, $\bs{R}_3^{j}$ can be rephrased as $8\times8$ matrices $A$ in terms of matrix representations of Majorana zero modes:
\begin{align}
\left.
\begin{aligned}
&A(\bs{R}_3^1)=\frac{1}{4}(\gamma_1'-\gamma_2')(\gamma_2'-\gamma_3)(\gamma_1-\gamma_2)(\gamma_2-\gamma_3')\\
&A(\bs{R}_3^2)=\frac{1}{4}(\gamma_1-\gamma_2')(\gamma_2'-\gamma_3)(\gamma_1'-\gamma_2)(\gamma_2-\gamma_3')\\
&A(\bs{R}_3^3)=\frac{1}{4}(\gamma_1-\gamma_2')(\gamma_2'-\gamma_3')(\gamma_1'-\gamma_2)(\gamma_2-\gamma_3)\\
&A(\bs{R}_3^4)=\frac{1}{4}(\gamma_1-\gamma_2)(\gamma_2-\gamma_3)(\gamma_1'-\gamma_2')(\gamma_2'-\gamma_3')
\end{aligned}
\right.
\end{align}
With group product $\bs{R}_3^1\bs{R}_3^2=\left(\bs{R}_3^3\right)^2$. Nevertheless, the representation $A$ of group $T$ satisfies:
\begin{align}
A(\bs{R}_3^1)A(\bs{R}_3^2)=-A\left[\left(\bs{R}_3^3\right)^2\right]
\end{align}
i.e., $A$ is a projective representation of the group $T$. Hence the Majorana zero modes $\gamma_{1,2,3}$ and $\gamma_{1,2,3}'$ cannot be gapped in a $T$-symmetric way, and the Majorana chain decoration on 1D block $\tau_3$ is \textit{obstructed}. 

\begin{figure}
\begin{tikzpicture}
\tikzstyle{sergio}=[rectangle,draw=none]
\draw[thick,color=green] (-0.5,-1) -- (-0.5,1);
\draw[thick,color=green] (0,-2.5) -- (0,-4.5);
\draw[thick,color=green] (-1,-1.5) -- (-3.5,-1.5);
\draw[thick,color=green] (1,-1.5) -- (3.5,-1.5);
\draw[thick,color=green] (1,-1) -- (2.5,0);
\draw[thick,color=green] (-0.5,-2.5) -- (-2,-3.5);
\filldraw[fill=black, draw=black] (-0.5,-1)circle (2pt);
\filldraw[fill=black, draw=black] (1,-1)circle (2pt);
\filldraw[fill=black, draw=black] (1,-1.5)circle (2pt);
\filldraw[fill=black, draw=black] (0,-2.5)circle (2pt);
\filldraw[fill=black, draw=black] (-0.5,-2.5)circle (2pt);
\filldraw[fill=black, draw=black] (-1,-1.5)circle (2pt);
\path (-0.75,-1) node [style=sergio] {$\gamma_1$};
\path (-1.25,-1.25) node [style=sergio] {$\gamma_2$};
\path (-0.5,-2.75) node [style=sergio] {$\gamma_3$};
\path (0.25,-2.75) node [style=sergio] {$\gamma_4$};
\path (1.25,-1.75) node [style=sergio] {$\gamma_5$};
\path (1.5,-1) node [style=sergio] {$\gamma_6$};
\path (-0.25,1.5) node [style=sergio] {$C_2^1$};
\path (-1.5,-3.5) node [style=sergio] {$C_2^2$};
\path (-3,-1.25) node [style=sergio] {$C_2^3$};
\draw[thick,color=green] (0,-1) -- (0,1);
\filldraw[fill=black, draw=black] (0,-1)circle (2pt);
\path (-0.2,-0.95) node [style=sergio] {$\gamma_1'$};
\draw[thick,color=green] (-1,-2) -- (-3.5,-2);
\filldraw[fill=black, draw=black] (-1,-2)circle (2pt);
\path (-1.25,-1.75) node [style=sergio] {$\gamma_2'$};
\draw[thick,color=green] (-1,-2.5) -- (-2.5,-3.5);
\filldraw[fill=black, draw=black] (-1,-2.5)circle (2pt);
\path (-1.5,-2.5) node [style=sergio] {$\gamma_3’$};
\draw[thick,color=green] (0.5,-2.5) -- (0.5,-4.5);
\filldraw[fill=black, draw=black] (0.5,-2.5)circle (2pt);
\path (0.75,-2.75) node [style=sergio] {$\gamma_4'$};
\draw[thick,color=green] (1,-2) -- (3.5,-2);
\filldraw[fill=black, draw=black] (1,-2)circle (2pt);
\path (1.25,-2.25) node [style=sergio] {$\gamma_5'$};
\draw[thick,color=green] (0.5,-1) -- (2,0);
\filldraw[fill=black, draw=black] (0.5,-1)circle (2pt);
\path (0.5,-0.7) node [style=sergio] {$\gamma_6'$};
\end{tikzpicture}
\caption{Double Majorana chains decoration on 1D blocks $\tau_3$, who leaves 12 dangling Majorana zero modes $\xi_1,\cdot\cdot\cdot,\xi_6$ and $\xi_1',\cdot\cdot\cdot,\xi_6'$.}
\label{tau3D}
\end{figure}
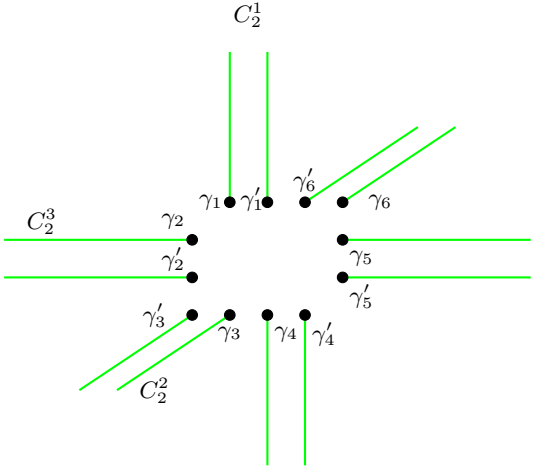

Consider the 1D fSPT phases decoration, leaving 12 dangling Majorana zero modes $\xi_j$ and $\xi_j'$ ($j=1,\cdot\cdot\cdot,6$) at $\mu$, see Fig. \ref{tau3D}. Define 6 complex fermions from these 12 Majorana zero modes:
\begin{align}
c_j^\dag=\frac{1}{2}(\xi_j+i\xi_j'),~~n_j=c_j^\dag c_j
\end{align}

These complex fermions span a 64-dimensional Hilbert space, and aforementioned Majorana zero modes $\xi_j$ and $\xi_j'$ can be represented as $64\times64$ matrices in this Hilbert space. Correspondingly, the symmetry operations $\bs{R}_2^k\in C_2^k$ ($k=1,2,3$, see Fig. \ref{T cell decomposition}) can be formulated in terms of matrix representations of Majorana zero modes $\xi_j$ and $\xi_j'$ as a representation $B$ of the point group $T$, as:
\begin{align}
B(\bs{R}_2^1)&=\frac{1}{8}\left(\xi_1-\xi_1'\right)\left(\xi_2-\xi_5\right)\left(\xi_2'-\xi_5'\right)\nonumber\\
&\cdot\left(\xi_3-\xi_6\right)\left(\xi_3'-\xi_6'\right)\left(\xi_4-\xi_4'\right)
\end{align}
\begin{align}
B(\bs{R}_2^2)&=\frac{1}{8}\left(\xi_2-\xi_2'\right)\left(\xi_1-\xi_4\right)\left(\xi_1'-\xi_4'\right)\nonumber\\
&\cdot\left(\xi_3-\xi_6\right)\left(\xi_3'-\xi_6'\right)\left(\xi_5-\xi_5'\right)
\end{align}
\begin{align}
B(\bs{R}_2^3)&=\frac{1}{8}\left(\xi_3-\xi_3'\right)\left(\xi_2-\xi_5\right)\left(\xi_2'-\xi_5'\right)\nonumber\\
&\cdot\left(\xi_1-\xi_4\right)\left(\xi_1'-\xi_4'\right)\left(\xi_6-\xi_6'\right)
\end{align}
We further consider another group element $\bs{R}_2^1\bs{R}_2^2\in T$. It is straightforward to verify that $B$ is a projective representation of $T$ because of the following relation:
\begin{align}
\begin{aligned}
&B(\bs{R}_2^1)B(\bs{R}_2^2)=-B(\bs{R}_2^1\bs{R}_2^2)\\
&B(\bs{R}_2^1)B(\bs{R}_2^3)=-B(\bs{R}_2^1\bs{R}_2^3)\\
&B(\bs{R}_2^2)B(\bs{R}_2^3)=-B(\bs{R}_2^2\bs{R}_2^3)
\end{aligned}
\end{align}
Hence the Majorana zero modes $\xi_j$ and $\xi_j'$ cannot be gapped in a $T$-symmetric way, and the corresponding 1D block state is \textit{obstructed}. Finally, for spinless fermions, there is no nontrivial 1D block state on $\tau_3$.

\paragraph{Spin-1/2 fermions}For spin-1/2 fermions, there is no root phase because nonzero $n_0$ and $n_1$ in Eq. (\ref{2-supercohomology}) is obstructed \cite{dihedral,wallpaper}, hence there is no nontrivial 1D block state on $\tau_3$. 

\begin{figure*}
\centering
\begin{tikzpicture}[scale=0.9]
\tikzstyle{sergio}=[rectangle,draw=none]
\filldraw[fill=white, draw=black, thick] (-11,-2)--(-8,-2)--(-7,-1)--(-10,-1)--cycle;
\filldraw[fill=white, draw=black, thick] (-11,-5)--(-8,-5)--(-7,-4);
\filldraw[fill=white, draw=black, thick, densely dashed] (-11,-5)--(-10,-4)--(-7,-4);
\filldraw[fill=white, draw=black, thick] (-11,-5)--(-11,-2);
\filldraw[fill=white, draw=black, thick, densely dashed] (-10,-4)--(-10,-1);
\filldraw[fill=white, draw=black, thick] (-8,-5)--(-8,-2);
\filldraw[fill=white, draw=black, thick] (-7,-4)--(-7,-1);
\draw[color=red, thick,->] (-11,-5)--(-9,-4.5);
\draw[color=red, thick] (-7,-4)--(-9,-4.5);
\draw[color=red, thick,->] (-7,-4)--(-7.5,-3);
\draw[color=red, thick] (-8,-2)--(-7.5,-3);
\draw[color=red, thick,->] (-8,-2)--(-9,-1.5);
\draw[color=red, thick] (-10,-1)--(-9,-1.5);
\draw[color=red, thick,->] (-10,-1)--(-10.5,-3);
\draw[color=red, thick] (-11,-5)--(-10.5,-3);
\filldraw[fill=white, draw=black, thick] (-5.5,-2)--(-2.5,-2)--(-1.5,-1)--(-4.5,-1)--cycle;
\filldraw[fill=white, draw=black, thick] (-5.5,-5)--(-2.5,-5)--(-1.5,-4);
\filldraw[fill=white, draw=black, thick, densely dashed] (-5.5,-5)--(-4.5,-4)--(-1.5,-4);
\filldraw[fill=white, draw=black, thick] (-5.5,-5)--(-5.5,-2);
\filldraw[fill=white, draw=black, thick, densely dashed] (-4.5,-4)--(-4.5,-1);
\filldraw[fill=white, draw=black, thick] (-2.5,-5)--(-2.5,-2);
\filldraw[fill=white, draw=black, thick] (-1.5,-4)--(-1.5,-1);
\draw[color=red, thick,->] (-1.5,-4)--(-3.5,-4.5);
\draw[color=red, thick] (-5.5,-5)--(-3.5,-4.5);
\draw[color=red, thick,->] (-4.5,-1)--(-3,-2.5);
\draw[color=red, thick] (-1.5,-4)--(-3,-2.5);
\draw[color=red, thick,->] (-2.5,-2)--(-3.5,-1.5);
\draw[color=red, thick] (-4.5,-1)--(-3.5,-1.5);
\draw[color=red, thick,->] (-5.5,-5)--(-4,-3.5);
\draw[color=red, thick] (-2.5,-2)--(-4,-3.5);
\filldraw[fill=white, draw=black, thick] (0,-2)--(3,-2)--(4,-1)--(1,-1)--cycle;
\filldraw[fill=white, draw=black, thick] (0,-5)--(3,-5)--(4,-4);
\filldraw[fill=white, draw=black, thick, densely dashed] (0,-5)--(1,-4)--(4,-4);
\filldraw[fill=white, draw=black, thick] (0,-5)--(0,-2);
\filldraw[fill=white, draw=black, thick, densely dashed] (1,-4)--(1,-1);
\filldraw[fill=white, draw=black, thick] (3,-5)--(3,-2);
\filldraw[fill=white, draw=black, thick] (4,-4)--(4,-1);
\draw[color=red, thick,->] (0,-5)--(0.5,-3);
\draw[color=red, thick] (0.5,-3)--(1,-1);
\draw[color=red, thick,->] (1,-1)--(2.5,-2.5);
\draw[color=red, thick] (4,-4)--(2.5,-2.5);
\draw[color=red, thick,->] (4,-4)--(3.5,-3);
\draw[color=red, thick] (3,-2)--(3.5,-3);
\draw[color=red, thick,->] (3,-2)--(1.5,-3.5);
\draw[color=red, thick] (0,-5)--(1.5,-3.5);
\end{tikzpicture}
\caption{Majorana chains with PBC surrounding $\mu$ by Majorana bubble construction on 2D blocks $\sigma_2$. The sub-digits $a=1,2,3,4$ depict the Majorana zero modes near the 1D blocks labeled by $a$.}
\label{T Majorana bubble}
\end{figure*}

\subsubsection{0D block states \label{T0D}}
The effective ``on-site'' symmetry on the 0D block $\mu$ is $A_4$, hence the candidate 0D block states are classified by the following two indices:
\begin{align}
\begin{aligned}
&n_0\in\mathcal{H}^0(A_4,\mathbb{Z}_2)=\mathbb{Z}_2\\
&\nu_1\in\mathcal{H}^1\left[A_4,U(1)\right]=\mathbb{Z}_3
\end{aligned}
\label{1-supercohomology T}
\end{align}
with the twisted cocycle condition:
\begin{align}
\mathrm{d}\nu_1=(-1)^{\omega_2\smile n_0}
\label{1-twisted T}
\end{align}
where $n_0$ depicts the parity of fermions, and $\nu_1$ depicts the 0D bSPT mode on the 0D block $\mu$. In particular, we demonstrate the representation of this 0D bSPT mode: we know that there are 3 eigenvalues of $C_3$ symmetry acting on the axis ($0$, $e^{2\pi i/3}$ and $e^{4\pi i/3}$); equivalently, it can be expressed by an index $n\in\mathbb{Z}_3$ ($e^{2\pi ni/3}$). For $T$ symmetry, there are four $C_3$ axes as illustrated in Fig. \ref{T cell decomposition} whose eigenvalues/eigenstates are characterized by four indices $n_j\in\mathbb{Z}_3$ ($j=1,2,3,4$). Therefore, the 0D bSPT modes with $A_4$ symmetry are characterized by the phase:
\[
\exp\left\{\frac{2\pi i}{3}(n_1+n_2+n_3+n_4)\right\}
\]
Equivalently, these 0D bSPT modes are characterized by a $\mathbb{Z}_3$-index defined as following:
\begin{align}
N=n_1+n_2+n_3+n_4~(\mathrm{mod}~3)\in\mathbb{Z}_3
\label{A4 index}
\end{align}
Subsequently, we demonstrate that this $\mathbb{Z}_3$-index can be trivialized by 1D bubble on $\tau_1/\tau_2$. We know the effective ``on-site'' symmetry on each 1D block $\tau_1/\tau_2$ is $\mathbb{Z}_3$, hence we can consider the 1D bubble(s) on $\tau_1/\tau_2$ which is just an interval on the 1D block $\tau_1$ and $\tau_2$ with one ending point residing $\mathbb{Z}_3$ eigenvalues $m_{1}$ and $m_2$ respectively and the other ending point residing $ m_{1}^*$ and $ m_2^*$.  We note that $m_i+m_i^*=0$ mod 3. We can enlarge these bubbles adiabatically and move the states with $m_1^*$ and $m_2^*$ eigenvalues to the infinite boundary and  $m_1$ and $m_2$ to the center, the 0D block $\mu$, in a symmetric way, which leaves the index at $\mu$:
$$4(m_1+m_2)\equiv{m_1+m_2}~(\mathrm{mod}~3)$$
For a specific $N$ [cf. Eq. (\ref{A4 index})], it can be trivialized by the above construction on 1D blocks if we choose $m_1+m_2\equiv N~(\mathrm{mod}~3)$. So, we can trivialize the 0D bSPT mode with $A_4$ symmetry on $\mu$.

Then consider the fermion parity of the 0D block $\mu$ for spinless fermions. In Refs. \cite{dihedral, wallpaper} we have demonstrated that a Majorana chain with periodic boundary condition (PBC) changes the fermion parity of the point it surrounded because the Majorana chain with PBC is not Kastyleyn oriented, and it can be constructed by bubble construction on 2D blocks. We discuss spinless and spin-1/2 fermions separately.


For spinless fermions, consider the Majorana bubble construction on each 2D block $\sigma_1$ that leaves 3 Majorana chains on each 1D block $\tau_2$. Directly discussing the Majorana bubble in bulk is rather complicated, hence we consider an alternative strategy: consider a 3D $T$-symmetric cubic lattice with an open boundary condition, and the Majorana bubbles on all 2D blocks $\sigma_1$ leave 2 Majorana chains along the diagonal line of each surface of the open cubic lattice as illustrated by the red lines in Fig. \ref{T Majorana bubble}, where red arrows label the directions of Majorana chains. On the one hand, if we reconnect all Majorana chains left by the bulk Majorana bubbles on all 2D blocks $\sigma_1$ towards three closed Majorana chains intertwining on the open cubic surface, each of them has an odd fermion parity. Therefore, we can deform the total fermion parity of the open cubic system without changing the topology of the system through Majorana bubble construction, which is equivalent to the fact that the complex fermion decoration on the 0D block $\mu$ is no longer a nontrivial block state, i.e., a trivialization has been established. 

For spin-1/2 fermions, the index $n_0$ in Eq. (\ref{1-supercohomology T}) representing the fermion parity of 0D block state on $\mu$ is \textit{obstructed}, hence there is no nontrivial 0D block state with odd fermion parity on the 0D block $\mu$. 

Finally, there is no nontrivial block state for the 3D system with $T$-symmetry, for both spinless and spin-1/2 fermions, and there is no nontrivial HO topological surface theory.

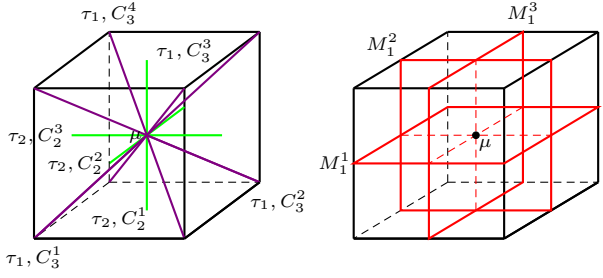
\begin{figure}
\begin{tikzpicture}[scale=0.5]
\tikzstyle{sergio}=[rectangle,draw=none]
\draw[thick] (8,1.5) -- (12,1.5);
\draw[thick] (8,1.5) -- (5.5,0);
\draw[thick] (12,1.5) -- (9.5,0);
\draw[thick] (5.5,0) -- (9.5,0);
\draw[thick] (5.5,0) -- (5.5,-4);
\draw[thick] (9.5,-4) -- (5.5,-4);
\draw[thick] (9.5,-4) -- (9.5,0);
\draw[thick] (12,-2.5) -- (12,1.5);
\draw[thick] (12,-2.5) -- (9.5,-4);
\draw[densely dashed] (8,-2.5) -- (8,1.5);
\draw[densely dashed] (8,-2.5) -- (12,-2.5);
\draw[densely dashed] (8,-2.5) -- (5.5,-4);
\draw[thick] (12,-2.5) -- (9.5,-4);
\draw[thick] (5.5,-4) -- (9.5,-4);
\path (9,-1.5) node [style=sergio] {\scriptsize$\mu$};
\draw[thick] (-1,1.5) -- (3,1.5);
\draw[thick] (-1,1.5) -- (-3,0);
\draw[thick] (3,1.5) -- (1,0);
\draw[thick] (-3,0) -- (1,0);
\draw[thick] (-3,0) -- (-3,-4);
\draw[thick] (1,-4) -- (-3,-4);
\draw[thick] (1,-4) -- (1,0);
\draw[thick] (3,-2.5) -- (3,1.5);
\draw[thick] (3,-2.5) -- (1,-4);
\draw[thick,color=green] (0,-1.25) -- (0,0.75);
\draw[thick,color=green] (0,-1.25) -- (0,-3.25);
\draw[thick,color=green] (0,-1.25) -- (-2,-1.25);
\draw[thick,color=green] (0,-1.25) -- (2,-1.25);
\draw[thick,color=green] (0,-1.25) -- (1,-0.5);
\draw[thick,color=green] (0,-1.25) -- (-1,-2);
\draw[densely dashed] (-1,-2.5) -- (-1,1.5);
\draw[densely dashed] (-1,-2.5) -- (3,-2.5);
\draw[densely dashed] (-1,-2.5) -- (-3,-4);
\draw[thick,color=violet] (3,1.5) -- (-3,-4);
\draw[thick,color=violet] (-1,1.5) -- (0,-1.25);
\draw[thick,color=violet] (-3,0) -- (3,-2.5);
\draw[thick,color=violet] (0,-1.25) -- (1,0);
\path (-3,-4.5) node [style=sergio] {\scriptsize$\tau_1,C_3^1$};
\path (3.5,-3) node [style=sergio] {\scriptsize$\tau_1,C_3^2$};
\path (1,1) node [style=sergio] {\scriptsize$\tau_1,C_3^3$};
\path (-1,2) node [style=sergio] {\scriptsize$\tau_1,C_3^4$};
\draw[thick,color=violet] (0,-1.25) -- (-3,-4);
\draw[thick,color=violet] (0,-1.25) -- (-1,-2.5);
\draw[thick,color=violet] (0,-1.25) -- (3,-2.5);
\draw[thick,color=violet] (0,-1.25) -- (1,-4);
\draw[thick] (3,-2.5) -- (1,-4);
\draw[thick] (-3,-4) -- (1,-4);
\path (-0.7269,-3.4699) node [style=sergio] {\scriptsize$\tau_2,C_2^1$};
\path (-2.9444,-1.257) node [style=sergio] {\scriptsize$\tau_2,C_2^3$};
\path (-1.8912,-1.9911) node [style=sergio] {\scriptsize$\tau_2,C_2^2$};
\path (-0.3,-1.3) node [style=sergio] {\scriptsize$\mu$};
\path (5,-2) node [style=sergio] {\scriptsize$M_1^1$};
\path (6.2647,1.1629) node [style=sergio] {\scriptsize$M_1^2$};
\path (10,2) node [style=sergio] {\scriptsize$M_1^3$};
\draw[draw=red, thick] (7.5,0)--(10,1.5)--(10,-2.5)--(7.5,-4)--cycle;
\draw[draw=red, thick] (9.5,-2)--(12,-0.5)--(8,-0.5)--(5.5,-2)--cycle;
\draw[draw=red, thick] (10.75,-3.25)--(10.75,0.75)--(6.75,0.75)--(6.75,-3.25)--cycle;
\draw[densely dashed,color=red] (6.75,-1.25) -- (10.75,-1.25);
\draw[densely dashed,color=red] (8.75,-3.25) -- (8.75,0.75);
\draw[densely dashed,color=red] (10,-0.5) -- (7.5,-2);
\filldraw[fill=black, draw=black] (8.75,-1.25)circle (2.5pt);
\end{tikzpicture}
\caption{The cell decomposition of the 3D system with point group symmetry $T_h$. There are four axes of 3-fold rotation symmetry labeled by $C_3^{1,2,3,4}$ and depicted by solid violet lines, across the center of the system as indicated by the solid dot (labeled by $\mu$); three axes of 2-fold rotation symmetry across the center, labeled by $C_2^{1,2,3}$ and depicted by solid green lines (see left panel); and three planes of reflection symmetry, labeled by $\bs{M}_1^{1,2,3}$. We plot cubic (black solid lines) for visual convenience.}
\label{Th cell decomposition}
\end{figure}

\subsection{$T_h$-symmetric lattice}
For $T_h$-symmetric cubic, by cell decomposition as illustrated in Figs. \ref{Th cell decomposition} and \ref{Th unit cell}, the ground-state wavefunction of the system can be decomposed to the direct products of wavefunctions of lower-dimensional blocks as:
\begin{align}
|\Psi\rangle=\bigotimes_{g\in T_h}|T_{g\lambda}\rangle\otimes\bigotimes\limits_{k=1}^3|\gamma_{g\sigma_k}\rangle\otimes\bigotimes\limits_{l=1}^3|\beta_{g\tau_l}\rangle\otimes|\alpha_\mu\rangle
\label{Th wavefunction}
\end{align}
where $|T_{g\lambda}\rangle$ is the wavefunction of 3D block state labeled by $g\lambda$ which is topologically trivial; $|\gamma_{g\sigma_1}\rangle$ is the wavefunction of 2D block state labeled by $g\sigma_1$, and $|\gamma_{g\sigma_2}\rangle/|\gamma_{g\sigma_3}\rangle$ is the wavefunction of 2D block state labeled by $g\sigma_2/g\sigma_3$ which is $\mathbb{Z}_2$-symmetric; $|\beta_{g\tau_1}\rangle$ is the wavefunction of 1D block state labeled by $\tau_1$ which is $\mathbb{Z}_3$-symmetric, and $|\beta_{g\tau_2}\rangle$ is the wavefunction of 1D block state labeled by $\tau_2$ which is $(\mathbb{Z}_2\times\mathbb{Z}_2)$-symmetric, and $|\beta_{g\tau_3}\rangle$ is the wavefunction of 1D block state labeled by $\tau_3$ which is $\mathbb{Z}_2$-symmetric; $|\alpha_\mu\rangle$ is the wavefunction of the 0D block state labeled by $\mu$ which is $(A_4\times\mathbb{Z}_2)$-symmetric.

\begin{figure}
\begin{tikzpicture}[scale=0.8]
\tikzstyle{sergio}=[rectangle,draw=none]
\draw[thick] (-3,-0.5) -- (1,-0.5);
\draw[densely dashed,thick] (-1,1) -- (3,1);
\draw[densely dashed,thick] (-1,1) -- (-3,-0.5);
\draw[thick] (3,1) -- (1,-0.5);
\draw[thick,color=violet] (-1,1) -- (0,2.25);
\draw[thick,color=violet] (1,-0.5) -- (0,2.25);
\draw[thick,color=violet] (-3,-0.5) -- (0,2.25);
\draw[thick,color=violet] (3,1) -- (0,2.25);
\draw[thick,color=green] (0,0.25) -- (0,2.25);
\filldraw[fill=black, draw=black] (0,2.25)circle (3pt);
\draw[densely dashed,color=red] (-2,0.25) -- (2,0.25);
\draw[densely dashed,color=red] (1,1) -- (-1,-0.5);
\draw[thick,color=red] (1,1) -- (0,2.25);
\draw[thick,color=red] (-1,-0.5) -- (0,2.25);
\draw[thick,color=red] (-2,0.25) -- (0,2.25);
\draw[thick,color=red] (2,0.25) -- (0,2.25);
\draw[thick] (4.25,-0.5) -- (6.25,-0.5);
\draw[densely dashed,color=red] (7.25,0.25) -- (6.25,-0.5);
\draw[densely dashed,color=red] (5.25,0.25) -- (4.25,-0.5);
\draw[densely dashed,color=red] (5.25,0.25) -- (7.25,0.25);
\draw[thick,color=green] (5.25,0.25) -- (5.25,2.25);
\draw[thick,color=red] (4.25,-0.5) -- (5.25,2.25);
\draw[thick,color=red] (7.25,0.25) -- (5.25,2.25);
\draw[thick,color=violet] (6.25,-0.5) -- (5.25,2.25);
\draw[thick,->] (4.9458,0.6294) -- (4.25,1.25);
\draw[thick,->] (5.9324,1.2136) -- (6.6459,1.5102);
\path (4,1.5) node [style=sergio] {$\sigma_2$};
\path (6.75,1.75) node [style=sergio] {$\sigma_3$};
\path (5.5032,0.5467) node [style=sergio] {$\sigma_1$};
\path (6.2984,0.5834) node [style=sergio] {$\sigma_1$};
\end{tikzpicture}
\caption{The unit cell of cell decomposition of the cubic lattice with point symmetry $T_h$. The left panel depicts the bottom rectangular pyramid of the cubic in Fig. \ref{Th cell decomposition}; the right panel illustrates the independent triangular pyramid $\lambda$, where $\sigma_1$ and $\sigma_2$ are two independent 2D blocks in the system. The red solid line in the right two panels are $\tau_3$. }
\label{Th unit cell}
\end{figure}

With the topological crystals, we consider the lower-dimensional block states and investigate the possible \textit{obstructions} and \textit{trivializations}.

\subsubsection{2D block states\label{Th2D}}
There is no effective ``on-site'' symmetry on all 2D blocks labeled by $\sigma_1$. The only possible root phase on 2D blocks is 2D $(p+ip)$-SC. If we decorate a 2D $(p+ip)$-SC with quantum number $n_1\in\mathbb{Z}$ on each 2D block $\sigma_1$, the chiral central charge on the 1D blocks labeled by $\tau_1$ is $3n_1/2$. Hence all nontrivial $(p+ip)$-SC decoration on 2D blocks is \textit{obstructed}.

Subsequently, we investigate the block state decorations on 2D blocks labeled by $\sigma_2$ and $\sigma_3$. The on-site symmetry of $\sigma_2$ is $\mathbb{Z}_2$ by reflection symmetry acting internally. First of all, we demonstrate that the block-state decoration on $\sigma_2$ and $\sigma_3$ should be identical: the shared border of $\sigma_2$ and $\sigma_3$ is $\tau_3$, with a $\Z_2$ symmetry (no enhanced symmetry). Therefore, if the decorated states on $\sigma_2$ and $\sigma_3$ are different, their shared border cannot be gapped out. For spinless fermions, there are two possible root phases of 2D block states on $\sigma_{2,3}$:
\begin{enumerate}[1.]
\item 2D $(p+ip)$-SC with $n_2\in\mathbb{Z}$ index;
\item 2D fermionic Levin-Gu state \cite{Gu-Levin} with $\nu_2\in\mathbb{Z}_8$ index.
\end{enumerate}
We note that for crystalline TSC, it is not necessary to treat Kitaev's $E_8$ state as an independent 2D block state, because Kitaev's $E_8$ state is equivalent to 16 layers of 2D $(p+ip)$-SC \cite{Cano_2014, You_2015}. We prove this issue in Supplementary Materials \cite{supplementary}. 

\begin{figure}
\begin{tikzpicture}
\tikzstyle{sergio}=[rectangle,draw=none]
\draw[draw=red, thick] (2,2)--(3,1)--(3,3)--(2,4)--cycle;
\draw[draw=red, thick] (0.5,3.5)--(1.5,2.5)--(1.5,4.5)--(0.5,5.5)--cycle;
\draw[draw=red, thick] (2.1519,2.4868)--(3.1314,3.4798)--(3.1314,5.4798)--(2.1519,4.4868)--cycle;
\draw[draw=red, thick] (0.3813,0.9911)--(1.3931,1.9355)--(1.3931,3.9355)--(0.3813,2.9911)--cycle;
\draw[densely dashed,color=red] (1.7845,4.6687) -- (1.7845,1.6687);
\path (0.5651,1.4663) node [style=sergio] {$1$};
\path (0.6965,4.9738) node [style=sergio] {$2$};
\path (2.9059,4.953) node [style=sergio] {$3$};
\path (2.7485,1.6685) node [style=sergio] {$4$};
\path (0.8877,2.3861) node [style=sergio] {$\sigma_2$};
\path (0.9281,4.1247) node [style=sergio] {$\sigma_3$};
\path (2.6263,4.1045) node [style=sergio] {$\sigma_2$};
\path (2.4763,2.3471) node [style=sergio] {$\sigma_3$};
\draw[->,ultra thick] (1.3931,2.32)--(1.3931,3.3293);
\draw[->,ultra thick] (2.1519,2.9908)--(2.1519,4);
\draw[->,ultra thick] (1.5,4)--(1.5,3);
\draw[->,ultra thick] (2,3.381)--(2,2.381);
\end{tikzpicture}
\caption{Chiralities of decorated 2D $(p+ip)$-SC on $\sigma_2$ near the 1D block $\tau_2$, depicted by black arrows.}
\label{edge theory}
\end{figure}
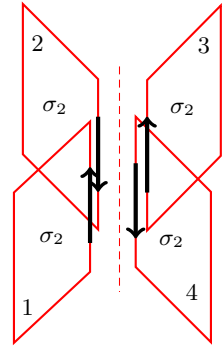

Consider the 2D $(p+ip)$-SC decorations: the chiralities of decorated phases are illustrated in Fig. \ref{edge theory} which indicate that the gapless modes left by these 2D block states should be nonchiral, and described by Eq. (\ref{Luttinger}). For monolayer $(p+ip)$-SC, there will be 4 chiral Majorana modes near the 1D block $\tau_2$ as their shared border, two of them $(\eta_1^\uparrow,\eta_2^\uparrow)$ move upward, and others $(\eta_1^\downarrow,\eta_2^\downarrow)$ move downward. Under reflection generators of $\mathbb{Z}_2\times\mathbb{Z}_2$, these chiral Majorana modes transform as:
\begin{align}
\begin{aligned}
&\bs{M}_1:~\left(\eta_1^\uparrow,\eta_1^\downarrow,\eta_2^\uparrow,\eta_2^\downarrow\right)\mapsto\left(\eta_2^\uparrow,\eta_1^\downarrow,\eta_1^\uparrow,\eta_2^\downarrow\right)\\
&\bs{M}_2:~\left(\eta_1^\uparrow,\eta_1^\downarrow,\eta_2^\uparrow,\eta_2^\downarrow\right)\mapsto\left(\eta_1^\uparrow,\eta_2^\downarrow,\eta_2^\uparrow,\eta_1^\downarrow\right)
\end{aligned}
\end{align}

We bosonize these chiral Majorana modes as:
\begin{align}
e^{i\phi_\uparrow}=\eta_1^\uparrow+i\eta_2^\uparrow,~~~e^{i\phi_\downarrow}=\eta_1^\downarrow+i\eta_2^\downarrow
\end{align}
subsequently, the chiral edge modes can be formulated by Eq. (\ref{Luttinger}) with $K=\sigma^z$. Under $\bs{M}_1$ and $\bs{M}_2$, the edge field $\Phi=(\phi_\uparrow,\phi_\downarrow)^T$ transforms as Eq. (\ref{K-matrix symmetry}), with
\begin{align}
\left\{
\begin{aligned}
&W^{\bs{M}_1}=-\sigma^z\\
&W^{\bs{M}_2}=\sigma^z
\end{aligned}
\right.,~~\left\{
\begin{aligned}
&\delta\Phi^{\bs{M}_1}=0\\
&\delta\Phi^{\bs{M}_2}=0
\end{aligned}
\right.
\end{align}
where we have shift the $\pi/2$ phase to zero by gauge transformation.
To investigate whether these modes can be fully gapped symmetrically, we calculate the anomaly indicators (\ref{anomaly indicator}) of two reflection generators $\bs{M}_1$ and $\bs{M}_2$. Nevertheless, both of them do not vanish,
\begin{align}
\begin{aligned}
&\nu_{\bs{M}_1}=1/4~(\mathrm{mod}~2)\\
&\nu_{\bs{M}_2}=-1/4~(\mathrm{mod}~2)
\end{aligned}
\end{align}
Hence, it is impossible to gap the above modes in a $\bs{M}_1$ and $\bs{M}_2$ symmetric way.

\begin{figure}
\begin{tikzpicture}[scale=0.9]
\tikzstyle{sergio}=[rectangle,draw=none]
\filldraw[fill=red!20,thick,draw=red] (2,-2.5) -- (1,-4) -- (1,-1) -- (2,0.5) -- (2,-2.5);
\draw[ultra thick,->,draw=red] (1,-1) -- (1.5,-0.25);
\draw[ultra thick,->,draw=red] (2,0.5) -- (2,-1);
\draw[ultra thick,->,draw=red] (2,-2.5) -- (1.5,-3.25);
\draw[ultra thick,->,draw=red] (1,-4) -- (1,-2.5);
\filldraw[fill=red!20,thick,draw=red] (-0.5,-2.5) -- (-1.5,-4) -- (-1.5,-1) -- (-0.5,0.5) -- (-0.5,-2.5);
\draw[ultra thick,->,draw=red] (-1.5,-1) -- (-1,-0.25);
\draw[ultra thick,->,draw=red] (-0.5,0.5) -- (-0.5,-1);
\draw[ultra thick,->,draw=red] (-0.5,-2.5) -- (-1,-3.25);
\draw[ultra thick,->,draw=red] (-1.5,-4) -- (-1.5,-2.5);
\draw[densely dashed] (0.75,-2.5) -- (-0.25,-4) -- (-0.25,-1) -- (0.75,0.5) -- (0.75,-2.5);
\path (0.3004,-1.84) node [style=sergio] {$M_1$};
\path (1.5,-1.6) node [style=sergio] {$p+ip$};
\path (1.5,-2) node [style=sergio] {$c=+$};
\path (-1,-1.6) node [style=sergio] {$p+ip$};
\path (-1,-2) node [style=sergio] {$c=+$};
\filldraw[fill=blue!20,thick,draw=blue] (3.5,-2.5) -- (2.5,-4) -- (2.5,-1) -- (3.5,0.5) -- (3.5,-2.5);
\draw[ultra thick,->,draw=blue] (3.5,0.5) -- (3,-0.25);
\draw[ultra thick,->,draw=blue] (2.5,-1) -- (2.5,-2.5);
\draw[ultra thick,->,draw=blue] (2.5,-4) -- (3,-3.25);
\draw[ultra thick,->,draw=blue] (3.5,-2.5) -- (3.5,-1);
\path (3,-1.6) node [style=sergio] {$p-ip$};
\path (3,-2) node [style=sergio] {$c=-$};
\filldraw[fill=blue!20,thick,draw=blue] (-2,-2.5) -- (-3,-4) -- (-3,-1) -- (-2,0.5) -- (-2,-2.5);
\draw[ultra thick,->,draw=blue] (-2,0.5) -- (-2.5,-0.25);
\draw[ultra thick,->,draw=blue] (-3,-1) -- (-3,-2.5);
\draw[ultra thick,->,draw=blue] (-3,-4) -- (-2.5,-3.25);
\draw[ultra thick,->,draw=blue] (-2,-2.5) -- (-2,-1);
\path (-2.5,-1.6) node [style=sergio] {$p-ip$};
\path (-2.5,-2) node [style=sergio] {$c=-$};
\path (-1,0.25) node [style=sergio] {$c_1^+$};
\path (1.5,0.25) node [style=sergio] {$c_2^+$};
\path (-2.8,0.25) node [style=sergio] {$d_L(d^+)$};
\path (2.7,0.25) node [style=sergio] {$d_R(d^-)$};
\end{tikzpicture}
\caption{2D block state deformed from 3D ``$p+ip$/Chern insulator'' bubble construction. The dashed red plate depicts the reflection plane and black arrowed solid plates represent the 2D $(p+ip)$-SCs/Chern insulators with their chiralities ($c$ depicts the Chern number of Chern insulator).}
\label{3D bubble}
\end{figure}
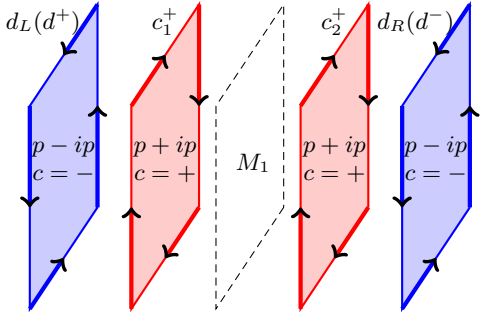

For bilayer $(p+ip)$-SCs, we demonstrate that near each $\sigma_2$, bilayer $(p+ip)$-SCs are equivalent to a 2D fermionic Levin-Gu state with quantum number $\nu=1$ ($\nu\in\mathbb{Z}_8$): As illustrated in Fig. \ref{3D bubble}, each bilayer $(p+ip)$-SCs leaves two chiral Majorana modes on its 1D border, with trivial $\mathbb{Z}_2$ symmetry action: $c_{1,2}^+\mapsto c_{1,2}^+$. Consider two 3D ``$(p-ip)$-SC'' bubbles sharing the 2D block $\sigma_2$ as their borders, leaving two $(p-ip)$-SCs at $\sigma_2$. At the 1D border of $\sigma_2$, 3D ``$(p-ip)$-SC'' bubbles leave another two chiral Majorana modes $d_{L,R}$ with opposite chirality, with $\mathbb{Z}_2$ symmetry acting as reflection: $d_L\leftrightarrow d_R$. Redefine these chiral Majorana modes:
\begin{align}
d^+=\frac{1}{2}\left(d_L+d_R\right),~~d^-=\frac{1}{2}\left(d_L-d_R\right)
\label{redefine}
\end{align}
With an alternative $\mathbb{Z}_2$ symmetry property: $d^{\pm}\mapsto\pm d^{\pm}$. Now we can assemble a $(p+ip)$-SC with a $(p-ip)$-SC, corresponding to the chiral Majorana modes labeled by $c_1^+$ and $d^+$, these chiral Majorana modes can be gapped in a $\mathbb{Z}_2$-symmetric way, and the corresponding assembly should be trivial; then another two layers with chiral Majorana modes $c_2^+$ and $d^-$ form a 2D fermionic Levin-Gu state with $\nu=1$. As a consequence, we can sort out the bilayer $(p+ip)$-SCs to Levin-Gu state decorations for the 2D block as reflection plane.

Subsequently, consider the decorations of 2D Levin-Gu states \cite{Gu-Levin} with a modulo 8 index $\nu$. The two nearby $\sigma_2$ and $\sigma_3$ in Fig. \ref{Th unit cell} (the right panel) should be decorated by the same $\nu$ state otherwise there is gapless along the red solid line. Then referring to Fig. \ref{edge theory}, if the shared index of 2D block labeled by 1 and 4 (corresponding to the two nearby $\sigma_2$ and $\sigma_3$ in Fig. \ref{Th unit cell}) is $\nu$, the shared index of 2D block labeled by 2 and 3  be $8-\nu$ under reflection symmetry and also be $\nu$ under 2-fold rotation, which requires that $8-\nu=\nu$ mod 8. Therefore, we only need to consider $\nu=4$ since other $\nu$ are incompatible with symmetry requirement.

For the $\nu=4$ case, there is a bosonic Levin-Gu state \cite{LevinGu} on each 2D block. Hence the 1D gapless mode at each 1D block $\tau_2$ has the Lagrangian (\ref{Luttinger}), with $K=(\sigma^x)^{\oplus4}$ and $\Phi=(\phi_1,\cdot\cdot\cdot,\phi_8)^T$. Under $\bs{R}$ and $\bs{M}$ as two generators of $\mathbb{Z}_2\times\mathbb{Z}_2$, $\Phi$ will be transformed as Eq. (\ref{K-matrix symmetry}), where
\begin{align}
W^{\bs{R}}=\left(
\begin{array}{cccc}
0 & 0 & 1 & 0\\
0 & 0 & 0 & 1\\
1 & 0 & 0 & 0\\
0 & 1 & 0 & 0\\
\end{array}
\right)\otimes\mathbbm{1}_{2\times2},~~\delta\phi^{\bs{R}}=0
\end{align}
\begin{align}
W^{\bs{M}}=\left(
\begin{array}{cccc}
0 & 0 & 1 & 0\\
0 & 1 & 0 & 0\\
1 & 0 & 0 & 0\\
0 & 0 & 0 & 1\\
\end{array}
\right)\otimes\mathbbm{1}_{2\times2},~~\delta\Phi^{\bs{M}}=\pi\chi
\end{align}
and $\chi=(1,1,1,1,1,1,1,1)^T$. The anomaly indicators (\ref{anomaly indicator}) of two reflection generators $\bs{M}_1=\bs{M}$ and $\bs{M}_2=\bs{M}\bs{R}$ vanish as $\nu_{\bs{M}_1}=\nu_{\bs{M}_2}=\nu_{\bs{R}}=0~(\mathrm{mod}~2)$, hence the corresponding 1D modes can definitely be gapped out symmetrically. We now introduce backscattering terms like Eq. (\ref{backscattering}) that gap out the edge without breaking the $\mathbb{Z}_2\times\mathbb{Z}_2$ symmetry, either explicitly or spontaneously. In order for $U(\Lambda_j)$ to be invariant under $\bs{M}$, we require that:
\begin{align}
\Lambda_j^T\chi\equiv0~(\mathrm{mod}~2)\Leftrightarrow\sum\limits_{k=1}^8\Lambda_j(k)\equiv0~(\mathrm{mod}~2)
\end{align}
Terms $U(\Lambda_j)$ can gap out the edge as long as the vectors $\{\Lambda_j\}$ satisfy the ``null-vector'' conditions (\ref{null-vector}). There are four linear independent solutions to this problem:
\begin{align}
\begin{aligned}
&\Lambda_1=\left(1,0,0,0,1,0,0,0\right)^T\\
&\Lambda_2=\left(0,1,0,0,0,-1,0,0\right)^T\\
&\Lambda_3=\left(0,0,1,0,0,0,1,0\right)^T\\
&\Lambda_4=\left(0,0,0,1,0,0,0,-1\right)^T
\end{aligned}
\label{Th backscattering}
\end{align}
and these 4 vectors correspond to 4 independent backscattering terms that can fully gap out the non-chiral Luttinger liquids near each 1D block $\tau_2$ without symmetry breaking according to the primitivity criteria in Sec.\ref{sec:blockdecoration}. As the consequence, the 2D bosonic Levin-Gu state decoration on $\sigma_2$ corresponds to an \textit{obstruction-free} block state.

Furthermore, we should consider if this 2D block state can be trivialized by 3D bubble construction. The only possible 3D bubble is formed by an enclosed 2D $(p+ip)$-SC, and we have discussed the effects of this bubble in Fig. \ref{3D bubble} that leads to the fact for spinless fermions, 2D block state from bilayer $(p+ip)$-SCs is equivalent to the Levin-Gu state decoration on each $\sigma_2$ and $\sigma_3$ that is \textit{trivialization-free}, and it is a nontrivial block state.

For spin-1/2 fermions, all 2D root phases on $\sigma_{1,2,3}$ are \textit{obstructed}. As the consequence, there is no nontrivial 2D block state for spin-1/2 fermions. 

\subsubsection{1D block states\label{Th1D}}
The effective ``on-site'' symmetry on the 1D block labeled by $\tau_{1}$ is $\mathbb{Z}_3$ by 3-fold rotation symmetry acting internally. Hence the candidate 1D block states are classified by group supercohomology (\ref{2-supercohomology}) with twisted cocycle conditions (\ref{2-twisted}) \cite{special,general1,general2}, with $G_{\mathrm{1D}}=\mathbb{Z}_3$ (and the spin of fermions is irrelevant). 

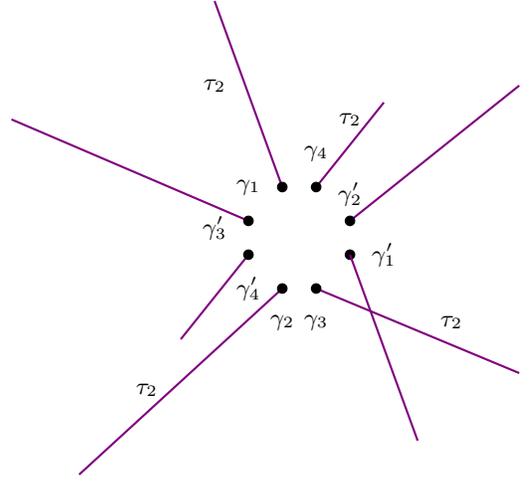
\begin{figure}
\begin{tikzpicture}[scale=0.9]
\tikzstyle{sergio}=[rectangle,draw=none]
\path (-1,0.5) node [style=sergio] {$\tau_2$};
\path (-2,-4) node [style=sergio] {$\tau_2$};
\path (1,0) node [style=sergio] {$\tau_2$};
\path (2.5,-3) node [style=sergio] {$\tau_2$};
\draw[thick,color=violet] (-1,1.75) -- (0,-1);
\draw[thick,color=violet] (0,-2.5) -- (-3,-5.25);
\draw[thick,color=violet] (0.5,-1) -- (1.5,0.25);
\draw[thick,color=violet] (0.5,-2.5) -- (3.5,-3.75);
\filldraw[fill=black, draw=black] (0,-1)circle (2pt);
\filldraw[fill=black, draw=black] (0.5,-1)circle (2pt);
\filldraw[fill=black, draw=black] (0,-2.5)circle (2pt);
\filldraw[fill=black, draw=black] (0.5,-2.5)circle (2pt);
\path (-0.5,-1) node [style=sergio] {$\gamma_1$};
\path (0,-3) node [style=sergio] {$\gamma_2$};
\path (0.5,-3) node [style=sergio] {$\gamma_3$};
\path (0.5,-0.5) node [style=sergio] {$\gamma_4$};
\draw[thick,color=violet] (-4,0) -- (-0.5,-1.5);
\path (-1,-1.6) node [style=sergio] {$\gamma_3'$};
\filldraw[fill=black, draw=black] (-0.5,-1.5)circle (2pt);
\draw[thick,color=violet] (-0.5,-2) -- (-1.5,-3.25);
\path (-0.5,-2.5) node [style=sergio] {$\gamma_4'$};
\filldraw[fill=black, draw=black] (-0.5,-2)circle (2pt);
\draw[thick,color=violet] (3.5,0.5) -- (1,-1.5);
\path (1.5,-2) node [style=sergio] {$\gamma_1'$};
\filldraw[fill=black, draw=black] (1,-2)circle (2pt);
\draw[thick,color=violet] (1,-2) -- (2,-4.75);
\path (1,-1.1) node [style=sergio] {$\gamma_2'$};
\filldraw[fill=black, draw=black] (1,-1.5)circle (2pt);
\end{tikzpicture}
\caption{Majorana chain decorations on 1D blocks labeled by $\tau_1$, which leaves 8 dangling Majorana zero modes $\gamma_j$ and $\gamma_j'$ ($j=1,2,3,4$) at the 0D block $\mu$.}
\label{tau1Th}
\end{figure}

Consider the Majorana chain decorations on 1D blocks labeled by $\tau_1$, as indicated in Fig. \ref{tau1Th} that leaves 8 dangling Majorana zero modes $\gamma_j$ and $\gamma_j'$ ($j=1,2,3,4$). The symmetry properties of these Majorana zero modes under 3-fold rotation are:
\begin{align}
\bs{R}_3:\left\{
\begin{aligned}
&\left(\gamma_1,\gamma_2,\gamma_3,\gamma_4\right)\mapsto\left(\gamma_1,\gamma_3,\gamma_4,\gamma_2\right)\\
&\left(\gamma_1',\gamma_2',\gamma_3',\gamma_4'\right)\mapsto\left(\gamma_1',\gamma_3',\gamma_4',\gamma_2'\right)
\end{aligned}
\right.
\label{R3Th}
\end{align}
Under 2-fold rotation $\bs{R}_2\in T_h$, these Majorana zero modes are transformed as:
\begin{align}
\bs{R}_2:\left\{
\begin{aligned}
&\left(\gamma_1,\gamma_2,\gamma_3,\gamma_4\right)\mapsto\left(\gamma_4,\gamma_3,\gamma_2,\gamma_1\right)\\
&\left(\gamma_1',\gamma_2',\gamma_3',\gamma_4'\right)\mapsto\left(\gamma_4',\gamma_3',\gamma_2',\gamma_1'\right)
\end{aligned}
\right.
\label{R2}
\end{align}
Under reflection $\bs{M}_1\in T_h$, these Majorana zero modes are transformed as:
\begin{align}
\bs{M}_1:~\left\{
\begin{aligned}
&\left(\gamma_1,\gamma_2,\gamma_3,\gamma_4\right)\mapsto\left(\gamma_2',\gamma_1',\gamma_4',\gamma_3'\right)\\
&\left(\gamma_1',\gamma_2',\gamma_3',\gamma_4'\right)\mapsto\left(\gamma_2,\gamma_1,\gamma_4,\gamma_3\right)
\end{aligned}
\right.
\label{M1}
\end{align}

Firstly we introduce an interacting Hamiltonian, including two 4-fermion interacting terms:
\begin{align}
H_U=U\left(\gamma_1\gamma_2\gamma_3\gamma_4+\gamma_1'\gamma_2'\gamma_3'\gamma_4'\right),~U>0
\end{align}
It is easy to verify that $H_U$ is symmetric under symmetry operations (\ref{R3Th})-(\ref{M1}). This Hamiltonian can open a Hubbard gap with 4-fold GSD that is characterized by:
\begin{align}
\gamma_1\gamma_2\gamma_3\gamma_4=\gamma_1'\gamma_2'\gamma_3'\gamma_4'=-1
\label{restrict Th}
\end{align}
To investigate whether this GSD can be lifted, we restrict the Hilbert space to the subspace constrained by relations (\ref{restrict Th}). In this subspace, we can further define two spin-1/2 degrees of freedom:
\begin{align}
S_x=\frac{i}{2}\gamma_1\gamma_2,~S_y=\frac{i}{2}\gamma_1\gamma_3,~S_z=\frac{i}{2}\gamma_1\gamma_4
\end{align}
\begin{align}
S_x'=\frac{i}{2}\gamma_1'\gamma_2',~S_y'=\frac{i}{2}\gamma_1'\gamma_3',~S_z'=\frac{i}{2}\gamma_1'\gamma_4'
\end{align}
and they satisfy the commutation relations of spin-1/2 degrees of freedom:
\begin{align}
\left[S_\mu,S_\nu\right]=i\epsilon_{\mu\nu\rho}S_\rho,~\left[S_\mu',S_\nu'\right]=i\epsilon_{\mu\nu\rho}S_\rho'
\end{align}
where $\epsilon_{\mu\nu\rho}$ is the Levi-Civita symbol and $\mu,\nu,\rho=x,y,z$. Furthermore, their symmetry properties are:
\begin{align}
\bs{R}_3:~\left\{
\begin{aligned}
&\left(S_x,S_y,S_z\right)\mapsto\left(S_y,S_z,S_x\right)\\
&\left(S_x',S_y',S_z'\right)\mapsto\left(S_y',S_z',S_x'\right)
\end{aligned}
\right.
\end{align}
\begin{align}
\bs{R}_2:~\left\{
\begin{aligned}
&\left(S_x,S_y,S_z\right)\mapsto-\left(S_x,S_y,S_z\right)\\
&\left(S_x',S_y',S_z'\right)\mapsto-\left(S_x',S_y',S_z'\right)
\end{aligned}
\right.
\end{align}
\begin{align}
\bs{M}_1:~\left\{
\begin{aligned}
&\left(S_x,S_y,S_z\right)\mapsto\left(-S_x',S_y',-S_z'\right)\\
&\left(S_x',S_y',S_z'\right)\mapsto\left(-S_x,S_y,-S_z\right)
\end{aligned}
\right.
\end{align}
Therefore, we can further add a Hamiltonian between these two spin-1/2 degrees of freedom:
\begin{align}
H_J=J\bs{S}\cdot\bs{S}',~~J>0
\end{align}
and $H_J$ splits the 4-fold degenerate ground states to a nondegenerate spin-singlet state with lower energy and 3-fold degenerate spin-triplet states. As the consequence, the dangling Majorana zero modes $\gamma_j$ and $\gamma_j'$ ($j=1,2,3,4$) can be fully gapped by Hamiltonian $H_U$ and $H_J$, and the Majorana chain decoration on 1D blocks labeled by $\tau_1$ is an obstruction-free block state. 

Subsequently, we consider the block state decoration on the 1D blocks labeled by $\tau_2$. The effective ``on-site'' symmetry on the 1D block labeled by $\tau_2$ is $\mathbb{Z}_2\times\mathbb{Z}_2$ by 2-fold rotation symmetry $C_2$ and reflection $\bs{M}_1$ acting internally. Hence the candidate 1D block states are classified by group supercohomology (\ref{2-supercohomology}) with twisted cocycle conditions (\ref{2-twisted}) \cite{special,general1,general2}, where $G_{\mathrm{1D}}=\mathbb{Z}_2\times\mathbb{Z}_2$. We discuss the spinless and spin-1/2 fermions separately. 

\paragraph{Spinless fermions}For spinless fermions, there are several possible invertible root phases on the 1D block $\tau_2$ whose symmetry is $\mathbb{Z}_2\times \mathbb{Z}_2\times \mathbb{Z}_2^f$:
\begin{enumerate}[1.]
\item Majorana chain, that needs only $\mathbb{Z}_2^f$ symmetry, denoted as $\mathsf{r}_0$;
\item two 1D fSPT phases, formed by double Majorana chains, protected jointly by $\mathbb{Z}_2\times \mathbb{Z}_2^f$ symmetry, denoted as $\mathsf{r}_1$ and $\mathsf{r}_2$ protected by $M_1$ and $M_2$ symmetry respectively,  and both the $\mathbb{Z}_2$ symmetry acts as exchanging the two majorana chain; 
\item 1D Haldane chain for integer spins, the bosonic SPT, protected by $\mathbb{Z}_2\times \mathbb{Z}_2$, denoted as $\mathsf{r}_3$, the edge mode from the projective representation of $\mathbb{Z}_2\times \mathbb{Z}_2$.  
\end{enumerate}

We can denote any phase by a four-component vector $\mathsf{r}=(\tilde{\mathsf{r}}_0,\tilde{\mathsf{r}}_1,\tilde{ \mathsf{r}}_2,\tilde{\mathsf{r}}_3)$  where $\tilde{\mathsf{r}}_i=0,1$. The above four root phases are labeled by $\mathsf{r}_0=(1,0,0,0)$, $\mathsf{r}_1=(0,1,0,0)$, $\mathsf{r}_2=(0,0,1,0)$ and $\mathsf{r}_3=(0,0,0,1)$.
We discuss them separately.

{
In Sec. \ref{T}, we have demonstrated that Majorana chain and double Majorana chains protected by either $M_1$ or $M_2$ symmetry (i.e., the root phase $\mathsf{r}_1$ or $\mathsf{r}_2$ ) decorated on 1D block $\tau_2$ are not compatible with $T$ as a subgroup of $T_h$, hence they are not compatible with $T_h$, either.   (We note that the state with 1d fSPT on $\tau_2$, either protected by $M_1$ or $M_2$, together with single Majorana chain on $\tau_3$ is obstruction-free, which however can be trivialized, for details see below.)}

Finally, Haldane chain decoration on each 1D block $\tau_2$ leaves 6 dangling spin-1/2 degrees of freedom $\bs{\tau}_j$ ($j=1,...,6$) near the 0D block $\mu$. Under $\bs{R}_2^1\in C_2^1$, these spin-1/2 degrees of freedom transform as:
\begin{align}
\left(\bs{\tau}_1,\bs{\tau}_2,\bs{\tau}_3,\bs{\tau}_4,\bs{\tau}_5,\bs{\tau}_6\right)\mapsto\left(\bs{\tau}_1,\bs{\tau}_3,\bs{\tau}_5,\bs{\tau}_4,\bs{\tau}_6,\bs{\tau}_2\right)
\end{align}
To gap out these spin-1/2 degrees of freedom, we simply introduce several symmetric Heisenberg interactions:
\begin{align}
H_J'=J'\left(\bs{\tau}_1\cdot\bs{\tau}_4+\bs{\tau}_2\cdot\bs{\tau}_5+\bs{\tau}_3\cdot\bs{\tau}_6\right)
\end{align}
And $H_J'$ leaves three spin-singlets as a non-degenerate ground state. As the consequence, the 1D Haldane chain decoration on 1D blocks labeled by $\tau_2$ is obstruction-free.

\paragraph{Spin-1/2 fermions}For spin-1/2 fermions, there is no nontrivial 1D SPT phase on $\tau_2$, hence there is also no nontrivial 1D block state.

Then we investigate the block state decoration on the 1D blocks $\tau_3$ with ``$\mathbb{Z}_2$ on-site symmetry'' by reflection symmetry $\bs{M}_1$ acting internally. Hence the candidate 1D block states are classified by group supercohomology (\ref{2-supercohomology}) with twisted cocycle conditions (\ref{2-twisted}) \cite{special,general1,general2}, where $G_{\mathrm{1D}}=\mathbb{Z}_2$. For spinless fermions, There are two possible root phases on each 1D block labeled by $\tau_3$:
\begin{enumerate}[1.]
\item Majorana chain;
\item 1D fSPT phase, formed by double Majorana chains.
\end{enumerate}
Majorana chain decoration on each 1D block $\tau_3$ leaves 12 dangling Majorana zero modes near 0D block $\mu$, and there are 4 dangling Majorana zero modes on each reflection plane. 3-fold rotations $\bs{R}_3^j\in C_3^j$ ($j=1,2,3,4$) permute these reflection planes ($xy$, $xz$ and $yz$ planes), and 2-fold rotations $\bs{R}_2^k\in C_2^k$ ($k=1,2,3$) are in-plane manipulations. We study the $xy$-plane as an example. 

\begin{figure}
\begin{tikzpicture}
\tikzstyle{sergio}=[rectangle,draw=none]
\draw[thick,color=red] (-0.5,0) -- (-1.5,1);
\draw[thick,color=red] (-0.5,-1) -- (-1.5,-2);
\draw[thick,color=red] (0.5,0) -- (1.5,1);
\draw[thick,color=red] (0.5,-1) -- (1.5,-2);
\draw[densely dashed,color=green] (0,1.5) -- (0,-2.5);
\draw[densely dashed,color=green] (-2,-0.5) -- (2,-0.5);
\filldraw[fill=black, draw=black] (-0.5,0)circle (2pt);
\filldraw[fill=black, draw=black] (0.5,0)circle (2pt);
\filldraw[fill=black, draw=black] (0.5,-1)circle (2pt);
\filldraw[fill=black, draw=black] (-0.5,-1)circle (2pt);
\path (0,-0.5) node [style=sergio] {$\bigotimes$};
\path (-0.4355,0.3493) node [style=sergio] {$\xi_1$};
\path (-0.7786,-0.8055) node [style=sergio] {$\xi_2$};
\path (0.3647,-1.2785) node [style=sergio] {$\xi_3$};
\path (0.7885,-0.1353) node [style=sergio] {$\xi_4$};
\path (-0.2267,-0.8351) node [style=sergio] {$R_2^1$};
\path (0.3037,1.5024) node [style=sergio] {$R_2^2$};
\path (1.7703,-0.1157) node [style=sergio] {$R_2^3$};
\end{tikzpicture}
\caption{Four Majorana zero modes near 0D block $\mu$ on the $xy$-plane, from Majorana chain decoration on 1D blocks $\tau_3$.}
\label{xy-plane}
\end{figure}
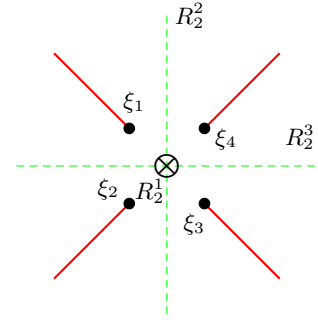

On the $xy$-plane, there are 4 dangling Majorana zero modes near 0D block $\mu$ from Majorana chain decorations, with the following rotation symmetry properties:
\begin{align}
\begin{aligned}
\bs{R}_2^1:~&\left(\xi_1,\xi_2,\xi_3,\xi_4\right)\mapsto\left(\xi_3,\xi_4,\xi_1,\xi_2\right)\\
\bs{R}_2^2:~&\left(\xi_1,\xi_2,\xi_3,\xi_4\right)\mapsto\left(\xi_4,\xi_3,\xi_2,\xi_1\right)\\
\bs{R}_2^3=\bs{R}_2^1\bs{R}_2^2:~&\left(\xi_1,\xi_2,\xi_3,\xi_4\right)\mapsto\left(\xi_2,\xi_1,\xi_4,\xi_3\right)
\end{aligned}
\end{align}
We define two complex fermions from $\xi_j$ ($j=1,2,3,4$):
\begin{align}
c_{13}^\dag=\frac{1}{2}(\xi_1+i\xi_3),~~c_{24}^\dag=\frac{1}{2}(\xi_2+i\xi_4)
\end{align}
They span a 4-dimensional Hilbert space: $|0\rangle$, $c_{13}^\dag|0\rangle$, $c_{24}^\dag|0\rangle$ and $c_{13}^\dag c_{24}^\dag|0\rangle$. Majorana zero modes $\xi_j$ can be represented as $4\times4$ matrices $A$ in this Hilbert space:
\begin{align}
\begin{aligned}
A(\bs{R}_2^1)&=\frac{1}{2}(\xi_1-\xi_3)(\xi_2-\xi_4)\\
A(\bs{R}_2^2)&=\frac{1}{2}(\xi_1-\xi_4)(\xi_2-\xi_3)\\
A(\bs{R}_2^3)=A(\bs{R}_2^1\bs{R}_2^2)&=\frac{1}{2}(\xi_1-\xi_2)(\xi_3-\xi_4)
\end{aligned}
\end{align}
It is straightforwardly to verify that $A$ is a projective representation of the $D_2^{xy}$ generated by $\bs{R}_2^k$ ($k=1,2,3$) as a subgroup of $T_h$ because of the following relation:
\begin{align}
A(\bs{R}_2^1)A(\bs{R}_2^2)=-A(\bs{R}_2^1\bs{R}_2^2)
\end{align}
Hence the Majorana zero modes $\xi_j$ ($j=1,2,3,4$) cannot be gapped in a $D_2^{xy}$-symmetric way. Equivalently, these Majorana zero modes can be treated as a spin-1/2 degree of freedom:
\begin{align}
S_x^{xy}=\frac{i}{2}\xi_1\xi_2,~S_y^{xy}=\frac{i}{2}\xi_1\xi_3,~S_z^{xy}=\frac{i}{2}\xi_1\xi_4
\end{align}
and all possible mass terms break $D_2^{xy}$ symmetry. Similarly, on the $xz$-plane and $yz$-plane, there is a spin-1/2 degree of freedom $\bs{S}^{xz}/\bs{S}^{yz}$ on each of them. Therefore, there are 3 spin-1/2 degrees of freedom near $\mu$ that cannot be gapped out in a $T_h$-symmetric way. As the consequence, Majorana chain decoration on $\tau_3$ is obstructed.

1D fSPT phase decoration leaves 24 dangling Majorana zero modes near 0D block $\mu$, and there are 8 dangling Majorana zero modes on each reflection plane. We investigate the $xy$-plane as an example.

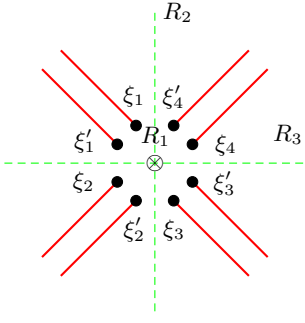
\begin{figure}
\begin{tikzpicture}[scale=1]
\tikzstyle{sergio}=[rectangle,draw=none]
\draw[thick,color=red] (-0.25,0) -- (-1.25,1);
\draw[thick,color=red] (-0.5,-0.25) -- (-1.5,0.75);
\draw[thick,color=red] (-0.5,-0.75) -- (-1.5,-1.75);
\draw[thick,color=red] (-0.25,-1) -- (-1.25,-2);
\draw[thick,color=red] (0.5,-0.25) -- (1.5,0.75);
\draw[thick,color=red] (0.25,0) -- (1.25,1);
\draw[thick,color=red] (0.25,-1) -- (1.25,-2);
\draw[thick,color=red] (0.5,-0.75) -- (1.5,-1.75);
\draw[densely dashed,color=green] (0,1.5) -- (0,-2.5);
\draw[densely dashed,color=green] (-2,-0.5) -- (2,-0.5);
\filldraw[fill=black, draw=black] (-0.25,0)circle (2pt);
\filldraw[fill=black, draw=black] (-0.5,-0.25)circle (2pt);
\filldraw[fill=black, draw=black] (0.5,-0.25) circle (2pt);
\filldraw[fill=black, draw=black] (0.25,0) circle (2pt);
\filldraw[fill=black, draw=black] (0.25,-1) circle (2pt);
\filldraw[fill=black, draw=black] (0.5,-0.75)circle (2pt);
\filldraw[fill=black, draw=black] (-0.5,-0.75)circle (2pt);
\filldraw[fill=black, draw=black] (-0.25,-1)circle (2pt);
\path (0,-0.5) node [style=sergio] {$\otimes$};
\path (-0.2759,0.3925) node [style=sergio] {$\xi_1$};
\path (-0.9283,-0.2026) node [style=sergio] {$\xi_1'$};
\path (-0.9461,-0.7563) node [style=sergio] {$\xi_2$};
\path (-0.2732,-1.3825) node [style=sergio] {$\xi_2'$};
\path (0.2563,-1.3968) node [style=sergio] {$\xi_3$};
\path (0.9246,-0.7689) node [style=sergio] {$\xi_3'$};
\path (0.9443,-0.2563) node [style=sergio] {$\xi_4$};
\path (0.2472,0.3798) node [style=sergio] {$\xi_4'$};
\path (0.0081,-0.1479) node [style=sergio] {$R_1$};
\path (0.3037,1.5024) node [style=sergio] {$R_2$};
\path (1.7703,-0.1157) node [style=sergio] {$R_3$};
\end{tikzpicture}
\caption{Eight Majorana zero modes near 0D block $\mu$ on the $xy$-plane, from 1D fSPT phase decoration on 1D blocks $\tau_3$.}
\label{xy-plane2}
\end{figure}

On the $xy$-plane, there are 8 dangling Majorana zero modes $\xi_j$ and $\xi_j'$ ($j=1,2,3,4$) near 0D block $\mu$, with the symmetry properties:
\begin{align}
\bs{R}_2^1:\left\{
\begin{aligned}
&\left(\xi_1,\xi_2,\xi_3,\xi_4\right)\mapsto\left(\xi_2,\xi_1,\xi_4,\xi_3\right)\\
&\left(\xi_1',\xi_2',\xi_3',\xi_4'\right)\mapsto\left(\xi_2',\xi_1',\xi_4',\xi_3'\right)
\end{aligned}
\right.
\end{align}
\begin{align}
\bs{R}_2^2:\left\{
\begin{aligned}
&\left(\xi_1,\xi_2,\xi_3,\xi_4\right)\mapsto\left(\xi_4,\xi_3,\xi_2,\xi_1\right)\\
&\left(\xi_1',\xi_2',\xi_3',\xi_4'\right)\mapsto\left(\xi_4',\xi_3',\xi_2',\xi_1'\right)
\end{aligned}
\right.
\end{align}
\begin{align}
\bs{M}_1^1:~\xi_j\leftrightarrow\xi_j'
\end{align}
We define four complex fermions from $\xi_j$ and $\xi_j'$:
\begin{align}
c_j^\dag=\frac{1}{2}(\xi_j+i\xi_j'),~j=1,2,3,4.
\label{Th fermion}
\end{align}
We firstly define a symmetric Hubbard interaction with the following Hamiltonian
\begin{align}
H_U=U\left[(n_1-\frac{1}{2})(n_3-\frac{1}{2})+(n_2-\frac{1}{2})(n_4-\frac{1}{2})\right],
\end{align}
where $n_{1,2,3,4}$ is the occupation number of fermions defined in Eq. \eqref{Th fermion}. In the ground state subspace of $H_U$, all degrees of freedom can be effectively described by two spin-1/2 degrees of freedom:
\begin{align}
\begin{gathered}
\tau_{13}^j=\left(c_1^\dag,c_3^\dag\right)\sigma^j\left(
\begin{array}{cc}
c_1 \\
c_3
\end{array}
\right)\\
\tau_{24}^j=\left(c_2^\dag,c_4^\dag\right)\sigma^j\left(
\begin{array}{cc}
c_2 \\
c_4
\end{array}
\right)
\end{gathered},
\end{align}
where $j=x,y,z$. And the ground state degeneracy of $H_U$ can be lifted by the following symmetric Heisenberg interaction,
\begin{align}
H_J=J\bs{\tau}_{13}\cdot\bs{\tau}_{24}.
\end{align}
Therefore, 1D fSPT decoration on $\tau_3$ is obstruction-free. 

For spin-1/2 fermions, there is no nontrivial 1D SPT phase on $\tau_3$.

Moreover, we need to consider whether these obstruction-free 1D block states can be trivialized. For systems with spinless fermions, we start by examining the 2D blocks labeled as $\sigma_1$. In this case, the only possible 2D bubble is the ``Majorana" bubble with anti-periodic boundary conditions (anti-PBC). Once these bubbles are enlarged and positioned near the borders, they result in 3 Majorana chains on each 1D block labeled as $\tau_1$ and 2 Majorana chains on each 1D block labeled as $\tau_3$  where the reflection symmetry acts as permutation symmetry of the double mojarana chain. Therefore, we conclude that the Majorana chain decoration on $\tau_1$ is equivalent to the 1D fSPT state decoration on $\tau_3$ by Majorana bubble on each $\sigma_1$.

Subsequently, we consider the 2D blocks labeled as $\sigma_2$ with a $\mathbb{Z}_2$ symmetry (denoted as $M_1$). Consequently, there are two possible 2D bubbles: the ``Majorana" bubble with anti-PBC and the ``1D fSPT" bubble (or saying ``double Majorana chain" bubble). Firstly, we explore the ``Majorana" bubble on $\sigma_2$, which results in 2 Majorana chains on each 1D block $\tau_2$ and 1 Majorana chains on each 1D block $\tau_3$.  {
Now we check how the symmetry acts on these majorana chains.  As there is  one Majorana chain on $\tau_3$,  we only need to consider the two Majorana chains on $\tau_2$. Under $M_1$, the two Majorana chains are invariant individually while under $M_2$, they are exchanged, which means it is a 1d fSPT protected by $M_2$ symmetry, corresponding to the phases labeled by $(\mathsf{r}_0,\mathsf{r}_1,\mathsf{r}_2,\mathsf{r}_3)=(0,0,1,0)$. Therefore, the ``Majorana bubble" on $\sigma_2$ trivializes the state with single Majorana chain on $\tau_3$ together with 1d fSPT protected by $M_2$ on $\tau_2$. 

Secondly, we consider the ``double Majorana" bubble on $\sigma_2$. We denote the two Majorana chains are built up by Majorana fermions $\gamma_1$ and $\gamma_2$, which, under $M_1$, get exchanged.  Further, under $M_2$, these two Majorana fermions transform into $\gamma_{1,2}'$, that comprise the ``double Majorana" bubble on another $\sigma_2$. The ``double Majorana" bubble on $\sigma_2$ finally results in a state with the 1d fSPT on $\tau_3$ together with four-Majorana-chain on $\tau_2$ built up by Majorana fermions $\gamma_{1,2}$ and $\gamma_{1,2}'$. To see which 1d phase the four-Majorana-chain belongs to, we check the symmetry: \begin{align}
    M_1: \,&\gamma_1\leftrightarrow \gamma_2\quad 
      \gamma_1'\leftrightarrow \gamma_2'\\
     M_2:\, &\gamma_1 \leftrightarrow \gamma_1'\quad 
     \gamma_2\leftrightarrow \gamma_2'
\end{align}
Then it is trivial 1d fSPT. Further we see whether it is Haldane phase. For this, we 
construct the corresponding operators in many-body Hilbert are given by
\begin{align}
    U(M_1)=\frac{1}{2}(\gamma_1-\gamma_2)(\gamma_1'-\gamma_2')\\
     U(M_2)=\frac{1}{2}(\gamma_1-\gamma_1')(\gamma_2-\gamma_2')
\end{align}
which anti-commute with each other, implying that the four-Majorana-chain state is the Haldane phase. Therefore, the ``double Majorana" bubble finally results in a state with 1d fSPT on $\tau_3$ together with Haldane phase on $\tau_2$. 

Similarly, we can consider the ``Majorana" bubble and ``double Majorana" bubble on $\sigma_3$, which result in trivialization: (1) the state with single Majorana chain on $\tau_3$ and the 1d fSPT protected by $M_1$ on $\tau_2$, (2) state with 1d fSPT on $\tau_3$ and Haldane phase on $\tau_2$. 
}


Therefore, for systems with spinless fermions, there is only one fundamental 1D block state: the Majorana chain decoration on 1D blocks $\tau_1$, or  1D fSPT state decoration on 1D blocks $\tau_3$, or Haldane phase decoration on 1D blocks $\tau_2$, all of which are equivalent through the 2D bubble equivalence.

For systems with spin-1/2 fermions, firstly we consider the 2D blocks $\sigma_1$, and the only possible 2D bubble is the ``Majorana'' bubble with anti-PBC. After enlarging these bubbles and proxy to the borders, they will leave 3 Majorana chains on each 1D block labeled by $\tau_1$ and 2 Majorana chains on each 1D block labeled by $\tau_3$. On each $\tau_3$, double Majorana chains can be trivialized for spin-1/2 fermions, hence the ``Majorana'' bubble changes nothing on $\tau_3$; on each $\tau_1$, 3 Majorana chains are equivalent to single Majorana chain, hence the Majorana chain decoration on 1D block labeled by $\tau_1$ is trivialized by ``Majorana'' bubble construction on 2D blocks labeled by $\sigma_1$.

Subsequently, we consider the 2D blocks $\sigma_2$. Due to the $\mathbb{Z}_2$ symmetry, there is no 2D bubble. Hence there is no nontrivial 1D block state for spin-1/2 fermions.

\subsubsection{0D block states\label{Th0D}}
The effective ``on-site'' symmetry on the 0D block $\mu$ is $A_4\times\mathbb{Z}_2$, hence the candidate 0D block states are classified by the following two indices:
\begin{align}
\begin{aligned}
&n_0\in\mathcal{H}^0(A_4\times\mathbb{Z}_2,\mathbb{Z}_2)=\mathbb{Z}_2\\
&\nu_1\in\mathcal{H}^1\left[A_4\times\mathbb{Z}_2,U(1)\right]=\mathbb{Z}_2\times\mathbb{Z}_3
\end{aligned}
\label{1-supercohomology Th}
\end{align}
with the twisted cocycle condition:
\begin{align}
\mathrm{d}\nu_1=(-1)^{\omega_2\smile n_0}
\label{1-twisted Th}
\end{align}
where $n_0$ depicts the parity of fermions, and $\nu_1$ depicts the 0D bSPT mode on the 0D block $\mu$ protected by $A_4\times\mathbb{Z}_2$ symmetry, characterizing the eigenvalue $\pm1$ of the product of $\bs{M}_1\in T_h$, and the product of eigenvalues $e^{i2n\pi/3}$ ($n=0,1,2$) of $\bs{R}_3\in T_h$. We demonstrate that these 0D bSPT mode can be trivialized by 1D bubble construction: consider the axes of 2-fold rotation generator $\bs{R}_2^j$ ($j=1,...,6$) of $T_h$, we decorate a complex fermion $c_j^\dag$ ($j=1,...,6$) on each of them. Near the 0D block $\mu$, these complex fermions form an atomic insulator:
\begin{align}
|\psi_\mu\rangle=c_1^\dag c_2^\dag c_3^\dag c_4^\dag c_5^\dag c_6^\dag|0\rangle
\label{Th atomic insulator}
\end{align}
Furthermore, these complex fermions will be transformed by reflection $\bs{M}_1^{1,2,3}$ as the following way:
\begin{align}
\begin{aligned}
&\bs{M}_1^1:\left(c_1^\dag, c_2^\dag, c_3^\dag, c_4^\dag, c_5^\dag, c_6^\dag\right)\mapsto\left(c_4^\dag, c_2^\dag, c_3^\dag, c_1^\dag, c_5^\dag, c_6^\dag\right)\\
&\bs{M}_1^2:\left(c_1^\dag, c_2^\dag, c_3^\dag, c_4^\dag, c_5^\dag, c_6^\dag\right)\mapsto\left(c_1^\dag, c_5^\dag, c_3^\dag, c_4^\dag, c_2^\dag, c_6^\dag\right)\\
&\bs{M}_1^3:\left(c_1^\dag, c_2^\dag, c_3^\dag, c_4^\dag, c_5^\dag, c_6^\dag\right)\mapsto\left(c_1^\dag, c_2^\dag, c_6^\dag, c_4^\dag, c_5^\dag, c_3^\dag\right)
\end{aligned}\nonumber
\end{align}
As the consequence, the atomic insulator $|\psi_\mu\rangle$ is an eigenstate of $\bs{M}_1^{1,2,3}$ with eigenvalue $-1$:
\begin{align}
\bs{M}_1^1|\psi_\mu\rangle=\bs{M}_1^2|\psi_\mu\rangle=\bs{M}_1^3|\psi_\mu\rangle=-|\psi_\mu\rangle
\end{align}
Therefore, the nontrivial 0D bSPT mode that characterizes the nontrivial reflection eigenvalue at $\mu$ can be rendered trivial through 1D bubble construction on $\tau_2$. Additionally, the trivialization of the $\mathbb{Z}_3$ index, derived from the linear representation of the $A_4$ symmetry, follows a similar pattern: incorporating 0D $\mathbb{Z}_3$ bSPT modes onto the 1D blocks labeled as $\tau_1$.

Now we examine the fermion parity of the 0D block $\mu$ in the case of spinless fermions. Previous references \cite{dihedral, wallpaper} have demonstrated that a Majorana chain with periodic boundary conditions (PBC) alters the fermion parity of the enclosed point. However, it is important to note that the Majorana chain is incompatible with reflection symmetry. As a result, unlike in the $T$-symmetric scenario, we cannot construct three Majorana chains with PBC that surround the 0D block $\mu$ and induce a change in fermion parity on $\mu$. Consequently, the presence of complex fermion decoration on the 0D block $\mu$ indicates a nontrivial block state.

For spin-1/2 fermions, the index $n_0$ in Eq. (\ref{1-supercohomology Th}) is considered ``obstructed". Furthermore, we repeatedly consider the atomic insulator $|\psi_\mu\rangle$, which remains an eigenstate of $\bs{M}_1^{1,2,3}$ but with an eigenvalue of +1 due to an additional minus sign resulting from the spin-1/2 nature of fermions. Consequently, the nontrivial eigenvalue of the reflection generator in the $T_h$ symmetry cannot be trivialized.

\subsubsection{Summary\label{summary Th}}
Finally, we summarize the results of classifications and corresponding block states. For systems with spinless fermions, the ultimate classification is $\mathbb{Z}_2^3$, with following three root block states:
\begin{enumerate}[1.]
\item Complex fermion decoration on 0D block $\mu$;
\item Majorana chain decoration on each 1D block $\tau_1$, {
which is equivalent to 1D fSPT chain decoration on each 1D block $\tau_3$ or Haldane phase deocration on each 1D block $\tau_2$};
\item 2D bosonic Levin-Gu state on each 2D block $\sigma_2$.
\end{enumerate}
For systems with spin-1/2 fermions, the ultimate classification is $\mathbb{Z}_2$, the root block state is 0D bSPT mode representing the eigenvalue $-1$ of reflection generator $\bs{M}_1^1\in T_h$ decoration on 0D block $\mu$. 

\subsubsection{Higher-order topological surface theory}
Equipped with specific block states, we can now delve into the corresponding higher-order (HO) topological surface theories through the prism of bulk-boundary correspondence.

In the case of spinless fermions, the classification of 3D $T_h$-symmetric topological phases is $\mathbb{Z}_2^3$, featuring three root phases as outlined in Section \ref{summary Th}. We investigate these root phases within the context of an open cubic structure, as depicted in Figure \ref{Th cell decomposition}.

The complex fermion decoration on the 0D block $\mu$ represents a 0D block state that does not leave any discernible features on the surface of an open cubic. Consequently, it does not constitute a nontrivial HO topological phase.

On the other hand, the Majorana chain decoration on each 1D block $\tau_1$ corresponds to a 1D block state that results in dangling Majorana zero modes at each corner of an open cubic. As a result, the nontrivial topological phase derived from this particular 1D block state manifests as a 3D fermionic third-order topological phase.

Similarly, the Haldane chain decoration on each 1D block $\tau_2$ represents a 1D block state that gives rise to a spin-1/2 degree of freedom at the center of each surface of an open cubic. Consequently, the nontrivial topological phase resulting from this 1D block state materializes as a 3D bosonic third-order topological phase.

\begin{figure}
\begin{tikzpicture}[scale=0.92]
\tikzstyle{sergio}=[rectangle,draw=none]
\draw[draw=black, thick] (-2,0.5)--(1,0.5)--(1,-2.5)--(-2,-2.5)--cycle;
\draw[densely dashed,color=red] (-0.5,0.5) -- (-0.5,-2.5);
\draw[densely dashed,color=red] (-2,-1) -- (1,-1);
\path (-0.2055,0.7887) node [style=sergio] {$M_1^3$};
\path (1.3561,-0.7194) node [style=sergio] {$M_1^2$};
\draw[thick,->](-0.5,-1)--(-0.2624,-0.5);
\path (0,-0.5) node [style=sergio] {$\boldsymbol{S}$};
\draw[draw=black, thick] (2.5,0.5)--(5.5,0.5)--(5.5,-2.5)--(2.5,-2.5)--cycle;
\draw[densely dashed,color=red] (4,0.5) -- (4,-2.5);
\draw[densely dashed,color=red] (2.5,-1) -- (5.5,-1);
\path (4.2945,0.7887) node [style=sergio] {$M_1^3$};
\path (5.8561,-0.7194) node [style=sergio] {$M_1^2$};
\filldraw[fill=black, draw=black] (4.25,-1.25)circle (2pt);
\filldraw[fill=black, draw=black] (4.25,-0.75)circle (2pt);
\filldraw[fill=black, draw=black] (3.75,-1.25)circle (2pt);
\filldraw[fill=black, draw=black] (3.75,-0.75)circle (2pt);
\path (3.7333,-0.4755) node [style=sergio] {$\gamma_1$};
\path (4.2699,-1.5368) node [style=sergio] {$\gamma_3$};
\path (4.27,-0.4755) node [style=sergio] {$\gamma_2$};
\path (3.7459,-1.5367) node [style=sergio] {$\gamma_4$};
\draw[thick,->] (3.5,-0.5)--(3,0);
\draw[thick,->] (3.5,-1.5)--(3,-2);
\draw[thick,->] (4.5,-0.5)--(5,0);
\draw[thick,->] (4.5,-1.5)--(5,-2);
\filldraw[fill=black, draw=black] (5.25,-2.25)circle (2pt);
\filldraw[fill=black, draw=black] (5.25,0.25)circle (2pt);
\filldraw[fill=black, draw=black] (2.75,-2.25)circle (2pt);
\filldraw[fill=black, draw=black] (2.75,0.25)circle (2pt);
\path (-2.5,0.75) node [style=sergio] {$(a)$};
\path (2,0.75) node [style=sergio] {$(b)$};
\path (3,-0.5) node [style=sergio] {$m_1$};
\path (5,-1.5) node [style=sergio] {$m_3$};
\path (5,-0.5) node [style=sergio] {$m_2$};
\path (3,-1.5) node [style=sergio] {$m_4$};
\end{tikzpicture}
\caption{Topological surface state of 3D $T_h$-symmetric third-order topological phase from 1D block state construction. (a): Dangling spin-1/2 degree of freedom at the top surface of an open cubic, from Haldane chain decoration on each 1D block $\tau_2$. (b): Dangling spin-1/2 degree of freedom formulated by 4 dangling Majorana zero modes $\gamma_{1,2,3,4}$ that can be smoothly deformed to the corner of the cubic.}
\label{surface}
\end{figure}
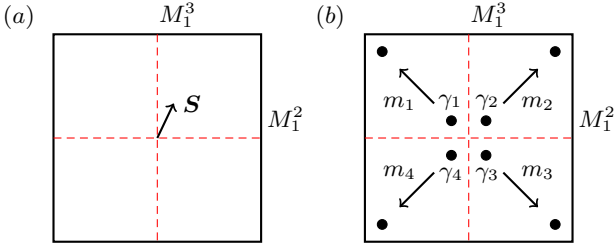

We have established that the 1D block states discussed in the preceding paragraphs are not independent when viewed from the perspective of the block states in the previous sections of this segment. To further clarify this point, we provide an alternative outlook from the boundary. Let us consider the third-order surface state of the 3D $T_h$-symmetric topological phase, which arises from decorating a Haldane chain on each 1D block $\tau_2$. Specifically, for the top surface of the open cubic, a spin-1/2 fermion resides at the center [see Fig. \ref{surface}(a)]. This spin-1/2 degree of freedom can be decomposed into a combination of four dangling Majorana zero modes, denoted as $\gamma_{1,2,3,4}$. The effective 2D crystalline symmetry on the top surface of the open cubic is governed by the 2-fold dihedral group $D_2$, generated by two reflection operators $\bs{M}_1^2$ and $\bs{M}1^3$. We assign specific symmetry properties to the four Majorana zero modes $\gamma{1,2,3,4}$, which are as follows:
\begin{align}
\begin{aligned}
&\bs{M}_1^2:~(\gamma_1,\gamma_2,\gamma_3,\gamma_4)\mapsto(\gamma_4,\gamma_3,\gamma_2,\gamma_1)\\
&\bs{M}_1^3:~(\gamma_1,\gamma_2,\gamma_3,\gamma_4)\mapsto(\gamma_2,\gamma_1,\gamma_4,\gamma_3)
\end{aligned}
\label{surface symmetry}
\end{align}
Firstly, we define a symmetric Hubbard interaction between these Majorana zero modes:
\begin{align}
H_s^U=U\gamma_1\gamma_2\gamma_3\gamma_4,~U>0
\end{align}
The ground states are characterized by $\gamma_1\gamma_2\gamma_3\gamma_4=-1$ with 2-fold degeneracy. Then define the operators:
\begin{align}
S_x=\frac{i}{2}\gamma_1\gamma_2,~S_y=\frac{i}{2}\gamma_1\gamma_3,~S_z=\frac{i}{2}\gamma_1\gamma_4
\end{align}
and they satisfy the commutation relations of spin-1/2 degrees of freedom ($\mu,\nu,\rho=x,y,z$):
\begin{align}
\left[S_\mu,S_\nu\right]=i\epsilon_{\mu\nu\rho}S_\rho
\end{align}
It can be easily justified that any mass terms involving the spin vector $\bs{S}=(S_x, S_y, S_z)$ would break the $D_2$ symmetry, thus confirming that $\bs{S}$ represents a dangling spin-1/2 degree of freedom. Similarly, a dangling spin-1/2 degree of freedom at the center of the top surface of an open cubic is equivalent to four dangling Majorana zero modes possessing the symmetry properties described in Equation (\ref{surface symmetry}). This equivalence holds true for the other surfaces of the cubic as well.

To demonstrate the smooth deformation of these Majorana zero modes to the corner of the cubic, we consider an open Majorana chain labeled by $m_1$ positioned as ``plates" on the boundary, as depicted in Fig. \ref{surface}(b). Near the surface center, the Majorana zero mode, serving as the edge mode of this Majorana chain, can be gapped out with the involvement of $\gamma_1$ without breaking any symmetry. Similarly, near the top-left corner of the surface, this Majorana chain gives rise to an additional Majorana zero mode as the edge mode on the opposite side. To ensure the construction is symmetric under $D_2$, we also incorporate Majorana chains labeled by $m_{2,3,4}$, resulting in the presence of four dangling Majorana zero modes at the surface corner (and the corner of the cubic as well). This deformation process is carried out for all other surfaces, ultimately leading to the smooth deformation of the dangling spin-1/2 degrees of freedom at the surface centers, arising from the Haldane chain decoration on 1D block $\tau_2$, into dangling Majorana zero modes at the corners of the cubic. Each corner features three equivalent dangling Majorana zero modes, which can be viewed as a single Majorana zero mode.

On the other hand, the 3D $T_h$-symmetric third-order topological phase constructed from the Majorana chain decoration on 1D block $\tau_1$ leaves dangling Majorana zero modes at each corner of the cubic, and the corresponding bulk states are topologically equivalent. The equivalence of bubbles in the bulk and the ``plate" equivalence on the boundary directly reflects the higher-order bulk-boundary correspondence of crystalline SPT phases.

In the case of the 2D bosonic Levin-Gu state decoration on each 2D block $\sigma_2$, non-chiral Luttinger liquids emerge on the surfaces, specifically at the vertical and horizontal links crossing the center of each surface of the cubic. As a result, the nontrivial topological phase constructed from this 2D block state represents a 3D bosonic second-order topological phase.

In the context of spin-1/2 fermions, the corresponding classification of 3D $T_h$-symmetric topological phases is $\mathbb{Z}_2$, with the root phase outlined in Section \ref{summary Th}. The 3D $T_h$-symmetric topological phase characterized by this 0D block state does not qualify as a nontrivial higher-order topological phase.

\begin{figure*}
\begin{tikzpicture}[scale=0.54]
\tikzstyle{sergio}=[rectangle,draw=none]
\path (17.0818,-0.7873) node [style=sergio] {$\mu$};
\draw[thick] (-1,1.5) -- (3,1.5);
\draw[thick] (-1,1.5) -- (-3,0);
\draw[thick] (3,1.5) -- (1,0);
\draw[thick] (-3,0) -- (1,0);
\draw[thick] (-3,0) -- (-3,-4);
\draw[thick] (1,-4) -- (-3,-4);
\draw[thick] (1,-4) -- (1,0);
\draw[thick] (3,-2.5) -- (3,1.5);
\draw[thick] (3,-2.5) -- (1,-4);
\draw[thick,color=green] (0,-1.25) -- (0,0.75);
\draw[thick,color=green] (0,-1.25) -- (0,-3.25);
\draw[thick,color=green] (0,-1.25) -- (-2,-1.25);
\draw[thick,color=green] (0,-1.25) -- (2,-1.25);
\draw[thick,color=green] (0,-1.25) -- (1,-0.5);
\draw[thick,color=green] (0,-1.25) -- (-1,-2);
\draw[densely dashed] (-1,-2.5) -- (-1,1.5);
\draw[densely dashed] (-1,-2.5) -- (3,-2.5);
\draw[densely dashed] (-1,-2.5) -- (-3,-4);
\draw[densely dashed,color=red] (3,1.5) -- (0,-1.25);
\draw[thick,color=violet] (-1,1.5) -- (0,-1.25);
\draw[densely dashed,color=red] (-3,0) -- (0,-1.25);
\draw[thick,color=violet] (0,-1.25) -- (1,0);
\path (-3.5,-4.5) node [style=sergio] {$\tau_1,C_3^1$};
\path (3.5,-3) node [style=sergio] {$\tau_1,C_3^2$};
\path (1,1) node [style=sergio] {$\tau_1,C_3^3$};
\path (-1.5,2) node [style=sergio] {$\tau_1,C_3^4$};
\draw[thick,color=violet] (0,-1.25) -- (-3,-4);
\draw[densely dashed,color=red] (0,-1.25) -- (-1,-2.5);
\draw[thick,color=violet] (0,-1.25) -- (3,-2.5);
\draw[densely dashed,color=red] (0,-1.25) -- (1,-4);
\draw[thick] (3,-2.5) -- (1,-4);
\draw[thick] (-3,-4) -- (1,-4);
\path (-1,-3.5) node [style=sergio] {$\tau_3,C_2^1$};
\path (-3,-1.5) node [style=sergio] {$\tau_3,C_2^3$};
\path (2,-0.5) node [style=sergio] {$\tau_3,C_2^2$};
\path (1.5,-4.5) node [style=sergio] {$\tau_2$};
\path (-0.2796,-2.2352) node [style=sergio] {$\tau_2$};
\path (-3.5,0) node [style=sergio] {$\tau_2$};
\path (3,2) node [style=sergio] {$\tau_2$};
\draw[thick] (24.25,1.5) -- (28.25,1.5);
\draw[thick] (24.25,1.5) -- (22.25,0);
\draw[thick] (28.25,1.5) -- (26.25,0);
\draw[thick] (22.25,0) -- (26.25,0);
\draw[thick] (22.25,0) -- (22.25,-4);
\draw[thick] (26.25,-4) -- (22.25,-4);
\draw[thick] (26.25,-4) -- (26.25,0);
\draw[thick] (28.25,-2.5) -- (28.25,1.5);
\draw[thick] (28.25,-2.5) -- (26.25,-4);
\draw[densely dashed] (24.25,-2.5) -- (24.25,1.5);
\draw[densely dashed] (24.25,-2.5) -- (28.25,-2.5);
\draw[densely dashed] (24.25,-2.5) -- (22.25,-4);
\draw[thick] (28.25,-2.5) -- (26.25,-4);
\draw[thick] (22.25,-4) -- (26.25,-4);
\path (-0.3,-1.3) node [style=sergio] {$\mu$};
\path (24.9296,0.4477) node [style=sergio] {$M_1^5$};
\path (26.4652,1.9832) node [style=sergio] {$M_1^6$};
\draw[draw=red, thick] (26.25,0)--(22.25,0)--(24.25,-2.5)--(28.25,-2.5)--cycle;
\draw[draw=red, thick] (24.25,1.5)--(28.25,1.5)--(26.25,-4)--(22.25,-4)--cycle;
\filldraw[fill=black, draw=black] (25.25,-1.25) circle (2.5pt);
\draw[densely dashed,color=red] (23.25,-1.25) -- (27.25,-1.25);
\draw[thick] (16,1.5) -- (20,1.5);
\draw[thick] (16,1.5) -- (13.5,0);
\draw[thick] (20,1.5) -- (17.5,0);
\draw[thick] (13.5,0) -- (17.5,0);
\draw[thick] (13.5,0) -- (13.5,-4);
\draw[thick] (17.5,-4) -- (13.5,-4);
\draw[thick] (17.5,-4) -- (17.5,0);
\draw[thick] (20,-2.5) -- (20,1.5);
\draw[thick] (20,-2.5) -- (17.5,-4);
\draw[densely dashed] (16,-2.5) -- (16,1.5);
\draw[densely dashed] (16,-2.5) -- (20,-2.5);
\draw[densely dashed] (16,-2.5) -- (13.5,-4);
\draw[thick] (20,-2.5) -- (17.5,-4);
\draw[thick] (13.5,-4) -- (17.5,-4);
\path (-0.3,-1.3) node [style=sergio] {$\mu$};
\path (17.75,0.75) node [style=sergio] {$M_1^4$};
\path (14.25,1.25) node [style=sergio] {$M_1^3$};
\draw[draw=red, thick] (16,-2.5)--(20,1.5)--(17.5,0)--(13.5,-4)--cycle;
\draw[draw=red, thick] (16,1.5)--(13.5,0)--(17.5,-4)--(20,-2.5)--cycle;
\filldraw[fill=black, draw=black] (16.75,-1.25) circle (2.5pt);
\draw[densely dashed,color=red] (15.5,-2) -- (18,-0.5);
\draw[thick] (7.5,1.5) -- (11.5,1.5);
\draw[thick] (7.5,1.5) -- (5.5,0);
\draw[thick] (11.5,1.5) -- (9.5,0);
\draw[thick] (5.5,0) -- (9.5,0);
\draw[thick] (5.5,0) -- (5.5,-4);
\draw[thick] (9.5,-4) -- (5.5,-4);
\draw[thick] (9.5,-4) -- (9.5,0);
\draw[thick] (11.5,-2.5) -- (11.5,1.5);
\draw[thick] (11.5,-2.5) -- (9.5,-4);
\draw[densely dashed] (7.5,-2.5) -- (7.5,1.5);
\draw[densely dashed] (7.5,-2.5) -- (11.5,-2.5);
\draw[densely dashed] (7.5,-2.5) -- (5.5,-4);
\draw[thick] (11.5,-2.5) -- (9.5,-4);
\draw[thick] (5.5,-4) -- (9.5,-4);
\path (-0.3,-1.3) node [style=sergio] {$\mu$};
\path (5,-2) node [style=sergio] {$M_1^1$};
\path (7,2) node [style=sergio] {$M_1^2$};
\draw[draw=red, thick] (5.5,0)--(11.5,1.5)--(11.5,-2.5)--(5.5,-4)--cycle;
\draw[draw=red, thick] (7.5,1.5)--(9.5,0)--(9.5,-4)--(7.5,-2.5)--cycle;
\filldraw[fill=black, draw=black] (8.5,-1.25) circle (2.5pt);
\draw[densely dashed,color=red] (8.5,0.75) -- (8.5,-3.25);
\end{tikzpicture}
\caption{The cell decomposition of the 3D system with point group symmetry $T_d$. There are four axes of 3-fold rotation symmetry labeled by $C_3^{1,2,3,4}$ and depicted by solid violet line segments ($\tau_1$) and dashed red line segments ($\tau_2$), across the center of the system (labeled by $\mu$); three axes of 2-fold rotation symmetry across the center, labeled by $C_2^{1,2,3}$ and depicted by solid green lines (see top left panel); and six planes of reflection symmetry, labeled by $\bs{M}_1^{1,2,3,4,5,6}$.}
\label{Td cell decomposition}
\end{figure*}
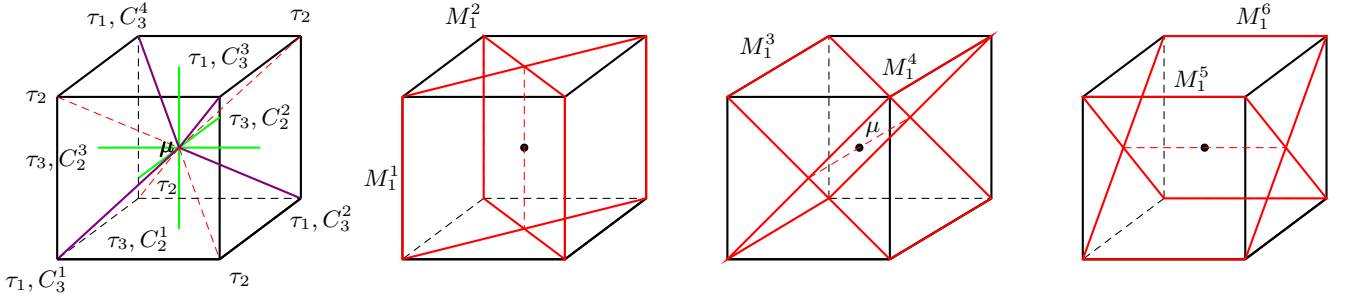

\subsection{$T_d$-symmetric lattice}
\begin{figure}
\begin{tikzpicture}[scale=0.78]
\tikzstyle{sergio}=[rectangle,draw=none]
\draw[thick] (-3,-0.5) -- (1,-0.5);
\draw[densely dashed,thick] (-1,1) -- (3,1);
\draw[densely dashed,thick] (-1,1) -- (-3,-0.5);
\draw[thick] (3,1) -- (1,-0.5);
\draw[densely dashed,color=red] (-1,1) -- (0,2.25);
\draw[densely dashed,color=red] (1,-0.5) -- (0,2.25);
\draw[thick,color=violet] (-3,-0.5) -- (0,2.25);
\draw[thick,color=violet] (3,1) -- (0,2.25);
\draw[thick,color=green] (0,0.25) -- (0,2.25);
\filldraw[fill=black, draw=black] (0,2.25)circle (3pt);
\draw[densely dashed] (-1,1) -- (1,-0.5);
\draw[densely dashed] (3,1) -- (-3,-0.5);
\draw[thick] (3.5,-0.5) -- (7.5,-0.5);
\draw[densely dashed] (6.5,0.25) -- (3.5,-0.5);
\draw[densely dashed] (6.5,0.25) -- (7.5,-0.5);
\draw[thick,color=green] (6.5,0.25) -- (6.5,2.25);
\draw[thick,color=violet] (3.5,-0.5) -- (6.5,2.25);
\draw[densely dashed,color=red] (7.5,-0.5) -- (6.5,2.25);
\draw[thick,->] (6,1) -- (5.3042,1.6206);
\draw[thick,->] (6.8213,0.6741) -- (7.3329,1.2556);
\path (5.0542,1.8706) node [style=sergio] {$\sigma_1$};
\path (6,0.5) node [style=sergio] {$\sigma_3$};
\path (7.5,1.5) node [style=sergio] {$\sigma_2$};
\end{tikzpicture}
\caption{The unit cell of cell decomposition of the cubic lattice with point symmetry $T_d$. The top panel depicts the bottom rectangular pyramid of the cubic in Fig. \ref{Td cell decomposition}; the bottom panel illustrates the independent triangular pyramid $\lambda$, where $\sigma_1$, $\sigma_2$ and $\sigma_3$ are three independent 2D blocks in the system.}
\label{Td unit cell}
\end{figure}
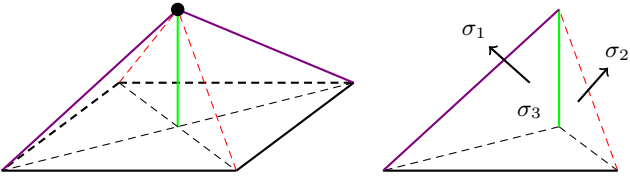

For $T_d$-symmetric cubic, by cell decomposition as illustrated in Figs. \ref{Td cell decomposition} and \ref{Td unit cell}, the ground-state wavefunction of the system can be decomposed to the direct products of wavefunctions of lower-dimensional blocks as:
\begin{align}
|\Psi\rangle=\bigotimes_{g\in T_d}|T_{g\lambda}\rangle\otimes\bigotimes\limits_{k=1}^3|\gamma_{g\sigma_k}\rangle\otimes\bigotimes\limits_{l=1}^3|\beta_{g\tau_l}\rangle\otimes|\alpha_\mu\rangle
\label{Td wavefunction}
\end{align}
where $|T_{g\lambda}\rangle$ is the wavefunction of 3D block state $g\lambda$ which is topologically trivial; $|\gamma_{g\sigma_{1,2,3}}\rangle$ is the wavefunction of 2D block state $g\sigma_{1,2,3}$ which is $\mathbb{Z}_2$-symmetric; $|\beta_{\tau_{1,2}}\rangle$ is the wavefunction of 1D block state $g\tau_{1,2}$ which is $(\mathbb{Z}_3\rtimes\mathbb{Z}_2)$-symmetric, and $|\beta_{\tau_3}\rangle$ is the wavefunction of 1D block state $g\tau_3$ which is $(\mathbb{Z}_2\times\mathbb{Z}_2)$-symmetric; $|\alpha_\mu\rangle$ is the wavefunction of 0D block state $\mu$ which is $S_4$-symmetric.

With the topological crystals, we decorate the lower-dimensional block states and investigate the possible \textit{obstructions} and \textit{trivializations}. 

\subsubsection{2D block states\label{Td2D}}
The effective on-site symmetry of 2D blocks labeled by $\sigma_j$ ($j=1,2,3$) is $\mathbb{Z}_2$ by reflection generator $\bs{M}_1^k$ ($k=1,...,6$) acting internally. For spinless fermions, there are two possible root phases:
\begin{enumerate}[1.]
\item 2D $(p+ip)$-SC;
\item 2D fermionic Levin-Gu state \cite{Gu-Levin} with $\nu\in\mathbb{Z}_8$ index.
\end{enumerate}

\begin{figure}
\begin{tikzpicture}
\tikzstyle{sergio}=[rectangle,draw=none]
\draw[thick] (-2.5,-4.5) -- (2,-4.5);
\draw[densely dashed] (0.5,-3.75) -- (-2.5,-4.5);
\draw[densely dashed] (0.5,-3.75) -- (2,-4.5);
\draw[thick,color=green] (0.5,-3.75) -- (0.5,-1.75);
\draw[thick,color=violet] (-2.5,-4.5) -- (0.5,-1.75);
\draw[densely dashed,color=red] (2,-4.5) -- (0.5,-1.75);
\draw[thick,->] (-0.9869,-3.7493) -- (0.0084,-2.8083);
\draw[thick,->] (0.2361,-2.7684) -- (0.2445,-3.4613);
\draw[thick,->,color=blue] (0.6559,-3.4505) -- (0.6443,-2.7492);
\draw[thick,->,color=blue] (0.7763,-2.9234) -- (1.088,-3.6056);
\draw[thick,->,color=red] (1.5911,-4.0107) -- (0.8017,-2.5131);
\draw[thick,->,color=red] (0.1357,-2.3865) -- (-1.6198,-3.9759);
\end{tikzpicture}
\caption{Chiralities of edge modes of decorated 2D $(p+ip)$-SC on 2D blocks $\sigma_j$ with indices $n_j$ ($j=1,2,3$), illustrated by black, blue, and red arrows, respectively.}
\label{Td chirality}
\end{figure}
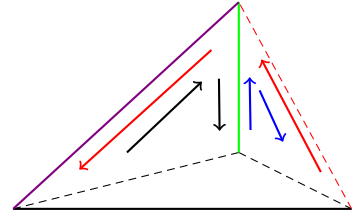

Firstly we consider the 2D $(p+ip)$-SC decorations on 2D blocks $\sigma_j$ with quantum number $n_j\in\mathbb{Z}$ ($j=1,2,3$), and their edge modes are illustrated in Fig. \ref{Td chirality}. As a consequence, the chiral central charges with decorated 2D $(p+ip)$-SCs on $\tau_k$ ($k=1,2,3$) are $3(n_1-n_3)/2$, $-3(n_2-n_3)/2$ and $-n_1+n_2$, respectively. To get a fully-gapped bulk, a necessary condition is that all these three quantities characterizing two times of chiral central charge of 1D blocks $\tau_{1,2,3}$ vanish:
\begin{align}
3n_1-3n_3=-3n_2+3n_3=-2n_1+2n_2=0
\end{align}
The solutions to these equations should satisfy $n_1=n_2=n_3=n$. Therefore, a necessary condition of obstruction-free block states from 2D $(p+ip)$-SC decoration on 2D blocks gives a $\mathbb{Z}$-index $n$, characterizing the number of decorated 2D $(p+ip)$-SCs on each 2D block $\sigma_j$ ($j=1,2,3$). Similar to Sec. \ref{Th2D}, the case of the monolayer is obstructed, and the case of the bilayer is trivialized.

Subsequently, we consider the 2D fermionic Levin-Gu state decorations on $\sigma_{1,2,3}$ with quantum number $\nu_{1,2,3}\in\mathbb{Z}_8$, their edge modes are similar to Fig. \ref{Td chirality} with a subtle difference: for 2D $(p+ip)$-SC, the arrows characterize the direction of chiral edge states; for 2D fermionic Levin-Gu state, the arrows characterize the direction of the current of $\mathbb{Z}_2$ charge. Therefore, the numbers of edge modes leaved by 2D fermionic Levin-Gu states on $\tau_{1,2,3}$ are $3\nu_1-3\nu_3$, $-3\nu_2+3\nu_3$ and $-2\nu_1+2\nu_2$, respectively. To get a fully-gapped bulk, a necessary condition is that all these three quantities are integer multiples of 8:
\begin{align}
\begin{aligned}
&3\nu_1-3\nu_3\equiv0~(\mathrm{mod}~8)\\
-&3\nu_2+3\nu_3\equiv0~(\mathrm{mod}~8)\\
-&2\nu_1+2\nu_2\equiv0~(\mathrm{mod}~8)
\end{aligned}
\end{align}
The solution to these equations should satisfy:
$$
\nu_1\equiv\nu_2\equiv\nu_3\equiv\nu~(\mathrm{mod}~8)
$$
Therefore, a necessary condition of obstruction-free block states from 2D fermionic Levin-Gu states decorations on 2D blocks gives a $\mathbb{Z}_8$-index $\nu$, characterizing the quantum number of decorated Levin-Gu states on each 2D block $\sigma_{1,2,3}$. Then we should study if the edge states near each 1D block $\tau_3$ left by 2D Levin-Gu states decorated on 2D blocks $\sigma_{1,2,3}$.

Similarly to $T_h$ symmetry, only $\nu=4$ state on $\sigma_{1,2,3}$ are compatible with the $T_d$ symmetry, which is  \textit{obstruction-free}. 

We should further investigate if these 2D block states can be trivialized. There is only one possible 3D bubble (see Fig. \ref{3D bubble}) because of the absence of on-site symmetry in all 3D blocks, which is irrelevant to 2D Levin-Gu state decorations. Therefore, there is no trivialization, and all nontrivial 2D block states form a $\mathbb{Z}_2$ group, composed by a 2D (bosonic) Levin-Gu state on each 2D block.

\subsubsection{1D block states}
The effective on-site symmetry of 1D blocks labeled by $\tau_{1,2}$ is $\mathbb{Z}_3\rtimes\mathbb{Z}_2$, by 3-fold rotation and reflection generators, $\bs{R}_3^{1,2,3}$ and $\bs{M}_1^{k}$ ($k=1,...,6$) acting internally. The effective on-site symmetry of 1D blocks labeled by $\tau_3$ is $\mathbb{Z}_2\times\mathbb{Z}_2$, by 2-fold rotation and reflection generators, $\bs{R}_2^{1,2,3}$ and $\bs{M}_1^{k}$ ($k=1,...,6$) acting internally. We discuss the spinless fermions and spin-1/2 fermions separately.

\paragraph{Spinless fermions}
For spinless fermions, there are two possible root phases on each $\tau_{1,2}$:
\begin{enumerate}[1.]
\item Majorana chain;
\item 1D fSPT phase, composed of double Majorana chains.
\end{enumerate}

Similar to the $T$-symmetric lattice, Majorana chain decoration solely on $\tau_1$ or $\tau_2$ is \textit{obstructed}, but jointly on $\tau_1$ and $\tau_2$ is \textit{obstruction-free}.

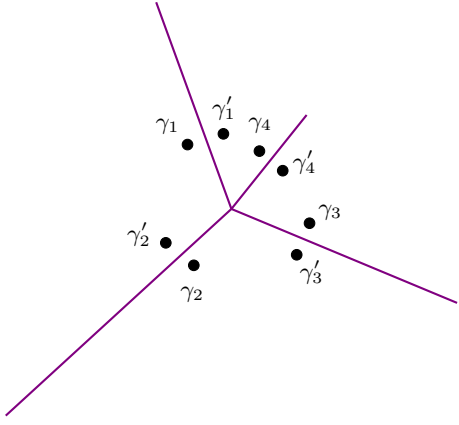
\begin{figure}
\begin{tikzpicture}
\tikzstyle{sergio}=[rectangle,draw=none]
\draw[thick,color=violet] (-1,1.5) -- (0,-1.25);
\draw[thick,color=violet] (0,-1.25) -- (1,0);
\draw[thick,color=violet] (0,-1.25) -- (-3,-4);
\draw[thick,color=violet] (0,-1.25) -- (3,-2.5);
\filldraw[fill=black, draw=black] (-0.5836,-0.3935)circle (2pt);
\filldraw[fill=black, draw=black] (-0.1064,-0.2508)circle (2pt);
\filldraw[fill=black, draw=black] (1.0385,-1.4399)circle (2pt);
\filldraw[fill=black, draw=black] (0.868,-1.8597)circle (2pt);
\filldraw[fill=black, draw=black] (0.3725,-0.4805)circle (2pt);
\filldraw[fill=black, draw=black] (0.6822,-0.741)circle (2pt);
\filldraw[fill=black, draw=black] (-0.5,-2)circle (2pt);
\filldraw[fill=black, draw=black] (-0.8722,-1.7008)circle (2pt);
\path (-0.8478,-0.1522) node [style=sergio] {$\gamma_1$};
\path (-0.0997,0.0903) node [style=sergio] {$\gamma_1'$};
\path (-0.5195,-2.3872) node [style=sergio] {$\gamma_2$};
\path (-1.2279,-1.587) node [style=sergio] {$\gamma_2'$};
\path (1.3133,-1.2983) node [style=sergio] {$\gamma_3$};
\path (1.0663,-2.0876) node [style=sergio] {$\gamma_3'$};
\path (0.3841,-0.1345) node [style=sergio] {$\gamma_4$};
\path (0.9723,-0.6279) node [style=sergio] {$\gamma_4'$};
\end{tikzpicture}
\caption{Double Majorana chains decoration on 1D blocks $\tau_1$, leaving 8 Majorana zero modes $\gamma_j$ and $\gamma_j'$ ($j=1,2,3,4$) near the 0D block $\mu$.}
\label{double}
\end{figure}

Subsequently, double Majorana chains decoration on 1D blocks $\tau_1$ leaves 8 Majorana zero modes at 0D block $\mu$ (see Fig. \ref{double}). These Majorana zero modes have the following symmetry properties:
\begin{align}
\bs{R}_3^1:~\left\{
\begin{aligned}
&\left(\gamma_1,\gamma_2,\gamma_3,\gamma_4\right)\mapsto\left(\gamma_1,\gamma_3,\gamma_4,\gamma_2\right)\\
&\left(\gamma_1',\gamma_2',\gamma_3',\gamma_4'\right)\mapsto\left(\gamma_1',\gamma_3',\gamma_4',\gamma_2'\right)
\end{aligned}
\right.
\end{align}
\begin{align}
\bs{R}_2^1:~\left\{
\begin{aligned}
&\left(\gamma_1,\gamma_2,\gamma_3,\gamma_4\right)\mapsto\left(\gamma_4,\gamma_3,\gamma_2,\gamma_1\right)\\
&\left(\gamma_1',\gamma_2',\gamma_3',\gamma_4'\right)\mapsto\left(\gamma_4',\gamma_3',\gamma_2',\gamma_1'\right)
\end{aligned}
\right.
\end{align}
\begin{align}
\bs{M}_1^2:~\left\{
\begin{aligned}
&\left(\gamma_1,\gamma_2,\gamma_3,\gamma_4\right)\mapsto\left(\gamma_1',\gamma_3',\gamma_2',\gamma_4'\right)\\
&\left(\gamma_1',\gamma_2',\gamma_3',\gamma_4'\right)\mapsto\left(\gamma_1,\gamma_3,\gamma_2,\gamma_4\right)
\end{aligned}
\right.
\end{align}

We first introduce an interacting Hamiltonian $H_U$ that is symmetric under $\bs{R}_3^1$, $\bs{R}_2^1$ and $\bs{M}_1^2$ ($U>0$):
\begin{align}
H_U=U\left(\gamma_1\gamma_2\gamma_3\gamma_4+\gamma_1'\gamma_2'\gamma_3'\gamma_4'\right)
\end{align}
$H_U$ can open a gap with 4-fold GSD characterized by:
\begin{align}
\gamma_1\gamma_2\gamma_3\gamma_4=\gamma_1'\gamma_2'\gamma_3'\gamma_4'=-1
\label{subspace}
\end{align}
We restrict the Hilbert space to the subspace constrained by Eq. (\ref{subspace}). In this subspace, we can further define two spin-1/2 degrees of freedom:
\begin{align}
S_x=\frac{i}{2}\gamma_1\gamma_2,~S_y=\frac{i}{2}\gamma_1\gamma_3,~S_z=\frac{i}{2}\gamma_1\gamma_4
\end{align}
and
\begin{align}
S_x'=\frac{i}{2}\gamma_1'\gamma_2',~S_y'=\frac{i}{2}\gamma_1'\gamma_3',~S_z'=\frac{i}{2}\gamma_1'\gamma_4'
\end{align}
and they satisfy the commutation relations of spin-1/2 degrees of freedom:
\begin{align}
\left[S_\mu,S_\nu\right]=i\epsilon_{\mu\nu\rho}S_\rho,~\left[S_\mu',S_\nu'\right]=i\epsilon_{\mu\nu\rho}S_\rho'
\end{align}
where $\epsilon_{\mu\nu\rho}$ is the Levi-Civita symbol and $\mu,\nu,\rho=x,y,z$. Furthermore, their symmetry properties are:
\begin{align}
\bs{R}_3^1:~\left\{
\begin{aligned}
&\left(S_x,S_y,S_z\right)\mapsto\left(S_y,S_z,S_x\right)\\
&\left(S_x',S_y',S_z'\right)\mapsto\left(S_y',S_z',S_x'\right)
\end{aligned}
\right.
\end{align}
\begin{align}
\bs{R}_2^1:~\left(\bs{S},\bs{S}'\right)\mapsto-\left(\bs{S},\bs{S}'\right)
\end{align}
\begin{align}
\bs{M}_1^2:~\left\{
\begin{aligned}
&\left(S_x,S_y,S_z\right)\mapsto\left(S_y',S_x',S_z'\right)\\
&\left(S_x',S_y',S_z'\right)\mapsto\left(S_y,S_x,S_z\right)
\end{aligned}
\right.
\end{align}
Therefore, we can further add a Hamiltonian between these two spin-1/2 degrees of freedom:
\begin{align}
H_J=J\bs{S}\cdot\bs{S}',~~J>0
\end{align}
and $H_J$ splits the 4-fold degenerate ground states to a nondegenerate spin-singlet state with lower energy and 3-fold degenerate spin-triplet states. As a consequence, the dangling Majorana zero modes $\gamma_j$ and $\gamma_j'$ ($j=1,2,3,4$) can be fully gapped by Hamiltonian $H_U$ and $H_J$, and the double Majorana chains decoration on $\tau_1$ is \textit{obstruction-free}. Similar for the double Majorana chains decoration on $\tau_2$. As a consequence, all obstruction-free 1D block states from $\tau_1$ and $\tau_2$ form a $\mathbb{Z}_2^3$ group with three generators:
\begin{enumerate}[1.]
\item Majorana chain decoration on both $\tau_1$ and $\tau_2$;
\item Double Majorana chains decoration on $\tau_1$ or $\tau_2$.
\end{enumerate}
For 1D blocks $\tau_3$, there are several possible root phases:
\begin{enumerate}[1.]
\item Majorana chain;
\item 1D fSPT phase, formed by double Majorana chains; \item 1D Haldane chain for integer spins.
\end{enumerate}
Similar to the $T_h$-symmetric systems, the first two root phases are \textit{obstructed} , but the last root phase is \textit{obstruction-free}. Therefore, all obstruction-free 1D block states form a $\{\mathrm{OFBS}\}_{T_d}^{\mathrm{1D}}=\mathbb{Z}_2^4$ group.

We should further investigate the possible trivializations. Recall the 2D blocks $\sigma_{1,2,3}$ with $\mathbb{Z}_2$ on-site symmetry, there are two possible 2D bubble constructions:
\begin{enumerate}[1.]
\item ``Majorana'' bubble construction;
\item ``Double Majorana'' bubble construction;
\end{enumerate}

``Majorana'' bubble on each 2D block $\sigma_3$ leaves 3 Majorana chains on each 1D block $\tau_1/\tau_2$ which is equivalent to a single Majorana chain. Equivalently, Majorana chain decoration on both 1D blocks $\tau_1$ and $\tau_2$ can be trivialized by the ``Majorana'' bubble on each 2D block $\sigma_3$. 

``Double Majorana'' bubble on each 2D block $\sigma_3$ leaves three 1D fSPT phases on each 1D block $\tau_1/\tau_2$ which is equivalent to double Majorana chains on each of them. Equivalently, double Majorana chains decoration on both 1D blocks $\tau_1$ and $\tau_2$ can be trivialized by 2D ``double Majorana'' bubble on each 2D block $\sigma_3$.

``Double Majorana'' bubble construction on each 2D block $\sigma_1$ leaves three 1D fSPT phases on each 1D block $\tau_1$ and two 1D fSPT phases on each 1D block $\tau_3$ which is equivalent to a Haldane chain on each $\tau_3$. Equivalently, the double Majorana chains decoration on each 1D block $\tau_1$ can be smoothly deformed to the Haldane chain decoration on each 1D block $\tau_3$, and these two 1D block states are topologically equivalent. Thus for spinless fermions, all obstruction-free that can be trivialized by 2D bubble constructions form the group $\{\mathrm{TBS}\}_{T_d}^{\mathrm{1D}}=\mathbb{Z}_2^3$, and all obstruction and trivialization free 1D block states form the quotient group of $\{\mathrm{OFBS}\}_{T_d}^{\mathrm{1D}}$ and $\{\mathrm{TBS}\}_{T_d}^{\mathrm{1D}}$:
\begin{align}
\mathcal{G}_{T_d}^{\mathrm{1D}}=\{\mathrm{OFBS}\}_{T_d}^{\mathrm{1D}}/\{\mathrm{TBS}\}_{T_d}^{\mathrm{1D}}=\mathbb{Z}_2
\end{align}

\paragraph{Spin-1/2 fermions}For spin-1/2 fermions, there is no root phase \cite{dihedral,wallpaper}, hence there is no nontrivial 1D block states.

\subsubsection{0D block states}
The effective ``on-site'' symmetry on the 0D block $\mu$ is $S_4$, hence the candidate 0D block states are classified by the following two indices:
\begin{align}
\begin{aligned}
&n_0\in\mathcal{H}^0(S_4,\mathbb{Z}_2)=\mathbb{Z}_2\\
&\nu_1\in\mathcal{H}^1\left[S_4,U(1)\right]=\mathbb{Z}_2
\end{aligned}
\label{1-supercohomology Td}
\end{align}
with the twisted cocycle condition:
\begin{align}
\mathrm{d}\nu_1=(-1)^{\omega_2\smile n_0}
\label{1-twisted Td}
\end{align}
where $n_0$ depicts the parity of fermions, and $\nu_1$ depicts the 0D bSPT mode on the 0D block $\mu$ protected by $S_4$ symmetry, characterizing the eigenvalues of the reflection generator $\bs{M}_1^{k}\in T_d$ ($k=1,...,6$), i.e., $\nu_1(\bs{M}_1^k)=-1$. For spinless fermions, we demonstrate that this 0D bSPT mode would be trivialized by 1D bubble construction: consider the 1D blocks labeled by $\tau_1$ and decorate a complex fermion on each of them, these complex fermions are labeled by $c_j^\dag$ ($j=1,2,3,4$). Near the 0D block $\mu$, these complex fermions form an atomic insulator:
\begin{align}
|\psi_\mu\rangle=c_1^\dag c_2^\dag c_3^\dag c_4^\dag|0\rangle
\label{Td atomic insulator}
\end{align}
Furthermore, these complex fermions will be transformed by reflection $\bs{M}_1^2$ as the following way:
\begin{align}
\bs{M}_1^2:\left(c_1^\dag, c_2^\dag, c_3^\dag, c_4^\dag\right)\mapsto\left(c_1^\dag, c_3^\dag, c_2^\dag, c_4^\dag\right)
\end{align}
As a consequence, the atomic insulator $|\psi_\mu\rangle$ is an eigenstate of $\bs{M}_1^2$ with eignevalue $-1$:
\begin{align}
\bs{M}_1^2|\psi_\mu\rangle=c_1^\dag c_3^\dag c_2^\dag c_4^\dag|0\rangle=-|\psi_\mu\rangle
\end{align}
Similarly, $|\psi_\mu\rangle$ is also an eigenstate of $\bs{M}_1^1$ with eigenvalue $-1$. Similar for all other reflection generators $\bs{M}_1^{3,4,5,6}$. Therefore, the nontrivial 0D bSPT block state decorated on $\mu$ would be trivialized by 1D bubble construction on $\tau_1$. 

Then consider the fermion parity of the 0D block $\mu$. Similar to the $T_h$-symmetric case, for spinless fermions, complex fermion decoration on 0D block $\mu$ is a nontrivial block state. 

For spin-1/2 fermions, the index $n_0$ in Eq. (\ref{1-supercohomology Td}) representing the fermion parity of 0D block state on $\mu$ is \textit{obstructed}. Furthermore, we repeatedly consider the atomic insulator $|\psi_\mu\rangle$, to satisfy the spin-1/2 condition (i.e., $\left(\bs{M}_1^2\right)^2=-1$), we should refine the symmetry properties of the complex fermions $c_{1,2,3,4}^\dag$ to:
\begin{align}
\bs{M}_1^2:\left(c_1^\dag, c_2^\dag, c_3^\dag, c_4^\dag\right)\mapsto\left(ic_1^\dag, c_3^\dag,-c_2^\dag, ic_4^\dag\right)
\end{align}
Then the atomic insulator $|\psi_\mu\rangle$ is an eigenstate of $\bs{M}_1^2$ with eigenvalue $-1$:
\begin{align}
\bs{M}_1^2|\psi_\mu\rangle=\left(ic_1^\dag\right)c_3^\dag\left(-c_2^\dag\right)\left(ic_4^\dag\right)|0\rangle=-|\psi_\mu\rangle
\end{align}
On the other hand, under $\bs{M}_1^1$, complex fermions $c_{1,2,3,4}^\dag$ transforms as:
\begin{align}
\bs{M}_1^1:\left(c_1^\dag, c_2^\dag, c_3^\dag, c_4^\dag\right)\mapsto\left(c_4^\dag, ic_3^\dag,-ic_2^\dag,-c_1^\dag\right)
\end{align}
Then the atomic insulator $|\psi_\mu\rangle$ is an eigenstate of $\bs{M}_1^1$ with eigenvalue $+1$. Similar for $\bs{M}_1^{3,4,5,6}$,
and there is no nontrivial 0D block state on $\mu$.

\subsubsection{Summary\label{summary Td}}
In this section, we summarize the classifications of 3D topological crystalline superconductors in the systems with $T_d$ point group symmetry, for both spinless and spin-1/2 fermions. For spinless fermions, the ultimate classification is:
\begin{align}
\mathcal{G}_{T_d}^0=\mathbb{Z}_2^3
\end{align}
with the following block states as the root phases:
\begin{enumerate}[1.]
\item 2D bosonic Levin-Gu states on $\sigma_1$ and $\sigma_2$;
\item Haldane chain decoration on each 1D block $\tau_3$, which is equivalent to double Majorana chains decoration on each $\tau_1$ or $\tau_2$ ($\mathbb{Z}_2$);
\item Complex fermion decoration on 0D block $\mu$ ($\mathbb{Z}_2$);
\end{enumerate}
And there is a nontrivial group extension between the last two root phases.

For spin-1/2 fermions, there is no nontrivial block state, hence the classification is trivial:
\begin{align}
\mathcal{G}_{T_d}^{1/2}=\mathbb{Z}_1
\end{align}

\subsubsection{Higher-order topological surface theory}
Having established concrete block states, we can now delve into the analysis of the corresponding higher-order (HO) topological surface theories through the lens of the bulk-boundary correspondence.

For spinless fermions, we examine these root phases on the open cubic depicted in Figure \ref{Td cell decomposition}. In the case of the 2D bosonic Levin-Gu state decoration on each 2D block, the resulting topological surface theory inherits an assembly of 1D nonchiral Luttinger liquids on the system's boundary. These nonchiral Luttinger liquids manifest themselves at the hinges of the cubic, including the diagonal and off-diagonal links on all surfaces. Consequently, the nontrivial 2D block states discussed in this paragraph constitute a 3D second-order topological phase.

Considering the 1D root phase arising from the Haldane chain decoration on each 1D block $\tau_3$, a spin-1/2 degree of freedom emerges at the center of each surface of the open cubic. Consequently, the nontrivial topological phase constructed from this 1D block state corresponds to a 3D bosonic third-order topological phase.

On the other hand, the complex fermion decoration on the 0D block $\mu$ represents a 0D block state that does not yield any observable features on the surface of the open cubic. Consequently, the 3D $T_d$-symmetric topological phase characterized by this 0D block state does not qualify as a nontrivial higher-order (HO) topological phase.

For systems involving spin-1/2 fermions, the corresponding classification of 3D $T_d$-symmetric topological phases is trivial. As a result, no higher-order (HO) topological surface states are present in this scenario.

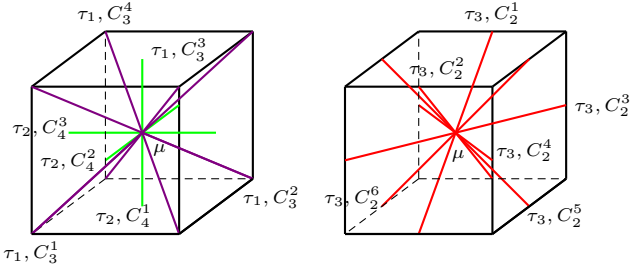
\begin{figure}
\begin{tikzpicture}[scale=0.49]
\tikzstyle{sergio}=[rectangle,draw=none]
\draw[thick] (7.5,1.5) -- (11.5,1.5);
\draw[thick] (7.5,1.5) -- (5.5,0);
\draw[thick] (11.5,1.5) -- (9.5,0);
\draw[thick] (5.5,0) -- (9.5,0);
\draw[thick] (5.5,0) -- (5.5,-4);
\draw[thick] (9.5,-4) -- (5.5,-4);
\draw[thick] (9.5,-4) -- (9.5,0);
\draw[thick] (11.5,-2.5) -- (11.5,1.5);
\draw[thick] (11.5,-2.5) -- (9.5,-4);
\draw[thick,color=red] (9.5,-2) -- (7.5,-0.5);
\draw[thick,color=red] (11.5,-0.5) -- (5.5,-2);
\draw[thick,color=red] (7.5,-4) -- (9.5,1.5);
\draw[thick,color=red] (10.5,-3.25) -- (6.5,0.75);
\draw[thick,color=red] (9.5,-2.5) -- (7.5,0);
\draw[thick,color=red] (6.5,-3.25) -- (10.5,0.75);
\draw[densely dashed] (7.5,-2.5) -- (7.5,1.5);
\draw[densely dashed] (7.5,-2.5) -- (11.5,-2.5);
\draw[densely dashed] (7.5,-2.5) -- (5.5,-4);
\draw[thick] (11.5,-2.5) -- (9.5,-4);
\draw[thick] (5.5,-4) -- (9.5,-4);
\path (9.5,2) node [style=sergio] {\scriptsize$\tau_3,C_2^1$};
\path (10.3733,-1.7172) node [style=sergio] {\scriptsize$\tau_3,C_2^4$};
\path (8,0.5) node [style=sergio] {\scriptsize$\tau_3,C_2^2$};
\path (12.5,-0.5) node [style=sergio] {\scriptsize$\tau_3,C_2^3$};
\path (11.2019,-3.5061) node [style=sergio] {\scriptsize$\tau_3,C_2^5$};
\path (5.7036,-3.0105) node [style=sergio] {\scriptsize$\tau_3,C_2^6$};
\path (8.5717,-1.8688) node [style=sergio] {\scriptsize$\mu$};
\draw[thick] (-1,1.5) -- (3,1.5);
\draw[thick] (-1,1.5) -- (-3,0);
\draw[thick] (3,1.5) -- (1,0);
\draw[thick] (-3,0) -- (1,0);
\draw[thick] (-3,0) -- (-3,-4);
\draw[thick] (1,-4) -- (-3,-4);
\draw[thick] (1,-4) -- (1,0);
\draw[thick] (3,-2.5) -- (3,1.5);
\draw[thick] (3,-2.5) -- (1,-4);
\draw[thick,color=green] (0,-1.25) -- (0,0.75);
\draw[thick,color=green] (0,-1.25) -- (0,-3.25);
\draw[thick,color=green] (0,-1.25) -- (-2,-1.25);
\draw[thick,color=green] (0,-1.25) -- (2,-1.25);
\draw[thick,color=green] (0,-1.25) -- (1,-0.5);
\draw[thick,color=green] (0,-1.25) -- (-1,-2);
\draw[densely dashed] (-1,-2.5) -- (-1,1.5);
\draw[densely dashed] (-1,-2.5) -- (3,-2.5);
\draw[densely dashed] (-1,-2.5) -- (-3,-4);
\draw[thick,color=violet] (3,1.5) -- (-3,-4);
\draw[thick,color=violet] (-1,1.5) -- (0,-1.25);
\draw[thick,color=violet] (-3,0) -- (3,-2.5);
\draw[thick,color=violet] (0,-1.25) -- (1,0);
\path (-3,-4.5) node [style=sergio] {\scriptsize$\tau_1,C_3^1$};
\path (3.5,-3) node [style=sergio] {\scriptsize$\tau_1,C_3^2$};
\path (1,1) node [style=sergio] {\scriptsize$\tau_1,C_3^3$};
\path (-1,2) node [style=sergio] {\scriptsize$\tau_1,C_3^4$};
\draw[thick,color=violet] (0,-1.25) -- (-3,-4);
\draw[thick,color=violet] (0,-1.25) -- (-1,-2.5);
\draw[thick,color=violet] (0,-1.25) -- (3,-2.5);
\draw[thick,color=violet] (0,-1.25) -- (1,-4);
\draw[thick] (3,-2.5) -- (1,-4);
\draw[thick] (-3,-4) -- (1,-4);
\path (-0.5,-3.5) node [style=sergio] {\scriptsize$\tau_2,C_4^1$};
\path (-2.7633,-1.1189) node [style=sergio] {\scriptsize$\tau_2,C_4^3$};
\path (-2,-2) node [style=sergio] {\scriptsize$\tau_2,C_4^2$};
\path (0.4668,-1.724) node [style=sergio] {\scriptsize$\mu$};
\end{tikzpicture}
\caption{The cell decomposition of the 3D system with point group symmetry $O$. There are four axes of 3-fold rotation symmetry labeled by $C_3^{1,2,3,4}$ across the center of the system (labeled by $\mu$); three axes of 4-fold rotation symmetry across the center, labeled by $C_4^{1,2,3}$ (see left panel); and six axes of 2-fold rotation symmetry across the center, labeled by $C_2^{1,2,3,4,5,6}$ (see right panel).}
\label{O cell decomposition}
\end{figure}

\subsection{$O$-symmetric lattice}
For $O$-symmetric cubic, by cell decomposition as illustrated in Figs. \ref{O cell decomposition} and \ref{O unit cell}, the ground-state wavefunction of the system can be decomposed to the direct product of wavefunctions of lower-dimensional blocks as:
\begin{align}
|\Psi\rangle=\bigotimes_{g\in O}|T_{g\lambda}\rangle\otimes\bigotimes\limits_{k=1}^2|T_{g\sigma_k}\rangle\otimes\bigotimes\limits_{l=1}^3|\beta_{g\tau_l}\rangle\otimes|\alpha_\mu\rangle
\label{O wavefunction}
\end{align}
where $|T_{g\lambda}\rangle$ is the wavefunction of 3D block state $g\lambda$ which is topologically trivial; $|T_{g\sigma_{1,2}}\rangle$ is the wavefunction of 2D block state $g\sigma_{1,2}$ which is topologically trivial; $|\beta_{g\tau_1}\rangle$ is the wavefunction of 1D block state $g\tau_1$ which is $\mathbb{Z}_3$-symmetric, $|\beta_{g\tau_2}\rangle$ is the wavefunction of 1D block state $g\tau_2$ which is $\mathbb{Z}_4$-symmetric, and $|\beta_{g\tau_3}\rangle$ is the wavefunction of 1D block state $g\tau_3$ which is $\mathbb{Z}_2$-symmetric; $|\alpha_\mu\rangle$ is the wavefunction of 0D block state $\mu$ which is $S_4$-symmetric.

With topological crystals, we decorate the lower-dimensional block states and investigate the possible \textit{obstructions} and \textit{trivializations}.

\subsubsection{2D block states\label{O2D}}
There is no effective ``on-site'' symmetry on all 2D blocks. The only possible root phase on 2D blocks is 2D $(p+ip)$-SC.  If we decorate a 2D $(p+ip)$-SC with quantum number $n_k\in\mathbb{Z}$ on each 2D block $\sigma_k$ ($k=1,2$), the chiral central charges on the 1D blocks labeled by $\tau_1$, $\tau_2$ and $\tau_3$ are $(3n_1+3n_2)/2$, $2n_2$ and $n_1$, respectively. Fully gapped bulk requires that all of these three quantities should vanish:
\[
3n_1+3n_2=4n_2=2n_1=0
\]
The only solution to these equations is $n_1=n_2=0$, hence all nontrivial $(p+ip)$-SC decoration on 2D blocks is \textit{obstructed}. 

\begin{figure}
\begin{tikzpicture}[scale=0.75]
\tikzstyle{sergio}=[rectangle,draw=none]
\draw[thick] (-3,-0.5) -- (1,-0.5);
\draw[densely dashed,thick] (-1,1) -- (3,1);
\draw[densely dashed,thick] (-1,1) -- (-3,-0.5);
\draw[thick] (3,1) -- (1,-0.5);
\draw[thick,color=violet] (-1,1) -- (0,2.25);
\draw[thick,color=violet] (1,-0.5) -- (0,2.25);
\draw[thick,color=violet] (-3,-0.5) -- (0,2.25);
\draw[thick,color=violet] (3,1) -- (0,2.25);
\draw[thick,color=green] (0,0.25) -- (0,2.25);
\filldraw[fill=black, draw=black] (0,2.25)circle (3pt);
\draw[densely dashed,thick] (-1,1) -- (1,-0.5);
\draw[densely dashed,thick] (3,1) -- (-3,-0.5);
\draw[thick,color=red] (1,1) -- (0,2.25);
\draw[thick,color=red] (-1,-0.5) -- (0,2.25);
\draw[thick,color=red] (-2,0.25) -- (0,2.25);
\draw[thick,color=red] (2,0.25) -- (0,2.25);
\draw[thick] (3.5,-0.5) -- (7.5,-0.5);
\draw[densely dashed,thick] (6.5,0.25) -- (7.5,-0.5);
\draw[densely dashed,thick] (6.5,0.25) -- (3.5,-0.5);
\draw[thick,color=violet] (6.5,2.25) -- (7.5,-0.5);
\draw[thick,color=violet] (6.5,2.25) -- (3.5,-0.5);
\draw[thick,color=green] (6.5,2.25) -- (6.5,0.25);
\draw[thick,color=red] (6.5,2.25) -- (5.5,-0.5);
\path (4.5,1.5) node [style=sergio] {$\sigma_2$};
\draw[thick,->] (5.5,0.5) -- (4.75,1.25);
\draw[thick,->] (6.7938,0.6294) -- (7.5176,1.2044);
\path (7.5,1.5) node [style=sergio] {$\sigma_2$};
\path (5.0784,0.141) node [style=sergio] {$\sigma_1$};
\path (6.4415,-0.1135) node [style=sergio] {$\sigma_1$};
\end{tikzpicture}
\caption{The unit cell of cell decomposition of the cubic lattice with point symmetry $O$. The left panel depicts the bottom rectangular pyramid of the cubic in Fig. \ref{O cell decomposition}; the right panel illustrates the independent triangular pyramid $\lambda$, where $\sigma_1$ and $\sigma_2$ are two independent 2D blocks in the system.}
\label{O unit cell}
\end{figure}

\subsubsection{1D block states}
The effective ``on-site'' symmetry on the 1D block labeled by $\tau_{1}$ is $\mathbb{Z}_3$ by 3-fold rotation symmetry acting internally. Hence the candidate 1D block states are classified by group supercohomology (\ref{2-supercohomology}) with twisted cocycle condition (\ref{2-twisted}) \cite{special,general1,general2}, and $G_{\mathrm{1D}}=\mathbb{Z}_3$. For 1D blocks $\tau_{1,2}$, $\mathcal{H}^2(\mathbb{Z}_3,\mathbb{Z}_2)=0$, hence the spin of fermions is irrelevant. Eqs. (\ref{2-supercohomology}) and (\ref{2-twisted}) indicate that the only possible nontrivial 1D block state is Majorana chain characterized by nonzero $n_0$ in Eq. (\ref{2-supercohomology}). 

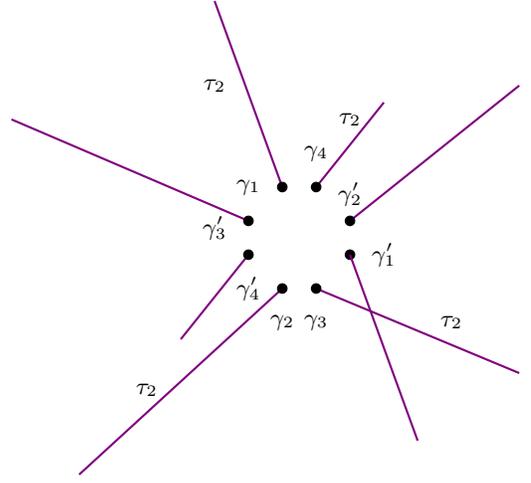
\begin{figure}
\begin{tikzpicture}[scale=0.9]
\tikzstyle{sergio}=[rectangle,draw=none]
\path (-1,0.5) node [style=sergio] {$\tau_2$};
\path (-2,-4) node [style=sergio] {$\tau_2$};
\path (1,0) node [style=sergio] {$\tau_2$};
\path (2.5,-3) node [style=sergio] {$\tau_2$};
\draw[thick,color=violet] (-1,1.75) -- (0,-1);
\draw[thick,color=violet] (0,-2.5) -- (-3,-5.25);
\draw[thick,color=violet] (0.5,-1) -- (1.5,0.25);
\draw[thick,color=violet] (0.5,-2.5) -- (3.5,-3.75);
\filldraw[fill=black, draw=black] (0,-1)circle (2pt);
\filldraw[fill=black, draw=black] (0.5,-1)circle (2pt);
\filldraw[fill=black, draw=black] (0,-2.5)circle (2pt);
\filldraw[fill=black, draw=black] (0.5,-2.5)circle (2pt);
\path (-0.5,-1) node [style=sergio] {$\gamma_1$};
\path (0,-3) node [style=sergio] {$\gamma_2$};
\path (0.5,-3) node [style=sergio] {$\gamma_3$};
\path (0.5,-0.5) node [style=sergio] {$\gamma_4$};
\draw[thick,color=violet] (-4,0) -- (-0.5,-1.5);
\path (-1,-1.6) node [style=sergio] {$\gamma_3'$};
\filldraw[fill=black, draw=black] (-0.5,-1.5)circle (2pt);
\draw[thick,color=violet] (-0.5,-2) -- (-1.5,-3.25);
\path (-0.5,-2.5) node [style=sergio] {$\gamma_4'$};
\filldraw[fill=black, draw=black] (-0.5,-2)circle (2pt);
\draw[thick,color=violet] (3.5,0.5) -- (1,-1.5);
\path (1.5,-2) node [style=sergio] {$\gamma_1'$};
\filldraw[fill=black, draw=black] (1,-2)circle (2pt);
\draw[thick,color=violet] (1,-2) -- (2,-4.75);
\path (1,-1.1) node [style=sergio] {$\gamma_2'$};
\filldraw[fill=black, draw=black] (1,-1.5)circle (2pt);
\end{tikzpicture}
\caption{Majorana chain decorations on 1D blocks labeled by $\tau_1$, which leaves 8 dangling Majorana zero modes $\gamma_j$ and $\gamma_j'$ ($j=1,2,3,4$) at the 0D block $\mu$.}
\label{tau1}
\end{figure}

Consider the Majorana chain decorations on 1D blocks $\tau_1$, as indicated in Fig. \ref{tau1} that leaves 8 dangling Majorana zero modes $\gamma_j$ and $\gamma_j'$ ($j=1,2,3,4$). The symmetry properties of these Majorana zero modes under 3-fold rotation are:
\begin{align}
\bs{R}_3:\left\{
\begin{aligned}
&\left(\gamma_1,\gamma_2,\gamma_3,\gamma_4\right)\mapsto\left(\gamma_1,\gamma_3,\gamma_4,\gamma_2\right)\\
&\left(\gamma_1',\gamma_2',\gamma_3',\gamma_4'\right)\mapsto\left(\gamma_1',\gamma_3',\gamma_4',\gamma_2'\right)
\end{aligned}
\right.
\end{align}
Under 4-fold rotation $\bs{R}_4\in O$, these Majorana zero modes are transformed as:
\begin{align}
\bs{R}_4:\left\{
\begin{aligned}
&\left(\gamma_1,\gamma_2,\gamma_3,\gamma_4\right)\mapsto\left(\gamma_3',\gamma_1',\gamma_4',\gamma_2'\right)\\
&\left(\gamma_1',\gamma_2',\gamma_3',\gamma_4'\right)\mapsto\left(\gamma_3,\gamma_1,\gamma_4,\gamma_2\right)
\end{aligned}
\right.
\end{align}

According to these 8 Majorana zero modes, we can define four complex fermions:
\begin{align}
c_j^\dag=\frac{1}{2}\left(\gamma_j+i\gamma_j'\right),~j=1,2,3,4.
\end{align}
And these complex fermions can form a 16-dimensional local Hilbert space, spanned by the following states (where $|0\rangle$ is the vacuum state):
\begin{align}
&|1\rangle=|0\rangle,~|2\rangle=c_1^\dag|0\rangle,~|3\rangle=c_2^\dag|0\rangle,~|4\rangle=c_3^\dag|0\rangle\nonumber\\
&|5\rangle=c_4^\dag|0\rangle,~|6\rangle=c_1^\dag c_2^\dag|0\rangle,~|7\rangle=c_1^\dag c_3^\dag|0\rangle\nonumber\\
&|8\rangle=c_1^\dag c_4^\dag|0\rangle,~|9\rangle=c_2^\dag c_3^\dag|0\rangle,~|10\rangle=c_2^\dag c_4^\dag|0\rangle\nonumber\\
&|11\rangle=c_3^\dag c_4^\dag|0\rangle,~|12\rangle=c_1^\dag c_2^\dag c_3^\dag|0\rangle,~|13\rangle=c_1^\dag c_2^\dag c_4^\dag|0\rangle\nonumber\\
&|14\rangle=c_1^\dag c_3^\dag c_4^\dag|0\rangle,~|15\rangle=c_2^\dag c_3^\dag c_4^\dag|0\rangle,~|16\rangle=c_1^\dag c_2^\dag c_3^\dag c_4^\dag|0\rangle\nonumber
\end{align}

In this Hilbert space, Majorana zero modes $\gamma_j$ and $\gamma_j'$ can be represented as $16\times16$ matrices. The symmetry operations $\bs{R}_3$ and $\bs{R}_4$ can be formulated in terms of matrix representations of Majorana zero modes $\gamma_j$ and $\gamma_j'$ as a representation $A$ of the group $O$, as:
\begin{align}
A(\bs{R}_3)=\frac{1}{4}\left(\gamma_2-\gamma_3\right)\left(\gamma_3-\gamma_4\right)\left(\gamma_2'-\gamma_3'\right)\left(\gamma_3'-\gamma_4'\right)
\end{align}
\begin{align}
A(\bs{R}_4)=&\frac{1}{8}\left(\gamma_1-\gamma_3'\right)\left(\gamma_3'-\gamma_4\right)\left(\gamma_4-\gamma_2'\right)\nonumber\\
&\cdot\left(\gamma_1'-\gamma_3\right)\left(\gamma_3-\gamma_4'\right)\left(\gamma_4'-\gamma_2\right)
\end{align}

Then we further consider another group element $\bs{R}_3\bs{R}_4\in O$. Under $\bs{R}_3\bs{R}_4$, aforementioned Majorana zero modes $\gamma_j$ and $\gamma_j'$ ($j=1,2,3,4$) are transformed as:
\begin{align}
\bs{R}_3\bs{R}_4:~\left\{
\begin{aligned}
&\left(\gamma_1,\gamma_2,\gamma_3,\gamma_4\right)\mapsto\left(\gamma_4',\gamma_1',\gamma_2',\gamma_3'\right)\\
&\left(\gamma_1',\gamma_2',\gamma_3',\gamma_4'\right)\mapsto\left(\gamma_4,\gamma_1,\gamma_2,\gamma_3\right)
\end{aligned}
\right.
\end{align}
We can also represent $\bs{R}_3\bs{R}_4$ in the aforementioned 16-dimensional Hilbert space in terms of the matrix representations of Majorana zero modes $\gamma_j$ and $\gamma_j'$:
\begin{align}
A(\bs{R}_3\bs{R}_4)=&\frac{1}{8}\left(\gamma_1-\gamma_4'\right)\left(\gamma_4'-\gamma_3\right)\left(\gamma_3-\gamma_2'\right)\nonumber\\
&\cdot\left(\gamma_1'-\gamma_4\right)\left(\gamma_4-\gamma_3'\right)\left(\gamma_3'-\gamma_2\right)
\end{align}
Nevertheless, we can straightforwardly check that $A$ is a projective representation of the point group $O$ because of the following relation:
\begin{align}
A(\bs{R}_3)A(\bs{R}_4)=-A(\bs{R}_3\bs{R}_4)
\end{align}
As a consequence, the Majorana zero modes $\gamma_j$ and $\gamma_j'$ cannot be gapped in an $O$-symmetric way without ground-state degeneracy.

Subsequently, we consider the block state decoration on the 1D blocks $\tau_2$. The effective ``on-site'' symmetry on the 1D block $\tau_2$ is $\mathbb{Z}_4$ by 4-fold rotation symmetry $C_4$ acting internally. Hence the candidate 1D block states are classified by group supercohomology (\ref{2-supercohomology}) with twisted cocycle conditions (\ref{2-twisted}), and $G_{\mathrm{1D}}=\mathbb{Z}_4$ \cite{special,general1,general2}. We discuss the spinless and spin-1/2 fermions separately.

\paragraph{Spinless fermions}For spinless fermions, there is $(\Z_2)^2$ classification, whose two root phases on the 1D block $\tau_2$:
\begin{enumerate}[1.]
\item Majorana chain;
\item 1D complex fermion decoration (fSPT) phase, formed by quadruple Majorana chains; $\mathbb{Z}_4$ symmetry action permutes them [from $(1234)$ to $(2341)$].
\end{enumerate}

Majorana chain decoration on each 1D block $\tau_3$ leaves 6 dangling Majorana zero modes $\gamma_j$ ($j=1,\cdot\cdot\cdot,6$) near the 0D block $\mu$, which is not compatible with $T$-symmetry as a subgroup of $O$, hence this 1D block state is \textit{obstructed}. 

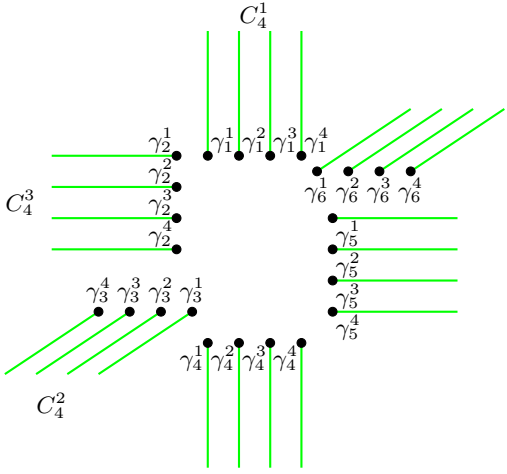
\begin{figure}
\begin{tikzpicture}[scale=0.83]
\tikzstyle{sergio}=[rectangle,draw=none]
\draw[thick,color=green] (-0.25,0.25) -- (-0.25,2.25);
\draw[thick,color=green] (-0.25,-2.75) -- (-0.25,-4.75);
\draw[thick,color=green] (-0.75,0.25) -- (-2.75,0.25);
\draw[thick,color=green] (1.75,-0.75) -- (3.75,-0.75);
\draw[thick,color=green] (2,0) -- (3.5,1);
\draw[thick,color=green] (2.5,0) -- (4,1);
\draw[thick,color=green] (3,0) -- (4.5,1);
\draw[thick,color=green] (-0.5,-2.25) -- (-2,-3.25);
\filldraw[fill=black, draw=black] (-0.25,0.25)circle (2pt);
\filldraw[fill=black, draw=black] (2,0)circle (2pt);
\filldraw[fill=black, draw=black] (3,0)circle (2pt);
\filldraw[fill=black, draw=black] (2.5,0)circle (2pt);
\filldraw[fill=black, draw=black] (1.75,-0.75)circle (2pt);
\filldraw[fill=black, draw=black] (-0.25,-2.75)circle (2pt);
\filldraw[fill=black, draw=black] (-0.5,-2.25)circle (2pt);
\filldraw[fill=black, draw=black] (-0.75,0.25)circle (2pt);
\path (0,0.5) node [style=sergio] {$\gamma_1^1$};
\path (-1,0.5) node [style=sergio] {$\gamma_2^1$};
\path (-0.5,-1.9) node [style=sergio] {$\gamma_3^1$};
\path (-0.5,-3) node [style=sergio] {$\gamma_4^1$};
\path (2,-1) node [style=sergio] {$\gamma_5^1$};
\path (2,-0.3) node [style=sergio] {$\gamma_6^2$};
\path (2.5,-0.3) node [style=sergio] {$\gamma_6^3$};
\path (3,-0.3) node [style=sergio] {$\gamma_6^4$};
\path (0.5,2.5) node [style=sergio] {$C_4^1$};
\path (-2.75,-3.75) node [style=sergio] {$C_4^2$};
\path (-3.25,-0.5) node [style=sergio] {$C_4^3$};
\draw[thick,color=green] (0.25,0.25) -- (0.25,2.25);
\draw[thick,color=green] (0.75,0.25) -- (0.75,2.25);
\draw[thick,color=green] (1.25,0.25) -- (1.25,2.25);
\filldraw[fill=black, draw=black] (0.25,0.25)circle (2pt);
\filldraw[fill=black, draw=black] (0.75,0.25)circle (2pt);
\filldraw[fill=black, draw=black] (1.25,0.25)circle (2pt);
\path (0.5,0.5) node [style=sergio] {$\gamma_1^2$};
\path (1,0.5) node [style=sergio] {$\gamma_1^3$};
\path (1.5,0.5) node [style=sergio] {$\gamma_1^4$};
\draw[thick,color=green] (-0.75,-0.25) -- (-2.75,-0.25);
\draw[thick,color=green] (-0.75,-0.75) -- (-2.75,-0.75);
\draw[thick,color=green] (-0.75,-1.25) -- (-2.75,-1.25);
\filldraw[fill=black, draw=black] (-0.75,-0.25)circle (2pt);
\filldraw[fill=black, draw=black] (-0.75,-1.25)circle (2pt);
\filldraw[fill=black, draw=black] (-0.75,-0.75)circle (2pt);
\path (-1,0) node [style=sergio] {$\gamma_2^2$};
\path (-1,-0.5) node [style=sergio] {$\gamma_2^3$};
\path (-1,-1) node [style=sergio] {$\gamma_2^4$};
\draw[thick,color=green] (-1,-2.25) -- (-2.5,-3.25);
\draw[thick,color=green] (-1.5,-2.25) -- (-3,-3.25);
\draw[thick,color=green] (-2,-2.25) -- (-3.5,-3.25);
\filldraw[fill=black, draw=black] (-1,-2.25)circle (2pt);
\filldraw[fill=black, draw=black] (-2,-2.25)circle (2pt);
\filldraw[fill=black, draw=black] (-1.5,-2.25)circle (2pt);
\path (-1,-1.9) node [style=sergio] {$\gamma_3^2$};
\path (-1.5,-1.9) node [style=sergio] {$\gamma_3^3$};
\path (-2,-1.9) node [style=sergio] {$\gamma_3^4$};
\draw[thick,color=green] (0.25,-2.75) -- (0.25,-4.75);
\draw[thick,color=green] (0.75,-2.75) -- (0.75,-4.75);
\draw[thick,color=green] (1.25,-2.75) -- (1.25,-4.75);
\filldraw[fill=black, draw=black] (0.25,-2.75)circle (2pt);
\filldraw[fill=black, draw=black] (1.25,-2.75)circle (2pt);
\filldraw[fill=black, draw=black] (0.75,-2.75)circle (2pt);
\path (0,-3) node [style=sergio] {$\gamma_4^2$};
\path (0.5,-3) node [style=sergio] {$\gamma_4^3$};
\path (1,-3) node [style=sergio] {$\gamma_4^4$};
\draw[thick,color=green] (1.75,-1.25) -- (3.75,-1.25);
\draw[thick,color=green] (1.75,-1.75) -- (3.75,-1.75);
\draw[thick,color=green] (1.75,-2.25) -- (3.75,-2.25);
\filldraw[fill=black, draw=black] (1.75,-1.25)circle (2pt);
\filldraw[fill=black, draw=black] (1.75,-1.75)circle (2pt);
\filldraw[fill=black, draw=black] (1.75,-2.25)circle (2pt);
\path (2,-1.5) node [style=sergio] {$\gamma_5^2$};
\path (2,-2) node [style=sergio] {$\gamma_5^3$};
\path (2,-2.5) node [style=sergio] {$\gamma_5^4$};
\draw[thick,color=green] (1.5,0) -- (3,1);
\filldraw[fill=black, draw=black] (1.5,0)circle (2pt);
\path (1.5,-0.3) node [style=sergio] {$\gamma_6^1$};
\end{tikzpicture}
\caption{Quadruple Majorana chains decoration on 1D blocks $\tau_2$, who leaves 24 dangling Majorana zero modes $\gamma_j^k$ ($j=1,...,6$ and $k=1,...,4$).}
\label{quadruple}
\end{figure}

1D fSPT phase decoration leaves 24 dangling Majorana zero modes $\gamma_j^{1,2,3,4}$ ($j=1,\cdot\cdot\cdot,6$) near the 0D block $\mu$ (see Fig. \ref{quadruple}). Define 12 complex fermions from these 24 Majorana zero modes, with their fermion number operator $n_{j,j+3}^k$:
\begin{align}
\left\{
\begin{aligned}
&(c_{j,j+3}^k)^\dag=\frac{1}{2}(\gamma_j^k+i\gamma_{j+3}^k)\\
&n_{j,j+3}^k=(c_{j,j+3}^k)^\dag c_{j,j+3}^k
\end{aligned}
\right.
\end{align}

These 12 complex fermions span a $2^{12}$-dimensional Hilbert space. Different from the double Majorana chains decoration case for $T$-symmetric cubic lattice, 24 Majorana zero modes near the 0D block $\mu$ form a linear representation of the point group $O$, and can be gapped in a symmetric way as the consequence: consider 4 Majorana zero modes $\gamma_1^1$, $\gamma_1^3$, $\gamma_4^1$, and $\gamma_4^3$ that forming a spin-1/2 degree of freedom at the low-energy subspace of the interaction $-\gamma_1^1\gamma_1^3\gamma_4^1\gamma_4^3$, namely
\begin{align}
S_1^x=\frac{i}{2}\gamma_1^1\gamma_1^3,~S_1^y=\frac{i}{2}\gamma_1^1\gamma_4^1,~S_1^z=\frac{i}{2}\gamma_1^1\gamma_4^3
\end{align}
Similarly, we can define 6 spin-1/2 degrees of freedom in total:
\begin{align}
\begin{gathered}
S_2^x=\frac{i}{2}\gamma_1^2\gamma_1^4,~S_2^y=\frac{i}{2}\gamma_1^2\gamma_4^2,~S_2^z=\frac{i}{2}\gamma_1^2\gamma_4^4\\
S_3^x=\frac{i}{2}\gamma_2^1\gamma_2^3,~S_3^y=\frac{i}{2}\gamma_2^1\gamma_5^1,~S_3^z=\frac{i}{2}\gamma_2^1\gamma_5^3\\
S_4^x=\frac{i}{2}\gamma_2^2\gamma_2^4,~S_4^y=\frac{i}{2}\gamma_2^2\gamma_5^2,~S_4^z=\frac{i}{2}\gamma_2^2\gamma_5^4\\
S_5^x=\frac{i}{2}\gamma_3^1\gamma_3^3,~S_5^y=\frac{i}{2}\gamma_3^1\gamma_6^1,~S_5^z=\frac{i}{2}\gamma_3^1\gamma_6^3\\
S_6^x=\frac{i}{2}\gamma_3^2\gamma_3^4,~S_6^y=\frac{i}{2}\gamma_3^2\gamma_6^2,~S_6^z=\frac{i}{2}\gamma_3^2\gamma_6^4
\end{gathered}
\end{align}
Then the following spin Hamiltonian can fully gap the junction at $\mu$:
\begin{align}
H=J\left(\bs{S}_1\cdot\bs{S}_2+\bs{S}_3\cdot\bs{S}_4+\bs{S}_5\cdot\bs{S}_6\right).
\end{align}

For spin-1/2 fermions, there is no difference from the spinless case on $\tau_1$ because there is no nontrivial extension between $\Z_3$ symmetry and fermion parity $\Z_2^f$; on $\tau_2/\tau_3$, the total on-site symmetry group is $\Z_8^f/\Z_4^f$, with trivial classification. Therefore, there is no nontrivial 1D block state for spin-1/2 case.

\subsubsection{0D block states}
The effective ``on-site'' symmetry on the 0D block $\mu$ is 4-fold symmetric group $S_4$, hence the candidate 0D block states are classified by the following two indices:
\begin{align}
\begin{aligned}
&n_0\in\mathcal{H}^0(S_4,\mathbb{Z}_2)=\mathbb{Z}_2\\
&\nu_1\in\mathcal{H}^1\left[S_4,U(1)\right]=\mathbb{Z}_2
\end{aligned}
\label{1-supercohomology O}
\end{align}
with the twisted cocycle condition:
\begin{align}
\mathrm{d}\nu_1=(-1)^{\omega_2\smile n_0}
\label{1-twisted O}
\end{align}
where $n_0$ depicts the parity of fermions, and $\nu_1$ depicts the 0D bSPT mode on the 0D block $\mu$ protected by $S_4$ symmetry, characterizing the eigenvalue $-1$ of the 4-fold rotation generator of the group $O$. We demonstrate that this 0D bSPT mode can be trivialized by 1D bubble construction: consider the axes of 4-fold rotation generator $\bs{R}_4$ of 3D point group $O$, we decorate a complex fermion $c_j^\dag$ ($j=1,...,6$) on each of them. Near the 0D block $\mu$, these complex fermions form an atomic insulator:
\begin{align}
|\psi_\mu\rangle=c_1^\dag c_2^\dag c_3^\dag c_4^\dag c_5^\dag c_6^\dag|0\rangle
\label{O atomic insulator}
\end{align}
Furthermore, these complex fermions will be transformed by 4-fold rotation $\bs{R}_4$ in the following way:
\begin{align}
\bs{R}_4:\left(c_1^\dag, c_2^\dag, c_3^\dag, c_4^\dag, c_5^\dag, c_6^\dag\right)\mapsto\left(c_1^\dag, c_3^\dag, c_5^\dag, c_4^\dag, c_6^\dag, c_2^\dag\right)
\end{align}
As the consequence, the atomic insulator $|\psi_\mu\rangle$ is an eigenstate of $\bs{R}_4$ with eignevalue $-1$:
\begin{align}
\bs{R}_4|\psi_\mu\rangle=c_1^\dag c_3^\dag c_5^\dag c_4^\dag c_6^\dag c_2^\dag|0\rangle=-|\psi_\mu\rangle
\label{R4 eigenvalue}
\end{align}
Hence, the nontrivial 0D block state decorated on $\mu$ in the bosonic symmetry-protected topological (bSPT) case can be rendered trivial by constructing a 1D bubble on $\tau_2$.

Now let us consider the fermion parity of the 0D block $\mu$. In the case of spinless fermions, similar to the scenario with $T$ symmetry, there exists no nontrivial 0D block state decoration on $\mu$ because the complex fermion decoration can be trivialized by the Majorana chain bubble on the 1D blocks $\tau_1$.

For spin-1/2 fermions, the index $n_0$ in Equation (\ref{1-supercohomology O}) that represents the fermion parity of the 0D block state on $\mu$ is obstructed. Furthermore, due to the fermions' half-integer spin, there is an additional factor of $-1$ in Equation (\ref{R4 eigenvalue}). Consequently, the eigenvalue of $-1$ for the 4-fold rotation cannot be trivialized for spin-1/2 fermions, and then the ultimate classification is $\Z_2$.

In summary, there is no nontrivial 0D block state for spinless fermions, and for spin-1/2 fermions, the nontrivial 0D block state is characterized by the eigenvalue of $-1$ under 4-fold rotation. Nevertheless, neither of these cases gives rise to a nontrivial higher-order (HO) topological surface theory.


\subsubsection{Summary}

To summarize, for spinless and spin one hafl fermions, the classification of $O$ symmetric phase are both $Z_2$, whose generators are the 1D FSPT decorated on $\tau_2$ and 0D bSPT decorated on the center $\mu$, respectively.

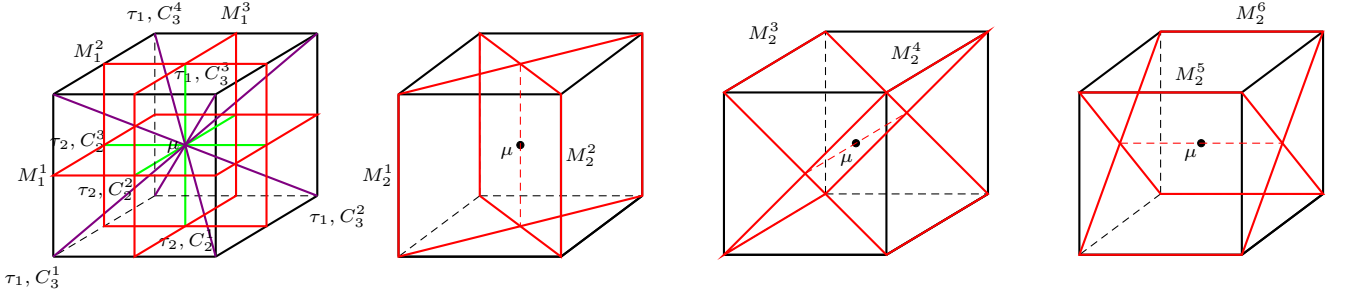
\begin{figure*}
\begin{tikzpicture}[scale=0.54]
\tikzstyle{sergio}=[rectangle,draw=none]
\draw[thick] (-0.5,1.5) -- (3.5,1.5);
\draw[thick] (-0.5,1.5) -- (-3,0);
\draw[thick] (3.5,1.5) -- (1,0);
\draw[thick] (-3,0) -- (1,0);
\draw[thick] (-3,0) -- (-3,-4);
\draw[thick] (1,-4) -- (-3,-4);
\draw[thick] (1,-4) -- (1,0);
\draw[thick] (3.5,-2.5) -- (3.5,1.5);
\draw[thick] (3.5,-2.5) -- (1,-4);
\draw[thick,color=green] (0.25,-1.25) -- (0.25,0.75);
\draw[thick,color=green] (0.25,-1.25) -- (0.25,-3.25);
\draw[thick,color=green] (0.25,-1.25) -- (-1.75,-1.25);
\draw[thick,color=green] (0.25,-1.25) -- (2.25,-1.25);
\draw[thick,color=green] (0.25,-1.25) -- (1.5,-0.5);
\draw[thick,color=green] (0.25,-1.25) -- (-1,-2);
\draw[densely dashed] (-0.5,-2.5) -- (-0.5,1.5);
\draw[densely dashed] (-0.5,-2.5) -- (3.5,-2.5);
\draw[densely dashed] (-0.5,-2.5) -- (-3,-4);
\draw[thick,color=violet] (3.5,1.5) -- (0.25,-1.25);
\draw[thick,color=violet] (-0.5,1.5) -- (0.25,-1.25);
\draw[thick,color=violet] (-3,0) -- (0.25,-1.25);
\draw[thick,color=violet] (0.25,-1.25) -- (1,0);
\path (-3.5,-4.5) node [style=sergio] {\scriptsize$\tau_1,C_3^1$};
\path (4,-3) node [style=sergio] {\scriptsize$\tau_1,C_3^2$};
\path (0.6702,0.5045) node [style=sergio] {\scriptsize$\tau_1,C_3^3$};
\path (-0.5,2) node [style=sergio] {\scriptsize$\tau_1,C_3^4$};
\draw[thick,color=violet] (0.25,-1.25) -- (-3,-4);
\draw[thick,color=violet] (0.25,-1.25) -- (-0.5,-2.5);
\draw[thick,color=violet] (0.25,-1.25) -- (3.5,-2.5);
\draw[thick,color=violet] (0.25,-1.25) -- (1,-4);
\path (0.2854,-3.6052) node [style=sergio] {\scriptsize$\tau_2,C_2^1$};
\path (-2.3949,-1.147) node [style=sergio] {\scriptsize$\tau_2,C_2^3$};
\path (-1.7058,-2.3338) node [style=sergio] {\scriptsize$\tau_2,C_2^2$};
\path (-0.05,-1.3) node [style=sergio] {\scriptsize$\mu$};
\path (-3.5,-2) node [style=sergio] {\scriptsize$M_1^1$};
\path (-2.1189,1.0691) node [style=sergio] {\scriptsize\scriptsize$M_1^2$};
\path (1.5,2) node [style=sergio] {\scriptsize$M_1^3$};
\draw[draw=red, thick] (-1,0)--(1.5,1.5)--(1.5,-2.5)--(-1,-4)--cycle;
\draw[draw=red, thick] (1,-2)--(3.5,-0.5)--(-0.5,-0.5)--(-3,-2)--cycle;
\draw[draw=red, thick] (2.25,-3.25)--(2.25,0.75)--(-1.75,0.75)--(-1.75,-3.25)--cycle;
\draw[thick] (24.2618,1.5458) -- (28.2618,1.5458);
\draw[thick] (24.2618,1.5458) -- (22.2618,0.0458);
\draw[thick] (28.2618,1.5458) -- (26.2618,0.0458);
\draw[thick] (22.2618,0.0458) -- (26.2618,0.0458);
\draw[thick] (22.2618,0.0458) -- (22.2618,-3.9542);
\draw[thick] (26.2618,-3.9542) -- (22.2618,-3.9542);
\draw[thick] (26.2618,-3.9542) -- (26.2618,0.0458);
\draw[thick] (28.2618,-2.4542) -- (28.2618,1.5458);
\draw[thick] (28.2618,-2.4542) -- (26.2618,-3.9542);
\draw[densely dashed] (24.2618,-2.4542) -- (24.2618,1.5458);
\draw[densely dashed] (24.2618,-2.4542) -- (28.2618,-2.4542);
\draw[densely dashed] (24.2618,-2.4542) -- (22.2618,-3.9542);
\draw[thick] (28.2618,-2.4542) -- (26.2618,-3.9542);
\draw[thick] (22.2618,-3.9542) -- (26.2618,-3.9542);
\path (25,-1.5) node [style=sergio] {\scriptsize$\mu$};
\path (25,0.5) node [style=sergio] {\scriptsize$M_2^5$};
\path (26.5,2) node [style=sergio] {\scriptsize$M_2^6$};
\draw[draw=red, thick] (26.2618,0.0458)--(22.2618,0.0458)--(24.2618,-2.4542)--(28.2618,-2.4542)--cycle;
\draw[draw=red, thick] (24.2618,1.5458)--(28.2618,1.5458)--(26.2618,-3.9542)--(22.2618,-3.9542)--cycle;
\filldraw[fill=black, draw=black] (25.2618,-1.2042) circle (2.5pt);
\draw[densely dashed,color=red] (23.2618,-1.2042) -- (27.2618,-1.2042);
\draw[thick] (16.0118,1.5458) -- (20.0118,1.5458);
\draw[thick] (16.0118,1.5458) -- (13.5118,0.0458);
\draw[thick] (20.0118,1.5458) -- (17.5118,0.0458);
\draw[thick] (13.5118,0.0458) -- (17.5118,0.0458);
\draw[thick] (13.5118,0.0458) -- (13.5118,-3.9542);
\draw[thick] (17.5118,-3.9542) -- (13.5118,-3.9542);
\draw[thick] (17.5118,-3.9542) -- (17.5118,0.0458);
\draw[thick] (20.0118,-2.4542) -- (20.0118,1.5458);
\draw[thick] (20.0118,-2.4542) -- (17.5118,-3.9542);
\draw[densely dashed] (16.0118,-2.4542) -- (16.0118,1.5458);
\draw[densely dashed] (16.0118,-2.4542) -- (20.0118,-2.4542);
\draw[densely dashed] (16.0118,-2.4542) -- (13.5118,-3.9542);
\draw[thick] (20.0118,-2.4542) -- (17.5118,-3.9542);
\draw[thick] (13.5118,-3.9542) -- (17.5118,-3.9542);
\path (16.5278,-1.6051) node [style=sergio] {\scriptsize$\mu$};
\path (18,1) node [style=sergio] {\scriptsize$M_2^4$};
\path (14.5,1.5) node [style=sergio] {\scriptsize$M_2^3$};
\draw[draw=red, thick] (16.0118,-2.4542)--(20.0118,1.5458)--(17.5118,0.0458)--(13.5118,-3.9542)--cycle;
\draw[draw=red, thick] (16.0118,1.5458)--(13.5118,0.0458)--(17.5118,-3.9542)--(20.0118,-2.4542)--cycle;
\filldraw[fill=black, draw=black] (16.7618,-1.2042) circle (2.5pt);
\draw[densely dashed,color=red] (15.5118,-1.9542) -- (18.0118,-0.4542);
\draw[thick] (7.5,1.5) -- (11.5,1.5);
\draw[thick] (7.5,1.5) -- (5.5,0);
\draw[thick] (11.5,1.5) -- (9.5,0);
\draw[thick] (5.5,0) -- (9.5,0);
\draw[thick] (5.5,0) -- (5.5,-4);
\draw[thick] (9.5,-4) -- (5.5,-4);
\draw[thick] (9.5,-4) -- (9.5,0);
\draw[thick] (11.5,-2.5) -- (11.5,1.5);
\draw[thick] (11.5,-2.5) -- (9.5,-4);
\draw[densely dashed] (7.5,-2.5) -- (7.5,1.5);
\draw[densely dashed] (7.5,-2.5) -- (11.5,-2.5);
\draw[densely dashed] (7.5,-2.5) -- (5.5,-4);
\draw[thick] (11.5,-2.5) -- (9.5,-4);
\draw[thick] (5.5,-4) -- (9.5,-4);
\path (8.1701,-1.4347) node [style=sergio] {\scriptsize$\mu$};
\path (5,-2) node [style=sergio] {\scriptsize$M_2^1$};
\path (10,-1.5) node [style=sergio] {\scriptsize$M_2^2$};
\draw[draw=red, thick] (5.5,0)--(11.5,1.5)--(11.5,-2.5)--(5.5,-4)--cycle;
\draw[draw=red, thick] (7.5,1.5)--(9.5,0)--(9.5,-4)--(7.5,-2.5)--cycle;
\filldraw[fill=black, draw=black] (8.5,-1.25) circle (2.5pt);
\draw[densely dashed,color=red] (8.5,0.75) -- (8.5,-3.25);
\end{tikzpicture}
\caption{The cell decomposition of the 3D system with point group symmetry $O_h$. There are four axes of 3-fold rotation symmetry labeled by $C_3^{1,2,3,4}$ and depicted by solid violet line segments, across the center of the system as indicated by the solid dot (labeled by $\mu$); three axes of 4-fold rotation symmetry across the center, labeled by $C_4^{1,2,3}$ and depicted by solid green lines (see left panel); and nine planes of reflection symmetry, labeled by $\bs{M}_1^k$ and $\bs{M}_2^l$, where $k=1,2,3$ and $l=1,...,6$.}
\label{Oh cell decomposition}
\end{figure*}

\subsection{$O_h$-symmetric lattice}
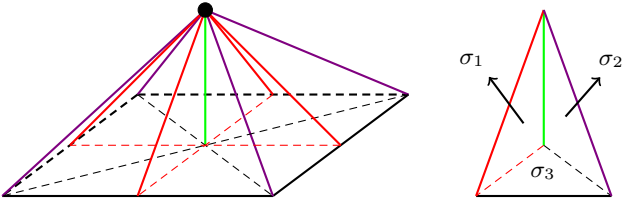
\begin{figure}
\begin{tikzpicture}[scale=0.9]
\tikzstyle{sergio}=[rectangle,draw=none]
\draw[thick] (-3,-0.5) -- (1,-0.5);
\draw[densely dashed,thick] (-1,1) -- (3,1);
\draw[densely dashed,thick] (-1,1) -- (-3,-0.5);
\draw[thick] (3,1) -- (1,-0.5);
\draw[thick,color=violet] (-1,1) -- (0,2.25);
\draw[thick,color=violet] (1,-0.5) -- (0,2.25);
\draw[thick,color=violet] (-3,-0.5) -- (0,2.25);
\draw[thick,color=violet] (3,1) -- (0,2.25);
\draw[thick,color=green] (0,0.25) -- (0,2.25);
\draw[densely dashed] (-1,1) -- (1,-0.5);
\draw[densely dashed] (3,1) -- (-3,-0.5);
\draw[thick] (4,-0.5) -- (6,-0.5);
\draw[densely dashed,color=red] (5,0.25) -- (4,-0.5);
\draw[densely dashed] (5,0.25) -- (6,-0.5);
\draw[thick,color=green] (5,0.25) -- (5,2.25);
\draw[thick,color=red] (4,-0.5) -- (5,2.25);
\draw[thick,color=violet] (6,-0.5) -- (5,2.25);
\draw[thick,->] (4.7013,0.5747) -- (4.1914,1.2597);
\draw[thick,->] (5.3213,0.6741) -- (5.8329,1.2556);
\path (3.9414,1.5097) node [style=sergio] {$\sigma_1$};
\path (4.9887,-0.158) node [style=sergio] {$\sigma_3$};
\path (6,1.5) node [style=sergio] {$\sigma_2$};
\draw[thick,color=red] (1,1) -- (0,2.25);
\draw[thick,color=red] (-1,-0.5) -- (0,2.25);
\draw[thick,color=red] (2,0.25) -- (0,2.25);
\draw[thick,color=red] (-2,0.25) -- (0,2.25);
\draw[densely dashed,color=red] (2,0.25) -- (-2,0.25);
\draw[densely dashed,color=red] (1,1) -- (-1,-0.5);
\filldraw[fill=black, draw=black] (0,2.25)circle (3pt);
\end{tikzpicture}
\caption{The unit cell of cell decomposition of the cubic lattice with point symmetry $O_h$. The top panel depicts the bottom rectangular pyramid of the cubic in Fig. \ref{Oh cell decomposition}; the bottom panel illustrates the independent triangular pyramid $\lambda$, where $\sigma_1$, $\sigma_2$ and $\sigma_3$ are three independent 2D blocks in the system.}
\label{Oh unit cell}
\end{figure}

For $O_h$-symmetric cubic, by cell decomposition as illustrated in Figs. \ref{Oh cell decomposition} and \ref{Oh unit cell}, the ground-state wavefunction of the system can be decomposed to the direct product of wavefunctions of lower-dimensional blocks as:
\begin{align}
|\Psi\rangle=\bigotimes_{g\in O_h}|T_{g\lambda}\rangle\otimes\bigotimes\limits_{k=1}^3|\gamma_{g\sigma_k}\rangle\otimes\bigotimes\limits_{l=1}^3|\beta_{g\tau_l}\rangle\otimes|\alpha_\mu\rangle
\label{Oh wavefunction}
\end{align}
where $|T_{g\lambda}\rangle$ is the wavefunction of 3D block state $g\lambda$ which is topologically trivial; $|\gamma_{\sigma_{1,2,3}}\rangle$ is the wavefunction of 2D block state $g\sigma_{1,2,3}$ which is $\mathbb{Z}_2$-symmetric; $|\beta_{\tau_1}\rangle$ is the wavefunction of 1D block state $g\tau_1$ which is $(\mathbb{Z}_3\rtimes\mathbb{Z}_2)$-symmetric, $|\beta_{\tau_2}\rangle$ is the wavefunction of 1D block state $g\tau_2$ which is $(\mathbb{Z}_4\times\mathbb{Z}_2)$-symmetric, and $|\beta_{\tau_3}\rangle$ is the wavefunction of 1D block state $g\tau_3$ which is $(\mathbb{Z}_2\times\mathbb{Z}_2)$-symmetric; $|\alpha_\mu\rangle$ is the wavefunction of 0D block state $\mu$ which is $(S_4\times\mathbb{Z}_2)$-symmetric.

With topological crystals, we decorate the lower-dimensional block states and investigate the possible \textit{obstructions} and \textit{trivializations}.

\subsubsection{2D block states\label{Oh2D}}
The effective on-site symmetry of 2D blocks labeled by $\sigma_j$ ($j=1,2,3$) is $\mathbb{Z}_2$ by reflection generator $\bs{M}_1^{1,2,3}$ or $\bs{M}_2^{k}$ ($k=1,...,6$) acting internally. We discuss spinless and spin-1/2 fermions separately.

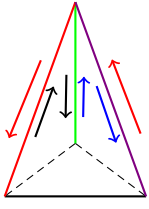
\begin{figure}
\begin{tikzpicture}[scale=0.94]
\tikzstyle{sergio}=[rectangle,draw=none];
\draw[thick] (-1,-4.75) -- (1,-4.75);
\draw[densely dashed] (0,-4) -- (-1,-4.75);
\draw[densely dashed] (0,-4) -- (1,-4.75);
\draw[thick,color=green] (0,-4) -- (0,-2);
\draw[thick,color=red] (-1,-4.75) -- (0,-2);
\draw[thick,color=violet] (1,-4.75) -- (0,-2);
\draw[thick,->,color=red] (-0.4887,-2.8239) -- (-0.9402,-3.9254);
\draw[thick,->,color=red] (0.9219,-3.8616) -- (0.5169,-2.8138);
\draw[thick,->] (-0.5677,-3.9097) -- (-0.3093,-3.1958);
\draw[thick,->] (-0.1275,-3.0294) -- (-0.1388,-3.6432);
\draw[thick,->,color=blue] (0.1095,-3.6275) -- (0.1207,-3.0564);
\draw[thick,->,color=blue] (0.3001,-3.189) -- (0.5755,-3.9902);
\end{tikzpicture}
\caption{Directions of edge modes of decorated 2D $(p+ip)$-SC on 2D blocks $\sigma_j$ with indices $n_j$ ($j=1,2,3$), illustrated by black, blue, and red arrows, respectively.}
\label{Oh chirality}
\end{figure}

\paragraph{Spinless fermions}For spinless fermions, there are two possible root phases:
\begin{enumerate}[1.]
\item 2D $(p+ip)$-SC;
\item 2D fermionic Levin-Gu state \cite{Gu-Levin} with $\nu\in\mathbb{Z}_8$ index.
\end{enumerate}
 We begin by examining the decorations of 2D $(p+ip)$-superconductors on the 2D blocks $\sigma_j$, where $n_j \in \mathbb{Z}$ denotes their corresponding phase indices ($j=1,2,3$). The edge modes resulting from these decorations are depicted in Figure \ref{Oh chirality}. Consequently, the chiral central charges on $\tau_k$ ($k=1,2,3$) are given by $(3n_2-3n_3)/2$, $2n_1-2n_2$, and $-n_1+n_3$ respectively. In order to achieve a fully-gapped bulk, it is necessary for all three quantities characterizing the chiral central charges of the 1D blocks $\tau_{1,2,3}$ to vanish, yielding the following equations:
\begin{align}
3n_2-3n_3=4n_1-4n_2=-2n_1+2n_3=0
\end{align}
The solutions to these equations are such that $n_1=n_2=n_3=n$. Furthermore, similar to the discussion in Section \ref{Th2D}, the case of monolayer/bilayer $(p+ip)$-superconductors is either obstructed or trivialized. Hence, there are no nontrivial 2D block states resulting from the decoration of 2D $(p+ip)$-superconductors.

Subsequently, we consider the decorations of 2D fermionic Levin-Gu states on $\sigma_{1,2,3}$, with $\nu_{1,2,3} \in \mathbb{Z}_8$ representing their  indices. The resulting edge modes are similar to those shown in Figure \ref{3D bubble}. Consequently, the number of edge modes left by the 2D fermionic Levin-Gu states on $\tau_{1,2,3}$ are $3\nu_2-3\nu_3$, $4\nu_1-4\nu_2$, and $-2\nu_1+2\nu_3$, respectively. To achieve a fully-gapped bulk, it is necessary for all three of these quantities to be integer multiples of 8.
\begin{align}
\begin{aligned}
&3\nu_2-3\nu_3\equiv0~(\mathrm{mod}~8)\\
&4\nu_1-4\nu_2\equiv0~(\mathrm{mod}~8)\\
-&2\nu_1+2\nu_3\equiv0~(\mathrm{mod}~8)
\end{aligned}
\end{align}
The solution to these equations should satisfy:
$$
\nu_2\equiv\nu_3~(\mathrm{mod}~8)~~\mathrm{and}~~\nu_1\equiv\nu_3~(\mathrm{mod}~4)
$$
A necessary condition of obstruction-free block states from 2D Levin-Gu states decorations gives a $\mathbb{Z}_8$-index $\nu$, characterizing the number of decorated Levin-Gu states on each 2D block $\sigma_{1,2,3}$. Furthermore, similar to Sec. \ref{Td2D}, we can only decorate 2D bosonic Levin-Gu state on $\sigma_1$ enforced by symmetry. As  consequence, the only possible non-vacuum 2D block states should be labeled by:
\begin{align}
(\nu_1,\nu_2,\nu_3)=(0,4,4),(4,0,0),(4,4,4)
\end{align}
i.e., the state decorated on each 2D block $\sigma_{1,2,3}$ should be a 2D bosonic Levin-Gu state \cite{LevinGu}. For the case with $(\nu_1,\nu_2,\nu_3)=(0,4,4)$, the 1D gapless theory on each $\tau_2$ is described by the non-chiral Luttinger liquid (\ref{Luttinger}), with $K=\left({\sigma^x}\right)^{\oplus4}$ and $\Phi=\left(\phi^1,\cdot\cdot\cdot,\phi^8\right)^T$. Under 4-fold rotation $\bs{R}$, the edge field $\Phi$ transformed as Eq. (\ref{K-matrix symmetry}), with
\begin{align}
W^{\bs{R}}=\left(
\begin{array}{cccc}
0 & 1 & 0 & 0\\
0 & 0 & 1 & 0\\
0 & 0 & 0 & 1\\
1 & 0 & 0 & 0\\
\end{array}
\right)\otimes\mathbbm{1}_{2\times2},~\delta\Phi^{\bs{R}}=0
\end{align}
Under reflection $\bs{M}$, the edge field transforms as:
\begin{align}
W^{\bs{M}}=\left(
\begin{array}{cccc}
0 & 0 & 1 & 0\\
0 & 1 & 0 & 0\\
1 & 0 & 0 & 0\\
0 & 0 & 0 & 1\\
\end{array}
\right)\otimes\mathbbm{1}_{2\times2},~\delta\Phi^{\bs{M}}=\pi\chi
\end{align}
and another reflection generator of $D_4$ group, $\bs{M}\bs{R}$, as
\begin{align}
W^{\bs{M}\bs{R}}=\left(
\begin{array}{cccc}
0 & 1 & 0 & 0\\
1 & 0 & 0 & 0\\
0 & 0 & 0 & 1\\
0 & 0 & 1 & 0\\
\end{array}
\right)\otimes\mathbbm{1}_{2\times2},~\delta\Phi^{\bs{M}\bs{R}}=\pi\chi
\end{align}
where $\chi=(1,1,1,1,1,1,1,1)^T$. On the one hand, the anomaly indicators (\ref{anomaly indicator}) of two reflection generators $\bs{M}$ and $\bs{MR}$ are vanishing as $\nu_{\bs{M}}=\nu_{\bs{MR}}=0~(\mathrm{mod}~2)$.

On the other hand, we construct backscattering terms (\ref{backscattering}) that gap out the edge without breaking the $\mathbb{Z}_4\rtimes\mathbb{Z}_2$ symmetry, either explicitly or spontaneously. There are four symmetric linear independent solutions to the ``null-vector'' problem:
\begin{align}
\begin{aligned}
&\Lambda_1=\left(1,1,0,0,1,1,0,0\right)^T\\
&\Lambda_2=\left(1,-1,0,0,1,-1,0,0\right)^T\\
&\Lambda_3=\left(0,0,1,1,0,0,1,1\right)^T\\
&\Lambda_4=\left(0,0,1,-1,0,0,1,-1\right)^T
\end{aligned}
\end{align}
which correspond to four independent backscattering terms which can fully gap out all 4 nonchiral Luttinger liquids near each 1D block $\tau_2$. Similar arguments can also be applied to 1D blocks $\tau_1$ (with $\mathbb{Z}_3\rtimes\mathbb{Z}_2$ symmetry) and $\tau_3$ (with $\mathbb{Z}_2\times\mathbb{Z}_2$ symmetry), and all nearby 1D theories from decorated Levin-Gu states can be fully gapped. As the consequence, 2D Levin-Gu state decoration on 2D blocks $\sigma_1$, $\sigma_2$ and $\sigma_3$ with quantum numbers $(\nu_1,\nu_2,\nu_3)=(0,4,4)$ is \textit{obstruction-free}. Furthermore, similar to Sec. \ref{Th2D}, the Levin-Gu state decoration on 2D blocks cannot be trivialized by 3D bubble constructions. Similar for cases with  $(\nu_1,\nu_2,\nu_3)=(4,0,0)$ and $(\nu_1,\nu_2,\nu_3)=(4,4,4)$, these 2D block states are all nontrivial, forming a $\mathbb{Z}_2^2$ group.

\paragraph{Spin-1/2 fermions}For spin-1/2 fermions, the effective on-site symmetry of each 2D block is $\mathbb{Z}_4^f$ as the nontrivial $\mathbb{Z}_2^f$ extension of $\mathbb{Z}_2$ symmetry by reflection generator acting internally. The classification of the corresponding 2D fSPT phases is trivial \cite{general2}, hence there is no nontrivial 2D block state for spin-1/2 fermions.

\subsubsection{1D block states}
The effective on-site symmetry of 1D blocks $\tau_{1}$ is $\mathbb{Z}_3\rtimes\mathbb{Z}_2$, by 3-fold rotation and reflection generators, $\bs{R}_3^{1,2,3}$ and $\bs{M}_2^{k}$ ($k=1,...,6$) acting internally; the effective on-site symmetry of 1D blocks labeled by $\tau_2$ is $\mathbb{Z}_4\rtimes\mathbb{Z}_2$, by 4-fold rotation and reflection generators, $\bs{R}_2^{1,2,3}$ and $\bs{M}_2^{k}$ ($k=1,...,6$) acting internally; the effective on-site symmetry of 1D blocks labeled by $\tau_3$ is $\tau_2$ is $\mathbb{Z}_2\times\mathbb{Z}_2$, by reflection generators $\bs{M}_{1,2}^k$ acting internally. We discuss the spinless fermions and spin-1/2 fermions separately.

\paragraph{Spinless fermions}For spinless fermions, there are two possible root phases on each $\tau_{1}$:
\begin{enumerate}[1.]
\item Majorana chain;
\item 1D fSPT phase, composed of double Majorana chains.
\end{enumerate}
Firstly, Majorana chain decoration on 1D blocks $\tau_1$ leaves 8 Majorana zero modes $\gamma_j$ and $\gamma_j'$ ($j=1,2,3,4$) at 0D block $\mu$. The symmetry properties of these Majorana zero modes under 3-fold rotation are:
\begin{align}
\bs{R}_3^4:\left\{
\begin{aligned}
&\left(\gamma_1,\gamma_2,\gamma_3,\gamma_4\right)\mapsto\left(\gamma_1,\gamma_3,\gamma_4,\gamma_2\right)\\
&\left(\gamma_1',\gamma_2',\gamma_3',\gamma_4'\right)\mapsto\left(\gamma_1',\gamma_3',\gamma_4',\gamma_2'\right)
\end{aligned}
\right.
\label{R3}
\end{align}
Under 4-fold rotation $\bs{R}_4^1\in O_h$, these Majorana zero modes are transformed as:
\begin{align}
\bs{R}_4^1:\left\{
\begin{aligned}
&\left(\gamma_1,\gamma_2,\gamma_3,\gamma_4\right)\mapsto\left(\gamma_3',\gamma_1',\gamma_4',\gamma_2'\right)\\
&\left(\gamma_1',\gamma_2',\gamma_3',\gamma_4'\right)\mapsto\left(\gamma_3,\gamma_1,\gamma_4,\gamma_2\right)
\end{aligned}
\right.
\label{R2}
\end{align}
Under reflection $\bs{M}_2^1\in O_h$, these Majorana zero modes are transformed as:
\begin{align}
\bs{M}_2^1:~\left\{
\begin{aligned}
&\left(\gamma_1,\gamma_2,\gamma_3,\gamma_4\right)\mapsto\left(\gamma_4,\gamma_2,\gamma_3,\gamma_1\right)\\
&\left(\gamma_1',\gamma_2',\gamma_3',\gamma_4'\right)\mapsto\left(\gamma_4',\gamma_2',\gamma_3',\gamma_1'\right)
\end{aligned}
\right.
\label{M21}
\end{align}
We can define 4 complex fermions from these Majorana zero modes: $c_j^\dag=(\gamma_j+i\gamma_j')/2$. and these complex fermions span a 16-dimensional Hilbert space. In this Hilbert space, Majorana zero modes $\gamma_{1,2,3,4}$ and $\gamma_{1,2,3,4}'$ can be represented as $16\times16$ matrices in this Hilbert space. Furthermore, the above symmetry generators can be represented as $16\times16$ matrices $A$ in terms of matrix representations of Majorana zero modes:
\begin{align}
A(\bs{R}_3^4)=\frac{1}{4}(\gamma_2-\gamma_3)(\gamma_3-\gamma_4)(\gamma_2'-\gamma_3')(\gamma_3'-\gamma_4')
\end{align}
\begin{align}
A(\bs{R}_4^1)=&\frac{1}{8}(\gamma_1-\gamma_3')(\gamma_3'-\gamma_4)(\gamma_4-\gamma_2')\nonumber\\
&\cdot(\gamma_1'-\gamma_3)(\gamma_3-\gamma_4')(\gamma_4'-\gamma_2)
\end{align}
\begin{align}
A(\bs{M}_2^1)=\frac{1}{2}(\gamma_1-\gamma_4)(\gamma_1'-\gamma_4')
\end{align}
Then we further consider another group element $\bs{R}_3^4\bs{R}_4^1\in O_h$. Under $\bs{R}_3^4\bs{R}_4^1$, aforementioned Majorana zero modes $\gamma_j$ and $\gamma_j'$ are transformed as:
\begin{align}
\bs{R}_3^4\bs{R}_4^1:~\left\{
\begin{aligned}
&\left(\gamma_1,\gamma_2,\gamma_3,\gamma_4\right)\mapsto\left(\gamma_4',\gamma_1',\gamma_2',\gamma_3'\right)\\
&\left(\gamma_1',\gamma_2',\gamma_3',\gamma_4'\right)\mapsto\left(\gamma_4,\gamma_1,\gamma_2,\gamma_3\right)
\end{aligned}
\right.
\end{align}
We can also represent $\bs{R}_3^4\bs{R}_4^1$ in the above 16-dimensional Hilbert space in terms of the matrix representations of Majorana zero modes $\gamma_j$ and $\gamma_j'$:
\begin{align}
A(\bs{R}_3^4\bs{R}_4^1)=&\frac{1}{8}\left(\gamma_1-\gamma_4'\right)\left(\gamma_4'-\gamma_3\right)\left(\gamma_3-\gamma_2'\right)\nonumber\\
&\cdot\left(\gamma_1'-\gamma_4\right)\left(\gamma_4-\gamma_3'\right)\left(\gamma_3'-\gamma_2\right)
\end{align}
We can straightforwardly check the following relation:
\begin{align}
A(\bs{R}_3^4)A(\bs{R}_4^1)=-A(\bs{R}_3^4\bs{R}_4^1)
\label{proj1}
\end{align}
Similarly, we have another two relations:
\begin{align}
\begin{aligned}
&A(\bs{R}_3^4)A(\bs{M}_2^1)=A(\bs{R}_3^4\bs{M}_2^1)\\
&A(\bs{R}_4^1)A(\bs{M}_2^1)=-A(\bs{R}_4^1\bs{M}_2^1)
\end{aligned}
\label{proj2}
\end{align}
From Eqs. (\ref{proj1}) and (\ref{proj2}), we conclude that $A$ is a projective representation of $O_h$. As the consequence, Majorana zero modes $\gamma_j$ and $\gamma_j'$ cannot be gapped in a $O_h$-symmetric way, and the 1D block state from Majorana chain decoration on $\tau_1$ is \textit{obstructed}.

Subsequently, double Majorana chains decoration on 1D blocks $\tau_1$ leaves 16 Majorana zero modes $\eta_j$, $\eta_j'$, $\xi_j$ and $\xi_j'$ ($j=1,2,3,4$) at 0D block $\mu$. The symmetry properties of these Majorana zero modes under 3-fold rotation $\bs{R}_3^4$ are:
\begin{align}
\bs{R}_3^4:\left\{
\begin{aligned}
&\left(\eta_1,\eta_2,\eta_3,\eta_4\right)\mapsto\left(\eta_1,\eta_3,\eta_4,\eta_2\right)\\
&\left(\eta_1',\eta_2',\eta_3',\eta_4'\right)\mapsto\left(\eta_1',\eta_3',\eta_4',\eta_2'\right)\\
&\left(\xi_1,\xi_2,\xi_3,\xi_4\right)\mapsto\left(\xi_1,\xi_3,\xi_4,\xi_2\right)\\
&\left(\xi_1',\xi_2',\xi_3',\xi_4'\right)\mapsto\left(\xi_1',\xi_3',\xi_4',\xi_2'\right)
\end{aligned}
\right.
\label{R32}
\end{align}
Under 4-fold rotation $\bs{R}_4^1\in O_h$, these Majorana zero modes are transformed as:
\begin{align}
\bs{R}_4^1:\left\{
\begin{aligned}
&\left(\eta_1,\eta_2,\eta_3,\eta_4\right)\mapsto\left(\xi_3,\xi_1,\xi_4,\xi_2\right)\\
&\left(\eta_1',\eta_2',\eta_3',\eta_4'\right)\mapsto\left(\xi_3',\xi_1',\xi_4',\xi_2'\right)\\
&\left(\xi_1,\xi_2,\xi_3,\xi_4\right)\mapsto\left(\eta_3,\eta_1,\eta_4,\eta_2\right)\\
&\left(\xi_1',\xi_2',\xi_3',\xi_4'\right)\mapsto\left(\eta_3',\eta_1',\eta_4',\eta_2'\right)
\end{aligned}
\right.
\label{R22}
\end{align}
Under reflection $\bs{M}_2^1\in O_h$, these Majorana zero modes are transformed as:
\begin{align}
\bs{M}_2^1:~\left\{
\begin{aligned}
&\left(\eta_1,\eta_2,\eta_3,\eta_4\right)\mapsto\left(\eta_4,\eta_2,\eta_3,\eta_1\right)\\
&\left(\eta_1',\eta_2',\eta_3',\eta_4'\right)\mapsto\left(\eta_4',\eta_2',\eta_3',\eta_1'\right)\\
&\left(\xi_1,\xi_2,\xi_3,\xi_4\right)\mapsto\left(\xi_4,\xi_2,\xi_3,\xi_1\right)\\
&\left(\xi_1',\xi_2',\xi_3',\xi_4'\right)\mapsto\left(\xi_4',\xi_2',\xi_3',\xi_1'\right)
\end{aligned}
\right.
\label{M22}
\end{align}
To gap out these Majorana zero modes, we introduce a symmetric interacting Hamiltonian:
\begin{align}
H_U=U\sum\limits_{j=1}^4\eta_j\eta_j'\xi_j\xi_j',~U>0
\end{align}
This Hamiltonian can open a Hubbard gap with 16-fold GSD that can be characterized by:
\begin{align}
\eta_j\eta_j'\xi_j\xi_j'=-1,~j=1,2,3,4.
\label{restrict}
\end{align}
To investigate whether this GSD can be lifted, we restrict the Hilbert space to the subspace constrained by relations (\ref{restrict}). In this subspace, we can further define four spin-1/2 degrees of freedom ($j=1,2,3,4$):
\begin{align}
S_j^x=\frac{i}{2}\eta_j\eta_j',~S_j^y=\frac{i}{2}\eta_j\xi_j,~S_j^z=\frac{i}{2}\eta_j\xi_j'
\end{align}
and they satisfy the commutation relations of spin-1/2 degrees of freedom ($\mu,\nu,\rho=x,y,z$):
\begin{align}
\left[S^\mu_j,S^\nu_j\right]=i\epsilon_{\mu\nu\rho}S^\rho_j,~j=1,2,3,4.
\end{align}
Furthermore, their symmetry properties are:
\begin{align}
\bs{R}_3^4:~\left(\bs{S}_1,\bs{S}_2,\bs{S}_3,\bs{S}_4\right)\mapsto\left(\bs{S}_1,\bs{S}_3,\bs{S}_4,\bs{S}_2\right)
\end{align}
\begin{align}
\bs{R}_4^1:~\left\{
\begin{aligned}
&\left(S_1^x,S_2^x,S_3^x,S_4^x\right)\mapsto\left(S_3^x,S_1^x,S_4^x,S_2^x\right)\\
&\left(S_1^y,S_2^y,S_3^y,S_4^y\right)\mapsto-\left(S_3^y,S_1^y,S_4^y,S_2^y\right)\\
&\left(S_1^z,S_2^z,S_3^z,S_4^z\right)\mapsto-\left(S_3^z,S_1^z,S_4^z,S_2^z\right)
\end{aligned}
\right.
\end{align}
\begin{align}
\bs{M}_2^1:~\left(\bs{S}_1,\bs{S}_2,\bs{S}_3,\bs{S}_4\right)\mapsto\left(\bs{S}_4,\bs{S}_2,\bs{S}_3,\bs{S}_1\right)
\end{align}
Therefore, we can further add a Heisenberg Hamiltonian between these four spin-1/2 degrees of freedom:
\begin{align}
H_J=J\left(\bs{S}_1\cdot\bs{S}_3+\bs{S}_2\cdot\bs{S}_4\right),~J>0
\end{align}
and $H_J$ splits the 16-fold degenerate ground states and obtains a non-degenerate ground state as the superposition of two spin-singlets formed by $(\bs{S}_1,\bs{S}_3)$ and $(\bs{S}_2,\bs{S}_4)$. As a consequence, the dangling Majorana zero modes near $\mu$ can be fully gapped by Hamiltonian $H_U$ and $H_J$, and the double Majorana chains decoration on 1D blocks $\tau_1$ is an \textit{obstruction-free} block state. 

Because of the $\mathbb{Z}_4\rtimes\mathbb{Z}_2$ on-site symmetry, there are three possible root phases on each $\tau_2$:
\begin{enumerate}[1.]
\item Majorana chain;
\item 1D fSPT phase (quadruple Majorana chains);
\item 1D bSPT phase (Haldane chain).
\end{enumerate}

Similar to the $O$-symmetric case, Majorana chain decoration is \textit{obstructed}, and quadruple Majorana chain decoration is \textit{obstruction-free}. Haldane chain decoration on $\tau_2$ leaves 6 dangling spin-1/2 degrees of freedom $\bs{\tau}_j$ ($j=1,...,6$), with the following symmetry properties:
\begin{align}
\begin{aligned}
\bs{R}_3^4:&\left(\bs{\tau}_1,\bs{\tau}_2,\bs{\tau}_3,\bs{\tau}_4,\bs{\tau}_5,\bs{\tau}_6\right)\mapsto\left(\bs{\tau}_2,\bs{\tau}_6,\bs{\tau}_4,\bs{\tau}_5,\bs{\tau}_3,\bs{\tau}_1\right)\\
\bs{R}_4^1:&\left(\bs{\tau}_1,\bs{\tau}_2,\bs{\tau}_3,\bs{\tau}_4,\bs{\tau}_5,\bs{\tau}_6\right)\mapsto\left(\bs{\tau}_1,\bs{\tau}_3,\bs{\tau}_5,\bs{\tau}_4,\bs{\tau}_6,\bs{\tau}_2\right)\\
\bs{M}_2^1:&\left(\bs{\tau}_1,\bs{\tau}_2,\bs{\tau}_3,\bs{\tau}_4,\bs{\tau}_5,\bs{\tau}_6\right)\mapsto\left(\bs{\tau}_1,\bs{\tau}_6,\bs{\tau}_5,\bs{\tau}_4,\bs{\tau}_3,\bs{\tau}_2\right)
\end{aligned}\nonumber
\end{align}
To gap out these spin-1/2 degrees of freedom, we can simply introduce several Heisenberg interactions:
\begin{align}
H_J'=J'\left(\bs{\tau}_1\cdot\bs{\tau}_4+\bs{\tau}_2\cdot\bs{\tau}_5+\bs{\tau}_3\cdot\bs{\tau}_6\right)
\end{align}
And $H_J'$ leaves three spin-singlets as a non-degenerate ground state. Hence the Haldane chain decoration on 1D blocks labeled by $\tau_2$ is \textit{obstruction-free}.

For 1D blocks labeled by $\tau_3$ with $\mathbb{Z}_2\times\mathbb{Z}_2$, there are three root phases:
\begin{enumerate}[1.]
\item Majorana chain;
\item 1D fSPT phase (double Majorana chains);
\item Haldane chain.
\end{enumerate}
Similar to the $T_h$-symmetric case, Majorana chain and double Majorana chain decorations are not compatible with the $T_h$ symmetry as a subgroup of $O_h$.

\begin{figure}
\begin{tikzpicture}
\tikzstyle{sergio}=[rectangle,draw=none]
\draw[thick,color=red] (-0.5,0) -- (-1.5,1);
\draw[thick,color=red] (-0.5,-1) -- (-1.5,-2);
\draw[thick,color=red] (0.5,0) -- (1.5,1);
\draw[thick,color=red] (0.5,-1) -- (1.5,-2);
\draw[densely dashed,color=green] (0,1.5) -- (0,-2.5);
\draw[densely dashed,color=green] (-2,-0.5) -- (2,-0.5);
\filldraw[fill=black, draw=black] (-0.5,0)circle (2pt);
\filldraw[fill=black, draw=black] (0.5,0)circle (2pt);
\filldraw[fill=black, draw=black] (0.5,-1)circle (2pt);
\filldraw[fill=black, draw=black] (-0.5,-1)circle (2pt);
\path (0,-0.5) node [style=sergio] {$\bigotimes$};
\path (-0.4355,0.3493) node [style=sergio] {$\bs{\tau}_1$};
\path (-0.7786,-0.8055) node [style=sergio] {$\bs{\tau}_2$};
\path (0.3647,-1.2785) node [style=sergio] {$\bs{\tau}_3$};
\path (0.7885,-0.1353) node [style=sergio] {$\bs{\tau}_4$};
\path (-0.2267,-0.8351) node [style=sergio] {$R_4^1$};
\path (0.3037,1.5024) node [style=sergio] {$R_4^2$};
\path (1.7703,-0.1157) node [style=sergio] {$R_4^3$};
\end{tikzpicture}
\caption{4 dangling spin-1/2 degrees of freedom $\bs{\tau}_{1,2,3,4}$ near 0D block $\mu$ on the $xy$-plane, from Haldane chain decoration on 1D blocks labeled by $\tau_3$.}
\label{xy-plane}
\end{figure}
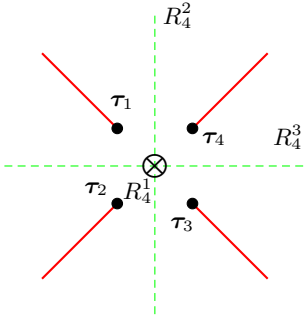

Haldane chain decoration on 1D blocks $\tau_3$ leaves 12 dangling spin-1/2 degrees of freedom near 0D block $\mu$, and there are 4 dangling spin-1/2 degrees of freedom on each reflection plane. 3-fold rotations $\bs{R}_3^j\in C_3^j$ ($j=1,2,3,4$) permute these reflection planes ($xy$, $xz$ and $yz$ planes), and 4-fold rotations $\bs{R}_4^k\in C_4^k$ ($k=1,2,3$) are in-plane manipulations. We study the $xy$-plane as an example (cf. Fig. \ref{xy-plane}).

The dangling spin-1/2 degrees of freedom on $xy$-plane are labeled by $\bs{\tau}_{1,2,3,4}$, with the following in-plane symmetry properties:
\begin{align}
\begin{aligned}
&\bs{R}_4^1:~(\bs{\tau}_1,\bs{\tau}_2,\bs{\tau}_3,\bs{\tau}_4)\mapsto(\bs{\tau}_2,\bs{\tau}_3,\bs{\tau}_4,\bs{\tau}_1)\\
&\bs{R}_4^2:~(\bs{\tau}_1,\bs{\tau}_2,\bs{\tau}_3,\bs{\tau}_4)\mapsto(\bs{\tau}_4,\bs{\tau}_3,\bs{\tau}_2,\bs{\tau}_1)\\
&\bs{R}_4^3:~(\bs{\tau}_1,\bs{\tau}_2,\bs{\tau}_3,\bs{\tau}_4)\mapsto(\bs{\tau}_2,\bs{\tau}_1,\bs{\tau}_4,\bs{\tau}_3)
\end{aligned}
\end{align}
To gap out these spin-1/2 degrees of freedom, we can introduce the symmetric Heisenberg Hamiltonian:
\begin{align}
H_J^{xy}=J^{xy}\left(\bs{\tau}_1\cdot\bs{\tau}_3+\bs{\tau}_2\cdot\bs{\tau}_4\right)
\end{align}
And $H_J^{xy}$ leaves two spin-singlets as a non-degenerate ground state. The problems in $xz$-plane and $yz$-plane are similar, hence the Haldane chain decoration on 1D blocks labeled by $\tau_2$ is \textit{obstruction-free}. We summarize all obstruction-free 1D block states:
\begin{enumerate}[1.]
\item Double Majorana chains decoration on 1D blocks $\tau_1$;
\item Quadruple Majorana chains decoration on each 1D block $\tau_2$;
\item Haldane chain decoration on 1D blocks $\tau_{2,3}$.
\end{enumerate}

Having all obstruction-free 1D block states, we should further investigate if they can be trivialized. The effective on-site symmetry of each 2D block is $\mathbb{Z}_2$, hence there are two possible bubble constructions:
\begin{enumerate}[1.]
\item Majorana chain with anti-PBC;
\item Double Majorana chains.
\end{enumerate}

Majorana bubble construction on each 2D block $\sigma_1$ leaves 4 Majorana chains on each 1D block $\tau_2$ and 2 Majorana chains on each 1D block $\tau_3$. Nevertheless, the corresponding 1D block states are all \textit{obstructed}, hence this type of Majorana bubble construction does not change the 1D block state. Similar to Majorana bubble construction on all other 2D blocks $\sigma_2/\sigma_3$. 

Double Majorana bubble construction on each 2D block $\sigma_1$ leaves 8 Majorana chains on each 1D block $\tau_2$ and 4 Majorana chains on each 1D block $\tau_3$ as a Haldane chain. Hence this bubble construction trivializes the Haldane chain decorations on $\tau_3$. Then double Majorana bubble construction on each 2D block $\sigma_2$ leaves 8 Majorana chains on each 1D block $\tau_2$ and 6 Majorana chains on each 1D block $\tau_3$ (equivalent to double Majorana chains). Hence the double Majorana chains decoration on $\tau_1$ is trivialized. 

As a consequence, all obstruction-free 1D block states are \textit{equivalent} by 2D bubble constructions, and the classification from 1D block states is $\mathbb{Z}_2$.

\paragraph{Spin-1/2 fermions}For spin-1/2 fermions, the effective on-site symmetry of each 1D block $\tau_1$ is $\mathbb{Z}_2\times_{\omega_2}(\mathbb{Z}_3\rtimes\mathbb{Z}_2)$, the effective on-site symmetry of each 1D block $\tau_2$ is $\mathbb{Z}_2^f\times_{\omega_2}(\mathbb{Z}_4\rtimes\mathbb{Z}_2)$, and the effective on-site symmetry of each 1D block $\tau_3$ is $\mathbb{Z}_2^f\times_{\omega_2}(\mathbb{Z}_2\times\mathbb{Z}_2)$. The classifications of the corresponding 1D fSPT phases on these 1D blocks are all trivial \cite{general2}, hence there is no nontrivial 1D block state for spin-1/2 fermions.

\subsubsection{0D block states}
The effective ``on-site'' symmetry on the 0D block $\mu$ is $S_4\times\mathbb{Z}_2$, hence the candidate 0D block states are classified by the following two indices:
\begin{align}
\begin{aligned}
&n_0\in\mathcal{H}^0(S_4\times\mathbb{Z}_2,\mathbb{Z}_2)=\mathbb{Z}_2\\
&\nu_1\in\mathcal{H}^1\left[S_4\times\mathbb{Z}_2,U(1)\right]=\mathbb{Z}_2^2
\end{aligned}
\label{1-supercohomology Oh}
\end{align}
with the twisted cocycle condition:
\begin{align}
\mathrm{d}\nu_1=(-1)^{\omega_2\smile n_0}
\label{1-twisted Oh}
\end{align}
where $n_0$ depicts the parity of fermions, and $\nu_1$ depicts the 0D bSPT mode on the 0D block $\mu$ protected by $S_4\times\mathbb{Z}_2$ symmetry, characterizing the eigenvalues $\pm1$ of reflection generators $\bs{M}_1^j$ and $\bs{M}_2^k$ ($j=1,2,3$ and $k=1,...,6$) of $O_h$. We study the spinless and spin-1/2 fermions separately.

\paragraph{Spinless fermions}For odd number of fermions at $\mu$, similar to $T_h$-symmetric systems, complex fermion decoration on 0D block $\mu$ is a nontrivial block state.

Then we investigate the 1D bubble constructions to seek the potential trivializations of 0D block states characterizing the nontrivial eigenvalues of symmetry generators. Consider 1D blocks labeled by $\tau_2$, we decorate a complex fermion $c_j^\dag$ ($j=1,...,6$) on each of them, which can be trivialized by smoothly deforming to infinite far away. Near the 0D block $\mu$, these complex fermions form an atomic insulator:
\begin{align}
|\psi_\mu\rangle=c_1^\dag c_2^\dag c_3^\dag c_4^\dag c_5^\dag c_6^\dag|0\rangle
\label{Oh atomic1}
\end{align}
Furthermore, these complex fermions will be transformed by reflection $\bs{M}_1^1$ as:
\begin{align}
\bs{M}_1^1:\left(c_1^\dag, c_2^\dag, c_3^\dag, c_4^\dag, c_5^\dag, c_6^\dag\right)\mapsto\left(c_4^\dag, c_2^\dag, c_3^\dag, c_1^\dag, c_5^\dag, c_6^\dag\right)\nonumber
\end{align}
As the consequence, the atomic insulator $|\psi_\mu\rangle$ is an eigenstate of $\bs{M}_1^1$ with eignevalue $-1$:
\begin{align}
\bs{M}_1^1|\psi_\mu\rangle=c_4^\dag c_2^\dag c_3^\dag c_1^\dag c_5^\dag c_6^\dag|0\rangle=-|\psi_\mu\rangle
\label{M1 trivialization}
\end{align}
Therefore, the nontrivial 0D bSPT block state decorated on $\mu$ can be trivialized by 1D bubble construction on $\tau_2$. On the other hand, these complex fermions will be transformed by reflection $\bs{M}_2^1$ as:
\begin{align}
\bs{M}_2^1:\left(c_1^\dag, c_2^\dag, c_3^\dag, c_4^\dag, c_5^\dag, c_6^\dag\right)\mapsto\left(c_1^\dag, c_6^\dag, c_5^\dag, c_4^\dag, c_3^\dag, c_2^\dag\right)\nonumber
\end{align}
As the consequence, the atomic insulator $|\psi_\mu\rangle$ is an eigenstate of $\bs{M}_2^1$ with eignevalue $1$:
\begin{align}
\bs{M}_2^1|\psi_\mu\rangle=c_1^\dag c_6^\dag c_5^\dag c_4^\dag c_3^\dag c_2^\dag|0\rangle=|\psi_\mu\rangle
\end{align}
Subsequently, we consider 1D blocks $\tau_1$ and decorate a complex fermion $a_j^\dag/{a_j'}^\dag$ ($j=1,2,3,4$) on each of them, which can be trivialized by smoothly deforming to infinite far away. Near the 0D block $\mu$, these complex fermions form an atomic insulator:
\begin{align}
|\phi_\mu\rangle=a_1^\dag a_2^\dag a_3^\dag a_4^\dag {a_1'}^\dag {a_2'}^\dag {a_3'}^\dag {a_4'}^\dag|0\rangle
\end{align}
Under $\bs{M}_1^1$, these complex fermions transform as:
\begin{align}
\bs{M}_1^1:~\left\{
\begin{aligned}
&\left(a_1^\dag,a_2^\dag,a_3^\dag,a_4^\dag\right)\mapsto\left({a_2'}^\dag,{a_1'}^\dag,{a_4'}^\dag,{a_3'}^\dag\right)\\
&\left({a_1'}^\dag,{a_2'}^\dag,{a_3'}^\dag,{a_4'}^\dag\right)\mapsto\left(a_2^\dag,a_1^\dag,a_4^\dag,a_3^\dag\right)
\end{aligned}
\right.
\end{align}
Hence the atomic insulator $|\phi_\mu\rangle$ is an eigenstate of $\bs{M}_1^1$ with eigenvalue $1$. Under $\bs{M}_2^1$, these complex fermions transform as:
\begin{align}
\bs{M}_2^1:\left\{
\begin{aligned}
&\left(a_1^\dag,a_2^\dag,a_3^\dag,a_4^\dag\right)\mapsto\left(a_4^\dag,a_2^\dag,a_3^\dag,a_1^\dag\right)\\
&\left({a_1'}^\dag,{a_2'}^\dag,{a_3'}^\dag,{a_4'}^\dag\right)\mapsto\left({a_4'}^\dag,{a_2'}^\dag,{a_3'}^\dag,{a_1'}^\dag\right)
\end{aligned}
\right.\nonumber
\end{align}
Hence the atomic insulator $|\phi_\mu\rangle$ is an eigenstate of $\bs{M}_2^1$ with eigenvalue $1$. 

We should further consider 1D bubble on $\tau_3$: the complex fermions on them are labeled by $\alpha_j^\dag$, $\beta_j^\dag$ and $\gamma_j^\dag$ ($j=1,2,3,4$), forming three atomic insulators on $xy$, $yz$ and $xz$ planes:
\begin{align}
\begin{aligned}
&|\psi_\mu^{xy}\rangle=\alpha_1^\dag\alpha_2^\dag\alpha_3^\dag\alpha_4^\dag|0\rangle\\
&|\psi_\mu^{yz}\rangle=\beta_1^\dag\beta_2^\dag\beta_3^\dag\beta_4^\dag|0\rangle\\
&|\psi_\mu^{xz}\rangle=\gamma_1^\dag\gamma_2^\dag\gamma_3^\dag\gamma_4^\dag|0\rangle\\
\end{aligned}
\label{Oh atomic2}
\end{align}
under $\bs{M}_2^1$, $|\psi_\mu^{yz}\rangle$ and $|\psi_\mu^{xz}\rangle$ exchange, while $|\psi_\mu^{xy}\rangle$ is transformed as:
\begin{align}
\bs{M}_2^1|\psi_\mu^{xy}\rangle=\alpha_1^\dag\alpha_4^\dag\alpha_3^\dag\alpha_2^\dag|0\rangle=-|\psi_\mu^{xy}\rangle
\end{align}
i.e., reflection eigenvalue $-1$ is trivialized by the above three atomic insulators.

Furthermore, on 1D blocks, we can also decorate a 0D bSPT mode on each of them that can be trivialized by smoothly deforming it to infinite far away. Nevertheless, all possible 1D ``bosonic'' bubble constructions will leave an even number of 0D bSPT modes near 0D block $\mu$. Hence only the 0D bSPT mode characterizing the eigenvalue $-1$ of $\bs{M}_1^1$ can be trivialized by 1D bubble construction, and all 0D block states form a $\mathbb{Z}_2$ group.

\paragraph{Spin-1/2 fermions}For spin-1/2 fermions, nonzero $n_0$ in Eq. (\ref{1-supercohomology Oh}) is obstructed by Eq. (\ref{1-twisted Oh}). Furthermore, there should be an additional $-1$ in Eq. (\ref{M1 trivialization}) because of the spin-1/2 nature of fermions, hence the 0D bSPT mode characterizing the eigenvalue $-1$ of $\bs{M}_1^1$ cannot be gapped by 1D bubble constructions, and all 0D block states form a $\mathbb{Z}_2^2$ group.

\subsubsection{Summary\label{summary Oh}}
In this section, we summarize all obstruction and trivialization-free block states with different dimensions, and the ultimate classification of 3D topological crystalline superconductors in the $O_h$-symmetric systems, with spinless and spin-1/2 fermions. 

For spinless fermions, the ultimate classification is $\mathbb{Z}_2^5$, with the following root phases:
\begin{enumerate}[1.]
\item 2D bosonic Levin-Gu state on each 2D block $\sigma_2$ \& $\sigma_3$;
\item 2D bosonic Levin-Gu state on each 2D block $\sigma_1$;
\item 1D block state: Haldane chain on $\tau_{2}$;
\item 1D block state: Quadruple Majorana chains on $\tau_{2}$;
\item Odd number of complex fermions on 0D block $\mu$;
\end{enumerate}

For spin-1/2 fermions, the ultimate classification of 3D topological crystalline superconductors in the $O_h$-symmetric system is $\mathbb{Z}_2^2$, with the following root phases:
\begin{enumerate}[1.]
\item 0D bSPT mode characterizing the eigenvalue $-1$ of the reflection generator $\bs{M}_1^1\in O_h$.
\item 0D bSPT mode characterizing the eigenvalue $-1$ of the reflection generator $\bs{M}_2^1\in O_h$.
\end{enumerate}

\subsubsection{Higher-order topological surface theory}
Now, equipped with specific block states, we are prepared to examine the corresponding higher-order (HO) topological surface theories through the framework of the bulk-boundary correspondence.

For systems without spin, the classification of 3D $O_h$-symmetric topological phases is $\mathbb{Z}_2^4$, featuring four root phases outlined in Section \ref{summary Oh}. we consider these root phases on the open cubic, as depicted in Figure \ref{Oh cell decomposition}.

In the case of the 2D bosonic Levin-Gu state on $\tau_{2,3}$, the topological surface theory inherits an assembly of 1D non-chiral Luttinger liquids on the system's boundary, precisely situated at all the hinges, as well as the diagonal and off-diagonal links of the surfaces of the cubic. Consequently, the resulting topological phase derived from the 2D block state in this scenario represents a 3D bosonic second-order topological phase.

Similarly, when employing the 2D bosonic Levin-Gu state on $\tau_1$, the surface exhibits several non-chiral Luttinger liquids situated along the vertical and horizontal links traversing the center of each surface of the cubic. Consequently, the resulting topological phase derived from this 2D block state corresponds to a 3D bosonic second-order topological phase.

In the case of the Haldane chain decoration, there exists a spin-1/2 degree of freedom at the center of each surface of the open cubic.

For systems with spin-1/2 fermions, the classification of 3D $O_h$-symmetric topological phases is $\mathbb{Z}_2^2$, consisting of two root phases that correspond to 0D block states, as summarized in Section \ref{summary Oh}. Analogous to the last two cases concerning spinless fermions, the 3D $O_h$-symmetric topological phases characterized by these 0D block states do not constitute nontrivial HO topological phases.

\section{Construction and classification of crystalline TI\label{TI classification}}
In Section \ref{TSC classification}, we developed and classified crystalline topological superconductors (TSCs) using explicit topological crystals. Now, in this section, we apply the same general framework to crystalline topological insulators (TIs) on inversion-symmetric and cubic lattices.

\subsection{Inversion-symmetric lattice\label{inversion TI}}
In systems with $U^f(1)$ charge conservation, we first argue that the fermion spin is irrelevant to the classification. In 1D, 2D, and 3D blocks, there are no on-site symmetries other than charge conservation. The on-site symmetry of the 0D block, under inversion symmetry, is $U^f(1)\times\mathbb{Z}_2$. The fermion spin is characterized by a factor system in the short exact sequence:

\begin{align}
0\rightarrow U^f(1)\rightarrow U^f(1)\times\mathbb{Z}_2\rightarrow\mathbb{Z}_2\rightarrow0
\end{align}

This factor system is classified by the group 2-cohomology $\mathcal{H}^2\left[\mathbb{Z}_2,U^f(1)\right]=\mathbb{Z}_1$. Therefore, the fermion spin is irrelevant in this case.

Next, we investigate the block states with different dimensions. For 2D blocks, there are two possible root phases: the Chern insulator and Kitaev's $E_8$ state. In the case of a monolayer Chern insulator on each $\sigma$ block, it leaves two chiral fermion modes near the 1D block $\tau$ with opposite chiralities. These modes can be gapped out by introducing a mass term. On the other hand, bilayer Chern insulators and Kitaev's $E_8$ state can be trivialized using 3D bubble equivalence. For example, a 3D "Chern insulator" bubble can change the layers of Chern insulators on $\sigma$ by two. Similar reasoning applies to Kitaev's $E_8$ state. Consequently, all 2D block states are classified by $\mathbb{Z}_2^2$.

There are no 1D root phases due to the absence of on-site symmetries.

For the 0D block, the root phases form the group $\mathbb{Z}\times\mathbb{Z}_2$, characterizing the $U^f(1)$ charge and the eigenvalue $-1$ under inversion symmetry. We can examine 1D bubble equivalence: by decorating a 0D fermionic mode with a $U^f(1)$ charge of 1 on each 1D block $\tau$, it can be smoothly deformed to infinity and trivialized. Near the 0D block, two such modes change the $U^f(1)$ charge of the 0D block state by two and the eigenvalue of inversion by $-1$. In other words, all trivial 0D block states form a group $2\mathbb{Z}$. Therefore, 0D block states are classified by the quotient group $\mathbb{Z}\times\mathbb{Z}_2/2\mathbb{Z}=\mathbb{Z}_4$, generated by 0D modes with odd $U^f(1)$ charge and eigenvalue $-1$ under inversion symmetry, with a nontrivial group extension.

We should further investigate if there are any group extensions. On the one hand, as previously demonstrated, bilayer Chern insulators on each $\sigma$ can be trivialized by 3D Chern insulator bubble construction; on the other hand, bilayer Chern insulators can be smoothly deformed to a spherical Chern insulator surrounding the 0D block which changes the $U^f(1)$ charge of 0D block state by 1. Consequently, there is a nontrivial group extension between 2D and 0D block-states. Finally, the ultimate classification of inversion-symmetric TI in 3D interacting fermionic systems is:
\begin{align}
    \mathcal{G}_{U(1)}^{S_2}=\Z_8\times \Z_2.
\end{align}
We further investigate the corresponding HO topological surface theories. The monolayer Chern insulator decoration on $\sigma$ leaves chiral fermions on the intersections of the open cubic and red plate in Fig. \ref{inversion cell decomposition}. 

Kitaev’s E8 state decoration on $\sigma$ leaves chiral bosons as the edge theory of E8 state on the intersections of the open cubic and red plate in Fig. \ref{inversion cell decomposition}. These two block-states correspond to two second-order topological surface theories.

\subsection{$T$-symmetric lattice}
For $T$-symmetric lattice, the cell decomposition is illustrated in Fig. \ref{T cell decomposition}, and the wavefunction has the form of Eq. (\ref{T wavefunction}). The decorated block states with different dimensions should be $U^f(1)$ charge conserved.

\subsubsection{2D block states}
There is no effective ``on-site'' symmetry on all 2D blocks, hence for both spinless and spin-1/2 fermions, there are two possible invertible topological phases on 2D blocks: integer quantum Hall insulator (Chern insulator) with chiral central charge $c_-=1$, and Kitaev's $E_8$ state \cite{supplementary} with chiral central charge $c_-=8$. If we decorate a Chern insulator/Kitaev's $E_8$ state with quantum number $n_k\in\mathbb{Z}$ on each 2D block $\sigma_k$ ($k=1,2$), the number of chiral central charges on the 1D blocks labeled by $\tau_1$, $\tau_2$ and $\tau_3$ are $3n_2/12n_2$, $3(n_1+n_2)/12(n_1+n_2)$ and $2n_1/8n_1$, respectively. Fully gapped bulk requires that all of these three quantities should vanish:
\[
12n_2=12(n_1+n_2)=8n_1=0
\]
The only solution to these equations is $n_1=n_2=0$, hence there is no nontrivial 2D block state.

\subsubsection{1D block states}
The effective ``on-site'' symmetry of 1D blocks $\tau_{1,2}/\tau_3$ is $\mathbb{Z}_3/\mathbb{Z}_2$, with the trivial classification of the corresponding root phases for both spinless and spin-1/2 fermions:
\begin{align}
\mathcal{H}^2\left[U^f(1)\times\mathbb{Z}_{3,2},U(1)\right]=\mathbb{Z}_1
\end{align}
Therefore, there is no nontrivial 1D block state.

\subsubsection{0D block states}
The effective ``on-site'' symmetry on 0D block $\mu$ is 4-fold alternating group $A_4$, with the classification:
\begin{align}
\begin{aligned}
n_0&\in\mathcal{H}^0(A_4,\mathbb{Z})=\mathbb{Z}\\
\nu_1&\in\mathcal{H}^1\left[A_4,U(1)\right]=\mathbb{Z}_3
\end{aligned}
\end{align}
where $n_0$ characterizes the sector of $U^f(1)$ charge, and $\nu_1$ represents the 0D bSPT modes. The twisted cocycle condition is defined for fSPT as:
\begin{align}
\mathrm{d}\nu_1=(-1)^{\omega_2\smile n_0},~\omega_2\in\mathcal{H}^2\left[A_4,U(1)\right]
\end{align}
The trivialization of $\mathbb{Z}_3$ is identical with crystalline TSC cases (see Sec. \ref{T0D}), for both spinless and spin-1/2 fermions. 

For the sector of $U^f(1)$ charge conservation, firstly we consider the 1D bubble equivalence. For $\tau_1$ or $\tau_2$, we decorate a 0D mode with $U^f(1)$ charge $e=n_{1,2}\in\mathbb{Z}$ on each of them which can be smoothly deformed to infinite far away and trivialized. Move these complex fermions proxy to the 0D block $\mu$, the $U^f(1)$ charge of 0D block will be changed by $4n_{1,2}$. Similar construction can be applied to 1D blocks $\tau_3$ and the the $U^f(1)$ charge of 0D block will be changed by $6n_{3}\in6\mathbb{Z}$. Combine these 1D bubble constructions lead to the change of $U^f(1)$ charge of $\mu$ by $4n_{1}+4n_2+6n_3$, and all even $U^f(1)$ charge of $\mu$ can be trivialized to 0 by above 1D bubble.

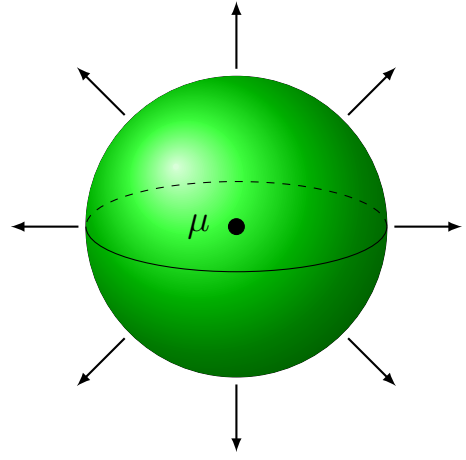
\begin{figure}
\begin{tikzpicture}[scale=1]
\usetikzlibrary{fadings,shadings}
\tikzstyle{sergio}=[rectangle,draw=none]
\draw (0,0) circle (2cm);
\shade[ball color=green] (0,0) circle (2cm);
\draw (-2,0) arc (180:360:2 and 0.6);
\draw[dashed] (2,0) arc (0:180:2 and 0.6);
\fill[fill=black] (0,0) circle (1pt);
\foreach \i in {0,45,...,315}
{
\begin{scope}[rotate=\i] 
\draw[-latex,thick] (2.1,0) -- (3,0);
\end{scope}
}
\filldraw[fill=black, draw=black] (0,0)circle (3pt);
\path (-0.5,0) node [style=sergio] {\Large$\mu$};
\end{tikzpicture}
\caption{Center of $T$-symmetric cubic $\mu$, surrounded by a monolayer Chern insulator with spherical geometry. Black arrows indicate the direction.}
\label{Chern insulator}
\end{figure}

Subsequently, a monolayer Chern insulator on a sphere can be shrunk to a point as a 0D fSPT mode with $U^f(1)$ charge $e=1$. We surround a Chern insulator with spherical geometry around $\mu$ that can be enlarged to infinite far and trivialized, see Fig. \ref{Chern insulator}. Shrink this Chern insulator to a point, it can change the $U^f(1)$ charge of $\mu$ by 1. Finally, the sector of $U^f(1)$ charge of $\mu$ is fully trivialized.

Furthermore, there is no nontrivial HO topological surface theory for $T$-symmetric crystalline TI.

\subsection{$T_h$-symmetric lattice}
For $T_h$-symmetric lattice, the cell decomposition is illustrated in Fig. \ref{Th cell decomposition}, and the wavefunction has the form of Eq. (\ref{Th wavefunction}). The decorated lower-dimensional states should be $U^f(1)$ charge conserved.

\subsubsection{2D block states\label{Th2D U(1)}}
There is no effective ``on-site'' symmetry on $\sigma_1$, with two possible root phases for both spinless and spin-1/2 fermions: Chern insulator and Kitaev's $E_8$ state. If we decorate a Chern insulator/Kitaev's $E_8$ state with index $n_1\in\mathbb{Z}$ on each $\sigma_1$, the chiral central charge of each 1D block $\tau_1$ is $3n_1/12n_1$ that should be vanishing. So there is no nontrivial 2D block state on $\sigma_1$.

The effective ``on-site'' symmetry on $\sigma_2$ is $\mathbb{Z}_2$, and the total symmetry group is the $U^f(1)\times\mathbb{Z}_2$, and the spin of fermions is irrelevant. There is an additional root phase other than Chern insulator and Kitaev's $E_8$ state: 2D $U^f(1)\times\mathbb{Z}_2$ fSPT phases, which is equivalent to fermionic Levin-Gu state with even index \cite{wang2021exactly}.

Consider the decorations of Chern insulators: the chiralities of decorated phases are illustrated in Fig. \ref{edge theory}, and the gapless mode on each $\tau_2$ is a nonchiral Luttinger liquid (\ref{Luttinger}). For the monolayer Chern insulator, the corresponding $K$-matrix is $K=\sigma^z\oplus\sigma^z$. There is a subtle point: we should identify the spin of fermions when dealing with the gapping problem of 1D modes of decorated 2D block states on $\tau_2$ as their shared border: the effective on-site symmetry on $\tau_2$ is $\mathbb{Z}_2\times\mathbb{Z}_2$, which can be nontrivially extended by $U^f(1)$ charge conservation, guaranteed by $\mathcal{H}^2\left[\mathbb{Z}_2\times\mathbb{Z}_2, U^f(1)\right]=\mathbb{Z}_2$.

For spinless fermions, under $\bs{M}_1$ and $\bs{M}_2$ as two generators of $\mathbb{Z}_2\times\mathbb{Z}_2$, the bosonic field $\Phi=(\phi_1,\phi_2,\phi_3,\phi_4)^T$ transforms as Eq. (\ref{K-matrix symmetry}), with
\begin{align}
W^{\bs{M}_1}=\left(
\begin{array}{cccc}
1 & 0 & 0 & 0\\
0 & 0 & 0 & 1\\
0 & 0 & 1 & 0\\
0 & 1 & 0 & 0
\end{array}
\right),~\delta\phi^{\bs{M}_1}=0
\end{align}
\begin{align}
W^{\bs{M}_2}=\left(
\begin{array}{cccc}
0 & 0 & 1 & 0\\
0 & 1 & 0 & 0\\
1 & 0 & 0 & 0\\
0 & 0 & 0 & 1
\end{array}
\right),~\delta\phi^{\bs{M}_2}=0
\end{align}
and $U^f(1)$ charge conservation [$\forall\theta\in U^f(1)$]:
\begin{align}
W^{U^f(1)}=\mathbbm{1}_{4\times4},~\delta\phi^{U^f(1)}=\theta(1,1,1,1)^T
\label{4-component U(1)}
\end{align}
On the one hand, the anomaly indicators (\ref{anomaly indicator}) of two reflection generators $\bs{M}_1$ and $\bs{M}_2$ are non-vanishing, as
\begin{align}
\begin{aligned}
&\nu_{\bs{M}_1}=\frac{1}{4}~(\mathrm{mod}~2)\\
&\nu_{\bs{M}_2}=\frac{1}{4}~(\mathrm{mod}~2)
\end{aligned}
\end{align}
Hence the monolayer Chern insulator on $\sigma_2$ is obstructed for spinless fermions.

On the other hand, there is only one linear independent solution to the ``null-vector'' problem (\ref{null-vector}):
\begin{align}
\Lambda=(1,1,1,1)^T
\end{align}
On the other hand, the corresponding edge theory $\Phi$ requires at least two independent backscattering terms (\ref{backscattering}) to fully gap them out. Hence we cannot symmetrically gap out these 1D modes, which confirms the results from the anomaly indicators.

For spin-1/2 fermions, under $\bs{M}_1$ and $\bs{M}_2$ as reflection generators of $\mathbb{Z}_2\times\mathbb{Z}_2$, the bosonic field $\Phi$ transforms as Eq. (\ref{K-matrix symmetry}), with:
\begin{align}
W^{\bs{M}_1}=\left(
\begin{array}{cccc}
0 & 0 & -1 & 0\\
0 & 1 & 0 & 0\\
-1 & 0 & 0 & 0\\
0 & 0 & 0 & 1
\end{array}
\right),~\delta\phi^{\bs{M}_1}=\frac{\pi}{2}\left(
\begin{array}{cccc}
1\\
-1\\
-1\\
-1
\end{array}
\right)
\end{align}
\begin{align}
W^{\bs{M}_2}=\left(
\begin{array}{cccc}
1 & 0 & 0 & 0\\
0 & 0 & 0 & -1\\
0 & 0 & 1 & 0\\
0 & -1 & 0 & 0
\end{array}
\right),~\delta\phi^{\bs{M}_2}=\frac{\pi}{2}\left(
\begin{array}{cccc}
-1\\
1\\
-1\\
-1
\end{array}
\right)
\end{align}
and $U^f(1)$ charge conservation:
\begin{align}
W^{U^f(1)}=\mathbbm{1}_{4\times4},~~\delta\phi^{U^f(1)}=\theta(1,1,-1,1)^T
\end{align}
On the one hand, the anomaly indicators (\ref{anomaly indicator}) of two reflection generators $\bs{M}_1$ and $\bs{M}_2$ vanish as $\nu_{\bs{M}_1}=\nu_{\bs{M}_2}=0~(\mathrm{mod}~2)$, which indicate that the monolayer Chern insulator on $\sigma_2$ is obstruction-free.

On the other hand, the ``null-vector'' problem (\ref{null-vector}) for the present case is quite subtle: there are two series of linearly independent solutions
\begin{align}
\begin{aligned}
&\left\{
\begin{aligned}
&\Lambda_1=(1,-1,1,1)^T\\
&\Lambda_1'=(1,1,1,-1)^T
\end{aligned}
\right.\\
&\left\{
\begin{aligned}
&\Lambda_2=(1,-1,-1,-1)^T\\
&\Lambda_2'=(1,1,-1,1)^T
\end{aligned}
\right.
\end{aligned}
\end{align}
It appears that $\Lambda_1$ and $\Lambda_1'$ correspond to two distinct backscattering terms (\ref{backscattering}). However, despite their independence, they lead to spontaneous symmetry breaking. Specifically, $\langle\phi^2-\phi^4\rangle$ exhibits two energy vacua: $0$ and $\pi$, which transform into each other through $\bs{M}_2$. A similar situation occurs with $\Lambda_2$ and $\Lambda_2'$, which spontaneously break $\bs{M}_1$ due to $\langle\phi^1-\phi^3\rangle$. By considering $\Lambda_{1,2}$ and $\Lambda_{1,2}'$ together, it is possible to restore these spontaneous symmetry breakings. Moreover, the corresponding backscattering terms can fully gap out the 1D gapless modes at each $\tau_2$, without breaking any symmetry explicitly or spontaneously. Consequently, the monolayer Chern insulator on $\sigma_2$ is obstruction-free for spin-1/2 fermions, which confirms the findings of the anomaly indicators.

Regarding bilayer Chern insulators, we can demonstrate their equivalence to the root phase of a 2D $U^f(1)\times\mathbb{Z}_2$ fSPT phase modulo 4, near each $\sigma_2$. By repeatedly referring to Fig. \ref{3D bubble}, we observe that each bilayer Chern insulator leaves two chiral edge modes on its 1D boundary, with a trivial $\mathbb{Z}_2$ symmetry action: $c_{1,2}^+\mapsto c_{1,2}^+$. Considering two 3D ``Chern insulator" bubbles that share the 2D block $\sigma_2$ as their boundary, we obtain two Chern insulators with chiral edge modes $d_L$ and $d_R$, which have opposite chirality at each $\sigma_2$, and the $\mathbb{Z}_2$ symmetry acts as reflection: $d_L\leftrightarrow d_R$. These chiral edge modes can be redefined as in Eq. (\ref{redefine}), leading to alternative $\mathbb{Z}_2$ symmetry properties: $d^{\pm}\mapsto\pm d^{\pm}$. Notably, the Chern insulators with $c_1^+$ and $d^+$ as their chiral edge modes can be assembled and gapped, while the other two Chern insulators with $c_2^+$ and $d^-$ as their chiral edge modes form the root phase of the 2D $U^f(1)\times\mathbb{Z}_2$ fSPT phase. As a result, we can categorize the bilayer Chern insulators as 2D fSPT decorations for the 2D block with the reflection plane, leading to a $\mathbb{Z}_8$ index.

Next, we consider Kitaev's $E_8$ state as a 2D block state on $\sigma_2$, with the chiralities illustrated in Fig. \ref{edge theory}. In the first step, by temporarily neglecting $U^f(1)$ charge conservation, the monolayer of Kitaev's $E_8$ state can be regarded as 16 layers of $(p+ip)$-superconductor \cite{supplementary}, which can be symmetrically gapped using appropriate backscattering terms (see Sec. \ref{Th2D}). Furthermore, the $U^f(1)$ charge conservation can be restored by assembling these backscattering terms. As a consequence, arbitrary layers of Kitaev's $E_8$ state are obstruction-free. 

We should further investigate whether this 2D block state can be rendered trivial through 3D bubble equivalence. Two possible 3D bubble constructions are considered: the ``Chern insulator bubble" and the ``Kitaev's $E_8$ state bubble". The Chern insulator bubble leads to a nontrivial extension of 2D block states composed of Chern insulators and $U^f(1)\times\mathbb{Z}_2$ fSPT phases. On the other hand, Kitaev's $E_8$ state bubble alters the layers of decorated Kitaev's $E_8$ states on $\sigma_2$ by 2 (refer to Fig. \ref{3D bubble}). As a result, the bilayer Kitaev's $E_8$ states on each $\sigma_2$ become trivialized through 3D bubble equivalence.

In conclusion, for spinless fermions, the nontrivial root phase 2D block states consist of monolayers of Kitaev's $E_8$ state and 2D Levin-Gu state \cite{LevinGu} on each $\sigma_2$, forming a $\mathbb{Z}_2^2$ group. For spin-1/2 fermions, the nontrivial root phase 2D block states consist of a monolayer of Kitaev's $E_8$ state and a Chern insulator on each $\sigma_2$, forming a $\mathbb{Z}_8\times\mathbb{Z}_2$ group.

\subsubsection{1D block states}
The 1D blocks $\tau_1$ have an effective on-site symmetry of $\mathbb{Z}_3$, and the corresponding 1D block states are characterized by the 2-cohomology of the total symmetry group $U^f(1)\times\mathbb{Z}_3$ for both spinless and spin-1/2 fermions. It is found that:
\begin{align}
\mathcal{H}^2\left[U^f(1)\times\mathbb{Z}_3,U(1)\right]=\mathbb{Z}_1
\end{align}
This implies that there are no nontrivial 1D block states on $\tau_1$.

On the other hand, the effective on-site symmetry of 1D blocks $\tau_2$ is $\mathbb{Z}_2\times\mathbb{Z}_2$, and the corresponding 1D block states are characterized by the 2-cohomology of the total symmetry group $U^f(1)\times(\mathbb{Z}_2\times\mathbb{Z}_2)$ for spinless fermions. Specifically:
\begin{align}
\mathcal{H}^2\left[U^f(1)\times(\mathbb{Z}_2\times\mathbb{Z}_2),U(1)\right]=\mathbb{Z}_2
\end{align}
This reveals that the nontrivial 1D root phase is the Haldane chain. As shown in Sec. \ref{Th1D}, the Haldane chain decoration on $\tau_2$ is a nontrivial 1D block state for spinless systems, and the arguments hold true for $U^f(1)$ charge-conserving systems.

However, for systems with spin-1/2 fermions, the 1D block states on $\tau_2$ are characterized by the 2-cohomology of the total symmetry group $U^f(1)\rtimes_{\rho_1,\omega_2}(\mathbb{Z}_2\times\mathbb{Z}_2)$:
\begin{align}
\mathcal{H}^2\left[U^f(1)\rtimes_{\rho_1,\omega_2}(\mathbb{Z}_2\times\mathbb{Z}_2),U(1)\right]=\mathbb{Z}_1
\end{align}
Here, the symbol ``$\rtimes_{\rho_1,\omega_2}$" represents the nontrivial group extension \cite{wang2021exactly}. Therefore, there are no nontrivial 1D block states on $\tau_2$ for spin-1/2 fermions.

To summarize, for spinless fermions, all 1D block states form a $\mathbb{Z}_2$ group, with the Haldane chain decoration on $\tau_2$ as the root phase. In contrast, for spin-1/2 fermions, there are no nontrivial 1D block states.

\subsubsection{0D block states}
The effective on-site symmetry on the 0D block $\mu$ is $A_4\times\mathbb{Z}_2$, with the classification data:
\begin{align}
\begin{aligned}
&n_0\in\mathcal{H}^0(A_4\times\mathbb{Z}_2,\mathbb{Z})=\mathbb{Z}\\
&\nu_1\in\mathcal{H}^1\left[A_4\times\mathbb{Z}_2,U(1)\right]=\mathbb{Z}_2\times\mathbb{Z}_3
\end{aligned}
\end{align}
The twisted cocycle condition (\ref{1-twisted Th}) is satisfied, where $n_0$ represents the $U^f(1)$ charge on $\mu$, and $\nu_1$ represents the 0D $A_4\times\mathbb{Z}_2$ bSPT modes. In Section \ref{Th0D}, we explained that the $\mathbb{Z}_3$ index can be rendered trivial through a 1D bubble construction on $\tau_1$, and these arguments hold true for $U^f(1)$ charge-conserving systems, including both spinless and spin-1/2 fermions.

For spinless fermions, let's consider an alternative 1D bubble construction. We decorate the fermionic 0D modes with $U^f(1)$ charges $n_1/n_2$ on each 1D block $\tau_1/\tau_2$ ($n_{1,2}\in\mathbb{Z}$). In particular, a ``particle" has a $U^f(1)$ charge of $+1$, and a ``hole" has a $U^f(1)$ charge of $-1$. When $n_2=1$, the 1D bubble on $\tau_2$ results in modes with a $U^f(1)$ charge of 6 on $\mu$, forming an atomic insulator (\ref{Th atomic insulator}) with a reflection eigenvalue of $-1$. On the other hand, when $n_1=1$, the 1D bubble on $\tau_1$ yields 0D modes with a $U^f(1)$ charge of 8 on $\mu$. We label the 0D mode on $\mu$ as $(n,\pm)$, where $n$ represents the $U^f(1)$ charge, and $\pm$ indicates the reflection eigenvalue. By employing these 1D bubble constructions, a vacuum 0D mode $(0,+)$ can be transformed into $(6n_1+8n_2,(-1)^{n_1})$, resulting in a $2\mathbb{Z}$ group:
\begin{align}
\left\{\left(6n_1+8n_2,(-1)^{n_1}\right)\big|n_1,n_2\in\mathbb{Z}\right\}=2\mathbb{Z}
\end{align}
We need to factor out these trivial phases from the bubble constructions, and all nontrivial 0D block states form a $\mathbb{Z}_4$ group, where $(2,+)$ is equivalent to $(0,-)$.

\paragraph{Spin-1/2 fermions}Firstly, 0D modes with odd $U^f(1)$ charge are obstructed by (\ref{1-twisted Th}). And for 1D block $\tau_2$ with $\mathbb{Z}_2\times\mathbb{Z}_2$ on-site symmetry, 0D modes with odd $U^f(1)$ charge are also obstructed. Then decorating fermionic 0D modes with $n_1/2n_2$ $U^f(1)$ charges on each 1D block $\tau_1/\tau_2$ ($n_{1,2}\in\mathbb{Z}$). For $n_1=1$, the 1D bubbles on $\tau_1$ leave 0D modes with $U^f(1)$ charge 8 on $\mu$; for $n_2=1$, the 1D bubble on $\tau_2$ leaves modes with $U^f(1)$ charge 6 on $\mu$, forming two copies of atomic insulator (\ref{Th atomic insulator}) with reflection eigenvalue $+1$. We label the 0D mode on $\mu$ by $(2n,\pm)$, where $2n$ characterizes the $U^f(1)$ charge and $\pm$ characterizes the reflection eigenvalue. By above bubble constructions, a vacuum 0D mode $(0,+)$ can be deformed to $(12n_1+8n_2,+)$, forming a $4\mathbb{Z}$ group:
\begin{align}
\left\{\left(12n_1+8n_2,+\right)\big|n_1,n_2\in\mathbb{Z}\right\}=4\mathbb{Z}
\end{align}
Quotient out these trivial phases and obtain that all nontrivial 0D block states form a $\mathbb{Z}_2^2$ group. We should further note that the Chern insulator surrounding $\mu$ illustrated in Fig. \ref{Chern insulator} is not compatible with reflection symmetry, hence there are no further trivializations for $T_h$-symmetric case.

\subsubsection{Summary\label{Th U(1) summary}}
In this section, we summarize all obstruction and trivialization-free block states and the ultimate classification of 3D topological crystalline insulators in $T_h$-symmetric systems, with spinless and spin-1/2 fermions.

For spinless fermions, the ultimate classification is $\mathbb{Z}_4\times\mathbb{Z}_2^3$, with the following root phases:
\begin{enumerate}[1.]
\item 2D $U^f(1)\times\mathbb{Z}_2$ bSPT phase decoration on each 2D block $\sigma_2$;
\item 2D Kitaev's $E_8$ state decoration on each 2D block $\sigma_2$;
\item Haldane chain decoration on each 1D block $\tau_2$;
\item 0D block states characterizing the parity of $U^f(1)$ charge and reflection eigenvalues $\pm$, with nontrivial extension. 
\end{enumerate}

For spin-1/2 fermions, the ultimate classification is $\mathbb{Z}_8\times\mathbb{Z}_2^3$, with the following root phases:
\begin{enumerate}[1.]
\item 2D monolayer Chern insulator and $U^f(1)\times\mathbb{Z}_2$ fSPT phases decoration on each 2D block $\sigma_2$, with nontrivial extension: bilayer Chern insulators should be extended to the root phase of 2D $U^f(1)\times\mathbb{Z}_2$ fSPT phases;
\item 2D Kitaev's $E_8$ state decoration on each 2D block $\sigma_2$;
\item 0D bSPT mode as block state;
\item 0D block state with $U^f(1)$ charge $n\equiv2~(\mathrm{mod}~4)$.
\end{enumerate}

\subsubsection{Higher-order topological surface theory}
With concrete block states, we are ready to investigate the corresponding HO topological surface theories by higher-order bulk-boundary correspondence. 

For $U^f(1)$ charge conserved systems with spinless fermions, the corresponding classification of 3D $T_h$-symmetric topological phases is $\mathbb{Z}_4\times\mathbb{Z}_2^3$, with root phases as summarized in Sec. \ref{Th U(1) summary}. 

For 2D block states, the classification is $\mathbb{Z}_2^2$ with two root phases: Kitaev's $E_8$ state and 2D bosonic Levin-Gu state on each $\sigma_2$. 

\paragraph{Kitaev's $E_8$ state}There are two chiral Luttinger liquids with $K$-matrix (\ref{K-matrix E8}) on each surface of the cubic (vertical and horizontal links across the center), each of them has chiral central charge $c_-=8$. 

\paragraph{2D bosonic Levin-Gu state}On each surface of the open cubic lattice, there exist two nonchiral Luttinger liquids. These Luttinger liquids have a $K$-matrix given by $K=\sigma^x$ and an on-site symmetry of $\mathbb{Z}_2$. Specifically, we have
\begin{align}
W^{\mathbb{Z}2}=\mathbbm{1}{2\times2},~\delta\phi^{\mathbb{Z}_2}=\pi(1,1)^T
\label{bosonic Levin-Gu}
\end{align}

When considering 1D blocks, the sole root phase present is the Haldane chain on each $\tau_2$. This configuration results in a spin-1/2 degree of freedom located at the center of each surface of the open cubic lattice.

For $U^f(1)$ charge-conserving systems with spin-1/2 fermions, the classification of 3D $T_h$-symmetric topological phases is $\mathbb{Z}_8\times\mathbb{Z}_2^3$, with root phases as summarized in Section \ref{Th U(1) summary}.

Regarding 2D block states, the classification is $\mathbb{Z}_8\times\mathbb{Z}2$, featuring two root phases: the monolayer Chern insulator and Kitaev's $E_8$ state on each $\sigma{1,2}$. Additionally, there is a nontrivial group extension between bilayer Chern insulators and the root phase of 2D $U^f(1)\times\mathbb{Z}_2$ fSPT states.

Let's focus on the Chern insulator. When decorating with monolayer Chern insulators, we observe the presence of two chiral Luttinger liquids on each surface of the open cubic lattice. These chiral Luttinger liquids have a chiral central charge of $c_-=1$.

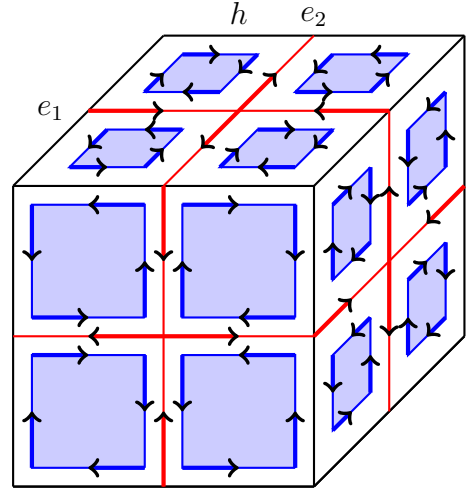
\begin{figure}
\begin{tikzpicture}[scale=1]
\tikzstyle{sergio}=[rectangle,draw=none]
\draw[thick] (8,2) -- (12,2);
\draw[thick] (8,2) -- (6,0);
\draw[thick] (12,2) -- (10,0);
\draw[thick] (6,0) -- (10,0);
\draw[thick] (6,0) -- (6,-4);
\draw[thick] (10,-4) -- (6,-4);
\draw[thick] (10,-4) -- (10,0);
\draw[thick] (12,-2) -- (12,2);
\draw[thick] (12,-2) -- (10,-4);
\draw[thick] (12,-2) -- (10,-4);
\draw[thick] (6,-4) -- (10,-4);
\draw[draw=red, thick] (8,0)--(10,2);
\draw[draw=red, thick] (10,-2)--(12,0);
\draw[draw=red, thick] (11,-3)--(11,1);
\draw[draw=red, thick] (8,-4)--(8,0);
\draw[draw=red, thick] (10,-2)--(6,-2);
\draw[draw=red, thick] (7,1)--(11,1);
\draw[draw=red,->, ultra thick] (7,1)--(8,1);
\draw[draw=red,->, ultra thick] (11,1)--(10,1);
\draw[draw=red,->, ultra thick] (9,1)--(9.5,1.5);
\draw[draw=red,->, ultra thick] (9,1)--(8.5,0.5);
\draw[draw=red,->, ultra thick] (8,0)--(8,-1);
\draw[draw=red,->, ultra thick] (8,-4)--(8,-3);
\draw[draw=red,->, ultra thick] (8,-2)--(9,-2);
\draw[draw=red,->, ultra thick] (8,-2)--(7,-2);
\draw[draw=red,->, ultra thick] (10,-2)--(10.5,-1.5);
\draw[draw=red,->, ultra thick] (12,0)--(11.5,-0.5);
\draw[draw=red,->, ultra thick] (11,-1)--(11,0);
\draw[draw=red,->, ultra thick] (11,-1)--(11,-2);
\filldraw[fill=blue!20,thick,draw=blue] (9.75,-3.75) -- (8.25,-3.75) -- (8.25,-2.25) -- (9.75,-2.25) -- (9.75,-3.75);
\draw[draw=blue,->, ultra thick] (9.75,-2.25)--(9,-2.25);
\draw[draw=blue,->, ultra thick] (8.25,-2.25)--(8.25,-3);
\draw[draw=blue,->, ultra thick] (8.25,-3.75)--(9,-3.75);
\draw[draw=blue,->, ultra thick] (9.75,-3.75)--(9.75,-3);
\filldraw[fill=blue!20,thick,draw=blue] (7.75,-3.75) -- (6.25,-3.75) -- (6.25,-2.25) -- (7.75,-2.25) -- (7.75,-3.75);
\draw[draw=blue,->, ultra thick] (6.25,-2.25)--(7,-2.25);
\draw[draw=blue,->, ultra thick] (6.25,-3.75)--(6.25,-3);
\draw[draw=blue,->, ultra thick] (7.75,-3.75)--(7,-3.75);
\draw[draw=blue,->, ultra thick] (7.75,-2.25)--(7.75,-3);
\filldraw[fill=blue!20,thick,draw=blue] (7.75,-1.75) -- (6.25,-1.75) -- (6.25,-0.25) -- (7.75,-0.25) -- (7.75,-1.75);
\draw[draw=blue,->, ultra thick] (7.75,-0.25)--(7,-0.25);
\draw[draw=blue,->, ultra thick] (6.25,-0.25)--(6.25,-1);
\draw[draw=blue,->, ultra thick] (6.25,-1.75)--(7,-1.75);
\draw[draw=blue,->, ultra thick] (7.75,-1.75)--(7.75,-1);
\filldraw[fill=blue!20,thick,draw=blue] (9.75,-1.75) -- (8.25,-1.75) -- (8.25,-0.25) -- (9.75,-0.25) -- (9.75,-1.75);
\draw[draw=blue,->, ultra thick] (8.25,-0.25)--(9,-0.25);
\draw[draw=blue,->, ultra thick] (8.25,-1.75)--(8.25,-1);
\draw[draw=blue,->, ultra thick] (9.75,-1.75)--(9,-1.75);
\draw[draw=blue,->, ultra thick] (9.75,-0.25)--(9.75,-1);
\filldraw[fill=blue!20,thick,draw=blue] (10.75,-1.75) -- (10.25,-2.25) -- (10.25,-3.25) -- (10.75,-2.75) -- (10.75,-1.75);
\draw[draw=blue,->, ultra thick] (10.25,-2.25)--(10.25,-2.75);
\draw[draw=blue,->, ultra thick] (10.25,-3.25)--(10.5,-3);
\draw[draw=blue,->, ultra thick] (10.75,-2.75)--(10.75,-2.25);
\draw[draw=blue,->, ultra thick] (10.75,-1.75)--(10.5,-2);
\filldraw[fill=blue!20,thick,draw=blue] (11.75,1.25) -- (11.25,0.75) -- (11.25,-0.25) -- (11.75,0.25) -- (11.75,1.25);
\draw[draw=blue,->, ultra thick] (11.25,0.75)--(11.25,0.25);
\draw[draw=blue,->, ultra thick] (11.25,-0.25)--(11.5,0);
\draw[draw=blue,->, ultra thick] (11.75,0.25)--(11.75,0.75);
\draw[draw=blue,->, ultra thick] (11.75,1.25)--(11.5,1);
\filldraw[fill=blue!20,thick,draw=blue] (11.75,-0.75) -- (11.25,-1.25) -- (11.25,-2.25) -- (11.75,-1.75) -- (11.75,-0.75);
\draw[draw=blue,->, ultra thick] (11.25,-2.25)--(11.25,-1.75);
\draw[draw=blue,->, ultra thick] (11.75,-1.75)--(11.5,-2);
\draw[draw=blue,->, ultra thick] (11.75,-0.75)--(11.75,-1.25);
\draw[draw=blue,->, ultra thick] (11.25,-1.25)--(11.5,-1);
\filldraw[fill=blue!20,thick,draw=blue] (10.75,0.25) -- (10.25,-0.25) -- (10.25,-1.25) -- (10.75,-0.75) -- (10.75,0.25);
\draw[draw=blue,->, ultra thick] (10.25,-1.25)--(10.25,-0.75);
\draw[draw=blue,->, ultra thick] (10.75,-0.75)--(10.5,-1);
\draw[draw=blue,->, ultra thick] (10.75,0.25)--(10.75,-0.25);
\draw[draw=blue,->, ultra thick] (10.25,-0.25)--(10.5,0);
\filldraw[fill=blue!20, draw=blue, thick] (9.25,0.75)--(10.25,0.75)--(9.75,0.25)--(8.75,0.25)--cycle;
\draw[draw=blue,->, ultra thick] (9.75,0.25)--(9.25,0.25);
\draw[draw=blue,->, ultra thick] (10.25,0.75)--(10,0.5);
\draw[draw=blue,->, ultra thick] (9.25,0.75)--(9.75,0.75);
\draw[draw=blue,->, ultra thick] (8.75,0.25)--(9,0.5);
\filldraw[fill=blue!20, draw=blue, thick] (10.25,1.75)--(11.25,1.75)--(10.75,1.25)--(9.75,1.25)--cycle;
\draw[draw=blue,->, ultra thick] (9.75,1.25)--(10.25,1.25);
\draw[draw=blue,->, ultra thick] (10.75,1.25)--(11,1.5);
\draw[draw=blue,->, ultra thick] (11.25,1.75)--(10.75,1.75);
\draw[draw=blue,->, ultra thick] (10.25,1.75)--(10,1.5);
\filldraw[fill=blue!20, draw=blue, thick] (7.25,0.75)--(8.25,0.75)--(7.75,0.25)--(6.75,0.25)--cycle;
\draw[draw=blue,->, ultra thick] (6.75,0.25)--(7.25,0.25);
\draw[draw=blue,->, ultra thick] (7.75,0.25)--(8,0.5);
\draw[draw=blue,->, ultra thick] (8.25,0.75)--(7.75,0.75);
\draw[draw=blue,->, ultra thick] (7.25,0.75)--(7,0.5);
\filldraw[fill=blue!20, draw=blue, thick] (8.25,1.75)--(9.25,1.75)--(8.75,1.25)--(7.75,1.25)--cycle;
\draw[draw=blue,->, ultra thick] (8.75,1.25)--(8.25,1.25);
\draw[draw=blue,->, ultra thick] (9.25,1.75)--(9,1.5);
\draw[draw=blue,->, ultra thick] (8.25,1.75)--(8.75,1.75);
\draw[draw=blue,->, ultra thick] (7.75,1.25)--(8,1.5);
\path (6.5,1) node [style=sergio] {\large$e_1$};
\path (10,2.25) node [style=sergio] {\large$e_2$};
\path (9,2.3) node [style=sergio] {\large$h$};
\end{tikzpicture}
\caption{Topological surface theory of 3D $T_h$-symmetric second-order topological phase from bilayer Chern insulators/the root phase of 2D $U^f(1)\times\mathbb{Z}_2$ fSPT states decoration on $\sigma_2$. each arrowed red segment labeled by $e_1/e_2$ depicts a chiral edge mode of bilayer Chern insulators with chiral central charge $c_-=2$; each arrowed blue plate depicts a monolayer Chern insulator with chiral central charge $c_-=1$. $h$ labels the hinges of the open cubic.}
\label{Th2D U(1) surface}
\end{figure}

In the previous section discussing bulk block states (refer to Section \ref{Th2D U(1)}), we have established the equivalence between bilayer Chern insulators and the root phase of 2D $U^f(1)\times\mathbb{Z}_2$ fSPT states on each $\sigma_2$ through the 3D "Chern insulator" bubble construction (see Fig. \ref{Chern insulator}). Now, in this section, we aim to demonstrate this equivalence on the boundary.

As depicted in Fig. \ref{Th2D U(1) surface}, each surface of the cubic lattice hosts two chiral Luttinger liquids with a chiral central charge of $c_-=2$. These chiral Luttinger liquids appear at the vertical and horizontal segments crossing the center. Additionally, on each surface, the red segments represent shared borders between adjacent 2D plates. To illustrate the equivalence, we decorate a monolayer Chern insulator on each plate (indicated by the blue plates in Fig. \ref{Th2D U(1) surface}).

In the vicinity of each red segment, the 1D physics resembles that shown in Fig. \ref{3D bubble}. By combining all the red and blue segments near each $e_1/e_2$ location, the total chiral central charge cancels out, and the corresponding 1D edge theory transforms into a nonchiral Luttinger liquid (\ref{Luttinger}). This nonchiral Luttinger liquid possesses a $K$-matrix given by $K=\sigma^z$ and an on-site symmetry of $\mathbb{Z}_2$. Specifically, we have:
\begin{align}
W^{\mathbb{Z}2}=\mathbbm{1}{2\times2},~\delta\phi^{\mathbb{Z}_2}=\pi(0,1)^T
\label{nu=2}
\end{align}
Remarkably, this nonchiral Luttinger liquid edge theory is identical to the 1D edge theory associated with the root phase of 2D $U^f(1)\times\mathbb{Z}_2$ fSPT states. At each hinge of the cubic lattice, two chiral Luttinger liquids with opposite chirality emerge from nearby 2D plates, which can be fully gapped. Consequently, the topological surface theory further supports the equivalence between bilayer Chern insulators and the root phase of $U^f(1)\times\mathbb{Z}_2$ fSPT state decorations on $\sigma_2$.

\paragraph{Kitaev's $E_8$ state}There are two chiral Luttinger liquids with $K$-matrix: 
\begin{align}
K_{E_8}=\left(
\begin{array}{cccccccc}
2 & -1 & 0 & 0 & 0 & 0 & 0 & 0\\
-1 & 2 & -1 & 0 & 0 & 0 & -1 & 0\\
0 & -1 & 2 & -1 & 0 & 0 & 0 & 0\\
0 & 0 & -1 & 2 & -1 & 0 & 0 & 0\\
0 & 0 & 0 & -1 & 2 & -1 & 0 & 0\\
0 & 0 & 0 & 0 & -1 & 2 & 0 & 0\\
0 & -1 & 0 & 0 & 0 & 0 & 2 & -1\\
0 & 0 & 0 & 0 & 0 & 0 & -1 & 2
\end{array}
\right)
\label{K-matrix E8}
\end{align}
on each surface of the cubic (vertical and horizontal links across the center), with chiral central charge $c_-=8$. 

\subsection{$T_d$-symmetric lattice}
For $T_d$-symmetric lattice, the cell decomposition is illustrated in Figs. \ref{Td cell decomposition} and \ref{Td unit cell}, and wavefunction has the form of Eq. (\ref{Td wavefunction}). The decorated block states should be $U^f(1)$ charge conserved. 

\subsubsection{2D block states}
The effective on-site symmetry of 2D blocks $\sigma_{1,2,3}$ is $\mathbb{Z}_2$. From Sec. \ref{Th2D U(1)}, there are three root phases:
\begin{enumerate}[1.]
\item Chern insulators, with $\mathbb{Z}$ classification;
\item Kitaev's $E_8$ state, with $\mathbb{Z}$ classification;
\item 2D $U^f(1)\times\mathbb{Z}_2$ fSPT phases, with $\mathbb{Z}_4$ classification;
\end{enumerate}

\paragraph{Chern insulator decoration} The chiralities of the decorated phases are depicted in Fig. \ref{Td chirality}, and the gapless mode on each $\tau_3$ surface can be described by a nonchiral Luttinger liquid (\ref{Luttinger}) with $\mathbb{Z}_2\times\mathbb{Z}_2$ symmetry, provided that the chiral central charges of the Chern insulators decorated on the 2D blocks $\sigma_{1,2,3}$ are equal. Similar to the discussion in Section \ref{Th2D U(1)}, the decoration of a monolayer Chern insulator is \textit{obstructed} for spinless fermions but \textit{obstruction-free} for spin-1/2 fermions. On the other hand, bilayer Chern insulators can be smoothly transformed into the root phase of 2D $U^f(1)\times\mathbb{Z}_2$ fSPT states through the 3D ``Chern insulator" bubble construction (see Fig. \ref{3D bubble}), resulting in a $\mathbb{Z}_8$ classification.

\paragraph{Kitaev's $E_8$ state decoration} The chiralities of the decorated phases are shown in Fig. \ref{Td chirality}. Similar to the discussion in Section \ref{Th2D U(1)}, arbitrary layers of Kitaev's $E_8$ states are \textit{obstruction-free}, but bilayer Kitaev's $E_8$ states can be trivialized by the 3D ``Kitaev $E_8$" bubble construction. Hence, for systems with either spinless or spin-1/2 fermions, the decorations of Kitaev's $E_8$ states on $\sigma_{1,2,3}$ yield a $\mathbb{Z}_2$ classification.

\paragraph{2D fSPT state decoration} In Section \ref{Td2D}, we have demonstrated that, for spinless fermions, only the 2D bosonic Levin-Gu state is \textit{obstruction-free}, and the same holds true for systems with $U^f(1)$ charge conservation.

\subsubsection{1D block states}
The effective on-site symmetry of the 1D blocks $\tau_{1,2}$ can be described by $\mathbb{Z}_3\rtimes\mathbb{Z}_2$. The spin of fermions is characterized by the short exact sequence:
\begin{align}
0\rightarrow U^f(1)\rightarrow G_f\rightarrow\mathbb{Z}_3\rtimes\mathbb{Z}_2\rightarrow0
\end{align}
The trivial factor system, ensured by group 2-cohomology $\mathcal{H}^2\left[\mathbb{Z}_3\rtimes\mathbb{Z}_2, U^f(1)\right]=\mathbb{Z}_1$, indicates that the spin of fermions is irrelevant when investigating the 1D block states on $\tau_{1,2}$. The 1D root phases are characterized by projective representations of the total symmetry group, which are classified by group 2-cohomology:
\begin{align}
\mathcal{H}^2\left[U^f(1)\times(\mathbb{Z}_3\rtimes\mathbb{Z}_2),U(1)\right]=\mathbb{Z}_1
\label{D3 1D U(1)}
\end{align}
This implies that there are no nontrivial 1D block states on $\tau_{1,2}$. On the other hand, the effective ``on-site" symmetry of $\tau_3$ is $\mathbb{Z}_2\times\mathbb{Z}_2$. The Haldane chain represents the only nontrivial root phase for spinless fermions, while there are no nontrivial root phases for spin-1/2 fermions (see Sec. \ref{Th2D U(1)}). Furthermore, the decoration of the Haldane chain on $\tau_3$ is both obstruction-free and free from trivialization.

\subsubsection{0D block states}
The effective on-site symmetry of the 0D block $\mu$ is $S_4$, with the classification data:
\begin{align}
\begin{aligned}
&n_0\in\mathcal{H}^0(S_4,\mathbb{Z})=\mathbb{Z}\\
&\nu_1\in\mathcal{H}^1\left[S_4,U(1)\right]=\mathbb{Z}_2
\end{aligned}
\end{align}
The twisted cocycle condition (\ref{1-twisted Td}) is satisfied in this case. Here, $n_0$ represents the $U^f(1)$ charge on $\mu$, and $\nu_1$ represents the 0D $S_4$ bSPT modes that characterize the reflection eigenvalues $\pm$. Similar to the $T_h$-symmetric case, the use of a Chern insulator surrounding $\mu$ (see Fig. \ref{Chern insulator}) is not applicable due to its incompatibility with reflection symmetry.

\paragraph{Spinless fermions} Let's consider the 1D bubble equivalence by decorating 0D modes with $n_{1,2,3}\in\mathbb{Z}$ $U^f(1)$ charges on each 1D block $\tau_{1,2,3}$. For $n_1/n_2=1$, the 1D bubble on $\tau_1/\tau_2$ results in modes with $U^f(1)$ charge 4 on $\mu$, forming an atomic insulator (\ref{Td atomic insulator}) with a reflection eigenvalue of $-1$. For $n_3=1$, the 1D bubble on $\tau_3$ leads to modes with $U^f(1)$ charge 6 on $\mu$. All trivial 0D block states that can be deformed from the vacuum $(0,+)$ by a 1D bubble form a group given by:
\begin{align}
\left\{\left(4n_1+4n_2+6n_3,(-1)^{n_1+n_2}\right)\big|n_{1,2,3}\in\mathbb{Z}\right\}=\mathbb{Z}
\end{align}
To characterize the nontrivial 0D block states, we need to quotient out these trivial phases, resulting in a $\mathbb{Z}_2$ group that characterizes the parity of the $U^f(1)$ charge.

\paragraph{Spin-1/2 fermions} Firstly, it is worth noting that 0D modes with an odd $U^f(1)$ charge are obstructed by (\ref{1-twisted Td}). Additionally, for 1D blocks $\tau_3$ with $\mathbb{Z}_2\times\mathbb{Z}_2$ symmetry, 0D modes with an odd $U^f(1)$ charge are also obstructed.

Now let's decorate the fermionic 0D modes with $n_{1,2}/2n_3$ $U^f(1)$ charges on each 1D block $\tau_{1,2}/\tau_3$ ($n_{1,2},n_3\in\mathbb{Z}$). For $n_{1,2}$, the 1D bubbles on $\tau_{1,2}$ result in modes with a $U^f(1)$ charge of 4 on $\mu$, forming an atomic insulator (\ref{Td atomic insulator}) with a reflection eigenvalue of $-1$. For $n_3=1$, the 1D bubbles on $\tau_3$ leave modes with a $U^f(1)$ charge of 12 on $\mu$.

All the trivial 0D block states that can be deformed from the vacuum $(0,+)$ by a 1D bubble form a group given by:
\begin{align}
\left\{\left(4(n_1+n_2+3n_3),(-1)^{n_1+n_2}\right)\big|n_{1,2,3}\in\mathbb{Z}\right\}=2\mathbb{Z}\nonumber
\end{align}

To characterize the nontrivial 0D block states, we need to quotient out these trivial phases, resulting in a $\mathbb{Z}_2$ group that characterizes the $U^f(1)$ charge $n\equiv2~(\mathrm{mod}~4)$.

\subsubsection{Summary\label{Td U(1) summary}}
In this section, we summarize all obstruction and trivialization-free block states with different dimensions.

For spinless fermions, the ultimate classification is $\mathbb{Z}_2^4$, with the following root phases:
\begin{enumerate}[1.]
\item 2D $U^f(1)\times\mathbb{Z}_2$ bSPT phase decoration on each 2D block $\sigma_{1,2,3}$;
\item Kitaev's $E_8$ state decoration on each 2D block $\sigma_{1,2,3}$;
\item Haldane chain decoration on each 1D block $\tau_3$;
\item 0D modes with odd $U^f(1)$ charge decoration on $\mu$.
\end{enumerate}

For spin-1/2 fermions, the ultimate classification is $\mathbb{Z}_8\times\mathbb{Z}_2^2$, with the following root phases:
\begin{enumerate}[1.]
\item 2D monolayer Chern insulator and $U^f(1)\times\mathbb{Z}_2$ fSPT phases decoration on each 2D block $\sigma_{1,2,3}$, with nontrivial extension: bilayer Chern insulators are extended to the root phase of 2D $U^f(1)\times\mathbb{Z}_2$ fSPT phases;
\item Kitaev's $E_8$ state decoration on each 2D block $\sigma_{1,2,3}$;
\item 0D block state with $U^f(1)$ charge $n\equiv2~(\mathrm{mod}~4)$.
\end{enumerate}

\subsubsection{Higher-order topological surface theory}
Having established the concrete block states, we can now explore the corresponding higher-order (HO) topological surface theory using the bulk-boundary correspondence.

For systems with spinless fermions and $U^f(1)$ charge conservation, the classification of 3D $T_d$-symmetric topological phases is $\mathbb{Z}_2^4$, with the root phases summarized in Sec. \ref{Td U(1) summary}.

Regarding the 2D block states, the classification is $\mathbb{Z}2^2$ and consists of two root phases: Kitaev's $E_8$ state and the 2D bosonic Levin-Gu state decorations on $\sigma{1,2,3}$.

\paragraph{Kitaev's $E_8$ state} Each surface of the open cubic exhibits two chiral Luttinger liquids with a $K$-matrix given by (\ref{K-matrix E8}). These chiral liquids possess a chiral central charge of $c_-=8$.

\paragraph{2D bosonic Levin-Gu state} On the surface of the open cubic, there exist two nonchiral Luttinger liquids characterized by $K=\sigma^x$ and an on-site $\mathbb{Z}_2$ symmetry (\ref{bosonic Levin-Gu}).

Regarding the 1D block states, the classification is $\mathbb{Z}_2$ with the root phase being the Haldane chain present on each $\tau_3$. This root phase gives rise to a spin-1/2 degree of freedom at the center of each surface of the open cubic.

In the case of $U^f(1)$ charge conserved systems with spin-1/2 fermions, the classification of 3D $T_d$-symmetric topological phases is $\mathbb{Z}_8\times\mathbb{Z}_2^2$, with the root phases summarized in Sec. \ref{Td U(1) summary}.

Regarding the 2D block states, the classification is $\mathbb{Z}_8\times\mathbb{Z}2$, featuring two root phases: the monolayer Chern insulator and Kitaev's $E_8$ state on each $\sigma{1,2,3}$, along with a nontrivial group extension connecting bilayer Chern insulators and the root phase of 2D $U^f(1)\times\mathbb{Z}_2$ fSPT states.

\paragraph{Monolayer Chern insulator} When considering monolayer Chern insulator decorations, each surface of the open cubic exhibits two chiral Luttinger liquids with a chiral central charge of $c_-=1$. These liquids are located at the diagonal and off-diagonal segments across the center.

\paragraph{Kitaev's $E_8$ state} On each surface of the cubic (vertical and horizontal links across the center), there are two chiral Luttinger liquids characterized by a $K$-matrix given by (\ref{K-matrix E8}). Each of these liquids possesses a chiral central charge of $c_-=8$.

\subsection{$O$-symmetric lattice}
For an $O$-symmetric lattice, the cell decomposition is depicted in Figs. \ref{O cell decomposition} and \ref{O unit cell}, and the wavefunction takes the form given in Eq. (\ref{O wavefunction}). It is important to ensure that the decorated block states conserve $U^f(1)$ charge.

\subsubsection{2D block states}
In the case of 2D block states, there is no on-site symmetry for any of the 2D blocks. Consequently, both spinless and spin-1/2 fermions allow for two possible 2D block states: the Chern insulator and Kitaev's $E_8$ state. However, as stated in Sec. \ref{O2D}, any nonvanishing invertible phase decoration on 2D blocks results in chiral edge modes on 1D blocks along their shared boundaries. Consequently, there are no nontrivial 2D block states.

\subsubsection{1D block states}
The effective on-site symmetries of the 1D blocks $(\tau_1, \tau_2, \tau_3)$ are $(\mathbb{Z}_3, \mathbb{Z}_4, \mathbb{Z}_2)$, and the spin of fermions is not relevant when investigating the 1D block states. The classification of 1D root phases is determined by the projective representations of the total symmetry group, as classified by:
\begin{align}
\mathcal{H}^2\left[U^f(1)\times\mathbb{Z}_{2,3,4},U(1)\right]=\mathbb{Z}_1,
\end{align}
implying that there are no nontrivial 1D block states.

\subsubsection{0D block states}
The effective on-site symmetry of the 0D block $\mu$ is $S_4$, and its classification is given by:
\begin{align}
\begin{aligned}
&n_0\in\mathcal{H}^0(S_4,\mathbb{Z})=\mathbb{Z},\
&\nu_1\in\mathcal{H}^1\left[S_4,U(1)\right]=\mathbb{Z}_2,
\end{aligned}
\end{align}
subject to the twisted cocycle condition (\ref{1-twisted O}), where $n_0$ characterizes the sector of $U^f(1)$ charge, and $\nu_1$ represents the 0D bSPT modes with a 4-fold rotation eigenvalue of $\pm1$.

\paragraph{Spinless fermions}For spinless fermions, there are three possible 1D bubble equivalences:
\begin{enumerate}[1.]
\item Fermionic modes with $U^f(1)$ charge $n_1$ on each $\tau_1$: changes the $U^f(1)$ charge on $\mu$ by $8n_1$;
\item Fermionic modes with $U^f(1)$ charge $n_2$ on each $\tau_2$: changes the $U^f(1)$ charge on $\mu$ by $6n_2$, forming $n_2$ copies of atomic insulator (\ref{O atomic insulator}) who changes the eigenvalue of 4-fold rotation by $(-1)^{n_2}$;
\item Fermionic modes with $U^f(1)$ charge $n_3$ on each $\tau_3$: changes the $U^f(1)$ charge on $\mu$ by $12n_3$, forming $n_3$ copies of atomic insulators:
\begin{align}
|\phi_\mu\rangle=\prod\limits_{j=1}^{12}c_j^\dag|0\rangle
\end{align}
with 4-fold rotation property:
\begin{align}
\bs{R}|\phi_\mu\rangle=\prod\limits_{j=0}^{2}c_{4j+2}^\dag c_{4j+3}^\dag c_{4j+4}^\dag c_{4j+1}^\dag=-|\phi_\mu\rangle
\end{align}
\end{enumerate}
By 1D bubbles, a vacuum 0D mode $(0,+)$ is deformed to $(8n_1+6n_2+12n_3,(-1)^{n_2+n_3})$, forming a $\mathbb{Z}$ group:
\begin{align}
\left\{(8n_1+6n_2+12n_3,(-1)^{n_2+n_3})\big|n_{1,2,3}\in\mathbb{Z}\right\}=\mathbb{Z}
\end{align}
Furthermore, we surround a Chern insulator $\mu$ as illustrated in Fig. \ref{Chern insulator} that changes the $U^f(1)$ charge of $\mu$ by $1$. Hence there is no nontrivial 0D block state.

\paragraph{Spin-1/2 fermions}The only difference between spinless and spin-1/2 fermions is that atomic insulators $|\phi_\mu\rangle$ and $|\phi_\mu\rangle$ cannot change the eigenvalues of 4-fold rotation because of the additional minus sign from spin-1/2 nature. Hence there is only one nontrivial 0D block state, characterizing the eigenvalue $-1$ of 4-fold rotation.

\subsubsection{Summary}
In this section we summarize all obstruction and trivialization free block states: for systems with spinless fermions, the ultimate classification is $\mathbb{Z}_1$; for systems with spin-1/2 fermions, the ultimate classification is $\mathbb{Z}_2$, the only nontrivial phase is 0D block state with eigenvalue $-1$ of 4-fold rotation.

Furthermore, there is no nontrivial HO topological surface theory because of the absence of 2D and 1D block state.

\subsection{$O_h$-symmetric lattice}
For $O_h$-symmetric lattice, the cell decomposition is illustrated in Figs. \ref{Oh cell decomposition} and Fig. \ref{Oh unit cell}, and the wavefunction has the form of Eq. (\ref{Oh wavefunction}). The decorated block states should be $U^f(1)$ charge conserved. 

\subsubsection{2D block states}
The effective on-site symmetry of 2D blocks $\sigma_{1,2,3}$ is $\mathbb{Z}_2$ by reflection symmetry acting internally. From Sec. \ref{Th2D U(1)}, there are three root phases: Chern insulators, Kitaev's $E_8$ state, and 2D $U^f(1)\times\mathbb{Z}_2$ fSPT phases. 

\paragraph{Chern insulator decoration}The chiralities of decorated phases are illustrated in Fig. \ref{Oh chirality}, and the gapless modes on each 1D block are described by nonchiral Luttinger liquid (\ref{Luttinger}) if the chiral central charges of Chern insulators decorated on all 2D blocks $\sigma_{1,2,3}$ are equal. Consider the monolayer Chern insulator decoration on each $\sigma_1$ and $\sigma_2$, leaving a nonchiral Luttinger liquid (\ref{Luttinger}) with $K$-matrix $K=(\sigma^z)^{\oplus4}$ and 8-component bosonic field $\Phi=(\phi_1,...,\phi_8)^T$ on each 1D block $\tau_2$. The effective on-site symmetry of $\tau_2$ is $\mathbb{Z}_4\rtimes\mathbb{Z}_2$ as the axis of 4-fold dihedral group symmetry $D_4=C_4\rtimes\mathbb{Z}_2^M$. 

For spinless fermions, under $\bs{R}\in C_4$ and $\bs{M}\in\mathbb{Z}_2^M$ as two generators of $D_4$, the bosonic field $\Phi$ transforms as Eq. (\ref{K-matrix symmetry}), with
\begin{align}
W^{\bs{R}}=\left(
\begin{array}{cccccccc}
0 & 0 & 1 & 0 & 0 & 0 & 0 & 0\\ 
0 & 0 & 0 & 1 & 0 & 0 & 0 & 0\\
0 & 0 & 0 & 0 & 1 & 0 & 0 & 0\\ 
0 & 0 & 0 & 0 & 0 & 1 & 0 & 0\\ 
0 & 0 & 0 & 0 & 0 & 0 & 1 & 0\\ 
0 & 0 & 0 & 0 & 0 & 0 & 0 & 1\\ 
1 & 0 & 0 & 0 & 0 & 0 & 0 & 0\\ 
0 & 1 & 0 & 0 & 0 & 0 & 0 & 0
\end{array}
\right),~~\delta\phi^{\bs{R}}=0
\end{align}
\begin{align}
W^{\bs{M}}=\left(
\begin{array}{cccccccc}
1 & 0 & 0 & 0 & 0 & 0 & 0 & 0\\ 
0 & 0 & 0 & 0 & 0 & 0 & 0 & 1\\
0 & 0 & 0 & 0 & 0 & 0 & 1 & 0\\ 
0 & 0 & 0 & 0 & 0 & 1 & 0 & 0\\ 
0 & 0 & 0 & 0 & 1 & 0 & 0 & 0\\ 
0 & 0 & 0 & 1 & 0 & 0 & 0 & 0\\ 
0 & 0 & 1 & 0 & 0 & 0 & 0 & 0\\ 
0 & 1 & 0 & 0 & 0 & 0 & 0 & 0
\end{array}
\right),~~\delta\phi^{\bs{M}}=0
\end{align}
and under another reflection generator $\bs{MR}$,
\begin{align}
W^{\bs{MR}}=\left(
\begin{array}{cccccccc}
0 & 0 & 0 & 0 & 0 & 0 & 1 & 0\\ 
0 & 0 & 0 & 0 & 0 & 1 & 0 & 0\\
0 & 0 & 0 & 0 & 1 & 0 & 0 & 0\\ 
0 & 0 & 0 & 1 & 0 & 0 & 0 & 0\\ 
0 & 0 & 1 & 0 & 0 & 0 & 0 & 0\\ 
0 & 1 & 0 & 0 & 0 & 0 & 0 & 0\\ 
1 & 0 & 0 & 0 & 0 & 0 & 0 & 0\\ 
0 & 0 & 0 & 0 & 0 & 0 & 0 & 1
\end{array}
\right),~\delta\phi^{\bs{MR}}=0
\end{align}
and $U^f(1)$ charge conservation [$\forall\theta\in U^f(1)$]:
\begin{align}
W^{U^f(1)}=\mathbbm{1}_{8\times8},~\delta\phi^{U^f(1)}=\theta(1,1,1,1,1,1,1,1)^T
\label{8-component U(1)}
\end{align}
On the one hand, the anomaly indicators (\ref{anomaly indicator}) of two reflection generators $\bs{M}$ and $\bs{MR}$ are non-vanishing, as
\begin{align}
\begin{aligned}
&\nu_{\bs{M}}=\frac{1}{2}~(\mathrm{mod}~2)\\
&\nu_{\bs{MR}}=\frac{3}{2}~(\mathrm{mod}~2)
\end{aligned}
\end{align}
Hence the monolayer Chern insulator on $\sigma_{1,2,3}$ is obstructed for spinless fermions.

On the other hand, there is only one linear independent solution to the ``null-vector'' problem (\ref{null-vector}):
\begin{align}
\Lambda=(1,1,1,1,1,1,1,1)^T
\end{align}
To fully gap out the 8-component bosonic field $\Phi$, we should introduce at least 4 symmetric independent backscattering terms (\ref{backscattering}). Hence the monolayer Chern insulator on $\sigma_{1,2,3}$ is obstructed for spinless fermions, which confirms the results of anomaly indicators.

For spin-1/2 fermions, under $\bs{R}$ and $\bs{M}$, the bosonic field $\Phi$ transforms as Eq. (\ref{K-matrix symmetry}), with:
\begin{align}
W^{\bs{R}}=\left(
\begin{array}{cccccccc}
0 & 0 & -1 & 0 & 0 & 0 & 0 & 0\\ 
0 & 0 & 0 & -1 & 0 & 0 & 0 & 0\\
0 & 0 & 0 & 0 & -1 & 0 & 0 & 0\\ 
0 & 0 & 0 & 0 & 0 & -1 & 0 & 0\\ 
0 & 0 & 0 & 0 & 0 & 0 & -1 & 0\\ 
0 & 0 & 0 & 0 & 0 & 0 & 0 & -1\\ 
-1 & 0 & 0 & 0 & 0 & 0 & 0 & 0\\ 
0 & -1 & 0 & 0 & 0 & 0 & 0 & 0
\end{array}
\right)
\end{align}
\begin{align}
W^{\bs{M}}=\left(
\begin{array}{cccccccc}
1 & 0 & 0 & 0 & 0 & 0 & 0 & 0\\ 
0 & 0 & 0 & 0 & 0 & 0 & 0 & -1\\
0 & 0 & 0 & 0 & 0 & 0 & -1 & 0\\ 
0 & 0 & 0 & 0 & 0 & -1 & 0 & 0\\ 
0 & 0 & 0 & 0 & 1 & 0 & 0 & 0\\ 
0 & 0 & 0 & -1 & 0 & 0 & 0 & 0\\ 
0 & 0 & -1 & 0 & 0 & 0 & 0 & 0\\ 
0 & -1 & 0 & 0 & 0 & 0 & 0 & 0
\end{array}
\right)
\end{align}
\begin{align}
W^{\bs{MR}}=\left(
\begin{array}{cccccccc}
0 & 0 & 0 & 0 & 0 & 0 & -1 & 0\\ 
0 & 0 & 0 & 0 & 0 & -1 & 0 & 0\\
0 & 0 & 0 & 0 & -1 & 0 & 0 & 0\\ 
0 & 0 & 0 & -1 & 0 & 0 & 0 & 0\\ 
0 & 0 & -1 & 0 & 0 & 0 & 0 & 0\\ 
0 & -1 & 0 & 0 & 0 & 0 & 0 & 0\\ 
-1 & 0 & 0 & 0 & 0 & 0 & 0 & 0\\ 
0 & 0 & 0 & 0 & 0 & 0 & 0 & 1
\end{array}
\right)
\end{align}
\begin{align}
W^{U^f(1)}=\mathbbm{1}_{8\times8}
\end{align}
and
\begin{align}
\begin{aligned}
&\delta\phi^{\bs{R}}=\pi(1,0,0,0,0,0,0,-1)^T\\
&\delta\phi^{\bs{M}}=\frac{\pi}{2}(-1,1,1,1,-1,-1,-1,-1)^T\\
&\delta\phi^{\bs{MR}}=\frac{\pi}{2}(1,1,1,1,-1,-1,-1,1)^T\\
&\delta\phi^{U^f(1)}=\theta(1,1,-1,1,1,1,-1,1)^T
\end{aligned}
\end{align}
On the one hand, the anomaly indicators (\ref{anomaly indicator}) of two reflection generators $\bs{M}$ and $\bs{MR}$ are vanishing as $\nu_{\bs{M}}=\nu_{\bs{MR}}=0~(\mathrm{mod}~2)$, hence the monolayer of Chern insulators decorated on $\sigma_{1,2}$ is \textit{obstruction-free} for spin-1/2 fermions.

On the other hand, there is four linear independent solutions to the ``null-vector'' problem (\ref{null-vector}) for this case, which corresponds to four independent backscattering terms:
\begin{align}
\begin{aligned}
&\Lambda_1=(1,-1,-1,-1,1,-1,-1,-1)^T\\
&\Lambda_2=(1,1,1,-1,-1,1,1,1)^T\\
&\Lambda_3=(-1,1,1,1,1,1,1,-1)^T\\
&\Lambda_4=(1,1,-1,1,1,-1,1,1)^T\\
\end{aligned}
\end{align}
Hence the monolayer of Chern insulators decorated on $\sigma_{1,2}$ is \textit{obstruction-free} for spin-1/2 fermions, which confirms the results of the anomaly indicators. Furthermore, bilayer Chern insulators can be smoothly deformed to the root phase of 2D $U^f(1)\times\mathbb{Z}_2$ fSPT by 3D ``Chern insulator'' bubble construction (see Fig. \ref{3D bubble}).

\paragraph{Kitaev's $E_8$ state decoration}The chirality of decorated phases are illustrated in Fig. \ref{Oh chirality}. The chiral central charges of decorated Kitaev's $E_8$ states on $\sigma_{1,2,3}$ should be equal. On the other hand, the edge theory of Kitaev's $E_8$ state is compatible with all symmetry, hence arbitrary layers of Kitaev's $E_8$ states are \textit{obstruction-free}. Furthermore, similar to Sec. \ref{Th2D U(1)}, bilayer Kitaev's $E_8$ states can be trivialized by 3D ``$E_8$'' bubble construction. 

\paragraph{2D fSPT state decoration}The 1D edge theory of the root phase of $U^f(1)\times\mathbb{Z}_2$ fSPT is described as a nonchiral Luttinger liquid (\ref{Luttinger}), with $K$-matrix $K=\sigma^z$ and $\mathbb{Z}_2$ symmetry action (\ref{K-matrix symmetry}), where
\begin{align}
W^{\mathbb{Z}_2}=\mathbbm{1}_{2\times2},~~\delta\phi^{\mathbb{Z}_2}=\pi(0,1)^T
\end{align}
For spinless fermions, similar to Sec. \ref{Oh2D}, suppose the indices of decorated 2D $U^f(1)\times\mathbb{Z}_2$ fSPT phases on $\sigma_{1,2,3}$ are $\zeta_{1,2,3}\in\mathbb{Z}_4$, satisfying the condition:
\begin{align}
3(\zeta_2-\zeta_3)\equiv4(\zeta_1-\zeta_2)\equiv2(\zeta_1-\zeta_3)~(\mathrm{mod}~4)
\end{align}
with solutions satisfying
\begin{align}
\zeta_2\equiv\zeta_3~(\mathrm{mod}~4)~\mathrm{and}~\zeta_1\equiv\zeta_3~(\mathrm{mod}~2)
\end{align}
For $\zeta_1=\zeta_2=\zeta_3=1$, the 1D gapless theory on each $\tau_3$ as the axis of $D_2$ point group is described by non-chiral Luttinger liquid (\ref{Luttinger}), with $K=(\sigma^z)^{\oplus4}$ and $\Phi=(\phi_1,\cdot\cdot\cdot,\phi_8)^T$. Under $\bs{M}_1$ and $\bs{M}_2$ as reflection generators of $D_2$, the bosonic field $\Phi$ transforms as Eq. (\ref{K-matrix symmetry}), with
\begin{align}
W^{\bs{M}_1}=\left(
\begin{array}{cccc}
0 & 0 & 1 & 0\\
0 & 1 & 0 & 0\\
1 & 0 & 0 & 0\\
0 & 0 & 0 & 1
\end{array}
\right)\otimes\mathbbm{1}_{2\times2},~\delta\phi^{\bs{M}_1}=\pi\chi
\end{align}
\begin{align}
W^{\bs{M}_2}=\left(
\begin{array}{cccc}
1 & 0 & 0 & 0\\
0 & 0 & 0 & 1\\
0 & 0 & 1 & 0\\
0 & 1 & 0 & 0
\end{array}
\right)\otimes\mathbbm{1}_{2\times2},~\delta\phi^{\bs{M}_2}=\pi\chi
\end{align}
and $U^f(1)$ charge conservation (\ref{8-component U(1)}), where $\chi=(0,1,0,1,0,1,0,1)^T$. One the one hand, the anomaly indicators of two reflection generators $\bs{M}_1$ and $\bs{M}_2$ are non-vanishing as $\nu_{\bs{M}_1}=\nu_{\bs{M}_2}=1~(\mathrm{mod}~2)$. Hence we cannot fully gap out the 1D gapless modes on $\tau_3$ in a symmetric way. 

On the other hand, without $U^f(1)$ charge conservation, there is four linear independent solutions to the ``null-vector'' problem (\ref{null-vector}) which correspond to four independent backscattering terms (\ref{backscattering}). Nevertheless, these backscattering terms are not compatible with $U^f(1)$ charge conservation, which confirms the results of the anomaly indicators. 

For $\zeta_1=1$ and $\zeta_{2,3}=3$, the 1D gapless theory on each $\tau_3$ is described by another non-chiral Luttinger liquid (\ref{Luttinger}), with $K=(\sigma^z)^{\oplus4}$ and $\Phi=(\phi_1,\cdot\cdot\cdot,\phi_8)^T$. Under $\bs{M}_{1,2}$, the bosonic field $\Phi$ transformed as Eq. (\ref{K-matrix symmetry}), with
\begin{align}
W^{\bs{M}_1}=\left(
\begin{array}{cccc}
0 & 0 & 1 & 0\\
0 & 1 & 0 & 0\\
1 & 0 & 0 & 0\\
0 & 0 & 0 & 1
\end{array}
\right)\otimes\mathbbm{1}_{2\times2},~\delta\phi^{\bs{M}_1}=\pi\eta
\end{align}
\begin{align}
W^{\bs{M}_2}=\left(
\begin{array}{cccc}
1 & 0 & 0 & 0\\
0 & 0 & 0 & 1\\
0 & 0 & 1 & 0\\
0 & 1 & 0 & 0
\end{array}
\right)\otimes\mathbbm{1}_{2\times2},~\delta\phi^{\bs{M}_2}=\pi\eta
\end{align}
and $U^f(1)$ charge conservation (\ref{8-component U(1)}), where $\eta=(0,1,1,0,0,1,1,0)^T$. On the one hand, the anomaly indicators (\ref{anomaly indicator}) of two reflection generators $\bs{M}_1$ and $\bs{M}_2$ are non-vanishing as $\nu_{\bs{M}_1}=\nu_{\bs{M}_2}=1~(\mathrm{mod}~2)$, hence we cannot fully gap out the 1D gapless modes on $\tau_3$ in a symmetric way.

On the other hand, there is no solution to the ``null-vector'' problem (\ref{null-vector}), which confirms the results of the anomaly indicators. 

Summarize above two gapping problem, 2D $U^f(1)\times\mathbb{Z}_2$ fSPT block states with $\zeta_1=1$ is \textit{obstructed} for spinless fermions. Similar for $\zeta_1=3$ cases. 

For $\zeta_1=2$ and $\zeta_{2,3}=0$, the 1D gapless theory on each $\tau_2$ is described by non-chiral Luttinger liquid (\ref{Luttinger}), with $K=(\sigma^x)^{\oplus4}$ and $\Phi=(\phi_1,\cdot\cdot\cdot,\phi_8)^T$. Under $\bs{M}_{1,2}'$ as reflection generators of $D_4$, the bosonic field $\Phi$ transformed as Eq. (\ref{K-matrix symmetry}), with
\begin{align}
W^{\bs{M}_1'}=\left(
\begin{array}{cccc}
0 & 0 & 1 & 0\\
0 & 1 & 0 & 0\\
1 & 0 & 0 & 0\\
0 & 0 & 0 & 1
\end{array}
\right)\otimes\mathbbm{1}_{2\times2},~\delta\phi^{\bs{M}_1'}=\pi\rho
\end{align}
\begin{align}
W^{\bs{M}_2'}=\left(
\begin{array}{cccc}
0 & 1 & 0 & 0\\
1 & 0 & 0 & 0\\
0 & 0 & 0 & 1\\
0 & 0 & 1 & 0
\end{array}
\right)\otimes\mathbbm{1}_{2\times2},~\delta\phi^{\bs{M}_2'}=\pi\rho
\end{align}
and $U^f(1)$ charge conservation (\ref{8-component U(1)}), where $\rho=(1,1,1,1,1,1,1,1)^T$. On the one hand, the anomaly indicators (\ref{anomaly indicator}) of two reflection generators $\bs{M}_1'$ and $\bs{M}_2'$ are vanishing as $\nu_{\bs{M}_1'}=\nu_{\bs{M}_2'}=0~(\mathrm{mod}~2)$. Hence the 1D gapless modes on $\tau_2$ can be fully gapped symmetrically.

On the other hand, there are four linear independent solutions to ``null-vector'' problem (\ref{null-vector}):
\begin{align}
\begin{aligned}
&\Lambda_1=(1,0,-1,0,1,0,-1,0)^T\\
&\Lambda_2=(0,1,0,-1,0,-1,0,1)^T\\
&\Lambda_3=(1,0,1,0,-1,0,-1,0)^T\\
&\Lambda_4=(0,-1,0,1,0,1,0,-1)^T
\end{aligned}
\label{Oh backscattering U(1)}
\end{align}
Hence the 1D gapless modes on $\tau_2$ can be fully gapped by corresponding backscattering terms, which confirms the results of the anomaly indicators. The 1D gapless theory on each $\tau_3$ is described by nonchiral Luttinger liquid (\ref{Luttinger}), with $K=(\sigma^x)^{\oplus2}$ and $\Phi=(\phi_1,\phi_2,\phi_3,\phi_4)^T$. Under $\bs{M}_{1,2}$, the bosonic field $\Phi$ transformed as Eq. (\ref{K-matrix symmetry}), with
\begin{align}
W^{\bs{M}_1}=\left(
\begin{array}{cccc}
0 & 0 & 1 & 0\\
0 & 0 & 0 & 1\\
1 & 0 & 0 & 0\\
0 & 1 & 0 & 0
\end{array}
\right),~\delta\phi^{\bs{M}_1}=\pi\left(
\begin{array}{cccc}
1\\
1\\
1\\
1
\end{array}
\right)
\end{align}
\begin{align}
W^{\bs{M}_2}=\mathbbm{1}_{4\times4},~\delta\phi^{\bs{M}_2}=\delta\phi^{\bs{M}_1}
\end{align}
and $U^f(1)$ charge conservation (\ref{4-component U(1)}). On the one hand, the anomaly indicators (\ref{anomaly indicator}) of two reflection generators $\bs{M}_1$ and $\bs{M}_2$ are vanishing as $\nu_{\bs{M}_1}=\nu_{\bs{M}_2}=0~(\mathrm{mod}~2)$. Hence the 1D gapless modes on $\tau_3$ can be fully gapped symmetrically.

On the other hand, there are two linearly independent solutions to the ``null-vector'' problem (\ref{null-vector}):
\begin{align}
\Lambda_1=(1,0,1,0)^T,~~\Lambda_2=(0,1,0,-1)^T
\end{align}
Therefore, 2D $U^f(1)\times\mathbb{Z}_2$ fSPT state decorations on $\sigma_{1,2,3}$ with $(\zeta_1,\zeta_2,\zeta_3)=(2,0,0)$ is \textit{obstruction-free}. Similarly, 2D $U^f(1)\times\mathbb{Z}_2$ fSPT state decorations on $\sigma_{1,2,3}$ with $(\zeta_1,\zeta_2,\zeta_3)=(0,2,2)$ is also \textit{obstruction-free}.

For spinless fermions, all nontrivial 2D block states form a $\mathbb{Z}_2^3$ group, with two root phases:
\begin{enumerate}[1.]
\item Kitaev's $E_8$ state on $\sigma_{1,2,3}$;
\item 2D $U^f(1)\times\mathbb{Z}_2$ bSPT state on $\sigma_{1}$/$\sigma_{2,3}$.
\end{enumerate}

For spin-1/2 fermions, we have proved that arbitrary layers of Chern insulators on $\sigma_{1,2,3}$ are obstruction-free. Furthermore, 2D $U^f(1)\times\mathbb{Z}_2$ fSPT state decorations can be smoothly deformed to Chern insulator decorations by 3D ``Chern insulator'' bubble constructions. 

Near 1D blocks $\tau_1$, the spin of fermions is irrelevant. As a consequence, the indices of decorated 2D $U^f(1)\times\mathbb{Z}_2$ fSPT states on $\sigma_{2,3}$ should be equal: $\zeta_2=\zeta_3$. 

For 2D block state with $(\zeta_1,\zeta_2,\zeta_3)=(1,0,0)$, consider the inverse procedure of 3D ``Chern insulator'' bubble constructions as illustrated in Fig. \ref{3D bubble}: on the one hand, a 2D $U^f(1)\times\mathbb{Z}_2$ fSPT state with $\zeta_1=1$ can be smoothly deformed to bilayer Chern insulators on each $\sigma_1$; on the other hand, by 3D ``Chern insulator'' bubble constructions, 2D blocks $\sigma_{2,3}$ with vacuum block states are smoothly deformed to bilayer Chern insulators on each of them. Above we have proved that for spin-1/2 fermions, arbitrary layers of Chern insulators on each 2D block $\sigma_{1,2,3}$ are obstruction-free, so do the 2D block state with $(\zeta_1,\zeta_2,\zeta_3)=(1,0,0)$. Similarly, the 2D block state with $(\zeta_1,\zeta_2,\zeta_3)=(0,1,1)$ is also \textit{obstruction-free}. Furthermore, there is no trivialization on the above two block states, consequently, for spin-1/2 fermions, all nontrivial 2D block states form a $\mathbb{Z}_8\times\mathbb{Z}_4\times\mathbb{Z}_2$ group, with the following root phases:
\begin{enumerate}[1.]
\item Monolayer Chern insulator on each 2D block $\sigma_{1,2,3}$;
\item 2D $U^f(1)\times\mathbb{Z}_2$ fSPT states with $(\zeta_1,\zeta_2,\zeta_3)=(1,0,0)$ or $(\zeta_1,\zeta_2,\zeta_3)=(0,1,1)$;
\item Kitaev's $E_8$ state on each 2D block $\sigma_{1,2,3}$.
\end{enumerate}
and there is a nontrivial group extension between the first two root phases. 

\subsubsection{1D block states}
The effective on-site symmetry of each 1D block $\tau_1$ is $\mathbb{Z}_3\rtimes\mathbb{Z}_2$ as the axis of $D_3$ symmetry, at which the spin of fermions is irrelevant. The corresponding 1D root phases are classified by Eq. (\ref{D3 1D U(1)}), i.e., trivial classification. 

The effective on-site symmetry of each 1D block $\tau_2/\tau_3$ is $\mathbb{Z}_4\rtimes\mathbb{Z}_2/\mathbb{Z}_2\times\mathbb{Z}_2$ as the axis of $D_4/D_2$ symmetry. For spinless fermions, the only nontrivial root phase is the Haldane chain, guaranteed by group 2-cohomology:
\begin{align}
\mathcal{H}^2\left[U^f(1)\times(\mathbb{Z}_{4,2}\rtimes\mathbb{Z}_2),U(1)\right]=\mathbb{Z}_2
\end{align}
For spin-1/2 fermions, the classification of the corresponding 1D root phase is trivial:
\begin{align}
\mathcal{H}^2\left[U^f(1)\rtimes_{\rho_1,\omega_2}(\mathbb{Z}_{4,2}\rtimes\mathbb{Z}_2),U(1)\right]=\mathbb{Z}_1
\end{align}
Therefore, for spinless fermions, all nontrivial 1D block states form a $\mathbb{Z}_2^2$ group, with the root phases: Haldane chain decorations on $\tau_2$ or $\tau_3$; for spin-1/2 fermions, there is no nontrivial 1D block state.

\subsubsection{0D block states}
The effective on-site symmetry of the 0D block $\mu$ is $S_4\times\mathbb{Z}_2$, with the classification data:
\begin{align}
\begin{aligned}
&n_0\in\mathcal{H}^0(S_4\times\mathbb{Z}_2,\mathbb{Z})=\mathbb{Z}\\
&\nu_1\in\mathcal{H}^1[S_4\times\mathbb{Z}_2,U(1)]=\mathbb{Z}_2^2
\end{aligned}
\end{align}
with twisted cocycle condition (\ref{1-twisted Oh}). $n_0$ depicts the $U^f(1)$ charge on $\mu$, $\nu_1$ depicts the 0D bSPT modes, characterizing the products of reflection eigenvalues $\pm1$ of $\bs{M}_1^{j}$ and $\bs{M}_2^k$, respectively ($j=1,2,3$ and $k=1,...,6$). 

\paragraph{Spinless fermions}Consider 1D bubble equivalences: decorate 0D modes with $n_{1,2,3}\in\mathbb{Z}$ $U^f(1)$ charges on each 1D block $\tau_{1,2,3}$ which can be smoothly deformed to infinite far away and trivialized. For $n_1=1$, the 1D bubbles on $\tau_1$ leave modes with $U^f(1)$ charge 8 on $\mu$; for $n_2=1$, the 1D bubbles on $\tau_2$ leave modes with $U^f(1)$ charge 6 on $\mu$, forming an atomic insulator (\ref{Oh atomic1}) with eigenvalue $-1$ of $M_{1}^{j}$; for $n_3=1$, the 1D bubbles on $\tau_3$ leave modes with $U^f(1)$ charge 12 on $\mu$, forming three atomic insulators (\ref{Oh atomic2}) who change the eigenvalue of $\bs{M}_2^k$, but their product is invariant. By 1D bubbles, all trivial 0D block state deformed from the vacuum 0D mode $(0,+,+)$ form the following group:
\begin{align}
\left\{\left(8n_1+6n_2+12n_3,(-1)^{n_2},+\right)\big|n_{1,2,3}\in\mathbb{Z}\right\}=2\mathbb{Z}\times\mathbb{Z}_2\nonumber
\end{align}
we quotient out these trivial phases, and all nontrivial 0D block states form a $\mathbb{Z}_4$ group, characterzing the parity of $U^f(1)$ charge and eigenvalue $-1$ of $\bs{M}_2^k$, with a nontrivial group extension. 

\paragraph{Spin-1/2 fermions}Firstly, 0D modes with odd $U^f(1)$ charge are obstructed by (\ref{1-twisted Oh}). On the other hand, 1D bubbles on $\tau_{2,3}$ with odd $U^f(1)$ charge are also obstructed. Hence for spin-1/2 fermions, all trivial 0D block state deformed from the vacuum 0D mode $(0,+,+)$ form an alternative group:
\begin{align}
&\left\{\left(8n_1+12n_2+24n_3,+,+\right)\big|n_{1,2,3}\in\mathbb{Z}\right\}=4\mathbb{Z}
\end{align}
after quotienting out these trivial phases, all nontrivial 0D block states form a $\mathbb{Z}_2^3$ group, with root phases:
\begin{enumerate}[1.]
\item 0D modes with $U^f(1)$ charge $n\equiv2~(\mathrm{mod}~4)$;
\item Eigenvalues $-1$ of $\bs{M}_1^j$ and $\bs{M}_2^k$.
\end{enumerate}

\subsubsection{Summary\label{Oh U(1) summary}}
In this section we summarize all obstruction and trivialization free block states: for systems with spinless fermions, the ultimate classification is $\mathcal{G}_{O_h,0}^{U(1)}=\mathbb{Z}_4\times\mathbb{Z}_2^4$, with the following root phases:
\begin{enumerate}[1.]
\item 2D monolayer Kitaev's $E_8$ state decoration on $\sigma_{1,2,3}$;
\item 2D $U^f(1)\times\mathbb{Z}_2$ bSPT phases decoration on $\sigma_{1}$;
\item 2D $U^f(1)\times\mathbb{Z}_2$ bSPT phases decoration on $\sigma_{2,3}$;
\item 1D Haldane chain decorations on $\tau_{2,3}$;
\item 0D mode with odd $U^f(1)$ charge and eigenvalue $-1$ of $\bs{M}_2^k$, with a nontrivial group extension.
\end{enumerate}

For systems with spin-1/2 fermions, the ultimate classification is $\mathcal{G}_{O_h,1/2}^{U(1)}=\mathbb{Z}_8\times\mathbb{Z}_4\times\mathbb{Z}_2^4$, with the following root phases:
\begin{enumerate}[1.]
\item 2D monolayer Chern insulator decoration on $\sigma_{1,2,3}$;
\item 2D $U^f(1)\times\mathbb{Z}_2$ fSPT phases decoration on $\sigma_{1,2,3}$, with quantum number $(\zeta_1,\zeta_2,\zeta_3)=(1,0,0)$ or $(\zeta_1,\zeta_2,\zeta_3)=(0,1,1)$;
\item 2D monolayer Kitaev's $E_8$ state decoration on $\sigma_{1,2,3}$;
\item 0D mode with $U^f(1)$ charge $n\equiv2~(\mathrm{mod}~4)$;
\item 0D mode with eigenvalues $-1$ of $\bs{M}_1^j/\bs{M}_2^k$ ($j=1,2,3$ and $k=1,2,3,4,5,6$).
\end{enumerate}
And there is a nontrivial group extension between the first two root phases.

\subsubsection{Higher-order topological surface theory}
With concrete block states, we are ready to investigate the corresponding HO topological surface theories by higher-order bulk-boundary correspondence. 

For $U^f(1)$ charge conserved systems with spinless fermions, the corresponding classification of 3D $O_h$-symmetric topological phases is $\mathbb{Z}_4\times\mathbb{Z}_2^4$, with root phases as summarized in Sec. \ref{Oh U(1) summary}.

For 2D block states, the classification is $\mathbb{Z}_2^2$ with two root phases: Kitaev's $E_8$ state and 2D bosonic Levin-Gu state decorations on $\sigma_{1,2,3}$.

\paragraph{Kitaev's $E_8$ state}There are four chiral Luttinger liquids with $K$-matrix (\ref{K-matrix E8}) on each surface of the open cubic (vertical, horizontal, diagonal, and off-diagonal segments across the center), each of them has chiral central charge $c_-=8$. 

\paragraph{2D bosonic Levin-Gu state}There are four nonchiral Luttinger liquids on the surface of the open cubic, with $K=sigma^x$ and $\mathbb{Z}_2$ on-site symmetry (\ref{bosonic Levin-Gu}).

For 1D block states, the classification is $\mathbb{Z}_2^2$, with root phases: Haldane chain decorations on $\tau_2/\tau_3$. 

\paragraph{Haldane chain on $\tau_2$}Leaving a spin-1/2 degree of freedom on the center of each surface of the open cubic.

\paragraph{Haldane chain on $\tau_3$}Leaving a spin-1/2 degree of freedom on the center of each hinge of the open cubic.

For $U^f(1)$ charge conserved systems with spin-1/2 fermions, the corresponding classification of 3D $O_h$-symmetric topological phases is $\mathbb{Z}_8\times\mathbb{Z}_4\times\mathbb{Z}_2^4$, with root phases as summarized in Sec. \ref{Oh U(1) summary}.

For 2D block states, the classification is $\mathbb{Z}_8\times\mathbb{Z}_4\times\mathbb{Z}_2$, with the following root phases:
\begin{enumerate}[1.]
\item Kitaev's $E_8$ state on each $\sigma_{1,2,3}$;
\item Monolayer Chern insulator on each $\sigma_{1,2,3}$;
\item 2D $U^f(1)\times\mathbb{Z}_2$ fSPT states on each $\sigma_1$/$\sigma_{2,3}$;
\end{enumerate}
With a nontrivial group extension (stacking) between bilayer Chern insulators and the root phase of 2D $U^f(1)\times\mathbb{Z}_2$ fSPT states decorations.

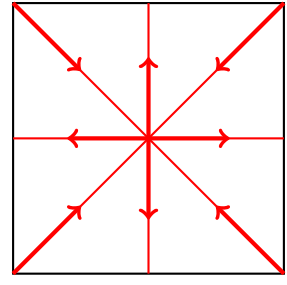
\begin{figure}
\begin{tikzpicture}[scale=1.2]
\tikzstyle{sergio}=[rectangle,draw=none]
\draw[draw=black, thick] (-2,0.5)--(1,0.5)--(1,-2.5)--(-2,-2.5)--cycle;
\draw[thick,color=red] (-0.5,0.5) -- (-0.5,-2.5);
\draw[thick,color=red] (-2,-1) -- (1,-1);
\draw[thick,color=red] (-2,0.5) -- (1,-2.5);
\draw[thick,color=red] (1,0.5) -- (-2,-2.5);
\draw[ultra thick,color=red,->] (1,0.5) -- (0.25,-0.25);
\draw[ultra thick,color=red,->] (-2,0.5) -- (-1.25,-0.25);
\draw[ultra thick,color=red,->] (-2,-2.5) -- (-1.25,-1.75);
\draw[ultra thick,color=red,->] (1,-2.5) -- (0.25,-1.75);
\draw[ultra thick,color=red,->] (-0.5,-1) -- (-0.5,-1.9);
\draw[ultra thick,color=red,->] (-0.5,-1) -- (0.4,-1);
\draw[ultra thick,color=red,->] (-0.5,-1) -- (-1.4,-1);
\draw[ultra thick,color=red,->] (-0.5,-1) -- (-0.5,-0.1);
\end{tikzpicture}
\caption{Chiralities of the second-order topological surface theory of monolayer Chern insulator/Kitaev's $E_8$ state decorations on each $\sigma_{1,2,3}$.}
\label{chiralities}
\end{figure}

\paragraph{Kitaev's $E_8$ state}There are four chiral Luttinger liquids with $K$-matrix (\ref{K-matrix E8}) on each surface of the open cubic, each of them has chiral central charge $c_-=8$. The chiralities of chiral Luttinger liquids on each surface are illustrated in Fig. \ref{chiralities}. 

\paragraph{Monolayer Chern insulator}There are four chiral Luttinger liquids with chiral central charge $c_-=1$ on each surface of the open cubic, with chiralities illustrated in Fig. \ref{chiralities}.

\paragraph{2D $U^f(1)\times\mathbb{Z}_2$ fSPT on $\sigma_1$}There are two nonchiral Luttinger liquids with $K=\sigma_z$ and $\mathbb{Z}_2$ symmetry transformation (\ref{nu=2}) on each surface of the open cubic, located at vertical and horizontal segments across the center. 

\paragraph{2D $U^f(1)\times\mathbb{Z}_2$ fSPT on $\sigma_{2,3}$}There are two nonchiral Luttinger liquids with $K=\sigma_z$ and $\mathbb{Z}_2$ symmetry transformation (\ref{nu=2}) on each surface of the open cubic, located at diagonal and off-diagonal segments across the center. 

\section{Generalized crystalline equivalence principle\label{equivalence}}
In this section, we discuss how to generalize the crystalline equivalence principle that has been conjectured and justified in interacting bosonic systems \cite{realspace} and 2D interacting fermionic systems \cite{rotation, dihedral, wallpaper}. By comparing the classification results of crystalline TSC and TI summarized in Tables \ref{TSC} and \ref{TI} with the classification results of 3D fSPT phases protected by corresponding internal symmetry groups \cite{mathematical}, we verify the fermionic version of the crystalline equivalence principle in 3D systems for all TSC and TI constructed in this paper, for both spinless and spin-1/2 fermions.

Slightly different from bosonic and 2D fermionic systems, we should map the space group symmetry to on-site symmetry according to the following rules:
\begin{enumerate}[1.]
\item $n$-fold rotation symmetry should be mapped to a $\mathbb{Z}_n$ on-site symmetry;
\item Reflection symmetry should be mapped to antiunitary time-reversal symmetry $\mathbb{Z}_2^T$;
\item The systems with spinless (spin-1/2) fermions should be mapped to spin-1/2 (spinless) fermions.
\end{enumerate}
With some exceptions: For $S_n$ symmetric systems, there are some subtleties for the correspondence: for $n=2,6$, the $\Z_2^f$ extensions should always be trivial no matter what the spin of fermion is; but for $n=4$, the spin of fermion should be twisted by the internal-crystalline correspondence. We give a physical argument of this subtlety: for $n=2,4,6$, an even-fold roto-reflection can be treated as the composition of a rotation with respect to the $z$-axis and a reflection with respect to the $xy$-plane, so for a fermion, if we apply $S_n$ symmetry by $n$ times, we effectively drive the fermion rotates by $n/2+1$ rounds. So for $n=2,6$, this number is even, which implies that the group extension between the fermion parity $\Z_2^f$ and $S_n$ is not sensitive to the real spin of fermion; but for $n=4$, this number is odd, which implies that the group extension is still important, and the spin twist should still be encountered in the internal-crystalline correspondence.

\section{Conclusion and discussion\label{conclusion}}
In this paper, we undertake a comprehensive study of crystalline topological superconductors (TSCs) and topological insulators (TIs) for interacting fermions on 3D lattices with specific point group symmetries. Our approach involves systematically constructing and classifying these phases. We begin by decomposing the 3D lattice into lower-dimensional blocks, each possessing its own effective ``on-site" symmetry due to the internal action of the point group symmetry. By employing cell decompositions, we decorate the lower-dimensional block states in accordance with their respective on-site symmetries.

For the 2D block states, it is possible that they give rise to one or more 1D gapless Luttinger liquids (\ref{Luttinger}) along their shared borders. To determine if these Luttinger liquids can be fully gapped out, we utilize the $K$-matrix formalism. If successful, we classify the corresponding 2D block states as ``obstruction-free"; otherwise, they are deemed "obstructed". In the case of 1D block states, they may generate several 0D gapless modes forming a local Hilbert space at the center of the point group as their shared border. The presence of a projective or linear representation of the total point group among these gapless modes serves as a sufficient condition indicating that they cannot be symmetrically gapped out. We classify such 1D block states as "obstructed", while those lacking this representation are labeled ``obstruction-free".

Based on the collection of obstruction-free block states, we further examine the possibility of ``trivializations" through bubble equivalences. A $d$-dimensional block should act as the shared border of multiple $(d+1)$-dimensional blocks, and the corresponding block states might undergo changes due to $d$-dimensional bubbles on the $(d+1)$-dimensional blocks. Block states that are both obstruction and trivialization-free correspond to nontrivial crystalline SPT phases. Following this framework, we provide complete classifications of 3D point group SPT phases in interacting fermionic systems. By comparing our results with mathematical classifications of 3D fSPT phases protected by corresponding on-site symmetries~\cite{mathematical,wang2021exactly}, we validate the ``crystalline equivalence principle" for 3D interacting fermionic systems. Furthermore, we make the significant prediction that almost all nontrivial 2D block states represent intriguing interacting topological phases.

Topological crystals not only yield a comprehensive classification of 3D crystalline fSPT phases but also provide a systematic understanding of their higher-order (HO) topological surface theories. In this paper, we undertake a systematic investigation of the HO topological surface theory of 3D crystalline fSPT phases using higher-order bulk-boundary correspondence. Our findings demonstrate that all 2D block states correspond to second-order topological surface theories, 1D block states correspond to third-order topological surface theories, and 0D block states do not give rise to nontrivial topological surface theories. Moreover, our analysis of block states in the bulk reveals that several nontrivial block states are equivalent and can be mutually deformed through bubble equivalences. This equivalence is also reflected in their HO topological surface theories, as the HO topological surface theories of two bubble-equivalent block states can be smoothly deformed into one another through "plates" constructions on the boundary (refer to Figs. \ref{surface} and \ref{Th2D U(1) surface}). The bubble equivalences in the bulk and ``plates'' equivalences on the boundary straightly manifest the bulk-boundary correspondence of 3D crystalline fSPT phases. 

It is important to note that our methodology is not limited to the cases discussed in this paper, but can be extended to more general scenarios, such as 3D systems with space group symmetries or systems protected by a combination of internal and crystalline symmetries. In the bulk, the block states must adhere to both the internally acting crystalline symmetry and the internal symmetry. Similarly, on the boundary, the higher-order (HO) topological surface theory can be readily observed through finite-size truncations, reinforcing the concept of bulk-boundary correspondence.

Furthermore, our systematic classification results and boundary manifestations offer excellent opportunities for experimental observations of 3D crystalline HO topological phases. Spectroscopic and transport measurements can be employed to detect the presence of HO topological surface theories in materials exhibiting crystalline symmetries. For instance, scanning tunneling microscopy (STM) can be utilized to determine the second- or third-order topological surface theory, characterized by the presence of 1D or 0D gapless modes on the surface, by measuring the local density of states in the open lattice's real-space configuration. Transport experiments can investigate the channel transport of second-order topological hinge modes along specific directions, providing further insight into the properties of these intriguing phases.

\begin{acknowledgements}
We thank Meng Cheng, Zhen Bi,  Ruochen Ma, and Liujun Zou for stimulating discussions. This work is supported by funding from Hong Kong’s Research Grants Council (GRF No. 14307621, ANR/RGC Joint Research Scheme No. A-CUHK402/18), and from the National Natural Science Foundation of China (Grand Nos. 12174068 and 11874115).
\end{acknowledgements}

\bibliography{apssamp}

\pagebreak

\clearpage

\appendix
\setcounter{equation}{0}
\newpage

\renewcommand{\thesection}{S-\arabic{section}} \renewcommand{\theequation}{S%
\arabic{equation}} \setcounter{equation}{0} \renewcommand{\thefigure}{S%
\arabic{figure}} \setcounter{figure}{0}

\onecolumngrid

\vskip0.2cm
\centerline{\large\textbf{Supplemental Materials of ``Topological crystals and higher-order topological surface}}
\vskip0.12cm
\centerline{\large\textbf{theories of three-dimensional crystalline symmetry-protected topological phases''}}

\vskip0.8cm
\twocolumngrid

\maketitle

\section{$K$-matrix formalism of fSPT phases}
In the main text, we use the $K$-matrix formalism to investigate the \textit{obstruction} of 2D block states, i.e., the gapping problem of the gapless edge states of 2D block states on the 1D blocks as their shared border. In this section we review the $K$-matrix formalism of fermionic symmetry-protected topological (fSPT) phases. A $U(1)$ Chern-Simons theory has the form:
\begin{align}
\mathcal{L}=\frac{K_{IJ}}{4\pi}\epsilon^{\mu\nu\lambda}a_\mu^I\partial_\nu a_\lambda^J+a_\mu^Ij_I^\mu+\cdot\cdot\cdot
\label{Chern-Simons}
\end{align}
where $K$ is a symmetric integral matrix, $\{a^I\}$ is a set of one-form gauge fields, and $\{j_I\}$ are the corresponding currents that couple to the gauge fields $a^I$. The symmetry is defined as: two theories $\mathcal{L}[a^I]$ and $\mathcal{L}[\tilde{a}^I]$ correspond to the same phase if there is an $n\times n$ integral unimodular matrix $W$ satisfying $\tilde{a}^I=W_{IJ}a^J$. 

The topological order described by Abelian Chern-Simons theory hosts Abelian anyon excitations. An anyon is labeled by an integer vector $l=(l_1,l_2,\cdot\cdot\cdot,l_n)^T$. The self and mutual statistics of anyons are:
\begin{align}
\begin{aligned}
\theta_l&=\pi l^TK^{-1}l\\
\theta_{l,l'}&=2\pi l^T K^{-1}l'
\end{aligned}
\end{align}

The total number of anyons and the ground-state degeneracy (GSD) on a torus are both given by $|\mathrm{det}K|$. For SPT phase, there is no GSD or anyon, hence we require $|\mathrm{det}K|=1$ for SPT phases. 

The $K$-matrix Chern-Simons theory has a well-known bulk-boundary correspondence \cite{Lu12,Ning21a}. In a system with open boundary, the edge thoery of (\ref{Chern-Simons}) has the form:
\begin{align}
\mathcal{L}_{\mathrm{edge}}=\frac{K_{IJ}}{4\pi}\left(\partial_x\phi^I\right)\left(\partial_t\phi^J\right)+\frac{V_{IJ}}{8\pi}\left(\partial_x\phi^I\right)\left(\partial_x\phi^J\right)
\label{LuttingerS}
\end{align}
where $\phi=(\{\phi^I\})^T$ are chiral bosonic fields on the edge and related to dynamical gauge field $a_\mu^I$ in the bulk by $a_\mu^I=\partial_\mu\phi^I$, and an anyon on the edge can be created by the operator $e^{il^T\phi}$.

\section{Equivalence between Kitaev's $E_8$ state and 16-layer 2D $(p+ip)$-SC}
In the main text, we have utilized the fact that Kitaev's $E_8$ state is equivalent to 16 layers of 2D $(p+ip)$-SC in the discussion of trivialization of Kitaev's $E_8$ state decoration on 2D blocks in TSC systems. In this section we prove this issue by an explicit $K$-matrix transformation. The $K$-matrix of Kitaev's $E_8$ state is:
\begin{align}
K_{E_8}=\left(
\begin{array}{cccccccc}
2 & -1 & 0 & 0 & 0 & 0 & 0 & 0\\
-1 & 2 & -1 & 0 & 0 & 0 & -1 & 0\\
0 & -1 & 2 & -1 & 0 & 0 & 0 & 0\\
0 & 0 & -1 & 2 & -1 & 0 & 0 & 0\\
0 & 0 & 0 & -1 & 2 & -1 & 0 & 0\\
0 & 0 & 0 & 0 & -1 & 2 & 0 & 0\\
0 & -1 & 0 & 0 & 0 & 0 & 2 & -1\\
0 & 0 & 0 & 0 & 0 & 0 & -1 & 2
\end{array}
\right)
\label{K-matrix E8S}
\end{align}
and the $K$-matrix of 16 layers of 2D $(p+ip)$-SC is $\mathbbm{1}_{8\times8}$. At the first glimpse, we call that Kitaev's $E_8$ state is equivalent to 16 layers of 2D $(p+ip)$-SC if we can find a $8\times8$ special linear matrix with integer elements $W_0\in SL(8,\mathbb{Z})$, such that $W_0^TK_{E_8}W_0=\mathbbm{1}_{8\times8}$ with chiral central charge $c_-=8$. Nevertheless, the absence of $W_0$ does not guarantee the inequivalence of two corresponding topological phases. Consider a $2\times2$ $K$-matrix $\sigma^z$: without symmetry, the corresponding phase is trivial because the 1D nonchiral Luttinger liquid on the edge can be gapped by backscattering term $\cos(\phi_1-\phi_2)$. We stack this trivial phase to Kitaev's $E_8$ state, with total $K$-matrix $K_{E_8}\oplus\sigma^z$. We find that $W^T(K_{E_8}\oplus\sigma^z)W=\mathbbm{1}_{8\times8}\oplus\sigma^z$, where
\begin{align}
W=\left(
\begin{array}{cccccccccc}
5 & 5 & 5 & 5 & 5 & 5 & 5 & 5 & 8 & 16\\
10 & 10 & 10 & 9 & 9 & 9 & 9 & 9 & 15 & 30\\
8 & 8 & 8 & 8 & 7 & 7 & 7 & 7 & 12 & 24\\
6 & 6 & 6 & 6 & 6 & 5 & 5 & 5 & 9 & 18\\
4 & 4 & 4 & 4 & 4 & 4 & 3 & 3 & 6 & 12\\
2 & 2 & 2 & 2 & 2 & 2 & 2 & 1 & 3 & 6\\
7 & 7 & 6 & 6 & 6 & 6 & 6 & 6 & 10 & 20\\
4 & 3 & 3 & 3 & 3 & 3 & 3 & 3 & 5 & 10\\
-1 & -1 & -1 & -1 & -1 & -1 & -1 & -1 & -3 & -4\\
2 & 2 & 2 & 2 & 2 & 2 & 2 & 2 & 4 & 7\\
\end{array}
\right)\nonumber
\end{align}
and $W\in SL(10,\mathbb{Z})$. Hence the following two phases are equivalent:
\begin{enumerate}[1.]
\item Kitaev's $E_8$ state stacked with a trivial phase described by $K_0=\sigma^z$;
\item 16 layers of 2D $(p+ip)$-SC stacked with a trivial phase described by $K_0=\sigma^z$
\end{enumerate}
Therefore, for systems without $U^f(1)$ charge conservation, Kitaev's $E_8$ state is equivalent to 16 layers of 2D $(p+ip)$-SC, and we do not treat the Kitaev's $E_8$ state as an independent 2D root phase. On the other hand, for systems with $U^f(1)$ charge conservation, we should treat the Kitaev's $E_8$ state as an independent 2D root phase, because we cannot deform it to eight layers of Chern insulator.

\section{Cell decomposition, classifications and HO topological surface theories}
In this section, we summarize all 3D crystalline fSPT phases protected by all other point group symmetries (in the main text, we have discussed six representative examples), for both spinless and spin-1/2 fermions, with and without $U^f(1)$ charge conservation. Including the cell decompositions of 3D point group symmetric lattice, the classifications, the corresponding root phases, and their HO topological surface theories.

\subsection{$C_2$-symmetric lattice}
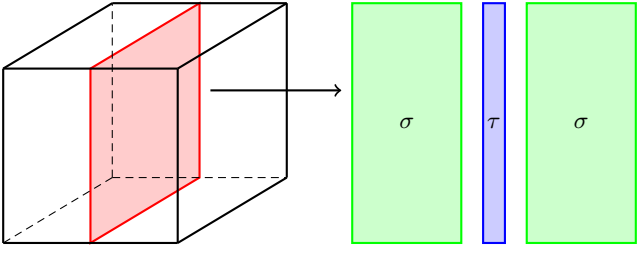
\begin{figure}
\begin{tikzpicture}[scale=0.58]
\tikzstyle{sergio}=[rectangle,draw=none]
\filldraw[fill=red!20, draw=red, thick] (-1,0)--(1.5,1.5)--(1.5,-2.5)--(-1,-4)--cycle;
\draw[thick] (-0.5,1.5) -- (3.5,1.5);
\draw[thick] (-0.5,1.5) -- (-3,0);
\draw[thick] (3.5,1.5) -- (1,0);
\draw[thick] (-3,0) -- (1,0);
\draw[thick] (-3,0) -- (-3,-4);
\draw[thick] (1,-4) -- (-3,-4);
\draw[thick] (1,-4) -- (1,0);
\draw[thick] (3.5,-2.5) -- (3.5,1.5);
\draw[thick] (3.5,-2.5) -- (1,-4);
\draw[densely dashed] (-0.5,-2.5) -- (-0.5,1.5);
\draw[densely dashed] (-0.5,-2.5) -- (3.5,-2.5);
\draw[densely dashed] (-0.5,-2.5) -- (-3,-4);
\filldraw[fill=green!20, draw=green, thick] (5,1.5)--(7.5,1.5)--(7.5,-4)--(5,-4)--cycle;
\filldraw[fill=green!20, draw=green, thick] (9,1.5)--(11.5,1.5)--(11.5,-4)--(9,-4)--cycle;
\filldraw[fill=blue!20, draw=blue, thick] (8.5,1.5)--(8,1.5)--(8,-4)--(8.5,-4)--cycle;
\path (6.25,-1.25) node [style=sergio] {$\sigma$};
\path (10.25,-1.25) node [style=sergio] {$\sigma$};
\path (8.25,-1.25) node [style=sergio] {$\tau$};
\draw[thick,->] (1.75,-0.5) -- (4.75,-0.5);
\end{tikzpicture}
\caption{The cell decomposition of 3D $C_2$-symmetric lattice. Red plate in the left panel is the equator; right panel shows the lower-dimensional blocks included in the equator, where green plates depict 2D blocks and blue segments depict 1D block.}
\label{C2 cell decomposition}
\end{figure}

For $C_2$-symmetric lattice with the cell decomposition in Fig. \ref{C2 cell decomposition}, the ground-state wavefunction of the system can be decomposed to the direct products of wavefunctions of lower-dimensional block states as:
\begin{align}
|\Psi\rangle=\bigotimes_{g\in C_2}|T_{g\lambda}\rangle\otimes|T_{g\sigma}\rangle\otimes|\alpha_{g\tau}\rangle
\label{C2 wavefunction}
\end{align}
where $|T_{g(\lambda,\sigma,\tau)}\rangle$ is the wavefunction of $d$D block state on $g(\lambda,\sigma,\tau)$ ($d=2,3$), which is topological trivial or invertible topological phase; $|\alpha_{g\tau}\rangle$ is the wavefunction of 1D block state which is $\mathbb{Z}_2$-symmetric. 

We summarize the classifications and corresponding root phases. For crystalline TSC with spinless fermions, we summarize all possible block states: 
\begin{enumerate}[1.]
\item 2D block $\sigma$: 2D $(p+ip)$-SC;
\item 1D block $\tau$: Majorana chain and 1D $\mathbb{Z}_2$ fSPT phase.
\end{enumerate}
2D $(p+ip)$-SC on $\sigma$ is obstructed because it leaves a chiral 1D mode on $\tau$ with chiral central charge $c_-=1$; 1D $\mathbb{Z}_2$ fSPT phase on $\tau$ is trivialized by 2D ``Majorana'' bubble equivalence on $\sigma$, and Majorana chain decoration on $\tau$ is trivialized by 2D anomalous $(p+ip)$-SC on the top surface of the open lattice. Hence the ultimate classification is trivial.

For crystalline TSC with spin-1/2 fermions, the only possible block state is 2D $(p+ip)$-SC on each $\sigma$ which is obstructed. Hence the ultimate classification is trivial.

For crystalline TI, the total group $G_f$ is the central extension of $C_2$ group with $U^f(1)$ charge conservation, characterized by the short exact sequence:
\begin{align}
0\rightarrow U^f(1)\rightarrow G_f\rightarrow C_2\rightarrow0
\end{align}
and the spin of fermions is characterized by the factor system of this short exact sequence, which is classified by group 2-cohomology $\mathcal{H}^2[C_2,U^f(1)]=\mathbb{Z}_1$. i.e., the spin of fermions is irrelevant for the $C_2$-symmetric case. All possible block states are located on 2D block $\sigma$: Chern insulators and Kitaev's $E_8$ states. Nevertheless, these chiral block states leave chiral 1D mode on $\tau$ which cannot be gapped out. Therefore, the ultimate classification is trivial.

\subsection{$C_{1h}$-symmetric lattice}
\begin{figure}
\begin{tikzpicture}[scale=0.8]
\tikzstyle{sergio}=[rectangle,draw=none]
\filldraw[fill=red!20, draw=red, thick] (-1,0)--(1.5,1.5)--(1.5,-2.5)--(-1,-4)--cycle;
\draw[thick] (-0.5,1.5) -- (3.5,1.5);
\draw[thick] (-0.5,1.5) -- (-3,0);
\draw[thick] (3.5,1.5) -- (1,0);
\draw[thick] (-3,0) -- (1,0);
\draw[thick] (-3,0) -- (-3,-4);
\draw[thick] (1,-4) -- (-3,-4);
\draw[thick] (1,-4) -- (1,0);
\draw[thick] (3.5,-2.5) -- (3.5,1.5);
\draw[thick] (3.5,-2.5) -- (1,-4);
\draw[densely dashed] (-0.5,-2.5) -- (-0.5,1.5);
\draw[densely dashed] (-0.5,-2.5) -- (3.5,-2.5);
\draw[densely dashed] (-0.5,-2.5) -- (-3,-4);
\path (1.7859,1.8085) node [style=sergio] {$M$};
\end{tikzpicture}
\caption{The cell decomposition of 3D $C_{1h}$-symmetric lattice. Red plate depicts the reflection plane.}
\label{C1h cell decomposition}
\end{figure}
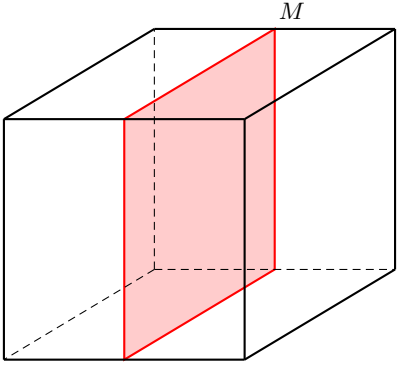

For $C_{1h}$-symmetric lattice with the cell decomposition in Fig. \ref{C1h cell decomposition}, the ground-state wavefunction of the system can be decomposed to the direct products of wavefunctions of lower-dimensional block states as:
\begin{align}
|\Psi\rangle=\bigotimes\limits_{g\in C_{1h}}|T_{g\lambda}\rangle\otimes|\alpha_{\sigma}\rangle
\end{align}
Where $|T_{g\lambda}\rangle$ is the wavefunction of 3D block state on $g\lambda$ which is topological trivial; $|\alpha_\sigma\rangle$ is the wavefunction of 2D block state which is $\mathbb{Z}_2$-symmetric.

We summarize the classifications and corresponding root phases. For crystalline TSC with spinless fermions, the ultimate classification is $\mathbb{Z}_{16}$, with the following two root phases:
\begin{enumerate}[1.]
\item Monolayer $(p+ip)$-SC on $\sigma$ ($\mathbb{Z}_2$);
\item $\mathbb{Z}_2$ fSPT phases on $\sigma$, with $\mathbb{Z}_8$ classification ($\mathbb{Z}_8$).
\end{enumerate}
And there is a nontrivial group extension between these two root phases: bilayer $(p+ip)$-SCs on $\sigma$ are equivalent to 2D $\mathbb{Z}_2$ fSPT phase with $\nu=1\in\mathbb{Z}_8$ on $\sigma$. Furthermore, by finite-size truncations, the HO topological surface theories are straightforwardly obtained. For the first root phase, the second-order topological surface theory is a chiral Majorana mode on the intersection of the open lattice and reflection plane in Fig. \ref{C1h cell decomposition}; for the second root phase, the second-order topological surface theory is a nonchiral Luttinger liquid with $K$-matrix $K=\sigma^z$ and $\mathbb{Z}_2$ symmetry property $W^{\mathbb{Z}_2}=\sigma^z$ and $\delta\phi^{\mathbb{Z}_2}=0$, on the intersection of the open lattice and reflection plane in Fig. \ref{C1h cell decomposition}.

For crystalline TSC with spin-1/2 fermions, the corresponding classification is trivial. 

For systems with $U^f(1)$ charge conservation, similar to inversion-symmetric case, the spin of fermions is irrelevant. The corresponding classification is $\mathbb{Z}_8\times\mathbb{Z}_2$, with three root phases:
\begin{enumerate}[1.]
\item Monolayer Chern insulator on $\sigma$ ($\mathbb{Z}_2$);
\item $U^f(1)\times\mathbb{Z}_2$ fSPT phases on $\sigma$, with $\mathbb{Z}_4$ classification ($\mathbb{Z}_4$);
\item Monolayer Kitaev's $E_8$ state on $\sigma$ ($\mathbb{Z}_2$).
\end{enumerate}
And there is a nontrivial group extension between first two root phases: bilayer Chern insulators on $\sigma$ are equivalent to 2D $U^f(1)\times\mathbb{Z}_2$ fSPT phase with $\nu=1\in\mathbb{Z}_4$ on $\sigma$. 

The second-order topological surface theory of different root phases on the intersection of the open lattice and reflection plane in Fig. \ref{C1h cell decomposition} are:
\begin{enumerate}[1.]
\item Monolayer Chern insulator: chiral fermion;
\item $U^f(1)\times\mathbb{Z}_2$ fSPT phase: 1D nonchiral Luttinger liquid with $K$-matrix $K=\sigma^z$ and $\mathbb{Z}_2$ symmetry property $W^{\mathbb{Z}_2}=\mathbbm{1}_{2\times2}$ and $\delta\phi^{\mathbb{Z}_2}=\pi(0,1)^T$;
\item Kitaev's $E_8$ state: 1D chiral Luttinger liquid with $K$-matrix in Eq. (\ref{K-matrix E8S}).
\end{enumerate}

\subsection{$C_{2h}$-symmetric lattice}
For $C_{2h}$-symmetric lattice with the cell decomposition in Fig. \ref{C2h cell decomposition}, the ground-state wavefunction of the system can be decomposed to the direct products of wavefunctions of lower-dimensional block states as:
\begin{align}
|\Psi\rangle=\bigotimes\limits_{g\in C_{2h}}|T_{g\lambda}\rangle\otimes\sum\limits_{k=1}^2|\gamma_{g\sigma_k}\rangle\otimes\sum\limits_{j=1}^2|\beta_{g\tau_j}\rangle\otimes|\alpha_\mu\rangle
\label{C_{2h} wavefunction}
\end{align}
where $|T_{g\lambda}\rangle$ is the wavefunction of 3D block state on $g\lambda$ which is topological trivial; $|\gamma_{g\sigma_1}\rangle$ is the wavefunction of 2D block state on $g\sigma_1$ which is topological trivial or invertible topological phase, $|\gamma_{g\sigma_2}\rangle$ is the $\mathbb{Z}_2$-symmetric wavefunction of 2D block state on $g\sigma_2$; $|\beta_{g\tau_{1,2}}\rangle$ is the $\mathbb{Z}_2$-symmetric wavefunction of 1D block state on $g\tau_{1,2}$; $|\alpha_\mu\rangle$ is the $(\mathbb{Z}_2\times\mathbb{Z}_2)$-symmetric wavefunction of 0D block state on $\mu$.

\begin{figure}
\begin{tikzpicture}[scale=0.56]
\tikzstyle{sergio}=[rectangle,draw=none]
\filldraw[fill=green!20, draw=green, thick] (5,1.5)--(7.5,1.5)--(7.5,-4)--(5,-4)--cycle;
\filldraw[fill=green!20, draw=green, thick] (9,1.5)--(11.5,1.5)--(11.5,-4)--(9,-4)--cycle;
\filldraw[fill=blue!20, draw=blue, thick] (8.5,1.5)--(8,1.5)--(8,-0.5)--(8.5,-0.5)--cycle;
\filldraw[fill=blue!20, draw=blue, thick] (8.5,-2)--(8,-2)--(8,-4)--(8.5,-4)--cycle;
\path (6.25,-1.25) node [style=sergio] {$\sigma_2$};
\path (10.25,-1.25) node [style=sergio] {$\sigma_2$};
\path (8.25,0.5) node [style=sergio] {\scriptsize$\tau_2$};
\path (8.25,-3) node [style=sergio] {\scriptsize$\tau_2$};
\filldraw[fill=white, draw=black] (8.25,-1.25)circle (13pt);
\path (8.25,-1.25) node [style=sergio] {$\mu$};
\filldraw[fill=red!20, draw=red, thick] (1,1.5)--(3.5,1.5)--(3.5,-4)--(1,-4)--cycle;
\filldraw[fill=red!20, draw=red, thick] (-3,1.5)--(-0.5,1.5)--(-0.5,-4)--(-3,-4)--cycle;
\filldraw[fill=green!20, draw=green, thick] (0,1.5)--(0.5,1.5)--(0.5,-0.5)--(0,-0.5)--cycle;
\filldraw[fill=green!20, draw=green, thick] (0,-2)--(0.5,-2)--(0.5,-4)--(0,-4)--cycle;
\filldraw[fill=blue!20, draw=blue, thick] (0,-1)--(0.5,-1)--(0.5,-1.5)--(0,-1.5)--cycle;
\path (-1.75,-1.25) node [style=sergio] {$\lambda$};
\path (2.25,-1.25) node [style=sergio] {$\lambda$};
\path (0.25,0.5) node [style=sergio] {\scriptsize$\sigma_1$};
\path (0.25,-3) node [style=sergio] {\scriptsize$\sigma_1$};
\path (0.25,-1.25) node [style=sergio] {\scriptsize$\tau_1$};
\path (6,0.5) node [style=sergio] {$M$};
\path (10.5,0.5) node [style=sergio] {$M$};
\path (0.75,-0.75) node [style=sergio] {\scriptsize$C_2$};
\path (0.25,-4.75) node [style=sergio] {\footnotesize North/south hemisphere};
\path (8.25,-4.75) node [style=sergio] {\footnotesize Equator};
\end{tikzpicture}
\caption{The cell decomposition of 3D $C_{2h}$-symmetric lattice. Left panel depicts the north/south hemisphere, including 3D blocks $\lambda$, 2D blocks $\sigma_1$ and 1D blocks $\tau_1$; right panel depicts the equator, including 2D blocks $\sigma_2$, 1D blocks $\tau_2$ and 0D block $\mu$. $C_2$ depicts the axis of 2-fold rotation, and $\bs{M}$ depicts the reflection plane.}
\label{C2h cell decomposition}
\end{figure}

We summarize the classifications and corresponding root phases. For crystalline TSC with spinless fermions, the possible root phases on different blocks are summarized as:
\begin{enumerate}[1.]
\item 2D block $\sigma_1$: 2D $(p+ip)$-SC;
\item 2D block $\sigma_2$: 2D $(p+ip)$-SC and 2D $\mathbb{Z}_2$ fSPT phases;
\item 1D block $\tau_1/\tau_2$: Majorana chain and 1D $\mathbb{Z}_2$ fSPT phases;
\item 0D block $\mu$: 0D modes with odd fermion parity, eigenvalues $-1$ of two generators of the $C_{2h}$ group.
\end{enumerate}
Then we consider all possible obstructions and trivializations. 2D $(p+ip)$-SCs on $\sigma_1$ with quantum number $n_1\in\mathbb{Z}$ leaves $2n_1$ chiral Majorana modes with chiral central charge $c_-=n_1$ on $\tau_1$ which cannot be gapped out, hence the corresponding block states are obstructed; odd layers of 2D $(p+ip)$-SC on $\sigma_2$ is obstructed because it is not compatible with the 2-fold in-plane rotation symmetry on the equator \cite{rotation}. Majorana chain decoration on each $\tau_1/\tau_2$ is obstructed because it leaves 2 dangling Majorana zero modes at $\mu$ which cannot be gapped out; 1D $\mathbb{Z}_2$ fSPT phase decoration on each $\tau_1/\tau_2$ is obstructed because it leaves 4 dangling Majorana zero modes at $\mu$ which cannot be gapped out in a $\mathbb{Z}_2\times\mathbb{Z}_2$ symmetric way \cite{dihedral}. Furthermore, if we consider $\tau_1$ and $\tau_2$ together and decorate a 1D $\mathbb{Z}_2$ fSPT phase on each of them, there are 8 dangling Majorana zero modes at $\mu$ which can be symmetrically gapped out.

If we decorate a complex fermion on each $\tau_1$ which can be adiabatically deformed to infinite far and trivialized, there will be an atomic insulator formed by two complex fermions at $\mu$, with eigenvalue $-1$ of the reflection with respect to the equator. Equivalently, the 0D block state characterizing the eigenvalue $-1$ of the reflection with respect to the equator is trivialized. Similar for the 0D block state characterizing the eigenvalue $-1$ of the 2-fold rotation. Furthermore, 2D ``Majorana'' bubble construction on each $\sigma_2$ can be smoothly deformed to a Majorana chain surrounding $\mu$ who changes the fermion parity of the 0D block state on $\mu$. Therefore, all possible 0D block states are trivialized. 

Then 2D ``Majorana'' bubble construction on each $\sigma_1$ leaves a 1D $\mathbb{Z}_2$ fSPT phase on each $\sigma_1$ and $\sigma_2$, and trivializes the corresponding 1D block state. Moreover, 3D ``$p+ip$'' bubble construction on each $\lambda$ will change the layers of $(p+ip)$-SCs on each $\sigma_2$ by two, and trivializes the corresponding block states.

Finally, the ultimate classification is $\mathbb{Z}_8$, the corresponding root phases are 2D $\mathbb{Z}_2$ fSPT phases on $\sigma_2$. The second-order topological surface theory of this block state is 1D nonchiral Luttinger liquid on the edge of the equator in Fig. \ref{C2h cell decomposition}, with $K$-matrix $K=\sigma^z$ and $\mathbb{Z}_2$ symmetry property $W^{\mathbb{Z}_2}=\sigma^z$ and $\delta\phi^{\mathbb{Z}_2}=0$. 

For crystalline TSC with spin-1/2 fermions, the nontrivial extension between $C_{2h}$ and $\Z_2^f$ is captured by the factor system $\omega_2$ of the following short exact sequence,
\begin{align}
1\rightarrow\Z_2^f\rightarrow G_f\rightarrow C_{2h}\rightarrow1,
\end{align}
for $g_{1/2}=(a_{1/2},b_{1/2})\in C_{2h}$ (where $a$ and $b$ are generators of rotation and reflection, respectively), the explicit expression of $\omega_2$ is
\begin{align}
\omega_2(g_1,g_2)=a_1a_2+b_1b_2~(\mathrm{mod}~2).
\end{align}
The physical meaning of this extension is that we have $\bs{R}^2=-1$ and $\bs{M}^2=-1$ as usual, but their composition, $\bs{I}=\bs{MR}$ as an inversion symmetry, satisfies that $(\bs{MR})^2=1$. A physical argument for this fact is that an inversion will rotate a spin-1/2 fermion by two rounds, which cannot be extended by $\Z_2^f$.

All possible block states are 2D $(p+ip)$-SC on $\sigma_1$ and 0D modes characterizing the odd fermion parity and eigenvalues $-1$ of two generators of the $C_{2h}$ group. Similar to the spinless fermions, 2D $(p+ip)$-SC on $\sigma_1$ is obstructed. For reflection eigenvalue $-1$ at $\mu$, we consider a mode with reflection eigenvalue $i$ on each $\tau_2$ that can be deformed to $\mu$ and trivializes the corresponding 0D block state; for rotation eigenvalue $-1$, similarly, we consider a mode with rotation eigenvalue $i$ on each $\tau_1$. For odd fermion parity at $\mu$, we consider the equator at which the inversion symmetry $I$ acts as an in-plane two-fold rotation with $\bs{I}^2=1$. Now we consider a Majorana bobble on each $\sigma_2$ that can be deformed to a Majorana chain with PBC surrounding $\mu$, which trivializes the odd fermion parity on it. Therefore, the ultimate classification is trivial. 

For crystalline TI with spinless fermions, all possible root phases on different blocks are summarized as:
\begin{enumerate}[1.]
\item 2D block $\sigma_1$: Chern insulator and Kitaev's $E_8$ state;
\item 2D block $\sigma_2$: Chern insulator, Kitaev's $E_8$ state and 2D $U^f(1)\times\mathbb{Z}_2$ fSPT phases;
\item 0D block $\mu$: 0D modes characterizing eigenvalues $-1$ of two generators of the $C_{2h}$ group, with different $U^f(1)$ charge.
\end{enumerate}
Similar to the crystalline TSC, all block states on $\sigma_1$ are obstructed because all of them are chiral, and 3D bubble equivalence on $\lambda$ can change the layers of decorated chiral phases on 2D blocks $\sigma_2$. Hence the corresponding classification is $\mathbb{Z}_8\times\mathbb{Z}_4\times\mathbb{Z}_2$, with the following root phases:
\begin{enumerate}[1.]
\item Monolayer Chern insulator on each $\sigma_2$ ($\mathbb{Z}_2$);
\item 2D $U^f(1)\times\mathbb{Z}_2$ fSPT phase on each $\sigma_2$ ($\mathbb{Z}_4$);
\item Monolayer Kitaev's $E_8$ state on each $\sigma_2$ ($\mathbb{Z}_2$);
\item 0D fermionic mode with odd $U^f(1)$ charge on each $\mu$ ($\mathbb{Z}_2$, characterizing the parity of $U^f(1)$ charge);
\item Eigenvalues $-1$ of rotation or reflection ($\mathbb{Z}_2$).
\end{enumerate}
Furthermore, there is a nontrivial extension between first two root phases: bilayer Chern insulators are equivalent to 2D $U^f(1)\times\mathbb{Z}_2$ fSPT phase with $\nu=1\in\mathbb{Z}_4$ on each $\sigma_2$; and there is another nontrivial extension between last two root phases. The second-order topological surface theories of different root phases on the edge of equator in Fig. \ref{C2h cell decomposition} are:
\begin{enumerate}[1.]
\item Monolayer Chern insulator: chiral fermion;
\item $U^f(1)\times\mathbb{Z}_2$ fSPT phase: 1D nonchiral Luttinger liquid with $K$-matrix $K=\sigma^z$ and $\mathbb{Z}_2$ symmetry property $W^{\mathbb{Z}_2}=\mathbbm{1}_{2\times2}$ and $\delta\phi^{\mathbb{Z}_2}=\pi(0,1)^T$;
\item Kitaev's $E_8$ state: 1D chiral Luttinger liquid with $K$-matrix in Eq. (\ref{K-matrix E8S}).
\end{enumerate}

For crystalline TI with spin-1/2 fermions, all possible root phases on different blocks are summarized as:
\begin{enumerate}[1.]
\item 2D block $\sigma_1$: Chern insulator and Kitaev's $E_8$ state;
\item 2D block $\sigma_2$: Chern insulator, Kitaev's $E_8$ state and 2D $U^f(1)\times\mathbb{Z}_2$ fSPT phases;
\item 0D block $\mu$: 0D modes characterizing eigenvalues $-1$ of two generators of the $C_{2h}$ group, with different $U^f(1)$ charge.
\end{enumerate}
All 2D block states on $\sigma_1$ are obstructed because all of them are chiral, and 3D bubble equivalence on $\lambda$ can change the layers of decorated chiral phases on 2D blocks $\sigma_2$. Hence the corresponding classification is $\mathbb{Z}_8\times\mathbb{Z}_4\times\mathbb{Z}_2$, with the following root phases:
\begin{enumerate}[1.]
\item Monolayer Chern insulator on each $\sigma_2$ ($\mathbb{Z}_2$);
\item 2D $U^f(1)\times\mathbb{Z}_2$ fSPT phase on each $\sigma_2$ ($\mathbb{Z}_4$);
\item Monolayer Kitaev's $E_8$ state on each $\sigma_2$ ($\mathbb{Z}_2$);
\item 0D fermionic mode with odd $U^f(1)$ charge on each $\mu$
($\mathbb{Z}_2$, characterizing the parity of $U^f(1)$ charge);
\item Eigenvalues $-1$ of rotation or reflection ($\mathbb{Z}_2$).
\end{enumerate}
There is a nontrivial extension between the first two root phases. The HO topological surface theories are identical to the systems with spinless fermions.

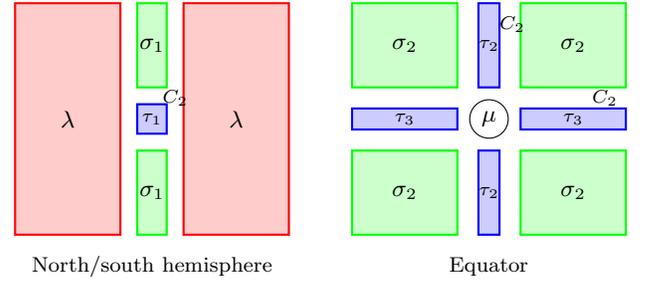
\begin{figure}
\begin{tikzpicture}[scale=0.56]
\tikzstyle{sergio}=[rectangle,draw=none]
\filldraw[fill=green!20, draw=green, thick] (5,1.5)--(7.5,1.5)--(7.5,-0.5)--(5,-0.5)--cycle;
\filldraw[fill=green!20, draw=green, thick] (9,1.5)--(11.5,1.5)--(11.5,-0.5)--(9,-0.5)--cycle;
\filldraw[fill=blue!20, draw=blue, thick] (8.5,1.5)--(8,1.5)--(8,-0.5)--(8.5,-0.5)--cycle;
\filldraw[fill=blue!20, draw=blue, thick] (8.5,-2)--(8,-2)--(8,-4)--(8.5,-4)--cycle;
\path (6.25,0.5) node [style=sergio] {$\sigma_2$};
\path (10.25,0.5) node [style=sergio] {$\sigma_2$};
\path (8.25,0.5) node [style=sergio] {\scriptsize$\tau_2$};
\path (8.25,-3) node [style=sergio] {\scriptsize$\tau_2$};
\filldraw[fill=white, draw=black] (8.25,-1.25)circle (13pt);
\path (8.25,-1.25) node [style=sergio] {$\mu$};
\filldraw[fill=red!20, draw=red, thick] (1,1.5)--(3.5,1.5)--(3.5,-4)--(1,-4)--cycle;
\filldraw[fill=red!20, draw=red, thick] (-3,1.5)--(-0.5,1.5)--(-0.5,-4)--(-3,-4)--cycle;
\filldraw[fill=green!20, draw=green, thick] (-0.1,1.5)--(0.6,1.5)--(0.6,-0.5)--(-0.1,-0.5)--cycle;
\filldraw[fill=green!20, draw=green, thick] (-0.1,-2)--(0.6,-2)--(0.6,-4)--(-0.1,-4)--cycle;
\filldraw[fill=blue!20, draw=blue, thick] (-0.1,-0.9)--(0.6,-0.9)--(0.6,-1.6)--(-0.1,-1.6)--cycle;
\path (-1.75,-1.25) node [style=sergio] {$\lambda$};
\path (2.25,-1.25) node [style=sergio] {$\lambda$};
\path (0.25,0.5) node [style=sergio] {$\sigma_1$};
\path (0.25,-3) node [style=sergio] {$\sigma_1$};
\path (0.25,-1.25) node [style=sergio] {\scriptsize$\tau_1$};
\filldraw[fill=green!20, draw=green, thick] (5,-2)--(7.5,-2)--(7.5,-4)--(5,-4)--cycle;
\filldraw[fill=green!20, draw=green, thick] (9,-2)--(11.5,-2)--(11.5,-4)--(9,-4)--cycle;
\path (6.25,-3) node [style=sergio] {$\sigma_2$};
\path (10.25,-3) node [style=sergio] {$\sigma_2$};
\filldraw[fill=blue!20, draw=blue, thick] (5,-1.5)--(5,-1)--(7.5,-1)--(7.5,-1.5)--cycle;
\filldraw[fill=blue!20, draw=blue, thick] (9,-1.5)--(9,-1)--(11.5,-1)--(11.5,-1.5)--cycle;
\path (6.25,-1.25) node [style=sergio] {\scriptsize$\tau_3$};
\path (10.25,-1.25) node [style=sergio] {\scriptsize$\tau_3$};
\path (11,-0.75) node [style=sergio] {\scriptsize$C_2$};
\path (8.8,1) node [style=sergio] {\scriptsize$C_2$};
\path (0.8,-0.75) node [style=sergio] {\scriptsize$C_2$};
\path (0.25,-4.75) node [style=sergio] {\footnotesize North/south hemisphere};
\path (8.25,-4.75) node [style=sergio] {\footnotesize Equator};
\end{tikzpicture}
\caption{The cell decomposition of 3D $D_2/V$-symmetric lattice. Left panel depicts the north/south hemisphere, including 3D blocks $\lambda$, 2D blocks $\sigma_1$ and 1D blocks $\tau_1$; right panel depicts the equator, including 2D blocks $\sigma_2$, 1D blocks $\tau_2$ and $\tau_3$, and 0D block $\mu$. $C_2$ depict the axes of 2-fold rotations.}
\label{D2 cell decomposition}
\end{figure}

\subsection{$D_2/V$-symmetric lattice}
For $D_2/V$-symmetric 3D lattice with the cell decomposition in Fig. \ref{D2 cell decomposition}, the ground-state wavefunction of the system can be decomposed to the direct products of wavefunctions of lower-dimensional block states as:
\begin{align}
|\Psi\rangle=\bigotimes\limits_{g\in D_2/V}|T_{g\lambda}\rangle\otimes\sum\limits_{k=1}^2|T_{g\sigma_k}\rangle\otimes\sum\limits_{j=1}^3|\beta_{g\tau_j}\rangle\otimes|\alpha_\mu\rangle
\label{D2 wavefunction}
\end{align}
where $|T_{g\lambda}\rangle/|T_{g\sigma_{1,2}}\rangle$ is the wavefunction of 3D/2D block state on $g\lambda/g\sigma_{1,2}$ which is topological trivial or invertible topological phase; $|\beta_{g\tau_{1,2,3}}\rangle$ is the $\mathbb{Z}_2$-symmetric wavefunction of 1D block state on $g\tau_{1,2,3}$; $|\alpha_\mu\rangle$ is the $(\mathbb{Z}_2\times\mathbb{Z}_2)$-symmetric wavefunction of 0D block state on $\mu$.

We summarize the classifications and corresponding root phases. For crystalline TSC with spinless fermions, all possible root phases are summarized as:
\begin{enumerate}[1.]
\item 2D blocks $\sigma_1/\sigma_2$: 2D $(p+ip)$-SC;
\item 1D blocks $\sigma_{1,2,3}$: Majorana chain and 1D $\mathbb{Z}_2$ fSPT phase;
\item 0D block $\mu$: 0D modes with odd fermion parity, eigenvalues $-1$ of two generators of the $D_2/V$ group.
\end{enumerate}
2D $(p+ip)$-SC on each $\sigma_1/\sigma_2$ is obstructed because it leaves several chiral modes on 1D blocks as the shared border of nearby 2D blocks. 

Majorana chain decoration on each $\tau_{1,2,3}$ is obstructed; 1D $\mathbb{Z}_2$ fSPT phase decoration on each $\tau_{1,2,3}$ is obstructed because it leaves 4 dangling Majorana zero modes at $\mu$ which cannot be gapped out in a $\mathbb{Z}_2\times\mathbb{Z}_2$ symmetric way \cite{dihedral}. Furthermore, if we consider $\tau_1$ and $\tau_2$ together and decorate a 1D $\mathbb{Z}_2$ fSPT phase on each of them, this 1D block state is obstruction-free. Hence all obstruction-free block states are summarized as:
\begin{enumerate}[1.]
\item Majorana chain decoration on each 1D block $(\tau_1,\tau_2)/(\tau_1,\tau_3)/(\tau_2,\tau_3)$ ($\mathbb{Z}_2^3$);
\item 0D modes with odd fermion parity, eigenvalues $-1$ of two generators of the $D_2/V$ group ($\mathbb{Z}_2^3$).
\end{enumerate}
On the other hand, Majorana chain decoration on $(\tau_1,\tau_2)$ can be trivialized by 2D ``Majorana'' bubble construction on each $\sigma_1$ \cite{dihedral}. Similar for $(\tau_1,\tau_3)$ and $(\tau_2,\tau_3)$. Hence all possible 1D block states are obstructed or trivialized.

If we decorate a complex fermion on each $\tau_1$ which can be adiabatically deformed to infinite far and trivialized, there will be an atomic insulator formed by two complex fermions at $\mu$, with eigenvalue $-1$ of the 2-fold rotation with respect to the $\tau_2$ and $\tau_3$ as its axis. Hence all possible 0D block states are trivialized, and the corresponding classification is trivial. 

For crystalline TSC with spin-1/2 fermions, the only possible root phases are 0D block states characterizing eigenvalues $-1$ of two generators of the $D_2/V$ group. There is no more obstruction and trivialization, and the corresponding classification is $\mathbb{Z}_2^2$, with two 0D root phases: eigenvalues $-1$ of two generators of the $D_2/V$ group.

For crystalline TI with spinless fermions, all possible block states are summarized as:
\begin{enumerate}[1.]
\item 2D blocks $\sigma_1/\sigma_2$: Chern insulators and Kitaev's $E_8$ states;
\item 0D blocks $\mu$: 0D modes characterizing the eigenvalues $-1$ of two generators of the $D_2/V$ group, with different $U^f(1)$ charge. 
\end{enumerate}
Similar to the crystalline TSC, all 2D block states are obstructed because all of these root phases are chiral. Furthermore, if we decorate a complex fermion on each $\tau_1$ which can be adiabatically deformed to infinite far and trivialized, there will be an atomic insulator with $U^f(1)$ charge 2 and eigenvalues $-1$ of 2-fold rotations with respect to $\tau_2$ and $\tau_3$. Similar construction is also applied to 1D blocks $\tau_2$ and $\tau_3$. Moreover, a Chern insulator enclosing the 0D block $\mu$ can change the $U^f(1)$ charge of $\mu$ by one, hence all possible 0D block states are trivialized. As the consequence, the classification of $D_2/V$-symmetric crystalline TI with spinless fermions is trivial.

For crystalline TI with spin-1/2 fermions, there are three 0D root phases:
\begin{enumerate}[1.]
\item 0D mode with odd $U^f(1)$ charge;
\item Eigenvalues $-1$ of two generators of $D_2/V$.
\end{enumerate}
and there is no more obstruction and trivialization. As the consequence, the corresponding classification is $\mathbb{Z}_2^3$. Because all nontrivial block states are 0D, there is no nontrivial HO topological surface theory.

\subsection{$C_{2v}$-symmetric lattice}
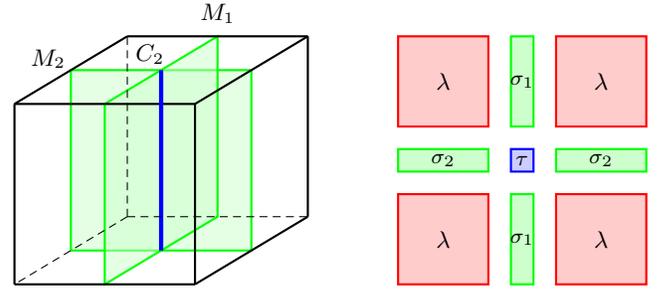
\begin{figure}
\begin{tikzpicture}[scale=0.6]
\tikzstyle{sergio}=[rectangle,draw=none]
\filldraw[fill=green!20, draw=green, thick,fill opacity=0.5] (-1.75,0.75)--(-1.75,-3.25)--(2.25,-3.25)--(2.25,0.75)--cycle;
\filldraw[fill=green!20, draw=green, thick,fill opacity=0.5] (-1,0)--(1.5,1.5)--(1.5,-2.5)--(-1,-4)--cycle;
\draw[thick] (-0.5,1.5) -- (3.5,1.5);
\draw[thick] (-0.5,1.5) -- (-3,0);
\draw[thick] (3.5,1.5) -- (1,0);
\draw[thick] (-3,0) -- (1,0);
\draw[thick] (-3,0) -- (-3,-4);
\draw[thick] (1,-4) -- (-3,-4);
\draw[thick] (1,-4) -- (1,0);
\draw[thick] (3.5,-2.5) -- (3.5,1.5);
\draw[thick] (3.5,-2.5) -- (1,-4);
\draw[draw=blue,ultra thick] (0.25,-3.25) -- (0.25,0.75);
\draw[densely dashed] (-0.5,-2.5) -- (-0.5,1.5);
\draw[densely dashed] (-0.5,-2.5) -- (3.5,-2.5);
\draw[densely dashed] (-0.5,-2.5) -- (-3,-4);
\filldraw[fill=red!20, draw=red, thick] (5.5,1.5)--(7.5,1.5)--(7.5,-0.5)--(5.5,-0.5)--cycle;
\filldraw[fill=red!20, draw=red, thick] (9,1.5)--(11,1.5)--(11,-0.5)--(9,-0.5)--cycle;
\filldraw[fill=green!20, draw=green, thick] (8.5,1.5)--(8,1.5)--(8,-0.5)--(8.5,-0.5)--cycle;
\filldraw[fill=green!20, draw=green, thick] (8.5,-2)--(8,-2)--(8,-4)--(8.5,-4)--cycle;
\filldraw[fill=green!20, draw=green, thick] (7.5,-1.5)--(7.5,-1)--(5.5,-1)--(5.5,-1.5)--cycle;
\filldraw[fill=green!20, draw=green, thick] (11,-1.5)--(11,-1)--(9,-1)--(9,-1.5)--cycle;
\filldraw[fill=red!20, draw=red, thick] (9,-2)--(11,-2)--(11,-4)--(9,-4)--cycle;
\filldraw[fill=red!20, draw=red, thick] (5.5,-2)--(7.5,-2)--(7.5,-4)--(5.5,-4)--cycle;
\path (6.5,0.5) node [style=sergio] {$\lambda$};
\path (6.5,-3) node [style=sergio] {$\lambda$};
\path (10,-3) node [style=sergio] {$\lambda$};
\path (10,0.5) node [style=sergio] {$\lambda$};
\path (1.5,2) node [style=sergio] {$M_1$};
\path (-2.25,1) node [style=sergio] {$M_2$};
\path (0,1.1) node [style=sergio] {$C_2$};
\path (8.25,0.5) node [style=sergio] {\footnotesize$\sigma_1$};
\path (8.25,-3) node [style=sergio] {\footnotesize$\sigma_1$};
\path (6.5,-1.25) node [style=sergio] {\footnotesize$\sigma_2$};
\path (10,-1.25) node [style=sergio] {\footnotesize$\sigma_2$};
\filldraw[fill=blue!20, draw=blue, thick] (8,-1)--(8.5,-1)--(8.5,-1.5)--(8,-1.5)--cycle;
\path (8.25,-1.25) node [style=sergio] {\footnotesize$\tau$};
\end{tikzpicture}
\caption{The cell decomposition of $C_{2v}$-symmetric lattice. Left panel depicts the whole lattice, right panel depicts the horizontal intersection, including 3D blocks $\lambda$, 2D blocks $\sigma_{1,2}$, and 1D block $\tau$. $C_2$ depicts the 2-fold rotation, $M_1$ and $M_2$ are reflection planes.}
\label{C2v cell decomposition}
\end{figure}

For $C_{2v}$-symmetric lattice, with the cell decomposition in Fig. \ref{C2v cell decomposition}, the ground-state wavefunction of the system can be decomposed to the direct products of wavefunctions of lower-dimensional block states as:
\begin{align}
|\Psi\rangle=\bigotimes\limits_{g\in C_{2v}}|T_{g\lambda}\rangle\otimes\sum\limits_{j=1}^{2}|\gamma_{g\sigma_j}\rangle\otimes|\beta_{\tau}\rangle
\label{C2v wavefunction}
\end{align}
where $|T_{g\lambda}\rangle$ is the wavefunction of 3D block state on $g\lambda$ which is topological trivial; $|\gamma_{g\sigma_{1,2}}\rangle$ is the $\mathbb{Z}_2$-symmetric wavefunction of 2D block state on $g\sigma_{1,2}$; $|\beta_{\tau}\rangle$ is the $(\mathbb{Z}_2\times\mathbb{Z}_2)$-symmetric wavefunction of 1D block state on $\tau$. 

We summarize the classifications and corresponding root phases. For crystalline TSC with spinless fermions, we summarize all possible root phases:
\begin{enumerate}[1.]
\item 2D blocks $\sigma_1$ and $\sigma_2$: 2D $(p+ip)$-SCs and 2D $\mathbb{Z}_2$ fSPT phases;
\item 1D blocks $\tau$: Majorana chain, 1D $\mathbb{Z}_2\times\mathbb{Z}_2$ fSPT phases and 1D Haldane chain.
\end{enumerate}
It is obviously that the only possible obstruction-free $p+ip$ block states require that the $p+ip$ blocks on $\sigma_1$ and $\sigma_2$ should have opposite chirality. Then we consider the ``Majorana'' bubble construction on the 2D blocks $\sigma_1$, it leaves a 1D $\mathbb{Z}_2$ fSPT phase on $\tau$; reversely, this 1D block state can be trivialized by 2D bubble equivalence. Similar for another root phase of 1D $\mathbb{Z}_2$ fSPT phase. Furthermore, if we consider the bubble formed by 1D $\mathbb{Z}_2$ fSPT phase on each $\sigma_1$, it leaves a Haldane chain on $\tau$. Consequently, only Majorana chain decoration on $\tau$ is obstruction-free, and the ultimate classification is $\mathbb{Z}_2^3$, with the following root phases:
\begin{enumerate}[1.]
\item 2D bosonic Levin-Gu state on each $\sigma_1/\sigma_2$ ($\mathbb{Z}_2^2$);
\item Majorana chain on each $\tau$ ($\mathbb{Z}_2$).
\end{enumerate}
The HO topological surface theories of different root phases are:
\begin{enumerate}[1.]
\item $2^{\mathrm{nd}}$-order: 1D nonchiral Luttinger liquid with $K$-matrix $K=\sigma^x$ and $\mathbb{Z}_2$ symmetry property: $W^{\mathbb{Z}_2}=\mathbbm{1}_{2\times2}$ and $\delta\phi^{\mathbb{Z}_2}=\pi(1,1)^T$, on the intersections between open lattice and 2D blocks $\sigma_1/\sigma_2$;
\item $3^{\mathrm{rd}}$-order: Dangling Majorana zero modes at the centers of top and bottom surfaces of the open lattice.
\end{enumerate}

For crystalline TSC with spin-1/2 fermions, the corresponding classification is trivial because there is no nontrivial block state, for both 2D and 1D blocks.

For crystalline TI with spinless fermions, all possible block states are summarized as:
\begin{enumerate}[1.]
\item 2D blocks $\sigma_1$ and $\sigma_2$: Chern insulators, Kitaev's $E_8$ states, and 2D $U^f(1)\times\mathbb{Z}_2$ fSPT phases;
\item 1D block $\tau$: Haldane chain.
\end{enumerate}
Similar arguments to the $T_h$-symmetric lattice in the main text, the only obstruction-free 2D block states are 2D $U^f(1)\times\mathbb{Z}_2$ bSPT phase on each $\sigma_1$ or $\sigma_2$ and Monolayer Kitaev's $E_8$ state on each $\sigma_1$ and $\sigma_2$, and the corresponding classification is $\mathbb{Z}_2^4$, with the following root phases:
\begin{enumerate}[1.]
\item 2D $U^f(1)\times\mathbb{Z}_2$ bSPT phase on each $\sigma_1$ or $\sigma_2$ ($\mathbb{Z}_2^2$);
\item Monolayer Kitaev's $E_8$ state on each $\sigma_1$ and $\sigma_2$ ($\mathbb{Z}_2$);
\item Haldane chain decoration on each $\tau$ ($\mathbb{Z}_2$).
\end{enumerate}
The HO topological surface theories of different root phases are:
\begin{enumerate}[1.]
\item $2^{\mathrm{nd}}$-order: 1D nonchiral Luttinger liquid with $K$-matrix $K=\sigma^x$ and $\mathbb{Z}_2$ symmetry property: $W^{\mathbb{Z}_2}=\mathbbm{1}_{2\times2}$ and $\delta\phi^{\mathbb{Z}_2}=\pi(1,1)^T$, on the intersections between open lattice and 2D blocks $\sigma_1/\sigma_2$;
\item $2^{\mathrm{nd}}$-order: 1D chiral Luttinger liquid with $K$-matrix (\ref{K-matrix E8S}), on the intersections between open lattice and 2D blocks $\sigma_1$ and $\sigma_2$;
\item $3^{\mathrm{rd}}$-order: dangling spin-1/2 degrees of freedom at the centers of top and bottom surfaces of the open lattice.
\end{enumerate}

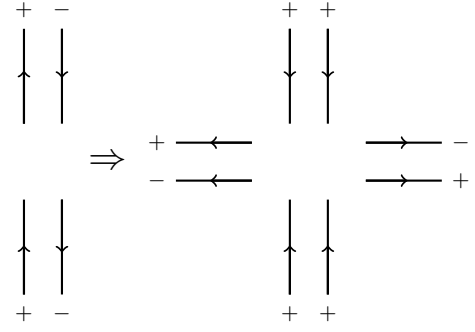
\begin{figure}
\begin{tikzpicture}[scale=1.01]
\tikzstyle{sergio}=[rectangle,draw=none]
\draw[draw=black,thick] (0,-0.5) -- (0,0.75);
\draw[draw=black,thick,->] (0,-0.5) -- (0,0.2);
\draw[draw=black,thick] (0.5,-0.5) -- (0.5,0.75);
\draw[draw=black,thick,->] (0.5,0.75) -- (0.5,0.1);
\draw[draw=black,thick] (0,-2.75) -- (0,-1.5);
\draw[draw=black,thick,->] (0,-2.75) -- (0,-2.1);
\draw[draw=black,thick] (0.5,-2.75) -- (0.5,-1.5);
\draw[draw=black,thick,->] (0.5,-1.5) -- (0.5,-2.2);
\path (0,1) node [style=sergio] {\footnotesize$+$};
\path (0.5,1) node [style=sergio] {\footnotesize$-$};
\path (0,-3) node [style=sergio] {\footnotesize$+$};
\path (0.5,-3) node [style=sergio] {\footnotesize$-$};
\path (1.1,-1) node [style=sergio] {\Large$\Rightarrow$};
\draw[draw=black,thick] (3.5,-0.5) -- (3.5,0.75);
\draw[draw=black,thick,->] (3.5,0.75) -- (3.5,0.1);
\draw[draw=black,thick] (4,-0.5) -- (4,0.75);
\draw[draw=black,thick,->] (4,0.75) -- (4,0.1);
\draw[draw=black,thick] (3.5,-2.75) -- (3.5,-1.5);
\draw[draw=black,thick,->] (3.5,-2.75) -- (3.5,-2.1);
\draw[draw=black,thick] (4,-2.75) -- (4,-1.5);
\draw[draw=black,thick,->] (4,-2.75) -- (4,-2.1);
\draw[draw=black,thick] (2,-0.75) -- (3,-0.75);
\draw[draw=black,thick,->] (3,-0.75) -- (2.45,-0.75);
\draw[draw=black,thick] (2,-1.25) -- (3,-1.25);
\draw[draw=black,thick,->] (4.5,-1.25) -- (5.05,-1.25);
\draw[draw=black,thick] (4.5,-0.75) -- (5.5,-0.75);
\draw[draw=black,thick,->] (4.5,-0.75) -- (5.05,-0.75);
\draw[draw=black,thick] (4.5,-1.25) -- (5.5,-1.25);
\draw[draw=black,thick,->] (3,-1.25) -- (2.45,-1.25);
\path (1.75,-0.75) node [style=sergio] {\footnotesize$+$};
\path (1.75,-1.25) node [style=sergio] {\footnotesize$-$};
\path (5.75,-0.75) node [style=sergio] {\footnotesize$-$};
\path (5.75,-1.25) node [style=sergio] {\footnotesize$+$};
\path (3.5,1) node [style=sergio] {\footnotesize$+$};
\path (4,-3) node [style=sergio] {\footnotesize$+$};
\path (3.5,-3) node [style=sergio] {\footnotesize$+$};
\path (4,1) node [style=sergio] {\footnotesize$+$};
\end{tikzpicture}
\caption{Equivalence between 2D $U^f(1)\times\mathbb{Z}_2$ fSPT phase on each $\sigma_1$ and bilayer Chern insulators on each $\sigma_1$ and $\sigma_2$, whose $\mathbb{Z}_2$ eigenvalues are depicted.}
\label{Chern bubble S}
\end{figure}

For crystalline TI with spin-1/2 fermions, all possible block states are on 2D blocks: Chern insulators, Kitaev's $E_8$ states and 2D $U^f(1)\times\mathbb{Z}_2$ fSPT phases. For chiral block states, nonchiral condition on $\tau$ requires the chiralities on $\sigma_1$ and $\sigma_2$ are opposite. Furthermore, bubble equivalences on 3D blocks can change the layers of chiral block states by two. Hence the corresponding classification is $\mathbb{Z}_8\times\mathbb{Z}_4\times\mathbb{Z}_2$, with the following root phases:
\begin{enumerate}[1.]
\item Monolayer Chern insulator on each $\sigma_1$ and $\sigma_2$ ($\mathbb{Z}_2$);
\item 2D $U^f(1)\times\mathbb{Z}_2$ fSPT phase with $\nu=1\in\mathbb{Z}_4$ on each $\sigma_1/\sigma_2$ ($\mathbb{Z}_4^2$);
\item Monolayer Kitaev's $E_8$ state on each $\sigma_1$ and $\sigma_2$ ($\mathbb{Z}_2$);
\end{enumerate}
And there is a nontrivial extensions between first two root phases: bilayer Chern insulators on each 2D block can be smoothly deformed to 2D $U^f(1)\times\mathbb{Z}_2$ fSPT phase with $\nu=1\in\mathbb{Z}_4$ on each $\sigma_1$ and $\nu=3\in\mathbb{Z}_4$ on each $\sigma_2$ by 3D ``Chern insulator'' bubble equivalence. We further demonstrate that the gapping problem of the second root phases is subtle: we cannot gap out the 1D nonchiral Luttinger liquid on $\tau$ leaved by 2D $U^f(1)\times\mathbb{Z}_2$ fSPT phase on $\sigma_1$ by direct $K$-matrix calculations, but there is a more sophisticated way: in the main text, we have demonstrated that 2D $U^f(1)\times\mathbb{Z}_2$ fSPT phase on $\sigma_1$ can be smoothly deformed to bilayer Chern insulators by 3D ``Chern insulator'' bubble equivalence on $\lambda$, with trivial $\mathbb{Z}_2$ symmetry property; on the other hand, 3D ``Chern insulator'' bubble equivalence leaves another bilayer Chern insulators on each $\sigma_2$, at which $\mathbb{Z}_2$ symmetry exchanges two layers of Chern insulators. Therefore, this root phase is smoothly deformed to bilayer Chern insulators on each $\sigma_1$ and $\sigma_2$ (see Fig. \ref{Chern bubble S}), which has been proved to be symmetrically gap out.
The second-order topological surface theories of these root phases are:
\begin{enumerate}[1.]
\item Monolayer Chern insulator on each $\sigma_1$ and $\sigma_2$: chiral fermions on the intersections between the open lattice, $\sigma_1$ and $\sigma_2$;
\item 2D $U^f(1)\times\mathbb{Z}_2$ fSPT phases with $\nu=1$ on each $\sigma_1$: 1D nonchiral Luttinger liquids with $K$-matrix $K=\sigma^z$ and $\mathbb{Z}_2$ symmetry property $W^{\mathbb{Z}_2}=\mathbbm{1}_{2\times2}$ and $\delta\phi^{\mathbb{Z}_2}=\pi(0,1)^T$ on the intersections between the open lattice and $\sigma_1$;
\item Monolayer Kitaev's $E_8$ state on each $\sigma_1$ and $\sigma_2$: 1D chiral Luttinger liquids with $K$-matrix (\ref{K-matrix E8S}) on the intersections between the open lattice, $\sigma_1$ and $\sigma_2$.
\end{enumerate}

\subsection{$D_{2h}/V_h$-symmetric lattice}
For $D_{2h}$-symmetric lattice, with the cell decomposition in Fig. \ref{D2h cell decomposition}, the ground-state wavefunction of the system can be decomposed to the direct products of wavefunctions of lower-dimensional block states as:
\begin{align}
|\Psi\rangle=\bigotimes\limits_{g\in D_{2h}}|T_{g\lambda}\rangle\otimes\sum\limits_{k=1}^3|\gamma_{g\sigma_{k}}\rangle\otimes\sum\limits_{j=1}^3|\beta_{g\tau_j}\rangle\otimes|\alpha_\mu\rangle
\label{D2h cell decomposition}
\end{align}
where $|T_{g\lambda}\rangle$ is the wavefunction of 3D block state on $g\lambda$ which is topological trivial; $|\gamma_{g\sigma_{1,2,3}}\rangle$ is the $\mathbb{Z}_2$-symmetric wavefunction of 2D block state on $g\sigma_{1,2,3}$; $|\beta_{g\tau_{1,2,3}}\rangle$ is the $(\mathbb{Z}_2\times\mathbb{Z}_2)$-symmetric wavefunction of 1D block state on $g\tau_{1,2,3}$; $|\alpha_\mu\rangle$ is the $(\mathbb{Z}_2\times\mathbb{Z}_2\times\mathbb{Z}_2)$-symmetric wavefunction of 0D block state on $\mu$.

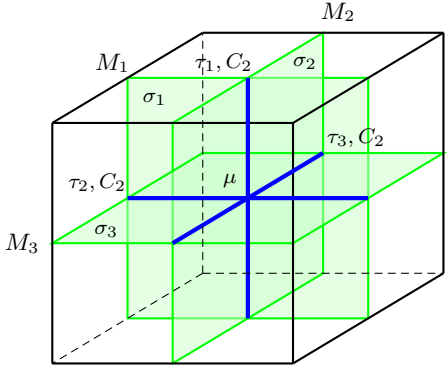
\begin{figure}
\begin{tikzpicture}[scale=0.8]
\tikzstyle{sergio}=[rectangle,draw=none]
\filldraw[fill=green!20, draw=green, thick,fill opacity=0.5] (-1.75,0.75)--(-1.75,-3.25)--(2.25,-3.25)--(2.25,0.75)--cycle;
\filldraw[fill=green!20, draw=green, thick,fill opacity=0.5] (-1,0)--(1.5,1.5)--(1.5,-2.5)--(-1,-4)--cycle;
\filldraw[fill=green!20, draw=green, thick,fill opacity=0.5] (-3,-2)--(-0.5,-0.5)--(3.5,-0.5)--(1,-2)--cycle;
\draw[thick] (-0.5,1.5) -- (3.5,1.5);
\draw[thick] (-0.5,1.5) -- (-3,0);
\draw[thick] (3.5,1.5) -- (1,0);
\draw[thick] (-3,0) -- (1,0);
\draw[thick] (-3,0) -- (-3,-4);
\draw[thick] (1,-4) -- (-3,-4);
\draw[thick] (1,-4) -- (1,0);
\draw[thick] (3.5,-2.5) -- (3.5,1.5);
\draw[thick] (3.5,-2.5) -- (1,-4);
\draw[draw=blue,ultra thick] (0.25,-3.25) -- (0.25,0.75);
\draw[draw=blue,ultra thick] (1.5,-0.5) -- (-1,-2);
\draw[draw=blue,ultra thick] (2.25,-1.25) -- (-1.75,-1.25);
\draw[densely dashed] (-0.5,-2.5) -- (-0.5,1.5);
\draw[densely dashed] (-0.5,-2.5) -- (3.5,-2.5);
\draw[densely dashed] (-0.5,-2.5) -- (-3,-4);
\path (-1.3,0.4) node [style=sergio] {\footnotesize$\sigma_1$};
\path (-2,1) node [style=sergio] {$M_1$};
\path (1.75,1.85) node [style=sergio] {$M_2$};
\path (-3.5,-2) node [style=sergio] {$M_3$};
\path (1.2,1) node [style=sergio] {\footnotesize$\sigma_2$};
\path (-2.1,-1.8) node [style=sergio] {\footnotesize$\sigma_3$};
\path (-0.15,1) node [style=sergio] {\footnotesize$\tau_1,C_2$};
\path (-2.25,-1) node [style=sergio] {\footnotesize$\tau_2,C_2$};
\path (2.05,-0.25) node [style=sergio] {\footnotesize$\tau_3,C_2$};
\path (-0.05,-0.95) node [style=sergio] {\footnotesize$\mu$};
\end{tikzpicture}
\caption{The cell decomposition of $D_{2h}/V_h$-symmetric lattice. Green plates depict 2D blocks $\sigma_{1,2,3}$, and blue segments represent 1D blocks $\tau_{1,2,3}$. $C_2$'s depict the axes of 2-fold rotations, and $\bs{M}_{1,2,3}$ are reflection planes.}
\label{D2h cell decomposition}
\end{figure}

We summarize the classifications and corresponding root phases. For crystalline TSC with spinless fermions, we summarize all possible block states as:
\begin{enumerate}[1.]
\item 2D blocks $\sigma_{1,2,3}$: 2D $(p+ip)$-SCs and 2D $\mathbb{Z}_2$ fSPT phases;
\item 1D blocks $\tau_{1,2,3}$: Majorana chain, 1D $\mathbb{Z}_2\times\mathbb{Z}_2$ fSPT phases, and Haldane chain;
\item 0D block $\mu$: 0D modes with odd fermion parity, eigenvalues $-1$ of three generators of the $D_{2h}/V_h$ group.
\end{enumerate}
Similar to the $T_h$-symmetric case in the main text, by $K$-matrix calculations, we conclude that except 2D bosonic Levin-Gu state, all other 2D block states are obstructed. 

For 1D block states, firstly it is easy to verify that all configurations of Majorana chain decorations are obstructed; the 1D $\mathbb{Z}_2\times\mathbb{Z}_2$ fSPT phases decorations solely on $\tau_1/\tau_2/\tau_3$ or all together are obstructed (see $T_h$-symmetric case in the main text). Furthermore, 2D ``Majorana'' bubble construction on $\sigma_1$ will be deformed to 1D $\mathbb{Z}_2\times\mathbb{Z}_2$ fSPT phases on $\tau_1$ and $\tau_2$. Reversely, the corresponding 1D block state is trivialized. Similar for $(\tau_1,\tau_3)$ and $(\tau_2,\tau_3)$. Moreover, if we consider the 1D $\mathbb{Z}_2$ fSPT phase as the 2D bubble on $\sigma_1$, it can be deformed to the Haldane chains on $\tau_1$ and $\tau_2$. Reversely, the corresponding 1D block state is trivialized, similar for $(\tau_1,\tau_3)$ and $(\tau_2,\tau_3)$. 

Then consider a complex fermion on each 1D block $\tau_1$ which can be adiabatically deformed to infinite far and trivialized, there will be an atomic insulator with two complex fermions and eigenvalue $-1$ of the reflection with respect to the horizontal plane at the 0D block $\mu$. Similar constructions can be applied to $\tau_2$ and $\tau_3$, hence all nontrivial eigenvalues of all three generators of the $D_{2h}/V_h$ group are trivialized. Therefore, the ultimate classification is $\mathbb{Z}_2^5$, with the following root phases:
\begin{enumerate}[1.]
\item 2D bosonic Levin-Gu state on each $\sigma_{1,2,3}$ ($\mathbb{Z}_2^3$);
\item 1D Haldane phase on each $\tau_1$, $\tau_2$ and $\tau_3$ ($\mathbb{Z}_2$);
\item 0D fermionic mode with odd fermion parity at $\mu$ ($\mathbb{Z}_2$).
\end{enumerate}
The HO topological surface theories of different root phases are:
\begin{enumerate}[1.]
\item $2^{\mathrm{nd}}$-order: 1D nonchiral Luttinger liquid with $K$-matrix $K=\sigma^x$ and $\mathbb{Z}_2$ symmetry property: $W^{\mathbb{Z}_2}=\mathbbm{1}_{2\times2}$ and $\delta\phi^{\mathbb{Z}_2}=\pi(1,1)^T$, on the intersections between cubic and 2D blocks $\sigma_{1,2,3}$;
\item $3^{\mathrm{rd}}$-order: a spin-1/2 degree of freedom at the center of each surface of the open lattice.
\end{enumerate}

For crystalline TSC with spin-1/2 fermions, all possible root phases are located at 0D block $\mu$, characterizing the eigenvalues $-1$ of all three generators of the $D_{2h}/V_h$ group, and there is no obstruction and trivialization. Hence the ultimate classification is $\mathbb{Z}_2^3$, whose root phases are 0D modes with eigenvalues $-1$ of three generators of $D_{2h}/V_h$ group at $\mu$. There is no HO topological surface theory because there is no block state with the dimension higher than zero.

For crystalline TI with spinless fermions, we summarize all possible block states as:
\begin{enumerate}[1.]
\item 2D blocks $\sigma_{1,2,3}$: Chern insulators, Kitaev's $E_8$ states, and 2D $U^f(1)\times\mathbb{Z}_2$ fSPT phases;
\item 1D blocks $\tau_{1,2,3}$: Haldane chain;
\item 0D block $\mu$: 0D modes characterizing the eigenvalues $-1$ of all three generators of the $D_{2h}/V_h$ group, with different $U^f(1)$ charges.
\end{enumerate}
Similar to the $T_h$-symmetric case in the main text, by $K$-matrix calculations, we conclude that except 2D bosonic Levin-Gu state and Kitaev's $E_8$ states on $\sigma_{1,2,3}$ with proper chiralities guaranteeing that all 1D blocks as their shared border should be nonchiral, all other 2D block states are obstructed. 

Then we consider the trivialization. Firstly, Kitaev's $E_8$ state as the 3D bubble on each $\lambda$ will change the layers of Kitaev's $E_8$ state on $\sigma_{1,2,3}$ by two, hence only monolayer will be nontrivial. Secondly consider a 0D fermionic mode with $U^f(1)$ charge $+1$ on each $\tau_1$, similar to crystalline TSCs, it will be deformed to an atomic insulator at $\mu$ with $U^f(1)$ charge $+1$ and eigenvalue $-1$ of the reflection with respect to the horizontal plane. Finally, the ultimate classification is $\mathbb{Z}_4\times\mathbb{Z}_2^7$, with the following root phases:
\begin{enumerate}[1.]
\item 2D $U^f(1)\times\mathbb{Z}_2$ bSPT phase on each $\sigma_{1,2,3}$ ($\mathbb{Z}_2^3$);
\item Monolayer Kitaev's $E_8$ state on each $\sigma_1$, $\sigma_2$ and $\sigma_3$ ($\mathbb{Z}_2$);
\item 1D Haldane phase on each $\tau_{1,2,3}$ ($\mathbb{Z}_2^3$);
\item 0D fermionic mode with odd $U^f(1)$ charge on $\mu$ ($\mathbb{Z}_2$);
\item The product of eigenvalues of three reflection generators of $D_{2h}/V_h$ group equals to $-1$ ($\mathbb{Z}_2$).
\end{enumerate}
And there is a nontrivial extension between last two root phases. The HO topological surface theories of different root phases are:
\begin{enumerate}[1.]
\item $2^{\mathrm{nd}}$-order: identical to crystalline TSC with spinless fermions;
\item $2^{\mathrm{nd}}$-order: 1D chiral Luttinger liquids with $K$-matrix (\ref{K-matrix E8S}), on the intersections of the open lattice and 2D blocks $\sigma_1$, $\sigma_2$ and $\sigma_3$;
\item $3^{\mathrm{rd}}$-order: a spin-1/2 degree of freedom at the center of top and bottom surfaces;
\item $3^{\mathrm{rd}}$-order: a spin-1/2 degree of freedom at the center of front and back surfaces;
\item $3^{\mathrm{rd}}$-order: a spin-1/2 degree of freedom at the center of left and right surfaces;
\end{enumerate}

For crystalline TI with spin-1/2 fermions, we summarize all possible block states as:
\begin{enumerate}[1.]
\item 2D blocks $\sigma_{1,2,3}$: Chern insulators, Kitaev's $E_8$ states, and 2D $U^f(1)\times\mathbb{Z}_2$ fSPT phases;
\item 0D block $\mu$: 0D modes characterizing the eigenvalues $-1$ of all three generators of the $D_{2h}/V_h$ group, with different but even $U^f(1)$ charges.
\end{enumerate}
There are some constraints on the chiral 2D block states because no-open-edge condition requires that the gapless modes leaved by these 2D block states near the 1D blocks as their shared border should be nonchiral. And the layers of chiral 2D block states can be changed by 3D bubble equivalences on $\lambda$. Hence the ultimate classification is $\mathbb{Z}_8\times\mathbb{Z}_4^2\times\mathbb{Z}_2^5$, with the following root phases:
\begin{enumerate}[1.]
\item Monolayer Chern insulator on each 2D block $\sigma_1$, $\sigma_2$ and $\sigma_3$ ($\mathbb{Z}_2$);
\item 2D $U^f(1)\times\mathbb{Z}_2$ fSPT phase on each 2D block $\sigma_1/\sigma_2/\sigma_3$ ($\mathbb{Z}_4^3$);
\item Monolayer Kitaev's $E_8$ state on each 2D block $\sigma_1$, $\sigma_2$ and $\sigma_3$ ($\mathbb{Z}_2$);
\item 0D mode with $U^f(1)$ charge $n\equiv2(\mathrm{mod}~4)$ on $\mu$ ($\mathbb{Z}_2$);
\item Eigenvalues $-1$ of three generators of $D_{2h}/V_h$ ($\mathbb{Z}_2^3$).
\end{enumerate}
and there is a nontrivial extensions between these root phases: bilayer Chern insulators on each $\sigma_1$, $\sigma_2$ and $\sigma_3$ are equivalent to 2D $U^f(1)\times\mathbb{Z}_2$ fSPT phases with $(\nu_1,\nu_2,\nu_3)=(1,3,1)$ on 2D blocks $\sigma_1$, $\sigma_2$ and $\sigma_3$. The second-order topological surface theories of different root phases are:
\begin{enumerate}[1.]
\item Chiral fermions on the intersections between the open lattice and all 2D blocks;
\item 1D nonchiral Luttinger liquids with $K$-matrix $K=\sigma^z$, and $\mathbb{Z}_2$ symmetry property $W^{\mathbb{Z}_2}=\mathbbm{1}_{2\times2}$ and $\delta\phi^{\mathbb{Z}_2}=\pi(0,1)^T$, on the intersections between the open lattice and 2D blocks decorated with a $U^f(1)\times\mathbb{Z}_2$ fSPT phase with $\nu=1$ on each of them;
\item 1D chiral Luttinger liquids with $K$-matrix (\ref{K-matrix E8S}), on the intersections between the open lattice and all 2D blocks.
\end{enumerate}

\begin{figure}
\begin{tikzpicture}[scale=0.6]
\tikzstyle{sergio}=[rectangle,draw=none]
\filldraw[fill=green!20, draw=green, thick,fill opacity=0.5] (-1.75,0.75)--(-1.75,-3.25)--(2.25,-3.25)--(2.25,0.75)--cycle;
\filldraw[fill=green!20, draw=green, thick,fill opacity=0.5] (-1,0)--(1.5,1.5)--(1.5,-2.5)--(-1,-4)--cycle;
\draw[thick] (-0.5,1.5) -- (3.5,1.5);
\draw[thick] (-0.5,1.5) -- (-3,0);
\draw[thick] (3.5,1.5) -- (1,0);
\draw[thick] (-3,0) -- (1,0);
\draw[thick] (-3,0) -- (-3,-4);
\draw[thick] (1,-4) -- (-3,-4);
\draw[thick] (1,-4) -- (1,0);
\draw[thick] (3.5,-2.5) -- (3.5,1.5);
\draw[thick] (3.5,-2.5) -- (1,-4);
\draw[draw=blue,ultra thick] (0.25,-3.25) -- (0.25,0.75);
\draw[densely dashed] (-0.5,-2.5) -- (-0.5,1.5);
\draw[densely dashed] (-0.5,-2.5) -- (3.5,-2.5);
\draw[densely dashed] (-0.5,-2.5) -- (-3,-4);
\filldraw[fill=red!20, draw=red, thick] (5.5,1.5)--(7.5,1.5)--(7.5,-0.5)--(5.5,-0.5)--cycle;
\filldraw[fill=red!20, draw=red, thick] (9,1.5)--(11,1.5)--(11,-0.5)--(9,-0.5)--cycle;
\filldraw[fill=green!20, draw=green, thick] (8.5,1.5)--(8,1.5)--(8,-0.5)--(8.5,-0.5)--cycle;
\filldraw[fill=green!20, draw=green, thick] (8.5,-2)--(8,-2)--(8,-4)--(8.5,-4)--cycle;
\filldraw[fill=green!20, draw=green, thick] (7.5,-1.5)--(7.5,-1)--(5.5,-1)--(5.5,-1.5)--cycle;
\filldraw[fill=green!20, draw=green, thick] (11,-1.5)--(11,-1)--(9,-1)--(9,-1.5)--cycle;
\filldraw[fill=red!20, draw=red, thick] (9,-2)--(11,-2)--(11,-4)--(9,-4)--cycle;
\filldraw[fill=red!20, draw=red, thick] (5.5,-2)--(7.5,-2)--(7.5,-4)--(5.5,-4)--cycle;
\path (6.5,0.5) node [style=sergio] {$\lambda$};
\path (6.5,-3) node [style=sergio] {$\lambda$};
\path (10,-3) node [style=sergio] {$\lambda$};
\path (10,0.5) node [style=sergio] {$\lambda$};
\path (0,1.1) node [style=sergio] {$C_4$};
\path (8.25,0.5) node [style=sergio] {\footnotesize$\sigma$};
\path (8.25,-3) node [style=sergio] {\footnotesize$\sigma$};
\path (6.5,-1.25) node [style=sergio] {\footnotesize$\sigma$};
\path (10,-1.25) node [style=sergio] {\footnotesize$\sigma$};
\filldraw[fill=blue!20, draw=blue, thick] (8,-1)--(8.5,-1)--(8.5,-1.5)--(8,-1.5)--cycle;
\path (8.25,-1.25) node [style=sergio] {\footnotesize$\tau$};
\end{tikzpicture}
\caption{The cell decomposition of $C_4$-symmetric lattice. The left panel depicts the whole lattice, the right panel depicts the horizontal intersection, including 3D blocks $\lambda$, 2D blocks $\sigma$, and 1D block $\tau$. $C_4$ depicts the axis of 4-fold rotation.}
\label{C4 cell decomposition}
\end{figure}
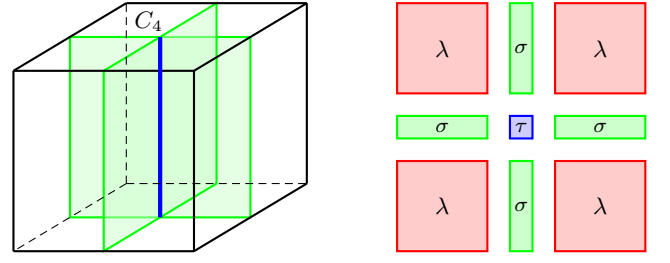

\begin{figure*}
\begin{tikzpicture}[scale=0.85]
\tikzstyle{sergio}=[rectangle,draw=none]
\filldraw[fill=green!20, draw=green, thick] (11.5,1.5)--(13.5,1.5)--(13.5,-0.5)--(11.5,-0.5)--cycle;
\filldraw[fill=green!20, draw=green, thick] (15,1.5)--(17,1.5)--(17,-0.5)--(15,-0.5)--cycle;
\filldraw[fill=blue!20, draw=blue, thick] (14.5,1.5)--(14,1.5)--(14,-0.5)--(14.5,-0.5)--cycle;
\filldraw[fill=blue!20, draw=blue, thick] (14.5,-2)--(14,-2)--(14,-4)--(14.5,-4)--cycle;
\path (12.5,0.5) node [style=sergio] {$\sigma_2$};
\path (16,0.5) node [style=sergio] {$\sigma_2$};
\path (14.25,0.5) node [style=sergio] {\scriptsize$\tau_2$};
\path (14.25,-3) node [style=sergio] {\scriptsize$\tau_2$};
\filldraw[fill=white, draw=black] (14.25,-1.25) circle (13pt);
\path (14.25,-1.25) node [style=sergio] {$\mu$};
\filldraw[fill=red!20, draw=red, thick] (1,1.5)--(3,1.5)--(3,-4)--(1,-4)--cycle;
\filldraw[fill=red!20, draw=red, thick] (-2.5,1.5)--(-0.5,1.5)--(-0.5,-4)--(-2.5,-4)--cycle;
\filldraw[fill=green!20, draw=green, thick] (0,1.5)--(0.5,1.5)--(0.5,-0.5)--(0,-0.5)--cycle;
\filldraw[fill=green!20, draw=green, thick] (0,-2)--(0.5,-2)--(0.5,-4)--(0,-4)--cycle;
\filldraw[fill=blue!20, draw=blue, thick] (0,-1)--(0.5,-1)--(0.5,-1.5)--(0,-1.5)--cycle;
\path (-1.5,-1.25) node [style=sergio] {$\lambda$};
\path (2,-1.25) node [style=sergio] {$\lambda$};
\path (0.25,0.5) node [style=sergio] {\scriptsize$\sigma_1$};
\path (0.25,-3) node [style=sergio] {\scriptsize$\sigma_1$};
\path (0.25,-1.25) node [style=sergio] {\scriptsize$\tau_1$};
\filldraw[fill=green!20, draw=green, thick] (11.5,-2)--(13.5,-2)--(13.5,-4)--(11.5,-4)--cycle;
\filldraw[fill=green!20, draw=green, thick] (15,-2)--(17,-2)--(17,-4)--(15,-4)--cycle;
\path (12.5,-3) node [style=sergio] {$\sigma_2$};
\path (16,-3) node [style=sergio] {$\sigma_2$};
\filldraw[fill=blue!20, draw=blue, thick] (11.5,-1.5)--(11.5,-1)--(13.5,-1)--(13.5,-1.5)--cycle;
\filldraw[fill=blue!20, draw=blue, thick] (15,-1.5)--(15,-1)--(17,-1)--(17,-1.5)--cycle;
\path (12.5,-1.25) node [style=sergio] {\scriptsize$\tau_2$};
\path (16,-1.25) node [style=sergio] {\scriptsize$\tau_2$};
\path (0.25,-4.75) node [style=sergio] {North hemisphere};
\path (14.25,-4.75) node [style=sergio] {Equator};
\filldraw[fill=red!20, draw=red, thick] (4.5,-2)--(10,-2)--(10,-4)--(4.5,-4)--cycle;
\filldraw[fill=red!20, draw=red, thick] (4.5,1.5)--(10,1.5)--(10,-0.5)--(4.5,-0.5)--cycle;
\filldraw[fill=green!20, draw=green, thick] (4.5,-1)--(4.5,-1.5)--(6.5,-1.5)--(6.5,-1)--cycle;
\filldraw[fill=green!20, draw=green, thick] (8,-1)--(8,-1.5)--(10,-1.5)--(10,-1)--cycle;
\filldraw[fill=blue!20, draw=blue, thick] (7,-1)--(7,-1.5)--(7.5,-1.5)--(7.5,-1)--cycle;
\path (7.25,0.5) node [style=sergio] {$\lambda$};
\path (7.25,-3) node [style=sergio] {$\lambda$};
\path (0.75,-0.75) node [style=sergio] {$S_4$};
\path (7.75,-0.75) node [style=sergio] {$S_4$};
\path (5.5,-1.25) node [style=sergio] {\footnotesize$\sigma_1$};
\path (9,-1.25) node [style=sergio] {\footnotesize$\sigma_1$};
\path (7.25,-1.25) node [style=sergio] {\footnotesize$\tau_1$};
\path (7.25,-4.75) node [style=sergio] {South hemisphere};
\end{tikzpicture}
\caption{The cell decomposition of $S_4$-symmetric lattice. Left panel depicts the north hemisphere, including 3D blocks $\lambda$, 2D blocks $\sigma_1$ and 1D blocks $\tau_1$; middle panel depicts the south hemisphere, including 3D blocks $\lambda$, 2D blocks $\sigma_1$ and 1D blocks $\tau_1$; right panel depicts the equator, including 2D blocks $\sigma_2$, 1D blocks $\tau_2$ and 0D block $\mu$. $S_4$ depicts the axis of the 4-fold rotoreflection.}
\label{S4 cell decomposition}
\end{figure*}
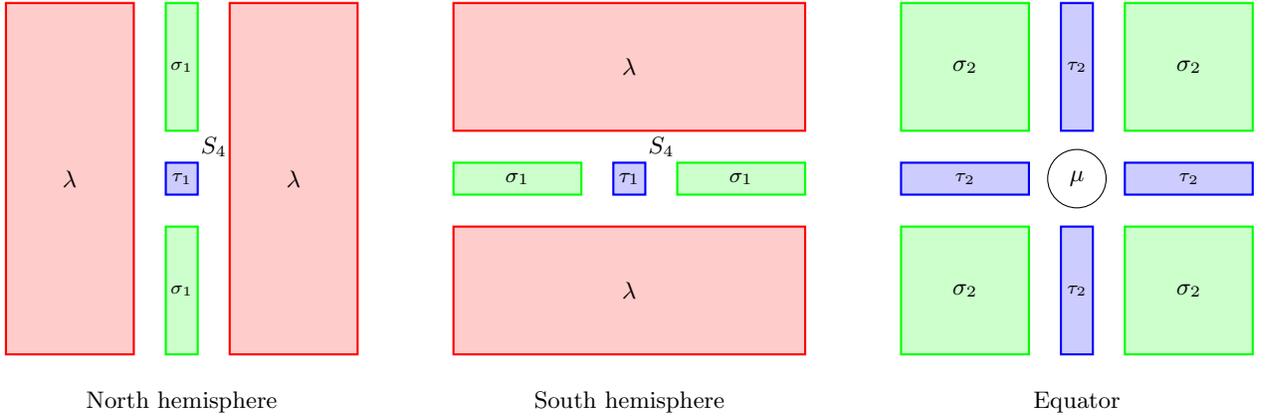

\subsection{$C_4$-symmetric lattice}
For $C_4$-symmetric lattice with the cell decomposition in Fig. \ref{C4 cell decomposition}, the ground-state wavefunction of the system can be decomposed to the direct products of wavefunctions of lower-dimensional block states as:
\begin{align}
|\Psi\rangle=\bigotimes_{g\in C_4}|T_{g\lambda}\rangle\otimes|T_{g\sigma}\rangle\otimes|\beta_{\tau}\rangle
\label{C4 wavefunction}
\end{align}
where $|T_{g\lambda}\rangle/|T_{g\sigma}\rangle$ is the wavefunction of 3D/2D block state on $g\lambda/g\sigma$ which is topological trivial or invertible topological phase; $|\beta_\tau\rangle$ is the $\mathbb{Z}_4$-symmetric wavefunction of 1D block state on $\tau$. 

We summarize the classifications and corresponding root phases. For crystalline TSC with spinless fermions, we summarize all possible root phases as:
\begin{enumerate}[1.]
\item 2D blocks $\sigma$: 2D $(p+ip)$-SCs;
\item 1D block $\tau$: Majorana chain and 1D $\mathbb{Z}_4$ fSPT phases.
\end{enumerate}
It is obviously that 2D $(p+ip)$-SCs on $\sigma$ is obstructed because it leaves chiral 1D mode on $\tau$, whose chiral central charge is $c_-=2$. Furthermore, 1D $\mathbb{Z}_4$ fSPT phase on $\tau$ is trivialized by 2D ``Majorana'' bubble construction on $\sigma$, and the Majorana chain decoration on $\tau$ is trivialized by anomalous $(p+ip)$-SC on the top surface of the open cubic \cite{rotation}. Hence the ultimate classifications are all trivial. 

For crystalline TSC with spin-1/2 fermions, the only possible root phase is 2D $(p+ip)$-SC on each $\sigma$ which is obstructed. Hence the classification is trivial. 

For crystalline TI with spinless and spin-1/2 fermions, the only possible root phases are Chern insulators and Kitaev's $E_8$ states on each $\sigma$, which is obstructed because these root phases are chiral. Hence the ultimate classifications are trivial.

\subsection{$S_4$-symmetric lattice}
For $S_4$-symmetric lattice with the cell decomposition in Fig. \ref{S4 cell decomposition}, the ground-state wavefunction of the system can be decomposed to the direct products of wavefunctions of lower-dimensional block states as:
\begin{align}
|\Psi\rangle=\bigotimes_{g\in S_4}|T_{g\lambda}\rangle\otimes\bigotimes\limits_{k=1}^2|T_{g\sigma_{k}}\rangle\otimes\bigotimes\limits_{j=1}^2|\beta_{\tau_j}\rangle\otimes|\alpha_\mu\rangle
\label{S4 wavefunction}
\end{align}
where $|T_{g\lambda}\rangle/|T_{g\sigma_{1,2}}\rangle$ is the wavefunction of 3D/2D block state on $g\lambda/g\sigma_{1,2}$ which is topological trivial or invertible topological phase; $|\beta_{g\tau_{1}}\rangle$ is the $\mathbb{Z}_2$-symmetric wavefunction of 1D block state on $g\tau_1$, $|\beta_{g\tau_2}\rangle$ is the wavefunction of 1D block state on $g\tau_2$ which is topological trivial or invertible topological phase; $|\alpha_\mu\rangle$ is the $\mathbb{Z}_4$-symmetric wavefunction of 0D block state on $\mu$.

We summarize the classifications and corresponding root phases. For crystalline TSC with spinless fermions, we summarize all possible block states as:
\begin{enumerate}[1.]
\item 2D blocks $\sigma_1$ and $\sigma_2$: 2D $(p+ip)$-SCs;
\item 1D blocks $\tau_1$: Majorana chain and 1D $\mathbb{Z}_2$ fSPT phase;
\item 1D blocks $\tau_2$: Majorana chain;
\item 0D block $\mu$: 0D modes with odd fermion parity, or characterizing the eigenvalues of 4-fold rotation.
\end{enumerate}
2D $(p+ip)$-SC on each $\sigma_1$ is obstructed because it leaves a chiral mode with chiral central charge $c_-=1$ on each $\tau_1$; 2D $(p+ip)$-SC on each $\sigma_2$ is also obstructed because the 2D $(p+ip)$-SC is not compatible with 4-fold rotation for spinless fermions \cite{rotation}. 

Majorana chain decoration on each $\tau_1$ leaves two Majorana fermions $\gamma_1$ and $\gamma_2$ near $\mu$, with the following property under the generator of the $S_4$ group:
\begin{align}
\bs{I}_r\in S_4:~\gamma_1\leftrightarrow\gamma_2
\end{align}
The fermion parity of these two Majorana fermions $P_f=i\gamma_1\gamma_2$ is odd under $\bs{I}_r$, hence they cannot be gapped out in a symmetric way, and the Majorana chain decoration on $\tau_1$ is obstructed. Similarly, Majorana chain decoration or 1D $\mathbb{Z}_2$ fSPT phase on $\tau_2$ is also obstructed. There is an exception: the 1D block state formed by Majorana chain decoration on $\tau_2$ and 1D $\mathbb{Z}_2$ fSPT phase on $\tau_1$ simultaneously is obstruction-free: We label the edge Majorana zero modes from the decorations on $\tau_2$ by $\gamma_{3,4,5,6}$, the $S_4$ symmetry properties of all Majorana zero modes near $\mu$ are
\begin{align}
\bs{I}_r:~(\gamma_1,\gamma_2,\gamma_3,\gamma_4,\gamma_5,\gamma_6)\mapsto(\gamma_2,\gamma_1,\gamma_4,\gamma_5,\gamma_6,\gamma_3),
\end{align}
then we see that the fermion parity $P_f=\prod_{j=1}^6\gamma_j$ is even under $\bs{I}_r$, i.e., Majorana zero modes $\gamma_{1,2,3,4,5,6}$ are compatible with the $S_4$ symmetry. Furthermore, We see that $S_4$ does not have nontrivial projective representation which is guaranteed by $\mathcal{H}^2[S_4, U(1)] = \mathbb{Z}_1$, hence we can always find a way to gap out all these 6 Majorana zero modes. Together with 0D modes with odd fermion parity or characterizing the eigenvalues of 4-fold rotation, we have obtained all obstruction-free block states.

Consider possible trivializations by bubble equivalences. A complex fermion on each $\tau_2$ which can be adiabatically deformed to infinite far and trivialized, which forms an atomic insulator with 4 complex fermions, with eigenvalue $-1$ of 4-fold rotation at $\mu$. Furthermore, the odd fermion parity of 0D block state at $\mu$ can be trivialized by ``Majorana'' bubble construction on each $\sigma_2$ that can be adiabatically deformed to a Majorana chain with PBC surrounding $\mu$ \cite{rotation}. Finally, the ultimate classification is $\mathbb{Z}_2^2$, whose root phases are
\begin{enumerate}[1.]
\item Majorana chain decoration on $\tau_1$ and $\tau_2$;
\item Eigenvalue $\pm i$ of 4-fold rotation on the equator ($S_4$ symmetry action is identical with a 4-fold rotation on the equator). 
\end{enumerate}

For crystalline TSC with spin-1/2 fermions, we summarize all possible root phases as:
\begin{enumerate}[1.]
\item 2D blocks $\sigma_1$ and $\sigma_2$: 2D $(p+ip)$-SCs;
\item 1D blocks $\tau_2$: Majorana chain;
\item 0D blocks $\mu$: 0D modes with odd fermion parity, or characterizing the eigenvalues of 4-fold rotation (at the equator). 
\end{enumerate}

2D $(p+ip)$-SC on each $\sigma_1$ is obstructed because it leaves a chiral mode with chiral central charge $c_-=1$ on each $\tau_1$; 2D $(p+ip)$-SC on each $\sigma_2$ is obstruction-free because 2D $(p+ip)$-SC is compatible with 4-fold rotation on the equator for spin-1/2 fermions.

Majorana chain decoration on each $\tau_2$ leaves 4 Majorana zero modes $\gamma_{3,4,5,6}$ near $\mu$, with the following $S_4$ symmetry properties
\begin{align}
\bs{I}_r:(\gamma_3,\gamma_4,\gamma_5,\gamma_6)\mapsto(\gamma_4,\gamma_5,\gamma_6,-\gamma_3),
\end{align}
hence they can be gapped out by the following symmetric Hamiltonian:
\begin{align}
H=i\gamma_3\gamma_5+i\gamma_4\gamma_6.
\end{align}
We summarize all obstruction-free block-states:
\begin{enumerate}[1.]
\item 2D $(p+ip)$-SC on each $\sigma_2$;
\item 1D Majorana chain on each $\tau_2$;
\item Complex fermion decoration on $\mu$;
\item 0D modes with $S_4$ eigenvalues $\{\pm i, \pm1\}$ on $\mu$.
\end{enumerate}

Then we consider possible trivializations. Firstly, we see that the Majorana bubble on each $\sigma_1$ will leave a Majorana chain on each $\tau_2$, hence the Majorana chain block state is trivialized. Then consider a complex fermion on each $\tau_1$, with the following $S_4$ properties:
\begin{align}
\bs{I}_r:(c_1^\dag,c_2^\dag)\mapsto(c_2^\dag,ic_1^\dag),
\end{align}
then we see that the 0D block state with $S_4$ eigenvalue $\pm i$ on $\mu$ is trivialized. Therefore, there are two root phases:
\begin{enumerate}[1.]
\item 2D $(p+ip)$-SC on each $\sigma_2$;
\item Complex fermion mode on $\mu$.
\end{enumerate}
Then we consider the most subtle point of this case: the above two root phases actually have nontrivial extension. Consider two layers of 2D $(p
+ip)$-SC on each $\sigma_2$, on the one hand, this 2D block state ought be trivialized by $(p+ip)$-SC bubble on each 3D block $\lambda$; on the other hand, bilayer $(p+ip)$-SCs on $\sigma_2$ leave 4 chiral Majorana zero modes $\eta_{1,2,3,4}$ at each vertical $\tau_2$, with the following properties under the symmetry action $S_4^2$,
\begin{align}
\left\{
\begin{aligned}
&\eta_1(x)\mapsto\eta_2(-x)\\
&\eta_2(x)\mapsto-\eta_1(-x)
\end{aligned}
\right.,~\left\{
\begin{aligned}
&\eta_3(x)\mapsto\eta_4(-x)\\
&\eta_4(x)\mapsto-\eta_3(-x)
\end{aligned}
\right.
\end{align}
Firstly, $\eta_1$ and $\eta_4$ can be gapped by the Hamiltonian
\begin{align}
H_{14}=\int\mathrm{d}x\cdot\eta_\1^T(i\sigma^z\partial_x+m\sigma^y)\eta_\1,
\end{align}
where $\eta_\1=(\eta_1,\eta_4)^T$. Enforced by symmetry, Majorana zero modes $\eta_2$ and $\eta_3$ (with $\eta_\2=(\eta_2,\eta_3)^T$) are gapped by the Hamiltonian,
\begin{align}
H_{23}=\int\mathrm{d}x\cdot\eta_\2^T(i\sigma^z\partial_x+m\sigma^y)\eta_\2.
\end{align}
Furthermore, bilayer $(p+ip)$-SCs on $\sigma_2$ leave 4 chiral Majorana zero modes $\eta_{5,6,7,8}$ at each horizontal $\tau_2$, with the following properties under the symmetry action $S_4^2$,
\begin{align}
\left\{
\begin{aligned}
&\eta_5(x)\mapsto\eta_6(-x)\\
&\eta_6(x)\mapsto-\eta_5(-x)
\end{aligned}
\right.,~\left\{
\begin{aligned}
&\eta_7(x)\mapsto\eta_8(-x)\\
&\eta_8(x)\mapsto-\eta_7 (-x)
\end{aligned}
\right.
\end{align}
Enforced by the $S_4$ symmetry, the gapping Hamiltonians along the horizontal $\tau_2$ should be 
\begin{align}
\begin{gathered}
H_{58}=\int\mathrm{d}x\cdot\eta_\3^T(i\sigma^z\partial_x-m\sigma^y)\eta_\3\\
H_{67}=\int\mathrm{d}x\cdot\eta_\4^T(i\sigma^z\partial_x-m\sigma^y)\eta_\4
\end{gathered},
\end{align}
where $\eta_\3=(\eta_5,\eta_8)^T$ and $\eta_\4=(\eta_6,\eta_7)^T$. Compare the Hamiltonians $H_{14}$, $H_{23}$, $H_{58}$, and $H_{67}$, we conclude that if $m>0$, there are two Majorana chains on each horizontal $\tau_2$; if $m<0$, there are two Majorana chains on each vertical $\tau_2$. We know that two Majorana chains along either direction can be deformed into a Majorana chain with periodic boundary conditions surrounding $\mu$, and change the fermion parity decorated on that. Therefore, we conclude that for this case, two block states have nontrivial extension towards a $\Z_4$ classification.

For systems with $U^f(1)$ charge conservation, similar to the $C_2$-symmetric case, the classifications and root phases are irrelevant to the spin of fermions because $\mathcal{H}^2[S_4, U(1)]=\mathbb{Z}_1$ that guarantees there is no nontrivial extension between $S_4$ and $U^f(1)$. We summarize all possible block states as
\begin{enumerate}[1.]
\item 2D blocks $\sigma_1$ and $\sigma_2$: Chern insulators and Kitaev's $E_8$ states;
\item 0D  block $\mu$: 0D modes characterizing the eigenvalues of 4-fold rotation, with different $U^f(1)$ charge.
\end{enumerate}
Chern insulators and Kitaev's $E_8$ states on $\sigma_1$ are obstructed because both of them leave chiral 1D mode at $\tau_1$ as their shared border. Furthermore, consider a 0D fermionic mode with $U^f(1)$ charge $+1$ on each $\tau_1$ which can be adiabatically deformed to infinite far and trivialized, they will form an atomic insulator $c_1^\dag c_2^\dag|0\rangle$, with eigenvalue $-1$ of $\bs{I}_r$ and $U^f(1)$ charge 2. Reversely, the corresponding 0D block state is trivialized. Therefore, the ultimate classification of crystalline TI is $\mathbb{Z}_2^4$, with the following root phases:
\begin{enumerate}[1.]
\item Monolayer Chern insulator on each $\sigma_2$ ($\mathbb{Z}_2$);
\item Monolayer Kitaev's $E_8$ state on each $\sigma_2$ ($\mathbb{Z}_2$);
\item 0D fermionic mode with odd $U^f(1)$ charge on $\mu$ ($\mathbb{Z}_2$);
\item 0D mode with eigenvalue $\pm i$ of 4-fold rotation on $\mu$ ($\mathbb{Z}_2$).
\end{enumerate}
The second-order topological surface theories of different root phases are:
\begin{enumerate}[1.]
\item 1D chiral fermions on the edge of equator;
\item 1D chiral Luttinger liquids with the $K$-matrix (\ref{K-matrix E8S}) on the edge of equator.
\end{enumerate}

\subsection{$C_{4h}$-symmetric lattice}
For $C_{4h}$-symmetric lattice with the cell decomposition in Fig. \ref{C4h cell decomposition}, the ground-state wavefunction of the system can be decomposed to the direct products of wavefunctions of lower-dimensional block states as:
\begin{align}
|\Psi\rangle=\bigotimes\limits_{g\in C_{4h}}|T_{g\lambda}\rangle\otimes\sum\limits_{k=1}^2|\gamma_{g\sigma_k}\rangle\otimes\sum\limits_{j=1}^2|\beta_{g\tau_j}\rangle\otimes|\alpha_\mu\rangle
\label{C4h cell decomposition}
\end{align}
where $|T_{g\lambda}\rangle$ is the wavefunction of 3D block state on $g\lambda$ which is topological trivial; $|\gamma_{g\sigma_1}\rangle$ is the wavefuntion of 2D block state on $g\sigma_1$ which is topological trivial or invertible topological phase; $|\gamma_{g\sigma_2}\rangle$ is the $\mathbb{Z}_2$-symmetric wavefunction of 2D block state on $g\sigma_2$; $|\beta_{g\tau_1}\rangle/|\beta_{g\tau_2}\rangle$ is the $\mathbb{Z}_4$-/$\mathbb{Z}_2$-symmetric wavefunction of 1D block state on $g\tau_1/g\tau_2$; $|\alpha_\mu\rangle$ is the $(\mathbb{Z}_4\times\mathbb{Z}_2)$-symmetric wavefunction of 0D block state on $\mu$.

\begin{figure}
\begin{tikzpicture}[scale=0.65]
\tikzstyle{sergio}=[rectangle,draw=none]
\filldraw[fill=green!20, draw=green, thick] (4.5,1.5)--(6.5,1.5)--(6.5,-0.5)--(4.5,-0.5)--cycle;
\filldraw[fill=green!20, draw=green, thick] (8,1.5)--(10,1.5)--(10,-0.5)--(8,-0.5)--cycle;
\filldraw[fill=blue!20, draw=blue, thick] (7.5,1.5)--(7,1.5)--(7,-0.5)--(7.5,-0.5)--cycle;
\filldraw[fill=blue!20, draw=blue, thick] (7.5,-2)--(7,-2)--(7,-4)--(7.5,-4)--cycle;
\path (5.5,0.5) node [style=sergio] {$\sigma_2$};
\path (9,0.5) node [style=sergio] {$\sigma_2$};
\path (7.25,0.5) node [style=sergio] {\scriptsize$\tau_2$};
\path (7.25,-3) node [style=sergio] {\scriptsize$\tau_2$};
\filldraw[fill=white, draw=black] (7.25,-1.25)circle (13pt);
\path (7.25,-1.25) node [style=sergio] {$\mu$};
\filldraw[fill=red!20, draw=red, thick] (1,1.5)--(3,1.5)--(3,-0.5)--(1,-0.5)--cycle;
\filldraw[fill=red!20, draw=red, thick] (-2.5,1.5)--(-0.5,1.5)--(-0.5,-0.5)--(-2.5,-0.5)--cycle;
\filldraw[fill=green!20, draw=green, thick] (0,1.5)--(0.5,1.5)--(0.5,-0.5)--(0,-0.5)--cycle;
\filldraw[fill=green!20, draw=green, thick] (0,-2)--(0.5,-2)--(0.5,-4)--(0,-4)--cycle;
\filldraw[fill=blue!20, draw=blue, thick] (0,-1)--(0.5,-1)--(0.5,-1.5)--(0,-1.5)--cycle;
\path (-1.5,0.5) node [style=sergio] {$\lambda$};
\path (2,0.5) node [style=sergio] {$\lambda$};
\path (0.25,0.5) node [style=sergio] {\scriptsize$\sigma_1$};
\path (0.25,-3) node [style=sergio] {\scriptsize$\sigma_1$};
\path (0.25,-1.25) node [style=sergio] {\scriptsize$\tau_1$};
\path (0.25,-4.75) node [style=sergio] {North/south hemisphere};
\path (7.25,-4.75) node [style=sergio] {Equator};
\filldraw[fill=red!20, draw=red, thick] (-2.5,-2)--(-0.5,-2)--(-0.5,-4)--(-2.5,-4)--cycle;
\path (-1.5,-3) node [style=sergio] {$\lambda$};
\filldraw[fill=red!20, draw=red, thick] (1,-2)--(3,-2)--(3,-4)--(1,-4)--cycle;
\path (2,-3) node [style=sergio] {$\lambda$};
\filldraw[fill=green!20, draw=green, thick] (-2.5,-1)--(-2.5,-1.5)--(-0.5,-1.5)--(-0.5,-1)--cycle;
\filldraw[fill=green!20, draw=green, thick] (1,-1)--(1,-1.5)--(3,-1.5)--(3,-1)--cycle;
\path (2,-1.25) node [style=sergio] {\scriptsize$\sigma_1$};
\path (-1.5,-1.25) node [style=sergio] {\scriptsize$\sigma_1$};
\filldraw[fill=green!20, draw=green, thick] (8,-2)--(10,-2)--(10,-4)--(8,-4)--cycle;
\filldraw[fill=green!20, draw=green, thick] (4.5,-2)--(6.5,-2)--(6.5,-4)--(4.5,-4)--cycle;
\path (5.5,-3) node [style=sergio] {$\sigma_2$};
\path (9,-3) node [style=sergio] {$\sigma_2$};
\path (0.75,-0.75) node [style=sergio] {\footnotesize$C_4$};
\path (9.5,1) node [style=sergio] {$M$};
\filldraw[fill=blue!20, draw=blue, thick] (4.5,-1.5)--(4.5,-1)--(6.5,-1)--(6.5,-1.5)--cycle;
\filldraw[fill=blue!20, draw=blue, thick] (8,-1.5)--(8,-1)--(10,-1)--(10,-1.5)--cycle;
\path (5.5,-1.25) node [style=sergio] {\scriptsize$\tau_2$};
\path (9,-1.25) node [style=sergio] {\scriptsize$\tau_2$};
\end{tikzpicture}
\caption{The cell decomposition of 3D $C_{4h}$-symmetric lattice. Left panel depicts the north/south hemisphere, including 3D blocks $\lambda$, 2D blocks $\sigma_1$ and 1D blocks $\tau_1$; right panel depicts the equator, including 2D blocks $\sigma_2$, 1D blocks $\tau_2$ and 0D block $\mu$. $C_4$ depicts the axis of 4-fold rotation, and $M$ depicts the reflection plane.}
\label{C4h cell decomposition}
\end{figure}

We summarize the classifications and corresponding root phases. For crystalline TSC with spinless fermions, we summarize all possible block states as:
\begin{enumerate}[1.]
\item 2D blocks $\sigma_1$: 2D $(p+ip)$-SCs;
\item 2D blocks $\sigma_2$: 2D $(p+ip)$-SCs and 2D $\mathbb{Z}_2$ fSPT phases;
\item 1D blocks $\tau_1$: Majorana chain and 1D $\mathbb{Z}_4$ fSPT phases;
\item 1D blocks $\tau_2$: Majorana chain and 1D $\mathbb{Z}_2$ fSPT phases;
\item 0D block $\mu$: 0D modes with odd fermion parity, eigenvalues of two generators of the $C_{4h}$ group.
\end{enumerate}
2D $(p+ip)$-SCs decoration on $\sigma_1$ is obstructed because it leaves 4 chiral Majorana modes on each $\tau_1$ with chiral central charge $c_-=2$ which cannot be gapped out; 2D $(p+ip)$-SCs decoration on $\sigma_2$ is also obstructed because for spinless fermions, 2D $(p+ip)$-SC is not compatible with the 4-fold rotation \cite{rotation}. 

Majorana chain decoration on $\tau_1$ leaves 2 Majorana zero modes $\gamma_1$ and $\gamma_2$, with the following reflection property:
\begin{align}
\bs{M}_z\in C_{4h}:~\gamma_1\leftrightarrow\gamma_2
\end{align}
The fermion parity defined from these two Majorana zero modes $P_f=i\gamma_1\gamma_2$ is odd under $\bs{M}_z$, hence they cannot be gapped out. Similar for Majorana chain decoration on $\tau_2$. For 1D $\mathbb{Z}_4$ fSPT phase decoration on each $\tau_1$, it leaves 8 Majorana zero modes at $\mu$, which form a projective representation of the $C_{4h}$ group. As a matter of fact, these Majorana zero modes cannot be gapped out, and the corresponding 1D block state is obstructed. Similarly, 1D $\mathbb{Z}_2$ fSPT phase decoration on each $\tau_2$ is also obstructed. We summarize all obstruction-free block states:
\begin{enumerate}[1.]
\item 2D $\mathbb{Z}_2$ fSPT phase decoration on each $\sigma_2$ ($\mathbb{Z}_8$);
\item 0D modes at $\mu$ with odd fermion parity, eigenvalues of two generators of the $C_{4h}$ group ($\mathbb{Z}_4\times\mathbb{Z}_2^2$).
\end{enumerate}
If we consider a complex fermion on each $\tau_1$ which can be adiabatically deformed to infinite far and trivialized, it forms an atomic insulator $c_1^\dag c_2^\dag|0\rangle$ with two complex fermions, with eigenvalue $-1$ of $\bs{M}_z$. Equivalently, the 0D mode characterizing the eigenvalue $-1$ of $\bs{M}_z$ is trivialized. Similarly, the eigenvalue $-1$ of the 4-fold rotation is also trivialized. Furthermore, odd fermion parity of $\mu$ is trivialized by ``Majorana'' bubble construction on $\sigma_2$ \cite{rotation}. Therefore, the ultimate classification is $\mathbb{Z}_8\times\mathbb{Z}_2$, with the following root phases:
\begin{enumerate}[1.]
\item 2D fermionic Levin-Gu state with the index $\nu=1\in\mathbb{Z}_8$ on each $\sigma_2$ ($\mathbb{Z}_8$);
\item 0D mode with eigenvalue $\pm i$ of 4-fold rotation on $\mu$ ($\mathbb{Z}_2$).
\end{enumerate}
The second-order topological surface theory of 2D root phase is 1D nonchiral Luttinger liquid with $K$-matrix $K=\sigma^z$ and $\mathbb{Z}_2$ symmetry property $W^{\mathbb{Z}_2}=\sigma^z$ and $\delta\phi^{\mathbb{Z}_2}=0$, on the edge of equator. 

For crystalline TSC with spin-1/2 fermions, all possible block states are summarized as follows:
\begin{enumerate}[1.]
\item 2D blocks $\sigma_1$: 2D $(p+ip)$-SCs;
\item 0D block $\mu$: 0D modes with odd fermion parity and characterizing the eigenvalues of two generators of the $C_{4h}$ group.
\end{enumerate}
Similar to the $C_{2h}$ case, the ultimate classification is $\Z_2$, and the nontrivial block state is the 0D mode with odd fermion parity on $\mu$.

For crystalline TI with spinless fermions, we summarize all possible block states as follows:
\begin{enumerate}[1.]
\item 2D blocks $\sigma_1$: Chern insulators and Kitaev's $E_8$ states;
\item 2D blocks $\sigma_2$: Chern insulators, Kitaev's $E_8$ states and 2D $U^f(1)\times\mathbb{Z}_2$ fSPT phases;
\item 0D block $\mu$: 0D modes characterizing the eigenvalues of two generators of the $C_{4h}$ group, with different $U^f(1)$ charge. 
\end{enumerate}
Similar to the crystalline TSC, all 2D block states on $\sigma_1$ are obstructed because they are chiral and leaves chiral 1D mode on each $\tau_1$ which cannot be gapped out. All obstruction-free block states are summarized as:
\begin{enumerate}[1.]
\item Chern insulator on each $\sigma_2$ ($\mathbb{Z}$);
\item Kitaev's $E_8$ state on each $\sigma_2$ ($\mathbb{Z}$);
\item 2D $U^f(1)\times\mathbb{Z}_2$ fSPT phase on each $\sigma_2$ ($\mathbb{Z}_4$);
\item 0D modes at $\mu$, characterizing the eigenvalues of two generators of the $C_{4h}$ group, with different $U^f(1)$ charge ($\mathbb{Z}\times\mathbb{Z}_4\times\mathbb{Z}_2$). 
\end{enumerate}
It is easy to see that 3D bubble equivalences on $\lambda$ will changes the layers of invertible topological phases on each $\sigma_2$ by two. And similar to the crystalline TSC, a 0D fermionic mode with $U^f(1)$ charge $+1$ at each $\tau_1$ changes the eigenvalue of $\bs{M}_z$ by $-1$ and the $U^f(1)$ charge by 2 at $\mu$, and a 0D fermionic mode with $U^f(1)$ charge $+1$ at each $\tau_2$ changes the eigenvalue of 4-fold rotation by $-1$ and the $U^f(1)$ charge by 4 at $\mu$. Hence the ultimate classification is $\mathbb{Z}_8\times\mathbb{Z}_4\times\mathbb{Z}_2^2$, with the following root phases:
\begin{enumerate}[1.]
\item Monolayer Chern insulator on each $\sigma_2$ ($\mathbb{Z}_2$);
\item 2D $U^f(1)\times\mathbb{Z}_2$ fSPT phase with $\nu=1\in\mathbb{Z}_4$ on each $\sigma_2$ ($\mathbb{Z}_4$);
\item Monolayer Kitaev's $E_8$ state on each $\sigma_2$ ($\mathbb{Z}_2$);
\item 0D mode with eigenvalue $\pm i$ of 4-fold rotation on $\mu$ ($\mathbb{Z}_2$);
\item 0D mode with odd $U^f(1)$ charge on $\mu$ ($\mathbb{Z}_2$);
\item 0D mode with eigenvalue $-1$ of reflection on $\mu$ ($\mathbb{Z}_2$).
\end{enumerate}
There is a nontrivial extension between the first two root phases and another nontrivial extension between the last two root phases. The second-order topological surface theories of 2D root phases are:
\begin{enumerate}[1.]
\item 1D chiral fermions on the edge of equator;
\item 1D nonchiral Luttinger liquid with $K$-matrix $K=\sigma^z$ and $\mathbb{Z}_2$ symmetry property $W^{\mathbb{Z}_2}=\mathbbm{1}_{2\times2}$ and $\delta\phi^{\mathbb{Z}_2}=\pi(0,1)^T$ on the edge of equator;
\item 1D chiral Luttinger liquid with $K$-matrix (\ref{K-matrix E8S}) on the edge of equator.
\end{enumerate}

For crystalline TI with spin-1/2 fermions, we summarize all possible block states as:
\begin{enumerate}[1.]
\item 2D blocks $\sigma_1$: Chern insulators and Kitaev's $E_8$ states;
\item 2D blocks $\sigma_2$: Chern insulators, Kitaev's $E_8$ states and 2D $U^f(1)\times\mathbb{Z}_2$ fSPT phases;
\item 0D block $\mu$: 0D modes characterizing the eigenvalues of two generators of the $C_{4h}$ group, with different even $U^f(1)$ charge. 
\end{enumerate}
Similar to the spinless fermions, all 2D block states on $\sigma_1$ are obstructed. All obstruction-free block states are summarized as:
\begin{enumerate}[1.]
\item Chern insulator on each $\sigma_2$ ($\mathbb{Z}$);
\item Kitaev's $E_8$ state on each $\sigma_2$ ($\mathbb{Z}$);
\item 2D $U^f(1)\times\mathbb{Z}_2$ fSPT phase on each $\sigma_2$ ($\mathbb{Z}_4$);
\item 0D modes at $\mu$, characterizing the eigenvalues of two generators of the $C_{4h}$ group, with different $U^f(1)$ charge ($\mathbb{Z}\times\mathbb{Z}_4\times\mathbb{Z}_2$).  
\end{enumerate}
Similar to the spinless fermions, eigenvalues $-1$ of two generators of the $C_{4h}$ group and even $U^f(1)$ charge are trivialized. Therefore, the ultimate classification is $\mathbb{Z}_8\times\mathbb{Z}_4\times\mathbb{Z}_2^2$, with the following root phases:
\begin{enumerate}[1.]
\item Monolayer Chern insulator on each $\sigma_2$ ($\mathbb{Z}_2$);
\item 2D $U^f(1)\times\mathbb{Z}_2$ fSPT phase with $\nu=1\in\mathbb{Z}_4$ on each $\sigma_2$ ($\mathbb{Z}_4$);
\item Monolayer Kitaev's $E_8$ state on each $\sigma_2$ ($\mathbb{Z}_2$);
\item 0D mode with eigenvalue $\pm i$ of 4-fold rotation on $\mu$ ($\mathbb{Z}_2$);
\item 0D mode with odd $U^f(1)$ charge on $\mu$ ($\mathbb{Z}_2$);
\item 0D mode with eigenvalue $-1$ of reflection on $\mu$ ($\mathbb{Z}_2$).
\end{enumerate}
The HO topological surface theories are identical to the systems with spinless fermions. 

\begin{figure}
\begin{tikzpicture}[scale=0.65]
\tikzstyle{sergio}=[rectangle,draw=none]
\filldraw[fill=green!20, draw=green, thick] (7,2)--(5.5,2)--(7,0.5)--cycle;
\filldraw[fill=green!20, draw=green, thick] (4.5,-0.5)--(4.5,1)--(6,-0.5)--cycle;
\filldraw[fill=green!20, draw=green, thick] (8,2)--(9.5,2)--(8,0.5)--cycle;
\filldraw[fill=green!20, draw=green, thick] (10.5,-0.5)--(10.5,1)--(9,-0.5)--cycle;
\filldraw[fill=blue!20, draw=blue, thick] (7.75,2)--(7.25,2)--(7.25,0.5)--(7.75,0.5)--cycle;
\filldraw[fill=blue!20, draw=blue, thick] (7.75,-2.5)--(7.25,-2.5)--(7.25,-4)--(7.75,-4)--cycle;
\path (6.5,1.5) node [style=sergio] {\scriptsize$\sigma_2$};
\path (5,0) node [style=sergio] {\scriptsize$\sigma_2$};
\path (10,0) node [style=sergio] {\scriptsize$\sigma_2$};
\path (8.5,1.5) node [style=sergio] {\scriptsize$\sigma_2$};
\path (7.5,1.25) node [style=sergio] {\scriptsize$\tau_2$};
\path (7.5,-3.25) node [style=sergio] {\scriptsize$\tau_2$};
\filldraw[fill=white, draw=black] (7.5,-1)circle (13pt);
\path (7.5,-1) node [style=sergio] {\scriptsize$\mu$};
\filldraw[fill=red!20, draw=red, thick] (1.25,1.75)--(3.25,1.75)--(3.25,-0.25)--(1.25,-0.25)--cycle;
\filldraw[fill=red!20, draw=red, thick] (-2.25,1.75)--(-0.25,1.75)--(-0.25,-0.25)--(-2.25,-0.25)--cycle;
\filldraw[fill=green!20, draw=green, thick] (0.25,1.75)--(0.75,1.75)--(0.75,-0.25)--(0.25,-0.25)--cycle;
\filldraw[fill=green!20, draw=green, thick] (0.25,-1.75)--(0.75,-1.75)--(0.75,-3.75)--(0.25,-3.75)--cycle;
\filldraw[fill=blue!20, draw=blue, thick] (0.25,-0.75)--(0.75,-0.75)--(0.75,-1.25)--(0.25,-1.25)--cycle;
\path (-1.25,0.75) node [style=sergio] {\scriptsize$\lambda$};
\path (2.25,0.75) node [style=sergio] {\scriptsize$\lambda$};
\path (0.5,0.75) node [style=sergio] {\scriptsize$\sigma_1$};
\path (0.5,-2.75) node [style=sergio] {\scriptsize$\sigma_1$};
\path (0.5,-1) node [style=sergio] {\scriptsize$\tau_1$};
\path (0.5,-4.75) node [style=sergio] {North/south hemisphere};
\path (7.75,-4.75) node [style=sergio] {Equator};
\filldraw[fill=red!20, draw=red, thick] (-2.25,-1.75)--(-0.25,-1.75)--(-0.25,-3.75)--(-2.25,-3.75)--cycle;
\path (-1.25,-2.75) node [style=sergio] {\scriptsize$\lambda$};
\filldraw[fill=red!20, draw=red, thick] (1.25,-1.75)--(3.25,-1.75)--(3.25,-3.75)--(1.25,-3.75)--cycle;
\path (2.25,-2.75) node [style=sergio] {\scriptsize$\lambda$};
\filldraw[fill=green!20, draw=green, thick] (-2.25,-0.75)--(-2.25,-1.25)--(-0.25,-1.25)--(-0.25,-0.75)--cycle;
\filldraw[fill=green!20, draw=green, thick] (1.25,-0.75)--(1.25,-1.25)--(3.25,-1.25)--(3.25,-0.75)--cycle;
\path (2.25,-1) node [style=sergio] {\scriptsize$\sigma_1$};
\path (-1.25,-1) node [style=sergio] {\scriptsize$\sigma_1$};
\filldraw[fill=green!20, draw=green, thick] (8,-2.5)--(9.5,-4)--(8,-4)--cycle;
\filldraw[fill=green!20, draw=green, thick] (4.5,-1.5)--(6,-1.5)--(4.5,-3)--cycle;
\path (5,-2) node [style=sergio] {\scriptsize$\sigma_2$};
\path (8.5,-3.5) node [style=sergio] {\scriptsize$\sigma_2$};
\filldraw[fill=blue!20, draw=blue, thick] (4.5,-1.25)--(4.5,-0.75)--(6,-0.75)--(6,-1.25)--cycle;
\filldraw[fill=blue!20, draw=blue, thick] (9,-1.25)--(9,-0.75)--(10.5,-0.75)--(10.5,-1.25)--cycle;
\path (5.25,-1) node [style=sergio] {\scriptsize$\tau_2$};
\path (9.75,-1) node [style=sergio] {\scriptsize$\tau_2$};
\filldraw[fill=blue!20, draw=blue, thick] (4.75,1.25)--(5.25,1.75)--(6.75,0.25)--(6.25,-0.25)--cycle;
\filldraw[fill=blue!20, draw=blue, thick] (10.25,1.25)--(9.75,1.75)--(8.25,0.25)--(8.75,-0.25)--cycle;
\filldraw[fill=blue!20, draw=blue, thick] (5.25,-3.75)--(4.75,-3.25)--(6.25,-1.75)--(6.75,-2.25)--cycle;
\filldraw[fill=blue!20, draw=blue, thick] (9.75,-3.75)--(10.25,-3.25)--(8.75,-1.75)--(8.25,-2.25)--cycle;
\filldraw[fill=green!20, draw=green, thick] (7,-4)--(7,-2.5)--(5.5,-4)--cycle;
\filldraw[fill=green!20, draw=green, thick] (9,-1.5)--(10.5,-3)--(10.5,-1.5)--cycle;
\path (6.5,-3.5) node [style=sergio] {\scriptsize$\sigma_2$};
\path (10,-2) node [style=sergio] {\scriptsize$\sigma_2$};
\path (1,-0.5) node [style=sergio] {$C_4$};
\path (7.5,0) node [style=sergio] {$C_2$};
\path (6.75,-1.75) node [style=sergio] {$C_2$};
\path (5.75,0.75) node [style=sergio] {\scriptsize$\tau_3$};
\path (9.25,0.75) node [style=sergio] {\scriptsize$\tau_3$};
\path (9.25,-2.75) node [style=sergio] {\scriptsize$\tau_3$};
\path (5.75,-2.75) node [style=sergio] {\scriptsize$\tau_3$};
\end{tikzpicture}
\caption{The cell decomposition of $D_4$-symmetric lattice. Left panel depicts the north/south hemisphere, including 3D blocks $\lambda$, 2D blocks $\sigma_1$ and 1D blocks $\tau_1$; right panel depicts the equator, including 2D blocks $\sigma_2$, 1D blocks $\tau_{2,3}$, and 0D block $\mu$. $C_4$ depicts the axis of 4-fold rotation, and $C_2$'s depict the axes of 2-fold rotations.}
\label{D4 cell decomposition}
\end{figure}

\subsection{$D_4$-symmetric lattice}
For $D_4$-symmetric lattice with the cell decomposition in Fig. \ref{D4 cell decomposition}, the ground-state wavefunction of the system can be decomposed to the direct products of wavefunctions of lower-dimensional block states as:
\begin{align}
|\Psi\rangle=\bigotimes\limits_{g\in D_4}|T_{g\lambda}\rangle\otimes\sum\limits_{k=1}^2|T_{g\sigma_k}\rangle\otimes\sum\limits_{j=1}^3|\beta_{g\tau_j}\rangle\otimes|\alpha_\mu\rangle
\label{D4 wavefunction}
\end{align}
where $|T_{g\lambda}\rangle/|T_{g\sigma_{1,2}}\rangle$ is the wavefunction of 3D/2D block state on $g\lambda/g\sigma_{1,2}$ which is topological trivial or invertible topological phase; $|\beta_{g\tau_1}\rangle$ is the $\mathbb{Z}_4$-symmetric wavefunction of 1D block state on $g\tau_1$, and $|\beta_{g\tau_{2,3}}\rangle$ is the $\mathbb{Z}_2$-symmetric wavefunction of 1D block state on $g\tau_{2,3}$; $|\alpha_\mu\rangle$ is the $(\mathbb{Z}_4\rtimes\mathbb{Z}_2)$-symmetric wavefunction of 0D block state on $\mu$.

We summarize the classifications and corresponding root phases. For crystalline TSC with spinless fermions, we summarize all possible root phases as:
\begin{enumerate}[1.]
\item 2D blocks $\sigma_1$ and $\sigma_2$: 2D $(p+ip)$-SCs;
\item 1D blocks $\tau_1$: Majorana chain and 1D $\mathbb{Z}_4$ fSPT phases;
\item 1D blocks $\tau_2$ and $\tau_3$: Majorana chain and 1D $\mathbb{Z}_2$ fSPT phases;
\item 0D block $\mu$: 0D modes with odd fermion parity, characterizing the eigenvalues of two generators of the $D_4$ group. 
\end{enumerate}
All 2D block states are obstructed because they leave chiral 1D modes on the 1D blocks as their shared border. Similar to the $C_{4h}$-symmetric case, Majorana chain decoration and 1D $\mathbb{Z}_4$ fSPT phase are obstructed. 

Then we focus on the equator, which is identical with 2D $D_4$-symmetric case. From Ref. \cite{dihedral} we know that the only obstruction and trivialization free block state is 1D $\mathbb{Z}_2$ fSPT phase decoration on $\tau_2$ or $\tau_3$. Hence the ultimate classification is $\mathbb{Z}_2$, with the root phase: 1D $\mathbb{Z}_2$ fSPT state (formed by double Majorana chains) on each $\tau_2$ or $\tau_3$. Its third-order topological surface theory is two dangling Majorana zero modes at the center of each vertical hinge of the open lattice. 

For crystalline TSC with spin-1/2 fermions, we summarize all possible block states as:
\begin{enumerate}[1.]
\item 2D blocks $\sigma_1$ and $\sigma_2$: 2D $(p+ip)$-SCs;
\item 0D block $\mu$: 0D modes characterizing the eigenvalues of two generators of the $D_4$ group. 
\end{enumerate}
Similar to the spinless fermions, all 2D block states are obstructed. And there is no more trivialization, hence the ultimate classification is $\mathbb{Z}_2^2$, the root phases are 0D modes with eigenvalues $-1$ of two rotation generators of $D_4$ group at $\mu$, and there is no HO topological surface theory.

For crystalline TI with spinless fermions, all possible block states are summarized as following:
\begin{enumerate}[1.]
\item 2D blocks $\sigma_1$ and $\sigma_2$: Chern insulators and Kitaev's $E_8$ states;
\item 0D block $\mu$: 0D modes characterizing the eigenvalues of two generators of the $D_4$ group, with different $U^f(1)$ charge.
\end{enumerate}
Similar to the crystalline TSC, all 2D block states are obstructed because they leave chiral modes on 1D blocks as their shared border. Furthermore, similar to the $D_2$-symmetric case, all possible 0D block states are trivialized by 1D bubble equivalences. Hence the ultimate classification is trivial. 

For crystalline TI with spin-1/2 fermions, all possible block states are summarized as following:
\begin{enumerate}[1.]
\item 2D blocks $\sigma_1$ and $\sigma_2$: Chern insulators and Kitaev's $E_8$ states;
\item 0D block $\mu$: 0D modes characterizing the eigenvalues of two generators of the $D_4$ group, with different even $U^f(1)$ charge.
\end{enumerate}
All possible 2D blocks-states are obstructed, and 0D modes with nonvanishing $U^f(1)$ charges are trivialized. Therefore, the ultimate classification is $\mathbb{Z}_2^2$, the root phases are 0D modes with eigenvalues $-1$ of two rotation generators of $D_4$ group at $\mu$, and there is no HO topological surface theory for both cases.

\subsection{$C_{4v}$-symmetric lattice}
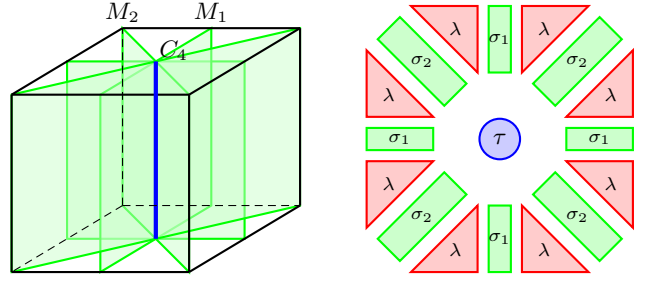
\begin{figure}
\begin{tikzpicture}[scale=0.59]
\tikzstyle{sergio}=[rectangle,draw=none]
\filldraw[fill=green!20, draw=green, thick,fill opacity=0.5] (-1.75,0.75)--(-1.75,-3.25)--(2.25,-3.25)--(2.25,0.75)--cycle;
\filldraw[fill=green!20, draw=green, thick,fill opacity=0.5] (-1,0)--(1.5,1.5)--(1.5,-2.5)--(-1,-4)--cycle;
\filldraw[fill=green!20, draw=green, thick,fill opacity=0.5] (1,0)--(-0.5,1.5)--(-0.5,-2.5)--(1,-4)--cycle;
\filldraw[fill=green!20, draw=green, thick,fill opacity=0.5] (-3,0)--(3.5,1.5)--(3.5,-2.5)--(-3,-4)--cycle;
\draw[thick] (-0.5,1.5) -- (3.5,1.5);
\draw[thick] (-0.5,1.5) -- (-3,0);
\draw[thick] (3.5,1.5) -- (1,0);
\draw[thick] (-3,0) -- (1,0);
\draw[thick] (-3,0) -- (-3,-4);
\draw[thick] (1,-4) -- (-3,-4);
\draw[thick] (1,-4) -- (1,0);
\draw[thick] (3.5,-2.5) -- (3.5,1.5);
\draw[thick] (3.5,-2.5) -- (1,-4);
\draw[draw=blue,ultra thick] (0.25,-3.25) -- (0.25,0.75);
\draw[densely dashed] (-0.5,-2.5) -- (-0.5,1.5);
\draw[densely dashed] (-0.5,-2.5) -- (3.5,-2.5);
\draw[densely dashed] (-0.5,-2.5) -- (-3,-4);
\filldraw[fill=red!20, draw=red, thick] (7.5,2)--(6,2)--(7.5,0.5)--cycle;
\filldraw[fill=red!20, draw=red, thick] (5,-0.5)--(5,1)--(6.5,-0.5)--cycle;
\filldraw[fill=red!20, draw=red, thick] (8.5,2)--(10,2)--(8.5,0.5)--cycle;
\filldraw[fill=red!20, draw=red, thick] (11,-0.5)--(11,1)--(9.5,-0.5)--cycle;
\filldraw[fill=green!20, draw=green, thick] (8.25,2)--(7.75,2)--(7.75,0.5)--(8.25,0.5)--cycle;
\filldraw[fill=green!20, draw=green, thick] (8.25,-2.5)--(7.75,-2.5)--(7.75,-4)--(8.25,-4)--cycle;
\path (7,1.5) node [style=sergio] {\scriptsize$\lambda$};
\path (5.5,0) node [style=sergio] {\scriptsize$\lambda$};
\path (10.5,0) node [style=sergio] {\scriptsize$\lambda$};
\path (9,1.5) node [style=sergio] {\scriptsize$\lambda$};
\path (8,1.25) node [style=sergio] {\scriptsize$\sigma_1$};
\path (8,-3.25) node [style=sergio] {\scriptsize$\sigma_1$};
\filldraw[fill=blue!20, draw=blue, thick] (8,-1)circle (13pt);
\path (8,-1) node [style=sergio] {$\tau$};
\filldraw[fill=red!20, draw=red, thick] (8.5,-2.5)--(10,-4)--(8.5,-4)--cycle;
\filldraw[fill=red!20, draw=red, thick] (5,-1.5)--(6.5,-1.5)--(5,-3)--cycle;
\path (5.5,-2) node [style=sergio] {\scriptsize$\lambda$};
\path (9,-3.5) node [style=sergio] {\scriptsize$\lambda$};
\filldraw[fill=green!20, draw=green, thick] (5,-1.25)--(5,-0.75)--(6.5,-0.75)--(6.5,-1.25)--cycle;
\filldraw[fill=green!20, draw=green, thick] (9.5,-1.25)--(9.5,-0.75)--(11,-0.75)--(11,-1.25)--cycle;
\path (5.75,-1) node [style=sergio] {\scriptsize$\sigma_1$};
\path (10.25,-1) node [style=sergio] {\scriptsize$\sigma_1$};
\filldraw[fill=green!20, draw=green, thick] (5.25,1.25)--(5.75,1.75)--(7.25,0.25)--(6.75,-0.25)--cycle;
\filldraw[fill=green!20, draw=green, thick] (10.75,1.25)--(10.25,1.75)--(8.75,0.25)--(9.25,-0.25)--cycle;
\filldraw[fill=green!20, draw=green, thick] (5.75,-3.75)--(5.25,-3.25)--(6.75,-1.75)--(7.25,-2.25)--cycle;
\filldraw[fill=green!20, draw=green, thick] (10.25,-3.75)--(10.75,-3.25)--(9.25,-1.75)--(8.75,-2.25)--cycle;
\filldraw[fill=red!20, draw=red, thick] (7.5,-4)--(7.5,-2.5)--(6,-4)--cycle;
\filldraw[fill=red!20, draw=red, thick] (9.5,-1.5)--(11,-3)--(11,-1.5)--cycle;
\path (7,-3.5) node [style=sergio] {\scriptsize$\lambda$};
\path (10.5,-2) node [style=sergio] {\scriptsize$\lambda$};
\path (6.25,0.75) node [style=sergio] {\scriptsize$\sigma_2$};
\path (9.75,0.75) node [style=sergio] {\scriptsize$\sigma_2$};
\path (9.75,-2.75) node [style=sergio] {\scriptsize$\sigma_2$};
\path (6.25,-2.75) node [style=sergio] {\scriptsize$\sigma_2$};
\path (0.65,1) node [style=sergio] {$C_4$};
\path (1.5,1.85) node [style=sergio] {$M_1$};
\path (-0.5,1.85) node [style=sergio] {$M_2$};
\end{tikzpicture}
\caption{The cell decomposition of $C_{4v}$-symmetric lattice. Left panel depicts the whole lattice, right panel depicts the horizontal intersection, including 3D blocks $\lambda$, 2D blocks $\sigma_{1,2}$ and 1D block $\tau$. $C_4$ depicts the axis of 4-fold rotation, and $M_{1,2}$ depict the reflection planes.}
\label{C4v cell decomposition}
\end{figure}

For $C_{4v}$-symmetric lattice with the cell decomposition in Fig. \ref{C4v cell decomposition}, the ground-state wavefunction of the system can be decomposed to the direct products of wavefunctions of lower-dimensional block states as:
\begin{align}
|\Psi\rangle=\bigotimes\limits_{g\in C_{4v}}|T_{g\lambda}\rangle\otimes\sum\limits_{j=1}^2|\gamma_{g\sigma_{j}}\rangle\otimes|\beta_\tau\rangle
\label{C4v wavefunction}
\end{align}
where $|T_{g\lambda}\rangle$ is the wavefunction of 3D block state on $g\lambda$ which is topological trivial; $|\gamma_{g\sigma_{1,2}}\rangle$ is the $\mathbb{Z}_2$-symmetric wavefunction of 2D block state on $g\sigma_{1,2}$; $|\beta_\tau\rangle$ is the $(\mathbb{Z}_4\rtimes\mathbb{Z}_2)$-symmetric wavefunction of 1D block state $\tau$.

We summarize the classifications and corresponding root phases. For crystalline TSC with spinless fermions, we summarize all possible block states as:
\begin{enumerate}[1.]
\item 2D blocks $\sigma_1$ and $\sigma_2$: 2D $(p+ip)$-SCs and 1D $\mathbb{Z}_2$ fSPT phases;
\item 1D block $\tau$: Majorana chain, 1D $\mathbb{Z}_4\rtimes\mathbb{Z}_2$ fSPT phases and Haldane chain.
\end{enumerate}
Similar to the $C_{2v}$-symmetric cases, all 2D block states except 2D bosonic Levin-Gu states on each $\sigma_1$ or $\sigma_2$ are obstructed, which can be checked by explicit $K$-matrix calculations. Then consider two root phases of 1D $\mathbb{Z}_4\rtimes\mathbb{Z}_2$ fSPT phases, they can be trivialized by 2D ``Majorana'' bubble constructions on $\sigma_1$ or $\sigma_2$. Therefore, the ultimate classification is $\mathbb{Z}_2^4$, with the following root phases:
\begin{enumerate}[1.]
\item 2D bosonic Levin-Gu state on each $\sigma_1/\sigma_2$ ($\mathbb{Z}_2^2$);
\item Majorana chain on $\tau$ ($\mathbb{Z}_2$);
\item Haldane chain on $\tau$ ($\mathbb{Z}_2$).
\end{enumerate}
The HO topological surface theories of different root phases are:
\begin{enumerate}[1.]
\item $2^{\mathrm{nd}}$-order: 1D nonchiral Luttinger liquids with $K$-matrix $K=\sigma^x$ and $\mathbb{Z}_2$ symmetry property $W^{\mathbb{Z}_2}=\mathbbm{1}_{2\times2}$ and $\delta\phi^{\mathbb{Z}_2}=\pi(1,1)^T$, on the intersections between the open lattice and 2D blocks $\sigma_1/\sigma_2$;
\item $3^{\mathrm{rd}}$-order: Dangling Majorana zero modes at the centers of top and bottom surfaces of the open lattice;
\item $3^{\mathrm{rd}}$-order: Spin-1/2 degrees of freedom at the centers of top and bottom surfaces of the open lattice.
\end{enumerate}

For crystalline TSC with spin-1/2 fermions, the corresponding classification is trivial because there is no nontrivial candidate block states on blocks with different dimensions.

\begin{figure*}
\begin{tikzpicture}[scale=0.75]
\tikzstyle{sergio}=[rectangle,draw=none]
\filldraw[fill=yellow!60, draw=yellow, thick,fill opacity=0.5] (-3.5,-2)--(-1,-0.5)--(3,-0.5)--(0.5,-2)--cycle;
\filldraw[fill=green!20, draw=green, thick,fill opacity=0.5] (0.5,0)--(-1,1.5)--(-1,-2.5)--(0.5,-4)--cycle;
\filldraw[fill=green!20, draw=green, thick,fill opacity=0.5] (-3.5,0)--(3,1.5)--(3,-2.5)--(-3.5,-4)--cycle;
\draw[thick] (-1,1.5) -- (3,1.5);
\draw[thick] (-1,1.5) -- (-3.5,0);
\draw[thick] (3,1.5) -- (0.5,0);
\draw[thick] (-3.5,0) -- (0.5,0);
\draw[thick] (-3.5,0) -- (-3.5,-4);
\draw[thick] (0.5,-4) -- (-3.5,-4);
\draw[thick] (0.5,-4) -- (0.5,0);
\draw[thick] (3,-2.5) -- (3,1.5);
\draw[thick] (3,-2.5) -- (0.5,-4);
\draw[draw=blue,ultra thick] (-0.25,-3.25) -- (-0.25,0.75);
\draw[draw=blue,ultra thick] (1,-0.5) -- (-1.5,-2);
\draw[draw=blue,ultra thick] (1.75,-1.25) -- (-2.25,-1.25);
\draw[draw=blue,ultra thick] (3,-0.5) -- (-3.5,-2);
\draw[draw=blue,ultra thick] (-1,-0.5) -- (0.5,-2);
\draw[densely dashed] (-1,-2.5) -- (-1,1.5);
\draw[densely dashed] (-1,-2.5) -- (3,-2.5);
\draw[densely dashed] (-1,-2.5) -- (-3.5,-4);
\filldraw[fill=green!20, draw=green, thick] (14.75,1.75)--(13.25,1.75)--(14.75,0.25)--cycle;
\filldraw[fill=green!20, draw=green, thick] (12.25,-0.75)--(12.25,0.75)--(13.75,-0.75)--cycle;
\filldraw[fill=green!20, draw=green, thick] (15.75,1.75)--(17.25,1.75)--(15.75,0.25)--cycle;
\filldraw[fill=green!20, draw=green, thick] (18.25,-0.75)--(18.25,0.75)--(16.75,-0.75)--cycle;
\filldraw[fill=blue!20, draw=blue, thick] (15.5,1.75)--(15,1.75)--(15,0.25)--(15.5,0.25)--cycle;
\filldraw[fill=blue!20, draw=blue, thick] (15.5,-2.75)--(15,-2.75)--(15,-4.25)--(15.5,-4.25)--cycle;
\path (14.25,1.25) node [style=sergio] {$\sigma_3$};
\path (12.75,-0.25) node [style=sergio] {$\sigma_3$};
\path (17.75,-0.25) node [style=sergio] {$\sigma_3$};
\path (16.25,1.25) node [style=sergio] {$\sigma_3$};
\path (15.25,1) node [style=sergio] {$\tau_2$};
\path (15.25,-3.5) node [style=sergio] {$\tau_2$};
\filldraw[fill=white!20, draw=black, thick] (15.25,-1.25)circle (13pt);
\path (15.25,-1.25) node [style=sergio] {$\mu$};
\filldraw[fill=green!20, draw=green, thick] (15.75,-2.75)--(17.25,-4.25)--(15.75,-4.25)--cycle;
\filldraw[fill=green!20, draw=green, thick] (12.25,-1.75)--(13.75,-1.75)--(12.25,-3.25)--cycle;
\path (12.75,-2.25) node [style=sergio] {$\sigma_3$};
\path (16.25,-3.75) node [style=sergio] {$\sigma_3$};
\filldraw[fill=blue!20, draw=blue, thick] (12.25,-1.5)--(12.25,-1)--(13.75,-1)--(13.75,-1.5)--cycle;
\filldraw[fill=blue!20, draw=blue, thick] (16.75,-1.5)--(16.75,-1)--(18.25,-1)--(18.25,-1.5)--cycle;
\path (13,-1.25) node [style=sergio] {$\tau_2$};
\path (17.5,-1.25) node [style=sergio] {$\tau_2$};
\filldraw[fill=blue!20, draw=blue, thick] (12.5,1)--(13,1.5)--(14.5,0)--(14,-0.5)--cycle;
\filldraw[fill=blue!20, draw=blue, thick] (18,1)--(17.5,1.5)--(16,0)--(16.5,-0.5)--cycle;
\filldraw[fill=blue!20, draw=blue, thick] (13,-4)--(12.5,-3.5)--(14,-2)--(14.5,-2.5)--cycle;
\filldraw[fill=blue!20, draw=blue, thick] (17.5,-4)--(18,-3.5)--(16.5,-2)--(16,-2.5)--cycle;
\filldraw[fill=green!20, draw=green, thick] (14.75,-4.25)--(14.75,-2.75)--(13.25,-4.25)--cycle;
\filldraw[fill=green!20, draw=green, thick] (16.75,-1.75)--(18.25,-3.25)--(18.25,-1.75)--cycle;
\path (14.25,-3.75) node [style=sergio] {$\sigma_3$};
\path (17.75,-2.25) node [style=sergio] {$\sigma_3$};
\path (13.5,0.5) node [style=sergio] {$\tau_3$};
\path (17,0.5) node [style=sergio] {$\tau_3$};
\path (17,-3) node [style=sergio] {$\tau_3$};
\path (13.5,-3) node [style=sergio] {$\tau_3$};
\filldraw[fill=red!20, draw=red, thick] (9.25,1.25)--(6.25,1.25)--(7.75,-0.25)--cycle;
\filldraw[fill=green!20, draw=green, thick] (5.5,0.5)--(6,1)--(7.5,-0.5)--(7,-1)--cycle;
\filldraw[fill=green!20, draw=green, thick] (10,0.5)--(9.5,1)--(8,-0.5)--(8.5,-1)--cycle;
\filldraw[fill=red!20, draw=red, thick] (5.25,-2.75)--(5.25,0.25)--(6.75,-1.25)--cycle;
\filldraw[fill=green!20, draw=green, thick] (6,-3.5)--(5.5,-3)--(7,-1.5)--(7.5,-2)--cycle;
\filldraw[fill=green!20, draw=green, thick] (9.5,-3.5)--(10,-3)--(8.5,-1.5)--(8,-2)--cycle;
\filldraw[fill=red!20, draw=red, thick] (9.25,-3.75)--(6.25,-3.75)--(7.75,-2.25)--cycle;
\filldraw[fill=red!20, draw=red, thick] (10.25,-2.75)--(10.25,0.25)--(8.75,-1.25)--cycle;
\filldraw[fill=blue!20, draw=blue, thick] (7.75,-1.25)circle (13pt);
\path (7.75,-1.25) node [style=sergio] {$\tau_1$};
\path (7.75,0.75) node [style=sergio] {$\lambda$};
\path (7.75,-3.25) node [style=sergio] {$\lambda$};
\path (9.75,-1.25) node [style=sergio] {$\lambda$};
\path (5.75,-1.25) node [style=sergio] {$\lambda$};
\path (6.5,0) node [style=sergio] {$\sigma_1$};
\path (9,-2.5) node [style=sergio] {$\sigma_1$};
\path (6.5,-2.5) node [style=sergio] {$\sigma_2$};
\path (9,0) node [style=sergio] {$\sigma_2$};
\path (0,1.1) node [style=sergio] {$S_4$};
\path (1,-0.2) node [style=sergio] {$C_2$};
\path (-2.5,-1) node [style=sergio] {$C_2$};
\path (-1,1.85) node [style=sergio] {$M_1$};
\path (3,1.85) node [style=sergio] {$M_2$};
\path (-0.25,-5) node [style=sergio] {Open lattice};
\path (7.75,-5) node [style=sergio] {North/south hemisphere};
\path (15.25,-5) node [style=sergio] {Equator};
\end{tikzpicture}
\caption{The cell decomposition of $D_{2d}$-symmetric lattice, with the yellow plate representing the plane of 4-fold improper rotation. Left panel depicts the whole lattice; middle panel depicts north/south hemisphere, including 3D blocks $\lambda$, 2D blocks $\sigma_{1,2}$ and 1D blocks $\tau_1$; right panel depicts the equator, including 2D blocks $\sigma_3$, 1D blocks $\tau_{2,3}$ and 0D block $\mu$. $S_4$ depicts the axis of the 4-fold rotoreflection, $C_2$'s depict the axes of the 2-fold rotations and $\bs{M}_{1,2}$ depict the reflection planes.}
\label{D2d cell decomposition}
\end{figure*}
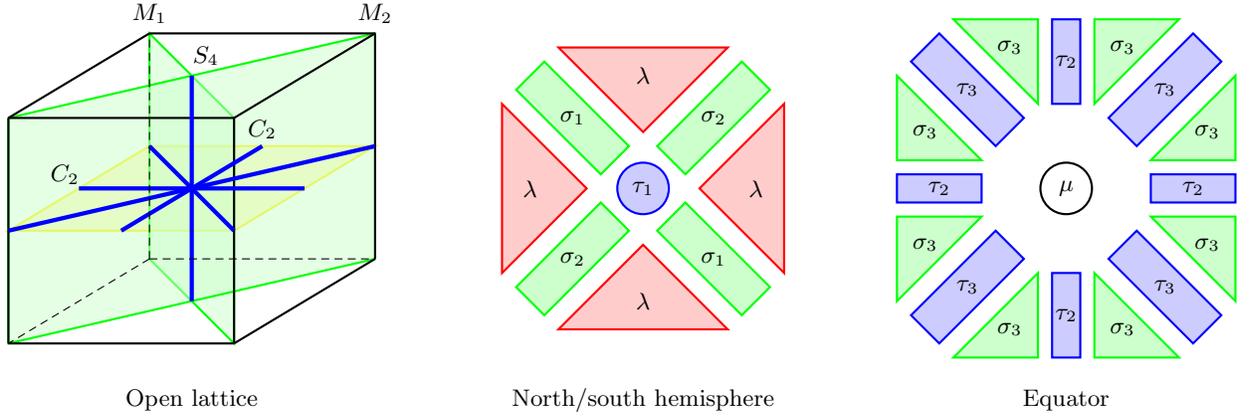

For crystalline TI with spinless fermions, we summarize all possible block states as:
\begin{enumerate}[1.]
\item 2D blocks $\sigma_1$ and $\sigma_2$: Chern insulators, Kitaev's $E_8$ states and 2D $U^f(1)\times\mathbb{Z}_2$ fSPT phases;
\item 1D block $\tau$: Haldane chain.
\end{enumerate}
By direct $K$-matrix calculations, all 2D block states except 2D $U^f(1)\times\mathbb{Z}_2$ bSPT phase on each $\sigma_1$ or $\sigma_2$ and Kitaev's $E_8$ states on $\sigma_1$ and $\sigma_2$ with opposite chiralities are all obstructed. Hence the ultimate classification is $\mathbb{Z}_2^4$, with the following root phases:
\begin{enumerate}[1.]
\item 2D $U^f(1)\times\mathbb{Z}_2$ bSPT phase on each $\sigma_1$ or $\sigma_2$ ($\mathbb{Z}_2^2$);
\item Monolayer Kitaev's $E_8$ state on each $\sigma_1$ and $\sigma_2$ ($\mathbb{Z}_2$);
\item Haldane chain decoration on $\tau$ ($\mathbb{Z}_2$).
\end{enumerate}
The HO topological surface theories of different root phases are:
\begin{enumerate}[1.]
\item $2^{\mathrm{nd}}$-order: 1D nonchiral Luttinger liquids with $K$-matrix $K=\sigma^x$ and $\mathbb{Z}_2$ symmetry property: $W^{\mathbb{Z}_2}=\mathbbm{1}_{2\times2}$ and $\delta\phi^{\mathbb{Z}_2}=\pi(1,1)^T$, on the intersections between the open lattice and 2D blocks $\sigma_1/\sigma_2$;
\item $2^{\mathrm{nd}}$-order: 1D chiral Luttinger liquids with $K$-matrix (\ref{K-matrix E8S}), on the intersections between the open lattice and 2D blocks $\sigma_1$ and $\sigma_2$;
\item $3^{\mathrm{rd}}$-order: dangling spin-1/2 degrees of freedom at the centers of top and bottom surfaces of the open lattice.
\end{enumerate}

For crystalline TI with spin-1/2 fermions, all possible block states are on 2D blocks: Chern insulators, Kitaev's $E_8$ states and 2D $U^f(1)\times\mathbb{Z}_2$ fSPT phases. We note that the chiralities of decorated chiral block states on $\sigma_1$ and $\sigma_2$ should be opposite to guarantee the nonchiral condition of 1D block $\tau$. Furthermore, the layers of chiral block states (Chern insulators and Kitaev's $E_8$ states) on $\sigma_1$ and $\sigma_2$ can be changed by 3D bubble equivalences on $\lambda$ by two. Hence the ultimate classification is $\mathbb{Z}_8\times\mathbb{Z}_4\times\mathbb{Z}_2$, with the following root phases:
\begin{enumerate}[1.]
\item Monolayer Chern insulator on each $\sigma_1$ and $\sigma_2$ ($\mathbb{Z}_2$);
\item 2D $U^f(1)\times\mathbb{Z}_2$ fSPT phase on each $\sigma_1/\sigma_2$ ($\mathbb{Z}_4^2$);
\item Monolayer Kitaev's $E_8$ state on each $\sigma_1$ and $\sigma_2$ ($\mathbb{Z}_2$);
\end{enumerate}
And there are several nontrivial extension between first two root phases: bilayer Chern insulators are equivalent to 2D $U^f(1)\times\mathbb{Z}_2$ fSPT phases with $(\nu_1,\nu_3)=(1,3)$ on each $\sigma_1$ and $\sigma_2$. The second-order topological surface theories of these root phases are:
\begin{enumerate}[1.]
\item Monolayer Chern insulator on each $\sigma_1$ and $\sigma_2$: chiral fermions on the intersections between the open lattice, $\sigma_1$ and $\sigma_2$;
\item 2D $U^f(1)\times\mathbb{Z}_2$ fSPT phase on each $\sigma_{1,2}$: 1D nonchiral Luttinger liquid with $K$-matrix $K=\sigma^z$ and $\mathbb{Z}_2$ symmetry property: $W^{\mathbb{Z}_2}=\mathbbm{1}_{2\times2}$ and $\delta\phi^{\mathbb{Z}_2}=\pi(0,1)^T/\pi(1,0)^T$ on the intersections between the open lattice and $\sigma_1/\sigma_2$;
\item Monolayer Kitaev's $E_8$ state on each $\sigma_1$ and $\sigma_2$: 1D chiral Luttinger liquids with $K$-matrix (\ref{K-matrix E8S}) on the intersections between the open lattice, $\sigma_1$ and $\sigma_2$.
\end{enumerate}

\subsection{$D_{2d}$-symmetric lattice}
For $D_{2d}$-symmetric lattice with the cell decomposition in Fig. \ref{D2d cell decomposition}, the ground-state wavefunction of the system can be decomposed to the direct products of wavefunctions of lower-dimensional block states as:
\begin{align}
|\Psi\rangle=\bigotimes\limits_{g\in D_{2d}}|T_{g\lambda}\rangle\otimes\sum\limits_{k=1}^3|\gamma_{g\sigma_k}\rangle\otimes\sum\limits_{j=1}^3|\beta_{g\tau_j}\rangle\otimes|\alpha_\mu\rangle
\label{D2d cell decomposition}
\end{align}
where $|T_{g\lambda}\rangle$ is the wavefunction of 3D block state on $g\lambda$ which is topological trivial; $|\gamma_{g\sigma_{1,2}}\rangle$ is the $\mathbb{Z}_2$-symmetric wavefunction of 2D block state on $g\sigma_{1,2}$, and $|\gamma_{g\sigma_3}\rangle$ is the wavefunction of 2D block state on $g\sigma_3$ which is topological trivial or invertible topological phase; $|\beta_{g\tau_1}\rangle$ is the $(\mathbb{Z}_2\times\mathbb{Z}_2)$-symmetric wavefunction of 1D block state on $g\tau_1$, and $|\beta_{g\tau_{2,3}}\rangle$ is the $\mathbb{Z}_2$-symmetric wavefunction of 1D block state on $g\tau_{2,3}$; $|\alpha_\mu\rangle$ is the $(\mathbb{Z}_4\rtimes\mathbb{Z}_2)$-symmetric wavefunction of 0D block state on $\mu$.

We summarize the classifications and corresponding root phases. For crystalline TSC with spinless fermions, we summarize all possible block states as:
\begin{enumerate}[1.]
\item 2D blocks $\sigma_1$ and $\sigma_2$: 2D $(p+ip)$-SCs and 2D $\mathbb{Z}_2$ fSPT phases; 
\item 2D blocks $\sigma_3$: 2D $(p+ip)$-SCs;
\item 1D blocks $\tau_1$: Majorana chain, 1D $\mathbb{Z}_2\times\mathbb{Z}_2$ fSPT phases and Haldane chain;
\item 1D blocks $\tau_2$ and $\tau_3$: Majorana chain and 1D $\mathbb{Z}_2$ fSPT phases;
\item 0D block $\mu$: 0D modes with odd fermion parity, characterizing the eigenvalues of two generators of the $D_{2d}$ group.
\end{enumerate}
By explicit $K$-matrix calculations, we see that all 2D block states except 2D bosonic Levin-Gu states on $\sigma_1$ or $\sigma_2$ are obstructed. 

Majorana chain decoration on $\tau_1$ leaves 2 Majorana zero modes $\gamma_1$ and $\gamma_2$ at $\mu$, with the following property under $\bs{I}_r\in S_4$:
\begin{align}
\bs{I}_r\in S_4:~\gamma_1\leftrightarrow\gamma_2
\end{align}
and their fermion parity $P_f=i\gamma_1\gamma_2$ is odd under $\bs{I}_r$. Reversely, the Majorana chain decoration on $\tau_1$ is obstructed. For 1D $\mathbb{Z}_2\times\mathbb{Z}_2$ fSPT phases decoration, it leaves 4 dangling Majorana zero modes $\xi_{1,2,3,4}$, with the following properties under $\bs{I}_r$:
\begin{align}
\bs{I}_r:~(\xi_1,\xi_2,\xi_3,\xi_4)\mapsto(\xi_2,\xi_3,\xi_4,\xi_1)
\end{align}
and their fermion parity $P_f'=-\xi_1\xi_2\xi_3\xi_4$ is odd under $\bs{I}_r$. Reversely, the 1D $\mathbb{Z}_2\times\mathbb{Z}_2$ fSPT phases decoration on $\tau_1$ are obstructed. 

Turn to the equator which is identical to the 2D $D_4$-symmetric case, from Ref. \cite{dihedral} we know that there is only one nontrivial 1D block state: 1D $\mathbb{Z}_2$ fSPT phases on $\tau_2$ or $\tau_3$. We further demonstrate that this phase is equivalent to the Haldane chain decoration on $\tau_1$: Consider 1D $\mathbb{Z}_2$ fSPT phase on each $\sigma_1$ as the 2D bubble, it will deform 1D $\mathbb{Z}_2$ fSPT phases on $\tau_3$ to the Haldane chain decoration on $\tau_1$. Moreover, from Ref. \cite{dihedral} we know that the 0D modes characterizing the eigenvalues $-1$ of two generators of the $D_{2d}$ group is trivialized by 1D bubble equivalences on $\tau_2$ and $\tau_3$. Therefore, the ultimate classification is $\mathbb{Z}_2^3$, with the following root phases:
\begin{enumerate}[1.]
\item 2D bosonic Levin-Gu state on each $\sigma_1$ and $\sigma_2$ ($\mathbb{Z}_2$);
\item 1D $\mathbb{Z}_2$ fSPT on each $\tau_2$ or $\tau_3$, which is equivalent to 1D Haldane chain on each $\tau_1$ ($\mathbb{Z}_2$);
\item 0D mode with odd fermion parity on $\mu$ ($\mathbb{Z}_2$).
\end{enumerate}
The HO topological surface theories of different root phases are:
\begin{enumerate}[1.]
\item $2^{\mathrm{nd}}$-order: 1D nonchiral Luttinger liquids with $K$-matrix $K=\sigma^x$ and $\mathbb{Z}_2$ symmetry property $W^{\mathbb{Z}_2}=\mathbbm{1}_{2\times2}$ and $\delta\phi^{\mathbb{Z}_2}=\pi(1,1)^T$, on the intersections between the open lattice and 2D blocks $\sigma_1/\sigma_2$;
\item $3^{\mathrm{rd}}$-order: Two Majorana zero modes at the center of each vertical hinge, which is equivalent to spin-1/2 degrees of freedom at the center of each horizontal surface of the open lattice.
\end{enumerate}

For crystalline TSC with spin-1/2 fermions, all possible block states are summarized as follows:
\begin{enumerate}[1.]
\item 2D block $\sigma_3$: 2D $(p+ip)$-SC;
\item 0D block $\mu$: 0D modes characterizing the eigenvalues of two generators of the $D_{2d}$ group.
\end{enumerate}
and similar to the spinless fermions, 2D $(p+ip)$-SC decoration on $\sigma_3$ is obstructed because it leaves chiral mode at each $\tau_2$ and $\tau_3$ with chiral central charge $c_-=1$ which cannot be gapped out. Therefore, the ultimate classification is $\mathbb{Z}_2$, with the root phase: 0D modes with eigenvalues $-1$ of the (improper) rotation generator of $D_{2d}$ group, which do not correspond to any HO topological surface theory. We note that an inversion operation is composed of a two-fold improper rotation and a two-fold rotation whose axis is on the equator, whose $-1$ eigenvalue can be trivialized by a complex fermion on each $\tau_1$.

For crystalline TI with spinless fermions, we summarize all possible block states as:
\begin{enumerate}[1.]
\item 2D blocks $\sigma_1$ and $\sigma_2$: Chern insulators, Kitaev's $E_8$ states and 2D $U^f(1)\times\mathbb{Z}_2$ fSPT phases;
\item 2D blocks $\sigma_3$: Chern insulators and Kitaev's $E_8$ states;
\item 1D block $\tau_1$: Haldane chain;
\item 0D block $\mu$: 0D modes characterizing eigenvalues of two generators of the $D_{2d}$ group, with different $U^f(1)$ charge. 
\end{enumerate}
From explicit $K$-matrix calculations, we conclude that except 2D $U^f(1)\times\mathbb{Z}_2$ bSPT phase decoration on each $\sigma_1$ or $\sigma_2$ and Kitaev'e $E_8$ states decoration on $\sigma_1$ and $\sigma_2$ with opposite chiralities, all other 2D block states are obstructed, and even layers of Kitaev's $E_8$ states can be trivialized by 3D bubble equivalence on each $\lambda$. Furthermore, by 1D bubble equivalences, 0D modes characterizing eigenvalues of two generators of the $D_{2d}$ group, with integer multiple of 4. Therefore, the ultimate classification is $\mathbb{Z}_4\times\mathbb{Z}_2^3$, with the following root phases:
\begin{enumerate}[1.]
\item Monolayer Kitaev's $E_8$ state on each $\sigma_1$ and $\sigma_2$ ($\mathbb{Z}_2$);
\item 2D $U^f(1)\times\mathbb{Z}_2$ bSPT phase on each $\sigma_1$ and $\sigma_2$ ($\mathbb{Z}_2$);
\item 1D Haldane chain on each $\tau_1$ ($\mathbb{Z}_2$);
\item 0D mode with $U^f(1)$ charge $n\equiv0,1,2,3~(\mathrm{mod}~4)$ on $\mu$ ($\mathbb{Z}_4$).
\end{enumerate}
The HO topological surface theories of different root phases are:
\begin{enumerate}[1.]
\item $2^{\mathrm{nd}}$-order: 1D chiral Luttinger liquids with $K$-matrix (\ref{K-matrix E8S}), on the intersections between the open lattice and 2D blocks $\sigma_1$ and $\sigma_2$;
\item $2^{\mathrm{nd}}$-order: 1D nonchiral Luttinger liquids with $K$-matrix $K=\sigma^x$ and $\mathbb{Z}_2$ symmetry property $W^{\mathbb{Z}_2}=\mathbbm{1}_{2\times2}$ and $\delta\phi^{\mathbb{Z}_2}=\pi(1,1)^T$, on the intersections between the open lattice and 2D blocks $\sigma_1/\sigma_2$;
\item $3^{\mathrm{rd}}$-order: A spin-1/2 degree of freedom at the center of each horizontal surface of the open lattice.
\end{enumerate}

For crystalline TI with spin-1/2 fermions, all possible block states are summarized as:
\begin{enumerate}[1.]
\item 2D blocks $\sigma_1$ and $\sigma_2$: Chern insulators, Kitaev's $E_8$ states and 2D $U^f(1)\times\mathbb{Z}_2$ fSPT phases;
\item 2D blocks $\sigma_3$: Chern insulators and Kitaev's $E_8$ states;
\item 0D block $\mu$: 0D modes characterizing eigenvalues of two generators of the $D_{2d}$ group, with different even $U^f(1)$ charge. 
\end{enumerate}
The chiral 2D block states should satisfy that they leave nonchiral 1D modes at their shared borders, hence all 2D block states on $\sigma_3$ are obstructed, and the chiralities of 2D chiral block states on $\sigma_1$ and $\sigma_2$ should be opposite. Furthermore, even layers of chiral 2D block states are trivialized by 3D bubble equivalences on $\lambda$, and the integer multiple of 4 $U^f(1)$ charges at $\mu$ are trivialized by 1D bubble equivalences on $\tau_{1,2,3}$. Therefore, the ultimate classification is $\mathbb{Z}_8\times\mathbb{Z}_2^5$, with the corresponding root phases:
\begin{enumerate}[1.]
\item Monolayer Chern insulator on each $\sigma_1$ and $\sigma_2$ ($\mathbb{Z}_2$);
\item 2D $U^f(1)\times\mathbb{Z}_2$ fSPT phase with on each $\sigma_1$ and $\sigma_2$, with quantum number $(\nu_1,\nu_2)=(1,3)$ ($\mathbb{Z}_4$);
\item 2D $U^f(1)\times\mathbb{Z}_2$ bSPT phase with solely on each $\sigma_1$ and $\sigma_2$ ($\mathbb{Z}_2$);
\item Monolayer Kitaev's $E_8$ state on each $\sigma_1$ and $\sigma_2$ ($\mathbb{Z}_2$);
\item 0D mode with $U^f(1)$ charge $n\equiv2(\mathrm{mod}~4)$ on $\mu$ ($\mathbb{Z}_2$);
\item 0D mode with eigenvalues $-1$ of two generators of $D_{2d}$ on $\mu$ ($\mathbb{Z}_2^2$).
\end{enumerate}
and there is a nontrivial extension between the first two root phases: bilayer Chern insulators on each 2D block $\sigma_1$ and $\sigma_2$ can be smoothly deformed to 2D $U^f(1)\times\mathbb{Z}_2$ fSPT phases on $\sigma_1$ and $\sigma_2$ with quantum numbers $(\nu_1,\nu_2)=(1,3)$, by 3D ``Chern insulator'' bubble equivalence. The second-order topological surface theories of different root phases are:
\begin{enumerate}[1.]
\item 1D chiral Luttinger liquids with $K$-matrix (\ref{K-matrix E8S}), on the intersections between the open lattice and 2D blocks $\sigma_1$ and $\sigma_2$;
\item Chiral fermions on the intersections of the open lattice and $\sigma_1$ and $\sigma_2$;
\item 1D nonchiral Luttinger liquids with $K$-matrix $K=\sigma^z$ and $\mathbb{Z}_2$ symmetry property $W^{\mathbb{Z}_2}=\mathbbm{1}_{2\times2}$ and $\delta\phi^{\mathbb{Z}_2}=\pi(0,1)^T$, on the intersections of the open lattice and $\sigma_1$ and $\sigma_2$.
\end{enumerate}

\subsection{$D_{4h}$-symmetric lattice}

\begin{figure*}
\begin{tikzpicture}[scale=0.75]
\tikzstyle{sergio}=[rectangle,draw=none]
\filldraw[fill=green!20, draw=green, thick,fill opacity=0.5] (-1.75,0.75)--(-1.75,-3.25)--(2.25,-3.25)--(2.25,0.75)--cycle;
\filldraw[fill=green!20, draw=green, thick,fill opacity=0.5] (-1,0)--(1.5,1.5)--(1.5,-2.5)--(-1,-4)--cycle;
\filldraw[fill=green!20, draw=green, thick,fill opacity=0.5] (1,0)--(-0.5,1.5)--(-0.5,-2.5)--(1,-4)--cycle;
\filldraw[fill=green!20, draw=green, thick,fill opacity=0.5] (-3,0)--(3.5,1.5)--(3.5,-2.5)--(-3,-4)--cycle;
\filldraw[fill=green!20, draw=green, thick,fill opacity=0.5] (-3,-2)--(-0.5,-0.5)--(3.5,-0.5)--(1,-2)--cycle;
\draw[thick] (-0.5,1.5) -- (3.5,1.5);
\draw[thick] (-0.5,1.5) -- (-3,0);
\draw[thick] (3.5,1.5) -- (1,0);
\draw[thick] (-3,0) -- (1,0);
\draw[thick] (-3,0) -- (-3,-4);
\draw[thick] (1,-4) -- (-3,-4);
\draw[thick] (1,-4) -- (1,0);
\draw[thick] (3.5,-2.5) -- (3.5,1.5);
\draw[thick] (3.5,-2.5) -- (1,-4);
\draw[draw=blue,ultra thick] (0.25,-3.25) -- (0.25,0.75);
\draw[draw=blue,ultra thick] (-1.75,-1.25) -- (2.25,-1.25);
\draw[draw=blue,ultra thick] (-1,-2) -- (1.5,-0.5);
\draw[draw=blue,ultra thick] (3.5,-0.5) -- (-3,-2);
\draw[draw=blue,ultra thick] (1,-2) -- (-0.5,-0.5);
\draw[densely dashed] (-0.5,-2.5) -- (-0.5,1.5);
\draw[densely dashed] (-0.5,-2.5) -- (3.5,-2.5);
\draw[densely dashed] (-0.5,-2.5) -- (-3,-4);
\filldraw[fill=green!20, draw=green, thick] (15,2)--(13.5,2)--(15,0.5)--cycle;
\filldraw[fill=green!20, draw=green, thick] (12.5,-0.5)--(12.5,1)--(14,-0.5)--cycle;
\filldraw[fill=green!20, draw=green, thick] (16,2)--(17.5,2)--(16,0.5)--cycle;
\filldraw[fill=green!20, draw=green, thick] (18.5,-0.5)--(18.5,1)--(17,-0.5)--cycle;
\filldraw[fill=blue!20, draw=blue, thick] (15.75,2)--(15.25,2)--(15.25,0.5)--(15.75,0.5)--cycle;
\filldraw[fill=blue!20, draw=blue, thick] (15.75,-2.5)--(15.25,-2.5)--(15.25,-4)--(15.75,-4)--cycle;
\path (14.5,1.5) node [style=sergio] {\scriptsize$\sigma_3$};
\path (13,0) node [style=sergio] {\scriptsize$\sigma_3$};
\path (18,0) node [style=sergio] {\scriptsize$\sigma_3$};
\path (16.5,1.5) node [style=sergio] {\scriptsize$\sigma_3$};
\path (15.5,1.25) node [style=sergio] {\scriptsize$\tau_2$};
\path (15.5,-3.25) node [style=sergio] {\scriptsize$\tau_2$};
\filldraw[fill=white!20, draw=black, thick] (15.5,-1)circle (13pt);
\path (15.5,-1) node [style=sergio] {$\mu$};
\filldraw[fill=green!20, draw=green, thick] (16,-2.5)--(17.5,-4)--(16,-4)--cycle;
\filldraw[fill=green!20, draw=green, thick] (12.5,-1.5)--(14,-1.5)--(12.5,-3)--cycle;
\path (13,-2) node [style=sergio] {\scriptsize$\sigma_3$};
\path (16.5,-3.5) node [style=sergio] {\scriptsize$\sigma_3$};
\filldraw[fill=blue!20, draw=blue, thick] (12.5,-1.25)--(12.5,-0.75)--(14,-0.75)--(14,-1.25)--cycle;
\filldraw[fill=blue!20, draw=blue, thick] (17,-1.25)--(17,-0.75)--(18.5,-0.75)--(18.5,-1.25)--cycle;
\path (13.25,-1) node [style=sergio] {\scriptsize$\tau_2$};
\path (17.75,-1) node [style=sergio] {\scriptsize$\tau_2$};
\filldraw[fill=blue!20, draw=blue, thick] (12.75,1.25)--(13.25,1.75)--(14.75,0.25)--(14.25,-0.25)--cycle;
\filldraw[fill=blue!20, draw=blue, thick] (18.25,1.25)--(17.75,1.75)--(16.25,0.25)--(16.75,-0.25)--cycle;
\filldraw[fill=blue!20, draw=blue, thick] (13.25,-3.75)--(12.75,-3.25)--(14.25,-1.75)--(14.75,-2.25)--cycle;
\filldraw[fill=blue!20, draw=blue, thick] (17.75,-3.75)--(18.25,-3.25)--(16.75,-1.75)--(16.25,-2.25)--cycle;
\filldraw[fill=green!20, draw=green, thick] (15,-4)--(15,-2.5)--(13.5,-4)--cycle;
\filldraw[fill=green!20, draw=green, thick] (17,-1.5)--(18.5,-3)--(18.5,-1.5)--cycle;
\path (14.5,-3.5) node [style=sergio] {\scriptsize$\sigma_3$};
\path (18,-2) node [style=sergio] {\scriptsize$\sigma_3$};
\path (13.75,0.75) node [style=sergio] {\scriptsize$\tau_3$};
\path (17.25,0.75) node [style=sergio] {\scriptsize$\tau_3$};
\path (17.25,-2.75) node [style=sergio] {\scriptsize$\tau_3$};
\path (13.75,-2.75) node [style=sergio] {\scriptsize$\tau_3$};
\filldraw[fill=red!20, draw=red, thick] (7.5,2)--(6,2)--(7.5,0.5)--cycle;
\filldraw[fill=red!20, draw=red, thick] (5,-0.5)--(5,1)--(6.5,-0.5)--cycle;
\filldraw[fill=red!20, draw=red, thick] (8.5,2)--(10,2)--(8.5,0.5)--cycle;
\filldraw[fill=red!20, draw=red, thick] (11,-0.5)--(11,1)--(9.5,-0.5)--cycle;
\filldraw[fill=green!20, draw=green, thick] (8.25,2)--(7.75,2)--(7.75,0.5)--(8.25,0.5)--cycle;
\filldraw[fill=green!20, draw=green, thick] (8.25,-2.5)--(7.75,-2.5)--(7.75,-4)--(8.25,-4)--cycle;
\path (7,1.5) node [style=sergio] {\scriptsize$\lambda$};
\path (5.5,0) node [style=sergio] {\scriptsize$\lambda$};
\path (10.5,0) node [style=sergio] {\scriptsize$\lambda$};
\path (9,1.5) node [style=sergio] {\scriptsize$\lambda$};
\path (8,1.25) node [style=sergio] {\scriptsize$\sigma_1$};
\path (8,-3.25) node [style=sergio] {\scriptsize$\sigma_1$};
\filldraw[fill=blue!20, draw=blue, thick] (8,-1)circle (13pt);
\path (8,-1) node [style=sergio] {$\tau_1$};
\filldraw[fill=red!20, draw=red, thick] (8.5,-2.5)--(10,-4)--(8.5,-4)--cycle;
\filldraw[fill=red!20, draw=red, thick] (5,-1.5)--(6.5,-1.5)--(5,-3)--cycle;
\path (5.5,-2) node [style=sergio] {\scriptsize$\lambda$};
\path (9,-3.5) node [style=sergio] {\scriptsize$\lambda$};
\filldraw[fill=green!20, draw=green, thick] (5,-1.25)--(5,-0.75)--(6.5,-0.75)--(6.5,-1.25)--cycle;
\filldraw[fill=green!20, draw=green, thick] (9.5,-1.25)--(9.5,-0.75)--(11,-0.75)--(11,-1.25)--cycle;
\path (5.75,-1) node [style=sergio] {\scriptsize$\sigma_1$};
\path (10.25,-1) node [style=sergio] {\scriptsize$\sigma_1$};
\filldraw[fill=green!20, draw=green, thick] (5.25,1.25)--(5.75,1.75)--(7.25,0.25)--(6.75,-0.25)--cycle;
\filldraw[fill=green!20, draw=green, thick] (10.75,1.25)--(10.25,1.75)--(8.75,0.25)--(9.25,-0.25)--cycle;
\filldraw[fill=green!20, draw=green, thick] (5.75,-3.75)--(5.25,-3.25)--(6.75,-1.75)--(7.25,-2.25)--cycle;
\filldraw[fill=green!20, draw=green, thick] (10.25,-3.75)--(10.75,-3.25)--(9.25,-1.75)--(8.75,-2.25)--cycle;
\filldraw[fill=red!20, draw=red, thick] (7.5,-4)--(7.5,-2.5)--(6,-4)--cycle;
\filldraw[fill=red!20, draw=red, thick] (9.5,-1.5)--(11,-3)--(11,-1.5)--cycle;
\path (7,-3.5) node [style=sergio] {\scriptsize$\lambda$};
\path (10.5,-2) node [style=sergio] {\scriptsize$\lambda$};
\path (6.25,0.75) node [style=sergio] {\scriptsize$\sigma_2$};
\path (9.75,0.75) node [style=sergio] {\scriptsize$\sigma_2$};
\path (9.75,-2.75) node [style=sergio] {\scriptsize$\sigma_2$};
\path (6.25,-2.75) node [style=sergio] {\scriptsize$\sigma_2$};
\path (0.3532,1.0733) node [style=sergio] {\footnotesize$C_4$};
\path (1.5,1.85) node [style=sergio] {$M_1$};
\path (-0.5,1.85) node [style=sergio] {$M_2$};
\path (-3.5,-2) node [style=sergio] {$M_3$};
\path (0.5,-5) node [style=sergio] {Open lattice};
\path (15.5,-5) node [style=sergio] {Equator};
\path (8,-5) node [style=sergio] {North/south hemisphere};
\end{tikzpicture}
\caption{The cell decomposition of $D_{4h}$-symmetric lattice. Left panel depicts the whole lattice; middle panel depicts the north/south hemisphere including 3D block $\lambda$ and 2D blocks $\sigma_{1,2}$; right panrl depicts the equator, including 2D blocks $\sigma_3$, 1D blocks $\tau_{2,3}$, and 0D blocks $\mu$. $C_4$ depicts the axis of 4-fold rotation, and $\bs{M}_{1,2,3}$ depict the reflection planes.}
\label{D4h cell decomposition}
\end{figure*}

For $D_{4h}$-symmetric lattice with the cell decomposition in Fig. \ref{D4h cell decomposition}, the ground-state wavefunction of the system can be decomposed to the direct products of wavefunctions of lower-dimensional block states as:
\begin{align}
|\Psi\rangle=\bigotimes\limits_{g\in D_{4h}}|T_{g\lambda}\rangle\otimes\sum\limits_{k=1}^3|\gamma_{g\sigma_k}\rangle\otimes\sum\limits_{j=1}^3|\beta_{g\tau_j}\rangle\otimes|\alpha_\mu\rangle
\label{D4h cell decomposition}
\end{align}
where $|T_{g\lambda}\rangle$ is the wavefunction of 3D block state on $g\lambda$ which is topological trivial; $|\gamma_{g\sigma_{1,2,3}}\rangle$ is the $\mathbb{Z}_2$-symmetric wavefunction of 1D block state on $g\sigma_{1,2,3}$; $|\beta_{g\tau_1}\rangle$ is the $(\mathbb{Z}_4\rtimes\mathbb{Z}_2)$-symmetric wavefunction of 1D block state on $g\tau_1$, and $|\beta_{g\tau_{2,3}}\rangle$ is the $(\mathbb{Z}_2\times\mathbb{Z}_2)$-symmetric wavefunction of 2D block state on $g\tau_{2,3}$; $|\alpha_{\mu}\rangle$ is the $\mathbb{Z}_2\times(\mathbb{Z}_4\rtimes\mathbb{Z}_2)$-symmetric wavefunction of 0D block state on $\mu$.

We summarize the classifications and corresponding root phases. For crystalline TSC with spinless fermions, we summarize all possible block states as:
\begin{enumerate}[1.]
\item 2D blocks $\sigma_1$, $\sigma_2$ and $\sigma_3$: 2D $(p+ip)$-SCs and 2D $\mathbb{Z}_2$ fSPT phases;
\item 1D block $\tau_1$: Majorana chain, 1D $\mathbb{Z}_4\rtimes\mathbb{Z}_2$ fSPT phases and Haldane chain;
\item 1D blocks $\tau_2$ and $\tau_3$: Majorana chains, 1D $\mathbb{Z}_2\times\mathbb{Z}_2$ fSPT phases and Haldane chains.
\item 0D block $\mu$: 0D modes with odd fermion parity, characterizing eigenvalues of all three generators of $D_{4h}$ the group.
\end{enumerate}
By explicit $K$-matrix calculations, we conclude that except 2D bosonic Levin-Gu state on each $\sigma_1/\sigma_2/\sigma_3$, all other 2D block states are obstructed. Similar to the $C_{4h}$-symmetric case, Majorana chain decoration and 1D $\mathbb{Z}_4\rtimes\mathbb{Z}_2$ fSPT phases decoration on $\tau_1$ are obstructed. 

Then focus on the equator, for the first root phase of 1D $\mathbb{Z}_2\times\mathbb{Z}_2$ fSPT phases, from Ref. \cite{dihedral}, the only obstruction and trivialization free block state is on $\tau_2$ or $\tau_3$; for the second root phase of 1D $\mathbb{Z}_2\times\mathbb{Z}_2$ fSPT phases, all of them are obstructed. At the 0D block $\mu$, it is easy to see that 0D modes characterizing eigenvalues $-1$ of all three generators of the $D_{4h}$ group are trivialized by 1D bubble equivalences. Therefore, the ultimate classification is $\mathbb{Z}_2^6$, with the corresponding root phases:
\begin{enumerate}[1.]
\item 2D bosonic Levin-Gu state on each $\sigma_1/\sigma_2/\sigma_3$ ($\mathbb{Z}_2^3$);
\item 1D Haldane phase on each $\tau_1$ ($\mathbb{Z}_2$);
\item The first root phase of 1D $\mathbb{Z}_2\times\mathbb{Z}_2$ fSPT phase on each $\tau_2$ or $\tau_3$ ($\mathbb{Z}_2$);
\item 0D mode with odd fermion parity on $\mu$ ($\mathbb{Z}_2$).
\end{enumerate}
The HO topological surface theories of different root phases are:
\begin{enumerate}[1.]
\item $2^{\mathrm{nd}}$-order: 1D nonchiral Luttinger liquids with $K$-matrix $K=\sigma^x$ and $\mathbb{Z}_2$ symmetry property: $W^{\mathbb{Z}_2}=\mathbbm{1}_{2\times2}$ and $\delta\phi^{\mathbb{Z}_2}=\pi(1,1)^T$, on the intersections between the open lattice and 2D blocks $\sigma_1/\sigma_2/\sigma_3$;
\item $3^{\mathrm{rd}}$-order: a spin-1/2 degree of freedom at the center of each surface of the open lattice;
\item $3^{\mathrm{rd}}$-order: double Majorana zero modes at the center of each vertical hinge of the open lattice.
\end{enumerate}

For crystalline TSC with spin-1/2 fermions, all possible block states are located at 0D block $\mu$: 0D modes characterizing eigenvalues $-1$ of three generators of $D_{4h}$ group. There is no further obstruction and trivialization, hence the ultimate classification is $\mathbb{Z}_2^3$.

For crystalline TI with spinless fermions, we summarize all possible block states as:
\begin{enumerate}[1.]
\item 2D blocks $\sigma_1$, $\sigma_2$ and $\sigma_3$: Chern insulators, Kitaev's $E_8$ states and 2D $U^f(1)\times\mathbb{Z}_2$ fSPT phases;
\item 1D blocks $\tau_1$, $\tau_2$ and $\tau_3$: Haldane chains;
\item 0D block $\mu$: 0D modes characterizing eigenvalues $-1$ of all three generators of the $D_{4h}$ group, with different $U^f(1)$ charge.
\end{enumerate}
From explicit $K$-matrix calculations, we conclude that except 2D $U^f(1)\times\mathbb{Z}_2$ bSPT phase decoration on each $\sigma_{1,2,3}$ and Kitaev's $E_8$ states decoration on $\sigma_1$, $\sigma_2$ and $\sigma_3$ with proper chiralities to guarantee the nonchiral 1D blocks, all other 2D block states are obstructed. Furthermore, the 0D modes characterizing the eigenvalues $-1$ of in-plane generators of the $D_{4h}$ group and even $U^f(1)$ charge are trivialized by 1D bubble equivalences. Hence the ultimate classification is $\mathbb{Z}_4\times\mathbb{Z}_2^7$, with the following root phases:
\begin{enumerate}[1.]
\item 2D bosonic Levin-Gu state on each $\sigma_1/\sigma_2/\sigma_3$ ($\mathbb{Z}_2^3$);
\item Monolayer Kitaev's $E_8$ state on each $\sigma_1$, $\sigma_2$ and $\sigma_3$ ($\mathbb{Z}_2$);
\item 1D Haldane phase on each $\tau_1/\tau_2/\tau_3$ ($\mathbb{Z}_2^3$);
\item 0D mode with odd $U^f(1)$ charge on $\mu$ ($\mathbb{Z}_2$);
\item 0D mode with eigenvalue $-1$ of horizontal reflection plane on $\mu$ ($\mathbb{Z}_2$).
\end{enumerate}
and there is a nontrivial extension between the last two root phases. The HO topological surface theories of different root phases are:
\begin{enumerate}[1.]
\item $2^{\mathrm{nd}}$-order: 1D nonchiral Luttinger liquids with $K$-matrix $K=\sigma^x$ and $\mathbb{Z}_2$ symmetry property: $W^{\mathbb{Z}_2}=\mathbbm{1}_{2\times2}$ and $\delta\phi^{\mathbb{Z}_2}=\pi(1,1)^T$, on the intersections between the open lattice and 2D blocks $\sigma_1/\sigma_2/\sigma_3$;
\item $2^{\mathrm{nd}}$-order: 1D chiral Luttinger liquids with $K$-matrix (\ref{K-matrix E8S}), on the intersections between the open lattice and 2D blocks $\sigma_1$, $\sigma_2$ and $\sigma_3$;
\item $3^{\mathrm{rd}}$-order: a spin-1/2 degree of freedom at the center of each horizontal surfaces;
\item $3^{\mathrm{rd}}$-order: a spin-1/2 degree of freedom at the center of each vertical surfaces;
\item $3^{\mathrm{rd}}$-order: a spin-1/2 degree of freedom at the center of each vertical hinges;
\end{enumerate}

For crystalline TI with spin-1/2 fermions, we summarize all possible block states as:
\begin{enumerate}[1.]
\item 2D blocks $\sigma_1$, $\sigma_2$ and $\sigma_3$: Chern insulators, Kitaev's $E_8$ states and 2D $U^f(1)\times\mathbb{Z}_2$ fSPT phases;
\item 0D block $\mu$: 0D modes characterizing eigenvalues $-1$ of all three generators of the $D_{4h}$ group, with different even $U^f(1)$ charge.
\end{enumerate}
There are some constraints on chiral 2D block states to guarantee that they leaves nonchiral 1D modes at their shared border, otherwise the chiral 2D block states are obstructed. Furthermore, the $U^f(1)$ charge of integer multiple of 4 is trivialized by 1D bubble equivalences. Therefore, the ultimate classification is $\mathbb{Z}_8\times\mathbb{Z}_4^2\times\mathbb{Z}_2^5$, with the following root phases:
\begin{enumerate}[1.]
\item Monolayer Chern insulator on each $\sigma_1$, $\sigma_2$ and $\sigma_3$ ($\mathbb{Z}_2$);
\item 2D $U^f(1)\times\mathbb{Z}_2$ fSPT phase on each $\sigma_1/\sigma_2/\sigma_3$ ($\mathbb{Z}_4^3$);
\item Monolayer Kitaev's $E_8$ state on each $\sigma_1$, $\sigma_2$ and $\sigma_3$ ($\mathbb{Z}_2$);
\item 0D mode with $U^f(1)$ charge $n\equiv2(\mathrm{mod}~4)$ on $\mu$ ($\mathbb{Z}_2$);
\item 0D mode with eigenvalues $-1$ of all three generators of the $D_{4h}$ group on $\mu$ ($\mathbb{Z}_2^3$).
\end{enumerate}
And there are several nontrivial extensions between these root phases: bilayer Chern insulators on each $\sigma_{1,2,3}$ are equivalent to 2D $U^f(1)\times\mathbb{Z}_2$ fSPT phases with $(\nu_1,\nu_2,\nu_3)=(1,3,1)$ on 2D blocks $\sigma_{1,2,3}$. The second-order topological surface theories of different root phases are:
\begin{enumerate}[1.]
\item Chiral fermions on the intersections between the open lattice and 2D blocks $\sigma_1$, $\sigma_2$ and $\sigma_3$;
\item 1D nonchiral Luttinger liquids with $K$-matrix $K=\sigma^z$ and $\mathbb{Z}_2$ symmetry property $W^{\mathbb{Z}_2}=\mathbbm{1}_{2\times2}$ and $\delta\phi^{\mathbb{Z}_2}=\pi(0,1)^T$, on the intersections between the open lattice and 2D blocks decorated with a $U^f(1)\times\mathbb{Z}_2$ fSPT phase with $\nu=1$ on each of them;
\item 1D chiral Luttinger liquids with $K$-matrix (\ref{K-matrix E8S}), on the intersections between the open lattice and 2D blocks $\sigma_1$, $\sigma_2$ and $\sigma_3$;
\end{enumerate}

\begin{figure}
\begin{tikzpicture}[scale=0.7]
\tikzstyle{sergio}=[rectangle,draw=none]
\draw[thick] (0,1.5) -- (-3,0);
\draw[thick] (0,1.5) -- (1,0);
\draw[thick] (-3,0) -- (1,0);
\draw[thick] (-3,0) -- (-3,-4);
\draw[thick] (1,-4) -- (-3,-4);
\draw[thick] (1,-4) -- (1,0);
\filldraw[fill=green!20, draw=green, thick, fill opacity=0.5] (-0.5,-3.5)--(-0.5,0.5)--(-3,0)--(-3,-4)--cycle;
\filldraw[fill=green!20, draw=green, thick, fill opacity=0.5] (-0.5,-3.5)--(-0.5,0.5)--(0,1.5)--(0,-2.5)--cycle;
\filldraw[fill=green!20, draw=green, thick, fill opacity=0.5] (-0.5,-3.5)--(-0.5,0.5)--(1,0)--(1,-4)--cycle;
\draw[densely dashed] (0,-2.5) -- (0,1.5);
\draw[densely dashed] (0,-2.5) -- (1,-4);
\draw[densely dashed] (0,-2.5) -- (-3,-4);
\draw[draw=blue,ultra thick] (-0.5,-3.5) -- (-0.5,0.5);
\filldraw[fill=red!20, draw=red, thick, fill opacity=0.5] (4.25,0.75)--(2.75,-2.25)--(4.25,-1.25)--cycle;
\filldraw[fill=green!20, draw=green, thick] (4.25,-1.75)--(4.65,-2.25)--(3.25,-3.25)--(2.85,-2.75)--cycle;
\filldraw[fill=red!20, draw=red, thick, fill opacity=0.5] (6.5,-3.5)--(3.5,-3.5)--(5,-2.5)--cycle;
\filldraw[fill=green!20, draw=green, thick] (5.75,-1.75)--(5.35,-2.25)--(6.75,-3.25)--(7.15,-2.75)--cycle;
\filldraw[fill=red!20, draw=red, thick, fill opacity=0.5] (5.75,0.75)--(7.25,-2.25)--(5.75,-1.25)--cycle;
\filldraw[fill=green!20, draw=green, thick] (5.3,-1.05)--(4.7,-1.05)--(4.7,0.75)--(5.3,0.75)--cycle;
\filldraw[fill=blue!20, draw=blue, thick] (5,-1.68)circle (13pt);
\path (5,-1.68) node [style=sergio] {$\tau$};
\path (3.75,-2.5) node [style=sergio] {$\sigma$};
\path (6.25,-2.5) node [style=sergio] {$\sigma$};
\path (5,-0.15) node [style=sergio] {$\sigma$};
\path (3.75,-1) node [style=sergio] {$\lambda$};
\path (6.25,-1) node [style=sergio] {$\lambda$};
\path (5,-3) node [style=sergio] {$\lambda$};
\path (-0.75,0.75) node [style=sergio] {$C_3$};
\end{tikzpicture}
\caption{The cell decomposition of $C_3$-symmetric lattice. Left panel depicts the whole lattice; right panel depicts the horizontal intersection of the lattice, including 3D blocks $\lambda$, 2D blocks $\sigma$ and 1D blocks $\tau$. $C_3$ depicts the axis of 3-fold rotation.}
\label{C3 cell decomposition}
\end{figure}
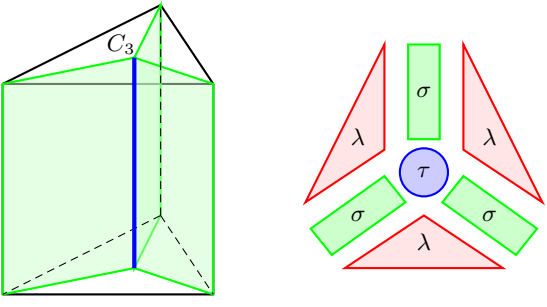

\subsection{$C_3$-symmetric lattice}
For $C_3$-symmetric lattice with the cell decomposition in Fig. \ref{C3 cell decomposition}, the ground-state wavefunction of the system can be decomposed to the direct products of wavefunctions of lower-dimensional block states as:
\begin{align}
|\Psi\rangle=\bigotimes\limits_{g\in C_3}|T_{g\lambda}\rangle\otimes|T_{g\sigma}\rangle\otimes|\beta_\tau\rangle
\label{C3 wavefunction}
\end{align}
where $|T_{g\lambda}\rangle/|T_{g\sigma}\rangle$ is the wavefunction of 3D/2D block state on $g\lambda/g\sigma$ which is topological trivial or invertible topological phase; $|\beta_\tau\rangle$ is the $\mathbb{Z}_3$-symmetric wavefunction of 1D block state on $\tau$.

We summarize the classifications and corresponding root phases. Similar to other rotational symmetric systems, the spin of fermions are irrelevant, for both crystalline TSC and TI. For crystalline TSC, we summarize all block states as:
\begin{enumerate}[1.]
\item 2D block $\sigma$: 2D $(p+ip)$-SC;
\item 1D block $\tau$: Majorana chain.
\end{enumerate}
2D $(p+ip)$-SC decoration on $\sigma$ leave a chiral mode on $\tau$ with chiral central charge $c_-=3/2$, which cannot be gapped out. Furthermore, Majorana chain decoration can be trivialized by 2D ``Majorana'' bubble construction on each $\sigma$. Hence the ultimate classifications are all trivial.

For crystalline TI, all block states are located on the 2D block $\sigma$: Chern insulators and Kitaev's $E_8$ states, who leave chiral 1D mode on $\tau$. As the consequence, all of them are obstructed, and the ultimate classification is trivial.

\subsection{$S_6$-symmetric lattice}
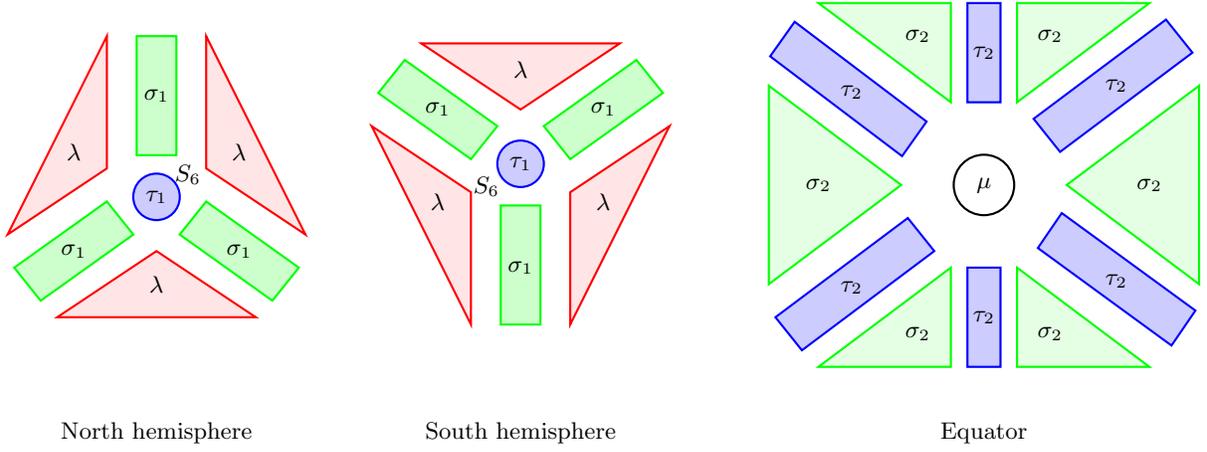
\begin{figure*}
\begin{tikzpicture}[scale=0.88]
\tikzstyle{sergio}=[rectangle,draw=none]
\filldraw[fill=red!20, draw=red, thick, fill opacity=0.5] (-5.5,0)--(-7,-3)--(-5.5,-2)--cycle;
\filldraw[fill=green!20, draw=green, thick] (-5.5,-2.5)--(-5.1,-3)--(-6.5,-4)--(-6.9,-3.5)--cycle;
\filldraw[fill=red!20, draw=red, thick, fill opacity=0.5] (-3.25,-4.25)--(-6.25,-4.25)--(-4.75,-3.25)--cycle;
\filldraw[fill=green!20, draw=green, thick] (-4,-2.5)--(-4.4,-3)--(-3,-4)--(-2.6,-3.5)--cycle;
\filldraw[fill=red!20, draw=red, thick, fill opacity=0.5] (-4,0)--(-2.5,-3)--(-4,-2)--cycle;
\filldraw[fill=green!20, draw=green, thick] (-4.45,-1.8)--(-5.05,-1.8)--(-5.05,0)--(-4.45,0)--cycle;
\filldraw[fill=blue!20, draw=blue, thick] (-4.75,-2.43)circle (10pt);
\path (-4.75,-2.43) node [style=sergio] {$\tau_1$};
\path (-6,-3.25) node [style=sergio] {$\sigma_1$};
\path (-3.5,-3.25) node [style=sergio] {$\sigma_1$};
\path (-4.75,-0.9) node [style=sergio] {$\sigma_1$};
\path (-6,-1.75) node [style=sergio] {$\lambda$};
\path (-3.5,-1.75) node [style=sergio] {$\lambda$};
\path (-4.75,-3.75) node [style=sergio] {$\lambda$};
\path (-4.2776,-2.0925) node [style=sergio] {$S_6$};
\path (0.2317,-2.2729) node [style=sergio] {$S_6$};
\filldraw[rotate around={180:(0.75,-1.93)},fill=red!20, draw=red, thick, fill opacity=0.5] (0,0.5)--(-1.5,-2.5)--(0,-1.5)--cycle;
\filldraw[rotate around={180:(0.75,-1.93)},fill=green!20, draw=green, thick] (0,-2)--(0.4,-2.5)--(-1,-3.5)--(-1.4,-3)--cycle;
\filldraw[rotate around={180:(0.75,-1.93)},fill=red!20, draw=red, thick, fill opacity=0.5] (2.25,-3.75)--(-0.75,-3.75)--(0.75,-2.75)--cycle;
\filldraw[rotate around={180:(0.75,-1.93)},fill=green!20, draw=green, thick] (1.5,-2)--(1.1,-2.5)--(2.5,-3.5)--(2.9,-3)--cycle;
\filldraw[rotate around={180:(0.75,-1.93)},fill=red!20, draw=red, thick, fill opacity=0.5] (1.5,0.5)--(3,-2.5)--(1.5,-1.5)--cycle;
\filldraw[rotate around={180:(0.75,-1.93)},fill=green!20, draw=green, thick] (1.05,-1.3)--(0.45,-1.3)--(0.45,0.5)--(1.05,0.5)--cycle;
\filldraw[rotate around={180:(0.75,-1.93)},fill=blue!20, draw=blue, thick] (0.75,-1.93)circle (10pt);
\path (0.75,-1.93) node [style=sergio] {$\tau_1$};
\path (-0.5,-1.1) node [style=sergio] {$\sigma_1$};
\path (2,-1.1) node [style=sergio] {$\sigma_1$};
\path (0.75,-3.5) node [style=sergio] {$\sigma_1$};
\path (-0.5,-2.5) node [style=sergio] {$\lambda$};
\path (2,-2.5) node [style=sergio] {$\lambda$};
\path (0.75,-0.5) node [style=sergio] {$\lambda$};
\filldraw[fill=green!20, draw=green, thick, fill opacity=0.5] (7.25,-5)--(5.25,-5)--(7.25,-3.5)--cycle;
\filldraw[fill=green!20, draw=green, thick, fill opacity=0.5] (8.25,-5)--(10.25,-5)--(8.25,-3.5)--cycle;
\filldraw[fill=green!20, draw=green, thick, fill opacity=0.5] (4.5,-3.75)--(4.5,-0.75)--(6.5,-2.25)--cycle;
\filldraw[fill=blue!20, draw=blue, thick] (8,-5)--(7.5,-5)--(7.5,-3.5)--(8,-3.5)--cycle;
\filldraw[fill=blue!20, draw=blue, thick] (4.6,-4.25)--(5,-4.75)--(7,-3.25)--(6.6,-2.75)--cycle;
\filldraw[fill=white!20, draw=black, thick] (7.75,-2.25)circle (13pt);
\filldraw[rotate around={286:(7.75,-2.5)},fill=blue!20, draw=blue, thick] (4.35,-4.5)--(4.75,-5)--(6.75,-3.5)--(6.35,-3)--cycle;
\filldraw[fill=green!20, draw=green, thick, fill opacity=0.5] (7.25,0.5)--(5.25,0.5)--(7.25,-1)--cycle;
\filldraw[fill=blue!20, draw=blue, thick] (8,-1)--(7.5,-1)--(7.5,0.5)--(8,0.5)--cycle;
\filldraw[fill=green!20, draw=green, thick, fill opacity=0.5] (8.25,0.5)--(10.25,0.5)--(8.25,-1)--cycle;
\filldraw[fill=blue!20, draw=blue, thick] (10.5,0.25)--(8.5,-1.25)--(8.9,-1.75)--(10.9,-0.25)--cycle;
\filldraw[rotate around={180:(7.75,-2.25)},fill=green!20, draw=green, thick, fill opacity=0.5] (4.5,-3.75)--(4.5,-0.75)--(6.5,-2.25)--cycle;
\filldraw[rotate around={287:(7.75,-2.25)},fill=blue!20, draw=blue, thick] (10.5,0.25)--(8.5,-1.25)--(8.9,-1.75)--(10.9,-0.25)--cycle;
\path (7.75,-2.25) node [style=sergio] {$\mu$};
\path (5.25,-2.25) node [style=sergio] {$\sigma_2$};
\path (10.25,-2.25) node [style=sergio] {$\sigma_2$};
\path (8.75,-4.5) node [style=sergio] {$\sigma_2$};
\path (6.75,-4.5) node [style=sergio] {$\sigma_2$};
\path (8.75,0) node [style=sergio] {$\sigma_2$};
\path (6.75,0) node [style=sergio] {$\sigma_2$};
\path (5.75,-0.85) node [style=sergio] {$\tau_2$};
\path (9.75,-0.75) node [style=sergio] {$\tau_2$};
\path (9.75,-3.7) node [style=sergio] {$\tau_2$};
\path (7.75,-4.25) node [style=sergio] {$\tau_2$};
\path (7.75,-0.25) node [style=sergio] {$\tau_2$};
\path (5.75,-3.8) node [style=sergio] {$\tau_2$};
\path (-4.75,-6) node [style=sergio] {North hemisphere};
\path (0.75,-6) node [style=sergio] {South hemisphere};
\path (7.75,-6) node [style=sergio] {Equator};
\end{tikzpicture}
\caption{The cell decomposition of $S_6$-symmetric lattice. Left panel depicts the north hemisphere, including 3D blocks $\lambda$, 2D blocks $\sigma_1$ and 1D blocks $\tau_1$; middle panel depicts the south hemisphere, including 3D blocks $\lambda$, 2D blocks $\sigma_1$ and 1D blocks $\tau_1$; right panel depicts the equator, including 2D blocks $\sigma_2$, 1D blocks $\tau_2$ and 0D block $\mu$. $S_6$ depicts the axis of the 6-fold rotoreflection.}
\label{S6 cell decomposition}
\end{figure*}

For $S_6$-symmetric lattice with the cell decomposition in Fig. \ref{S6 cell decomposition}, the ground-state wavefunction of the system can be decomposed to the direct products of wavefunctions of lower-dimensional block states as:
\begin{align}
|\Psi\rangle=\bigotimes_{g\in S_6}|T_{g\lambda}\rangle\otimes\sum\limits_{k=1}^2|T_{g\sigma_{k}}\rangle\otimes\sum\limits_{j=1}^2|\beta_{\tau_j}\rangle\otimes|\alpha_\mu\rangle
\label{S6 wavefunction}
\end{align}
where $|T_{g\lambda}\rangle/|T_{g\sigma_{1,2}}\rangle$ is the wavefunction of 3D/2D block state on $g\lambda/g\sigma_{1,2}$ which is topological trivial or invertible topological phase; $|\beta_{g\tau_{1}}\rangle$ is the $\mathbb{Z}_3$-symmetric wavefunction of 1D block state on $g\tau_1$, $|\beta_{g\tau_2}\rangle$ is the wavefunction of 1D block state on $g\tau_2$ which is topological trivial or invertible topological phase; $|\alpha_\mu\rangle$ is the $\mathbb{Z}_6$-symmetric wavefunction of 0D block state on $\mu$.

We summarize the classifications and corresponding root phases. For crystalline TSC with spinless fermions, all possible block states are summarized as following:
\begin{enumerate}[1.]
\item 2D blocks $\sigma_1$ and $\sigma_2$: 2D $(p+ip)$-SC;
\item 1D block $\tau_1$ and $\tau_2$: Majorana chain;
\item 0D block $\mu$: 0D mode characterizing different eigenvalues of $\bs{I}_r\in S_6$ as the generator of the $S_6$ group, with odd/even fermion parity.
\end{enumerate}
2D $(p+ip)$-SC on $\sigma_1$ is obstructed because it leaves a chiral 1D mode on each $\tau_1$ with chiral central charge $c_-=3/2$; 2D $(p+ip)$-SC on $\sigma_2$ is also obstructed because $(p+ip)$-SC is not compatible with in-plane 6-fold rotation. 

Majorana chain decoration on each $\tau_1$ leaves 2 Majorana zero modes $\gamma_1$ and $\gamma_2$, with the following symmetry property:
\begin{align}
\bs{I}_r\in S_6:~\gamma_1\leftrightarrow\gamma_2
\end{align}
with the fermion parity $P_f=i\gamma_1\gamma_2$ which is odd under $\bs{I}_r\in S_6$. Hence the Majorana chain decoration on $\tau_1$ is obstructed. Similar for Majorana chain decoration on $\tau_2$. Furthermore, from Ref. \cite{rotation} we know that 0D mode with odd fermion parity is trivialized by 2D ``Majorana'' bubble construction, and eigenvalue $-1$ of $\bs{I}_r$ is trivialized by 1D bubble equivalence on $\tau_2$. Moreover, we decorate a 0D $\mathbb{Z}_3$ SPT mode characterized by a phase factor $e^{2\pi i/3}$ which can be adiabatically deformed to infinite far and trivialized, this 1D bubble construction changes the phase factor of $\mu$ by $e^{4\pi i/3}$, and the corresponding 0D block state is trivialized. Hence the ultimate classification is trivial.

For crystalline TI, similar to the $C_2$-symmetric case, we argue that the spin of fermions is irrelevant. We summarize all possible block states as:
\begin{enumerate}[1.]
\item 2D blocks $\sigma_1$ and $\sigma_2$: Chern insulators and Kitaev's $E_8$ states;
\item 1D block $\tau_1$ and $\tau_2$: Majorana chain;
\item 0D block $\mu$: 0D modes characterizing eigenvalues of $\bs{I}_r\in S_6$, with different $U^f(1)$ charge.
\end{enumerate}
All block states on 2D block $\sigma_1$ are obstructed because they leave chiral 1D modes on 1D block $\tau_1$ as their shared border. Furthermore, similar to crystalline TSC, the phase factors $e^{2n\pi i/3}$ ($n=0,1,2$) are trivialized by 1D bubble equivalences on $\tau_1$. Therefore, the ultimate classification is $\mathbb{Z}_4\times\mathbb{Z}_2^2$, with the following root phases:
\begin{enumerate}[1.]
\item Monolayer Chern insulator on each $\sigma_2$ ($\mathbb{Z}_2$);
\item Monolayer Kitaev's $E_8$ state on each $\sigma_2$ ($\mathbb{Z}_2$);
\item 0D mode with odd $U^f(1)$ charge on $\mu$ ($\mathbb{Z}_2$);
\item 0D mode with eigenvalue $-1$ of the generator of $S_6$ on $\mu$ ($\mathbb{Z}_2$).
\end{enumerate}
and there is a nontrivial extension between last two root phases. The second-order topological surface theories of these root phases are:
\begin{enumerate}[1.]
\item Chiral fermions on the edge of the equator;
\item 1D chiral Luttinger liquids with $K$-matrix (\ref{K-matrix E8S}) on the edge of the equator.
\end{enumerate}

\subsection{$D_3$-symmetric lattice}
\begin{figure}
\begin{tikzpicture}[scale=0.65]
\tikzstyle{sergio}=[rectangle,draw=none]
\filldraw[fill=red!20, draw=red, thick, fill opacity=0.5] (-5.5,0)--(-7,-3)--(-5.5,-2)--cycle;
\filldraw[fill=green!20, draw=green, thick] (-5.5,-2.5)--(-5.1,-3)--(-6.5,-4)--(-6.9,-3.5)--cycle;
\filldraw[fill=red!20, draw=red, thick, fill opacity=0.5] (-3.25,-4.25)--(-6.25,-4.25)--(-4.75,-3.25)--cycle;
\filldraw[fill=green!20, draw=green, thick] (-4,-2.5)--(-4.4,-3)--(-3,-4)--(-2.6,-3.5)--cycle;
\filldraw[fill=red!20, draw=red, thick, fill opacity=0.5] (-4,0)--(-2.5,-3)--(-4,-2)--cycle;
\filldraw[fill=green!20, draw=green, thick] (-4.45,-1.8)--(-5.05,-1.8)--(-5.05,0)--(-4.45,0)--cycle;
\filldraw[fill=blue!20, draw=blue, thick] (-4.75,-2.43)circle (10pt);
\path (-4.75,-2.43) node [style=sergio] {$\tau_1$};
\path (-4.25,-2.1) node [style=sergio] {\scriptsize$C_3$};
\path (2.25,-1.4) node [style=sergio] {$C_2$};
\path (1.584,-2.8429) node [style=sergio] {$C_2$};
\path (-6,-3.25) node [style=sergio] {\scriptsize $\sigma_1$};
\path (-3.5,-3.25) node [style=sergio] {\scriptsize $\sigma_1$};
\path (-4.75,-0.9) node [style=sergio] {\scriptsize $\sigma_1$};
\path (-6,-1.75) node [style=sergio] {$\lambda$};
\path (-3.5,-1.75) node [style=sergio] {$\lambda$};
\path (-4.75,-3.75) node [style=sergio] {$\lambda$};
\filldraw[fill=green!20, draw=green, thick, fill opacity=0.5] (1.75,-5)--(-0.25,-5)--(1.75,-3.5)--cycle;
\filldraw[fill=green!20, draw=green, thick, fill opacity=0.5] (2.75,-5)--(4.75,-5)--(2.75,-3.5)--cycle;
\filldraw[fill=green!20, draw=green, thick, fill opacity=0.5] (-1,-3.75)--(-1,-0.75)--(1,-2.25)--cycle;
\filldraw[fill=blue!20, draw=blue, thick] (2.5,-5)--(2,-5)--(2,-3.5)--(2.5,-3.5)--cycle;
\filldraw[fill=blue!20, draw=blue, thick] (-0.9,-4.25)--(-0.5,-4.75)--(1.5,-3.25)--(1.1,-2.75)--cycle;
\filldraw[fill=white!20, draw=black, thick] (2.25,-2.25)circle (13pt);
\filldraw[rotate around={286:(2.25,-2.5))},fill=blue!20, draw=blue, thick] (-1.15,-4.5)--(-0.75,-5)--(1.25,-3.5)--(0.85,-3)--cycle;
\filldraw[fill=green!20, draw=green, thick, fill opacity=0.5] (1.75,0.5)--(-0.25,0.5)--(1.75,-1)--cycle;
\filldraw[fill=blue!20, draw=blue, thick] (2.5,-1)--(2,-1)--(2,0.5)--(2.5,0.5)--cycle;
\filldraw[fill=green!20, draw=green, thick, fill opacity=0.5] (2.75,0.5)--(4.75,0.5)--(2.75,-1)--cycle;
\filldraw[fill=blue!20, draw=blue, thick] (5,0.25)--(3,-1.25)--(3.4,-1.75)--(5.4,-0.25)--cycle;
\filldraw[rotate around={180:(2.25,-2.25)},fill=green!20, draw=green, thick, fill opacity=0.5] (-1,-3.75)--(-1,-0.75)--(1,-2.25)--cycle;
\filldraw[rotate around={287:(2.25,-2.25)},fill=blue!20, draw=blue, thick] (5,0.25)--(3,-1.25)--(3.4,-1.75)--(5.4,-0.25)--cycle;
\path (2.25,-2.25) node [style=sergio] {$\mu$};
\path (-0.25,-2.25) node [style=sergio] {$\sigma_2$};
\path (4.75,-2.25) node [style=sergio] {$\sigma_2$};
\path (3.25,-4.5) node [style=sergio] {$\sigma_2$};
\path (1.25,-4.5) node [style=sergio] {$\sigma_2$};
\path (3.25,0) node [style=sergio] {$\sigma_2$};
\path (1.25,0) node [style=sergio] {$\sigma_2$};
\path (0.25,-0.85) node [style=sergio] {\scriptsize$\tau_3$};
\path (4.25,-0.75) node [style=sergio] {\scriptsize$\tau_3$};
\path (4.25,-3.7) node [style=sergio] {\scriptsize$\tau_2$};
\path (2.25,-4.25) node [style=sergio] {\scriptsize$\tau_3$};
\path (2.25,-0.25) node [style=sergio] {\scriptsize$\tau_2$};
\path (0.25,-3.8) node [style=sergio] {\scriptsize$\tau_2$};
\path (-4.75,-6) node [style=sergio] {North/south hemisphere};
\path (2.25,-6) node [style=sergio] {Equator};
\end{tikzpicture}
\caption{The cell decomposition of $D_3$-symmetric lattice. Left panel depicts the north/south hemisphere, including 3D blocks $\lambda$, 2D blocks $\sigma_1$ and 1D blocks $\tau_1$; right panel depicts the equator, including 2D blocks $\sigma_2$, 1D blocks $\tau_{2,3}$ and 0D block $\mu$. $C_3$ depicts the axis of 3-fold rotation, and $C_2$'s depicts the axes of 2-fold rotations.}
\label{D3 cell decomposition}
\end{figure}
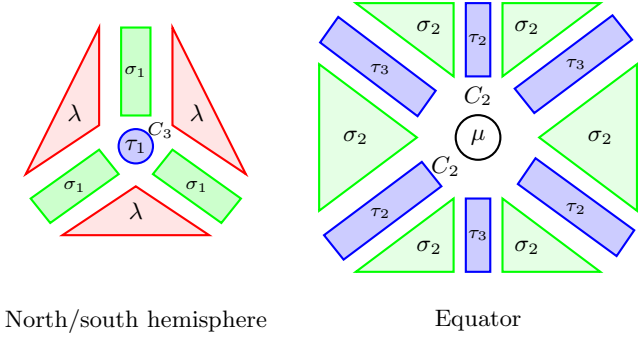

For $D_3$-symmetric lattice with the cell decomposition in Fig. \ref{D3 cell decomposition}, the ground-state wavefunction of the system can be decomposed to the direct products of wavefunctions of lower-dimensional block states as:
\begin{align}
|\Psi\rangle=\bigotimes\limits_{g\in D_3}|T_{g\lambda}\rangle\otimes\sum\limits_{k=1}^2|T_{g\sigma_k}\rangle\otimes\sum\limits_{j=1}^3|\beta_{g\tau_j}\rangle\otimes|\alpha_\mu\rangle
\label{D3 wavefunction}
\end{align}
where $|T_{g\lambda}\rangle/|T_{g\sigma_{1,2}}\rangle$ is the wavefunction of 3D/2D block state on $g\lambda/g\sigma_{1,2}$ which is topological trivial or invertible topological phase; $|\beta_{g\tau_1}\rangle$ is the $\mathbb{Z}_3$-symmetric wavefunction of 1D block state on $g\tau_1$, and $|\beta_{g\tau_{2,3}}\rangle$ is the $\mathbb{Z}_2$-symmetric wavefunction of 1D block state on $g\tau_{2,3}$; $|\alpha_\mu\rangle$ is the $(\mathbb{Z}_3\rtimes\mathbb{Z}_2)$-symmetric wavefunction of 0D block state on $\mu$.

We summarize the classifications and corresponding root phases. For crystalline TSC with spinless fermions, we summarize all possible block states as:
\begin{enumerate}[1.]
\item 2D blocks $\sigma_1$ and $\sigma_2$: 2D $(p+ip)$-SCs;
\item 1D block $\tau_1$: Majorana chain;
\item 1D blocks $\tau_2$ and $\tau_3$: Majorana chains and 1D $\mathbb{Z}_2$ fSPT phases;
\item 0D block $\mu$: 0D modes characterizing the eigenvalue $-1$ of $\mathbb{Z}_2\subset\mathbb{Z}_3\rtimes\mathbb{Z}_2$, with odd/even fermion parity.
\end{enumerate}
2D $(p+ip)$-SCs decoration on $\sigma_1$ and $\sigma_2$ are obstructed because they leave chiral 1D modes on $\tau_{1,2,3}$. Majorana chain decoration on $\tau_1$ is not compatible with 2-fold rotation whose axis is located on the equator, and Majorana chain decoration solely on $\tau_2/\tau_3$ is obstructed because it leaves three Majorana zero modes at the center of the system which cannot be gapped out. If we consider $\tau_2$ and $\tau_3$ together and decorate a Majorana chain on each of them, there will be 6 dangling Majorana zero modes $\gamma_{j}$ ($j=1,\cdot\cdot\cdot,6$) at the 0D block $\mu$, with the following properties under two rotation generators of the $D_3$ group:
\begin{align}
\begin{aligned}
&\bs{R}_{\tau_1}:~(\gamma_1,\gamma_2,\gamma_3,\gamma_4,\gamma_5,\gamma_6)\mapsto(\gamma_3,\gamma_4,\gamma_5,\gamma_6,\gamma_1,\gamma_2)\\
&\bs{R}_{\tau_2}:~(\gamma_1,\gamma_2,\gamma_3,\gamma_4,\gamma_5,\gamma_6)\mapsto(\gamma_1,\gamma_6,\gamma_5,\gamma_4,\gamma_3,\gamma_2)
\end{aligned}\nonumber
\end{align}
We can symmetrically gap them out by introducing the following Hamiltonian:
\begin{align}
H=i\gamma_1\gamma_4+i\gamma_2\gamma_5-i\gamma_3\gamma_6
\end{align}
Nevertheless, this block state is trivialized by anomalous $(p+ip)$-SC with six vortices on the open surface of the lattice \cite{rotation}. Furthermore, the 0D mode with odd fermion parity is trivialized by 1D bubble equivalence on $\tau_2/\tau_3$ and 0D mode characterizing eigenvalue $-1$ of 2-fold rotation generator of the $D_3$ group is trivialized by 1D bubble equivalence on $\tau_1$. Therefore, the ultimate classifications is trivial. 

For crystalline TSC with spin-1/2 fermions, we summarize all possible block states as:
\begin{enumerate}[1.]
\item 2D blocks $\sigma_1$ and $\sigma_2$: 2D $(p+ip)$-SCs;
\item 1D block $\tau_1$: Majorana chain;
\item 0D block $\mu$: 0D modes characterizing the eigenvalue $-1$ of $\mathbb{Z}_2\subset\mathbb{Z}_3\rtimes\mathbb{Z}_2$, with odd/even fermion parity.
\end{enumerate}
Similar to the spinless fermions, all 2D block states are obstructed because they violate the nonchiral condition. Majorana chain decoration on $\tau_1$ is obstruction-free but can be trivialized by 2D ``Majorana'' bubble construction on $\sigma_1$. Furthermore, the 0D modes characterizing the eigenvalue $-1$ of the 2-fold rotation generator of $D_3$ group with both even and odd fermion parities are trivialized by 1D bubble equivalence on $\tau_2/\tau_3$. Hence the ultimate classification is trivial. 

For crystalline TI, similar to the $C_2$-symmetric case, we argue that the spin of fermions is irrelevant. We summarize all possible block states as:
\begin{enumerate}[1.]
\item 2D blocks $\sigma_1$ and $\sigma_2$: Chern insulators and Kitaev's $E_8$ states;
\item 0D block $\mu$: 0D modes characterizing the eigenvalue $-1$ of $\mathbb{Z}_2\subset\mathbb{Z}_3\rtimes\mathbb{Z}_2$, with different $U^f(1)$ charge.
\end{enumerate}
And all 2D block states are obstructed because they violate the nonchiral condition; and all 0D block states are trivialized by 1D bubble equivalences. Hence the ultimate classification is trivial.

\subsection{$C_{3v}$-symmetric lattice}
For $C_{3v}$-symmetric lattice with the cell decomposition in Fig. \ref{C3v cell decomposition}, the ground-state wavefunction of the system can be decomposed to the direct products of wavefunctions of lower-dimensional block states as:
\begin{align}
|\Psi\rangle=\bigotimes\limits_{g\in C_{4v}}|T_{g\lambda}\rangle\otimes\sum\limits_{j=1}^2|\gamma_{g\sigma_{j}}\rangle\otimes|\beta_\tau\rangle
\label{C3v wavefunction}
\end{align}
where $|T_{g\lambda}\rangle$ is the wavefunction of 3D block state on $g\lambda$ which is topological trivial; $|\gamma_{g\sigma_{1,2}}\rangle$ is the $\mathbb{Z}_2$-symmetric wavefunction of 2D block state on $g\sigma_{1,2}$; $|\beta_\tau\rangle$ is the $(\mathbb{Z}_3\rtimes\mathbb{Z}_2)$-symmetric wavefunction of 1D block state $\tau$.

\begin{figure}
\begin{tikzpicture}[scale=0.6]
\tikzstyle{sergio}=[rectangle,draw=none]
\draw[thick] (-6.5,1) -- (-8.5,0);
\draw[thick] (-6.5,1) -- (-4.5,1);
\draw[thick] (-3.5,0) -- (-4.5,1);
\draw[thick] (-3.5,0) -- (-5.5,-1);
\draw[thick] (-7.5,-1) -- (-5.5,-1);
\draw[thick] (-8.5,0) -- (-7.5,-1);
\draw[thick] (-8.5,0) -- (-8.5,-4);
\draw[thick] (-7.5,-5) -- (-8.5,-4);
\draw[thick] (-7.5,-5) -- (-5.5,-5);
\draw[thick] (-3.5,-4) -- (-5.5,-5);
\draw[thick] (-3.5,-4) -- (-3.5,0);
\draw[densely dashed,thick] (-3.5,-4) -- (-4.5,-3);
\draw[densely dashed,thick] (-6.5,-3) -- (-4.5,-3);
\draw[densely dashed,thick] (-6.5,-3) -- (-8.5,-4);
\draw[densely dashed,thick] (-6.5,-3) -- (-6.5,1);
\draw[densely dashed,thick] (-4.5,-3) -- (-4.5,1);
\draw[thick] (-5.5,-1) -- (-5.5,-5);
\draw[thick] (-7.5,-1) -- (-7.5,-5);
\filldraw[fill=green!20, draw=green, thick, fill opacity=0.5] (-5.5,-5)--(-6.5,-3)--(-6.5,1)--(-5.5,-1)--cycle;
\filldraw[fill=green!20, draw=green, thick, fill opacity=0.5] (-4.5,-3)--(-7.5,-5)--(-7.5,-1)--(-4.5,1)--cycle;
\filldraw[fill=green!20, draw=green, thick, fill opacity=0.5] (-3.5,-4)--(-8.5,-4)--(-8.5,0)--(-3.5,0)--cycle;
\draw[color=blue,ultra thick] (-6,0) -- (-6,-4);
\filldraw[fill=red!20, draw=red, thick, fill opacity=0.5] (1.5,-4.75)--(-0.5,-4.75)--(1.5,-3.25)--cycle;
\filldraw[fill=red!20, draw=red, thick, fill opacity=0.5] (2.5,-4.75)--(4.5,-4.75)--(2.5,-3.25)--cycle;
\filldraw[fill=red!20, draw=red, thick, fill opacity=0.5] (-1.25,-3.5)--(-1.25,-0.5)--(0.75,-2)--cycle;
\filldraw[fill=green!20, draw=green, thick] (2.25,-4.75)--(1.75,-4.75)--(1.75,-3.25)--(2.25,-3.25)--cycle;
\filldraw[fill=green!20, draw=green, thick] (-1.15,-4)--(-0.75,-4.5)--(1.25,-3)--(0.85,-2.5)--cycle;
\filldraw[fill=blue!20, draw=blue, thick] (2,-2)circle (13pt);
\filldraw[rotate around={286:(2,-2.25))},fill=green!20, draw=green, thick] (-1.4,-4.25)--(-1,-4.75)--(1,-3.25)--(0.6,-2.75)--cycle;
\filldraw[fill=red!20, draw=red, thick, fill opacity=0.5] (1.5,0.75)--(-0.5,0.75)--(1.5,-0.75)--cycle;
\filldraw[fill=green!20, draw=green, thick] (2.25,-0.75)--(1.75,-0.75)--(1.75,0.75)--(2.25,0.75)--cycle;
\filldraw[fill=red!20, draw=red, thick, fill opacity=0.5] (2.5,0.75)--(4.5,0.75)--(2.5,-0.75)--cycle;
\filldraw[fill=green!20, draw=green, thick] (4.75,0.5)--(2.75,-1)--(3.15,-1.5)--(5.15,0)--cycle;
\filldraw[rotate around={180:(2,-2)},fill=red!20, draw=red, thick, fill opacity=0.5] (-1.25,-3.5)--(-1.25,-0.5)--(0.75,-2)--cycle;
\filldraw[rotate around={287:(2,-2)},fill=green!20, draw=green, thick] (4.75,0.5)--(2.75,-1)--(3.15,-1.5)--(5.15,0)--cycle;
\path (2,-2) node [style=sergio] {$\mu$};
\path (-0.5,-2) node [style=sergio] {$\lambda$};
\path (4.5,-2) node [style=sergio] {$\lambda$};
\path (3,-4.25) node [style=sergio] {$\lambda$};
\path (1,-4.25) node [style=sergio] {$\lambda$};
\path (3,0.25) node [style=sergio] {$\lambda$};
\path (1,0.25) node [style=sergio] {$\lambda$};
\path (0,-0.6) node [style=sergio] {\scriptsize$\sigma_2$};
\path (4,-0.5) node [style=sergio] {\scriptsize$\sigma_2$};
\path (4,-3.45) node [style=sergio] {\scriptsize$\sigma_1$};
\path (2,-4) node [style=sergio] {\scriptsize$\sigma_2$};
\path (2,0) node [style=sergio] {\scriptsize$\sigma_1$};
\path (0,-3.55) node [style=sergio] {\scriptsize$\sigma_1$};
\path (-5.806,0.5155) node [style=sergio] {$C_3$};
\path (-4.5,1.35) node [style=sergio] {$M_1$};
\path (-6.5,1.35) node [style=sergio] {$M_2$};
\end{tikzpicture}
\caption{The cell decomposition of $C_{3v}$-symmetric lattice. Left panel depicts the whole lattice, right panel depicts the horizontal intersection, including 3D blocks $\lambda$, 2D blocks $\sigma_{1,2}$ and 1D block $\tau$. $C_3$ depicts the axis of 3-fold rotation, and $\bs{M}_{1,2}$ depicts the reflection planes.}
\label{C3v cell decomposition}
\end{figure}

\begin{figure*}
\begin{tikzpicture}[scale=0.85]
\tikzstyle{sergio}=[rectangle,draw=none]
\draw[thick] (-13,1) -- (-15,0);
\draw[thick] (-13,1) -- (-11,1);
\draw[thick] (-10,0) -- (-11,1);
\draw[thick] (-10,0) -- (-12,-1);
\draw[thick] (-14,-1) -- (-12,-1);
\draw[thick] (-15,0) -- (-14,-1);
\draw[thick] (-15,0) -- (-15,-4);
\draw[thick] (-14,-5) -- (-15,-4);
\draw[thick] (-14,-5) -- (-12,-5);
\draw[thick] (-10,-4) -- (-12,-5);
\draw[thick] (-10,-4) -- (-10,0);
\draw[densely dashed,thick] (-10,-4) -- (-11,-3);
\draw[densely dashed,thick] (-13,-3) -- (-11,-3);
\draw[densely dashed,thick] (-13,-3) -- (-15,-4);
\draw[densely dashed,thick] (-13,-3) -- (-13,1);
\draw[densely dashed,thick] (-11,-3) -- (-11,1);
\draw[thick] (-12,-1) -- (-12,-5);
\draw[thick] (-14,-1) -- (-14,-5);
\filldraw[fill=green!20, draw=green, thick, fill opacity=0.5] (-12,-5)--(-13,-3)--(-13,1)--(-12,-1)--cycle;
\filldraw[fill=green!20, draw=green, thick, fill opacity=0.5] (-11,-3)--(-14,-5)--(-14,-1)--(-11,1)--cycle;
\filldraw[fill=green!20, draw=green, thick, fill opacity=0.5] (-10,-4)--(-15,-4)--(-15,0)--(-10,0)--cycle;
\draw[color=blue,ultra thick] (-12.5,0) -- (-12.5,-4);
\draw[color=blue,ultra thick] (-12,-1) -- (-13,-3);
\draw[color=blue,ultra thick] (-11,-2.5) -- (-14,-1.5);
\draw[color=blue,ultra thick] (-10.5,-1.5) -- (-14.5,-2.5);
\draw[color=blue,ultra thick] (-14,-3) -- (-11,-1);
\draw[color=blue,ultra thick] (-13,-1) -- (-12,-3);
\draw[color=blue,ultra thick] (-15,-2) -- (-10,-2);
\filldraw[fill=yellow!20, draw=yellow, thick, fill opacity=0.5] (-10,-2)--(-11,-1)--(-13,-1)--(-15,-2)--(-14,-3)--(-12,-3)--cycle;
\filldraw[fill=white!20, draw=black, thick] (2,-2)circle (13pt);
\filldraw[fill=blue!20, draw=blue, thick] (2.25,-0.75)--(1.75,-0.75)--(1.75,1.25)--(2.25,1.25)--cycle;
\filldraw[rotate around={30:(2,-2)},fill=blue!20, draw=blue, thick] (2.25,-0.75)--(1.75,-0.75)--(1.75,1.25)--(2.25,1.25)--cycle;
\filldraw[rotate around={60:(2,-2)},fill=blue!20, draw=blue, thick] (2.25,-0.75)--(1.75,-0.75)--(1.75,1.25)--(2.25,1.25)--cycle;
\filldraw[rotate around={90:(2,-2)},fill=blue!20, draw=blue, thick] (2.25,-0.75)--(1.75,-0.75)--(1.75,1.25)--(2.25,1.25)--cycle;
\filldraw[rotate around={120:(2,-2)},fill=blue!20, draw=blue, thick] (2.25,-0.75)--(1.75,-0.75)--(1.75,1.25)--(2.25,1.25)--cycle;
\filldraw[rotate around={150:(2,-2)},fill=blue!20, draw=blue, thick] (2.25,-0.75)--(1.75,-0.75)--(1.75,1.25)--(2.25,1.25)--cycle;
\filldraw[rotate around={180:(2,-2)},fill=blue!20, draw=blue, thick] (2.25,-0.75)--(1.75,-0.75)--(1.75,1.25)--(2.25,1.25)--cycle;
\filldraw[rotate around={210:(2,-2)},fill=blue!20, draw=blue, thick] (2.25,-0.75)--(1.75,-0.75)--(1.75,1.25)--(2.25,1.25)--cycle;
\filldraw[rotate around={240:(2,-2)},fill=blue!20, draw=blue, thick] (2.25,-0.75)--(1.75,-0.75)--(1.75,1.25)--(2.25,1.25)--cycle;
\filldraw[rotate around={270:(2,-2)},fill=blue!20, draw=blue, thick] (2.25,-0.75)--(1.75,-0.75)--(1.75,1.25)--(2.25,1.25)--cycle;
\filldraw[rotate around={300:(2,-2)},fill=blue!20, draw=blue, thick] (2.25,-0.75)--(1.75,-0.75)--(1.75,1.25)--(2.25,1.25)--cycle;
\filldraw[rotate around={330:(2,-2)},fill=blue!20, draw=blue, thick] (2.25,-0.75)--(1.75,-0.75)--(1.75,1.25)--(2.25,1.25)--cycle;
\path (2,-2) node [style=sergio] {$\mu$};
\path (2,0.25) node [style=sergio] {\scriptsize$\tau_2$};
\path[rotate around={30:(2,-2)}] (2,0.25) node [style=sergio] {\scriptsize$\tau_3$};
\path[rotate around={60:(2,-2)}] (2,0.25) node [style=sergio] {\scriptsize$\tau_2$};
\path[rotate around={90:(2,-2)}] (2,0.25) node [style=sergio] {\scriptsize$\tau_3$};
\path[rotate around={120:(2,-2)}] (2,0.25) node [style=sergio] {\scriptsize$\tau_2$};
\path[rotate around={150:(2,-2)}] (2,0.25) node [style=sergio] {\scriptsize$\tau_3$};
\path[rotate around={180:(2,-2)}] (2,0.25) node [style=sergio] {\scriptsize$\tau_2$};
\path[rotate around={210:(2,-2)}] (2,0.25) node [style=sergio] {\scriptsize$\tau_3$};
\path[rotate around={240:(2,-2)}] (2,0.25) node [style=sergio] {\scriptsize$\tau_2$};
\path[rotate around={270:(2,-2)}] (2,0.25) node [style=sergio] {\scriptsize$\tau_3$};
\path[rotate around={300:(2,-2)}] (2,0.25) node [style=sergio] {\scriptsize$\tau_2$};
\path[rotate around={330:(2,-2)}] (2,0.25) node [style=sergio] {\scriptsize$\tau_3$};
\filldraw[fill=green!20, draw=green, thick, fill opacity=0.5] (1.5,1.25)--(0.75,1)--(1.5,-0.25)--cycle;
\filldraw[rotate around={30:(2,-2)},fill=green!20, draw=green, thick, fill opacity=0.5] (1.5,1.25)--(0.75,1)--(1.5,-0.25)--cycle;
\filldraw[rotate around={60:(2,-2)},fill=green!20, draw=green, thick, fill opacity=0.5] (1.5,1.25)--(0.75,1)--(1.5,-0.25)--cycle;
\filldraw[rotate around={90:(2,-2)},fill=green!20, draw=green, thick, fill opacity=0.5] (1.5,1.25)--(0.75,1)--(1.5,-0.25)--cycle;
\filldraw[rotate around={120:(2,-2)},fill=green!20, draw=green, thick, fill opacity=0.5] (1.5,1.25)--(0.75,1)--(1.5,-0.25)--cycle;
\filldraw[rotate around={150:(2,-2)},fill=green!20, draw=green, thick, fill opacity=0.5] (1.5,1.25)--(0.75,1)--(1.5,-0.25)--cycle;
\filldraw[rotate around={180:(2,-2)},fill=green!20, draw=green, thick, fill opacity=0.5] (1.5,1.25)--(0.75,1)--(1.5,-0.25)--cycle;
\filldraw[rotate around={210:(2,-2)},fill=green!20, draw=green, thick, fill opacity=0.5] (1.5,1.25)--(0.75,1)--(1.5,-0.25)--cycle;
\filldraw[rotate around={240:(2,-2)},fill=green!20, draw=green, thick, fill opacity=0.5] (1.5,1.25)--(0.75,1)--(1.5,-0.25)--cycle;
\filldraw[rotate around={270:(2,-2)},fill=green!20, draw=green, thick, fill opacity=0.5] (1.5,1.25)--(0.75,1)--(1.5,-0.25)--cycle;
\filldraw[rotate around={300:(2,-2)},fill=green!20, draw=green, thick, fill opacity=0.5] (1.5,1.25)--(0.75,1)--(1.5,-0.25)--cycle;
\filldraw[rotate around={330:(2,-2)},fill=green!20, draw=green, thick, fill opacity=0.5] (1.5,1.25)--(0.75,1)--(1.5,-0.25)--cycle;
\path (1.2,0.75) node [style=sergio] {\scriptsize$\sigma_3$};
\path[rotate around={30:(2,-2)}] (1.2,0.75) node [style=sergio] {\scriptsize$\sigma_3$};
\path[rotate around={60:(2,-2)}] (1.2,0.75) node [style=sergio] {\scriptsize$\sigma_3$};
\path[rotate around={90:(2,-2)}] (1.2,0.75) node [style=sergio] {\scriptsize$\sigma_3$};
\path[rotate around={120:(2,-2)}] (1.2,0.75) node [style=sergio] {\scriptsize$\sigma_3$};
\path[rotate around={150:(2,-2)}] (1.2,0.75) node [style=sergio] {\scriptsize$\sigma_3$};
\path[rotate around={180:(2,-2)}] (1.2,0.75) node [style=sergio] {\scriptsize$\sigma_3$};
\path[rotate around={210:(2,-2)}] (1.2,0.75) node [style=sergio] {\scriptsize$\sigma_3$};
\path[rotate around={240:(2,-2)}] (1.2,0.75) node [style=sergio] {\scriptsize$\sigma_3$};
\path[rotate around={270:(2,-2)}] (1.2,0.75) node [style=sergio] {\scriptsize$\sigma_3$};
\path[rotate around={300:(2,-2)}] (1.2,0.75) node [style=sergio] {\scriptsize$\sigma_3$};
\path[rotate around={330:(2,-2)}] (1.2,0.75) node [style=sergio] {\scriptsize$\sigma_3$};
\filldraw[fill=blue!20, draw=blue, thick] (-5.5,-2)circle (13pt);
\filldraw[fill=green!20, draw=green, thick] (-5.25,-1)--(-5.75,-1)--(-5.75,1)--(-5.25,1)--cycle;
\filldraw[rotate around={60:(-5.5,-2)},fill=green!20, draw=green, thick] (-5.25,-1)--(-5.75,-1)--(-5.75,1)--(-5.25,1)--cycle;
\filldraw[rotate around={120:(-5.5,-2)},fill=green!20, draw=green, thick] (-5.25,-1)--(-5.75,-1)--(-5.75,1)--(-5.25,1)--cycle;
\filldraw[rotate around={180:(-5.5,-2)},fill=green!20, draw=green, thick] (-5.25,-1)--(-5.75,-1)--(-5.75,1)--(-5.25,1)--cycle;
\filldraw[rotate around={240:(-5.5,-2)},fill=green!20, draw=green, thick] (-5.25,-1)--(-5.75,-1)--(-5.75,1)--(-5.25,1)--cycle;
\filldraw[rotate around={300:(-5.5,-2)},fill=green!20, draw=green, thick] (-5.25,-1)--(-5.75,-1)--(-5.75,1)--(-5.25,1)--cycle;
\filldraw[fill=red!20, draw=red, thick] (-7.9,-0)--(-6,1)--(-6,-1.1)--cycle;
\filldraw[rotate around={60:(-5.5,-2)},fill=red!20, draw=red, thick] (-7.9,-0)--(-6,1)--(-6,-1.1)--cycle;
\filldraw[rotate around={120:(-5.5,-2)},fill=red!20, draw=red, thick] (-7.9,-0)--(-6,1)--(-6,-1.1)--cycle;
\filldraw[rotate around={180:(-5.5,-2)},fill=red!20, draw=red, thick] (-7.9,-0)--(-6,1)--(-6,-1.1)--cycle;
\filldraw[rotate around={240:(-5.5,-2)},fill=red!20, draw=red, thick] (-7.9,-0)--(-6,1)--(-6,-1.1)--cycle;
\filldraw[rotate around={300:(-5.5,-2)},fill=red!20, draw=red, thick] (-7.9,-0)--(-6,1)--(-6,-1.1)--cycle;
\path (-4.25,0) node [style=sergio] {\scriptsize$\lambda$};
\path[rotate around={60:(-5.5,-2)}] (-4.25,0) node [style=sergio] {\scriptsize$\lambda$};
\path[rotate around={120:(-5.5,-2)}] (-4.25,0) node [style=sergio] {\scriptsize$\lambda$};
\path[rotate around={180:(-5.5,-2)}] (-4.25,0) node [style=sergio] {\scriptsize$\lambda$};
\path[rotate around={240:(-5.5,-2)}] (-4.25,0) node [style=sergio] {\scriptsize$\lambda$};
\path[rotate around={300:(-5.5,-2)}] (-4.25,0) node [style=sergio] {\scriptsize$\lambda$};
\path (-5.5,0) node [style=sergio] {\scriptsize$\sigma_1$};
\path[rotate around={60:(-5.5,-2)}] (-5.5,0) node [style=sergio] {\scriptsize$\sigma_2$};
\path[rotate around={120:(-5.5,-2)}] (-5.5,0) node [style=sergio] {\scriptsize$\sigma_1$};
\path[rotate around={180:(-5.5,-2)}] (-5.5,0) node [style=sergio] {\scriptsize$\sigma_2$};
\path[rotate around={240:(-5.5,-2)}] (-5.5,0) node [style=sergio] {\scriptsize$\sigma_1$};
\path[rotate around={300:(-5.5,-2)}] (-5.5,0) node [style=sergio] {\scriptsize$\sigma_2$};
\path (-5.5,-2) node [style=sergio] {$\tau_1$};
\path (-12.4,0.5) node [style=sergio] {$S_6$};
\path (-13,1.35) node [style=sergio] {$M_1$};
\path (-11,1.35) node [style=sergio] {$M_2$};
\path (-14.4,-1.4) node [style=sergio] {$C_2$};
\path (-13.3347,-0.8395) node [style=sergio] {$C_2$};
\path (-11.7898,-0.7377) node [style=sergio] {$C_2$};
\path (-10.6822,-0.8058) node [style=sergio] {$C_2$};
\path (-10.2408,-1.315) node [style=sergio] {$C_2$};
\path (-9.6169,-1.994) node [style=sergio] {$C_2$};
\path (-12.5,-6) node [style=sergio] {Open lattice};
\path (-5.5,-6) node [style=sergio] {North/south hemisphere};
\path (2,-6) node [style=sergio] {Equator};
\end{tikzpicture}
\caption{The cell decomposition of $D_{3d}$-symmetric lattice, with the yellow plate representing the plane of 6-fold improper rotation. Left panel depicts the whole lattice; middle panel depicts north/south hemisphere, including 3D blocks $\lambda$, 2D blocks $\sigma_{1,2}$ and 1D blocks $\tau_1$; right panel depicts the equator, including 2D blocks $\sigma_3$, 1D blocks $\tau_{2,3}$ and 0D block $\mu$. $S_6$ depicts the axis of the 6-fold rotoreflection, $\bs{M}_{1,2}$ depict the reflection planes and $C_2$'s depict the axes of the 2-fold rotations.}
\label{D3d cell decomposition}
\end{figure*}

We summarize the classifications and corresponding root phases. For crystalline TSC with spinless fermions, we summarize all possible block states as:
\begin{enumerate}[1.]
\item 2D blocks $\sigma_1$ and $\sigma_2$: 2D $(p+ip)$-SCs and 2D $\mathbb{Z}_2$ fSPT phases;
\item 1D block $\tau$: Majorana chain and 1D $\mathbb{Z}_3\rtimes\mathbb{Z}_2$ fSPT phase.
\end{enumerate}
The chiralities of the decorated 2D $(p+ip)$-SCs on $\sigma_1$ and $\sigma_2$ should be opposite to guarantee the nonchiral condition on $\tau$. It is easy to see that Majorana chain decoration on $\tau$ can be trivialized by ``Majorana'' bubble construction on each $\sigma_1$, and 1D $\mathbb{Z}_3\rtimes\mathbb{Z}_2$ fSPT phase on $\tau$ can be trivialized by  1D $\mathbb{Z}_2$ fSPT phase as 2D bubble construction on each $\sigma_1$, and even layers of 2D $(p+ip)$-SCs is trivialized by 3D bubble equivalence on $\lambda$. Therefore, the ultimate classification is $\mathbb{Z}_{16}$, with two root phases:
\begin{enumerate}[1.]
\item Monolayer $(p+ip)$-SC on each $\sigma_1$ and $\sigma_2$ ($\mathbb{Z}_2$);
\item 2D fermionic Levin-Gu state with $\nu=1\in\mathbb{Z}_8$ on each $\sigma_1$ and $\sigma_2$ ($\mathbb{Z}_8$).
\end{enumerate}
and there is a nontrivial extension between these two root phases. The second-order topological surface theories of different root phases are:
\begin{enumerate}[1.]
\item Chiral Majorana modes on the intersections between the open lattice and 2D blocks $\sigma_1$ and $\sigma_2$ (i.e., verticle hinges of the open hexagonal prism in Fig. \ref{C3v cell decomposition});
\item 1D nonchiral Luttinger liquids with $K$-matrix $K=\sigma^z$ and $\mathbb{Z}_2$ symmetry property $W^{\mathbb{Z}_2}=\sigma^z$ and $\delta\phi^{\mathbb{Z}_2}=0$, on the verticle hinges of the open hexagonal prism in Fig. \ref{C3v cell decomposition}.
\end{enumerate}

For crystalline TSC with spin-1/2 fermions, the corresponding classification is trivial because there is no nontrivial possible block state.

For crystalline TI, similar to the $C_2$-symmetric lattice, the spin of fermions is irrelevant. All possible block states are on the 2D blocks $\sigma_1$ and $\sigma_2$: Chern insulators, Kitaev's $E_8$ states and 2D $U^f(1)\times\mathbb{Z}_2$ fSPT phases. The chiral block states on $\sigma_1$ and $\sigma_2$ should have opposite chiralities to warrant the nonchiral condition on $\tau$. Furthermore, even layers of chiral block states are trivialized by 3D bubble equivalences on $\lambda$. Therefore, the ultimate classification is $\mathbb{Z}_8\times\mathbb{Z}_2$, with the following root phases:
\begin{enumerate}[1.]
\item Monolayer Chern insulator on each $\sigma_1$ and $\sigma_2$ ($\mathbb{Z}_2$);
\item 2D $U^f(1)\times\mathbb{Z}_2$ fSPT phase with $\nu=1\in\mathbb{Z}_4$ on each $\sigma_1$ and $\sigma_2$ ($\mathbb{Z}_4$);
\item Monolayer Kitaev's $E_8$ state on each $\sigma_1$ and $\sigma_2$ ($\mathbb{Z}_2$).
\end{enumerate}
And there is a nontrivial extension between the first two root phases: bilayer Chern insulators on each $\sigma_1$ and $\sigma_2$ can be smoothly deformed to 2D $U^f(1)\times\mathbb{Z}_2$ fSPT phases on $\sigma_1$ and $\sigma_2$ with quantum numbers $(\nu_1,\nu_2)=(1,3)$. The second-order topological surface theories of these root phases are:
\begin{enumerate}[1.]
\item Chiral fermions on the verticle hinges of the open hexagonal prism in Fig. \ref{C3v cell decomposition};
\item 1D nonchiral Luttinger liquids with $K$-matrix $K=\sigma^z$ and $\mathbb{Z}_2$ symmetry property: $W^{\mathbb{Z}_2}=\mathbbm{1}_{2\times2}$ and $\delta\phi^{\mathbb{Z}_2}=\pi(0,1)^T/\pi(1,0)^T$, on the verticle hinges of the open hexagonal prism in Fig. \ref{C3v cell decomposition};
\item 1D chiral Luttinger liquids with $K$-matrix (\ref{K-matrix E8S}) on the verticle hinges of the open hexagonal prism in Fig. \ref{C3v cell decomposition}.
\end{enumerate}

\subsection{$D_{3d}$-symmetric lattice}
For $D_{3d}$-symmetric lattice with the cell decomposition in Fig. \ref{D3d cell decomposition}, the ground-state wavefunction of the system can be decomposed to the direct products of wavefunctions of lower-dimensional block states as:
\begin{align}
|\Psi\rangle=\bigotimes\limits_{g\in D_{2d}}|T_{g\lambda}\rangle\otimes\sum\limits_{k=1}^3|\gamma_{g\sigma_k}\rangle\otimes\sum\limits_{j=1}^3|\beta_{g\tau_j}\rangle\otimes|\alpha_\mu\rangle
\label{D3d cell decomposition}
\end{align}
where $|T_{g\lambda}\rangle$ is the wavefunction of 3D block state on $g\lambda$ which is topological trivial; $|\gamma_{g\sigma_{1,2}}\rangle$ is the $\mathbb{Z}_2$-symmetric wavefunction of 2D block state on $g\sigma_{1,2}$, and $|\gamma_{g\sigma_3}\rangle$ is the wavefunction of 2D block state on $g\sigma_3$ which is topological trivial or invertible topological phase; $|\beta_{g\tau_1}\rangle$ is the $(\mathbb{Z}_3\rtimes\mathbb{Z}_2)$-symmetric wavefunction of 1D block state on $g\tau_1$, and $|\beta_{g\tau_{2,3}}\rangle$ is the $\mathbb{Z}_2$-symmetric wavefunction of 1D block state on $g\tau_{2,3}$; $|\alpha_\mu\rangle$ is the $(\mathbb{Z}_6\rtimes\mathbb{Z}_2)$-symmetric wavefunction of 0D block state on $\mu$.

We summarize the classifications and corresponding root phases. For crystalline TSC with spinless fermions, we summarize all possible block states as:
\begin{enumerate}[1.]
\item 2D blocks $\sigma_1$ and $\sigma_2$: 2D $(p+ip)$-SCs and 2D $\mathbb{Z}_2$ fSPT phases;
\item 2D blocks $\sigma_3$: 2D $(p+ip)$-SC;
\item 1D block $\tau_1$: Majorana chain and 1D $\mathbb{Z}_3\rtimes\mathbb{Z}_2$ fSPT phase;
\item 1D blocks $\tau_2$ and $\tau_3$: Majorana chains and 1D $\mathbb{Z}_2$ fSPT phases.
\item 0D block $\mu$: 0D modes characterizing eigenvalues of two generators of the $D_{3d}$ group, with even/odd fermion parity.
\end{enumerate}
By explicit $K$-matrix calculations, except 2D bosonic Levin-Gu state on each $\sigma_1$ and $\sigma_2$, all other 2D block states are obstructed. For 1D block $\tau_1$, both Majorana chain decoration and 1D $\mathbb{Z}_3\rtimes\mathbb{Z}_2$ fSPT phase decoration are obstructed: Majorana chain decoration leaves two Majorana zero modes at $\mu$, whose fermion parity is odd under $\bs{I}_r\in S_6\subset D_{3d}$. There is an exception: Consider the Majorana chain decoration on $\tau_1$ and $\tau_2$ simultaneously, it will leaves 8 Majorana zero modes near $\mu$, with the following symmetry properties:
\begin{align}
S_6:~
\begin{gathered}
\gamma_1\leftrightarrow\gamma_2\\
\gamma_{3,4,5,6,7,8}\mapsto\gamma_{4,5,6,7,8,3}
\end{gathered},
\end{align}
\begin{align}
C_2:~
\begin{gathered}
\gamma_{1}\leftrightarrow\gamma_2\\
\gamma_{3,4,5,6,7,8}\leftrightarrow\gamma_{3,8,7,6,5,4}
\end{gathered},
\end{align}
and they form a linear representation of the $D_{3d}\times\Z_2^f$ group. Therefore, Majorana chain decoration on $\tau_1$ and $\tau_2$ is a valid block-state. 1D $\mathbb{Z}_3\rtimes\mathbb{Z}_2$ fSPT phase decoration leaves four Majorana zero modes at $\mu$ that can be treated as a spin-1/2 degree of freedom, who is exactly the projective representation of the $D_{3d}$ group at $\mu$. From Ref. \cite{dihedral} we know that for the equator, $D_{3d}$ acts identical to the 2D $D_{6}$ group, and there is no nontrivial in-plane 1D block state. Furthermore, 0D modes characterizing eigenvalues $-1$ of two generators of the $D_{3d}$ group are trivialized by 1D bubble equivalences on $\tau_2$ and $\tau_3$, respectively. Hence the ultimate classification is $\mathbb{Z}_2^3$, with the following root phases:
\begin{enumerate}[1.]
\item 2D bosonic Levin-Gu state on each $\sigma_1$ and $\sigma_2$ ($\mathbb{Z}_2$);
\item Majorana chain on each $\tau_1$ and $\tau_2$ ($\mathbb{Z}_2$);
\item 0D mode with odd fermion parity on $\mu$ ($\mathbb{Z}_2$).
\end{enumerate}
The second-order topological surface theories of different root phases are 1D nonchiral Luttinger liquids with $K$-matrix $K=\sigma^x$ and $\mathbb{Z}_2$ symmetry property $W^{\mathbb{Z}_2}=\mathbbm{1}_{2\times2}$ and $\delta\phi^{\mathbb{Z}_2}=\pi(1,1)^T$, on the intersections between the open lattice and 2D blocks $\sigma_1/\sigma_2$. The third order topological surface states of Majorana chain decoration are dangling Majorana zero mode at each center of top and bottom surface, and the intersections between $\tau_2$ and the open boundary.

For crystalline TSC with spin-1/2 fermions, we summarize all possible block states:
\begin{enumerate}[1.]
\item 2D block $\sigma_3$: 2D $(p+ip)$-SC;
\item 0D block $\mu$: 0D modes characterizing the eigenvalues $-1$ of two generators of the $D_{3d}$ group.
\end{enumerate}
The 2D block state is obstructed because of the violation of the nonchiral condition on $\tau_2$ and $\tau_3$, and the 0D block states can be trivialized by some 1D bubble construction: firstly, the $-1$ eigenvalue of the inversion which is composed of a 2-fold rotation and another 2-fold rotation whose axis is on the equator can be trivialized by a complex fermion on each $\tau_1$, and the eigenvalues of 6-fold rotation can be trivialized by a generator of $\mathbb{Z}_{12}^f$. Therefore, the ultimate classification is trivial.

For crystalline TI with spinless fermions, we summarize all possible block states as:
\begin{enumerate}[1.]
\item 2D blocks $\sigma_1$ and $\sigma_2$: Chern insulators, Kitaev's $E_8$ states and 2D $U^f(1)\times\mathbb{Z}_2$ fSPT phases;
\item 2D blocks $\sigma_3$: Chern insulators and Kitaev's $E_8$ states;
\item 0D block $\mu$: 0D modes characterizing the eigenvalues $-1$ of two generators of the $D_{3d}$ group, with different $U^f(1)$ charge.
\end{enumerate}
By explicit $K$-matrix calculations, except 2D bosonic Levin-Gu state on each $\sigma_1$ and $\sigma_2$, and Kitaev's $E_8$ states decoration on $\sigma_1$ and $\sigma_2$ with opposite chiralities, all other 2D block states are obstructed. For 0D block $\mu$, 0D modes characterizing the eigenvalues $-1$ of two generators of the $D_{3d}$ group, together with $4n$ $U^f(1)$ charge ($n\in\mathbb{Z}$), are trivialized by 1D bubble equivalences. Therefore, the ultimate classification is $\mathbb{Z}_4\times\mathbb{Z}_2^2$, with the following root phases:
\begin{enumerate}[1.]
\item Monolayer Kitaev's $E_8$ state on each $\sigma_1$ and $\sigma_2$ ($\mathbb{Z}_2$);
\item 2D $U^f(1)\times\mathbb{Z}_2$ bSPT phase on each $\sigma_1$ and $\sigma_2$ ($\mathbb{Z}_2$);
\item 0D mode with $U^f(1)$ charge $n\equiv0,1,2,3~(\mathrm{mod}~4)$ on $\mu$ ($\mathbb{Z}_4$).
\end{enumerate}
The second-order topological surface theories of these root phases are:
\begin{enumerate}[1.]
\item 1D chiral Luttinger liquids with $K$-matrix (\ref{K-matrix E8S}) on all verticle hinges of the open lattice in Fig. \ref{D3d cell decomposition};
\item 1D nonchiral Luttinger liquids with $K$-matrix $K=\sigma^x$ and $\mathbb{Z}_2$ symmetry property $W^{\mathbb{Z}_2}=\mathbbm{1}_{2\times2}$ and $\delta\phi^{\mathbb{Z}_2}=\pi(1,1)^T$, on the intersections between the open lattice and 2D blocks $\sigma_1/\sigma_2$.
\end{enumerate}

For crystalline TI with spin-1/2 fermions, all possible block states are summarized as following: 
\begin{enumerate}[1.]
\item 2D blocks $\sigma_1$ and $\sigma_2$: Chern insulators, Kitaev's $E_8$ states and 2D $U^f(1)\times\mathbb{Z}_2$ fSPT phases;
\item 2D blocks $\sigma_3$: Chern insulators and Kitaev's $E_8$ states;
\item 0D block $\mu$: 0D modes characterizing eigenvalues $-1$ of two generators of the $D_{3d}$ group, with different even $U^f(1)$ charge.
\end{enumerate}
The chiral block states on $\sigma_1$ and $\sigma_2$ should have opposite chiralities to guarantee the non-chirality of the 1D block $\tau_1$, and all block states on $\sigma_3$ are obstructed. Furthermore, even layers of chiral block states are trivialized by 3D bubble equivalences on $\lambda$, and 0D modes with even $U^f(1)$ charge are trivialized by 1D bubble equivalences. Therefore, the ultimate classification is $\mathbb{Z}_8\times\mathbb{Z}_2^3$, with the following root phases:
\begin{enumerate}[1.]
\item Monolayer Chern insulator on each $\sigma_1$ and $\sigma_2$ ($\mathbb{Z}_2$);
\item 2D $U^f(1)\times\mathbb{Z}_2$ fSPT phase with $(\nu_1,\nu_2)=(1,3)\in\mathbb{Z}_4^2$ on each $\sigma_1$ and $\sigma_2$ ($\mathbb{Z}_4$);
\item Monolayer Kitaev's $E_8$ state on each $\sigma_1$ and $\sigma_2$ ($\mathbb{Z}_2$);
\item 0D modes with eigenvalues $-1$ of two generators of $D_6$ symmetry on the equator ($\mathbb{Z}_2^2$).
\end{enumerate}
and there is a nontrivial extension between the first two root phases. The second-order topological surface theories of these root phases are:
\begin{enumerate}[1.]
\item Chiral fermions on all verticle hinges of the open lattice;
\item 1D nonchiral Luttinger liquids with $K$-matrix $K=\sigma^z$ and $\mathbb{Z}_2$ symmetry property $W^{\mathbb{Z}_2}=\mathbbm{1}_{2\times2}$ and $\delta\phi^{\mathbb{Z}_2}=\pi(0,1)^T$, on all verticle hinges of the open lattice;
\item 1D chiral Luttinger liquids with $K$-matrix (\ref{K-matrix E8S}) on all verticle hinges of the open lattices.
\end{enumerate}

\begin{figure}
\begin{tikzpicture}[scale=0.65]
\tikzstyle{sergio}=[rectangle,draw=none]
\draw[thick] (-13,1) -- (-15,0);
\draw[thick] (-13,1) -- (-11,1);
\draw[thick] (-10,0) -- (-11,1);
\draw[thick] (-10,0) -- (-12,-1);
\draw[thick] (-14,-1) -- (-12,-1);
\draw[thick] (-15,0) -- (-14,-1);
\draw[thick] (-15,0) -- (-15,-4);
\draw[thick] (-14,-5) -- (-15,-4);
\draw[thick] (-14,-5) -- (-12,-5);
\draw[thick] (-10,-4) -- (-12,-5);
\draw[thick] (-10,-4) -- (-10,0);
\draw[densely dashed,thick] (-10,-4) -- (-11,-3);
\draw[densely dashed,thick] (-13,-3) -- (-11,-3);
\draw[densely dashed,thick] (-13,-3) -- (-15,-4);
\draw[densely dashed,thick] (-13,-3) -- (-13,1);
\draw[densely dashed,thick] (-11,-3) -- (-11,1);
\draw[thick] (-12,-1) -- (-12,-5);
\draw[thick] (-14,-1) -- (-14,-5);
\filldraw[fill=green!20, draw=green, thick, fill opacity=0.5] (-12,-5)--(-13,-3)--(-13,1)--(-12,-1)--cycle;
\filldraw[fill=green!20, draw=green, thick, fill opacity=0.5] (-11,-3)--(-14,-5)--(-14,-1)--(-11,1)--cycle;
\filldraw[fill=green!20, draw=green, thick, fill opacity=0.5] (-10,-4)--(-15,-4)--(-15,0)--(-10,0)--cycle;
\draw[color=blue,ultra thick] (-12.5,0) -- (-12.5,-4);
\filldraw[fill=blue!20, draw=blue, thick] (-5.5,-2)circle (13pt);
\filldraw[fill=green!20, draw=green, thick] (-5.25,-1)--(-5.75,-1)--(-5.75,1)--(-5.25,1)--cycle;
\filldraw[rotate around={60:(-5.5,-2)},fill=green!20, draw=green, thick] (-5.25,-1)--(-5.75,-1)--(-5.75,1)--(-5.25,1)--cycle;
\filldraw[rotate around={120:(-5.5,-2)},fill=green!20, draw=green, thick] (-5.25,-1)--(-5.75,-1)--(-5.75,1)--(-5.25,1)--cycle;
\filldraw[rotate around={180:(-5.5,-2)},fill=green!20, draw=green, thick] (-5.25,-1)--(-5.75,-1)--(-5.75,1)--(-5.25,1)--cycle;
\filldraw[rotate around={240:(-5.5,-2)},fill=green!20, draw=green, thick] (-5.25,-1)--(-5.75,-1)--(-5.75,1)--(-5.25,1)--cycle;
\filldraw[rotate around={300:(-5.5,-2)},fill=green!20, draw=green, thick] (-5.25,-1)--(-5.75,-1)--(-5.75,1)--(-5.25,1)--cycle;
\filldraw[fill=red!20, draw=red, thick] (-7.9,-0)--(-6,1)--(-6,-1.1)--cycle;
\filldraw[rotate around={60:(-5.5,-2)},fill=red!20, draw=red, thick] (-7.9,-0)--(-6,1)--(-6,-1.1)--cycle;
\filldraw[rotate around={120:(-5.5,-2)},fill=red!20, draw=red, thick] (-7.9,-0)--(-6,1)--(-6,-1.1)--cycle;
\filldraw[rotate around={180:(-5.5,-2)},fill=red!20, draw=red, thick] (-7.9,-0)--(-6,1)--(-6,-1.1)--cycle;
\filldraw[rotate around={240:(-5.5,-2)},fill=red!20, draw=red, thick] (-7.9,-0)--(-6,1)--(-6,-1.1)--cycle;
\filldraw[rotate around={300:(-5.5,-2)},fill=red!20, draw=red, thick] (-7.9,-0)--(-6,1)--(-6,-1.1)--cycle;
\path (-4.25,0) node [style=sergio] {\scriptsize$\lambda$};
\path[rotate around={60:(-5.5,-2)}] (-4.25,0) node [style=sergio] {\scriptsize$\lambda$};
\path[rotate around={120:(-5.5,-2)}] (-4.25,0) node [style=sergio] {\scriptsize$\lambda$};
\path[rotate around={180:(-5.5,-2)}] (-4.25,0) node [style=sergio] {\scriptsize$\lambda$};
\path[rotate around={240:(-5.5,-2)}] (-4.25,0) node [style=sergio] {\scriptsize$\lambda$};
\path[rotate around={300:(-5.5,-2)}] (-4.25,0) node [style=sergio] {\scriptsize$\lambda$};
\path (-5.5,0) node [style=sergio] {\scriptsize$\sigma$};
\path[rotate around={60:(-5.5,-2)}] (-5.5,0) node [style=sergio] {\scriptsize$\sigma$};
\path[rotate around={120:(-5.5,-2)}] (-5.5,0) node [style=sergio] {\scriptsize$\sigma$};
\path[rotate around={180:(-5.5,-2)}] (-5.5,0) node [style=sergio] {\scriptsize$\sigma$};
\path[rotate around={240:(-5.5,-2)}] (-5.5,0) node [style=sergio] {\scriptsize$\sigma$};
\path[rotate around={300:(-5.5,-2)}] (-5.5,0) node [style=sergio] {\scriptsize$\sigma$};
\path (-5.5,-2) node [style=sergio] {$\tau$};
\path (-12.35,0.5) node [style=sergio] {$C_6$};
\path (-12.5,-6) node [style=sergio] {Open lattice};
\path (-5.5,-6) node [style=sergio] {Intersection};
\end{tikzpicture}
\caption{The cell decomposition of $C_6$-symmetric lattice. Left panel depicts the whole lattice, right panel depicts the horizontal intersection, including 3D blocks $\lambda$, 2D blocks $\sigma$ and 1D block $\tau$. $C_6$ depicts the axis of the 6-fold rotation.}
\label{C6 cell decomposition}
\end{figure}
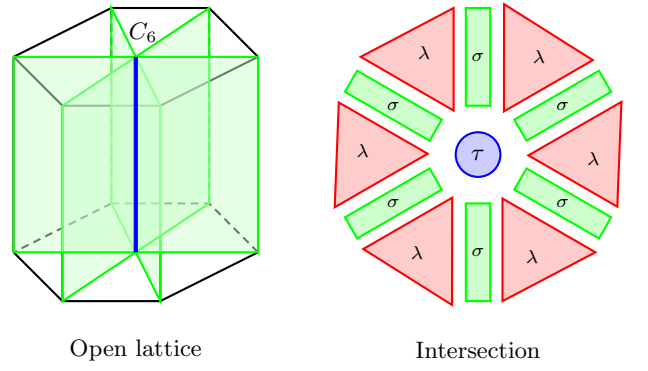

\subsection{$C_6$-symmetric lattice}
For $C_6$-symmetric lattice with the cell decomposition in Fig. \ref{C6 cell decomposition}, the ground-state wavefunction of the system can be decomposed to the direct products of wavefunctions of lower-dimensional block states as:
\begin{align}
|\Psi\rangle=\bigotimes_{g\in C_6}|T_{g\lambda}\rangle\otimes|T_{g\sigma}\rangle\otimes|\beta_{\tau}\rangle
\label{C4 wavefunction}
\end{align}
where $|T_{g\lambda}\rangle/|T_{g\sigma}\rangle$ is the wavefunction of 3D/2D block state on $g\lambda/g\sigma$ which is topological trivial or invertible topological phase; $|\beta_\tau\rangle$ is the $\mathbb{Z}_6$-symmetric wavefunction of 1D block state on $\tau$. 

We summarize the classifications and corresponding root phases. For crystalline TSC with spinless fermions, we summarize all possible block states as:
\begin{enumerate}[1.]
\item 2D block $\sigma$: 2D $(p+ip)$-SC;
\item 1D block $\tau$: Majorana chain and 1D $\mathbb{Z}_6$ fSPT phase.
\end{enumerate}
2D block state is obstructed because it leaves a chiral 1D mode at $\mu$; Majorana chain decoration on $\tau$ is trivialized by anomalous $(p+ip)$-SC on the top surface of the open lattice \cite{rotation}; 1D $\mathbb{Z}_6$ fSPT phase decoration on $\tau$ is trivialized by 2D ``Majorana'' bubble construction on each $\sigma$. Hence the ultimate classifications are trivial.

For crystalline TSC with spin-1/2 fermions, the only possible block state is 2D $(p+ip)$-SC on each $\sigma$ which is obstructed. Hence the corresponding classification is trivial. 

For crystalline TI, similar to the $C_2$-symmetric case, we argue that the spin of fermions is irrelevant. All possible block states are located at 2D block $\sigma$: Chern insulator and Kitaev's $E_8$ state, and all of them are obstructed because of the violation of the non-chirality on $\tau$. Hence the ultimate classification is trivial.

\subsection{$C_{3h}$-symmetric lattice}
For $C_{3h}$-symmetric lattice with the cell decomposition in Fig. \ref{C3h cell decomposition}, the ground-state wavefunction of the system can be decomposed to the direct products of wavefunctions of lower-dimensional block states as:
\begin{align}
|\Psi\rangle=\bigotimes\limits_{g\in C_{3h}}|T_{g\lambda}\rangle\otimes\sum\limits_{k=1}^2|\gamma_{g\sigma_k}\rangle\otimes\sum\limits_{j=1}^2|\beta_{g\tau_j}\rangle\otimes|\alpha_\mu\rangle
\label{C3h cell decomposition}
\end{align}
where $|T_{g\lambda}\rangle$ is the wavefunction of 3D block state on $g\lambda$ which is topological trivial; $|\gamma_{g\sigma_1}\rangle$ is the wavefuntion of 2D block state on $g\sigma_1$ which is topological trivial or invertible topological phase; $|\gamma_{g\sigma_2}\rangle$ is the $\mathbb{Z}_2$-symmetric wavefunction of 2D block state on $g\sigma_2$; $|\beta_{g\tau_1}\rangle/|\beta_{g\tau_2}\rangle$ is the $\mathbb{Z}_3$/$\mathbb{Z}_2$-symmetric wavefunction of 1D block state on $g\tau_1/g\tau_2$; $|\alpha_\mu\rangle$ is the $(\mathbb{Z}_3\times\mathbb{Z}_2)$-symmetric wavefunction of 0D block state on $\mu$.

\begin{figure}
\begin{tikzpicture}[scale=0.75]
\tikzstyle{sergio}=[rectangle,draw=none]
\filldraw[fill=green!20, draw=green, thick, fill opacity=0.5] (0.5,0)--(-1,-3)--(0.5,-2)--cycle;
\filldraw[fill=blue!20, draw=blue, thick] (0.5,-2.5)--(0.9,-3)--(-0.5,-4)--(-0.9,-3.5)--cycle;
\filldraw[fill=green!20, draw=green, thick, fill opacity=0.5] (2.75,-4.25)--(-0.25,-4.25)--(1.25,-3.25)--cycle;
\filldraw[fill=blue!20, draw=blue, thick] (2,-2.5)--(1.6,-3)--(3,-4)--(3.4,-3.5)--cycle;
\filldraw[fill=green!20, draw=green, thick, fill opacity=0.5] (2,0)--(3.5,-3)--(2,-2)--cycle;
\filldraw[fill=blue!20, draw=blue, thick] (1.55,-1.8)--(0.95,-1.8)--(0.95,0)--(1.55,0)--cycle;
\filldraw[fill=white!20, draw=black, thick] (1.25,-2.43)circle (13pt);
\path (1.25,-2.43) node [style=sergio] {$\mu$};
\path (0,-3.25) node [style=sergio] {$\tau_2$};
\path (2.5,-3.25) node [style=sergio] {$\tau_2$};
\path (1.25,-0.9) node [style=sergio] {$\tau_2$};
\path (0,-1.75) node [style=sergio] {$\sigma_2$};
\path (2.5,-1.75) node [style=sergio] {$\sigma_2$};
\path (1.25,-3.75) node [style=sergio] {$\sigma_2$};
\path (-4.75,-6) node [style=sergio] {North/south hemisphere};
\path (1.25,-6) node [style=sergio] {Equator};
\filldraw[fill=red!20, draw=red, thick, fill opacity=0.5] (-5.5,0)--(-7,-3)--(-5.5,-2)--cycle;
\filldraw[fill=green!20, draw=green, thick] (-5.5,-2.5)--(-5.1,-3)--(-6.5,-4)--(-6.9,-3.5)--cycle;
\filldraw[fill=red!20, draw=red, thick, fill opacity=0.5] (-3.25,-4.25)--(-6.25,-4.25)--(-4.75,-3.25)--cycle;
\filldraw[fill=green!20, draw=green, thick] (-4,-2.5)--(-4.4,-3)--(-3,-4)--(-2.6,-3.5)--cycle;
\filldraw[fill=red!20, draw=red, thick, fill opacity=0.5] (-4,0)--(-2.5,-3)--(-4,-2)--cycle;
\filldraw[fill=green!20, draw=green, thick] (-4.45,-1.8)--(-5.05,-1.8)--(-5.05,0)--(-4.45,0)--cycle;
\filldraw[fill=blue!20, draw=blue, thick] (-4.75,-2.43)circle (10pt);
\path (-4.75,-2.43) node [style=sergio] {$\tau_1$};
\path (-6,-3.25) node [style=sergio] {$\sigma_1$};
\path (-3.5,-3.25) node [style=sergio] {$\sigma_1$};
\path (-4.75,-0.9) node [style=sergio] {$\sigma_1$};
\path (-6,-1.75) node [style=sergio] {$\lambda$};
\path (-3.5,-1.75) node [style=sergio] {$\lambda$};
\path (-4.75,-3.75) node [style=sergio] {$\lambda$};
\path (-4.2627,-2.0639) node [style=sergio] {\scriptsize$C_3$};
\path (-0.1,-0.5) node [style=sergio] {$M$};
\end{tikzpicture}
\caption{The cell decomposition of $C_{3h}$-symmetric lattice. Left panel depicts the north/south hemisphere, including 3D blocks $\lambda$, 2D blocks $\sigma_1$ and 1D blocks $\tau_1$; right panel depicts the equator, including 2D blocks $\sigma_2$, 1D blocks $\tau_2$ and 0D block $\mu$. $C_3$ depicts the axis of the 3-fold rotation and $\bs{M}$ depicts the reflection plane.}
\label{C3h cell decomposition}
\end{figure}

We summarize the classifications and corresponding root phases. For crystalline TSC with spinless fermions, we summarize all possible block states as:
\begin{enumerate}[1.]
\item 2D block $\sigma_1$: 2D $(p+ip)$-SC;
\item 2D block $\sigma_2$: 2D $(p+ip)$-SC and 2D $\mathbb{Z}_2$ fSPT phases;
\item 1D block $\tau_1$: Majorana chain;
\item 1D block $\tau_2$: Majorana chain and 1D $\mathbb{Z}_2$ fSPT phase;
\item 0D block $\mu$: 0D modes characterizing eigenvalues of two generators of the $C_{3h}$ group, with even/odd fermion parity.
\end{enumerate}
2D $(p+ip)$-SC decoration on $\sigma_1$ is obstructed because it leaves chiral mode on each $\sigma_1$; 2D $(p+ip)$-SC decoration on $\sigma_2$ is obstructed by 3-fold rotation on the equator. Majorana chain decoration on $\tau_1$ leaves two Majorana zero modes $\gamma_{1,2}$ at $\mu$, whose fermion parity is odd under reflection with respect to the equator. On the equator, both 1D block states on $\tau_2$ are obstructed. Put a complex fermion on each $\tau_2$ which can be adiabatically deformed to infinite far and trivialized, it can change the fermion parity of $\mu$. Furthermore, 0D mode with eigenvalue $-1$ of reflection is trivialized by an atomic insulator constructed by 1D bubble equivalence on $\tau_1$; 0D modes with eigenvalues $e^{2n\pi i/3}$ ($n=0,1,2$) are trivialized by 1D bubble equivalence on $\tau_1$ with a 0D modes with eigenvalues $e^{2m\pi i/3}$ ($m=0,1,2$) on each of them. Hence the classification is $\mathbb{Z}_8$, with the root phase: 2D fermionic Levin-Gu state on each $\sigma_2$. Its second-order topological surface theory is 1D nonchiral Luttinger liquids with $K$-matrix $K=\sigma^z$ and $\mathbb{Z}_2$ symmetry property $W^{\mathbb{Z}_2}=\sigma^z$ and $\delta\phi^{\mathbb{Z}_2}=0$, on the edge of the equator.

For crystalline TSC with spin-1/2 fermions, we summarize all possible block states as:
\begin{enumerate}[1.]
\item 2D block $\sigma_1$: 2D $(p+ip)$-SC;
\item 0D block $\mu$: 0D modes characterizing eigenvalues of two generators of the $C_{3h}$ group, with even/odd fermion parity.
\end{enumerate}
Similar to the spinless fermions, the 2D block state is obstructed and all 0D block states are trivialized. Hence the 
ultimate classification is trivial.

For crystalline TI, similar to the $C_2$-symmetric case, the spin of fermions is irrelevant to the classification. All possible block states are summarized as following:
\begin{enumerate}[1.]
\item 2D block $\sigma_1$: Chern insulators and Kitaev's $E_8$ states;
\item 2D block $\sigma_2$: Chern insulators, Kitaev's $E_8$ states and 2D $U^f(1)\times\mathbb{Z}_2$ fSPT phases;
\item 0D block $\mu$: 0D modes characterizing eigenvalues of two generators of the $C_{3h}$ group, with different $U^f(1)$ charge. 
\end{enumerate}
All 2D block states are obstructed because of the violation of the nonchiral condition on each $\tau_1$. And similar to the crystalline TSC, all 0D block states are trivialized, and even layers of chiral block states on $\sigma_2$ are trivialized by 3D bubble equivalences on $\lambda$. Therefore, the ultimate classification is $\mathbb{Z}_8\times\mathbb{Z}_2$, with the following root phases:
\begin{enumerate}[1.]
\item Monolayer Chern insulator on each $\sigma_2$ ($\mathbb{Z}_2$);
\item 2D $U^f(1)\times\mathbb{Z}_2$ fSPT phase with $\nu=1\in\mathbb{Z}_4$ on each $\sigma_2$ ($\mathbb{Z}_4$);
\item Monolayer Kitaev's $E_8$ state on each $\sigma_2$ ($\mathbb{Z}_2$).
\end{enumerate}
and there is a nontrivial extension between the first two root phases. Their second-order topological surface theories are:
\begin{enumerate}[1.]
\item Chiral fermions on the edge of the equator;
\item 1D nonchiral Luttinger liquids with $K$-matrix $K=\sigma^z$ and $\mathbb{Z}_2$ symmetry property $W^{\mathbb{Z}_2}=\mathbbm{1}_{2\times2}$ and $\delta\phi^{\mathbb{Z}_2}=\pi(0,1)^T$, on the edge of the equator;
\item 1D chiral Luttinger liquids with $K$-matrix (\ref{K-matrix E8S}) on the edge of the equator.
\end{enumerate}

\subsection{$C_{6h}$-symmetric lattice}
For $C_{6h}$-symmetric lattice with the cell decomposition in Fig. \ref{C6h cell decomposition}, the ground-state wavefunction of the system can be decomposed to the direct products of wavefunctions of lower-dimensional block states as:
\begin{align}
|\Psi\rangle=\bigotimes\limits_{g\in C_{3h}}|T_{g\lambda}\rangle\otimes\sum\limits_{k=1}^2|\gamma_{g\sigma_k}\rangle\otimes\sum\limits_{j=1}^2|\beta_{g\tau_j}\rangle\otimes|\alpha_\mu\rangle
\label{C6h cell decomposition}
\end{align}
where $|T_{g\lambda}\rangle$ is the wavefunction of 3D block state on $g\lambda$ which is topological trivial; $|\gamma_{g\sigma_1}\rangle$ is the wavefuntion of 2D block state on $g\sigma_1$ which is topological trivial or invertible topological phase; $|\gamma_{g\sigma_2}\rangle$ is the $\mathbb{Z}_2$-symmetric wavefunction of 2D block state on $g\sigma_2$; $|\beta_{g\tau_1}\rangle/|\beta_{g\tau_2}\rangle$ is the $\mathbb{Z}_6$/$\mathbb{Z}_2$-symmetric wavefunction of 1D block state on $g\tau_1/g\tau_2$; $|\alpha_\mu\rangle$ is the $(\mathbb{Z}_6\times\mathbb{Z}_2)$-symmetric wavefunction of 0D block state on $\mu$.

\begin{figure}
\begin{tikzpicture}[scale=0.6]
\tikzstyle{sergio}=[rectangle,draw=none]
\filldraw[fill=blue!20, draw=blue, thick] (-5.5,-2)circle (13pt);
\filldraw[fill=green!20, draw=green, thick] (-5.25,-1)--(-5.75,-1)--(-5.75,1)--(-5.25,1)--cycle;
\filldraw[rotate around={60:(-5.5,-2)},fill=green!20, draw=green, thick] (-5.25,-1)--(-5.75,-1)--(-5.75,1)--(-5.25,1)--cycle;
\filldraw[rotate around={120:(-5.5,-2)},fill=green!20, draw=green, thick] (-5.25,-1)--(-5.75,-1)--(-5.75,1)--(-5.25,1)--cycle;
\filldraw[rotate around={180:(-5.5,-2)},fill=green!20, draw=green, thick] (-5.25,-1)--(-5.75,-1)--(-5.75,1)--(-5.25,1)--cycle;
\filldraw[rotate around={240:(-5.5,-2)},fill=green!20, draw=green, thick] (-5.25,-1)--(-5.75,-1)--(-5.75,1)--(-5.25,1)--cycle;
\filldraw[rotate around={300:(-5.5,-2)},fill=green!20, draw=green, thick] (-5.25,-1)--(-5.75,-1)--(-5.75,1)--(-5.25,1)--cycle;
\filldraw[fill=red!20, draw=red, thick] (-7.9,-0)--(-6,1)--(-6,-1.1)--cycle;
\filldraw[rotate around={60:(-5.5,-2)},fill=red!20, draw=red, thick] (-7.9,-0)--(-6,1)--(-6,-1.1)--cycle;
\filldraw[rotate around={120:(-5.5,-2)},fill=red!20, draw=red, thick] (-7.9,-0)--(-6,1)--(-6,-1.1)--cycle;
\filldraw[rotate around={180:(-5.5,-2)},fill=red!20, draw=red, thick] (-7.9,-0)--(-6,1)--(-6,-1.1)--cycle;
\filldraw[rotate around={240:(-5.5,-2)},fill=red!20, draw=red, thick] (-7.9,-0)--(-6,1)--(-6,-1.1)--cycle;
\filldraw[rotate around={300:(-5.5,-2)},fill=red!20, draw=red, thick] (-7.9,-0)--(-6,1)--(-6,-1.1)--cycle;
\path (-4.25,0) node [style=sergio] {\scriptsize$\lambda$};
\path[rotate around={60:(-5.5,-2)}] (-4.25,0) node [style=sergio] {\scriptsize$\lambda$};
\path[rotate around={120:(-5.5,-2)}] (-4.25,0) node [style=sergio] {\scriptsize$\lambda$};
\path[rotate around={180:(-5.5,-2)}] (-4.25,0) node [style=sergio] {\scriptsize$\lambda$};
\path[rotate around={240:(-5.5,-2)}] (-4.25,0) node [style=sergio] {\scriptsize$\lambda$};
\path[rotate around={300:(-5.5,-2)}] (-4.25,0) node [style=sergio] {\scriptsize$\lambda$};
\path (-5.5,0) node [style=sergio] {\scriptsize$\sigma_1$};
\path[rotate around={60:(-5.5,-2)}] (-5.5,0) node [style=sergio] {\scriptsize$\sigma_1$};
\path[rotate around={120:(-5.5,-2)}] (-5.5,0) node [style=sergio] {\scriptsize$\sigma_1$};
\path[rotate around={180:(-5.5,-2)}] (-5.5,0) node [style=sergio] {\scriptsize$\sigma_1$};
\path[rotate around={240:(-5.5,-2)}] (-5.5,0) node [style=sergio] {\scriptsize$\sigma_1$};
\path[rotate around={300:(-5.5,-2)}] (-5.5,0) node [style=sergio] {\scriptsize$\sigma_1$};
\path (-5.5,-2) node [style=sergio] {$\tau_1$};
\filldraw[fill=white!20, draw=black, thick] (2.25,-2)circle (13pt);
\filldraw[fill=blue!20, draw=blue, thick] (2.5,-1)--(2,-1)--(2,1)--(2.5,1)--cycle;
\filldraw[rotate around={60:(2.25,-2)},fill=blue!20, draw=blue, thick] (2.5,-1)--(2,-1)--(2,1)--(2.5,1)--cycle;
\filldraw[rotate around={120:(2.25,-2)},fill=blue!20, draw=blue, thick] (2.5,-1)--(2,-1)--(2,1)--(2.5,1)--cycle;
\filldraw[rotate around={180:(2.25,-2)},fill=blue!20, draw=blue, thick] (2.5,-1)--(2,-1)--(2,1)--(2.5,1)--cycle;
\filldraw[rotate around={240:(2.25,-2)},fill=blue!20, draw=blue, thick] (2.5,-1)--(2,-1)--(2,1)--(2.5,1)--cycle;
\filldraw[rotate around={300:(2.25,-2)},fill=blue!20, draw=blue, thick] (2.5,-1)--(2,-1)--(2,1)--(2.5,1)--cycle;
\filldraw[fill=green!20, draw=green, thick] (-0.15,0)--(1.75,1)--(1.75,-1.1)--cycle;
\filldraw[rotate around={60:(2.25,-2)},fill=green!20, draw=green, thick] (-0.15,0)--(1.75,1)--(1.75,-1.1)--cycle;
\filldraw[rotate around={120:(2.25,-2)},fill=green!20, draw=green, thick] (-0.15,0)--(1.75,1)--(1.75,-1.1)--cycle;
\filldraw[rotate around={180:(2.25,-2)},fill=green!20, draw=green, thick] (-0.15,0)--(1.75,1)--(1.75,-1.1)--cycle;
\filldraw[rotate around={240:(2.25,-2)},fill=green!20, draw=green, thick] (-0.15,0)--(1.75,1)--(1.75,-1.1)--cycle;
\filldraw[rotate around={300:(2.25,-2)},fill=green!20, draw=green, thick] (-0.15,0)--(1.75,1)--(1.75,-1.1)--cycle;
\path (3.5,0) node [style=sergio] {\scriptsize$\sigma_2$};
\path[rotate around={60:(2.25,-2)}] (3.5,0) node [style=sergio] {\scriptsize$\sigma_2$};
\path[rotate around={120:(2.25,-2)}] (3.5,0) node [style=sergio] {\scriptsize$\sigma_2$};
\path[rotate around={180:(2.25,-2)}] (3.5,0) node [style=sergio] {\scriptsize$\sigma_2$};
\path[rotate around={240:(2.25,-2)}] (3.5,0) node [style=sergio] {\scriptsize$\sigma_2$};
\path[rotate around={300:(2.25,-2)}] (3.5,0) node [style=sergio] {\scriptsize$\sigma_2$};
\path (2.25,0) node [style=sergio] {\scriptsize$\tau_2$};
\path[rotate around={60:(2.25,-2)}] (2.25,0) node [style=sergio] {\scriptsize$\tau_2$};
\path[rotate around={120:(2.25,-2)}] (2.25,0) node [style=sergio] {\scriptsize$\tau_2$};
\path[rotate around={180:(2.25,-2)}] (2.25,0) node [style=sergio] {\scriptsize$\tau_2$};
\path[rotate around={240:(2.25,-2)}] (2.25,0) node [style=sergio] {\scriptsize$\tau_2$};
\path[rotate around={300:(2.25,-2)}] (2.25,0) node [style=sergio] {\scriptsize$\tau_2$};
\path (2.25,-2) node [style=sergio] {$\mu$};
\path (-5.5,-6) node [style=sergio] {North/south hemisphere};
\path (2.25,-6) node [style=sergio] {Equator};
\path (-5,-1.5) node [style=sergio] {\scriptsize$C_6$};
\path (4,1) node [style=sergio] {$M$};
\end{tikzpicture}
\caption{The cell decomposition of $C_{6h}$-symmetric lattice. Left panel depicts the north/south hemisphere, including 3D blocks $\lambda$, 2D blocks $\sigma_1$ and 1D blocks $\tau_1$; right panel depicts the equator, including 2D blocks $\sigma_2$, 1D blocks $\tau_2$ and 0D block $\mu$. $C_6$ depicts the axis of the 6-fold rotation and $\bs{M}$ depicts the reflection plane.}
\label{C6h cell decomposition}
\end{figure}

We summarize the classifications and corresponding root phases of crystalline SPT phases. For crystalline TSC with spinless fermions, we summarize all possible block states as:
\begin{enumerate}[1.]
\item 2D block $\sigma_1$: 2D $(p+ip)$-SCs;
\item 2D block $\sigma_2$: 2D $(p+ip)$-SCs and 2D $\mathbb{Z}_2$ fSPT phases;
\item 1D block $\tau_1$: Majorana chain and 1D $\mathbb{Z}_6$ fSPT phase;
\item 1D block $\tau_2$: Majorana chain and 1D $\mathbb{Z}_2$ fSPT phase;
\item 0D block $\mu$: 0D modes characterizing eigenvalues of two generators of the $C_{6h}$ group, with even/odd fermion parity.
\end{enumerate}
2D $(p+ip)$-SCs on $\sigma_1$ are obstructed because they leaves chiral 1D mode on each $\tau_1$; and 2D $(p+ip)$-SCs on $\sigma_2$ because it is not compatible with 6-fold rotation on the equator. Majorana chain decoration on $\tau_1$ is obstructed because they leave two Majorana zero modes at $\mu$ whose fermion parity is odd under the reflection with respect to the equator; Majorana chain decoration on $\tau_2$ is obstructed because they leave six Majorana zero modes at $\mu$ whose fermion parity is odd under the 6-fold rotation. Consider a complex fermion on each $\tau_1$ which can be adiabatically deformed to infinite far and trivialized, it will form an atomic insulator with two complex fermions $c_1^\dag c_2^\dag|0\rangle$ whose eigenvalue of reflection with respect to the equator is $-1$. Equivalently, it trivializes the 0D mode characterizing eigenvalue $-1$ of the reflection. Similar to the eigenvalue $-1$ of 6-fold rotation. Other eigenvalues of 6-fold rotation can be trivialized by a 0D mode with 6-fold rotation eigenvalue $e^{\pi i/3}$ on each $\tau_1$. Furthermore, 0D mode with odd fermion parity is trivialized by 2D ``Majorana'' bubble construction on $\sigma_2$. Hence the ultimate classification is $\mathbb{Z}_8$, and the only root phase is the 2D fermionic Levin-Gu state on each 2D block $\sigma_2$. Its second-order topological surface theory is 1D Luttinger liquids with $K$-matrix $K=\sigma^z$ and $\mathbb{Z}_2$ symmetry property $W^{\mathbb{Z}_2}=\sigma^z$ and $\delta\phi^{\mathbb{Z}_2}=0$, on the edge of the equator.

For crystalline TSC with spin-1/2 fermions, we summarize all possible block states as:
\begin{enumerate}[1.]
\item 2D block $\sigma_1$: 2D $(p+ip)$-SCs;
\item 0D block $\mu$: 0D modes with odd fermion parity and characterizing eigenvalues of two generators of the $C_{6h}$ group.
\end{enumerate}
Similar to the $C_{2h}$ case, the ultimate classification should be trivial. 

For crystalline TI with spinless fermions, we summarize all possible block states as:
\begin{enumerate}[1.]
\item 2D block $\sigma_1$: Chern insulators and Kitaev's $E_8$ states;
\item 2D block $\sigma_2$: Chern insulators, Kitaev's $E_8$ states and 2D $U^f(1)\times\mathbb{Z}_2$ fSPT phases;
\item 0D block $\mu$: 0D modes characterizing eigenvalues of two generators of the $C_{6h}$ group, with different $U^f(1)$ charge.
\end{enumerate}
All 2D block states on $\sigma_1$ are obstructed because of the violation of the nonchiral condition on $\tau_1$, and even layers of chiral block states on $\sigma_2$ are trivialized by 3D bubble equivalences on $\lambda$. Similar to the crystalline TSC, 0D modes characterizing eigenvalues of 6-fold rotations and even $U^f(1)$ charge are all trivialized. Therefore, the ultimate classification is $\mathbb{Z}_8\times\mathbb{Z}_4\times\mathbb{Z}_2$, with the following root phases:
\begin{enumerate}[1.]
\item Monolayer Chern insulator on each $\sigma_2$ ($\mathbb{Z}_2$);
\item 2D $U^f(1)\times\mathbb{Z}_2$ fSPT phase with $\nu=1\in\mathbb{Z}_4$ on each $\sigma_2$ ($\mathbb{Z}_4$);
\item Monolayer Kitaev's $E_8$ state on each $\sigma_2$ ($\mathbb{Z}_2$);
\item 0D mode with odd $U^f(1)$ charge on $\mu$ ($\mathbb{Z}_2$);
\item 0D mode with eigenvalue $-1$ of reflection with respect to the equator on $\mu$ ($\mathbb{Z}_2$).
\end{enumerate}
and there are two nontrivial extensions: one extension is between the first two root phases, another extension is between the last two root phases. Their second-order topological surface theories on the edge of the equator are summarized below:
\begin{enumerate}[1.]
\item Chiral fermions on the edge of the equator;
\item 1D nonchiral Luttinger liquids with $K$-matrix $K=\sigma^z$ and $\mathbb{Z}_2$ symmetry property $W^{\mathbb{Z}_2}=\mathbbm{1}_{2\times2}$ and $\delta\phi^{\mathbb{Z}_2}=\pi(0,1)^T$, on the edge of the equator;
\item 1D chiral Luttinger liquids with $K$-matrix (\ref{K-matrix E8S}) on the edge of the equator.
\end{enumerate}

For crystalline TI with spin-1/2 fermions, we summarize all possible block states as:
\begin{enumerate}[1.]
\item 2D block $\sigma_1$: Chern insulators and Kitaev's $E_8$ states;
\item 2D block $\sigma_2$: Chern insulators, Kitaev's $E_8$ states and 2D $U^f(1)\times\mathbb{Z}_2$ fSPT phases;
\item 0D block $\mu$: 0D modes characterizing eigenvalues of two generators of the $C_{6h}$ group, with different even $U^f(1)$ charge.
\end{enumerate}
The obstructions and trivializations are identical to the spinless fermions except that 0D mode characterizing eigenvalue $-1$ of 6-fold rotation remains nontrivial. Hence the ultimate classification is $\mathbb{Z}_8\times\mathbb{Z}_2^3$, with the following root phases:
\begin{enumerate}[1.]
\item Monolayer Chern insulator on each $\sigma_2$ ($\mathbb{Z}_2$);
\item 2D $U^f(1)\times\mathbb{Z}_2$ fSPT phase with $\nu=1\in\mathbb{Z}_4$ on each $\sigma_2$ ($\mathbb{Z}_4$);
\item Monolayer Kitaev's $E_8$ state on each $\sigma_2$ ($\mathbb{Z}_2$);
\item 0D modes with eigenvalues $-1$ of two generators of $C_{6h}$ on $\mu$ ($\mathbb{Z}_2^2$).
\end{enumerate}
with a nontrivial extension between the first two root phases. Their second-order topological surface theories on the edge of the equator are:
\begin{enumerate}[1.]
\item Chiral fermions on the edge of the equator;
\item 1D nonchiral Luttinger liquids with $K$-matrix $K=\sigma^z$ and $\mathbb{Z}_2$ symmetry property $W^{\mathbb{Z}_2}=\mathbbm{1}_{2\times2}$ and $\delta\phi^{\mathbb{Z}_2}=\pi(0,1)^T$ on the edge of the equator;
\item 1D chiral Luttinger liquids with $K$-matrix (\ref{K-matrix E8S}) on the edge of the equator.
\end{enumerate}

\begin{figure}
\begin{tikzpicture}[scale=0.6]
\tikzstyle{sergio}=[rectangle,draw=none]
\filldraw[fill=white!20, draw=black, thick] (2,-2)circle (13pt);
\filldraw[fill=blue!20, draw=blue, thick] (2.25,-0.75)--(1.75,-0.75)--(1.75,1.25)--(2.25,1.25)--cycle;
\filldraw[rotate around={30:(2,-2)},fill=blue!20, draw=blue, thick] (2.25,-0.75)--(1.75,-0.75)--(1.75,1.25)--(2.25,1.25)--cycle;
\filldraw[rotate around={60:(2,-2)},fill=blue!20, draw=blue, thick] (2.25,-0.75)--(1.75,-0.75)--(1.75,1.25)--(2.25,1.25)--cycle;
\filldraw[rotate around={90:(2,-2)},fill=blue!20, draw=blue, thick] (2.25,-0.75)--(1.75,-0.75)--(1.75,1.25)--(2.25,1.25)--cycle;
\filldraw[rotate around={120:(2,-2)},fill=blue!20, draw=blue, thick] (2.25,-0.75)--(1.75,-0.75)--(1.75,1.25)--(2.25,1.25)--cycle;
\filldraw[rotate around={150:(2,-2)},fill=blue!20, draw=blue, thick] (2.25,-0.75)--(1.75,-0.75)--(1.75,1.25)--(2.25,1.25)--cycle;
\filldraw[rotate around={180:(2,-2)},fill=blue!20, draw=blue, thick] (2.25,-0.75)--(1.75,-0.75)--(1.75,1.25)--(2.25,1.25)--cycle;
\filldraw[rotate around={210:(2,-2)},fill=blue!20, draw=blue, thick] (2.25,-0.75)--(1.75,-0.75)--(1.75,1.25)--(2.25,1.25)--cycle;
\filldraw[rotate around={240:(2,-2)},fill=blue!20, draw=blue, thick] (2.25,-0.75)--(1.75,-0.75)--(1.75,1.25)--(2.25,1.25)--cycle;
\filldraw[rotate around={270:(2,-2)},fill=blue!20, draw=blue, thick] (2.25,-0.75)--(1.75,-0.75)--(1.75,1.25)--(2.25,1.25)--cycle;
\filldraw[rotate around={300:(2,-2)},fill=blue!20, draw=blue, thick] (2.25,-0.75)--(1.75,-0.75)--(1.75,1.25)--(2.25,1.25)--cycle;
\filldraw[rotate around={330:(2,-2)},fill=blue!20, draw=blue, thick] (2.25,-0.75)--(1.75,-0.75)--(1.75,1.25)--(2.25,1.25)--cycle;
\path (2,-2) node [style=sergio] {$\mu$};
\path (2,0.25) node [style=sergio] {\scriptsize$\tau_2$};
\path[rotate around={30:(2,-2)}] (2,0.25) node [style=sergio] {\scriptsize$\tau_3$};
\path[rotate around={60:(2,-2)}] (2,0.25) node [style=sergio] {\scriptsize$\tau_2$};
\path[rotate around={90:(2,-2)}] (2,0.25) node [style=sergio] {\scriptsize$\tau_3$};
\path[rotate around={120:(2,-2)}] (2,0.25) node [style=sergio] {\scriptsize$\tau_2$};
\path[rotate around={150:(2,-2)}] (2,0.25) node [style=sergio] {\scriptsize$\tau_3$};
\path[rotate around={180:(2,-2)}] (2,0.25) node [style=sergio] {\scriptsize$\tau_2$};
\path[rotate around={210:(2,-2)}] (2,0.25) node [style=sergio] {\scriptsize$\tau_3$};
\path[rotate around={240:(2,-2)}] (2,0.25) node [style=sergio] {\scriptsize$\tau_2$};
\path[rotate around={270:(2,-2)}] (2,0.25) node [style=sergio] {\scriptsize$\tau_3$};
\path[rotate around={300:(2,-2)}] (2,0.25) node [style=sergio] {\scriptsize$\tau_2$};
\path[rotate around={330:(2,-2)}] (2,0.25) node [style=sergio] {\scriptsize$\tau_3$};
\filldraw[fill=green!20, draw=green, thick, fill opacity=0.5] (1.5,1.25)--(0.75,1)--(1.5,-0.25)--cycle;
\filldraw[rotate around={30:(2,-2)},fill=green!20, draw=green, thick, fill opacity=0.5] (1.5,1.25)--(0.75,1)--(1.5,-0.25)--cycle;
\filldraw[rotate around={60:(2,-2)},fill=green!20, draw=green, thick, fill opacity=0.5] (1.5,1.25)--(0.75,1)--(1.5,-0.25)--cycle;
\filldraw[rotate around={90:(2,-2)},fill=green!20, draw=green, thick, fill opacity=0.5] (1.5,1.25)--(0.75,1)--(1.5,-0.25)--cycle;
\filldraw[rotate around={120:(2,-2)},fill=green!20, draw=green, thick, fill opacity=0.5] (1.5,1.25)--(0.75,1)--(1.5,-0.25)--cycle;
\filldraw[rotate around={150:(2,-2)},fill=green!20, draw=green, thick, fill opacity=0.5] (1.5,1.25)--(0.75,1)--(1.5,-0.25)--cycle;
\filldraw[rotate around={180:(2,-2)},fill=green!20, draw=green, thick, fill opacity=0.5] (1.5,1.25)--(0.75,1)--(1.5,-0.25)--cycle;
\filldraw[rotate around={210:(2,-2)},fill=green!20, draw=green, thick, fill opacity=0.5] (1.5,1.25)--(0.75,1)--(1.5,-0.25)--cycle;
\filldraw[rotate around={240:(2,-2)},fill=green!20, draw=green, thick, fill opacity=0.5] (1.5,1.25)--(0.75,1)--(1.5,-0.25)--cycle;
\filldraw[rotate around={270:(2,-2)},fill=green!20, draw=green, thick, fill opacity=0.5] (1.5,1.25)--(0.75,1)--(1.5,-0.25)--cycle;
\filldraw[rotate around={300:(2,-2)},fill=green!20, draw=green, thick, fill opacity=0.5] (1.5,1.25)--(0.75,1)--(1.5,-0.25)--cycle;
\filldraw[rotate around={330:(2,-2)},fill=green!20, draw=green, thick, fill opacity=0.5] (1.5,1.25)--(0.75,1)--(1.5,-0.25)--cycle;
\path (1.2,0.75) node [style=sergio] {\scriptsize$\sigma_2$};
\path[rotate around={30:(2,-2)}] (1.2,0.75) node [style=sergio] {\scriptsize$\sigma_2$};
\path[rotate around={60:(2,-2)}] (1.2,0.75) node [style=sergio] {\scriptsize$\sigma_2$};
\path[rotate around={90:(2,-2)}] (1.2,0.75) node [style=sergio] {\scriptsize$\sigma_2$};
\path[rotate around={120:(2,-2)}] (1.2,0.75) node [style=sergio] {\scriptsize$\sigma_2$};
\path[rotate around={150:(2,-2)}] (1.2,0.75) node [style=sergio] {\scriptsize$\sigma_2$};
\path[rotate around={180:(2,-2)}] (1.2,0.75) node [style=sergio] {\scriptsize$\sigma_2$};
\path[rotate around={210:(2,-2)}] (1.2,0.75) node [style=sergio] {\scriptsize$\sigma_2$};
\path[rotate around={240:(2,-2)}] (1.2,0.75) node [style=sergio] {\scriptsize$\sigma_2$};
\path[rotate around={270:(2,-2)}] (1.2,0.75) node [style=sergio] {\scriptsize$\sigma_2$};
\path[rotate around={300:(2,-2)}] (1.2,0.75) node [style=sergio] {\scriptsize$\sigma_2$};
\path[rotate around={330:(2,-2)}] (1.2,0.75) node [style=sergio] {\scriptsize$\sigma_2$};
\filldraw[fill=blue!20, draw=blue, thick] (-5.5,-2)circle (13pt);
\filldraw[fill=green!20, draw=green, thick] (-5.25,-1)--(-5.75,-1)--(-5.75,1)--(-5.25,1)--cycle;
\filldraw[rotate around={60:(-5.5,-2)},fill=green!20, draw=green, thick] (-5.25,-1)--(-5.75,-1)--(-5.75,1)--(-5.25,1)--cycle;
\filldraw[rotate around={120:(-5.5,-2)},fill=green!20, draw=green, thick] (-5.25,-1)--(-5.75,-1)--(-5.75,1)--(-5.25,1)--cycle;
\filldraw[rotate around={180:(-5.5,-2)},fill=green!20, draw=green, thick] (-5.25,-1)--(-5.75,-1)--(-5.75,1)--(-5.25,1)--cycle;
\filldraw[rotate around={240:(-5.5,-2)},fill=green!20, draw=green, thick] (-5.25,-1)--(-5.75,-1)--(-5.75,1)--(-5.25,1)--cycle;
\filldraw[rotate around={300:(-5.5,-2)},fill=green!20, draw=green, thick] (-5.25,-1)--(-5.75,-1)--(-5.75,1)--(-5.25,1)--cycle;
\filldraw[fill=red!20, draw=red, thick] (-7.9,-0)--(-6,1)--(-6,-1.1)--cycle;
\filldraw[rotate around={60:(-5.5,-2)},fill=red!20, draw=red, thick] (-7.9,-0)--(-6,1)--(-6,-1.1)--cycle;
\filldraw[rotate around={120:(-5.5,-2)},fill=red!20, draw=red, thick] (-7.9,-0)--(-6,1)--(-6,-1.1)--cycle;
\filldraw[rotate around={180:(-5.5,-2)},fill=red!20, draw=red, thick] (-7.9,-0)--(-6,1)--(-6,-1.1)--cycle;
\filldraw[rotate around={240:(-5.5,-2)},fill=red!20, draw=red, thick] (-7.9,-0)--(-6,1)--(-6,-1.1)--cycle;
\filldraw[rotate around={300:(-5.5,-2)},fill=red!20, draw=red, thick] (-7.9,-0)--(-6,1)--(-6,-1.1)--cycle;
\path (-4.25,0) node [style=sergio] {\scriptsize$\lambda$};
\path[rotate around={60:(-5.5,-2)}] (-4.25,0) node [style=sergio] {\scriptsize$\lambda$};
\path[rotate around={120:(-5.5,-2)}] (-4.25,0) node [style=sergio] {\scriptsize$\lambda$};
\path[rotate around={180:(-5.5,-2)}] (-4.25,0) node [style=sergio] {\scriptsize$\lambda$};
\path[rotate around={240:(-5.5,-2)}] (-4.25,0) node [style=sergio] {\scriptsize$\lambda$};
\path[rotate around={300:(-5.5,-2)}] (-4.25,0) node [style=sergio] {\scriptsize$\lambda$};
\path (-5.5,0) node [style=sergio] {\scriptsize$\sigma_1$};
\path[rotate around={60:(-5.5,-2)}] (-5.5,0) node [style=sergio] {\scriptsize$\sigma_1$};
\path[rotate around={120:(-5.5,-2)}] (-5.5,0) node [style=sergio] {\scriptsize$\sigma_1$};
\path[rotate around={180:(-5.5,-2)}] (-5.5,0) node [style=sergio] {\scriptsize$\sigma_1$};
\path[rotate around={240:(-5.5,-2)}] (-5.5,0) node [style=sergio] {\scriptsize$\sigma_1$};
\path[rotate around={300:(-5.5,-2)}] (-5.5,0) node [style=sergio] {\scriptsize$\sigma_1$};
\path (-5.5,-2) node [style=sergio] {$\tau_1$};
\path (-5.5,-6) node [style=sergio] {North/south hemisphere};
\path (2,-6) node [style=sergio] {Equator};
\path (-5,-1.5) node [style=sergio] {\scriptsize$C_6$};
\path (2,-1.2) node [style=sergio] {$C_2$};
\path (1.35,-2.5) node [style=sergio] {$C_2$};
\end{tikzpicture}
\caption{The cell decomposition of $D_6$-symmetric lattice. Left panel depicts the north/south hemisphere, including 3D blocks $\lambda$, 2D blocks $\sigma_1$ and 1D blocks $\tau_1$; right panel depicts the equator, including 2D blocks $\sigma_2$, 1D blocks $\tau_{2,3}$, and 0D block $\mu$. $C_6$ depicts the axis of the 6-fold rotation, and $C_2$'s depict the axes of the 2-fold rotations.}
\label{D6 cell decomposition}
\end{figure}
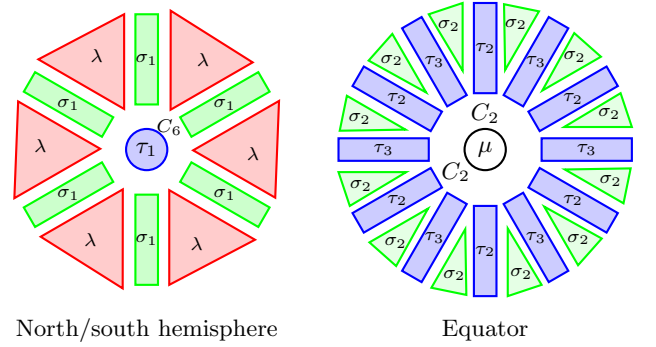

\subsection{$D_6$-symmetric lattice}
For $D_6$-symmetric lattice with the cell decomposition in Fig. \ref{D6 cell decomposition}. the ground-state wavefunction of the system can be decomposed to the direct products of wavefunctions of lower-dimensional block states as:
\begin{align}
|\Psi\rangle=\bigotimes\limits_{g\in D_6}|T_{g\lambda}\rangle\otimes\sum\limits_{k=1}^2|T_{g\sigma_k}\rangle\otimes\sum\limits_{j=1}^3|\beta_{g\tau_j}\rangle\otimes|\alpha_\mu\rangle
\label{D6 wavefunction}
\end{align}
where $|T_{g\lambda}\rangle/|T_{g\sigma_{1,2}}\rangle$ is the wavefunction of 3D/2D block state on $g\lambda/g\sigma_{1,2}$ which is topological trivial or invertible topological phase; $|\beta_{g\tau_1}\rangle$ is the $\mathbb{Z}_6$-symmetric wavefunction of 1D block state on $g\tau_1$, and $|\beta_{g\tau_{2,3}}\rangle$ is the $\mathbb{Z}_2$-symmetric wavefunction of 1D block state on $g\tau_{2,3}$; $|\alpha_\mu\rangle$ is the $(\mathbb{Z}_6\rtimes\mathbb{Z}_2)$-symmetric wavefunction of 0D block state on $\mu$.

We summarize the classifications and corresponding root phases. For crystalline TSC with spinless fermions, we summarize all possible block states as:
\begin{enumerate}[1.]
\item 2D blocks $\sigma_1$ and $\sigma_2$: 2D $(p+ip)$-SCs;
\item 1D block $\tau_1$: Majorana chain and 1D $\mathbb{Z}_6$ fSPT phases;
\item 1D blocks $\tau_2$ and $\tau_3$: Majorana chain and 1D $\mathbb{Z}_2$ fSPT phases;
\item 0D block $\mu$: 0D modes characterizing eigenvalues $-1$ of two generators of the $D_6$ group, with even/odd fermion parity.
\end{enumerate}
Similar to the $D_2$-symmetric cases, all these candidate block states are obstructed or trivialized. Hence the ultimate classification is trivial. 

For crystalline TSC with spin-1/2 fermions, all possible block states are summarized as following:
\begin{enumerate}[1.]
\item 2D blocks $\sigma_1$ and $\sigma_2$: 2D $(p+ip)$-SCs;
\item 0D block $\mu$: 0D modes characterizing eigenvalues $-1$ of two generators of the $D_6$ group.
\end{enumerate}
It is straightforwardly to see that all 2D block states are obstructed because the violation of the nonchiral condition at nearby 1D blocks. Hence the ultimate classification is $\mathbb{Z}_2^2$, the root phases are 0D modes with eigenvalues $-1$ of two rotation generators of $D_6$ group at $\mu$, and there is no HO topological surface theory.

Similar to the $D_2$-symmetric case, for crystalline TI with spinless fermions, the classification is trivial. For crystalline TI with spin-1/2 fermions, the classification is $\mathbb{Z}_2^3$, with three root phases:
\begin{enumerate}[1.]
\item 0D mode with odd $U^f(1)$ charge;
\item 0D modes with eigenvalues $-1$ of two rotation generators of $D_6$.
\end{enumerate}
and there is no HO topological surface theory.

\subsection{$C_{6v}$-symmetric lattice}
For $C_{6v}$-symmetric lattice with the cell decomposition in Fig. \ref{C6v cell decomposition}, the ground-state wavefunction of the system can be decomposed to the direct products of wavefunctions of lower-dimensional block states as:
\begin{align}
|\Psi\rangle=\bigotimes\limits_{g\in C_{6v}}|T_{g\lambda}\rangle\otimes\sum\limits_{j=1}^2|\gamma_{g\sigma_{j}}\rangle\otimes|\beta_\tau\rangle
\label{C6v wavefunction}
\end{align}
where $|T_{g\lambda}\rangle$ is the wavefunction of 3D block state on $g\lambda$ which is topological trivial; $|\gamma_{g\sigma_{1,2}}\rangle$ is the $\mathbb{Z}_2$-symmetric wavefunction of 2D block state on $g\sigma_{1,2}$; $|\beta_\tau\rangle$ is the $(\mathbb{Z}_6\rtimes\mathbb{Z}_2)$-symmetric wavefunction of 1D block state $\tau$.

\begin{figure}
\begin{tikzpicture}[scale=0.61]
\tikzstyle{sergio}=[rectangle,draw=none]
\draw[thick] (-5.5,1) -- (-7.5,0);
\draw[thick] (-5.5,1) -- (-3.5,1);
\draw[thick] (-2.5,0) -- (-3.5,1);
\draw[thick] (-2.5,0) -- (-4.5,-1);
\draw[thick] (-6.5,-1) -- (-4.5,-1);
\draw[thick] (-7.5,0) -- (-6.5,-1);
\draw[thick] (-7.5,0) -- (-7.5,-4);
\draw[thick] (-6.5,-5) -- (-7.5,-4);
\draw[thick] (-6.5,-5) -- (-4.5,-5);
\draw[thick] (-2.5,-4) -- (-4.5,-5);
\draw[thick] (-2.5,-4) -- (-2.5,0);
\draw[densely dashed,thick] (-2.5,-4) -- (-3.5,-3);
\draw[densely dashed,thick] (-5.5,-3) -- (-3.5,-3);
\draw[densely dashed,thick] (-5.5,-3) -- (-7.5,-4);
\draw[densely dashed,thick] (-5.5,-3) -- (-5.5,1);
\draw[densely dashed,thick] (-3.5,-3) -- (-3.5,1);
\draw[thick] (-4.5,-1) -- (-4.5,-5);
\draw[thick] (-6.5,-1) -- (-6.5,-5);
\filldraw[fill=green!20, draw=green, thick, fill opacity=0.5] (-4.5,-5)--(-5.5,-3)--(-5.5,1)--(-4.5,-1)--cycle;
\filldraw[fill=green!20, draw=green, thick, fill opacity=0.5] (-3.5,-3)--(-6.5,-5)--(-6.5,-1)--(-3.5,1)--cycle;
\filldraw[fill=green!20, draw=green, thick, fill opacity=0.5] (-2.5,-4)--(-7.5,-4)--(-7.5,0)--(-2.5,0)--cycle;
\filldraw[fill=green!20, draw=green, thick, fill opacity=0.5] (-3.5,-4.5)--(-6.5,-3.5)--(-6.5,0.5)--(-3.5,-0.5)--cycle;
\filldraw[fill=green!20, draw=green, thick, fill opacity=0.5] (-5.5,-5)--(-4.5,-3)--(-4.5,1)--(-5.5,-1)--cycle;
\filldraw[fill=green!20, draw=green, thick, fill opacity=0.5] (-7,-4.5)--(-3,-3.5)--(-3,0.5)--(-7,-0.5)--cycle;
\draw[color=blue,ultra thick] (-5,0) -- (-5,-4);
\filldraw[fill=blue!20, draw=blue, thick] (2,-2)circle (13pt);
\filldraw[fill=green!20, draw=green, thick] (2.25,-0.75)--(1.75,-0.75)--(1.75,1.25)--(2.25,1.25)--cycle;
\filldraw[rotate around={30:(2,-2)},fill=green!20, draw=green, thick] (2.25,-0.75)--(1.75,-0.75)--(1.75,1.25)--(2.25,1.25)--cycle;
\filldraw[rotate around={60:(2,-2)},fill=green!20, draw=green, thick] (2.25,-0.75)--(1.75,-0.75)--(1.75,1.25)--(2.25,1.25)--cycle;
\filldraw[rotate around={90:(2,-2)},fill=green!20, draw=green, thick] (2.25,-0.75)--(1.75,-0.75)--(1.75,1.25)--(2.25,1.25)--cycle;
\filldraw[rotate around={120:(2,-2)},fill=green!20, draw=green, thick] (2.25,-0.75)--(1.75,-0.75)--(1.75,1.25)--(2.25,1.25)--cycle;
\filldraw[rotate around={150:(2,-2)},fill=green!20, draw=green, thick] (2.25,-0.75)--(1.75,-0.75)--(1.75,1.25)--(2.25,1.25)--cycle;
\filldraw[rotate around={180:(2,-2)},fill=green!20, draw=green, thick] (2.25,-0.75)--(1.75,-0.75)--(1.75,1.25)--(2.25,1.25)--cycle;
\filldraw[rotate around={210:(2,-2)},fill=green!20, draw=green, thick] (2.25,-0.75)--(1.75,-0.75)--(1.75,1.25)--(2.25,1.25)--cycle;
\filldraw[rotate around={240:(2,-2)},fill=green!20, draw=green, thick] (2.25,-0.75)--(1.75,-0.75)--(1.75,1.25)--(2.25,1.25)--cycle;
\filldraw[rotate around={270:(2,-2)},fill=green!20, draw=green, thick] (2.25,-0.75)--(1.75,-0.75)--(1.75,1.25)--(2.25,1.25)--cycle;
\filldraw[rotate around={300:(2,-2)},fill=green!20, draw=green, thick] (2.25,-0.75)--(1.75,-0.75)--(1.75,1.25)--(2.25,1.25)--cycle;
\filldraw[rotate around={330:(2,-2)},fill=green!20, draw=green, thick] (2.25,-0.75)--(1.75,-0.75)--(1.75,1.25)--(2.25,1.25)--cycle;
\path (2,-2) node [style=sergio] {\Large $\tau$};
\path (2,0.25) node [style=sergio] {\scriptsize$\sigma_1$};
\path[rotate around={30:(2,-2)}] (2,0.25) node [style=sergio] {\scriptsize$\sigma_2$};
\path[rotate around={60:(2,-2)}] (2,0.25) node [style=sergio] {\scriptsize$\sigma_1$};
\path[rotate around={90:(2,-2)}] (2,0.25) node [style=sergio] {\scriptsize$\sigma_2$};
\path[rotate around={120:(2,-2)}] (2,0.25) node [style=sergio] {\scriptsize$\sigma_1$};
\path[rotate around={150:(2,-2)}] (2,0.25) node [style=sergio] {\scriptsize$\sigma_2$};
\path[rotate around={180:(2,-2)}] (2,0.25) node [style=sergio] {\scriptsize$\sigma_1$};
\path[rotate around={210:(2,-2)}] (2,0.25) node [style=sergio] {\scriptsize$\sigma_2$};
\path[rotate around={240:(2,-2)}] (2,0.25) node [style=sergio] {\scriptsize$\sigma_1$};
\path[rotate around={270:(2,-2)}] (2,0.25) node [style=sergio] {\scriptsize$\sigma_2$};
\path[rotate around={300:(2,-2)}] (2,0.25) node [style=sergio] {\scriptsize$\sigma_1$};
\path[rotate around={330:(2,-2)}] (2,0.25) node [style=sergio] {\scriptsize$\sigma_2$};
\filldraw[fill=red!20, draw=red, thick, fill opacity=0.5] (1.5,1.25)--(0.75,1)--(1.5,-0.25)--cycle;
\filldraw[rotate around={30:(2,-2)},fill=red!20, draw=red, thick, fill opacity=0.5] (1.5,1.25)--(0.75,1)--(1.5,-0.25)--cycle;
\filldraw[rotate around={60:(2,-2)},fill=red!20, draw=red, thick, fill opacity=0.5] (1.5,1.25)--(0.75,1)--(1.5,-0.25)--cycle;
\filldraw[rotate around={90:(2,-2)},fill=red!20, draw=red, thick, fill opacity=0.5] (1.5,1.25)--(0.75,1)--(1.5,-0.25)--cycle;
\filldraw[rotate around={120:(2,-2)},fill=red!20, draw=red, thick, fill opacity=0.5] (1.5,1.25)--(0.75,1)--(1.5,-0.25)--cycle;
\filldraw[rotate around={150:(2,-2)},fill=red!20, draw=red, thick, fill opacity=0.5] (1.5,1.25)--(0.75,1)--(1.5,-0.25)--cycle;
\filldraw[rotate around={180:(2,-2)},fill=red!20, draw=red, thick, fill opacity=0.5] (1.5,1.25)--(0.75,1)--(1.5,-0.25)--cycle;
\filldraw[rotate around={210:(2,-2)},fill=red!20, draw=red, thick, fill opacity=0.5] (1.5,1.25)--(0.75,1)--(1.5,-0.25)--cycle;
\filldraw[rotate around={240:(2,-2)},fill=red!20, draw=red, thick, fill opacity=0.5] (1.5,1.25)--(0.75,1)--(1.5,-0.25)--cycle;
\filldraw[rotate around={270:(2,-2)},fill=red!20, draw=red, thick, fill opacity=0.5] (1.5,1.25)--(0.75,1)--(1.5,-0.25)--cycle;
\filldraw[rotate around={300:(2,-2)},fill=red!20, draw=red, thick, fill opacity=0.5] (1.5,1.25)--(0.75,1)--(1.5,-0.25)--cycle;
\filldraw[rotate around={330:(2,-2)},fill=red!20, draw=red, thick, fill opacity=0.5] (1.5,1.25)--(0.75,1)--(1.5,-0.25)--cycle;
\path (1.2,0.75) node [style=sergio] {\scriptsize$\lambda$};
\path[rotate around={30:(2,-2)}] (1.2,0.75) node [style=sergio] {\scriptsize$\lambda$};
\path[rotate around={60:(2,-2)}] (1.2,0.75) node [style=sergio] {\scriptsize$\lambda$};
\path[rotate around={90:(2,-2)}] (1.2,0.75) node [style=sergio] {\scriptsize$\lambda$};
\path[rotate around={120:(2,-2)}] (1.2,0.75) node [style=sergio] {\scriptsize$\lambda$};
\path[rotate around={150:(2,-2)}] (1.2,0.75) node [style=sergio] {\scriptsize$\lambda$};
\path[rotate around={180:(2,-2)}] (1.2,0.75) node [style=sergio] {\scriptsize$\lambda$};
\path[rotate around={210:(2,-2)}] (1.2,0.75) node [style=sergio] {\scriptsize$\lambda$};
\path[rotate around={240:(2,-2)}] (1.2,0.75) node [style=sergio] {\scriptsize$\lambda$};
\path[rotate around={270:(2,-2)}] (1.2,0.75) node [style=sergio] {\scriptsize$\lambda$};
\path[rotate around={300:(2,-2)}] (1.2,0.75) node [style=sergio] {\scriptsize$\lambda$};
\path[rotate around={330:(2,-2)}] (1.2,0.75) node [style=sergio] {\scriptsize$\lambda$};
\path (-5,-6) node [style=sergio] {Open lattice};
\path (2,-6) node [style=sergio] {Intersection};
\path (-5,0.6) node [style=sergio] {\scriptsize$C_6$};
\path (-4.5,1.35) node [style=sergio] {\scriptsize$M_1$};
\path (-5.5,1.35) node [style=sergio] {\scriptsize$M_2$};
\end{tikzpicture}
\caption{The cell decomposition of $C_{6v}$-symmetric lattice.  Left panel depicts the whole lattice, while the right panel depicts the horizontal intersection, including 3D blocks $\lambda$, 2D blocks $\sigma_1$ and $\sigma_2$, and 1D block $\tau$. $C_6$ depicts the axis of the 6-fold rotation, and $\bs{M}_{1,2}$ depict the reflection planes.}
\label{C6v cell decomposition}
\end{figure}

We summarize the classifications and corresponding root phases of crystalline topological phases. For crystalline TSC with spinless fermions, we summarize all possible block states as:
\begin{enumerate}[1.]
\item 2D blocks $\sigma_1$ and $\sigma_2$: 2D $(p+ip)$-SCs and 2D $\mathbb{Z}_2$ fSPT phases;
\item 1D block $\tau$: Majorana chain, 1D $\mathbb{Z}_6\rtimes\mathbb{Z}_2$ fSPT phases and Haldane chain. 
\end{enumerate}
By explicit $K$-matrix calculations, we see that except 2D bosonic Levin-Gu states on $\sigma_1/\sigma_2$, all other 2D block states are obstructed. For two root phases of 1D $\mathbb{Z}_6\rtimes\mathbb{Z}_2$ fSPT phases on $\tau$, the first root phase can be trivialized by ``Majorana'' bubble construction on $\sigma_1$ and the second root phase can be trivialized by ``Majorana'' bubble construction on both $\sigma_1$ and $\sigma_2$. The Haldane chain on $\tau$ can be trivialized by 1D $\mathbb{Z}_2$ fSPT phase on each $\sigma_1$ as 2D bubble equivalence. Therefore, the ultimate classification is $\mathbb{Z}_2^3$, with the following root phases:
\begin{enumerate}[1.]
\item 2D bosonic Levin-Gu state on each $\sigma_1/\sigma_2$ ($\mathbb{Z}_2^2$);
\item Majorana chain on $\tau$ ($\mathbb{Z}_2$).
\end{enumerate}
The HO topological surface theories of these different root phases are:
\begin{enumerate}[1.]
\item $2^{\mathrm{nd}}$-order: 1D nonchiral Luttinger liquids with $K$-matrix $K=\sigma^x$ and $\mathbb{Z}_2$ symmetry property $W^{\mathbb{Z}_2}=\mathbbm{1}_{2\times2}$ and $\delta\phi^{\mathbb{Z}_2}=\pi(1,1)^T$, on the intersections between the open lattice and 2D blocks $\sigma_1/\sigma_2$;
\item $3^{\mathrm{rd}}$-order: Dangling Majorana zero modes at the centers of top and bottom surfaces of the open lattice.
\end{enumerate}

For crystalline TSC with spin-1/2 fermions, the ultimate classification is trivial because there is no nontrivial block states on lower-dimensional blocks, even we do not concentrate on the obstruction and trivialization.

For crystalline TI with spinless fermions, we summarize all possible block states as:
\begin{enumerate}[1.]
\item 2D blocks $\sigma_1$ and $\sigma_2$: Chern insulators, Kitaev's $E_8$ states and 2D $U^f(1)\times\mathbb{Z}_2$ fSPT phases;
\item 1D block $\tau$: Haldane chain.
\end{enumerate}
By explicit $K$-matrix calculations, we conclude that except 2D $U^f(1)\times\mathbb{Z}_2$ bSPT phase on each $\sigma_1/\sigma_2$ and Kitaev's $E_8$ states decoration on $\sigma_1$ and $\sigma_2$ with opposite chiralities (chiral central charges $c_-=\pm8$), all other 2D block states are obstructed. Hence the ultimate classification is $\mathbb{Z}_2^4$, with the following root phases:
\begin{enumerate}[1.]
\item 2D $U^f(1)\times\mathbb{Z}_2$ bSPT phase on each $\sigma_1/\sigma_2$ ($\mathbb{Z}_2^2$);
\item Monolayer Kitaev's $E_8$ state on each $\sigma_1$ and $\sigma_2$ ($\mathbb{Z}_2$);
\item Haldane chain decoration on $\tau$ ($\mathbb{Z}_2$).
\end{enumerate}
The HO topological surface theories of different root phases are:
\begin{enumerate}[1.]
\item $2^{\mathrm{nd}}$-order: 1D nonchiral Luttinger liquids with $K$-matrix $K=\sigma^x$ and $\mathbb{Z}_2$ symmetry property $W^{\mathbb{Z}_2}=\mathbbm{1}_{2\times2}$ and $\delta\phi^{\mathbb{Z}_2}=\pi(1,1)^T$, on the intersections between the open lattice and 2D blocks $\sigma_1/\sigma_2$;
\item $2^{\mathrm{nd}}$-order: 1D chiral Luttinger liquids with $K$-matrix (\ref{K-matrix E8S}), on all verticle hinges of the open lattice;
\item $3^{\mathrm{rd}}$-order: dangling spin-1/2 degrees of freedom at the centers of top and bottom surfaces of the open lattice.
\end{enumerate}

For crystalline TI with spin-1/2 fermions, all possible block states are located on the 2D blocks $\sigma_1$ and $\sigma_2$: Chern insulators, Kitaev's $E_8$ states and 2D $U^f(1)\times\mathbb{Z}_2$ fSPT phases. All chiral block states on $\sigma_1$ and $\sigma_2$ should have opposite chiralities to guarantee a nonchiral 1D mode on $
\tau$, and even layers of chiral block states on $\sigma_1$ and $\sigma_2$ are trivialized by 3D bubble equivalences on $\lambda$. Therefore, the ultimate classification is $\mathbb{Z}_8\times\mathbb{Z}_4\times\mathbb{Z}_2$, with the following root phases:
\begin{enumerate}[1.]
\item Monolayer Chern insulator on each $\sigma_1$ and $\sigma_2$ ($\mathbb{Z}_2$);
\item 2D $U^f(1)\times\mathbb{Z}_2$ fSPT phase on each $\sigma_1/\sigma_2$ ($\mathbb{Z}_4^2$);
\item Monolayer Kitaev's $E_8$ state on each $\sigma_1$ and $\sigma_2$ ($\mathbb{Z}_2$);
\end{enumerate}
And there are several nontrivial extensions between these root phases: bilayer Chern insulators on each 2D block can be smoothly deformed to 2D $U^f(1)\times\mathbb{Z}_2$ fSPT phase on $\sigma_1$ and $\sigma_2$ with $(\nu_1,\nu_2)=(1,3)$, by 3D ``Chern insulator'' bubble equivalence. The second-order topological surface theories of these root phases are:
\begin{enumerate}[1.]
\item Monolayer Chern insulator on each $\sigma_1$ and $\sigma_2$: chiral fermions on the intersections between the open lattice, $\sigma_1$ and $\sigma_2$;
\item 2D $U^f(1)\times\mathbb{Z}_2$ fSPT phase on each $\sigma_{1,2}$: 1D nonchiral Luttinger liquid with $K$-matrix $K=\sigma^z$ and $\mathbb{Z}_2$ symmetry property: $W^{\mathbb{Z}_2}=\mathbbm{1}_{2\times2}$ and $\delta\phi^{\mathbb{Z}_2}=\pi(0,1)^T/\pi(1,0)^T$ on the intersections between the open lattice and $\sigma_1/\sigma_2$;
\item Monolayer Kitaev's $E_8$ state on each $\sigma_1$ and $\sigma_2$: 1D chiral Luttinger liquids with $K$-matrix (\ref{K-matrix E8S}) on the intersections between the open lattice, $\sigma_1$ and $\sigma_2$.
\end{enumerate}

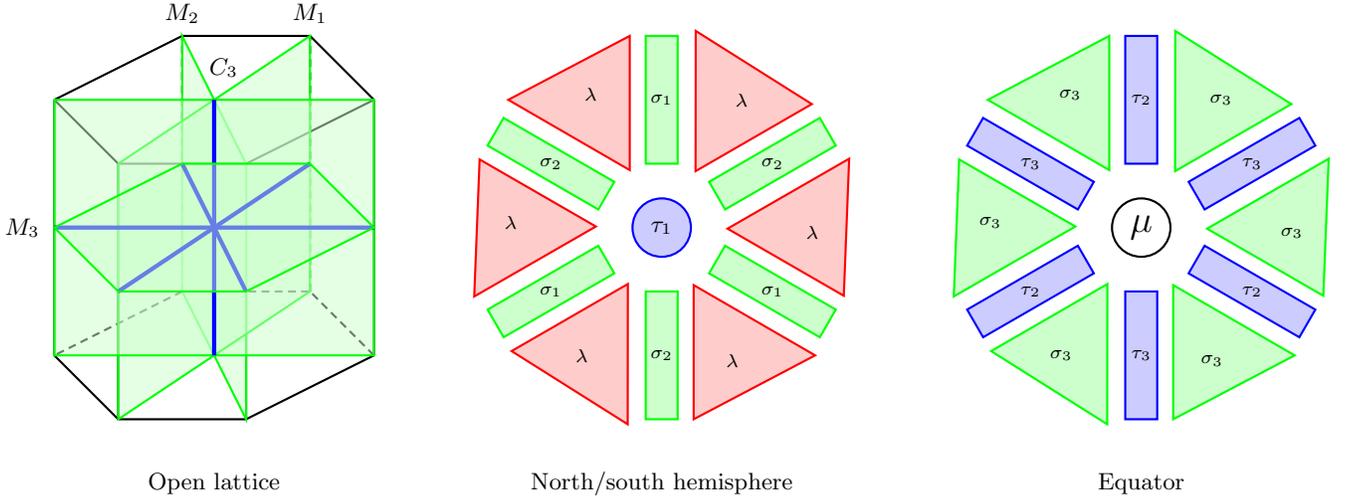
\begin{figure*}
\begin{tikzpicture}[scale=0.85]
\tikzstyle{sergio}=[rectangle,draw=none]
\draw[thick] (-13,1) -- (-15,0);
\draw[thick] (-13,1) -- (-11,1);
\draw[thick] (-10,0) -- (-11,1);
\draw[thick] (-10,0) -- (-12,-1);
\draw[thick] (-14,-1) -- (-12,-1);
\draw[thick] (-15,0) -- (-14,-1);
\draw[thick] (-15,0) -- (-15,-4);
\draw[thick] (-14,-5) -- (-15,-4);
\draw[thick] (-14,-5) -- (-12,-5);
\draw[thick] (-10,-4) -- (-12,-5);
\draw[thick] (-10,-4) -- (-10,0);
\draw[densely dashed,thick] (-10,-4) -- (-11,-3);
\draw[densely dashed,thick] (-13,-3) -- (-11,-3);
\draw[densely dashed,thick] (-13,-3) -- (-15,-4);
\draw[densely dashed,thick] (-13,-3) -- (-13,1);
\draw[densely dashed,thick] (-11,-3) -- (-11,1);
\draw[thick] (-12,-1) -- (-12,-5);
\draw[thick] (-14,-1) -- (-14,-5);
\filldraw[fill=green!20, draw=green, thick, fill opacity=0.5] (-12,-5)--(-13,-3)--(-13,1)--(-12,-1)--cycle;
\filldraw[fill=green!20, draw=green, thick, fill opacity=0.5] (-11,-3)--(-14,-5)--(-14,-1)--(-11,1)--cycle;
\filldraw[fill=green!20, draw=green, thick, fill opacity=0.5] (-10,-4)--(-15,-4)--(-15,0)--(-10,0)--cycle;
\draw[color=blue,ultra thick] (-12.5,0) -- (-12.5,-4);
\draw[color=blue,ultra thick] (-14,-3) -- (-11,-1);
\draw[color=blue,ultra thick] (-13,-1) -- (-12,-3);
\draw[color=blue,ultra thick] (-15,-2) -- (-10,-2);
\filldraw[fill=green!20, draw=green, thick, fill opacity=0.5] (-10,-2)--(-11,-1)--(-13,-1)--(-15,-2)--(-14,-3)--(-12,-3)--cycle;
\filldraw[fill=white!20, draw=black, thick] (2,-2)circle (13pt);
\filldraw[fill=blue!20, draw=blue, thick] (2.25,-1)--(1.75,-1)--(1.75,1)--(2.25,1)--cycle;
\filldraw[rotate around={60:(2,-2)},fill=blue!20, draw=blue, thick] (2.25,-1)--(1.75,-1)--(1.75,1)--(2.25,1)--cycle;
\filldraw[rotate around={120:(2,-2)},fill=blue!20, draw=blue, thick] (2.25,-1)--(1.75,-1)--(1.75,1)--(2.25,1)--cycle;
\filldraw[rotate around={180:(2,-2)},fill=blue!20, draw=blue, thick] (2.25,-1)--(1.75,-1)--(1.75,1)--(2.25,1)--cycle;
\filldraw[rotate around={240:(2,-2)},fill=blue!20, draw=blue, thick] (2.25,-1)--(1.75,-1)--(1.75,1)--(2.25,1)--cycle;
\filldraw[rotate around={300:(2,-2)},fill=blue!20, draw=blue, thick] (2.25,-1)--(1.75,-1)--(1.75,1)--(2.25,1)--cycle;
\filldraw[fill=green!20, draw=green, thick] (-0.4,0)--(1.5,1)--(1.5,-1.1)--cycle;
\filldraw[rotate around={60:(2,-2)},fill=green!20, draw=green, thick] (-0.4,0)--(1.5,1)--(1.5,-1.1)--cycle;
\filldraw[rotate around={120:(2,-2)},fill=green!20, draw=green, thick] (-0.4,0)--(1.5,1)--(1.5,-1.1)--cycle;
\filldraw[rotate around={180:(2,-2)},fill=green!20, draw=green, thick] (-0.4,0)--(1.5,1)--(1.5,-1.1)--cycle;
\filldraw[rotate around={240:(2,-2)},fill=green!20, draw=green, thick] (-0.4,0)--(1.5,1)--(1.5,-1.1)--cycle;
\filldraw[rotate around={300:(2,-2)},fill=green!20, draw=green, thick] (-0.4,0)--(1.5,1)--(1.5,-1.1)--cycle;
\path (3.25,0) node [style=sergio] {\scriptsize$\sigma_3$};
\path[rotate around={60:(2,-2)}] (3.25,0) node [style=sergio] {\scriptsize$\sigma_3$};
\path[rotate around={120:(2,-2)}] (3.25,0) node [style=sergio] {\scriptsize$\sigma_3$};
\path[rotate around={180:(2,-2)}] (3.25,0) node [style=sergio] {\scriptsize$\sigma_3$};
\path[rotate around={240:(2,-2)}] (3.25,0) node [style=sergio] {\scriptsize$\sigma_3$};
\path[rotate around={300:(2,-2)}] (3.25,0) node [style=sergio] {\scriptsize$\sigma_3$};
\path (2,0) node [style=sergio] {\scriptsize$\tau_2$};
\path[rotate around={60:(2,-2)}] (2,0) node [style=sergio] {\scriptsize$\tau_3$};
\path[rotate around={120:(2,-2)}] (2,0) node [style=sergio] {\scriptsize$\tau_2$};
\path[rotate around={180:(2,-2)}] (2,0) node [style=sergio] {\scriptsize$\tau_3$};
\path[rotate around={240:(2,-2)}] (2,0) node [style=sergio] {\scriptsize$\tau_2$};
\path[rotate around={300:(2,-2)}] (2,0) node [style=sergio] {\scriptsize$\tau_3$};
\path (2,-2) node [style=sergio] {\Large$\mu$};
\path (-12.5,-6) node [style=sergio] {Open lattice};
\path (-5.5,-6) node [style=sergio] {North/south hemisphere};
\path (2,-6) node [style=sergio] {Equator};
\filldraw[fill=blue!20, draw=blue, thick] (-5.5,-2)circle (13pt);
\filldraw[fill=green!20, draw=green, thick] (-5.25,-1)--(-5.75,-1)--(-5.75,1)--(-5.25,1)--cycle;
\filldraw[rotate around={60:(-5.5,-2)},fill=green!20, draw=green, thick] (-5.25,-1)--(-5.75,-1)--(-5.75,1)--(-5.25,1)--cycle;
\filldraw[rotate around={120:(-5.5,-2)},fill=green!20, draw=green, thick] (-5.25,-1)--(-5.75,-1)--(-5.75,1)--(-5.25,1)--cycle;
\filldraw[rotate around={180:(-5.5,-2)},fill=green!20, draw=green, thick] (-5.25,-1)--(-5.75,-1)--(-5.75,1)--(-5.25,1)--cycle;
\filldraw[rotate around={240:(-5.5,-2)},fill=green!20, draw=green, thick] (-5.25,-1)--(-5.75,-1)--(-5.75,1)--(-5.25,1)--cycle;
\filldraw[rotate around={300:(-5.5,-2)},fill=green!20, draw=green, thick] (-5.25,-1)--(-5.75,-1)--(-5.75,1)--(-5.25,1)--cycle;
\filldraw[fill=red!20, draw=red, thick] (-7.9,-0)--(-6,1)--(-6,-1.1)--cycle;
\filldraw[rotate around={60:(-5.5,-2)},fill=red!20, draw=red, thick] (-7.9,-0)--(-6,1)--(-6,-1.1)--cycle;
\filldraw[rotate around={120:(-5.5,-2)},fill=red!20, draw=red, thick] (-7.9,-0)--(-6,1)--(-6,-1.1)--cycle;
\filldraw[rotate around={180:(-5.5,-2)},fill=red!20, draw=red, thick] (-7.9,-0)--(-6,1)--(-6,-1.1)--cycle;
\filldraw[rotate around={240:(-5.5,-2)},fill=red!20, draw=red, thick] (-7.9,-0)--(-6,1)--(-6,-1.1)--cycle;
\filldraw[rotate around={300:(-5.5,-2)},fill=red!20, draw=red, thick] (-7.9,-0)--(-6,1)--(-6,-1.1)--cycle;
\path (-4.25,0) node [style=sergio] {\scriptsize$\lambda$};
\path[rotate around={60:(-5.5,-2)}] (-4.25,0) node [style=sergio] {\scriptsize$\lambda$};
\path[rotate around={120:(-5.5,-2)}] (-4.25,0) node [style=sergio] {\scriptsize$\lambda$};
\path[rotate around={180:(-5.5,-2)}] (-4.25,0) node [style=sergio] {\scriptsize$\lambda$};
\path[rotate around={240:(-5.5,-2)}] (-4.25,0) node [style=sergio] {\scriptsize$\lambda$};
\path[rotate around={300:(-5.5,-2)}] (-4.25,0) node [style=sergio] {\scriptsize$\lambda$};
\path (-5.5,0) node [style=sergio] {\scriptsize$\sigma_1$};
\path[rotate around={60:(-5.5,-2)}] (-5.5,0) node [style=sergio] {\scriptsize$\sigma_2$};
\path[rotate around={120:(-5.5,-2)}] (-5.5,0) node [style=sergio] {\scriptsize$\sigma_1$};
\path[rotate around={180:(-5.5,-2)}] (-5.5,0) node [style=sergio] {\scriptsize$\sigma_2$};
\path[rotate around={240:(-5.5,-2)}] (-5.5,0) node [style=sergio] {\scriptsize$\sigma_1$};
\path[rotate around={300:(-5.5,-2)}] (-5.5,0) node [style=sergio] {\scriptsize$\sigma_2$};
\path (-5.5,-2) node [style=sergio] {$\tau_1$};
\path (-12.35,0.5) node [style=sergio] {$C_3$};
\path (-11,1.35) node [style=sergio] {$M_1$};
\path (-13,1.35) node [style=sergio] {$M_2$};
\path (-15.5,-2) node [style=sergio] {$M_3$};
\end{tikzpicture}
\caption{The cell decomposition of $D_{3h}$-symmetric lattice. Left panel depicts the whole open lattice; middle panel depicts the north/south hemisphere, including 3D blocks $\lambda$, 2D blocks $\sigma_1$ \& $\sigma_2$, and 1D blocks $\tau_1$; right panel depicts the equator, including 2D blocks $\sigma_3$, 1D blocks $\tau_2$ \& $\tau_3$, and 0D block $\mu$. $C_3$ depicts the axis of the 3-fold rotation, and $\bs{M}_{1,2,3}$ depict the reflection planes.}
\label{D3h cell decomposition}
\end{figure*}

\subsection{$D_{3h}$-symmetric lattice}
For $D_{3h}$-symmetric lattice with the cell decomposition in Fig. \ref{D3h cell decomposition}, the ground-state wavefunction of the system can be decomposed to the direct products of wavefunctions of lower-dimensional block states as:
\begin{align}
|\Psi\rangle=\bigotimes\limits_{g\in D_{3h}}|T_{g\lambda}\rangle\otimes\sum\limits_{k=1}^3|\gamma_{g\sigma_k}\rangle\otimes\sum\limits_{j=1}^3|\beta_{g\tau_j}\rangle\otimes|\alpha_\mu\rangle
\label{D3h cell decomposition}
\end{align}
where $|T_{g\lambda}\rangle$ is the wavefunction of 3D block state on $g\lambda$ which is topological trivial; $|\gamma_{g\sigma_{1,2,3}}\rangle$ is the $\mathbb{Z}_2$-symmetric wavefunction of 2D block state on $g\sigma_{1,2,3}$; $|\beta_{g\tau_1}\rangle$ is the $(\mathbb{Z}_3\rtimes\mathbb{Z}_2)$-symmetric wavefunction of 2D block state on $g\tau_1$, and $|\beta_{g\tau_{2,3}}\rangle$ is the $(\mathbb{Z}_2\times\mathbb{Z}_2)$-symmetric wavefunction of 2D block state on $g\tau_{2,3}$; $|\alpha_{\mu}\rangle$ is the $\mathbb{Z}_2\times(\mathbb{Z}_3\rtimes\mathbb{Z}_2)$-symmetric wavefunction of 0D block state on $\mu$.

We summarize the classifications and corresponding root phases of $D_{3h}$-symmetric crystalline topological phases. For crystalline TSC with spinless fermions, we summarize all possible block states as:
\begin{enumerate}[1.]
\item 2D blocks $\sigma_1$, $\sigma_2$ and $\sigma_3$: 2D $(p+ip)$-SCs and 2D $\mathbb{Z}_2$ fSPT phases;
\item 1D block $\tau_1$: Majorana chain and 1D $\mathbb{Z}_3\rtimes\mathbb{Z}_2$ fSPT phase;
\item 1D blocks $\tau_2$ and $\tau_3$: Majorana chains, 1D $\mathbb{Z}_2\times\mathbb{Z}_2$ fSPT phases and Haldane chains;
\item 0D block $\mu$: 0D modes characterizing eigenvalues $-1$ of two generators of the $D_{3h}$ group, with even/odd fermion parity.
\end{enumerate}
By explicit $K$-matrix calculations, we know that except 2D bosonic Levin-Gu state on $(\sigma_1,\sigma_2)$ and $\sigma_3$, all other 2D block states are obstructed. 

Similar to the $D_{3d}$-symmetric case, Majorana chain decoration and 1D $\mathbb{Z}_3\rtimes\mathbb{Z}_2$ fSPT phases are obstructed. For 1D blocks $\tau_2$ and $\tau_3$ on the equator, the first root phase of 1D $\mathbb{Z}_2\times\mathbb{Z}_2$ fSPT phases solely on $\tau_2$ or $\tau_3$ are obstructed, and if we consider $\tau_2$ and $\tau_3$ together, the corresponding 1D block state is trivialized by 2D ``Majorana'' bubble construction on each $\sigma_3$; the second root phase of 1D $\mathbb{Z}_2\times\mathbb{Z}_2$ fSPT phases on $\tau_2$ or $\tau_3$ are trivialized by 2D ``Majorana'' bubble constructions on $\sigma_1$ or $\sigma_2$, respectively. Furthermore, Haldane chain decoration on $\tau_2$ and $\tau_3$ can be trivialized by 1D $\mathbb{Z}_2$ fSPT phase as the 2D bubble equivalence on each $\sigma_1$. 

For the 0D block states on $\mu$, consider a complex fermion on each $\tau_1$ which can be adiabatically deformed to infinite far and trivialized, it leaves an atomic insulator with two complex fermions who has eigenvalue $-1$ under the reflection with respect to the equator. On the equator, similar to the 2D $D_3$-symmetric case, all 0D block states are trivialized. 

Therefore, the ultimate classification is $\mathbb{Z}_2^3$, with the following root phases:
\begin{enumerate}[1.]
\item 2D bosonic Levin-Gu state on each 2D block $\sigma_1$ and $\sigma_2$ ($\mathbb{Z}_2$);
\item 2D bosonic Levin-Gu state on each 2D block $\sigma_3$ ($\mathbb{Z}_2$);
\item Majorana chain decoration on each $\tau_2$ and $\tau_3$ ($\mathbb{Z}_2$).
\end{enumerate}
Their HO topological surface theories are listed as following:
\begin{enumerate}[1.]
\item $2^{\mathrm{nd}}$-order: 1D nonchiral Luttinger liquids with $K$-matrix $K=\sigma^x$ and $\mathbb{Z}_2$ symmetry property $W^{\mathbb{Z}_2}=\mathbbm{1}_{2\times2}$ and $\delta\phi^{\mathbb{Z}_2}=\pi(1,1)^T$, on each verticle hinges of the open lattice;
\item $2^{\mathrm{nd}}$-order: 1D nonchiral Luttinger liquids with $K$-matrix $K=\sigma^x$ and $\mathbb{Z}_2$ symmetry property $W^{\mathbb{Z}_2}=\mathbbm{1}_{2\times2}$ and $\delta\phi^{\mathbb{Z}_2}=\pi(1,1)^T$, on the edge of the equator.
\end{enumerate}

For crystalline TSC with spin-1/2 fermions, all possible block states are located on the 0D block: 0D modes characterizing eigenvalues $-1$ of two generators of the $D_{3h}$ group, which are all trivialized according to the way we have discussed in the case with spinless fermions. Therefore, the ultimate classification is trivial.

For crystalline TI with spinless  fermions, we summarize all possible block states as:
\begin{enumerate}[1.]
\item 2D blocks $\sigma_1$, $\sigma_2$ and $\sigma_3$: Chern insulators, Kitaev's $E_8$ states and 2D $U^f(1)\times\mathbb{Z}_2$ fSPT phases;
\item 1D blocks $\tau_2$ and $\tau_3$: Haldane chain;
\item 0D block $\mu$: 0D modes characterizing eigenvalues $-1$ of two generators of the $D_{3h}$ group, with different $U^f(1)$ charges.
\end{enumerate}
By explicit $K$-matrix calculations, we conclude that except 2D $U^f(1)\times\mathbb{Z}_2$ bSPT phase on each 2D block $(\sigma_1,\sigma_2)$ and $\sigma_3$, and monolayer Kitaev's $E_8$ states on $(\sigma_1,\sigma_2,\sigma_3)$ with chiralities $(+,-,+)$, all other 2D block states are obstructed. Furthermore, 0D block state characterizing eigenvalue $-1$ of the reflection with respect to the equator is trivialized by 1D bubble equivalence on $\tau_1$, hence the ultimate classification is $\mathbb{Z}_2^5$, with the following root phases:
\begin{enumerate}[1.]
\item 2D $U^f(1)\times\mathbb{Z}_2$ bSPT phase on each 2D block $\sigma_1$ and $\sigma_2$ ($\mathbb{
Z}_2$);
\item 2D $U^f(1)\times\mathbb{Z}_2$ bSPT phase on each 2D block $\sigma_3$ ($\mathbb{Z}_2$);
\item Monolayer Kitaev's $E_8$ state on each 2D block $\sigma_1$, $\sigma_2$ and $\sigma_3$ ($\mathbb{Z}_2$). 
\item Haldane chain on each $\tau_2$ and $\tau_3$ ($\mathbb{Z}_2$);
\item 0D mode characterizing eigenvalue $-1$ of the vertical reflection ($\mathbb{Z}_2$).
\end{enumerate}
Their HO topological surface theories are listed as following:
\begin{enumerate}[1.]
\item $2^{\mathrm{nd}}$-order: 1D nonchiral Luttinger liquids with $K$-matrix $K=\sigma^x$ and $\mathbb{Z}_2$ symmetry property $W^{\mathbb{Z}_2}=\mathbbm{1}_{2\times2}$ and $\delta\phi^{\mathbb{Z}_2}=\pi(1,1)^T$, on each verticle hinges of the open lattice;
\item $2^{\mathrm{nd}}$-order: 1D nonchiral Luttinger liquids with $K$-matrix $K=\sigma^x$ and $\mathbb{Z}_2$ symmetry property $W^{\mathbb{Z}_2}=\mathbbm{1}_{2\times2}$ and $\delta\phi^{\mathbb{Z}_2}=\pi(1,1)^T$, on the edge of the equator;
\item $2^{\mathrm{nd}}$-order: 1D chiral Luttinger liquids with $K$-matrix (\ref{K-matrix E8S}) on all verticle hinges of the open lattice and the edge of the equator;
\item $3^{\mathrm{rd}}$-order: A dangling spin-1/2 degree of freedom at the center of each hinge.
\end{enumerate}

For crystalline TI with spin-1/2 fermions, we summarize all possible block states as:
\begin{enumerate}[1.]
\item 2D blocks $\sigma_1$, $\sigma_2$ and $\sigma_3$: Chern insulators, Kitaev's $E_8$ states and 2D $U^f(1)\times\mathbb{Z}_2$ fSPT phases;
\item 0D block $\mu$: 0D modes characterizing eigenvalues $-1$ of two generators of the $D_{3h}$ group, with different even $U^f(1)$ charges.
\end{enumerate}
In order to obtain obstruction-free block states, the chiralities of all chiral 2D block states should satisfy $(+,-,+)$ to guarantee that they do not leave chiral 1D mode on their shared border, and even layers of chiral block states are trivialized by 3D bubble equivalences on $\lambda$. Therefore, the ultimate classification is $\mathbb{Z}_8\times\mathbb{Z}_4\times\mathbb{Z}_2^3$, with the following root phases:
\begin{enumerate}[1.]
\item Monolayer Chern insulator on each 2D block $\sigma_1$, $\sigma_2$ and $\sigma_3$ ($\mathbb{Z}_2$);
\item 2D $U^f(1)\times\mathbb{Z}_2$ fSPT phase on each 2D blocks $\sigma_1$ and $\sigma_2$ ($\mathbb{Z}_4$);
\item 2D $U^f(1)\times\mathbb{Z}_2$ fSPT phase on each 2D blocks $\sigma_3$ ($\mathbb{Z}_4$);
\item Monolayer Kitaev's $E_8$ state on each 2D block $\sigma_1$, $\sigma_2$ and $\sigma_3$ ($\mathbb{Z}_2$);
\item 0D modes with eigenvalues $-1$ of verticle and horizontal reflection planes ($\mathbb{Z}_2^2$).
\end{enumerate}
And there are several nontrivial extensions between these root phases: bilayer Chern insulators on each 2D block can be smoothly deformed to 2D $U^f(1)\times\mathbb{Z}_2$ fSPT phase with $(\nu_1,\nu_2,\nu_3)=(1,3,1)$, by 3D ``Chern insulator'' bubble equivalence. The HO topological surface theories of these root phases are:
\begin{enumerate}[1.]
\item $2^{\mathrm{nd}}$-order: Chiral fermions on all verticle hinges and the edge of the equator;
\item $2^{\mathrm{nd}}$-order: 1D nonchiral Luttinger liquids with $K$-matrix $K=\sigma^z$ and $\mathbb{Z}_2$ symmetry property $W^{\mathbb{Z}_2}=\mathbbm{1}_{2\times2}$ and $\delta\phi^{\mathbb{Z}_2}=\pi(0,1)^T/\pi(1,0)^T$, on the intersections between the open lattice and 2D blocks decorated with a $U^f(1)\times\mathbb{Z}_2$ fSPT phase with $\nu=1/\nu=3$ on each of them;
\item $2^{\mathrm{nd}}$-order: 1D chiral Luttinger liquids with $K$-matrix (\ref{K-matrix E8S}) on all verticle hinges and the edge of the equator.
\end{enumerate}

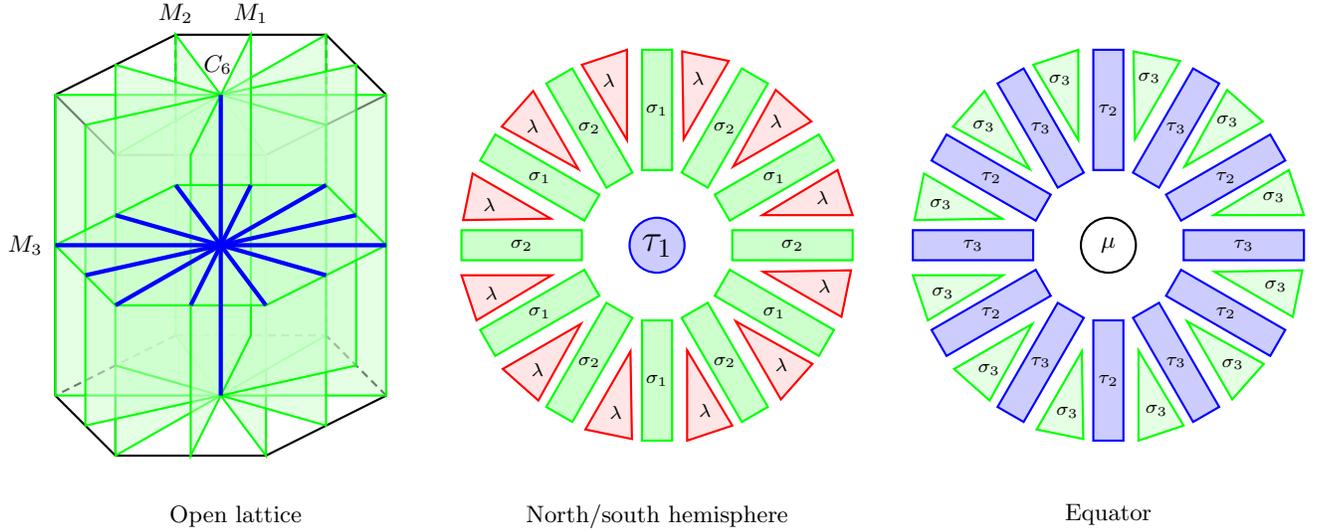
\begin{figure*}
\begin{tikzpicture}[scale=0.8]
\tikzstyle{sergio}=[rectangle,draw=none]
\draw[thick] (-13.5,1.5) -- (-15.5,0.5);
\draw[thick] (-13.5,1.5) -- (-11,1.5);
\draw[thick] (-10,0.5) -- (-11,1.5);
\draw[thick] (-10,0.5) -- (-12,-0.5);
\draw[thick] (-14.5,-0.5) -- (-12,-0.5);
\draw[thick] (-15.5,0.5) -- (-14.5,-0.5);
\draw[thick] (-15.5,0.5) -- (-15.5,-4.5);
\draw[thick] (-14.5,-5.5) -- (-15.5,-4.5);
\draw[thick] (-14.5,-5.5) -- (-12,-5.5);
\draw[thick] (-10,-4.5) -- (-12,-5.5);
\draw[thick] (-10,-4.5) -- (-10,0.5);
\draw[densely dashed,thick] (-10,-4.5) -- (-11,-3.5);
\draw[densely dashed,thick] (-13.5,-3.5) -- (-11,-3.5);
\draw[densely dashed,thick] (-13.5,-3.5) -- (-15.5,-4.5);
\draw[densely dashed,thick] (-13.5,-3.5) -- (-13.5,1.5);
\draw[densely dashed,thick] (-11,-3.5) -- (-11,1.5);
\draw[thick] (-12,-0.5) -- (-12,-5.5);
\draw[thick] (-14.5,-0.5) -- (-14.5,-5.5);
\filldraw[fill=green!20, draw=green, thick, fill opacity=0.5] (-12,-5.5)--(-13.5,-3.5)--(-13.5,1.5)--(-12,-0.5)--cycle;
\filldraw[fill=green!20, draw=green, thick, fill opacity=0.5] (-11,-3.5)--(-14.5,-5.5)--(-14.5,-0.5)--(-11,1.5)--cycle;
\filldraw[fill=green!20, draw=green, thick, fill opacity=0.5] (-10,-4.5)--(-15.5,-4.5)--(-15.5,0.5)--(-10,0.5)--cycle;
\filldraw[fill=green!20, draw=green, thick, fill opacity=0.5] (-11,-5)--(-14.5,-4)--(-14.5,1)--(-11,0)--cycle;
\filldraw[fill=green!20, draw=green, thick, fill opacity=0.5] (-10.5,-4)--(-15,-5)--(-15,0)--(-10.5,1)--cycle;
\filldraw[fill=green!20, draw=green, thick, fill opacity=0.5] (-12.25,-3.5)--(-13.25,-5.5)--(-13.25,-0.5)--(-12.25,1.5)--cycle;
\filldraw[fill=green!20, draw=green, thick, fill opacity=0.5] (-10,-2)--(-11,-1)--(-13.5,-1)--(-15.5,-2)--(-14.5,-3)--(-12,-3)--cycle;
\draw[color=blue,ultra thick] (-12.75,0.5) -- (-12.75,-4.5);
\draw[color=blue,ultra thick] (-14.5,-3) -- (-11,-1);
\draw[color=blue,ultra thick] (-13.5,-1) -- (-12,-3);
\draw[color=blue,ultra thick] (-15.5,-2) -- (-10,-2);
\draw[color=blue,ultra thick] (-14.5,-1.5) -- (-11,-2.5);
\draw[color=blue,ultra thick] (-15,-2.5) -- (-10.5,-1.5);
\draw[color=blue,ultra thick] (-12.25,-1) -- (-13.25,-3);
\filldraw[fill=blue!20, draw=blue, thick] (-5.5,-2)circle (13pt);
\filldraw[fill=green!20, draw=green, thick] (-5.25,-0.75)--(-5.75,-0.75)--(-5.75,1.25)--(-5.25,1.25)--cycle;
\filldraw[rotate around={30:(-5.5,-2)},fill=green!20, draw=green, thick] (-5.25,-0.75)--(-5.75,-0.75)--(-5.75,1.25)--(-5.25,1.25)--cycle;
\filldraw[rotate around={60:(-5.5,-2)},fill=green!20, draw=green, thick] (-5.25,-0.75)--(-5.75,-0.75)--(-5.75,1.25)--(-5.25,1.25)--cycle;
\filldraw[rotate around={90:(-5.5,-2)},fill=green!20, draw=green, thick] (-5.25,-0.75)--(-5.75,-0.75)--(-5.75,1.25)--(-5.25,1.25)--cycle;
\filldraw[rotate around={120:(-5.5,-2)},fill=green!20, draw=green, thick] (-5.25,-0.75)--(-5.75,-0.75)--(-5.75,1.25)--(-5.25,1.25)--cycle;
\filldraw[rotate around={150:(-5.5,-2)},fill=green!20, draw=green, thick] (-5.25,-0.75)--(-5.75,-0.75)--(-5.75,1.25)--(-5.25,1.25)--cycle;
\filldraw[rotate around={180:(-5.5,-2)},fill=green!20, draw=green, thick] (-5.25,-0.75)--(-5.75,-0.75)--(-5.75,1.25)--(-5.25,1.25)--cycle;
\filldraw[rotate around={210:(-5.5,-2)},fill=green!20, draw=green, thick] (-5.25,-0.75)--(-5.75,-0.75)--(-5.75,1.25)--(-5.25,1.25)--cycle;
\filldraw[rotate around={240:(-5.5,-2)},fill=green!20, draw=green, thick] (-5.25,-0.75)--(-5.75,-0.75)--(-5.75,1.25)--(-5.25,1.25)--cycle;
\filldraw[rotate around={270:(-5.5,-2)},fill=green!20, draw=green, thick] (-5.25,-0.75)--(-5.75,-0.75)--(-5.75,1.25)--(-5.25,1.25)--cycle;
\filldraw[rotate around={300:(-5.5,-2)},fill=green!20, draw=green, thick] (-5.25,-0.75)--(-5.75,-0.75)--(-5.75,1.25)--(-5.25,1.25)--cycle;
\filldraw[rotate around={330:(-5.5,-2)},fill=green!20, draw=green, thick] (-5.25,-0.75)--(-5.75,-0.75)--(-5.75,1.25)--(-5.25,1.25)--cycle;
\path (-5.5,-2) node [style=sergio] {\Large$\tau_1$};
\path (-5.5,0.25) node [style=sergio] {\scriptsize$\sigma_1$};
\path[rotate around={30:(-5.5,-2)}] (-5.5,0.25) node [style=sergio] {\scriptsize$\sigma_2$};
\path[rotate around={60:(-5.5,-2)}] (-5.5,0.25) node [style=sergio] {\scriptsize$\sigma_1$};
\path[rotate around={90:(-5.5,-2)}] (-5.5,0.25) node [style=sergio] {\scriptsize$\sigma_2$};
\path[rotate around={120:(-5.5,-2)}] (-5.5,0.25) node [style=sergio] {\scriptsize$\sigma_1$};
\path[rotate around={150:(-5.5,-2)}] (-5.5,0.25) node [style=sergio] {\scriptsize$\sigma_2$};
\path[rotate around={180:(-5.5,-2)}] (-5.5,0.25) node [style=sergio] {\scriptsize$\sigma_1$};
\path[rotate around={210:(-5.5,-2)}] (-5.5,0.25) node [style=sergio] {\scriptsize$\sigma_2$};
\path[rotate around={240:(-5.5,-2)}] (-5.5,0.25) node [style=sergio] {\scriptsize$\sigma_1$};
\path[rotate around={270:(-5.5,-2)}] (-5.5,0.25) node [style=sergio] {\scriptsize$\sigma_2$};
\path[rotate around={300:(-5.5,-2)}] (-5.5,0.25) node [style=sergio] {\scriptsize$\sigma_1$};
\path[rotate around={330:(-5.5,-2)}] (-5.5,0.25) node [style=sergio] {\scriptsize$\sigma_2$};
\filldraw[fill=red!20, draw=red, thick, fill opacity=0.5] (-6,1.25)--(-6.75,1)--(-6,-0.25)--cycle;
\filldraw[rotate around={30:(-5.5,-2)},fill=red!20, draw=red, thick, fill opacity=0.5] (-6,1.25)--(-6.75,1)--(-6,-0.25)--cycle;
\filldraw[rotate around={60:(-5.5,-2)},fill=red!20, draw=red, thick, fill opacity=0.5] (-6,1.25)--(-6.75,1)--(-6,-0.25)--cycle;
\filldraw[rotate around={90:(-5.5,-2)},fill=red!20, draw=red, thick, fill opacity=0.5] (-6,1.25)--(-6.75,1)--(-6,-0.25)--cycle;
\filldraw[rotate around={120:(-5.5,-2)},fill=red!20, draw=red, thick, fill opacity=0.5] (-6,1.25)--(-6.75,1)--(-6,-0.25)--cycle;
\filldraw[rotate around={150:(-5.5,-2)},fill=red!20, draw=red, thick, fill opacity=0.5] (-6,1.25)--(-6.75,1)--(-6,-0.25)--cycle;
\filldraw[rotate around={180:(-5.5,-2)},fill=red!20, draw=red, thick, fill opacity=0.5] (-6,1.25)--(-6.75,1)--(-6,-0.25)--cycle;
\filldraw[rotate around={210:(-5.5,-2)},fill=red!20, draw=red, thick, fill opacity=0.5] (-6,1.25)--(-6.75,1)--(-6,-0.25)--cycle;
\filldraw[rotate around={240:(-5.5,-2)},fill=red!20, draw=red, thick, fill opacity=0.5] (-6,1.25)--(-6.75,1)--(-6,-0.25)--cycle;
\filldraw[rotate around={270:(-5.5,-2)},fill=red!20, draw=red, thick, fill opacity=0.5] (-6,1.25)--(-6.75,1)--(-6,-0.25)--cycle;
\filldraw[rotate around={300:(-5.5,-2)},fill=red!20, draw=red, thick, fill opacity=0.5] (-6,1.25)--(-6.75,1)--(-6,-0.25)--cycle;
\filldraw[rotate around={330:(-5.5,-2)},fill=red!20, draw=red, thick, fill opacity=0.5] (-6,1.25)--(-6.75,1)--(-6,-0.25)--cycle;
\path (-6.3,0.75) node [style=sergio] {\scriptsize$\lambda$};
\path[rotate around={30:(-5.5,-2)}] (-6.3,0.75) node [style=sergio] {\scriptsize$\lambda$};
\path[rotate around={60:(-5.5,-2)}] (-6.3,0.75) node [style=sergio] {\scriptsize$\lambda$};
\path[rotate around={90:(-5.5,-2)}] (-6.3,0.75) node [style=sergio] {\scriptsize$\lambda$};
\path[rotate around={120:(-5.5,-2)}] (-6.3,0.75) node [style=sergio] {\scriptsize$\lambda$};
\path[rotate around={150:(-5.5,-2)}] (-6.3,0.75) node [style=sergio] {\scriptsize$\lambda$};
\path[rotate around={180:(-5.5,-2)}] (-6.3,0.75) node [style=sergio] {\scriptsize$\lambda$};
\path[rotate around={210:(-5.5,-2)}] (-6.3,0.75) node [style=sergio] {\scriptsize$\lambda$};
\path[rotate around={240:(-5.5,-2)}] (-6.3,0.75) node [style=sergio] {\scriptsize$\lambda$};
\path[rotate around={270:(-5.5,-2)}] (-6.3,0.75) node [style=sergio] {\scriptsize$\lambda$};
\path[rotate around={300:(-5.5,-2)}] (-6.3,0.75) node [style=sergio] {\scriptsize$\lambda$};
\path[rotate around={330:(-5.5,-2)}] (-6.3,0.75) node [style=sergio] {\scriptsize$\lambda$};
\filldraw[fill=white!20, draw=black, thick] (2,-2)circle (13pt);
\filldraw[fill=blue!20, draw=blue, thick] (2.25,-0.75)--(1.75,-0.75)--(1.75,1.25)--(2.25,1.25)--cycle;
\filldraw[rotate around={30:(2,-2)},fill=blue!20, draw=blue, thick] (2.25,-0.75)--(1.75,-0.75)--(1.75,1.25)--(2.25,1.25)--cycle;
\filldraw[rotate around={60:(2,-2)},fill=blue!20, draw=blue, thick] (2.25,-0.75)--(1.75,-0.75)--(1.75,1.25)--(2.25,1.25)--cycle;
\filldraw[rotate around={90:(2,-2)},fill=blue!20, draw=blue, thick] (2.25,-0.75)--(1.75,-0.75)--(1.75,1.25)--(2.25,1.25)--cycle;
\filldraw[rotate around={120:(2,-2)},fill=blue!20, draw=blue, thick] (2.25,-0.75)--(1.75,-0.75)--(1.75,1.25)--(2.25,1.25)--cycle;
\filldraw[rotate around={150:(2,-2)},fill=blue!20, draw=blue, thick] (2.25,-0.75)--(1.75,-0.75)--(1.75,1.25)--(2.25,1.25)--cycle;
\filldraw[rotate around={180:(2,-2)},fill=blue!20, draw=blue, thick] (2.25,-0.75)--(1.75,-0.75)--(1.75,1.25)--(2.25,1.25)--cycle;
\filldraw[rotate around={210:(2,-2)},fill=blue!20, draw=blue, thick] (2.25,-0.75)--(1.75,-0.75)--(1.75,1.25)--(2.25,1.25)--cycle;
\filldraw[rotate around={240:(2,-2)},fill=blue!20, draw=blue, thick] (2.25,-0.75)--(1.75,-0.75)--(1.75,1.25)--(2.25,1.25)--cycle;
\filldraw[rotate around={270:(2,-2)},fill=blue!20, draw=blue, thick] (2.25,-0.75)--(1.75,-0.75)--(1.75,1.25)--(2.25,1.25)--cycle;
\filldraw[rotate around={300:(2,-2)},fill=blue!20, draw=blue, thick] (2.25,-0.75)--(1.75,-0.75)--(1.75,1.25)--(2.25,1.25)--cycle;
\filldraw[rotate around={330:(2,-2)},fill=blue!20, draw=blue, thick] (2.25,-0.75)--(1.75,-0.75)--(1.75,1.25)--(2.25,1.25)--cycle;
\path (2,-2) node [style=sergio] {$\mu$};
\path (2,0.25) node [style=sergio] {\scriptsize$\tau_2$};
\path[rotate around={30:(2,-2)}] (2,0.25) node [style=sergio] {\scriptsize$\tau_3$};
\path[rotate around={60:(2,-2)}] (2,0.25) node [style=sergio] {\scriptsize$\tau_2$};
\path[rotate around={90:(2,-2)}] (2,0.25) node [style=sergio] {\scriptsize$\tau_3$};
\path[rotate around={120:(2,-2)}] (2,0.25) node [style=sergio] {\scriptsize$\tau_2$};
\path[rotate around={150:(2,-2)}] (2,0.25) node [style=sergio] {\scriptsize$\tau_3$};
\path[rotate around={180:(2,-2)}] (2,0.25) node [style=sergio] {\scriptsize$\tau_2$};
\path[rotate around={210:(2,-2)}] (2,0.25) node [style=sergio] {\scriptsize$\tau_3$};
\path[rotate around={240:(2,-2)}] (2,0.25) node [style=sergio] {\scriptsize$\tau_2$};
\path[rotate around={270:(2,-2)}] (2,0.25) node [style=sergio] {\scriptsize$\tau_3$};
\path[rotate around={300:(2,-2)}] (2,0.25) node [style=sergio] {\scriptsize$\tau_2$};
\path[rotate around={330:(2,-2)}] (2,0.25) node [style=sergio] {\scriptsize$\tau_3$};
\filldraw[fill=green!20, draw=green, thick, fill opacity=0.5] (1.5,1.25)--(0.75,1)--(1.5,-0.25)--cycle;
\filldraw[rotate around={30:(2,-2)},fill=green!20, draw=green, thick, fill opacity=0.5] (1.5,1.25)--(0.75,1)--(1.5,-0.25)--cycle;
\filldraw[rotate around={60:(2,-2)},fill=green!20, draw=green, thick, fill opacity=0.5] (1.5,1.25)--(0.75,1)--(1.5,-0.25)--cycle;
\filldraw[rotate around={90:(2,-2)},fill=green!20, draw=green, thick, fill opacity=0.5] (1.5,1.25)--(0.75,1)--(1.5,-0.25)--cycle;
\filldraw[rotate around={120:(2,-2)},fill=green!20, draw=green, thick, fill opacity=0.5] (1.5,1.25)--(0.75,1)--(1.5,-0.25)--cycle;
\filldraw[rotate around={150:(2,-2)},fill=green!20, draw=green, thick, fill opacity=0.5] (1.5,1.25)--(0.75,1)--(1.5,-0.25)--cycle;
\filldraw[rotate around={180:(2,-2)},fill=green!20, draw=green, thick, fill opacity=0.5] (1.5,1.25)--(0.75,1)--(1.5,-0.25)--cycle;
\filldraw[rotate around={210:(2,-2)},fill=green!20, draw=green, thick, fill opacity=0.5] (1.5,1.25)--(0.75,1)--(1.5,-0.25)--cycle;
\filldraw[rotate around={240:(2,-2)},fill=green!20, draw=green, thick, fill opacity=0.5] (1.5,1.25)--(0.75,1)--(1.5,-0.25)--cycle;
\filldraw[rotate around={270:(2,-2)},fill=green!20, draw=green, thick, fill opacity=0.5] (1.5,1.25)--(0.75,1)--(1.5,-0.25)--cycle;
\filldraw[rotate around={300:(2,-2)},fill=green!20, draw=green, thick, fill opacity=0.5] (1.5,1.25)--(0.75,1)--(1.5,-0.25)--cycle;
\filldraw[rotate around={330:(2,-2)},fill=green!20, draw=green, thick, fill opacity=0.5] (1.5,1.25)--(0.75,1)--(1.5,-0.25)--cycle;
\path (1.2,0.75) node [style=sergio] {\scriptsize$\sigma_3$};
\path[rotate around={30:(2,-2)}] (1.2,0.75) node [style=sergio] {\scriptsize$\sigma_3$};
\path[rotate around={60:(2,-2)}] (1.2,0.75) node [style=sergio] {\scriptsize$\sigma_3$};
\path[rotate around={90:(2,-2)}] (1.2,0.75) node [style=sergio] {\scriptsize$\sigma_3$};
\path[rotate around={120:(2,-2)}] (1.2,0.75) node [style=sergio] {\scriptsize$\sigma_3$};
\path[rotate around={150:(2,-2)}] (1.2,0.75) node [style=sergio] {\scriptsize$\sigma_3$};
\path[rotate around={180:(2,-2)}] (1.2,0.75) node [style=sergio] {\scriptsize$\sigma_3$};
\path[rotate around={210:(2,-2)}] (1.2,0.75) node [style=sergio] {\scriptsize$\sigma_3$};
\path[rotate around={240:(2,-2)}] (1.2,0.75) node [style=sergio] {\scriptsize$\sigma_3$};
\path[rotate around={270:(2,-2)}] (1.2,0.75) node [style=sergio] {\scriptsize$\sigma_3$};
\path[rotate around={300:(2,-2)}] (1.2,0.75) node [style=sergio] {\scriptsize$\sigma_3$};
\path[rotate around={330:(2,-2)}] (1.2,0.75) node [style=sergio] {\scriptsize$\sigma_3$};
\path (-12.5,-6.5) node [style=sergio] {Open lattice};
\path (-5.5,-6.5) node [style=sergio] {North/south hemisphere};
\path (2,-6.4756) node [style=sergio] {Equator};
\path (-12.8,1) node [style=sergio] {$C_6$};
\path (-12.25,1.85) node [style=sergio] {$M_1$};
\path (-13.5,1.85) node [style=sergio] {$M_2$};
\path (-16,-2) node [style=sergio] {$M_3$};
\end{tikzpicture}
\caption{The cell decomposition of $D_{6h}$-symmetric lattice. Left panel depicts the whole open lattice; middle panel depicts the north/south hemisphere, including 3D blocks $\lambda$, 2D blocks $\sigma_1$ \& $\sigma_2$, and 1D blocks $\tau_1$; right panel depicts the equator, including 2D blocks $\sigma_3$, 1D blocks $\tau_2$ \& $\tau_3$, and 0D block $\mu$. $C_{6}$ depicts the axis of the 6-fold rotation, and $\bs{M}_{1,2,3}$ depict the reflection planes.}
\label{D6h cell decomposition}
\end{figure*}

\subsection{$D_{6h}$-symmetric lattice}
For $D_{6h}$-symmetric lattice with the cell decomposition in Fig. \ref{D6h cell decomposition}, the ground-state wavefunction of the system can be decomposed to the direct products of wavefunctions of lower-dimensional block states as:
\begin{align}
|\Psi\rangle=\bigotimes\limits_{g\in D_{6h}}|T_{g\lambda}\rangle\otimes\sum\limits_{k=1}^3|\gamma_{g\sigma_k}\rangle\otimes\sum\limits_{j=1}^3|\beta_{g\tau_j}\rangle\otimes|\alpha_\mu\rangle
\label{D6h cell decomposition}
\end{align}
where $|T_{g\lambda}\rangle$ is the wavefunction of 3D block state on $g\lambda$ which is topological trivial; $|\gamma_{g\sigma_{1,2,3}}\rangle$ is the $\mathbb{Z}_2$-symmetric wavefunction of 2D block state on $g\sigma_{1,2,3}$; $|\beta_{g\tau_1}\rangle$ is the $(\mathbb{Z}_6\rtimes\mathbb{Z}_2)$-symmetric wavefunction of 2D block state on $g\tau_1$, and $|\beta_{g\tau_{2,3}}\rangle$ is the $(\mathbb{Z}_2\times\mathbb{Z}_2)$-symmetric wavefunction of 2D block state on $g\tau_{2,3}$; $|\alpha_{\mu}\rangle$ is the $\mathbb{Z}_2\times(\mathbb{Z}_6\rtimes\mathbb{Z}_2)$-symmetric wavefunction of 0D block state on $\mu$.

We summarize the classifications and corresponding root phases of $D_{6h}$-symmetric crystalline topological phases. For crystalline TSC with spinless fermions, we summarize all possible block states as:
\begin{enumerate}[1.]
\item 2D blocks $\sigma_1$, $\sigma_2$ and $\sigma_3$: 2D $(p+ip)$-SCs and 2D $\mathbb{Z}_2$ fSPT phases;
\item 1D blocks $\tau_1$: Majorana chain, 1D $\mathbb{Z}_6\rtimes\mathbb{Z}_2$ fSPT phases and Haldane chain;
\item 1D blocks $\tau_2$ and $\tau_3$: Majorana chain, 1D $\mathbb{Z}_2\times\mathbb{Z}_2$ fSPT phases and Haldane chain;
\item 0D block $\mu$: 0D modes characterizing the eigenvalues $-1$ of all three generators of the $D_{6h}$ group, with even/odd fermion parity. 
\end{enumerate}
By explicit $K$-matrix calculation, we conclude that except 2D bosonic Levin-Gu state on each 2D block $\sigma_1/\sigma_2/\sigma_3$, all other 2D block states are obstructed. 

Similar to the $D_{2h}$-symmetric case, Majorana chain decorations on $\tau_{1,2,3}$ and all possible combinations are obstructed; 1D $\mathbb{Z}_6\rtimes\mathbb{Z}_2$ fSPT phases on $\tau_1$ are obstructed, and for 1D $\mathbb{Z}_2\times\mathbb{Z}_2$ fSPT phases on $\tau_2$ and $\tau_3$, only the root phase corresponding to the reflection with respect to the equator on both $\tau_2$ and $\tau_3$ are obstruction-free, which is trivialized by 2D ``Majorana'' bubble construction on each $\sigma_3$. For 1D Haldane chain, we see that if we construct a 1D $\mathbb{Z}_2$ fSPT phase on each $\sigma_1$ as 2D bubble equivalence, Haldane chain decoration on $\tau_1$ and $\tau_2$ is trivialized. Similar for Haldane chain decorations on $(\tau_1,\tau_3)$ and $(\tau_2,\tau_3)$. 

For 0D block $\mu$, complex fermion bubble on $\tau_1/\tau_2/\tau_3$ trivializes the 0D mode characterizing eigenvalue $-1$ of the reflection with respect to the equator/$\tau_3$/$\tau_2$.  Therefore, the ultimate classification is $\mathbb{Z}_2^5$, with the folllowing root phases:
\begin{enumerate}[1.]
\item 2D bosonic Levin-Gu state on each 2D block $\sigma_1/\sigma_2/\sigma_3$ ($\mathbb{Z}_2^3$);
\item Haldane chain on each 1D block $\tau_1$, $\tau_2$ and $\tau_3$ ($\mathbb{Z}_2$);
\item 0D mode with odd fermion parity ($\mathbb{Z}_2$).
\end{enumerate}
The HO topological surface theories of these root phases are listed as following:
\begin{enumerate}[1.]
\item $2^{\mathrm{nd}}$-order: 1D nonchiral Luttinger liquids with $K$-matrix $K=\sigma^x$ and $\mathbb{Z}_2$ symmetry property $W^{\mathbb{Z}_2}=\mathbbm{1}_{2\times2}$ and $\delta\phi^{\mathbb{Z}_2}=\pi(1,1)^T$, on the intersections between the open lattice and 2D blocks $\sigma_1/\sigma_2/\sigma_3$;
\item $3^{\mathrm{rd}}$-order: A dangling spin-1/2 degree of freedom at the center of each verticle hinges of the open lattice, and centers and top and bottom surface of the open lattice.
\end{enumerate}

For crystalline TSC with spin-1/2 fermions, the ultimate classification is $\mathbb{Z}_2^3$ because there is no nontrivial block states on 2D and 1D blocks, the root phases are 0D modes with eigenvalues $-1$ of three generators of $D_{6h}$ group, and there is no nontrivial HO topological surface theory.

For crystalline TI with spinless fermions, we summarize all possible block states as:
\begin{enumerate}[1.]
\item 2D blocks $\sigma_1$, $\sigma_2$ and $\sigma_3$: Chern insulators, Kitaev's $E_8$ states and 2D $U^f(1)\times\mathbb{Z}_2$ fSPT phases;
\item 1D blocks $\tau_1$, $\tau_2$ and $\tau_3$: Haldane chain;
\item 0D block $\mu$: 0D modes characterizing the eigenvalues $-1$ of all three generators of the $D_{6h}$ group, with different $U^f(1)$ charge. 
\end{enumerate}
By explicit $K$-matrix calculations, we conclude that except 2D $U^f(1)\times\mathbb{Z}_2$ bSPT phase on each $\sigma_1/\sigma_2/\sigma_3$ and monolayer Kitaev's $E_8$ state on each $\sigma_1$, $\sigma_2$ and $\sigma_3$ with chiralities $(+,-,+)$, all other 2D block states are obstructed. Furthermore, similar to the crystalline TSC, 0D modes characterizing eigenvalues $-1$ of two vertical reflection generators of the $D_{6h}$ group, with even $U^f(1)$ charges are trivialized. Therefore, the ultimate classification is $\mathbb{Z}_4\times\mathbb{Z}_2^7$, with the following root phases:
\begin{enumerate}[1.]
\item 2D $U^f(1)\times\mathbb{Z}_2$ bSPT phase on each 2D block $\sigma_1/\sigma_2/\sigma_3$ ($\mathbb{Z}_2^3$);
\item Monolayer Kitaev's $E_8$ state on each $\sigma_1$, $\sigma_2$ and $\sigma_3$, with chiralities $(+,-,+)$ ($\mathbb{Z}_2$);
\item Haldane chain on each 1D block $\tau_1/\tau_2/\tau_3$ ($\mathbb{Z}_2^3$);
\item 0D mode with odd $U^f(1)$ charge on $\mu$ ($\mathbb{Z}_2$);
\item 0D mode with eigenvalue $-1$ of reflection with respect to the equator ($\mathbb{Z}_2$).
\end{enumerate}
and there is a nontrivial extension between the last two root phases: a 0D mode with $U^f(1)$ charge $n\equiv2(\mathrm{mod}~4)$ is extended to a 0D mode with eigenvalue $-1$ if reflection operation with respect to the equator. The HO topological surface theories of these root phases are listed as following:
\begin{enumerate}[1.]
\item $2^{\mathrm{nd}}$-order: 1D nonchiral Luttinger liquids with $K$-matrix $K=\sigma^x$ and $\mathbb{Z}_2$ symmetry property $W^{\mathbb{Z}_2}=\mathbbm{1}_{2\times2}$ and $\delta\phi^{\mathbb{Z}_2}=\pi(1,1)^T$, on the intersections between the open lattice and 2D blocks $\sigma_1/\sigma_2/\sigma_3$;
\item $3^{\mathrm{rd}}$-order: A dangling spin-1/2 degree of freedom at the intersections between 1D blocks $\tau_1/\tau_2/\tau_3$ and the open lattice.
\end{enumerate}

For crystalline TI with spin-1/2 fermions, we summarize all possible block states as:
\begin{enumerate}[1.]
\item 2D blocks $\sigma_1$, $\sigma_2$ and $\sigma_3$: Chern insulators, Kitaev's $E_8$ states and 2D $U^f(1)\times\mathbb{Z}_2$ fSPT phases;
\item 0D block $\mu$: 0D modes characterizing the eigenvalues $-1$ of all three generators of the $D_{6h}$ group, with different even $U^f(1)$ charge. \end{enumerate}
We note that 2D chiral block states should satisfy some conditions to guarantee that there is no chiral 1D modes leaved by 2D block states on their shared borders: the chiralities should be $(+,-,+)$, and even layers of chiral block states are trivialized by 3D bubble equivalences on the $\lambda$. Furthermore, 0D modes with $U^f(1)$ charge as integer multiple of 4 is trivialized by 1D bubble equivalences. Hence the ultimate classification is $\mathbb{Z}_8\times\mathbb{Z}_4^2\times\mathbb{Z}_2^5$, with the following root phases:
\begin{enumerate}[1.]
\item Monolayer Chern insulator on each $\sigma_1$, $\sigma_2$ and $\sigma_3$, with chiralities $(+,-,+)$ ($\mathbb{Z}_2$);
\item 2D $U^f(1)\times\mathbb{Z}_2$ fSPT phase on each $\sigma_1/\sigma_2/\sigma_3$ ($\mathbb{Z}_4^3$);
\item Monolayer Kitaev's $E_8$ state on each $\sigma_1$, $\sigma_2$ and $\sigma_3$, with chiralities $(+,-,+)$ ($\mathbb{Z}_2$);
\item 0D mode with $U^f(1)$ charge $n\equiv2(\mathrm{mod}~4)$ on $\mu$ ($\mathbb{Z}_2$);
\item 0D mode with eigenvalue $-1$ of three generators of the $D_{6h}$ group on $\mu$ ($\mathbb{Z}_2^3$).
\end{enumerate}
And there are several nontrivial extensions between these root phases: bilayer Chern insulators on each $\sigma_{1,2,3}$ can be smoothly deformed to 2D $U^f(1)\times\mathbb{Z}_2$ fSPT phases with $(\nu_1,\nu_2,\nu_3)=(1,3,1)$ on 2D blocks $\sigma_{1,2,3}$, by 3D ``Chern insulator'' bubble equivalence. The second-order topological surface theories of different root phases are:
\begin{enumerate}[1.]
\item Chiral fermions on the intersections between the open lattice and 2D blocks $\sigma_1$, $\sigma_2$ and $\sigma_3$;
\item 1D nonchiral Luttinger liquids with $K$-matrix $K=\sigma^z$ and $\mathbb{Z}_2$ symmetry property $W^{\mathbb{Z}_2}=\mathbbm{1}_{2\times2}$ and $\delta\phi^{\mathbb{Z}_2}=\pi(0,1)^T$, on the intersections between the open lattice and 2D blocks decorated with a $U^f(1)\times\mathbb{Z}_2$ fSPT phase with $\nu=1$ on each of them;
\item 1D chiral Luttinger liquids with $K$-matrix (\ref{K-matrix E8S}), on the intersections between the open lattice and 2D blocks $\sigma_1$, $\sigma_2$ and $\sigma_3$;
\end{enumerate}

\end{document}